%% file: ms.tex
\documentclass[10pt,italian,english,a5paper,twoside,openright]{book}
\pdfoutput=1 
\title{Model checking: the interval way}
\author{Alberto Molinari}
\date{}
 
\input{config.tex}

\begin{document}

\frontmatter

\input{FrontBackMatter/frontespizio.tex}


\cleardoublepage
\input{FrontBackMatter/date_copyr.tex}

\cleardoublepage
\input{FrontBackMatter/cit.tex}

\cleardoublepage
\input{FrontBackMatter/abstract.tex}

\cleardoublepage
\input{FrontBackMatter/ringraziamenti.tex}
 
\cleardoublepage
\dominitoc 
\tableofcontents


\mainmatter

\part{Introduction}\label{part:intro}
\include{Chaps/Intro/introFOR}

\part{MC for $\HS$ under homogeneity}\label{part:HShomo}
\include{Chaps/Intro/MCfullHShomo}
\include{Chaps/ICALP_D/ICALPmain}
\include{Chaps/TCS17/TCS17main}

\include{Chaps/IC17/IC17main}

\include{Chaps/TOCL17/TOCL17main}

\part{MC for $\HS$ relaxing homogeneity}\label{part:HSrelaxhomo}
\include{Chaps/Gandalf17RIVISTA/Gand17main}

\part{Interval-based system models}\label{part:timelines}
\include{Chaps/Timelines/TimelinesMain}
\include{Chaps/Concl/conclMain}

\appendix
\part{Appendices}
\include{Chaps/Appendices/appendixICALP_D}
\include{Chaps/Appendices/appendixTCS17}
\include{Chaps/Appendices/appendixIC17}
\include{Chaps/Appendices/appendixTOCL17}
\include{Chaps/Appendices/appendixGand17}
\include{Chaps/Appendices/appendixTimelines}

\backmatter
\cleardoublepage

\phantomsection
 \addcontentsline{toc}{chapter}{\bibname}
 \bibliographystyle{alphaurl}
 \bibliography{biblioEXTR.bib}


\end{document}

%% file: config.tex
\usepackage[a-1b]{pdfx}
\usepackage[utf8]{inputenc}
\usepackage{babel}
\usepackage[babel=true,final]{microtype}
\usepackage{cmap}
\usepackage[tt=false, type1=true]{libertine}
\usepackage[libertine]{newtxmath}
\usepackage{lettrine}

\usepackage[T1]{fontenc}

\usepackage[inner=1.8cm,outer=1.5cm,top=1.8cm,bottom=1.5cm,heightrounded]{geometry}
\usepackage{emptypage} 
\usepackage{nopageno} 

\usepackage{graphicx,rotating}
\usepackage[table]{xcolor}
\usepackage{tikz,tikz-qtree}\usetikzlibrary{positioning,arrows,automata,patterns,decorations.pathmorphing,decorations.pathreplacing,fit, patterns, calc,shapes}
\usepackage{xspace}

\usepackage[chapter]{algorithm}
\usepackage[noend]{algpseudocode} 
\algrenewcommand{\algorithmiccomment}[1]{\hfill$\triangleleft$ {\small #1}}
\algrenewcommand\algorithmicindent{1em}
\algloopdefx{Return}[1]{\textbf{return} #1}
\algblock{Case}{EndCase}
\algcblock[Case]{Case}{case}{EndCase}

\usepackage{multirow,multicol}
\usepackage{hhline}

\usepackage[autostyle=true]{csquotes}

\usepackage[pdfa]{hyperref}
\hypersetup{
    unicode=true, colorlinks=true, allcolors=blue, hypertexnames=false, plainpages=false}

\usepackage{sectsty}
\usepackage[Bjornstrup]{fncychap}
\usepackage{fancyhdr}
\ChTitleVar{\raggedleft\LARGE\sffamily\bfseries}

\allsectionsfont{\sffamily}
\sectionfont{
	\sffamily
	\sectionrule{0pt}{0pt}{-5pt}{0.5pt}
}

\usepackage[labelfont={bf}, font={small,sf}]{caption}

\usepackage[tight,english]{minitoc} 
\usepackage{enumitem}
\setlist{parsep=0pt plus 1pt, itemsep=2.0pt plus 1.0pt minus 1.0pt, topsep=3pt plus 1pt minus 1 pt}
\setlist[enumerate]{font=\sffamily\bfseries}
\setlist[enumerate,1]{label=\arabic*.}
\setlist[enumerate,2]{label=\alph*.}
\setlist[enumerate,3]{label=\roman*.}

\usepackage{tcolorbox}
\tcbuselibrary{skins,breakable}

\input{macro_mate}

\newenvironment{chapref}
    {\vspace*{-\baselineskip}\hfill\small}
    {}

%% file: macro_mate.tex
\usepackage{amsmath,amsthm,amssymb}
\usepackage{mathtools}

\allowdisplaybreaks

\DeclareMathAlphabet{\mathpzc}{OT1}{pzc}{m}{it}

\theoremstyle{plain}
    \newtheorem{proposition}{Proposition}[section]
    \newtheorem*{proposition*}{Proposition}
    \newtheorem{theorem}[proposition]{Theorem}
    \newtheorem*{theorem*}{Theorem}
    \newtheorem{lemma}[proposition]{Lemma}
    \newtheorem*{lemma*}{Lemma}
    \newtheorem{corollary}[proposition]{Corollary}

\theoremstyle{definition}
    \newtheorem{definition}[proposition]{Definition}
    \newtheorem{example}[proposition]{Example}
    \newtheorem{property}[proposition]{Property}

\theoremstyle{remark}
    \newtheorem{remark}[proposition]{Remark}
    \newtheorem{claim}[proposition]{Claim}

\tcolorboxenvironment{proposition}{enhanced jigsaw,colback=gray!30!white,colframe=white,arc=0mm,breakable,before skip=5pt plus 0.2em,after skip=5pt plus 0.2em}
\tcolorboxenvironment{proposition*}{enhanced jigsaw,colback=gray!30!white,colframe=white,arc=0mm,breakable,before skip=5pt plus 0.2em,after skip=5pt plus 0.2em}
\tcolorboxenvironment{theorem}{enhanced jigsaw,colback=gray!30!white,colframe=white,arc=0mm,breakable,before skip=5pt plus 0.2em,after skip=5pt plus 0.2em}
\tcolorboxenvironment{theorem*}{enhanced jigsaw,colback=gray!30!white,colframe=white,arc=0mm,breakable,before skip=5pt plus 0.2em,after skip=5pt plus 0.2em}
\tcolorboxenvironment{lemma}{enhanced jigsaw,colback=gray!30!white,colframe=white,arc=0mm,breakable,before skip=5pt plus 0.2em,after skip=5pt plus 0.2em}
\tcolorboxenvironment{lemma*}{enhanced jigsaw,colback=gray!30!white,colframe=white,arc=0mm,breakable,before skip=5pt plus 0.2em,after skip=5pt plus 0.2em}
\tcolorboxenvironment{corollary}{enhanced jigsaw,colback=gray!30!white,colframe=white,arc=0mm,breakable,before skip=5pt plus 0.2em,after skip=5pt plus 0.2em}
\tcolorboxenvironment{definition}{enhanced jigsaw,colback=gray!30!white,colframe=white,arc=0mm,breakable,before skip=5pt plus 0.2em,after skip=5pt plus 0.2em}
\tcolorboxenvironment{example}{enhanced jigsaw,colback=white,colframe=black,boxrule=0.5pt,arc=1mm,breakable,before skip=8pt plus 0.2em,after skip=8pt plus 0.2em}
\tcolorboxenvironment{property}{enhanced jigsaw,colback=gray!30!white,colframe=white,arc=0mm,breakable,before skip=5pt plus 0.2em,after skip=5pt plus 0.2em}
\tcolorboxenvironment{remark}{enhanced jigsaw,colback=gray!30!white,colframe=white,arc=0mm,breakable,before skip=5pt plus 0.2em,after skip=5pt plus 0.2em}
\tcolorboxenvironment{claim}{enhanced jigsaw,colback=gray!30!white,colframe=white,arc=0mm,breakable,before skip=5pt plus 0.2em,after skip=5pt plus 0.2em}

\makeatletter
\renewenvironment{proof}[1][\proofname]{%
\par
\pushQED{\qed}%
\normalfont 
\noindent{\bfseries #1\@addpunct{.}}\hspace\labelsep\ignorespaces
}{%
\popQED
\@endpefalse
\par\smallskip
}
\makeatother

\DeclareMathAlphabet{\mathpzc}{OT1}{pzc}{m}{it}

\newcommand{\Ku}{\ensuremath{\mathpzc{K}}}
\newcommand{\Prop}{\ensuremath{\mathpzc{AP}}}
\newcommand{\AP}{\Prop}
\newcommand{\States}{S}
\newcommand{\Edges}{\ensuremath{\textit{R}}}
\newcommand{\sinit}{s_0}
\newcommand{\Lab}{\mu}
\newcommand{\KuDef}{(\Prop,\States,\Edges,\Lab,\sinit)}

\newcommand{\Nat}{\mathbb{N}}

\DeclareMathOperator{\true}{\top}
\DeclareMathOperator{\false}{\bot}

\newcommand{\Length}{\textit{length}}

\newcommand{\tpl}[1]{(#1)}
\newcommand{\tupleof}[1]{(#1)}

\DeclareMathOperator{\Trk}{Trc}
\DeclareMathOperator{\Pref}{Pref}
\DeclareMathOperator{\Suff}{Suff}
\DeclareMathOperator{\states}{states}
\DeclareMathOperator{\intstates}{intstates}
\DeclareMathOperator{\lst}{lst}
\DeclareMathOperator{\fst}{fst}

\DeclareMathOperator{\nestbe}{\mathsf{Nest}_{BE}}
\DeclareMathOperator{\nestb}{\mathsf{Nest}_B}
\DeclareMathOperator{\neste}{\mathsf{Nest}_E}

\newcommand{\nnf}{\textsf{NNF}}

\DeclareMathOperator{\Root}{root}
\DeclareMathOperator{\DV}{\mathcal{V}}
\DeclareMathOperator{\DE}{\mathcal{E}}

\newcommand{\RE}{\mathsf{RE}}
\newcommand{\Lang}{{\mathpzc{L}}}
\newcommand{\lang}{\Lang}
\newcommand{\SPEC}{\textsf{Spec}}
\newcommand{\NFA}{\text{\sffamily NFA}}
\newcommand{\DFA}{\text{\sffamily DFA}}
\newcommand{\Au}{\ensuremath{\mathcal{A}}}
\newcommand{\Du}{\ensuremath{\mathcal{D}}}

\DeclareMathOperator{\hsA}{\langle A\rangle}
\DeclareMathOperator{\hsL}{\langle L\rangle}
\DeclareMathOperator{\hsB}{\langle B\rangle}
\DeclareMathOperator{\hsE}{\langle E\rangle}
\DeclareMathOperator{\hsD}{\langle D\rangle}
\DeclareMathOperator{\hsO}{\langle O\rangle}
\DeclareMathOperator{\hsX}{\langle X\rangle}
\DeclareMathOperator{\hsG}{\langle G\rangle}
\DeclareMathOperator{\hsAt}{\langle \overline{A}\rangle}
\DeclareMathOperator{\hsLt}{\langle \overline{L}\rangle}
\DeclareMathOperator{\hsBt}{\langle \overline{B}\rangle}
\DeclareMathOperator{\hsEt}{\langle \overline{E}\rangle}
\DeclareMathOperator{\hsDt}{\langle \overline{D}\rangle}
\DeclareMathOperator{\hsOt}{\langle \overline{O}\rangle}
\DeclareMathOperator{\hsXt}{\langle \overline{X}\rangle}
\DeclareMathOperator{\hsAu}{\mathopen[ A \mathclose]}

\DeclareMathOperator{\hsBu}{\mathopen[ B \mathclose]}
\DeclareMathOperator{\hsEu}{\mathopen[ E \mathclose]}
\DeclareMathOperator{\hsDu}{\mathopen[ D \mathclose]}

\DeclareMathOperator{\hsXu}{\mathopen[ X \mathclose]}
\DeclareMathOperator{\hsGu}{\mathopen[ G \mathclose]}
\DeclareMathOperator{\hsAtu}{\mathopen[ \overline{A} \mathclose]}

\DeclareMathOperator{\hsBtu}{\mathopen[ \overline{B} \mathclose]}
\DeclareMathOperator{\hsEtu}{\mathopen[ \overline{E} \mathclose]}

\DeclareMathOperator{\hsXtu}{\mathopen[ \overline{X} \mathclose]}

\newcommand{\A}{\mathsf{A}}
\renewcommand{\AE}{\mathsf{AE}}
\newcommand{\AB}{\mathsf{AB}}
\newcommand{\Abar}{\mathsf{\overline{A}}}
\newcommand{\AAbar}{\mathsf{A\overline{A}}}
\newcommand{\AAbarB}{\mathsf{A\overline{A}B}}
\newcommand{\AbarB}{\mathsf{\overline{A}B}}
\newcommand{\AAbarE}{\mathsf{A\overline{A}E}}
\newcommand{\AAbarBbarEbar}{\mathsf{A\overline{A}\overline{B}\overline{E}}}
\newcommand{\AAbarBBbar}{\mathsf{A\overline{A}B\overline{B}}}
\newcommand{\AAbarEEbar}{\mathsf{A\overline{A}E\overline{E}}}
\newcommand{\ABbar}{\mathsf{A\overline{B}}}
\newcommand{\Bbar}{\mathsf{\overline{B}}}
\newcommand{\Ebar}{\mathsf{\overline{E}}}
\renewcommand{\B}{\mathsf{B}}
\newcommand{\E}{\mathsf{E}}
\renewcommand{\L}{\mathsf{L}}
\newcommand{\D}{\mathsf{D}}
\newcommand{\BE}{\mathsf{BE}}
\newcommand{\AbarE}{\mathsf{\overline{A}E}}
\newcommand{\BEbar}{\mathsf{B\overline{E}}}
\newcommand{\ABBbar}{\mathsf{AB\overline{B}}}
\newcommand{\AAbarBBbarEbar}{\mathsf{A\overline{A}B\overline{B}\overline{E}}}

\newcommand{\AAbarEBbarEbar}{\mathsf{A\overline{A}E\overline{B}\overline{E}}}

\newcommand{\HSprop}{\mathsf{Prop}}

\newcommand{\HS}{\text{\sffamily HS}}
\newcommand{\PITL}{\text{\sffamily PITL}}
\newcommand{\DC}{\text{\sffamily DC}}
\newcommand{\CDT}{\text{\sffamily CDT}}
\newcommand{\FO}{\text{\sffamily FO}}
\newcommand{\CTLStar}{\text{\sffamily CTL$^{*}$}}
\newcommand{\CTLStarLP}{\text{\sffamily CTL$^{*}_{lp}$}}
\newcommand{\LTL}{\text{\sffamily LTL}}
\newcommand{\LTLP}{\text{\sffamily LTL$_p$}}
\newcommand{\CTL}{\text{\sffamily CTL}}
\newcommand{\Downarrowx}{\text{$\downarrow$$x$}}
\newcommand{\Downarrowy}{\text{$\downarrow$$y$}}
\newcommand{\until}{\textsf{U}}
\newcommand{\Next}{\textsf{X}}
\newcommand{\Always}{\textsf{G}}
\newcommand{\Eventually}{\textsf{F}}
\newcommand{\EQ}{\exists}
\newcommand{\EQF}{\exists_f}
\newcommand{\AQ}{\forall}
\newcommand{\EQSubf}{\exists\textit{SubF}}
\newcommand{\present}{\textit{present}}

\newcommand{\PTIME}{\textbf{P}}
\newcommand{\NP}{\textbf{NP}}
\newcommand{\PSPACE}{\textbf{PSPACE}}
\newcommand{\Psp}{\PSPACE}
\newcommand{\NPsp}{\textbf{NPSPACE}}
\newcommand{\LOGSPACE}{\textbf{LOGSPACE}}
\newcommand{\NLOGSPACE}{\textbf{NLOGSPACE}}
\newcommand{\NLOGSP}{\NLOGSPACE}
\newcommand{\EXPSPACE}{\textbf{EXPSPACE}}
\newcommand{\NEXPTIME}{\textbf{NEXPTIME}}
\newcommand{\EXPTIME}{\textbf{EXPTIME}}
\newcommand{\AEXP}{\textbf{AEXP}}
\newcommand{\LINAEXPTIME}{\mathbf{AEXP_{pol}}}
\newcommand{\PH}{\textbf{PH}}
\newcommand{\co}{\textbf{co-}}
\newcommand{\Th}{\PTIME^{\NP[O(\log n)]}}
\newcommand{\Thsq}{\PTIME^{\NP[O(\log^2 n)]}}
\newcommand{\Thpar}{\PTIME^{\NP}_{\parallel}}
\newcommand{\Thparlogn}{\PTIME^{\NP}_{\parallel O(\log n)}}
\newcommand{\Tower}{\mathsf{Tower}}
\newcommand{\TOWER}{\textbf{TOWER}}

%% file: FrontBackMatter/frontespizio.tex
\thispagestyle{empty}
\tikz[remember picture,overlay] \node[inner sep=0pt,opacity=0.2] at 
(current page.center){\includegraphics[width=\paperwidth,height=\paperheight
]{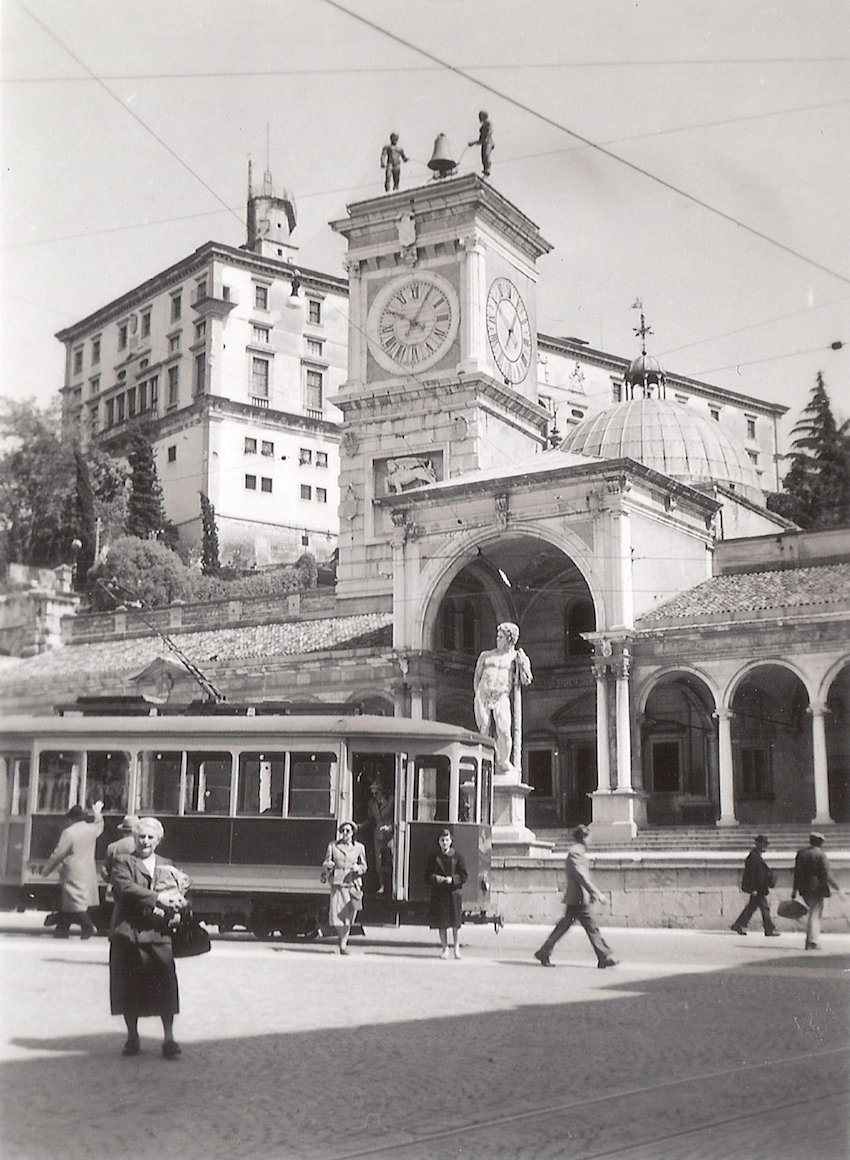}};

\tikz[remember picture,overlay] \node[inner sep=0pt] at 
(current page.center){
\begin{minipage}[c][0.8\paperheight][c]{0.8\paperwidth}

\begin{center}
\textsc{
{\huge Università degli Studi di Udine}
\bigskip 
\hrule
\bigskip
\begin{Large}
Dipartimento di Matematica, Informatica e Fisica\\\bigskip
Corso di Dottorato in\\ Informatica e Scienze Matematiche e Fisiche\\\bigskip
Ciclo XXXI
\end{Large}
}

\vfill

{\Huge \textsc{Model Checking:}}
\\ \vspace{0.3cm}
{\Huge \textsc{the Interval Way}}
\\ \vspace{1cm}
{\huge Tesi di Dottorato}

\vfill

\begin{Large}
\begin{flushleft}
Relatore \\ 
prof.\ Angelo MONTANARI\\\bigskip
Correlatore\\
prof.\ Adriano PERON
\end{flushleft}
\begin{flushright}
Dottorando \\
Alberto MOLINARI
\end{flushright}
\end{Large}
%
%
\end{center}

\end{minipage}
};

%% file: FrontBackMatter/date_copyr.tex
\thispagestyle{empty}

\begin{center}
	\begin{tabular}{rl}
	\textbf{Submission:}& October 31, 2018 \\ 
	\textbf{Defense:}& February 28, 2019\\ 
	\end{tabular} 
\end{center}

\vfill

\hrulefill

\vfill

\noindent
\textbf{Author’s e-mail:} \\\href{mailto:molinari.alberto@gmail.com}{molinari.alberto@gmail.com}

\bigskip

\noindent
\textbf{Author’s address:} \\
via del Prà delle Molle, 13 \\
IT-31029 Vittorio Veneto (TV) \\
Italy

\vfill

\begin{center}
This is an arXived version of the thesis. The \textbf{official version} can be 
found at:
\medskip

\url{http://albertom.altervista.org/Th.pdf}.
\end{center}

\vfill

\begin{center}
\includegraphics[scale=0.25]{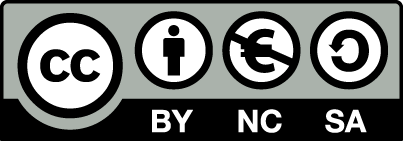} 
\medskip

\textcopyright{} 2019 by Alberto Molinari
\end{center}

\noindent
This work is licensed under the Creative Commons Attribution-NonCommercial-ShareAlike 4.0 International License.
To view a copy of this license, please visit
\url{http://creativecommons.org/licenses/by-nc-sa/4.0/}
or send a letter to Creative Commons, PO Box 1866, Mountain View, CA 94042, USA.

%% file: FrontBackMatter/cit.tex
\thispagestyle{empty}

\vspace*{0.3\textheight}

\begin{flushright}
	{\itshape
		\lq\lq Multa non quia difficilia sunt non audemus, sed quia non audemus 
		sunt difficilia.\rq\rq}\\
	\medskip
		\textsc{Seneca}, Epistulae morales ad Lucilium, 104, 26
\end{flushright}

%% file: FrontBackMatter/abstract.tex
\thispagestyle{empty}

\begin{center}
\bfseries \abstractname 
\end{center}

\emph{Formal methods} are structured methodologies that support the development of critical systems---whose safety and reliability are fundamental requirements---with the aim of 
establishing system correctness with mathematical rigor,
providing effective verification techniques and tools, and reducing verification time while simultaneously increasing coverage.

\emph{Model checking} (MC) is a family of formal methods that have been accepted by industry and are becoming integral part of standards and of system development cycles. 
In MC, some properties of a finite-state transition system are expressed in suitable specification languages and then verified over a model of the system itself (usually a finite Kripke structure) through \emph{exhaustive enumeration of all the reachable states}. This technique is \emph{fully automatic} and every time the design violates a desired property, a \emph{counterexample} is produced, which illustrates a behavior falsifying such a property: this is extremely useful for debugging.

The most famous MC techniques---just to mention a few, \emph{partial order reduction}, \emph{symbolic} and \emph{bounded} MC---were developed starting from the late 80s, bearing in mind the well-known ``point-based'' temporal logics \LTL\ and \CTL. However, while the expressiveness of such logics is beyond doubt,
there are some properties we may want to check that are inherently ``interval-based'' and thus cannot be expressed by point-based temporal logics, e.g., ``the proposition $p$ has to hold in at least an average number of system states in a given computation sector''. Here \emph{interval temporal logics} (ITLs) come into play, providing an alternative setting for 
reasoning about time. Such logics deal with intervals, instead of points, as their primitive entities: this feature gives them the ability of 
expressing temporal properties, such as actions with duration, accomplishments, and temporal aggregations, which cannot be dealt with in standard point-based logics.

The \emph{Halpern
and Shoham's modal logic of time intervals} (\HS, for short) is one of the most famous ITLs:
it features one modality for each of the 13 possible ordering
relations between pairs of intervals, apart from equality.
In this thesis we focus our attention on MC based on \HS , in the role of property specification language,
for which a little work has been done if compared to MC for point-based temporal logics.
The idea is to evaluate \HS\ formulas on finite Kripke structures, making it
possible to check the correctness of the behavior of systems with respect to 
meaningful interval properties.
To this end, we interpret each one of the (possibly infinitely many) finite paths of a Kripke structure as an interval, 
and we define its atomic properties on the basis of the properties of the states composing it,
at first assuming the \emph{homogeneity principle}: the latter enforces an atomic property to hold over an interval if and only if it holds over all its subintervals.
%
We prove that MC for \HS\ interpreted over finite Kripke structures is a \emph{decidable} problem (whose computational complexity has a nonelementary upper bound), and
then we show it to be \EXPSPACE-hard.

Since the problem provably admits no polynomial-time decision procedure, we also focus on \HS\ fragments, which 
feature considerably better complexities---from \EXPSPACE, down to \Psp\ and to low levels of the polynomial hierarchy---yet retaining the ability to capture meaningful interval properties of state transition systems.
Several MC algorithms are presented, tailored to the specific fragments being considered, and founded on concepts and techniques different from each other.

Moreover, we study the expressive power of $\HS$ in MC, in comparison with that of the standard point-based logics $\LTL$, $\CTL$ and $\CTLStar$, still under the homogeneity principle, which is then relaxed showing how this impacts on the complexity of MC for \HS\ and its fragments, and on the expressiveness of the logic.

Finally, we consider a possible replacement of Kripke structures by a more expressive model, which allows us to directly  describe systems in terms of their \emph{interval-based behaviour and properties}, thus paving the way for a more general interval-based MC.

\bigskip
\noindent\hrulefill
\bigskip

\begin{center}
\begin{tabular}{rl}
    \textbf{Keywords:} & Model checking, interval temporal logics, \\
    & timelines, computational complexity \\ 
    \rule[-1ex]{0pt}{4.5ex} \textbf{2010 MSC:} & 03B70, 68Q60 \\ 
    \rule[-1ex]{0pt}{4.5ex} \textbf{ACM classes:} & F.4.1, D.2.4 \\ 
\end{tabular}
\end{center}

\cleardoublepage
\thispagestyle{empty}
\selectlanguage{italian}
\begin{center}
\bfseries \abstractname 
\end{center}

I \emph{metodi formali} sono metodologie strutturate che supportano lo sviluppo di sistemi critici---la sicurezza ed affidabilità dei quali sono requisiti fondamentali---allo scopo di dimostrare la correttezza di tali sistemi con rigore matematico, fornendo tecniche e strumenti di verifica efficaci, e riducendo il tempo del processo di verifica, aumentando contemporaneamente il grado di copertura.

Il \emph{model checking} (MC)
è una famiglia di metodi formali che sono stati accettati dal mondo dell'industria e stanno diventando parte integrante di standard e dei cicli di sviluppo dei sistemi.
Nel contesto del MC, alcune proprietà di un sistema di transizione a stati finiti vengono espresse mediante linguaggi di specifica e, successivamente, queste sono verificate su un modello del sistema stesso (di solito una struttura di Kripke finita), tramite \emph{l'enumerazione completa di tutti gli stati raggiungibili}.
Questa tecnica è \emph{totalmente automatica} ed ogni volta che viene violata una proprietà desiderata, viene fornito un \emph{controesempio} che illustra un comportamento che falsifica tale proprietà: ciò è estremamente utile per il processo di debugging.

Le più famose tecniche di MC, come la \emph{partial order reduction}, il \emph{symbolic} ed il \emph{bounded} MC,
furono sviluppate verso la fine degli anni 80 prendendo in considerazione le famose logiche temporali \LTL\ e \CTL, che sono basate su punti. Tuttavia, nonostante l'indubbia espressività di tali logiche, esistono alcune proprietà che potremmo voler verificare che hanno inerentemente una semantica intervallare e quindi non possono essere espresse da logiche puntuali, per esempio: ``la proposizione $p$ deve valere in almeno un dato numero medio di stati del sistema, in un settore di computazione specifico''.
Le logiche temporali intervallari entrano in gioco in questi casi, permettendoci di ragionare su aspetti temporali in modo diverso: esse adottano gli intervalli, invece dei punti, come loro entità primitive.
Questa caratteristica dà loro l'abilità di esprimere proprietà intervallari, come azioni con durata, conseguimenti di obiettivi e aggregazioni temporali, che non possono essere trattate nelle logiche puntuali standard.

La \emph{logica modale degli intervalli temporali di Halpern e Shoham} (\HS\  in breve) è una delle più famose logiche intervallari: essa possiede una modalità per ognuna delle 13 possibili relazioni di ordinamento fra coppie di intervalli, eccetto l'uguaglianza.
In questa tesi viene considerato il problema del MC basato su \HS , come linguaggio di specifica delle proprietà, il quale ha ricevuto ben poca attenzione in letteratura in confronto al MC per logiche temporali puntuali.
L'idea è quella di valutare formule di \HS\  su strutture di Kripke finite, per riuscire a verificare la correttezza del comportamento di un sistema rispetto a proprietà intervallari.
A questo scopo, ognuno dei percorsi finiti di una struttura di Kripke (i quali possono essere presenti in quantità infinita) è interpretato come un intervallo, e le proprietà atomiche che valgono su quest'ultimo sono definite sulla base di quelle degli stati che lo costituiscono, inizialmente secondo il \emph{principio di omogeneità}: esso prevede che una proprietà atomica valga su un intervallo se e solo se vale su tutti i suoi sottointervalli. 
Dimostriamo innanzitutto che il MC per \HS\  interpretata su strutture di Kripke finite è un problema \emph{decidibile} (la sua complessità computazionale ha un upper bound non-elementare); poi mostriamo che esso è \EXPSPACE-hard.

Poiché il problema non ammette procedure di decisione di complessità polinomiale, 
consideriamo anche frammenti di \HS, i quali si caratterizzano per complessità notevolmente migliori---da
\EXPSPACE, giù fino a \Psp\ e a livelli bassi della gerarchia polinomiale---pur tuttavia mantenendo l'abilità di esprimere
proprietà intervallari significative dei sistemi di transizione. Presentiamo svariati algoritmi di MC, costruiti ad-hoc per gli specifici frammenti considerati, e fondati su concetti e tecniche diversi fra loro.

Inoltre studiamo il potere espressivo di \HS\ in confronto a quello delle logiche puntuali standard $\LTL$, $\CTL$ e $\CTLStar$, sempre sotto l'ipotesi di omogeneità, la quale viene poi rilassata mostrando quali implicazioni ha questo sulla complessità del MC per \HS\ ed i suoi frammenti, e sull'espressività della logica stessa.

Infine consideriamo una possibile alternativa alle strutture di Kripke: studiamo un modello di sistemi più espressivo, che ci permette di descrivere gli stessi direttamente in termini delle loro proprietà intervallari. Questo apre la strada a un MC basato su intervalli più generale.

\bigskip
\noindent\hrulefill
\bigskip

\begin{center}
\begin{tabular}{rl}
    \textbf{Parole chiave:} & Model checking, logiche temporali intervallari, \\
    & timeline, complessità computazionale \\ 
    \rule[-1ex]{0pt}{4.5ex} \textbf{2010 MSC:} & 03B70, 68Q60 \\ 
    \rule[-1ex]{0pt}{4.5ex} \textbf{Classi ACM:} & F.4.1, D.2.4 \\ 
\end{tabular}
\end{center}

\selectlanguage{english}

%% file: FrontBackMatter/ringraziamenti.tex
\thispagestyle{empty}

\begin{center}
\bfseries Acknowledgements (and some trifles)
\end{center}

{\itshape
This thesis is the result of a three-year period of research on automatic system verification, which actually started with my master thesis (and even before graduating!).
This ``journey'' called PhD
has given me the opportunity of enriching myself (from the cultural point of view, clearly!), of knowing outstanding people from this research area, and traveling to places (most of them beautiful!) where, probably, I would have never gone in other situations.

I would like to thank for this experience some people that I affectionately want to call my ``travel mates'':
in primis, Angelo Montanari and Adriano Peron---also for helping me during my (seemingly periodic) ``reflection periods''---then Pietro Sala,
Laura Bozzelli and Gabriele Puppis.
I also express gratitude to the external referees, Alessio Lomuscio and 
Stéphane Demri, for the reviews of my 
thesis.

I thank my parents, my grandma, my aunt and uncle for\dots{} everything!
I bet my words would not be enough to express what I feel like saying,
but I am sure they know how grateful I am to them. 
And, I was forgetting\dots{} thank you also to my brother for taking me to the station so many times :)
}

%% file: Chaps/Intro/introFOR.tex
\chapter{Introduction}
\label{chap:introFOR}
\minitoc\mtcskip

\lettrine[lines=3]{A}{s the years pass by,} information and communication technology (ICT) systems 
are becoming more and more pervasive in our lives, having 
a crucial role to play in countless industry, communication, security, cybernetics areas and applications.
Embedded systems are employed for critical purposes, 
such as air traffic and railway control, telephone networks and nuclear plants monitoring.
Security protocols are at the basis of e-commerce websites and services,
and are exploited in all applications committed to ensure user privacy.
Biomedical instruments and equipment are endowed with
automatic or proactive functionalities, 
and are supposed to help humans and prevent human error.

The essential requirements of safety, reliability and correctness for these systems suggest
(or, rather, compel)
to support their development, throughout the phases of the life cycle, 
by structured methodologies founded on \emph{formal methods}, 
with the aim to provide effective verification techniques and tools, and reduce the verification time, while simultaneously increasing coverage.

The typical techniques of system engineering for system validation, namely, \emph{simulation} and \emph{testing}, 
are clearly not sufficient alone nowadays. They are dynamic techniques: 
the former involves performing some experiments with a restrictive set of scenarios or over a model, 
the latter actually running the system, whose correctness is determined by forcing it to traverse a set of execution paths. However, an exhaustive consideration of all execution paths is practically unfeasible, and thus testing and simulation can never be complete, that is, they can only show the presence of errors, but \emph{not their absence}.
Moreover, their effectiveness decreases dramatically as the complexity of design grows, 
they often discover errors and unpredictable behaviors in systems at late stages of development (or even when they have already been deployed), and are not effective at determining the more subtle and hidden bugs.
Just to mention a few of them, some famed examples of bugs which were discovered too late---so to speak---were those affecting, with catastrophic economic, functional, security or life consequences, the Intel Pentium CPU floating-point division unit~\cite{pratt1995}, the Ariane 5 rocket~\cite{jazequel1997}, the radiation therapy machine Therac 25~\cite{thomas1994}, the Needham-Schroeder authentication protocol~\cite{Lowe1996} and the ISDN protocol~\cite{holzmann1994}.

\emph{Formal methods}
are complementary to simulation and testing.
They can be considered as ``the applied mathematics for modeling and analyzing ICT systems''~\cite{Baier2008}, being their aim to establish system correctness with mathematical rigor. As a matter of fact,
\emph{all the possible states and scenarios of a system} are considered during formal verification, in order to prove that the system features some desired properties such as deadlock freedom, data integrity, liveness, safety, fairness, responsiveness, interference freedom, and so on.
The two most famous approaches to formal verification are \emph{axiomatic reasoning}  and \emph{model checking}.

Axiomatic reasoning involves specifying the desired properties of a system by means of \emph{formulas} of some logic; then a \emph{proof system}, consisting of axioms and rules, allows one to formally prove that the system meets the expected behavior. A well-known example is Hoare's tuple-based proof system. 
However, axiomatic reasoning has several limitations:
the proof rules are designed only for an \emph{a posteriori} verification of existing software, not for its systematic development; 
it is very time-consuming, cumbersome, and can be performed only by experts. As a consequence, it is mostly employed for (parts of) critical systems or security protocols.%
\footnote{Or to verify model checkers!}
Additionally, such techniques were developed having in mind  \emph{transformational} systems/programs:
in the 70s most early computer programs were designed to \emph{compute} something (i.e., transforming initial data to final ones), and correctness was to be assessed by showing that, given some preconditions on the input, desired post-conditions hold on the output of the program. The first tools of \emph{computer-aided verification}, namely, theorem provers, made their debut then, with the objective of automating mathematical logical reasoning. The language of such tools was often indeed Hoare-like. However, proofs could not be completely automated, and thus theorem provers were rather proof assistants and checkers.

In the 80s the focus shifted to \emph{reactive systems}, whose goal is
continuously interacting with the environment within
which they operate, instead of terminating their execution returning some value
(e.g., operating systems, communication protocols, programs
for process control, embedded systems). 
One of the most successful techniques for the verification of such systems is \emph{model checking}, on which research actively started during the early 80s, and originated from the independent work of Clarke and Emerson~\cite{CE81} and Queille and Sifakis~\cite{Queille1982}.

\section{An overview on MC}
In model checking (MC)~\cite{CE81,CGP02,VW86b,Queille1982} some properties of a 
system are expressed in a \emph{temporal logic} (e.g., $\CTL$ or $\LTL$) and then verified over a model of the system itself (usually a labeled state-transition graph or Kripke structure) through \emph{exhaustive}---implicit or explicit---enumeration of all the reachable states. This technique is \emph{fully automatic}: MC algorithms
either terminate positively (proving that all properties are met), or produce a \emph{counterexample} (namely, a behavior that falsifies a property), which is extremely useful for debugging purposes.

Typical system properties that are to be checked over system models by MC algorithms are:
\begin{itemize}
	\item \emph{functional correctness} (``the system does what it is supposed to do'');
	\item \emph{safety} (``something bad never happens'');
	\item \emph{liveness} (``something good will eventually happen'');
	\item \emph{fairness} (``under certain conditions, an event occurs repeatedly'');
	\item absence of \emph{deadlock states} (``the system never reaches a state in which it remains indefinitely'');
	\item \emph{real-time properties} (``the system satisfies some specific timing constraints during its operation'').
\end{itemize}

The whole process of MC consists of three major phases~\cite{Baier2008}.
\begin{enumerate}
    \item The \emph{modeling} phase: $(i)$~a system is modeled in an accurate and unambiguous way in terms of  a labeled state-transition graph or using a \emph{model description language} (e.g., Promela, VHDL, Verilog). Then, $(ii)$~the properties of the system meant to be checked are formalized unambiguously in a suitable \emph{property specification language}, typically by formulas of some \emph{temporal logic}. 
    \item The \emph{running} phase: the model checker is executed on the system model and on the property specifications, checking the validity of the latter over the former, that is---in mathematical terms---the model checker verifies whether the system model is a model of the specifications/formulas (hence the name ``model checking'').
    \item The \emph{analysis} phase: the outcome of the running phase is analyzed. If a property is violated, this may depend on several causes:
    \begin{itemize}
        \item the presence of a \emph{modeling} error (i.e., the model does not faithfully reflect the behavior/design of the system, thus needs to be corrected);
        \item if there is no modeling error, then a \emph{design} error may have been discovered (i.e., the design has to be changed, along with the resulting model), or
        \item a \emph{property} error has taken place (i.e., the requirements to be validated have not been correctly ``translated'' into the specifications).
    \end{itemize}
    It is worth pointing out that, whenever the model gets changed, verification has to be restarted from scratch: the design is verified only when all properties have successfully been checked with respect to the ``final version'' of the system model.
\end{enumerate}

We now list the strengths of the MC approach.
\begin{itemize}
    \item  First of all, it is a \emph{general} verification approach that has been employed in innumerable areas, such as:
    
    \begin{itemize}
        \item planning~\cite{DBLP:conf/ecp/GiunchigliaT99}, 
        \item communication and security protocols~\cite{holzmann1994,Lowe1996,4271662,basin2015model}, 
        \item embedded reactive systems~\cite{Cimatti2001}, 
        \item computer device drivers~\cite{Witkowski:2007}, 
        \item database-backed web applications~\cite{Gligoric2013}, 
        \item concurrency control and transaction atomicity~\cite{CPE:CPE1876}, 
        \item automated verification of UML design of applications~\cite{DONINI200619},
        \item verification of air traffic control systems~\cite{708566},
        \item testing of railway control systems~\cite{BGGMM17,Nardone2016}, 
        \item analysis of complex circuits~\cite{Burch:1991},
        \item design and development of (components of) CPUs~\cite{fix2008}, and verification of their microcode~\cite{Arons2005}, and even
        \item verification of clinical guidelines~\cite{Giordano06}.
    \end{itemize}
    \item It supports \emph{partial specification}, i.e., no complete requirements/system specifications are needed before information can be obtained on its correctness.
    \item  It supports \emph{partial verification}, i.e., MC allows us to verify also \emph{subsets} of properties at a time; thus one may choose to temporarily neglect properties of secondary importance. 
    \item As we already mentioned, MC is \emph{exhaustive}, meaning that all possible system behaviors are considered, and all (reachable) states are enumerated/visited.
	\item It provides \emph{diagnostic information} whenever a property is invalidated; this is useful for debugging.
	\item It has been accepted by industry, commercial model checkers have been available for a long time, and MC-based verification is becoming integral part of standards and of system development cycles.
\end{itemize}

Conversely, we now highlight weaknesses and limitations of MC.
\begin{itemize}
    \item  First of all, MC algorithms \emph{verify a system model}, and not the actual system! Therefore \lq\lq any obtained result is as good as the system model\rq\rq{}~\cite{Baier2008}. 
    \item MC techniques suffer from the so-called \emph{state space explosion problem}:
    in order to accurately represent complex systems, models with a huge number of states may be necessary, sometimes possibly exceeding the available computation resources (memory and/or processing time). 
    The parallel composition of two concurrent components is modeled by taking the product of the corresponding state spaces. This means that the global state space of a concurrent system has size exponential in the number of concurrent processes running. 
    We will shortly list some effective methods to deal with this problem; 
    however,
	models of realistic systems may sometimes still be too large to be processed.
	In such situations MC can be used to find bugs (but not for an exhaustive verification).
	\item It is worth observing that modeling is a task far from being trivial and automatable: on one hand the model has to capture all the salient features and aspects of the system to verify; on the other hand, details which are useless for the correctness proof have to be removed, as they would just weight verification down, increasing the number of states of the model. However, if the model is too coarse, MC may fail to find bugs existing in the real system. Conversely, if it is too fine-grained, spurious states/counterexamples may be generated, which are not reachable in the real system.
	Thus, expertise is required in finding suitable abstractions to obtain smaller, yet faithful, system models. 
    \item Simulation and testing are still needed, in support of MC, to find fabrication faults (for hardware) or coding errors (for software), as these arise during implementation, thus falling outside of the application of MC.
    \item  Only stated and formalized requirements can be verified (\emph{coverage} problem);
	again, complementary techniques, such as testing, may reveal additional problems, that were not considered and checked. 
	\item MC is mainly suited for \emph{control-intensive applications}, and less for data-intensive ones, as the former can usually be represented as finite state systems, whereas data typically feature infinite domains.
\end{itemize}

MC techniques and tools can also be employed in ``reverse way'', that is,
not during the specification and design phases, but
to check already existing systems, by ``extracting'' a model from them and verifying it. 
This approach is known as ``MC at implementation level''.
The automated generation of models amenable to MC from programs
written in C, C++ or Java~\cite{Hatcliff2001,Godefroid:1997} 
has been studied, for instance,
at Microsoft~\cite{Ball2001} and at NASA~\cite{Havelund2000}.

As we have already said, 
MC techniques have to cope with the problem of the \emph{state space explosion},
which becomes particularly serious when the system being verified is very complex or many components are working in parallel; this is the case, in particular, of software, which is extremely flexible and tends to get more and more complex (as one of Nathan's \lq\lq laws\rq\rq\ of software states). 
\begin{itemize}
\item In this respect, \emph{symbolic MC}~\cite{bur90,cou90,mcm93} allows an exhaustive implicit enumeration of a huge quantity of states (even more than $10^{120}$): ordered binary decision diagrams (OBDDs), particular data structures for representing Boolean functions, are used to get concise representations of (sets of states of) transition systems and to efficiently manipulate them through fix-point algorithms. A very successful model checker was developed based on OBDDs, SMV, nowadays exploited as the basis for several commercial model checkers (e.g., CadenceSMV~\cite{cadence} and NuSMV~\cite{Cimatti2000nusmv}).

\item
\emph{Partial order reduction}~\cite{pored93} tries to reduce the size of the state space by making computations that differ only in the ordering of independently executed actions collapse, as they are indistinguishable by the specification (i.e., they can be considered equivalent) and thus only one for each group needs to be tested. This approach works particularly well when applied to, for instance, communication protocols and distributed algorithms, in which several loosely coupled components are working in parallel. The model checker SPIN makes use of partial order reduction.

\item
Another traditional MC methodology is \emph{bounded MC} \cite{biere2003bounded}: proposed in~\cite{biere1999symbolic}, its basic idea is searching for a counterexample in computations whose length is bounded by a fixed integer $k$. Then, either a bug is found, or one can increase $k$ until the problem becomes intractable, or the so-called \emph{completeness threshold} is reached (i.e., for high enough values of $k$, this technique is guaranteed to find any existing counterexample). 
In bounded MC both the specifications of the system and the properties to check have to be translated into a Boolean formula. In this way, it is possible to employ SAT-solvers (extremely efficient procedures that can decide the satisfiability
of Boolean formulas) in MC, which are less sensitive to the state space explosion problem than OBDD-based solvers. However, this method is in general \emph{incomplete} if the bound is not high enough, hence it is used as a complementary technique to OBDD-based symbolic MC: the former is usually exploited for falsification, i.e., finding counterexamples and bugs, while the latter for verification.

\item
\emph{IC3} (acronym of the expression ``Incremental Construction of Inductive Clauses of Indubitable Correctness'')~\cite{Bradley2011,Bradley2012} is another successful SAT-based MC algorithm,
mainly used for the verification of safety properties (but it has been adapted to liveness and fairness as well). 
IC3 keeps a list of formulas $F_i$ representing an over-approximation of the system states reachable in up to $i$ steps from the initial ones. These formulas are iteratively enriched with additional clauses, that strengthen the approximation of the reachable states, and are determined with the help of a SAT-solver. Eventually, either a pair of formulas in the list, say $(F_k,F_{k+1})$, become identical---which means, by the correctness of IC3, that the property holds---or a counterexample is found. Throughout the process, the \emph{incrementality} feature of SAT-solvers is exploited: IC3 generates many similar formulas, following a common structural pattern; these can be solved quickly by re-using knowledge acquired in previous runs of the SAT-solver.

\item
\emph{CEGAR} (CounterExample-Guided Abstraction Refinement), 
a technique typically used for safety/universal properties,
minimizes the state space by \emph{hiding system variables}
that are irrelevant with respect to the property to check.
To this aim, an \emph{abstraction} of the system is produced, which must be \emph{conservative}, namely,
if a universal property is true in the abstract system, 
it will also be true in the concrete system model but, conversely,
even if the abstract model invalidates the
specification, the concrete one may still satisfy it.
In such a case, intuitively, the produced abstraction is too
coarse to validate the specification, as there are some abstract computation paths, reaching error states, not corresponding to
any real system paths.
Thus, CEGAR algorithms iteratively refine the abstraction based on unfeasible counterexamples, until either a real
counterexample is obtained (which is reported to the user), or no more spurious
ones are generated (hence the property holds in the real system model as well).

\emph{Lazy abstraction}~\cite{Henzinger02,McMillan2006} is an evolution of the CEGAR approach in which 
refinement of the abstraction $(i)$~is performed on-the-fly during MC, 
and $(ii)$~it does not affect already explored parts
of the abstract state space, which saves a high computation time.
\end{itemize}

The techniques seen so far were born to deal with
\emph{finite-state} systems. However, while 
the finite-state assumption is often realistic for
hardware systems,
this is not the case of software, whose state space can either be really infinite, or it is convenient to assume so. 
Being able to deal with infinite-state systems is thus essential, but raises the obvious issue
of the impossibility of any exhaustive exploration.
Such systems are defined in terms of first-order logic predicates and formulas, typically with an underlying theory 
(e.g., the theory of real arithmetic), and not by Boolean formulas. As a consequence, in this context,
SAT-solvers are replaced by Satisfiability Modulo Theory
(SMT) solvers, namely, procedures for deciding the satisfiability of formulas of (decidable fragments of) first-order logic.

The \emph{predicate abstraction} technique~\cite{daniel_et_al}
reduces infinite-state MC to finite-state MC, by representing the infinite state space as an
abstract one.
Abstract computation paths are overapproximations of real paths, thus
spurious counterexamples can be generated: 
abstraction refinement techniques are needed to deal with this.
As an example, IC3 has been adapted to handle infinite-state systems by ``embedding'' predicate abstraction~\cite{Cimatti2012,Cimatti2014,Cimatti2016}.

\section{Logics for MC}
The previous section demonstrates that a thorough investigation into MC has been performed over the years:
nowadays
MC techniques are mature and
MC tools have gained acceptance in industry, becoming routinely used for many applications in several companies (e.g., IBM, Motorola, Intel, Nokia,\dots).

Two temporal logics, $\LTL$ and $\CTL$, take the lion's share as the property specification language in MC tools.
In 1977, Pnueli~\cite{Pnu77} proposed the use of the linear temporal logic $\LTL$ for program verification.
$\LTL$ allows one to reason about changes in the truth value of formulas in system models over a linearly-ordered temporal domain, where each moment in time has a unique possible future.
More precisely, one has to consider all possible paths in a model and to analyze, for each of them, how atomic properties (proposition letters), labeling the states, change from one state to the next one along the path.
The MC problem for $\LTL$ turns out to be $\PSPACE$-complete~\cite{CGP02,Sistla:1985}. This logic has also been investigated with respect to the \emph{satisfiability} (SAT) problem (useful, for instance, in planning and as a \emph{sanity check} for formulas in system verification) which is, again, $\PSPACE$-complete~\cite{Sistla:1985}.

Four years later, in 1981, Clarke and Emerson invented the branching time logic $\CTL$~\cite{CE81}, whose model of time is a tree, i.e., the future is not determined, as there are different paths in the future, any one of which may be realized. 
The MC problem for $\CTL$ is in $\PTIME$~\cite{Clarke:1986}, while its SAT is $\EXPTIME$-complete~\cite{FL79}.
$\CTL$ and $\LTL$ are somewhat complementary, as there are properties expressible in only one of them~\cite{EH86}. 

Since then, several extensions of these logics have been investigated, as well as many of their fragments. Just to mention some,
we recall the extension of $\LTL$ with 
\emph{promptness}, that makes it 
possible to bound the delay with which a liveness request is fulfilled~\cite{DBLP:journals/fmsd/KupfermanPV09}, the \emph{metric} extension of $\LTL$ called $\mathsf{MTL}$~\cite{Ouaknine08}, and
the fragments of $\CTL$ investigated in~\cite{MEIER2008201}.

Algorithms for other logics, for instance, $\CTLStar$, $\mu$-calculus and first-order extensions thereof, have been developed, but those which find application in industrial-level MC are mostly $\LTL$ and $\CTL$.

\section{Interval temporal logics and $\HS$}

All the aforementioned logics are \emph{``point-based''}, meaning that they can only predicate properties of system/computation states.
Conversely,
there are some properties we might want to check that are \emph{inherently ``interval-based''} and thus cannot be expressed by point-based temporal logics, e.g., 
``For every time interval $I$ during which the green light is on for the vehicles on either road at the intersection, 
for some time interval beginning strictly before and ending strictly after $I$,
 the green
light must be continuously off and the red light on
for the vehicles on the other intersecting road''~\cite{DBLP:journals/eatcs/MonicaGMS11}.
Here interval temporal logics (ITLs) come into play, providing an alternative setting for 
reasoning about time~\cite{HS91,chopping_intervals,venema1990}. Such logics feature intervals, instead of points, as their basic ontological temporal entities; this choice gives them the ability to 
express temporal properties, such as actions with duration, accomplishments, and temporal aggregations, which cannot be dealt with by standard (point-based) logics.

ITLs have been applied in a variety of computer
science fields, including:  
\begin{itemize}
    \item \emph{computational linguistics} (for the study of progressive tenses in natural language~\cite{DBLP:journals/ai/Pratt-Hartmann05}),
    \item \emph{artificial intelligence} (for qualitative reasoning, reasoning about action and
change, planning, 
configuration, multi-agent systems~\cite{DBLP:journals/logcom/BowmanT03,DBLP:conf/ecp/GiunchigliaT99,DBLP:conf/tacas/LomuscioR06}),
    \item \emph{theoretical computer science} (in formal verification~\cite{DBLP:series/eatcs/ChaochenH04,digital_circuits_thesis}), and
    \item \emph{temporal databases} (for the specification of the expected behavior of concurrency control and transaction scheduling~\cite{roadmap_intervals}).
\end{itemize}

Moszkowski's Propositional Interval Temporal Logic ($\PITL$)~\cite{digital_circuits_thesis}
is one of the first ITLs, introduced in 1983 with the aim of specifying and verifying hardware components.
In the 90s, Duration Calculus ($\DC$)---an extension of $\PITL$ featuring the concept of duration of an event on a time interval---was studied~\cite{CHAOCHEN1991269} and, since then, developed and applied to the specification and design of time-critical systems~\cite{DBLP:series/eatcs/ChaochenH04,Hansen2007}.
However, its semantics is point-based, being proposition letters interpreted only over time points.
In~\cite{chopping_intervals}, Venema introduced the ITL $\CDT$ featuring the binary modality C (chop)---the natural \emph{chopping} operation on an interval involves bisecting it into two consecutive parts/subintervals---and its two residual modalities D and T.

A prominent position among ITLs is occupied by \emph{Halpern
and Shoham's modal logic of time intervals} ($\HS$, for short)~\cite{HS91}.
$\HS$ features one modality for each of the 13 possible ordering
relations between pairs of intervals (the so-called \emph{Allen's
relations}~\cite{All83}), apart from equality.
As an example, the statement ``the current interval meets an interval over
which $p$ holds'' can be expressed in $\HS$ by the formula $\hsA p$,
where $\hsA$ is the (existential) $\HS$ modality for Allen's relation
\emph{meet}.

\begin{sidewaystable}
\renewcommand{\arraystretch}{1.4}
\centering
\caption{Allen's relations and corresponding $\HS$ modalities.}\label{allen}
\begin{tabular}{lllc}
\hline
\rule[-1ex]{0pt}{3.5ex} Allen relation & $\HS$ modality & Definition w.r.t. interval structures &  Example\\
\hline

&   &   & \multirow{7}{*}{\input{Draw/allensRels.tex}}\\

\textsc{meets} & $\hsA$ (after) & $[x,y]\mathpzc{R}_A[v,z]\iff y=v$ &\\

\textsc{before} & $\hsL$ (later) & $[x,y]\mathpzc{R}_L[v,z]\iff y<v$ &\\

\textsc{started-by} & $\hsB$ (begins) & $[x,y]\mathpzc{R}_B[v,z]\iff x=v\wedge z<y$ &\\

\textsc{finished-by} & $\hsE$ (ends) & $[x,y]\mathpzc{R}_E[v,z]\iff y=z\wedge x<v$ &\\

\textsc{contains} & $\hsD$ (during) & $[x,y]\mathpzc{R}_D[v,z]\iff x<v\wedge z<y$ &\\

\textsc{overlaps} & $\hsO$ (overlaps) & $[x,y]\mathpzc{R}_O[v,z]\iff x<v<y<z$ &\\

\hline
\end{tabular}
\end{sidewaystable}
Table~\ref{allen} depicts 6 of the 13 Allen's relations
together with the corresponding $\HS$ (existential) modalities. 
The other 7 are equality and the 6 inverse (symmetrical) relations 
(given a binary relation $\mathpzc{R}$, its inverse $\overline{\mathpzc{R}}$ is such that $b \overline{\mathpzc{R}} a$ iff $a \mathpzc{R} b$). 

The language of $\HS$ features a set of proposition letters $\Prop$, the Boolean connectives $\neg$ and $\vee$, and a temporal modality for each of the (non trivial) Allen's relations, namely, $\hsA$, $\hsL$, $\hsB$, $\hsE$, $\hsD$, $\hsO$, $\hsAt$, $\hsLt$, $\hsBt$, $\hsEt$, $\hsDt$ and $\hsOt$.
$\HS$ formulas are defined by the following grammar:
\begin{equation*}
    \psi ::= p \;\vert\; \neg\psi \;\vert\; \psi \vee \psi \;\vert\; \hsX\psi \;\vert\; \hsXt\psi, \ \ \mbox{ with } p\in\mathpzc{AP},\; X\in\{A,L,B,E,D,O\}.
\end{equation*}
Throughout the thesis, we will make use of the standard abbreviations of propositional logic, e.g., we write $\psi \wedge \phi$ for $\neg(\neg\psi \vee \neg\phi)$, $\psi \rightarrow \phi$ for $\neg \psi \vee \phi$, $\psi \leftrightarrow \phi$ for $\left(\psi \rightarrow \phi\right)\wedge\left(\phi \rightarrow \psi\right)$, $\true$ (true) for $p\vee\neg p$, and $\false$ (false) for $\neg\true$.

For all $X\in\{A,L,B,E,D,O\}$, the (dual) universal modalities $\hsXu\psi$ and $\hsXtu\psi$ are defined as $\neg\hsX\neg\psi$ and $\neg\hsXt\neg\psi$, respectively. 
We denote by $\mathsf{X_1\cdots X_n}$ the fragment of $\HS$, closed under Boolean connectives, that features (existential and universal) modalities for $\langle X_1\rangle,\ldots, \langle X_n\rangle $ only.

W.l.o.g., we assume the \emph{non-strict semantics} of $\HS$, which admits intervals consisting of a single point.\footnote{All the results we prove in this thesis hold for the strict semantics as well, which disallows point-intervals.} Under such an assumption, all $\HS$ modalities can be expressed in terms of
$\hsB, \hsE, \hsBt$ and $\hsEt$~\cite{HS91}.
As an example, $\hsA$ (resp., $\hsAt$) can be expressed by $\hsE$ and $\hsBt$ (resp., $\hsB$ and $\hsEt$) as 
\[\hsA \varphi= (\hsEu\bot \wedge (\varphi \vee \hsBt \varphi)) \vee \hsE (\hsEu\bot \wedge (\varphi \vee \hsBt \varphi))\] 
\[(\text{resp., } \hsAt \varphi= (\hsBu\bot \wedge (\varphi \vee \hsEt \varphi)) \vee \hsB (\hsBu\bot \wedge (\varphi \vee \hsEt \varphi))).\]
As for all other modalities, the next equivalences hold:
\begin{equation*}
\begin{array}{cc}
\hsL\psi=\hsA\hsA\psi & \qquad \hsLt\psi=\hsAt\hsAt\psi \\ 
\hsD\psi=\hsB\hsE\psi= \hsE\hsB\psi & \qquad \hsDt\psi=\hsBt\hsEt\psi =\hsEt\hsBt\psi \\ 
\hsO\psi=\hsE\hsBt\psi & \qquad \hsOt\psi=\hsB\hsEt\psi .
\end{array}
\end{equation*}
We also use the auxiliary operator $\hsG$ of $\HS$ (and its dual $\hsGu$), which allows one to select  arbitrary subintervals of a given interval, and is defined as: $\hsG\psi= \psi \vee \hsB\psi \vee \hsE\psi \vee \hsB\hsE\psi$.

$\HS$ can thus be seen as a multi-modal logic with   
four primitive modalities
and its semantics can be defined over a multi-modal structure, called \emph{abstract interval model}, where intervals are treated as atomic objects and Allen's relations as 
binary relations between pairs of intervals.
Since later we will focus on some $\HS$ fragments not including some of the primitive modalities, we consider explicitly both $\hsA$ and $\hsAt$ in addition to $\hsB, \hsE, \hsBt$ and $\hsEt$.

\begin{definition}[Abstract interval model]\label{def:AIM}
An \emph{abstract interval model} is a tuple \[\mathpzc{A}=(\mathpzc{AP},\mathbb{I},A_\mathbb{I},B_\mathbb{I},E_\mathbb{I},\sigma),\] where:
\begin{itemize}
    \item $\mathpzc{AP}$ is a finite set of proposition letters;
    \item $\mathbb{I}$ is a possibly infinite set of atomic objects (worlds);
    \item $A_\mathbb{I}$, $B_\mathbb{I}$, $E_\mathbb{I}$ are three binary relations over $\mathbb{I}$;
    \item $\sigma:\mathbb{I}\mapsto 2^{\mathpzc{AP}}$ is a (total) labeling function, which assigns a set of proposition letters to each world.
\end{itemize}
\end{definition}
Intuitively, in the interval setting, $\mathbb{I}$ is a set of intervals, $A_\mathbb{I}$, $B_\mathbb{I}$, and $E_\mathbb{I}$ are interpreted as Allen's interval relations $A$ (\emph{meets}), $B$
(\emph{started-by}), and $E$ (\emph{finished-by}), respectively, and $\sigma$ assigns to each interval the set of proposition letters that hold over it.

\begin{definition}[$\HS$ semantics]\label{def:satisfaction}
Given an abstract interval model $\mathpzc{A}=(\mathpzc{AP},\mathbb{I},A_\mathbb{I},B_\mathbb{I},E_\mathbb{I}, \sigma)$
and a world $I\in\mathbb{I}$, the truth of an $\HS$ formula over $I$ is defined by induction on the structural complexity of the formula as:
\begin{itemize}
    \item $\mathpzc{A},I\models p$ iff $p\in \sigma(I)$, for any proposition letter $p\in\mathpzc{AP}$;
    \item $\mathpzc{A},I\models \neg\psi$ iff it is not true that $\mathpzc{A},I\models \psi$ (also denoted as $\mathpzc{A},I\not\models \psi$);
        \item $\mathpzc{A},I\models \psi \vee \varphi$ iff $\mathpzc{A},I\models \psi$ or $\mathpzc{A},I\models \varphi$;
    \item $\mathpzc{A},I\models \hsX\psi$, for $X \in\{A,B,E\}$, iff there exists $J\in\mathbb{I}$ such that $I\, X_\mathbb{I}\, J$ and $\mathpzc{A},J\models \psi$;
    \item $\mathpzc{A},I\models \hsXt\psi$, for $\overline{X} \in\{\overline{A},\overline{B},\overline{E}\}$, iff there exists $J\in\mathbb{I}$ such that $J\, X_\mathbb{I}\, I$ and $\mathpzc{A},J\models \psi$.
\end{itemize}
\end{definition}

\emph{Satisfiability} (SAT) and \emph{validity} are defined in the usual way: an $\HS$ formula $\psi$ is satisfiable if there exists an interval model $\mathpzc{A}$ and a world (interval) $I$ such that $\mathpzc{A},I\models \psi$. Moreover, $\psi$ is valid if $\mathpzc{A},I\models \psi$ for all worlds (intervals) $I$ of every interval model $\mathpzc{A}$.


In~\cite{HS91}  
it was shown that the \emph{validity problem} for $\HS$ interpreted
over any class of linear orders that contains at least one linear order with an
infinite ascending or descending sequence of points,
 is \emph{undecidable} (more precisely, r.e.\ hard).
For example, it is undecidable over $\mathbb{N}$ (the natural numbers), $\mathbb{Z}$ (the integers), $\mathbb{Q}$ (the rationals) and $\mathbb{R}$ (the reals), 
as well as over the classes of all linear orders, all dense linear orders, and all discrete linear orders.
Moreover, over any class of \emph{Dedekind complete} ordered structures containing at least an order with an infinite ascending sequence,
it is $\Pi_1^1$-hard (i.e., not recursively axiomatizable). This is true, e.g., over $\mathbb{N}$, $\mathbb{Z}$,  and $\mathbb{R}$, being Dedekind complete. 

The authors of~\cite{DBLP:journals/eatcs/MonicaGMS11} state that:
\begin{quote}
[\dots] so general undecidability results about $\HS$ are given in~\cite{HS91}
that for a long time it was considered unsuitable for practical applications and
attracted little interest among computer scientists.  
\end{quote}

Since then, a lot of work has been done on the SAT
problem for \emph{fragments} of $\HS$, which showed that undecidability rules also over
them~\cite{Bresolin2008,DBLP:conf/time/BresolinMGMS09,DBLP:journals/amai/BresolinMGMS14,DBLP:conf/asian/Lodaya00,DBLP:journals/fuin/MarcinkowskiM14}. As an example, the fragment $\BE$ (i.e., formulas of $\HS$ with $\hsB$ and $\hsE$ modalities only) is undecidable, again, over the class of all linear orders, over all dense linear orders, discrete linear orders and finite linear orders~\cite{DBLP:conf/asian/Lodaya00}.

The reason for these undecidability results must be ascribed to the very nature of purely interval-based temporal reasoning, where 
proposition letters are interpreted on 
intervals (and not on points): 
the set-theoretic interpretation of an
$\HS$ formula in a model is a set of abstract intervals, that is, a set of pairs of points (a binary relation over the underlying linear order).
Automata-based methods, founded for instance on Büchi and Rabin theorems (implying decidability of MSO theories of various linear orders and trees), do not apply here, as SAT and validity for ITLs are
dyadic, and not monadic. New approaches for proving decidability of $\HS$ fragments with genuinely interval-based semantics were needed.

Meanwhile, several modifications or restrictions, essentially reducing the interval-based semantics to
a point-based one,  were proposed to remedy the problem and obtain decidable systems. As an example, already in~\cite{digital_circuits_thesis} Moszkowski showed that the decidability of the (otherwise undecidable) logic $\PITL$ can be recovered by constraining  proposition letters to be
point-wise and defining truth of an interval as truth on its initial point (\emph{locality principle}). 

The bleak picture started lightening up later, with the discovery of several non-trivial cases of decidable fragments of $\HS$~\cite{DBLP:journals/logcom/BresolinGMS10, DBLP:journals/apal/BresolinGMS09, BMSS11, MPS10,MPSS10,DBLP:conf/time/MontanariS12INS, DBLP:conf/tableaux/BresolinMSS11INS}, including the \emph{interval logic of temporal
neighbourhood} $\AAbar$ and the \emph{temporal logic of sub-intervals} $\D$. %
In particular, the former was shown to be decidable over the class of 
all linear orders, of all dense, of all discrete, and of all finite linear orders;%
\footnote{It is also worth mentioning the decidable metric extension of the fragment $\AAbar$ called MPNL, studied in~\cite{DBLP:journals/sosym/BresolinMGMS13,DBLP:conf/ecai/BresolinMGMS10,DBLP:journals/corr/abs-1106-1241}, that features 
special atomic propositions representing integer constraints (equalities  and  inequalities) on the length of the intervals over which they are predicated.}
as for the latter, the situation is more involved:
$\D$ is \emph{decidable} over all dense linear orders, \emph{undecidable} over discrete linear orders and finite linear orders~\cite{DBLP:journals/fuin/MarcinkowskiM14}, 
and it is not known whether it is decidable or not in the case of all linear orders.
Some other fragments, such as $\B\Bbar$ and $\E\Ebar$, have actually a \emph{point-based} semantics: one of the endpoints of every interval related to the current one can ``move'', but the other always remains fixed; as a consequence they can be polynomially translated into a basic point-based temporal logic (namely, the fragment of $\LTL$ with only the future $\Eventually$ and past $\mathsf{P}$ operators): any interval is mimicked by its only moving endpoint; it follows that $\B\Bbar$ and $\E\Ebar$ are decidable over the class of all linear orders and, in particular, $\NP$-complete~\cite{roadmap_intervals}.
On the other hand, another fragment, $\AAbarBBbar$, is decidable over finite linear orders, $\mathbb{Q}$, as well as over the class of
all linear orders, but undecidable over Dedekind-complete linear orders (in particular, $\mathbb{N}$ and $\mathbb{R}$)~\cite{DBLP:conf/mfcs/MontanariPS14Ratio}.
These snapshots on $\HS$ SAT make us conclude, with the authors of~\cite{DBLP:journals/eatcs/MonicaGMS11}, that
\begin{quote}
[\dots] the trade-off
between expressiveness and computational affordability in the family of $\HS$ fragments
is rather subtle and sometimes unpredictable, with the border between decidability and undecidability cutting right
across the core of that family.
\end{quote}

\section{The problem of MC for \HS}

Whereas, as we have seen, darkness reigns in the realm of SAT with few exceptions,
fortunately a dim light brightens the whole land of MC for $\HS$:
in this thesis, we will consider this \emph{decidable} problem,
for which a little work has been done, if compared to SAT of $\HS$ and especially to MC of point-based temporal logics.

For quite a long time, no studies on MC for $\HS$ were undertaken: the problem was finding 
suitable system models over which to interpret such an interval-based property specification language 
(clearly, this was a novel issue to face, as the SAT problem is defined more simply on linear orders).
In 2011, Dario Della Monica concluded his PhD thesis about decidability and undecidability results on SAT for $\HS$ and its fragments in this way~\cite{DarioDM:phd_thesis}:
\begin{quote}
Even if this topic [MC] has been extensively studied and successfully applied to real-world domains in the context of point-based logics, it is still quite
unexplored in the interval setting. The major difficulty concerns the problem of finding a convenient way to (finitely) represent the models to be checked.
\end{quote}

The first idea we explore in this thesis is evaluating $\HS$ formulas on (very standard) system models, consisting essentially in \emph{labeled graphs}, called finite \emph{Kripke structures} (which will be formally defined later), making it
possible to check the correctness of the behavior of systems with respect to 
meaningful interval properties. 
In order to verify interval properties of computations, one needs to collect 
information about the states of a system into computation stretches.
To this end, each finite path (i.e., a \emph{trace}) in a Kripke structure is interpreted as an interval, 
and its labeling is defined on the basis of the labeling of the states composing it, initially
according to the \emph{homogeneity assumption}~\cite{Roe80}, which enforces a proposition letter $p$ to hold on an interval $I$ if and only if $p$ holds on all the subintervals of $I$.
Such an assumption is fairly natural for many applications; moreover, it easily enables us to 
interpret an \emph{interval-based} property specification language on a model which is \emph{point-based} (as a matter of fact, Kripke structures make explicit how a system evolves \emph{state-by-state}, and describe which are the atomic properties that hold true at every single state).

Then we devote our efforts to 
relaxing homogeneity.
For the purpose, we investigate the possibility of using regular expressions to define the labeling of proposition letters over intervals in terms of the component states (as we will see, homogeneity can be trivially encoded by regular expressions, as a particular case). This impacts on the complexity of MC for \HS\ and its fragments, and on the expressiveness of the logic as well. 
Another possibility is completely abandoning Kripke structures in favour of different, \emph{inherently interval-based} system models: \emph{timelines}, structures that have been fruitfully exploited in temporal planning for quite a long time, are employed in this thesis in a different way: more expressive than Kripke structures, they 
allow us to directly describe systems in terms of their \emph{interval-based behaviour and properties},
paving the way for a more general interval-based MC.

\section{Organization of the thesis}
Let us now take a deeper look at the contents of the thesis, while presenting its organization.
\begin{enumerate}
    \item In the next chapter---which starts Part~\ref{part:HShomo}---we show that finite Kripke structures can be suitably mapped into
abstract interval models, 
whose worlds are all the possible traces of the former.
However, since Kripke structures may have loops and thus infinitely many traces, abstract interval models feature, in 
general, an infinite domain. 
In order to devise a MC procedure for $\HS$ over finite Kripke
structures---hence showing the decidability of MC for $\HS$---we prove a small model theorem stating that, given an $\HS$ 
formula $\psi$ and a finite Kripke structure $\Ku$, there exists a \emph{finite}
interval model which is equivalent to the one induced by $\Ku$ with respect 
to the satisfiability of $\psi$.
The main technical ingredients are $(i)$~the definition of a suitable equivalence relation 
over traces of $\Ku$, which is parametric in the nesting 
depth of modalities $\hsB$ and $\hsE$ in $\psi$, and $(ii)$~the proof that 
the resulting quotient structure is finite and equivalent to the one induced by 
$\Ku$ with respect to the satisfiability of $\psi$.

The chapter is concluded showing that the problem does not provably admit any polynomial-time decision procedure (as it is \EXPSPACE-hard). For this reason, 
\item
in the subsequent chapters of Part~\ref{part:HShomo} we focus on fragments of \HS , which 
feature considerably better complexities---from \EXPSPACE, down to \Psp\ and to low levels of the polynomial hierarchy. Nevertheless, they have the ability to capture meaningful interval properties of state transition systems.
We describe several MC algorithms, tailored to the specific fragments being considered, which feature as building blocks concepts and techniques different from one another (e.g., small model properties, trace contractions, Boolean circuits with oracles, finite state automata,\ldots). Most of the times, we also prove lower bounds matching the complexity of MC algorithms.

Moreover, we study the expressive power of $\HS$ in MC, in comparison with that of the standard point-based logics $\LTL$, $\CTL$ and $\CTLStar$, still under homogeneity.
Additionally, we introduce some \emph{semantic variants} of \HS\  and evaluate their expressiveness and succinctness with respect to the aforementioned point-based logics.

\item
The homogeneity assumption is relaxed in Part~\ref{part:HSrelaxhomo}: as we have anticipated, at first we still consider Kripke structures as system models, but we 
use regular expressions to define the labeling of proposition letters over intervals in terms of the component states.
In particular, MC for \HS\ extended with regular expressions remains still decidable: this result is proved via a (non completely standard) automata-theoretic approach.

As for the fragments of \HS, the MC problem for \lq\lq small\rq\rq\ ones features an increased (\Psp) complexity, whereas formulas of other, more expressive fragments, can be checked at no extra computational cost (w.r.t.\ the homogeneous case).
Such complexity results are achieved by heavily exploiting \emph{non-deterministic} and \emph{alternating} algorithms.

\item
Finally, in Part~\ref{part:timelines} we
consider the \emph{timeline-based MC problem},
where system models are given by the aforementioned \emph{timelines},
and the property specification language is the logic $\mathsf{MITL}$: the choice of replacing $\HS$ with the latter is justified by the need for a \emph{timed} logic, appropriate to be interpreted over timelines, where the dimension of time is essential. 

Most of Part~\ref{part:timelines} will be actually devoted to the study of the so-called \emph{timeline-based planning problem} over \emph{dense temporal domains}:
first, such problem can be considered as a necessary condition for MC, the former playing the role of a \lq\lq feasibility check\rq\rq\ of the system description (as a matter of fact, whereas Kripke structures are \lq\lq extensive\rq\rq\ system descriptions, timelines are not, given the presence of constraint rules which must be checked for satisfiability). 
Moreover, dense time domains (as opposed to discrete ones, which are standard in planning) are necessary to avoid discreteness in system descriptions, that can be abstracted at a higher level, enabling us to express really interval-based properties of systems.
Since, in its full generality, timeline-based planning over dense domains is, unfortunately, undecidable, suitable restrictions on timelines are identified at the beginning of Part~\ref{part:timelines} in order to recover its decidability, and finally deal with timeline-based MC.

\end{enumerate}

%% file: Draw/allensRels.tex
\begin{tikzpicture}[scale=1.2]
\draw[draw=none,use as bounding box](-0.3,0.2) rectangle (3.3,-3.1);
\coordinate [label=left:\textcolor{red}{$x$}] (A0) at (0,0);
\coordinate [label=right:\textcolor{red}{$y$}] (B0) at (1.5,0);
\draw[red] (A0) -- (B0);
\fill [red] (A0) circle (2pt);
\fill [red] (B0) circle (2pt);

\coordinate [label=left:$v$] (A) at (1.5,-0.5);
\coordinate [label=right:$z$] (B) at (2.5,-0.5);
\draw[black] (A) -- (B);
\fill [black] (A) circle (2pt);
\fill [black] (B) circle (2pt);

\coordinate [label=left:$v$] (A) at (2,-1);
\coordinate [label=right:$z$] (B) at (3,-1);
\draw[black] (A) -- (B);
\fill [black] (A) circle (2pt);
\fill [black] (B) circle (2pt);

\coordinate [label=left:$v$] (A) at (0,-1.5);
\coordinate [label=right:$z$] (B) at (1,-1.5);
\draw[black] (A) -- (B);
\fill [black] (A) circle (2pt);
\fill [black] (B) circle (2pt);

\coordinate [label=left:$v$] (A) at (0.5,-2);
\coordinate [label=right:$z$] (B) at (1.5,-2);
\draw[black] (A) -- (B);
\fill [black] (A) circle (2pt);
\fill [black] (B) circle (2pt);

\coordinate [label=left:$v$] (A) at (0.5,-2.5);
\coordinate [label=right:$z$] (B) at (1,-2.5);
\draw[black] (A) -- (B);
\fill [black] (A) circle (2pt);
\fill [black] (B) circle (2pt);

\coordinate [label=left:$v$] (A) at (1.3,-3);
\coordinate [label=right:$z$] (B) at (2.3,-3);
\draw[black] (A) -- (B);
\fill [black] (A) circle (2pt);
\fill [black] (B) circle (2pt);

\coordinate (A1) at (0,-3);
\coordinate (B1) at (1.5,-3);
\draw[dotted] (A0) -- (A1);
\draw[dotted] (B0) -- (B1);
\end{tikzpicture}

%% file: Chaps/Intro/MCfullHShomo.tex
\chapter{MC for full $\HS$ under homogeneity}\label{chap:MCfullHShomo}
\begin{chapref}
The references for this chapter are~\cite{MMMPP15,ijcar16,BOZZELLI2018}.
\end{chapref}

\minitoc\mtcskip

\lettrine[lines=3]{I}{n this chapter} we prove two fundamental results. On one hand we show that MC for $\HS$ under homogeneity over finite Kripke structures is decidable.
On the other we prove a complexity lower bound for the problem, namely, its $\EXPSPACE$-hardness. In the following description of the contents of the next sections, we detail how these results are achieved.

\paragraph{Organization of the chapter.} 
\begin{itemize}
    \item In Section~\ref{sec:KrAndindAIM}, after introducing Kripke structures, we show how these system models can be mapped into abstract interval models, where each world corresponds to a trace (finite path) of the structure: this allows us to interpret $\HS$ formulas over Kripke structures, and then to define interval-based MC. Since a Kripke structure may feature loops and thus infinitely many traces, the domain of the induced abstract interval model is infinite.
    Thus, to prove decidability of MC,
    \item in Section~\ref{sec:descr} 
    we introduce BE-descriptors, that are trees built starting from the traces of a Kripke structure $\Ku$, whose nodes have labels over states of $\Ku$, which allow us to define an equivalence relation of finite index over traces of $\Ku$, parametric in the nesting depth of modalities $\hsB$ and $\hsE$ in the formula $\psi$ to check.
    \item We prove in Section~\ref{sec:decidProof} that such a relation represents a sufficient condition for two (related) traces to be indistinguishable w.r.t.\ the truth value of any $\HS$ formula with nesting depth of $\hsB$ and $\hsE$ bounded by a value equal to the depth of the considered BE-descriptors. Decidability of $\HS$ MC (under homogeneity) follows as a result of such a relation (and by minor technical ingredients). We can finally devise a \emph{nonelementary complexity} decision procedure for the problem.
    \item In Section~\ref{sec:BEhard}, we prove the $\EXPSPACE$-hardness of MC for the $\HS$ fragment $\BE$. Since MC for full $\HS$ is clearly at least as hard as MC for $\BE$, such a lower bound immediately propagates to full $\HS$. The section concludes with some remarks on the complexity gap (nonelementary--$\EXPSPACE$) deriving from the results proved for the problem.
    \item In Section~\ref{sec:overvHomo} we give an overview of the complexity of MC for several $\HS$ fragments studied under homogeneity, which will be considered in the next chapters.
    \item We conclude with Section~\ref{sec:LOMrelated}, in which we shortly review some research papers (not of the present authors) dealing with $\HS$ MC (under different semantic assumptions).
\end{itemize}

\section{Preliminaries: Kripke structures and induced abstract interval models}\label{sec:KrAndindAIM}
In the context of MC, finite state systems are usually modelled as \emph{finite Kripke structures}, 
namely, graphs where nodes represent the states of the system, edges represent transitions between states, and there is a labelling function that associates every node with a set of (atomic) properties that hold true in the corresponding system state. 
Systems described in terms of many modelling languages (e.g., Promela) are translated (usually, at the cost of an exponential blow-up) into Kripke structures, before the MC running phase begins. 

\begin{definition}[Kripke structure]\label{def:kripkestructure}
A Kripke structure is a tuple \[\Ku=\KuDef,\] where $\Prop$ is a finite set of proposition letters, $\States$ is a set of states (\emph{worlds}), $\Edges\subseteq \States\times \States$ is a left-total relation%
\footnote{A relation $\Edges\subseteq \States\times \States$ is left-total if, for all $s\in \States$, there exists at least one $s'\in \States$ such that $(s,s')\in\Edges$.}
between pairs of states (\emph{accessibility/transition} relation), $\Lab:\States\to 2^\mathpzc{\Prop}$ is a total labelling function, and $\sinit\in \States$ is called the \emph{initial} state.
We say that $\Ku$ is finite if $\States$ is finite.
\end{definition}
For all $s\in \States$, $\Lab(s)$ captures the set of proposition letters that hold at that state;
$\Edges$ constrains the evolution of the system over time, namely, it specifies how the system can evolve from one state to another; it is left-total because the paths of $\Ku$ are meant to represent 
system computations. 
If $(s,s')\in\Edges$, we say that $s'$ is a \emph{successor} of $s$, and $s$ a \emph{predecessor} of $s'$. By $\Edges(s)$ we denote the set of successors of $s$.

A simple Kripke structure, consisting of two states only, is reported in the following 
example. We will use it as a running example throughout the thesis.

\begin{example} \label{ex:kripke1}
Figure~\ref{K2} depicts a two-state Kripke structure 
$\Ku_2$ (the initial state is identified by a double circle).
Despite its simplicity, it features an \emph{infinite number} of different (finite) 
paths. 
$\Ku_2$ is defined by the 
quintuple:
\[(\{p,q\},\{s_0,s_1\},\{(s_0,s_0),(s_0,s_1),(s_1,s_0),(s_1,s_1)\},\Lab,\sinit),\]
where $\mu(s_0)=\{p\}$ and $\mu(s_1)=\{q\}$.
\begin{figure}[H]
\centering
\begin{tikzpicture}[->,>=stealth,thick,shorten >=1pt,auto,node distance=2cm,every node/.style={circle,draw}]
    \node [style={double}](v0) {$\stackrel{s_0}{p}$};
    \node (v1) [right of=v0] {$\stackrel{s_1}{q}$};
    \draw (v0) to [bend right] (v1);
    \draw (v1) to [bend right] (v0);
    \draw (v0) to [loop left] (v0);
    \draw (v1) to [loop right] (v1);
\end{tikzpicture}
\caption{The finite Kripke structure $\Ku_2$.}\label{K2}
\end{figure}
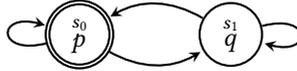
\end{example}

The next definition formalizes the notion of path in a Kripke structure.
\begin{definition}[Trace over $\Ku$]
A trace $\rho$ over a Kripke structure $\Ku=\KuDef$ is a finite sequence of states $s_1\cdots s_n$, with $n\geq 1$, such that $(s_i,s_{i+1})\in \Edges$ for all $i\in\{1,\ldots ,n-1\}$.
\end{definition}

Let $\Trk_\Ku$ be the (infinite) set of all traces over a finite/infinite Kripke structure $\Ku$. For any trace $\rho=s_1\cdots s_n \in \Trk_\Ku$, we define:
\begin{itemize}
    \item $|\rho|=n$ and, for $1\leq i\leq |\rho|$, $\rho(i)=s_i$ (we also say that $i$ is a \lq\lq $\rho$-position\rq\rq );
    \item $\states(\rho)=\{s_1,\ldots,s_n\}\subseteq \States$;
    \item $\intstates(\rho)=\{s_2,\ldots,s_{n-1}\}\subseteq \States$;
    \item $\fst(\rho)=s_1$ and $\lst(\rho)=s_n$;
    \item $\rho(i,j)=s_i\cdots s_j$, for $1\leq i \leq j\leq |\rho|$, is a subtrace of $\rho$;
    \item $\Pref(\rho)=\{\rho(1,i) \mid 1\leq i\leq |\rho|-1\}$ is the set of all proper prefixes of $\rho$;
    \item $\Suff(\rho)=\{\rho(i,|\rho|) \mid 2\leq i\leq |\rho|\}$ is the set of all proper suffixes of $\rho$. 
\end{itemize}
If $\fst(\rho)=s_0$---where $s_0$ is the initial state of $\Ku$---$\rho$ is said to be an \emph{initial trace}. 

Given $\rho,\rho' \in \Trk_\mathpzc{K}$, we denote by $\rho\cdot\rho'$ the concatenation of the traces $\rho$ and $\rho'$.
Moreover, if $\lst(\rho)=\fst(\rho')$, $\rho\star \rho'$ denotes $\rho(1,|\rho|-1)\cdot \rho'$ ($\star$-concatenation). In the following, when we write $\rho\star\rho'$, we implicitly assume that
$\lst(\rho)=\fst(\rho')$.

An abstract interval model (over $\Trk_\Ku$)---recall Definition~\ref{def:AIM}---can be naturally associated with a finite Kripke structure by interpreting every trace as an interval bounded by its first and last states.
\begin{definition}[Abstract interval model induced by $\Ku$]\label{def:inducedmodel}
The abstract interval model induced by a finite Kripke structure $\Ku=\KuDef$ is the abstract interval model $\mathpzc{A}_\Ku=(\Prop,\mathbb{I},A_\mathbb{I},B_\mathbb{I},E_\mathbb{I},\sigma)$, where:
    \begin{itemize}
        \item $\mathbb{I}=\Trk_\Ku$,
        \item $A_\mathbb{I}=\left\{(\rho,\rho')\in\mathbb{I}\times\mathbb{I}\mid \lst(\rho)=\fst(\rho')\right\}$,
        \item $B_\mathbb{I}=\left\{(\rho,\rho')\in\mathbb{I}\times\mathbb{I}\mid \rho'\in\Pref(\rho)\right\}$,
        \item $E_\mathbb{I}=\left\{(\rho,\rho')\in\mathbb{I}\times\mathbb{I}\mid \rho'\in\Suff(\rho)\right\}$,
        \item $\sigma:\mathbb{I}\to 2^\Prop$ is such that, for all $\rho\in\mathbb{I}$, 
        \begin{equation*}
            \sigma(\rho)=\bigcap_{s\in\states(\rho)}\Lab(s).
        \end{equation*}
    \end{itemize}
\end{definition}
In Definition \ref{def:inducedmodel}, relations $A_\mathbb{I},B_\mathbb{I}$, and $E_\mathbb{I}$ are interpreted as Allen's interval relations $A$, $B$ and $E$, respectively. Moreover, according to the definition of $\sigma$, a proposition letter $p\in\Prop$ holds over $\rho=s_1\cdots s_n$ if and only if it holds over all the states $s_1, \ldots , s_n$ of $\rho$. This conforms to the \emph{homogeneity principle}~\cite{Roe80}, according to which a proposition letter holds over an interval if and only if it holds over all of its subintervals.

Satisfiability of an $\HS$ formula over a finite Kripke structure can now be given in terms of induced abstract interval models.
\begin{definition}[Satisfiability of $\HS$ formulas over finite Kripke structures]\label{def:satkripke}
Let $\Ku$ be a finite Kripke structure, $\rho$ be a trace in $\Trk_\Ku$,
$\psi$ be an $\HS$ formula. We say that the pair $(\Ku,\rho)$ satisfies $\psi$, denoted by $\Ku,\rho\models \psi$, if and only if it holds that $\mathpzc{A}_\Ku,\rho\models \psi$.
\end{definition}
We are now ready to formally state the \emph{MC problem} for $\HS$ over finite Kripke structures: it is the problem of deciding whether or not  $\Ku\models \psi$.
\begin{definition}[Model checking]\label{def:MCkripke}
Let $\Ku$ be a finite Kripke structure and $\psi$ be an $\HS$ formula. We say that
$\Ku$ models $\psi$, denoted by $\Ku\models \psi$, if and only if, 
for all \emph{initial} traces $\rho\in\Trk_\Ku$, it holds that $\Ku,\rho\models \psi$.
\end{definition}
It is worth pointing out that every finite Kripke structure $\Ku$ induces an abstract interval model, and that only interval models arising from finite Kripke structures are considered in the MC problem. 

Given a finite $\Ku=\KuDef$, being $\Edges$ left-total and $\States$ finite,
$\Ku$ has to feature some loops, and thus
an infinite number of traces; as a consequence, 
the \emph{MC problem is not trivially decidable}.

We conclude this section by giving some examples of meaningful properties of Kripke structures and traces that can be expressed in $\HS$. 

\begin{example}\label{example:length}
The formula $\hsBu\bot$ can be used to select all and only the traces of length $1$. Indeed, given any $\rho$ with $|\rho|=1$, independently of $\Ku$, it holds that $\Ku,\rho\models \hsBu\bot$, because $\rho$ has no proper prefixes. On the other hand, it holds that $\Ku,\rho\models \hsB\top$ if (and only if) $|\rho| > 1$.

Modality $\hsB$ can actually be used to constrain the length of an interval to be greater than, less than,  or equal to any value $k$. Let us denote $k$ nested applications of $\hsB$ by $\hsB^k$. It holds that 
$\Ku,\rho\models \hsB^k\top$ if and only if $|\rho|\geq k+1$. Analogously, $\Ku,\rho\models \hsBu^k\bot$ if and only if $|\rho|\leq k$. 
Let $\Length_k$ be a shorthand for $\hsBu^{k}\bot \wedge \hsB^{k-1}\top$. It holds that $\Ku,\rho\models \Length_k$ if and only if $|\rho|=k$ (this formula will be used several times in the next chapters).
\end{example}

\begin{example}
Let us consider again the finite Kripke structure $\Ku_2$ of Example~\ref{ex:kripke1}, depicted in Figure~\ref{K2}. For the sake of brevity, for any trace $\rho$, we denote by $\rho^n$
the trace obtained by concatenating $n$ copies of $\rho$.
The truth of the following statements can be easily checked:
    ($i$)~$\Ku_2,(s_0s_1)^2\models \hsA q$;
    ($ii$)~$\Ku_2,s_0s_1s_0\not\models \hsA q$;
    ($iii$)~$\Ku_2,(s_0s_1)^2\models \hsAt p$;
    ($iv$)~$\Ku_2,s_1s_0s_1\not\models \hsAt p$.
The above statements show that modalities $\hsA$ and $\hsAt$ can be used to distinguish between traces that \emph{start or end at different states}. 

We would like to draw attention to the \emph{branching} semantics of modalities $\hsA$ and $\hsAt$ 
(in our experience, at the beginning, this causes confusion for the reader\dots): 
$\hsA$ (resp., $\hsAt$) allows one to ``move'' to \emph{any} trace branching on the right/future (resp., left/past) of the considered one, e.g., if $\rho=s_1s_0$, then 
$\rho\, A_\mathbb{I}\, s_0$,
$\rho\, A_\mathbb{I}\, s_0s_0$, 
$\rho\, A_\mathbb{I}\, s_0s_1$, 
$\rho\, A_\mathbb{I}\, s_0s_0s_0$, 
$\rho\, A_\mathbb{I}\, s_0s_0s_1$, 
$\rho\, A_\mathbb{I}\, s_0s_1s_0s_1$, and so on.
Analogously,
$s_1\, A_\mathbb{I}\, \rho$,
$s_0s_1\, A_\mathbb{I}\, \rho$, 
$s_1s_1\, A_\mathbb{I}\, \rho$,\dots 

Figure~\ref{fig:exaAAt} illustrates the \lq\lq behaviour\rq\rq{} of $\hsA$ and $\hsAt$.

\begin{figure}[H]
    \centering
    \begin{tikzpicture}
			\filldraw [gray] (4,-1) circle (2pt)
				(2,-1) circle (2pt)
				(0,0) circle (2pt)
				(0,-1) circle (2pt)
				(0,-2) circle (2pt)
				(6,0) circle (2pt)
				(6,-1) circle (2pt)
				(6,-2) circle (2pt)	;
				\draw [black]  (0,-1) -- (6,-1);
				\draw [black] (4,-1) -- (6,0);
				\draw [black] (4,-1) -- (6,-2);
					\draw [black] (0,0) -- (2,-1);
				\draw [black] (0,-2) -- (2,-1);
				\draw [dashed, orange] (0,0.2) -> (2,-0.8) -> (4,-0.8);
				\draw [dashed, orange] (2,-1.2) -> (4,-1.2) -> (6,-2.2);
				
				\node [orange] at (1,0) {$\varphi_1$};	
				\node [red] at (2.5,-0.5) {$\hsAt\varphi_1$};
			\node [orange] at (5,-2) {$\varphi_2$};	
				\node [red] at (2.5,-1.5) {$\hsA \varphi_2$};
						
			\end{tikzpicture}
    \caption{The branching semantics of modalities $\hsA$ and $\hsAt$}
    \label{fig:exaAAt}
\end{figure}

Modalities $\hsB$ and $\hsE$ can be exploited to distinguish between traces encompassing a different number of iterations of a given loop. This is the case, for instance, with the following statements:
\begin{itemize}    
    \item $\Ku_2,(s_1s_0)^3 s_1\models \hsB \big(\hsA p \wedge \hsB \left(\hsA p \wedge \hsB\hsA p\right)\big)$;
    \item $\Ku_2,(s_1s_0)^2 s_1\not\models \hsB \big(\hsA p \wedge \hsB \left(\hsA p \wedge \hsB\hsA p\right)\big)$.
\end{itemize}

$\HS$ makes it possible to distinguish between traces $\rho_1=s_0^3s_1s_0$ and $\rho_2=s_0s_1s_0^3$, which involve the same number of iterations of the same loops, but differ in the order of loop occurrences: $\Ku_2,\rho_1\models \hsB\big(\hsA q \wedge \hsB(\hsA p\wedge \hsB\true)\big)$, but $\Ku_2,\rho_2\not\models \hsB\big(\hsA q \wedge \hsB(\hsA p\wedge \hsB\true)\big)$.

Also $\hsBt$ and $\hsEt$, the inverses of $\hsB$ and $\hsE$, are branching---in the future and in the past, respectively---just like $\hsA$ and $\hsAt$. See Figure~\ref{fig:exaBtEt}.

\begin{figure}[H]
    \centering
    \begin{tikzpicture}
				\filldraw [gray] (0,2) circle (2pt)
				(2,2) circle (2pt)
				(4,3) circle (2pt)
				(4,2) circle (2pt)
				(4,1) circle (2pt);
				\draw [black]  (0,2) -- (4,2);

				\draw [black] (2,2) -- (4,3);
				\draw [black] (2,2) -- (4,1);

				\draw [dashed, orange] (0,2.2) -> (2,2.2) -> (4,3.2);
				\draw [dashed, red] (0,1.8) -> (2,1.8);

				\node [orange] at (0.5,2.5) {$\varphi_1$};	
				\node [red] at (0.5,1.5) {$\hsBt\varphi_1$};
			
						
			\end{tikzpicture}
\hspace{1.1cm}
			\begin{tikzpicture}
				\filldraw [gray] (4,-1) circle (2pt)
				(2,-1) circle (2pt)
				(0,0) circle (2pt)
				(0,-1) circle (2pt)
				(0,-2) circle (2pt);
				\draw [black]  (0,-1) -- (4,-1);
				\draw [black] (0,0) -- (2,-1);
				\draw [black] (0,-2) -- (2,-1);
				\draw [dashed, orange] (0,0.2) -> (2,-0.8) -> (4,-0.8);
				\draw [dashed, red] (2,-1.2) -> (4,-1.2);
				\node [orange] at (2.5,-0.5) {$\varphi_1$};	
				\node [red] at (2.5,-1.5) {$\hsEt\varphi_1$};
			\end{tikzpicture}
    \caption{The branching semantics of modalities $\hsBt$ and $\hsEt$}
    \label{fig:exaBtEt}
\end{figure}
\end{example}

\begin{example}\label{example:Ksched}
In Figure~\ref{KSched}, we give an example of a finite Kripke structure $\Ku_{Sched}$ that models the behaviour of a scheduler serving three processes which are continuously requesting the use of a common resource. 

The initial state (denoted by a double circle) 
is $s_0$: no process is served in that state. In any other state $s_i$ and $\overline{s}_i$, with $i \in \{1,2,3\}$, the $i$-th process is served (this is denoted by the fact that $p_i$ holds in those 
states). For the sake of readability, edges are marked either by $r_i$, for $\mathsf{request}(i)$, or by $u_i$, for $\mathsf{unlock}(i)$. However, edge labels do not have a semantic value, i.e., they are neither part of the structure definition, nor proposition letters; they are simply used to ease reference to edges. 

Process $i$ is served in state $s_i$, then, after ``some time'', a transition $u_i$ from $s_i$ to $\overline{s}_i$ is taken; subsequently, process $i$ cannot be served again immediately, as $s_i$ is not directly reachable from $\overline{s}_i$ (the scheduler cannot serve the same process twice in two successive rounds). A transition $r_j$, with $j\neq i$, from $\overline{s}_i$ to $s_j$ is then taken and process $j$ is served. This structure can be easily generalised to a higher number of processes.

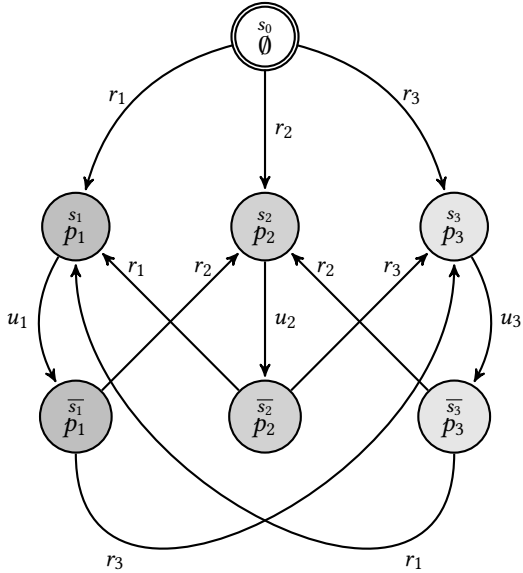
\begin{figure}[H]
\centering
\begin{tikzpicture}[->,>=stealth',shorten >=1pt,auto,node distance=2.5cm,thick,main node/.style={circle,draw}]
  \node[main node,style={double}] (1) {$\stackrel{s_0}{\emptyset}$};
  \node[main node,fill=gray!35] (3) [below of=1] {$\stackrel{s_2}{p_2}$};
  \node[main node,fill=gray!50] (2) [left of=3] {$\stackrel{s_1}{p_1}$};
  \node[main node,fill=gray!20] (4) [right of=3] {$\stackrel{s_3}{p_3}$};
  \node[main node,fill=gray!50] (5) [below of=2] {$\stackrel{\overline{s_1}}{p_1}$};
  \node[main node,fill=gray!35] (6) [below of=3] {$\stackrel{\overline{s_2}}{p_2}$};
  \node[main node,fill=gray!20] (7) [below of=4] {$\stackrel{\overline{s_3}}{p_3}$};

  \path[every node/.style={font=\small}]
    (1) edge [bend right] node[left] {$r_1$} (2)
        edge node {$r_2$} (3)
        edge [bend left] node[right] {$r_3$} (4)
    (2) edge [bend right] node [left] {$u_1$} (5)
    (3) edge node {$u_2$} (6)
    (4) edge [bend left] node [right] {$u_3$} (7)
    (5) edge node[very near end,left] {$r_2$} (3)
    (5) edge [out=270,in=270,looseness=1.3] node [near start,swap] {$r_3$} (4)
    (6) edge node[very near end,right] {$r_1$} (2)
    (6) edge node[very near end,left] {$r_3$} (4)
    (7) edge [out=270,in=270,looseness=1.3] node [near start] {$r_1$} (2)
    (7) edge node[very near end,right] {$r_2$} (3)
    ;
\end{tikzpicture}
\vspace{-1.4cm}
\caption{The finite Kripke structure $\Ku_{Sched}$.}\label{KSched}
\end{figure}

We now show how some meaningful properties to be checked over 
$\Ku_{Sched}$ 
can be expressed in $\HS$. 
In all formulas, we force the validity of the considered property over all legal computation sub-intervals by using modality $\hsEu$ (all computation sub-intervals are suffixes of at least one initial trace). Moreover, we will use the shorthand $wit_{\geq 2}(\{p_1,p_2,p_3\})$ for the formula
\begin{equation*}
(\hsD p_1\wedge \hsD p_2)\vee (\hsD p_1\wedge \hsD p_3)\vee (\hsD p_2\wedge \hsD p_3),
\end{equation*}
which states that there exist at least two sub-intervals such that $p_i$ holds over the former and $p_j$ over the latter, with $i,j\in\{1,2,3\}$ and $j\neq i$ (such a formula can be easily generalised to an arbitrary set of proposition letters and to any natural number $k$).

The truth of the following statements can be easily checked:
\begin{itemize}
    \item $\Ku_{Sched}\models\hsEu\left(\hsB^4\top\rightarrow wit_{\geq 2}(\{p_1,p_2,p_3\})\right)$;
    \item $\Ku_{Sched}\not\models\hsEu\left(\hsB^{10}\top\rightarrow \hsD p_3\right)$;
    \item $\Ku_{Sched}\not\models\hsEu\left(\hsB^6\top\rightarrow \hsD p_1\wedge \hsD p_2\wedge \hsD p_3\right)$.
\end{itemize}

The first formula states that in any suffix of an initial trace of length greater than or equal to 5 at least 2 proposition letters are witnessed. $\Ku_{Sched}$ satisfies the formula since a process cannot be executed twice in a row. 

The second formula states that in any suffix of an initial trace of length at least 11, process 3 is executed at least once in some internal states (\emph{non starvation}). $\Ku_{Sched}$ does not satisfy the formula since the scheduler can avoid executing a process ad libitum. 

The third formula states that in any suffix of an initial trace of length greater than or equal to 7, $p_1$, $p_2$, $p_3$ are all witnessed. The only way to satisfy this property is to constrain the scheduler to execute the three processes in a strictly periodic manner (\emph{strict alternation}), that is, $p_i p_j p_k p_i p_j p_k p_i p_j p_k\cdots$, for $i,j,k \in \{1,2,3\}$ and $i \neq  j \neq  k \neq i$, but $\Ku_{Sched}$ does not meet such a
requirement.

This example will be referred to also in the next chapters.
\end{example}

\section{The fundamental notion of $BE_k$-descriptor}\label{sec:descr}
In the previous section we have shown that, for any given finite Kripke structure $\Ku$, one can find a corresponding induced abstract interval model $\mathpzc{A}_\Ku$, featuring one interval for each trace of $\Ku$. Since $\Ku$ has loops (each state must have at least one successor, as the transition relation $\Edges$ is left-total), the number of its traces, and thus the number of intervals of $\mathpzc{A}_\Ku$, is infinite.

In this section we prove that, given a finite Kripke structure $\Ku$ and an $\HS$ formula $\varphi$, there exists a \emph{finite} representation for $\mathpzc{A}_\Ku$, equivalent to $\mathpzc{A}_\Ku$ with respect to the satisfiability of $\varphi$ (in fact, of a class of formulas including $\varphi$).

We start with the definition of some basic notions. The first one is the \mbox{BE-nesting} depth of an $\HS$ formula.

\begin{definition}[BE-nesting depth of an $\HS$ formula]\label{defnest}
Let $\psi$ be an $\HS$ formula. The BE-nesting depth of $\psi$, denoted by $\nestbe(\psi)$, is defined by induction on the structure of the formula as follows:
    \begin{itemize}
        \item $\nestbe(p)=0$ for any proposition letter $p\in\Prop$;
        \item $\nestbe(\neg\psi)=\nestbe(\psi)$;
        \item $\nestbe(\psi\wedge\varphi)=\max\{\nestbe(\psi),\nestbe(\varphi)\}$;
        \item $\nestbe(\hsB\psi)=\nestbe(\hsE\psi)=1+\nestbe(\psi)$;
        \item $\nestbe(\hsX\psi)=\nestbe(\psi)$, for $X\in\{A, \overline{A}, \overline{B},                           \overline{E}\}$.
    \end{itemize}
\end{definition}
In the following, we denote by $\nestb(\psi)$ the \lq\lq restriction\rq\rq\ of $\nestbe(\psi)$ to $\hsB$ modality only (i.e., $\nestb(\psi)$ accounts only for the nesting depth of $\hsB$ and disregards $\hsE$). Clearly $\nestb(\psi)=\nestbe(\psi)$ if $\psi$ is devoid of occurrences of $\hsE$. The analogous for $\neste(\psi)$.

Making use of the notion of BE-nesting depth of a formula, we can define a relation of $k$-equivalence over traces.
\begin{definition}[$k$-equivalence]\label{def:k-equivalence}
Let $\Ku$ be a finite Kripke structure and $\rho$ and $\rho'$ be two traces in $\Trk_\Ku$. We say that $\rho$ and $\rho'$ are $k$-equivalent if and only if, for every $\HS$ formula $\psi$ with $\nestbe(\psi)=k$, we have $\Ku,\rho\models \psi$ if and only if $\Ku,\rho'\models \psi$.
\end{definition}
It can be easily proved that $k$-equivalence \lq\lq propagates downwards\rq\rq .
\begin{proposition}
Let $\Ku$ be a finite Kripke structure and $\rho$ and $\rho'$ be two traces in $\Trk_\Ku$. If $\rho$ and $\rho'$ are $k$-equivalent, then they are $h$-equivalent, 
for all $0\leq h\leq k$.
\end{proposition}
 \begin{proof}
 Let us assume that $\Ku,\rho\models \psi$, with $0\leq \nestbe(\psi)\leq k$. Consider the formula $\hsB^k\top$, whose BE-nesting depth is equal to $k$. It trivially holds that either $\Ku,\rho\models \hsB^k\top$ or $\Ku,\rho\models \neg\hsB^k\top$. In the first case, we have that $\Ku,\rho\models \hsB^k\top \wedge \psi$. Since $\nestbe\left(\hsB^k\top \wedge \psi\right)=k$, from the hypothesis, it follows that $\Ku,\rho'\models \hsB^k\top \wedge \psi$, and thus $\Ku,\rho'\models \psi$. The other case is symmetric.
 \end{proof}

We are now ready to introduce the notion of \emph{descriptor}, which will play a fundamental role in the definition of finite abstract interval models.

\begin{definition}[$B$-descriptor and $E$-descriptor] \label{def:descr}
Let $\Ku=\KuDef$ be a finite Kripke structure. 
A $B$-descriptor (resp., $E$-descriptor) is a labelled tree $\mathpzc{D}=(\DV,\DE,\lambda)$, where $\DV$ is a finite set of vertices, $\DE\subseteq \DV\times \DV$ is a set of edges, and $\lambda:\DV\to \States\times 2^\States\times \States$ is a node labelling function, that satisfies the following conditions:
    \begin{enumerate}
        \item for all $(v,v')\!\in\! \DE$, with $\lambda(v)\!=\!(s_{in},A,s_{fin})$ and $\lambda(v')\! =\! (s_{in}',A', s_{fin}')$, it holds that $A'\subseteq A$, $s_{in}=s_{in}'$, and $s_{fin}'\in A$
            (resp., $A'\subseteq A$, $s_{fin}=s_{fin}'$, and $s_{in}'\in A$);
        \item for all pairs of edges $(v,v'), (v,v'')\in \DE$, 
        if the subtree rooted in $v'$ is isomorphic to the subtree rooted in $v''$, then $v'=v''$
        (here and in the following, we write subtree for maximal subtree).
    \end{enumerate}
\end{definition}
(2.) of Definition \ref{def:descr} simply states that no two subtrees, whose roots are siblings, 
can be isomorphic (note that $\lambda$ is taken into account).

For $X\in\{B,E\}$, the \emph{depth} of an $X$-descriptor $(\DV,\DE,\lambda)$ is the depth of the tree $(\DV,\DE)$. We call an $X$-descriptor of depth $k\in\Nat$ an $X_k$-descriptor. An $X_0$-descriptor $\mathpzc{D}$ consists of its root only, which is denoted by $\Root(\mathpzc{D})$. 
A label of a node will be referred to as a \emph{descriptor element}.
%
Hereafter, two descriptors will be considered \emph{equal up to isomorphism}. 

The following proposition holds.
\begin{proposition}\label{finitedescr}
Given a finite Kripke structure $\Ku=\KuDef$, for all $k\in\Nat$ there exists a \emph{finite number} of possible $B_k$-descriptors (resp., $E_k$-descriptors).
\end{proposition}
\begin{proof}
We consider the case of $B_k$-descriptors (the case of $E_k$-descriptors is analogous).
For $k=0$, there are at most $|S|\cdot 2^{|S|} \cdot |S|$ pairwise distinct $B_0$-descriptors. 
As for the inductive step, let us assume $h$ to be the number of pairwise distinct $B$-descriptors of depth at most $k$. The number of $B_{k+1}$-descriptors is at most $|S| \cdot 2^{|S|}\cdot |S| \cdot 2^h$ (there are at most $|S|\cdot 2^{|S|}\cdot |S|$ possible choices for the root, which can have any subset of the $h$ $B$-descriptors of depth at most $k$ as subtrees). By K\"onig's lemma, they are all finite as their depth is $k+1$ and the root has a finite number of children (no two subtrees of the root can be isomorphic).
\end{proof}

Proposition~\ref{finitedescr} provides an upper bound to the number of distinct $B_k$-descriptors (resp., $E_k$-descriptors), and thus to the number of nodes of each $B_{k+1}$-descriptor (resp., $E_{k+1}$-descriptors), for $k\in\Nat$, which is \emph{not elementary} with respect to $|S|$ and $k$, being $|S|$ the exponent and $k$ the height of the exponential tower. As a matter of fact, this is a very rough upper bound, since some descriptors may not have depth $k+1$ and some of the ``generated'' trees might not even fulfill the definition of descriptor.

We show now how $B$-descriptors and $E$-descriptors can be used to extract relevant information from the traces of a finite Kripke structure to use in MC. 

Let $\Ku$ be a finite Kripke structure and $\rho$ be a trace in $\Trk_\Ku$, with $|\rho|\geq 2$.
For any $k \geq 0$, the label of the root of both the $B_k$-descriptor and $E_k$-descriptor for
$\rho$ is the triple $(\fst(\rho),\intstates(\rho),\lst(\rho))$. The root of the $B_k$-descriptor has a child for each prefix $\rho'$, with $|\rho'|\geq 2$, of $\rho$, labelled with $(\fst(\rho'),\intstates(\rho'),\lst(\rho'))$.
Such a construction is then iteratively applied to the children of the root until either depth $k$ is reached or a trace of length 2 is being considered on a node. 
The length-1 prefix of $\rho$ can be recovered as a special case, being just $\fst(\rho)$. We should associate with it a descriptor element $(\fst(\rho),\emptyset,\bot)$ as a child of the root, where $\bot$ is just a marker for \lq\lq no state\rq\rq. However, in the rest of this section, for a more uniform and elegant characterization of descriptors, we assume \emph{traces, suffixes and prefixes to have length at least 2}, as those having length 1 can always be treated as special (trivial) cases.
The $E_k$-descriptor is built in a similar way by considering the suffixes of $\rho$.

In general $B$- and $E$-descriptors do not convey enough information to determine which trace they were built from (this will be clear shortly). However, they can be exploited to determine which $\HS$ formulas are satisfied by the trace from which they have been built:
    $(i)$~to check satisfiability of proposition letters, they keep information about initial, final, and internal states of the trace;
    $(ii)$~for $\hsA\psi$ and $\hsAt\psi$ formulas, they store the final and initial states of the trace;
    $(iii)$~for $\hsB\psi$ formulas, the $B$-descriptor keeps information about all the prefixes of the trace; 
    $(iv)$~for $\hsE\psi$ formulas, the $E$-descriptor keeps information about all the suffixes of the trace;
    $(v)$~no additional information is needed for $\hsBt\psi$ and $\hsEt\psi$ formulas.

Let $\Ku$ be a finite Kripke structure. The $B_k$-descriptor (resp., $E_k$-descriptor) for a trace $\rho$ in $\Trk_\Ku$ is formally defined as follows.
\begin{definition}[B-/E-descriptor for a trace] \label{def:tracedescr}
Let $\Ku$ be a finite Kripke structure, $\rho$ be a trace in $\Trk_\Ku$, and $k\in\Nat$. The $B_k$-descriptor (resp., $E_k$-descriptor) for $\rho$ is inductively defined as follows:
    \begin{itemize}
        \item for $k=0$, the $B_k$-descriptor (resp., $E_k$-descriptor) for $\rho$ is the tree $\mathpzc{D} = (\Root(\mathpzc{D}),\emptyset,$ $\lambda)$, where 
        \begin{equation*}
            \lambda(\Root(\mathpzc{D}))=(\fst(\rho),\intstates(\rho), \lst(\rho));
        \end{equation*}                
        
        \item for $k>0$, the $B_k$-descriptor (resp., $E_k$-descriptor) for $\rho$ is the tree $\mathpzc{D} = (\DV,\DE,\lambda)$, where 
        \begin{equation*}
            \lambda(\Root(\mathpzc{D}))=(\fst(\rho),\intstates(\rho),\lst(\rho)),
        \end{equation*}                
        which satisfies the following conditions:
            \begin{enumerate}
                \item for each prefix (resp., suffix) $\rho'$ of $\rho$, there exists $v\in \DV$ such that $(\Root(\mathpzc{D}),v)\in \DE$ and the subtree rooted in $v$ is the $B_{k-1}$-descriptor (resp., $E_{k-1}$-descriptor) for $\rho'$;
                \item for each vertex $v\in \DV$ such that $(\Root(\mathpzc{D}),v)\in \DE$, there exists a prefix (resp., suffix) $\rho'$ of $\rho$ such that the subtree rooted in $v$ is the $B_{k-1}$-descriptor (resp., $E_{k-1}$-descriptor) for $\rho'$;
            \item for all pairs of edges $(\Root(\mathpzc{D}),v'), (\Root(\mathpzc{D}),v'')\in \DE$, if the subtree rooted in $v'$ is isomorphic to the subtree rooted in $v''$, then $v'=v''$.
            \end{enumerate}
    \end{itemize}
\end{definition}
Any $B_k$-descriptor (resp., $E_k$-descriptor) for some trace satisfies the 
conditions of Definition~\ref{def:descr} (in particular (1.)), but not vice 
versa. See the next example.

\enlargethispage{2.5\baselineskip}

\begin{example}
Consider, for instance, the $B_1$-descriptor reported in Figure~\ref{fig:wrongdesc}.
It is built on a set of states $\States$ including at least states $s_0$, $s_1$, $s_2$ and $s_3$, and it
satisfies both conditions of Definition~\ref{def:descr}. However, no trace of a finite Kripke structure can be described by it, as no trace may feature two prefixes to be associated with the first two children of the root.

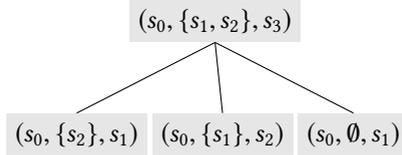
\begin{figure}[H]
\centering
\begin{tikzpicture}[level distance=15mm,every node/.style={fill=gray!20}]
\Tree [.$(s_0,\{s_1,s_2\},s_3)$
    $(s_0,\{s_2\},s_1)$
    $(s_0,\{s_1\},s_2)$
    $(s_0,\emptyset,s_1)$
] 
\end{tikzpicture}
\caption{$B_1$-descriptor devoid of a corresponding trace (in any Kripke structure).}\label{fig:wrongdesc}
\end{figure}
\end{example}

\begin{example}
In Figure~\ref{Keqtr1} and \ref{Keqtr2}, we depict the $B_2$- and $E_2$-descriptors for the trace $s_0s_1s_0s_0s_1$ of the Kripke structure $\Ku_2$ of Figure~\ref{K2}.

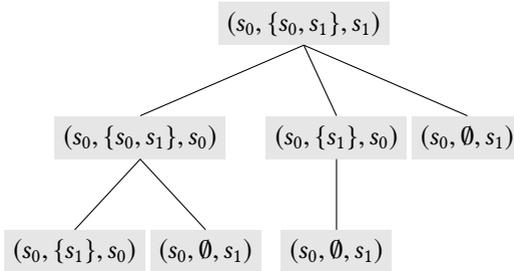
\begin{figure}[H]
\centering
\begin{tikzpicture}[level distance=15mm,every node/.style={fill=gray!20}]
\Tree [.$(s_0,\{s_0,s_1\},s_1)$
	[.$(s_0,\{s_0,s_1\},s_0)$
		$(s_0,\{s_1\},s_0)$
		$(s_0,\emptyset,s_1)$
]	[.$(s_0,\{s_1\},s_0)$
		$(s_0,\emptyset,s_1)$
]	$(s_0,\emptyset,s_1)$
]
\end{tikzpicture}
\caption{$B_2$-descriptor 
for the trace  $s_0s_1s_0s_0s_1$ of $\Ku_2$.}\label{Keqtr1}
\end{figure}

\begin{figure}[H]
\centering
\begin{tikzpicture}[level distance=15mm,every node/.style={fill=gray!20}]
\Tree [.$(s_0,\{s_0,s_1\},s_1)$
	[.$(s_1,\{s_0\},s_1)$
		$(s_0,\{s_0\},s_1)$
		$(s_0,\emptyset,s_1)$
]	[.$(s_0,\{s_0\},s_1)$
		$(s_0,\emptyset,s_1)$
]	$(s_0,\emptyset,s_1)$
]
\end{tikzpicture}
\caption{$E_2$-descriptor 
for the trace $s_0s_1s_0s_0s_1$ of $\Ku_2$.}\label{Keqtr2}
\end{figure}
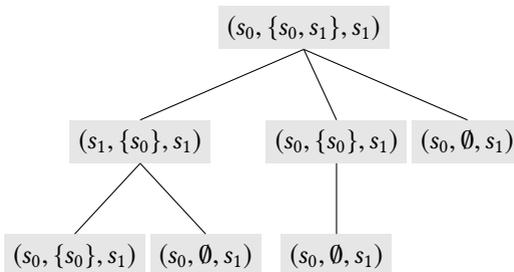
\end{example}

\begin{example}
In Figure~\ref{removeisom} we show the $B_2$-descriptor for the trace $\rho = s_0s_1s_0s_0s_0s_0s_1$ of $\Ku_{2}$. It is worth noticing that there exist two distinct prefixes of the trace $\rho$, that is, the traces $\rho'=s_0s_1s_0s_0s_0s_0$ and $\rho''=s_0s_1s_0s_0s_0$, which have the same $B_1$-descriptor. Since, according to Definition~\ref{def:tracedescr}, no tree can occur more than once as a subtree of the same node (in this example, the root), in the $B_2$-descriptor for $\rho$, the prefixes $\rho'$ and $\rho''$ are represented by the same tree (the first subtree of the root on the left). This shows that, in general, the root of a descriptor for a trace with $h$ proper prefixes may have less than $h$ children.

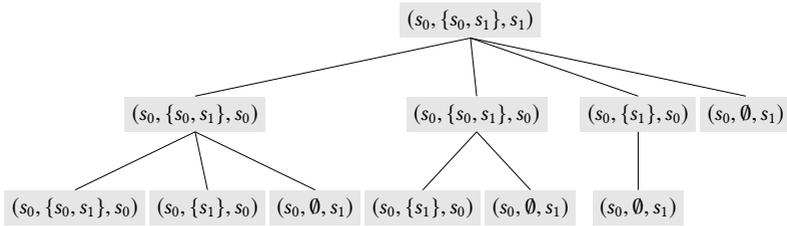
\begin{figure}[H] 
\centering
\resizebox{\textwidth}{!}{
\begin{tikzpicture}[level distance=15mm,every node/.style={fill=gray!20}]
\Tree [.$(s_0,\{s_0,s_1\},s_1)$
	[.$(s_0,\{s_0,s_1\},s_0)$
		$(s_0,\{s_0,s_1\},s_0)$
		$(s_0,\{s_1\},s_0)$
		$(s_0,\emptyset,s_1)$
]	[.$(s_0,\{s_0,s_1\},s_0)$
		$(s_0,\{s_1\},s_0)$
		$(s_0,\emptyset,s_1)$
]	[.$(s_0,\{s_1\},s_0)$
		$(s_0,\emptyset,s_1)$
]	$(s_0,\emptyset,s_1)$
]
\end{tikzpicture}}
\caption{The $B_2$-descriptor for the trace $s_0s_1s_0s_0s_0s_0s_1$ of $\Ku_{2}$.}\label{removeisom}
\end{figure}
\end{example}

\begin{example}
This example shows that not all of the $B_k$-descriptors that can be generated from the set of states of a given finite Kripke structure are $B_k$-descriptors for some trace of that structure (the same fact is true for $E_k$-descriptors).

Let us consider the finite Kripke structure $\Ku$ in Figure~\ref{akripke} and the $B_1$-descriptor $\mathpzc{D}_{B_1}$
in Figure~\ref{akripketr}. 
By inspecting $\mathpzc{D}_{B_1}$, it can be easily checked that it can be the $B_1$-descriptor for traces of the form $s_0s_1^hs_3^2$, with $h \geq 2$, only. However, no trace of this form can be obtained by unravelling $\Ku$.

\begin{figure}[H]
\centering
\begin{tikzpicture}[->,>=stealth',shorten >=1pt,auto,node distance=1.5cm,thick,main node/.style={circle,draw}]
    \node (5) {};    
    \node[main node] (1) [above of=5] {$s_1$};
    \node[main node,style={double}] (0) [left of=5] {$s_0$};
    \node[main node] (2) [below of=5] {$s_2$};
    \node[main node] (3) [right of=5] {$s_3$};

  \path[every node/.style={font=\small}]
    (0) edge (1)
        edge (2)
    (1) edge [bend left] (2)
        edge (3)
    (2) edge [bend left] (1)
        edge (3)
    (3) edge [loop right] (3)
    ;
\end{tikzpicture}
\caption{A finite Kripke structure $\Ku$.}\label{akripke}
\end{figure}

\begin{figure}[H]
\centering
\begin{tikzpicture}[level distance=15mm,every node/.style={fill=gray!20}]
\Tree [.$(s_0,\{s_1,s_3\},s_3)$
    $(s_0,\{s_1\},s_3)$
    $(s_0,\{s_1\},s_1)$
    $(s_0,\emptyset,s_1)$
] 
\end{tikzpicture}
\caption{$\mathpzc{D}_{B_1}$: a $B_1$-descriptor not corresponding to any of the traces of $\Ku$ in Figure \ref{akripke}.}\label{akripketr}
\end{figure}
\end{example}

To check an $\HS$ formula against a given finite Kripke structure we actually need to account for both the \emph{started-by} ($B$) and \emph{finished-by} ($E$) relations at the same time. To this end, we introduce $BE_k$-descriptors for traces. Given a finite Kripke structure $\Ku$ and a trace $\rho$ in $\Trk_\Ku$,
the $BE_k$-descriptor for $\rho$ can be obtained from a suitable merging of its $B_k$-descriptor and $E_k$-descriptor. It can be viewed as a sort of ``product'' of the $B_k$-descriptor and the $E_k$-descriptor for $\rho$, and it is formally defined as follows:

\begin{definition}[BE-descriptor for a trace]\label{def:BEdescr}
Let $\Ku=\KuDef$ be a finite Kripke structure, $\rho$ be a trace in $\Trk_\Ku$,
and $k \in \Nat$.
The $BE_k$-descriptor for $\rho$ is a labelled tree $\mathpzc{D}=(\DV,\DE,\lambda)$, where $\DV$ is 
a finite set of vertices, $\DE=\DE_B\cup \DE_E$, with $\DE_B\subseteq \DV\times \DV$ the set of ``$B$-edges'',
$\DE_E\subseteq \DV\times \DV$ the set of ``$E$-edges'', and $\DE_B\cap \DE_E=\emptyset$, and $\lambda:\DV\to \States\times 2^\States\times \States$, which is inductively defined on $k\in\Nat$ as follows:
    \begin{itemize}
        \item for $k=0$, the $BE_k$-descriptor for $\rho$ is $\mathpzc{D}=(\Root(\mathpzc{D}),\emptyset,\lambda)$, where 
        \begin{equation*}
            \lambda(\Root(\mathpzc{D}))=(\fst(\rho),\intstates(\rho),\lst(\rho)).
        \end{equation*}        
        
        \item for $k>0$, the $BE_k$-descriptor for $\rho$ is $\mathpzc{D}=(\DV,\DE,\lambda)$ with 
        \begin{equation*}
            \lambda(\Root(\mathpzc{D}))=(\fst(\rho),\intstates(\rho),\lst(\rho))
        \end{equation*}                
         which satisfies the following conditions:
            \begin{enumerate}
                \item[1a.] for each prefix $\rho'$ of $\rho$, there exists $v\in \DV$ such that $(\Root(\mathpzc{D}),v)\in \DE_B$ and the subtree rooted in $v$ is the $BE_{k-1}$-descriptor for $\rho'$;
                \item[1b.] for each vertex $v\in \DV$ such that $(\Root(\mathpzc{D}),v)\in \DE_B$, there exists a prefix $\rho'$ of $\rho$ such that the subtree rooted in $v$ is the $BE_{k-1}$-descriptor for $\rho'$;
                \item[1c.] for all pairs of edges $(\Root(\mathpzc{D}),v'), (\Root(\mathpzc{D}),v'') \in \DE_B$, if the subtree rooted in $v'$ is isomorphic to the  subtree rooted in $v''$, then $v'=v''$;
                \item[2a.] for each suffix $\rho''$ of $\rho$, there exists $v\in \DV$ such that $(\Root(\mathpzc{D}),v)\in \DE_E$ and the subtree rooted in $v$ is the $BE_{k-1}$-descriptor for $\rho''$;
                \item[2b.] for each vertex $v\in \DV$ such that $(\Root(\mathpzc{D}),v)\in \DE_E$, there exists a suffix $\rho''$ of $\rho$ such that the subtree rooted in $v$ is the $BE_{k-1}$-descriptor for $\rho''$;
                \item[2c.] for all pairs of edges $(\Root(\mathpzc{D}),v'), (\Root(\mathpzc{D}),v'')\in \DE_E$, if the subtree rooted in $v'$ is isomorphic to the subtree rooted in $v''$, then $v'=v''$.
            \end{enumerate}
    \end{itemize}
\end{definition}
From Definition~\ref{def:BEdescr}, it easily follows that for all $(v,v')\in \DE_B$, with $\lambda(v)=(s_{in},A,s_{fin})$ and $\lambda(v')=(s_{in}',A',s_{fin}')$, we have $A'\subseteq A$, $s_{in}=s_{in}'$, and $s_{fin}'\in A$, and  for all $(v,v')\in \DE_E$, with $\lambda(v)=(s_{in},A,s_{fin})$ and $\lambda(v')=(s_{in}',A',s_{fin}')$, we have $A'\subseteq A$, $s_{fin}=s_{fin}'$ and $s_{in}'\in A$.

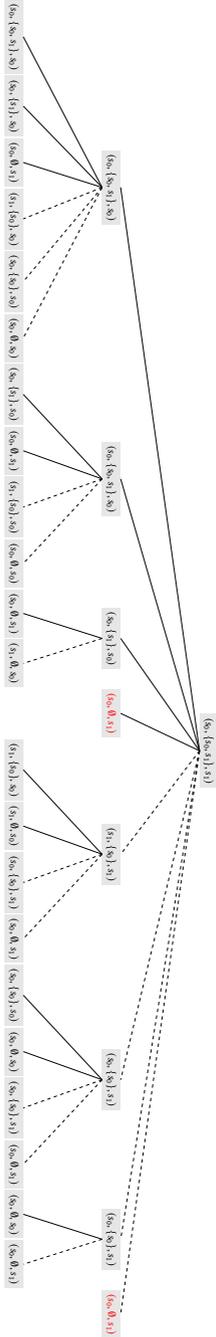
\begin{sidewaysfigure}
\centering
\resizebox{\textheight}{!}{
\begin{tikzpicture}[level distance=30mm,every node/.style={fill=gray!20}]
\Tree [.$(s_0,\{s_0,s_1\},s_1)$
	[.$(s_0,\{s_0,s_1\},s_0)$
		$(s_0,\{s_0,s_1\},s_0)$
		$(s_0,\{s_1\},s_0)$
		$(s_0,\emptyset,s_1)$
		\edge[dashed]; $(s_1,\{s_0\},s_0)$
		\edge[dashed]; $(s_0,\{s_0\},s_0)$
		\edge[dashed]; $(s_0,\emptyset,s_0)$
]	[.$(s_0,\{s_0,s_1\},s_0)$
		$(s_0,\{s_1\},s_0)$
		$(s_0,\emptyset,s_1)$
		\edge[dashed]; $(s_1,\{s_0\},s_0)$
		\edge[dashed]; $(s_0,\emptyset,s_0)$
]	[.$(s_0,\{s_1\},s_0)$
		$(s_0,\emptyset,s_1)$
		\edge[dashed]; $(s_1,\emptyset,s_0)$
]	\textcolor{red}{$(s_0,\emptyset,s_1)$}
    \edge[dashed]; [.$(s_1,\{s_0\},s_1)$
		$(s_1,\{s_0\},s_0)$
		$(s_1,\emptyset,s_0)$
		\edge[dashed]; $(s_0,\{s_0\},s_1)$
		\edge[dashed]; $(s_0,\emptyset,s_1)$
]	\edge[dashed]; [.$(s_0,\{s_0\},s_1)$
        $(s_0,\{s_0\},s_0)$
		$(s_0,\emptyset,s_0)$
		\edge[dashed]; $(s_0,\{s_0\},s_1)$
		\edge[dashed]; $(s_0,\emptyset,s_1)$
]	\edge[dashed]; [.$(s_0,\{s_0\},s_1)$
        $(s_0,\emptyset,s_0)$
		\edge[dashed]; $(s_0,\emptyset,s_1)$
]	\edge[dashed]; \textcolor{red}{$(s_0,\emptyset,s_1)$}
]
\end{tikzpicture}}
\caption{An example of $BE_2$-descriptor.}\label{treqdescr}
\end{sidewaysfigure}

\begin{example}
In Figure~\ref{treqdescr} at page~\pageref{treqdescr}, with reference to the finite Kripke structure $\Ku_{2}$ of Figure~\ref{K2}, we give an example of a $BE_2$-descriptor. $B$-edges are represented by solid lines, while $E$-edges are represented by dashed lines. It is worth pointing out that the $BE_2$-descriptor of Figure~\ref{treqdescr} turns out to be the $BE_2$-descriptor for both the trace $\rho=s_0s_1s_0^3s_1$ and the trace $\rho'=s_0s_1s_0^4s_1$ (and many others). As we will see very soon, this is not an exception, but the rule: different traces of a finite Kripke structure are described by the same $BE$-descriptor.
Notice also that it features two isomorphic subtrees for the same node (the root). They both consist of a single node, labelled with $(s_0,\emptyset,s_1)$. However, this does not violate Definition~\ref{def:BEdescr} since one of them is connected to the parent via a $B$-edge and the other via an $E$-edge.
\end{example}

It can be easily checked that the $BE_{k-1}$-descriptor $\mathpzc{D}_{BE_{k-1}}$ for a trace $\rho$ can be obtained from the 
$BE_k$-descriptor $\mathpzc{D}_{BE_k}$ for $\rho$ by removing the nodes at depth $k$ (if any) and the isomorphic subtrees possibly resulting from such a removal (see (1c.) of Definition~\ref{def:BEdescr}). 

$B_k$ and $E_k$-descriptors can be recovered from $BE_k$ ones. The $B_k$-descriptor $\mathpzc{D}_{B_k}$ for a trace $\rho$ can be obtained from the $BE_k$-descriptor $\mathpzc{D}_{BE_k}$ for $\rho$ by pruning it in such a way that only those vertices of $\mathpzc{D}_{BE_k}$ which are connected to the root via paths consisting of $B$-edges only are maintained (the set of edges of $\mathpzc{D}_{B_k}$ and its labelling function can be obtained by restricting those of $\mathpzc{D}_{BE_k}$ to the nodes of $\mathpzc{D}_{B_k}$). The $E_k$-descriptor $\mathpzc{D}_{E_k}$ of $\rho$ can be obtained in a similar way.

We focus now our attention on the relationships between the traces obtained from the unravelling of a finite Kripke structure and their $BE_k$-descriptors. A key observation is that, even though the number of traces of a finite Kripke structure $\Ku$ is infinite, for any $k\in\Nat$ the set of $BE_k$-descriptors for its traces is finite. This is an immediate consequence of Definition~\ref{def:BEdescr} and Proposition~\ref{finitedescr}. Thus, \emph{at least one $BE_k$-descriptor must be the $BE_k$-descriptor for infinitely many traces}. 

$BE_k$-descriptors naturally induce an \emph{equivalence relation of finite index} over the set of traces of a finite Kripke structure, called $k$-descriptor equivalence relation.

\begin{definition}[$k$-descriptor equivalence]
Let $\Ku$ be a finite Kripke structure, $\rho,\rho'$ be two traces in $\Trk_\Ku$,
and $k \in \Nat$. We say that $\rho$ and $\rho'$ are \emph{$k$-descriptor equivalent}, denoted by $\rho\sim_k\rho'$, if and only if the $BE_k$-descriptors for $\rho$ and $\rho'$ coincide.
\end{definition}

In the next section we will see that, for any given pair of traces $\rho,\rho'\in\Trk_\Ku$, 
if $\rho\sim_k\rho'$, then $\rho$ and $\rho'$ are $k$-equivalent (see Definition \ref{def:k-equivalence}).

\section{The decidability proof}\label{sec:decidProof}

In this section we 
will see the main results behind the decidability of the MC problem for $\HS$ formulas over finite Kripke structures.
We refer to~\cite{MMMPP15} for further details and missing proofs.

As a preliminary step, we state a right extension property. Let $\Ku$ be a finite Kripke structure, $k\in\Nat$, and $\rho$ and $\rho'$ be two traces in $\Trk_\Ku$ with the same $BE_k$-descriptor (and thus, in particular, $\lst(\rho)=\lst(\rho')$). The property states that if we extend $\rho$ and $\rho'$ ``to the right'' with the same trace $\overline{\rho}$ in $\Trk_\Ku$, with $\left(\lst(\rho),\fst(\overline{\rho})\right)\in\Edges$, then the resulting traces $\rho\cdot \overline{\rho}$ and $\rho'\cdot\overline{\rho}$ (both belonging to $\Trk_\Ku$) have the same $BE_k$-descriptor as well. 
An analogous property holds for the extension of the two traces $\rho$ and $\rho'$ ``to the left'', which guarantees that $\overline{\rho}\cdot \rho$ and $\overline{\rho}\cdot \rho'$ have the same $BE_k$-descriptor (left extension property).

\begin{proposition}[Right extension property]\label{extLemma}
Let $\Ku=\KuDef$ be a finite Kripke structure and let $\rho$ and $\rho'$ be two traces in $\Trk_\Ku$ with $\rho\sim_k\rho'$. For any trace $\overline{\rho}$ in $\Trk_\Ku$, with $\left(\lst(\rho),\fst(\overline{\rho})\right)\in\Edges$, the two traces $\rho\cdot \overline{\rho}$ and $\rho'\cdot\overline{\rho}$ belong to $\Trk_\Ku$ and $\rho\cdot\overline{\rho}\sim_k\rho'\cdot\overline{\rho}$.
\end{proposition}

The next theorem proves that, for any pair of traces $\rho,\rho'\in\Trk_\Ku$, if $\rho\sim_k\rho'$, then $\rho$ and $\rho'$ are $k$-equivalent (see Definition \ref{def:k-equivalence}). 
\begin{theorem}\label{satPresB}
Let $\Ku$ be a finite Kripke structure, $\rho$ and $\rho'$ be two traces in $\Trk_\Ku$, and $\psi$ be a $\HS$ formula with $\nestbe(\psi)=k$. If $\rho\sim_k\rho'$ then 
$\Ku,\rho\models\psi\iff \Ku,\rho'\models\psi$.
\end{theorem}

Since the set of $BE_k$-descriptors for the traces of a finite Kripke structure $\Ku$ is finite, i.e., the equivalence relation $\sim_k$ has a finite index, there always exists a finite number of $BE_k$-descriptors that ``satisfy'' a $\HS$ formula $\psi$ with $\nestbe(\psi)= k$ (this can be formally proved by a quotient construction \cite{MMMPP15}). 

Thus we can reduce the MC problem for $\HS$ over finite Kripke structures to MC for multi-modal finite Kripke structures, whose nodes are all possible \emph{witnessed} descriptors with depth up to $k$, and there is a distinct accessibility relation for each one of the $\HS$ modalities $A$, $B$, $E$, $\overline{A}$, $\overline{B}$ and $\overline{E}$.
Since the MC problem for multi-modal finite Kripke structures is decidable (in polynomial time with respect to the size of the multi-modal Kripke structure and to the length of the formula~\cite{Gabbay87,Lan06}),
decidability of the MC problem for $\HS$ against finite Kripke structures follows, provided that we can effectively determine which are the descriptors witnessed (by traces) in the structure, as shown in the proof of the next theorem.
\begin{theorem} 
The MC problem for $\HS$ formulas over finite Kripke structures is decidable (with \emph{nonelementary} complexity).
\end{theorem}
\begin{proof}
Let $\Ku$ be a finite Kripke structure and let $\varphi$ be the $\HS$ formula to check, with $\nestbe(\varphi) = k$. 

We first prove that, in order to select the $BE_h$-descriptors, with $0\leq h\leq k$, witnessed by some trace in $\Ku$,
we can restrict ourselves to traces devoid of prefixes associated with the same $BE_k$-descriptor.
Let $\rho \in \Trk_\Ku$ and let $\rho', \rho''$ be two prefixes of $\rho$, with 
$|\rho''|<|\rho'| \leq |\rho|$ (notice that we allow $\rho'$ to coincide with $\rho$).
Moreover, let $\rho = \rho' \cdot \tilde{\rho}$, for some $\tilde{\rho}$ with 
$|\tilde{\rho}| \geq 1$ (in case $|\rho| = |\rho'|$, $\rho = \rho'$).
If the $BE_k$-descriptors for $\rho'$ and $\rho''$ are the same then, by Proposition \ref{extLemma},
it holds that the $BE_k$-descriptor for $\rho'' \cdot \tilde{\rho}$ is equal to the one for 
$\rho' \cdot \tilde{\rho} = \rho$. Hence, we can safely replace $\rho$ by the $k$-descriptor equivalent 
shorter trace $\rho'' \cdot \tilde{\rho}$. 
We can iterate such a contraction process until there are no more pairs of prefixes associated with the same 
$BE_k$-descriptor.\footnote{As a matter of fact, the same argument can be given by referring to
suffixes instead of prefixes. Anyway, as one can easily see, making use of both the right extension 
and the left extension properties does not allow us to improve the claimed bound.}

Now, Proposition~\ref{finitedescr} provides a nonelementary upper bound to the number $\alpha$ of distinct 
$BE_h$-descriptors, with $0\leq h\leq k$ (as well as to their size), with 
respect to the size of $\Ku$ and $k$. 
A bound on the length of the traces in $\Trk_\Ku$ that we need to consider in order
to determine the witnessed $BE_h$-descriptors
in an effective way immediately follows: it is $1+\alpha$. 

Hence, in order to generate all the witnessed $BE_h$-descriptors, with $0\leq h\leq k$, it suffices
to list, for all states $s$ of $\Ku$, all the traces starting from $s$, ordered by length,
until the above bound is reached, and then to build the corresponding $BE_h$-descriptors, with 
$0\leq h\leq k$.

We conclude that the derived MC problem for multi-modal finite Kripke 
structures has to be solved over a model whose size has a nonelementary upper bound.
\end{proof}


\input{Chaps/Intro/LowerBoundBE.tex}

\input{Chaps/Intro/generalPict.tex}

%% file: Chaps/Intro/LowerBoundBE.tex
\newcommand{\Instance}{\mathcal{I}}
\newcommand{\Beg}{\textit{beg}}
\newcommand{\End}{\textit{end}}
\newcommand{\Left}{\textit{left}}
\newcommand{\Right}{\textit{right}}
\newcommand{\Down}{\textit{down}}
\newcommand{\Up}{\textit{up}}
\newcommand{\Init}{\textit{init}}
\newcommand{\Final}{\textit{final}}

\section{\EXPSPACE-hardness of MC for $\BE$}\label{sec:BEhard}

In the previous section we have seen that MC for $\HS$ can be decided in nonelementary time. 
Obviously, this does \emph{not} imply that such problem is \emph{provably} nonelementary, meaning that no algorithm with elementary complexity may exist for it. Currently, \emph{the existence of such an algorithm is still an open issue}.

Conversely, the best complexity lower bound known to date is $\EXPSPACE$-hardness. This derives from the $\HS$ fragment $\BE$, whose modalities can express properties of both interval prefixes and suffixes
simultaneously: 
we now prove that the MC problem for $\BE$
is $\EXPSPACE$-hard; since MC for full $\HS$ is clearly at least as hard as MC for $\BE$, such a lower bound immediately propagates to full $\HS$. 
The result is obtained by a polynomial-time reduction from a \emph{domino-tiling problem for grids with rows of single exponential length}~\cite{harel92} to MC for $\BE$. 

We start with the definition of the domino-tiling problem.
An instance $\Instance$ of a domino-tiling problem for grids with rows of single exponential length is a tuple $\Instance =\tupleof{C,\Delta,n,d_\Init,d_\Final}$, where $C$ is a finite set of colors, $\Delta \subseteq C^{4}$ is a set of tuples $\tupleof{c_\Down,c_\Left,c_\Up,c_\Right}$ of four colors, called \emph{domino-types}, $n>0$ is a  natural number encoded in \emph{unary},
and $d_\Init,d_\Final\in\Delta$ are two distinguished domino-types (respectively, the initial and final domino-types).
The \emph{size} of $\Instance$ is defined as $|C|+|\Delta|+n$.

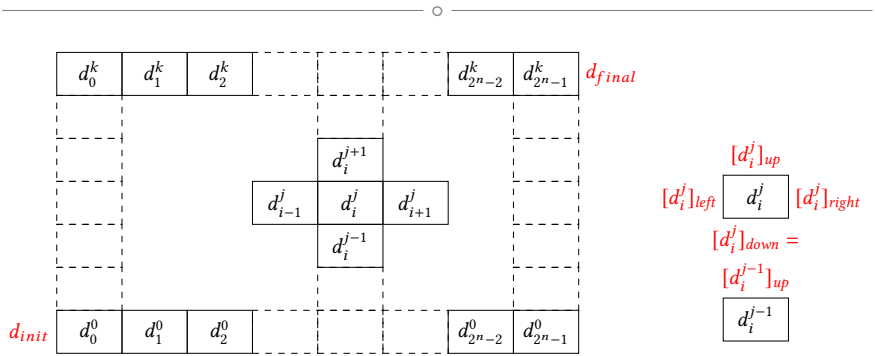
\begin{figure}[b]
    \centering
    \resizebox{\linewidth}{!}{\input{Chaps/Intro/tilingGraph.tex}}
    \caption{A (generic) instance of the domino-tiling problem, where $d^i_j$ denotes $f(i,j)$.}\label{fig:til}
\end{figure}

Intuitively, a tiling of a grid is a color labelling of the edges of each cell (see Figure~\ref{fig:til}).
Formally, a \emph{tiling of $\Instance$}  is a mapping $f:[0,k]\times [0,2^{n}-1] \rightarrow \Delta$, for some $k\geq 0$, that satisfies the following constraints:
\begin{itemize}
  \item two adjacent cells in a row have the same color on the shared edge, namely, for all $(i,j)\in [0,k]\times [0,2^{n}-2]$,
   $[f(i,j)]_{\Right}=[f(i,j+1)]_{\Left}$ (\emph{horizontal requirement});
  \item two adjacent cells in a column have the same color on the shared edge, namely, for all $(i,j)\in [0,k-1]\times [0,2^{n}-1]$,
   $[f(i,j)]_{\Up}=[f(i+1,j)]_{\Down}$ (\emph{vertical requirement});
  \item $f(0,0)=d_\Init$ (\emph{initialization requirement}) and $f(k,2^{n}-1)=d_\Final$ (\emph{acceptance requirement}).
\end{itemize}

Checking the existence (respectively, non-existence) of a tiling of $\Instance$ is an $\EXPSPACE$-complete problem~\cite{harel92}.

%

We now show how the domino-tiling problem can be reduced in polynomial time to the MC problem for $\BE$.
In particular, we show how to build in polynomial time a finite Kripke structure $\Ku_\Instance$ and a $\BE$ formula $\varphi_\Instance$ such that there exists an initial trace of $\Ku_\Instance$ satisfying $\varphi_\Instance$ if and only if there exists a tiling of $\Instance$. Hence, $\Ku_\Instance\models\neg\varphi_\Instance$ if and only if there is no tiling of $\Instance$.

\begin{figure}[tb]
    \centering
    \resizebox{\linewidth}{!}{\input{Chaps/Intro/TilingEncoding.tex}}
    \vspace{-0.4cm}
    \caption{Encoding of a tiling as a word, where $d^i_j$ denotes $f(i,j)$.}\label{fig:row}
\end{figure}
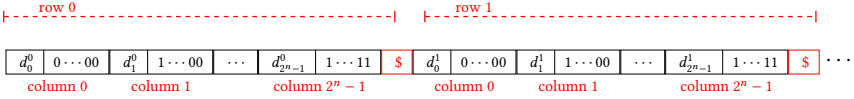

The encoding of tilings exploits the set of proposition letters $\Prop = \Delta \cup \{\$,0,1\}$.
%
Proposition letters in $\{0,1\}$  are used for the binary encoding of the value of an $n$-bit counter numbering the cells of a row of a  tiling, while the proposition letter $\$$ is used as a separator.
In particular, a cell with content $d\in\Delta$ and column number $j\in [0,2^{n}-1]$ is encoded by the word of length $n+1$ over $\Prop$
given by $d \,b_1\cdots b_n$,
 where $b_1 \cdots b_n$ is the binary encoding of the column number $j$ ($b_n$ being the most significant bit). A row is then represented by the word listing the encodings of cells from left to right, and a tiling $f$  consisting of  $k+1$ rows is encoded by the finite word $r_0 \$ r_1 \cdots \$ r_k$, where $r_i$ is the encoding of the
  $i$-th row of $f$, for all $i\in [0,k]$. See Figure~\ref{fig:row} for a graphical account of a word encoding of a tiling.
  
The Kripke structure $\Ku_\Instance$ is trivially defined as
 \[
 \Ku_\Instance = \tpl{\Prop, \Prop, \Prop\times \Prop, \Lab,d_\Init}
 ,\]
 where $\Lab(p)=\{p\}$, for each $p\in\Prop$. Thus, the initial traces of $\Ku_\Instance$ correspond to the finite words over $\Prop$ which start with the initial domino type $d_\Init$.
 
In order to build the $\BE$ formula $\varphi_\Instance$, we use some auxiliary formulas, namely, $\Length_i$, $\Beg(p)$, $\End(p)$, $\phi_{\textit{cell}}$, and $\theta_j(b,b')$, where $i\in [1,2n+2]$, $j\in [2,n+1]$, $p\in \Prop$, and $b,b'\in \{0,1\}$.

\begin{figure}[t]
    \centering
    \scalebox{0.4}{\includegraphics{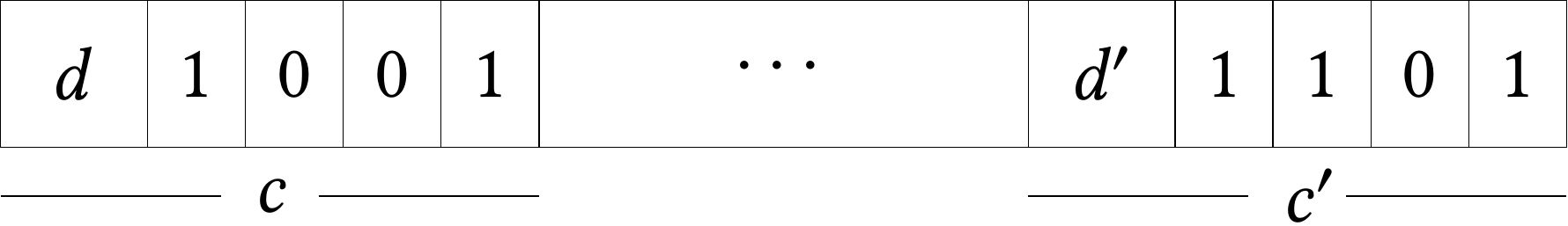}}
    \caption{Encoding of a trace $\rho$ starting with a cell $c=(d\, 1001)$ and ending with a cell $c'=(d'\, 1101)$ (here, $n=4$). The formula $\theta_2(1,1)$ is satisfied by $\rho$, while $\theta_3(1,0)$ is not.}
    \label{fig:thetaj}
\end{figure}

The formula  $\Length_i$, already presented in Example~\ref{example:length}, has size linear in $i$ and characterizes the traces having length $i$: 
\[
\Length_i= \hsBu^{i} \bot \wedge \hsB^{i-1} \top.
\]
The formula $\Beg(p)$ (resp., $\End(p)$) captures the traces of $\Ku$ which start (resp., end) in the state $p$:
\[
\Beg(p)= (p\wedge \Length_1) \vee \hsB(p\wedge \Length_1),
\quad
\End(p)= (p\wedge \Length_1) \vee \hsE(p\wedge \Length_1).
\]
The formula $\phi_{\textit{cell}}$ captures the traces of $\Ku_\Instance$ which encode cells:
\[
\phi_{\textit{cell}}= \Length_{n+1}\wedge \Big(\displaystyle{\bigvee_{d\in \Delta} \Beg(d)}\Big)\wedge \hsEu(\Beg(0)\vee \Beg(1)).
\]
Finally, for all $j\in [2,n+1]$ and $b,b'\in \{0,1\}$, the formula $\theta_j(b,b')$ is defined as:
\[
\theta_j(b,b')= \hsB(\Length_j \wedge \End(b))\wedge \hsE(\Length_{n-j+2} \wedge \Beg(b')).
\]
It is satisfied by a trace $\rho$ if $|\rho|\geq j+1$, $|\rho|\geq n-j+3$, $\rho(j)= b$, and $\rho(|\rho|-n+j-1)=b'$. In particular, for a trace $\rho$ starting with a cell $c$ and ending with a cell $c'$, $\theta_j(b,b')$ is satisfied by $\rho$ if the $(j-1)$-th bit of $c$ is $b$ and the $(j-1)$-th bit of $c'$ is $b'$. See Figure \ref{fig:thetaj} for an example.

Additionally, we use the derived operator $\hsG$ and its dual $\hsGu$, which allow us to select  arbitrary subtraces of the given trace, including the trace itself:
\[
\hsG\psi= \psi \vee \hsB\psi \vee \hsE\psi \vee \hsB\hsE\psi.
\]

The formula $\varphi_\Instance$ is defined as follows:
\[
\varphi_\Instance= \varphi_{\textit{b}} \wedge \varphi_{\textit{req}}\wedge \varphi_{\textit{inc}}\wedge \varphi_{\textit{rr}}\wedge \varphi_{\textit{rc}}.
\]
The conjunct $\varphi_{\textit{b}}$ checks that the given trace starts with a cell with content $d_\Init$ and column number $0$, and ends with a cell with content $d_\Final$
and column number $2^{n}-1$:
\[
\varphi_{\textit{b}} = \hsB\phi_{\textit{cell}}\wedge \Beg(d_\Init) \wedge  \hsE(\phi_{\textit{cell}} \wedge \Beg(d_\Final)) \wedge
\displaystyle{\bigwedge_{j=2}^{n+1}}\theta_j(0,1). 
\]
The conjunct $\varphi_{\textit{req}}$ ensures the following two requirements:
$(i)$ each occurrence of $\$$ in the given trace is followed by a cell with column number $0$ and
$(ii)$ each cell $c$ in the given trace is followed either by another cell, or by the separator $\$$, and in the latter case $c$ has column number $2^{n}-1$.
The first requirement is encoded by the formula: 
\[
\hsGu\bigl((\Length_{n+2} \wedge \Beg(\$)) \longrightarrow \hsE(\phi_{\textit{cell}}\wedge \hsEu\Beg(0))\bigr);
\]
the second one by the formula:
\begin{multline*}
\hsGu\Bigl\{(\Length_{n+2} \wedge \displaystyle{\bigvee_{d\in \Delta}} \Beg(d)) \longrightarrow\\
\Bigl(\hsB\phi_{\textit{cell}} \,\wedge\,  (\End(\$)\vee \displaystyle{\bigvee_{d\in \Delta}} \End(d))\, \wedge\, (\End(\$) \longrightarrow \hsEu(\Beg(\$)\vee \Beg(1)))\Bigr)\Bigr\}.
\end{multline*}
The conjunct $\varphi_{\textit{inc}}$  checks that adjacent cells along the given trace have consecutive columns numbers:
\[
\varphi_{\textit{inc}} = \hsGu\Bigl( \phi_{\textit{two\_cells}} \longrightarrow \displaystyle{\bigvee_{j=2}^{n+1}}\bigl[\theta_j(0,1)\wedge \bigwedge_{h=2}^{j-1}\theta_h(1,0) \wedge \bigwedge_{h=j+1}^{n+1}\bigvee_{b\in\{0,1\}}\theta_h(b,b)\bigr]\Bigr ),
\]
where $\phi_{\textit{two\_cells}}$ is given by
$
\Length_{2n+2}\wedge  \hsB\phi_{\textit{cell}}\wedge  \hsE\phi_{\textit{cell}}
$.
Note that $\varphi_{\textit{req}}$ and $\varphi_{\textit{inc}}$ ensure that column numbers are correctly encoded.

The conjunct $\varphi_{\textit{rr}}$  checks that adjacent cells in a row   have the same color on the shared edge:
\[
\varphi_{\textit{rr}} = \hsGu\Bigl( \phi_{\textit{two\_cells}} \longrightarrow\quad \smashoperator{\bigvee_{(d,d')\in \Delta\times \Delta\mid d_{\Right}= d'_{\Left}}} \quad (\Beg(d) \wedge \hsE(\Length_{n+1}\wedge \Beg(d')))\Bigr ) .
\]
Finally, the conjunct   $\varphi_{\textit{rc}}$  checks that adjacent cells in a column  have the same color on the shared edge.
For this, it suffices to require that
the following condition holds:
\begin{itemize}
  \item for each subtrace of the given one containing exactly one occurrence of $\$$, starting with a cell $c$, and ending with a cell $c'$, if $c$ and $c'$ have the same column number, then $d_{\Up}= d'_{\Down}$, where $d$ (respectively, $d'$) is the content of $c$ (respectively, $c'$).
\end{itemize}
Accordingly, the formula $\varphi_{\textit{rc}}$ is defined as follows, where we use the formulas $\theta_j(b,b)$, with $j\in [2,n+1]$ and $b\in \{0,1\}$, to express that $c$ and $c'$ have the same column number:
\begin{multline*}
  \varphi_{\textit{rc}} = \hsGu\Bigl\{\,\, \Big(\phi_{\textit{one}}(\$) \,\,\wedge \,\, \hsB\phi_{\textit{cell}} \,\,\wedge \,\, \hsE\phi_{\textit{cell}} \,\,\wedge \,\, \displaystyle{\bigwedge_{j=2}^{n+1}\bigvee_{b\in \{0,1\}}}\theta_j(b,b)\,\,\Bigr)\\
  \longrightarrow \smashoperator[r]{\bigvee_{(d,d')\in \Delta\times \Delta\mid d_{\Up}= d'_{\Down}}}\quad (\Beg(d) \wedge \hsE(\Length_{n+1}\wedge \Beg(d')))\,\,\,\Bigr\},
\end{multline*}
where $\phi_{\textit{one}}(\$)$ is defined as
\[
(\hsB\End(\$))\,\,\wedge\,\, \neg(\hsB(\End(\$) \wedge \hsB\End(\$)))
.\]

The formula $\varphi_\Instance$ has length polynomial in the size of $\Instance$. 
By construction, a trace $\rho$ of $\Ku_\Instance$ satisfies $\varphi_\Instance$ if and only if $\rho$ encodes a tiling.
Since the initial traces of $\Ku_\Instance$ are the finite words over $\Prop$ starting with $d_\Init$, it follows that there exists a tiling of $\Instance$ if and only if there exists an initial trace
of $\Ku_\Instance$ which satisfies $\varphi_\Instance$. 

The given reduction proves the following theorem. 

\begin{theorem}\label{theorem:lowerBoundBE} The MC problem for $\BE$ formulas over finite Kripke structures is \EXPSPACE-hard (under polynomial-time reductions).
\end{theorem}

Before concluding the section,
we would like to comment on the complexity gap deriving from the upper and lower bounds proved for $\HS$ (and $\BE$) MC.

Let us start by introducing \emph{star-free regular expressions} over a finite alphabet $\Sigma$, with $|\Sigma|\geq 2$, that are defined by the grammar
\[r::= \emptyset \mid a \mid r \cdot r \mid r \cup r \mid \neg r ,\]
where $a\in\Sigma$.
Every regular expression defines a language $\Lang(r)$ of finite words over $\Sigma$ by induction on its structural complexity
as follows:
   $\Lang(\emptyset)=\emptyset$,
   $\Lang(a)=\{a\}$,
   $\Lang(r_1\cdot r_2)=\Lang(r_1)\cdot \Lang(r_2)$,
   $\Lang(r_1\cup r_2)=\Lang(r_1)\cup \Lang(r_2)$ and
   $\Lang(\neg r)=\Sigma^*\setminus\Lang(r)$.\footnote{%
   As a standard notation, the dot $\cdot$ represents the concatenation of words, and $\Sigma^*$ (resp., $\Sigma^+$) the set of all possible \emph{finite} (resp., finite and non-empty) words over $\Sigma$.
   } 
It is well-known that the \emph{language-emptiness problem for star-free regular expressions is (provably) nonelementary}~\cite{Stockmeyer:1973,stockmeyer1974} (more properly, \TOWER-complete in the notation of~\cite{Schmitz:2016}), being negation the \lq\lq difficult case\rq\rq.

Let us now slightly modify the definition of these expressions. We consider:
\[r::= \emptyset \mid a \mid r \cdot \Sigma^+ \mid \Sigma^+\cdot r \mid r \cup r \mid \neg r .\]
Again, negation is present and there is no Kleene star (as $\Sigma^+$ can be \lq\lq rewritten\rq\rq{} as $\neg\emptyset$); \emph{concatenation is now weakened}: in fact, $r \cdot \Sigma^+$ and $\Sigma^+\cdot r$ represent right-/left-extensions of a word in $\Lang(r)$ with any (non-empty) word.
To the best of our knowledge, nothing is known about the precise complexity of this variant of star-free regular expressions.

The reader may now be wondering why we care so much about these regular expressions. The answer is given by the following fact:
\emph{there is a trivial reduction from language emptiness for this variant of expressions to $\BE$ MC, and vice versa}.
Intuitively, $r \cdot \Sigma^+$ (resp., $\Sigma^+\cdot r$) can be \lq\lq simulated\rq\rq{} by $\hsB$ (resp., $\hsE$).%
\footnote{It is also worth noting that the \lq\lq general\rq\rq{} concatenation $r\cdot r$ cannot be simulated by $\BE$: the \emph{chop} operator would be needed, which bisects an interval into two consecutive parts/subintervals: $\psi_1\langle C\rangle \psi_2$ predicates $\psi_1$ over the first subinterval and $\psi_2$ over the second.}
As an immediate consequence, an improvement on the known (upper or lower) bounds for any of the two problems would immediately propagate to the other.

Unfortunately, the nonelementary lower bound proved for (standard) star-free regular expressions by Stockmeyer in his PhD thesis \cite{stockmeyer1974} cannot be adapted to the variant, failing to encode the so-called \lq\lq \emph{movable rulers}\rq\rq, 
namely, regular expressions that are \lq\lq satisfied\rq\rq{} only by (all) words of a precise length.
Conversely, it is not clear how the (nonelementary) complexity of the standard automata-based algorithm for deciding the emptiness of regular expressions could possibly be lowered by exploiting the weakened concatenation; analogously, the nonelementary algorithm for full $\HS$ MC represents, at the moment, the best algorithm also for $\BE$ MC.

%% file: Chaps/Intro/tilingGraph.tex
\begin{tikzpicture}[node distance=0 cm,outer sep = 0pt]


\tikzstyle{cella}=[draw, rectangle,  minimum height=0.7cm, minimum width={width("dddddd")},anchor=south west];
\tikzstyle{dcella}=[draw, rectangle, dashed, minimum height=0.7cm, minimum width={width("dddddd")},anchor=south west];
\node[cella] (k-0) at (0,300) {$d^k_0$};
\node[cella] (k-1) [right = of k-0] {$d^k_1$};
\node[cella] (k-2) [right = of k-1] {$d^k_2$};
\node[dcella] (k-3) [right = of k-2] {};
\node[dcella] (k-4) [right = of k-3] {};
\node[dcella] (k-5) [right = of k-4] {};
\node[cella] (k-6) [right = of k-5] {$d^k_{2^n-2}$};
\node[cella] (k-7) [right = of k-6] {$d^k_{2^n-1}$};
\node[dcella] (k1-4) [below = of k-4] {};
\node[cella] (k2-4) [below = of k1-4] {$d^{j+1}_i$};
\node[cella] (k3-4) [below = of k2-4] {$d^{j}_i$};
\node[cella] (k4-4) [below = of k3-4] {$d^{j-1}_i$};
\node[cella] (k3-3) [left = of k3-4] {$d^{j}_{i-1}$};
\node[cella] (k3-5) [right = of k3-4] {$d^{j}_{i+1}$};
\node[dcella] (k5-4) [below = of k4-4] {};
%
\node[dcella] (1-4) [below = of k5-4] {};
\node[dcella] (1-5) [right = of 1-4] {};
\node[dcella] (1-3) [left = of 1-4] {};
\node[cella] (1-2) [left = of 1-3] {$d^0_2$};
\node[cella] (1-1) [left = of 1-2] {$d^0_1$};
\node[cella] (1-0) [left = of 1-1] {$d^0_0$};
\node[cella] (1-6) [right = of 1-5] {$d^0_{2^n-2}$};
\node[cella] (1-7) [right = of 1-6] {$d^0_{2^n-1}$};
 \node [align=left,red](13) [left = of 1-0] {$d_{init}$};
 \node [align=left,red](13) [right = of k-7] {$d_{final}$};
\node[cella] (k-k) at (10.9,298) {$d^j_i$};
\node [align=left,red] [left = of k-k] {$[d^j_i]_\Left$};
\node [align=left,red] [right = of k-k] {$[d^j_i]_\Right$};
\node [align=left,red] [below = of k-k] {$[d^j_i]_\Down =$};
\node [align=left,red] [above = of k-k] {$[d^j_i]_\Up$};
\node[cella] (k-k-1) at (10.9,296) {$d^{j-1}_i$};
\node [align=left,red] [above = of k-k-1] {$[d^{j-1}_i]_\Up$};
\node[dcella] (i-0) [below = of k-0] {};
\node[dcella] (i-1) [below = of i-0] {};
\node[dcella] (i-2) [below = of i-1] {};
\node[dcella] (i-3) [below = of i-2] {};
\node[dcella] (i-4) [below = of i-3] {};
\node[dcella] (j-0) [below = of k-7] {};
\node[dcella] (j-1) [below = of j-0] {};
\node[dcella] (j-2) [below = of j-1] {};
\node[dcella] (j-3) [below = of j-2] {};
\node[dcella] (j-4) [below = of j-3] {};

\end{tikzpicture}

%% file: Chaps/Intro/TilingEncoding.tex
\newcommand{\cellTwo}[2]{
    \begin{tabular}{c|c}
        \rule[-1ex]{0pt}{3.5ex}
    #1 & #2 \\
    \end{tabular}}

\newcommand{\cellOne}[1]{
    \begin{tabular}{c}
		\rule[-1ex]{0pt}{3.5ex}
		#1
	\end{tabular}}

\begin{tikzpicture}[node distance=0 cm,->,>=stealth',shorten >=1pt,auto,semithick,main node/.style={rectangle,draw, inner sep=0pt}]  

\tikzstyle{gray node}=[fill=gray!30]
 \node [main node](0) at (-4,0) {\cellTwo{$d^0_0$}{$0\cdots 00$}};
 \node [main node](1) [right = of 0] {\cellTwo{$d^0_1$}{$1\cdots 00$}};
\node [main node](11) [right = of 1] {\cellOne{$\cdots$}};
 \node [main node](2) [right = of 11] {\cellTwo{$d^0_{2^n-1}$}{$1\cdots 11$}};
 \node [main node, red](3) [right = of 2]  {\cellOne{\$}};
  \node [main node](4) [right = of 3]  {\cellTwo{$d^1_0$}{$0\cdots 00$}};
   \node [main node](5) [right = of 4]  {\cellTwo{$d^1_1$}{$1\cdots 00$}};
\node [main node](55) [right = of 5] {\cellOne{$\cdots$}};
   \node [main node](6) [right = of 55]  {\cellTwo{$d^1_{2^n-1}$}{$1\cdots 11$}};
   \node [main node, red](7) [right = of 6]  {\cellOne{\$}};
   \node (8) [right = of 7] {\Large $\cdots$};

   \node [align=left,red](10) [below = of 0] {column $0$};
   \node [align=left,red](11) [below = of 1] {column $1$};
   \node [align=left,red](12) [below = of 2] {column $2^n - 1$};
   \node [align=left,red](13) [below = of 4] {column $0$};
   \node [align=left,red](14) [below = of 5] {column $1$};
   \node [align=left,red](15) [below = of 6] {column $2^n - 1$};
 \node [align=left,red](20) at (-4,1.2) {row $0$};
  \node [align=left,red](20) at (5.2,1.2) {row $1$};
 
\draw [{|-|},dashed,red,thick] (-5.2,1) -- (3.5,1);
\draw [{|-|},dashed,red,thick] (4.1,1) -- (12.8,1);

%
%

\end{tikzpicture}

%% file: Chaps/Intro/generalPict.tex
\section{An overview on MC for $\HS$ and its fragments under homogeneity}
\label{sec:overvHomo}

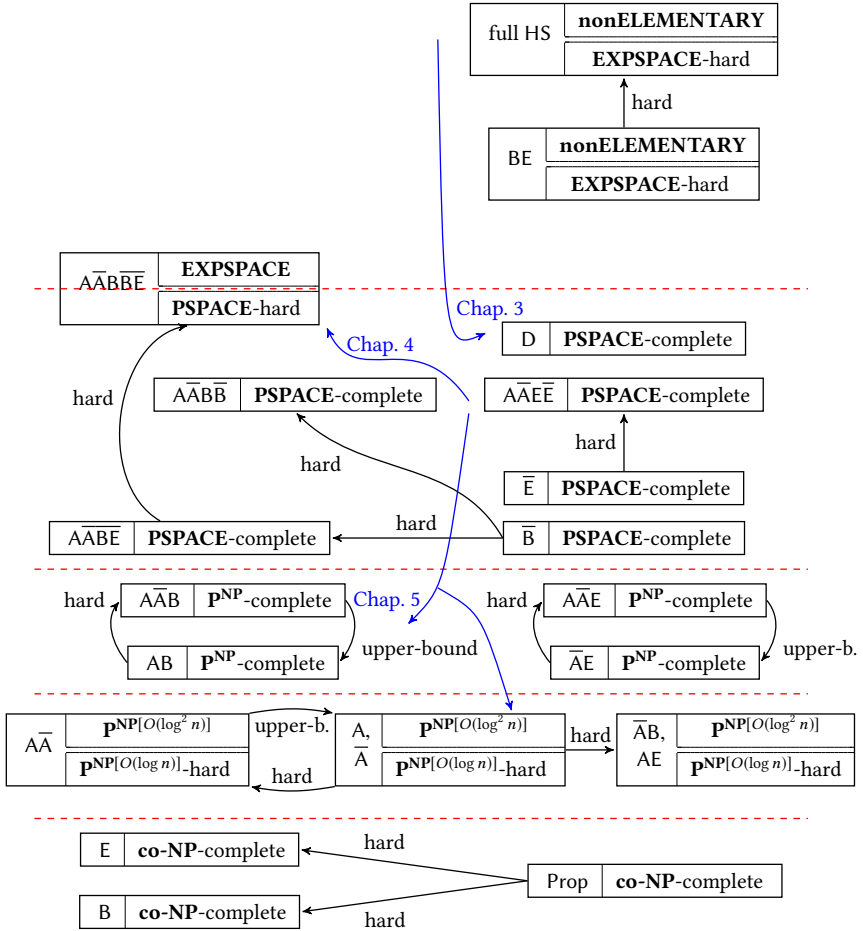
\begin{figure}[p]
\centering
    \input{Chaps/Intro/overGraph}
    \caption{Complexity of the MC problem for $\HS$ and its fragments. Red lines separate different complexity classes. Blue arrows depict the contents of the next three chapters.}
    \label{fGr}
\end{figure}

As the $\EXPSPACE$-hardness of the fragment $\BE$ clearly propagates to full $\HS$, $\HS$ MC represents a \emph{provably intractable problem} (here and in the following, we say \lq\lq intractable\rq\rq---by borrowing the terminology from Meyer and Stockmeyer~\cite{Stockmeyer:1973}---when a problem can not be provably solved in polynomial time).
The reasons which originated a systematic, in-depth investigation on complexity and expressiveness of $\HS$ fragments should then appear obvious: lowering the computational complexity of MC has crucial importance; at the same time, identifying a good \emph{trade-off between complexity and expressiveness} of a logic is fundamental as well. In the next chapters we start considering the issue of complexity which, for $\HS$ fragments where properties of prefixes (namely, $\hsB$) and suffixes ($\hsE$) of intervals are dealt with separately, is markedly lower than that of full $\HS$ (see Figure~\ref{fGr} for a graphical overview). Expressiveness will be addressed in Chapter~\ref{chap:TOCL17}: there we will make
a comparison of different semantic variants of $\HS$; 
in addition we will compare such variants with the standard point-based temporal logics $\LTL$, $\CTL$, and $\CTLStar$.

Since the combined use of modalities for prefixes $\hsB$ and suffixes $\hsE$ is critical,
the first fragments taken into consideration were 
$\AAbarBBbarEbar$ and $\AAbarEBbarEbar$:
these (syntactically maximal) fragments are obtained from full $\HS$ (i.e., $\AAbar\BE\Bbar\Ebar$) in an obvious way, that is, by removing either $\B$ or $\E$.
In~\cite{MMP15} we devised, for both of them, an $\EXPSPACE$ MC algorithm which finds, for each trace of the input Kripke structure, a satisfaction-preserving trace of bounded exponential length, i.e., a \emph{trace representative}. In this way, the algorithm needs to check only trace representatives instead of traces of unbounded length. In the same paper, it was proved that formulas satisfying a constant bound on the B-nesting depth (resp., E-nesting depth) can be checked in polynomial working space. As a consequence MC for $\AAbar\Bbar\Ebar$ is in $\PSPACE$ (its formulas do not feature $\hsB$/$\hsE$, hence $\nestb(\psi)=\neste(\psi)=0$).
The techniques employed in~\cite{MMP15} will be summarized at the beginning of Section~\ref{sec:AAbarBBbarEbar}.
However, the proof of existence of trace representatives is rather involved and it exploits very technical arguments. In this thesis, we will prove---in a (hopefully!) much more understandable and compact way---membership to $\EXPSPACE$ of MC for $\AAbarBBbarEbar$ and $\AAbarEBbarEbar$ (Section~\ref{sec:AAbarBBbarEbar}) by having recourse to completely different notions.

Still in Chapter~\ref{chap:TCS17} (Section~\ref{sec:AAbarEEbar}), we prove that MC for the $\HS$ fragment $\AAbarBBbar$ (resp., $\AAbarEEbar$) 
is in $\PSPACE$.
Since MC for the $\HS$ fragment featuring only one modality for right (resp., left) interval extensions $\Bbar$ (resp., $\Ebar$) is $\PSPACE$-hard (see Appendix~\ref{sect:BbarHard}),%
\footnote{The $\PSPACE$-hardness of MC for $\Bbar$/$\Ebar$ gives the best (unmatching) complexity lower bound also for $\AAbarBBbarEbar$ and $\AAbarEBbarEbar$ MC.}
$\PSPACE$-completeness immediately follows: 
the complexity turns out be the same as the one of $\LTL$ MC (known to be $\PSPACE$-complete~\cite{Sistla:1985}).
As a \lq\lq byproduct\rq\rq , we show that MC for the one-modality fragments $\B$ and $\E$ ($\B$ and $\E$ dealt with separately!) turns out to be $\co\NP$-complete.%
\footnote{The $\co\NP$-hardness derives from that of the fragment $\HSprop$ (the purely propositional fragment of $\HS$), as proved in~\cite{MMP15B}.}
These results are achieved by means of a \emph{small-model property}: intuitively, given a trace $\rho$ in a finite Kripke structure and a formula $\varphi$ of $\AAbarBBbar$/$\AAbarEEbar$, we prove that, by iteratively contracting $\rho$ it is always possible to build another trace
whose length is \emph{polynomially bounded} in the size of the formula and of the Kripke structure, which preserves the satisfiability of  $\varphi$  with respect to  $\rho$. 

In Chapter~\ref{chap:IC17}, we analyze the sub-fragments of $\AAbarBBbar$ (respectively, $\AAbarEEbar$), which are still expressive enough to capture meaningful interval properties of state transition systems and whose MC problem has a computational complexity markedly lower than that of full $\HS$,
namely, $\A$, $\Abar$, $\AAbar$, $\AB$, $\AbarB$, $\AE$, $\AbarE$, $\AAbarB$, and  $\AAbarE$.
All these have a similar computational complexity, as their MC problem settles in one of the lowest levels of the \emph{polynomial-time hierarchy}, $\PTIME^{\NP}$, or below. Such a class consists of the set of problems decided by a deterministic polynomial-time bounded Turing machine, with the \lq\lq support\rq\rq{} of an oracle for the class $\NP$,
that is, a tool which decides, in one computation step, whether an instance of a problem belonging to $\NP$ is positive or not. 
$\PTIME^{\NP}$ is also referred to as $\PTIME$ \emph{relative to} $\NP$ (relativization).
However, though the fragments in the considered set are similar, some differences can be marked. In particular, the fragments $\A$, $\Abar$, $\AAbar$, $\AbarB$, and $\AE$ are actually \lq\lq easier\rq\rq{} than the other ones, since they require the $\PTIME$ Turing machine to perform just $O(\log^2 n)$ queries to the $\NP$ oracle, for an input size $n$, instead of $O(n^k)$ queries, for some constant $k\geq 0$, as it is allowed in the general case for a  polynomial running time machine. The MC problem for these fragments witnesses
a \lq\lq non-standard\rq\rq{} complexity class in the polynomial-time hierarchy, called \emph{bounded-query} class, that will be presented in detail in Section~\ref{sub:compl}.
More precisely, 
we devise a $\PTIME^{\NP}$ MC algorithm for $\AAbarB$ and $\AAbarE$ in Section~\ref{sec:AAbarBalgo}, and then prove a matching complexity lower bound (Section~\ref{sec:ABhard});
then we show that MC for $\A$, $\Abar$, $\AAbar$, $\AbarB$ and $\AE$   
is still in $\PTIME^{\NP}$, but only $O(\log^2 n)$ queries to the $\NP$ oracle are necessary (Section~\ref{sect:AAbarAlg}).
This result is achieved by a reduction \emph{to} the problem TB(SAT)~\cite{schnoebelen2003}, whose instances are complex 
circuits where some of the gates are endowed with $\NP$ oracles. 
Finally, we identify a lower bound, which shows that \emph{at least $\log n$ queries} are needed to solve the problem (Section~\ref{sect:AHard}). Unfortunately, such bound does not exactly match the upper bound, leaving open the question whether the problem can be solved by $o(\log^2 n)$ (i.e., strictly less than $O(\log^2 n)$) queries to an $\NP$ oracle, or a tighter lower bound can be proved (or both).
 
In the next Chapter~\ref{chap:ICALP}, we focus on another $\HS$ fragment that has been studied, namely, $\D$---%
also known as \lq\lq the logic of sub-intervals\rq\rq---whose MC on finite Kripke structures turns out to be $\Psp$-complete (the same complexity result holds also for its SAT over finite linear orders, assuming homogeneity). 
Modality $\hsD$ can be easily defined by $\hsB$ and $\hsE$, as
$\hsD \varphi = \hsB \hsE \varphi = \hsE \hsB \varphi$,
hence $\D$ is a fragment of $\BE$. However
$\Psp$-completeness of $\D$ MC and SAT strongly contrasts with the case of $\BE$ (whose MC and SAT are $\EXPSPACE$-hard under homogeneity whereas, without homogeneity, SAT is undecidable over the class of finite and discrete linear orders~\cite{DBLP:journals/fuin/MarcinkowskiM14}).%
\footnote{We point out that homogeneity changes the status of the SAT problem for $\HS$ and its fragments. We will show in Chapter~\ref{chap:TOCL17} that, when interpreted over the (infinite) fullpaths of a finite Kripke structure (which is not the way we interpret $\HS$ here), $\LTL$ and $\HS$ have the same expressive power under homogeneity, but the latter is provably exponentially more succinct. As a byproduct, the SAT problem for full $\HS$, with such a semantics, turns out to be decidable. Therefore, \emph{under homogeneity}, the relevant issue for SAT of $\HS$ becomes its complexity, rather than its decidability.}

We now summarize the contents of the next three chapters, by a kind of \lq\lq journey\rq\rq{} among complexity classes, which is depicted by the blue arrows of Figure~\ref{fGr}.

\paragraph{Organization of the next chapters.}
\begin{itemize}
	\item  We start with Chapter~\ref{chap:ICALP}, presenting $\D$ MC and SAT. Whereas MC and SAT for $\BE$ are nonelementarily decidable and $\EXPSPACE$-hard, the situation is much better if we restrict to $\BE$'s fragment $\D$: we will prove $\PSPACE$ membership of both problems for it under homogeneity, and comment on their $\PSPACE$-hardness.
	\item In Chapter~\ref{chap:TCS17}, at first we remain in $\PSPACE$: we prove that MC for $\AAbarBBbar$ (resp., $\AAbarEEbar$) belongs to that class. Since MC for $\Bbar$ (resp., $\Ebar$) is $\PSPACE$-hard (this is proved in Appendix~\ref{sect:BbarHard}), $\PSPACE$-completeness of all fragments \lq\lq in between\rq\rq{} $\Bbar$ and $\AAbarBBbar$ (resp., $\Ebar$ and $\AAbarEEbar$) follows. It is surprising that $\Bbar$ and $\AAbarBBbar$ have exactly the same complexity, being the latter much more expressive than the former, as it can also express properties of prefixes, and of intervals in the future as well as in the past.
If we then \emph{add} $\Ebar$ to $\AAbarBBbar$ getting $\AAbarBBbarEbar$ (resp., $\Bbar$ to $\AAbarEEbar$ getting $\AAbarEBbarEbar$) we show that the complexity \lq\lq raises\rq\rq{} to $\EXPSPACE$ (Section~\ref{sec:AAbarBBbarEbar});%
 \footnote{The fragments $\AAbarBBbarEbar$/$\AAbarEBbarEbar$ are presented after $\AAbarBBbar$/$\AAbarEEbar$ for technical reasons.} 
 conversely,  
 \item in Chapter~\ref{chap:IC17}, we show the results achieved by \emph{removing} $\Bbar$ from $\AAbarBBbar$  (resp., $\Ebar$ from $\AAbarEEbar$). As we said, this leads to a \lq\lq fall\rq\rq{} towards a low level of the polynomial hierarchy, namely $\PTIME^{\NP}$, where MC for $\A$, $\Abar$, $\AAbar$, $\AB$, $\AbarB$, $\AE$, $\AbarE$, $\AAbarB$ and  $\AAbarE$ is.
\end{itemize}

\input{Chaps/Intro/LOMrelated.tex}

%% file: Chaps/Intro/overGraph.tex
\newcommand{\cellThree}[3]{
\begin{tabular}{c|c}
\rule[-1ex]{0pt}{3.5ex}
\multirow{2}{*}{#1} & #2 \\ 
\hhline{~=}\rule[-1ex]{0pt}{3.5ex}
 & #3 
\end{tabular}}

\newcommand{\cellTwo}[2]{\begin{tabular}{c|c}
\rule[-1ex]{0pt}{3.5ex}
#1 & #2 \\
\end{tabular}}

\newcommand{\cellOne}[1]{\begin{tabular}{c}
		\rule[-1ex]{0pt}{3.5ex}
		#1
	\end{tabular}}

\resizebox{\linewidth}{!}{
\begin{tikzpicture}[->,>=stealth',shorten >=1pt,auto,semithick,main node/.style={rectangle,draw, inner sep=0pt}]  

\tikzstyle{gray node}=[fill=gray!30]

    \node [main node](0) at (-4,0) {\cellTwo{$\AAbarBbarEbar$}{$\Psp$-complete}};
    \node [main node](1) at (3,0)  {\cellTwo{$\Bbar$}{$\Psp$-complete}};
    \node [main node](21) at (3,0.8)  {\cellTwo{$\Ebar$}{$\Psp$-complete}};
    \node [main node](31) at (3,2.3)  {\cellTwo{$\AAbarEEbar$}{$\Psp$-complete}};
    \node [main node](31a) at (3,3.2)  {\cellTwo{$\D$}{$\Psp$-complete}};
    \node [main node](32) at (-2.3,2.3)  {\cellTwo{$\AAbarBBbar$}{$\Psp$-complete}};
    \node [main node](2) at (-5.0,-3.4) {\cellThree{$\AAbar$}{$\Thsq$}{$\Th$-hard }};
    \node [main node](22) at (0.2,-3.4) {\cellThree{\hspace*{-6pt}$\begin{array}{c}\A, \\ \Abar \end{array}$\hspace*{-6pt}}{$\Thsq$}{$\Th$-hard }};
    \node [main node](92) at (4.8,-3.4) {\cellThree{\hspace*{-6pt}$\begin{array}{c}\Abar\B, \\ \A\E \end{array}$\hspace*{-6pt}}{$\Thsq$}{$\Th$-hard}};
    
    \node [main node](72) at (-3.3,-1.0)  {\cellTwo{$\AAbarB$}{$\PTIME^{\NP}$-complete}};
    \node [main node](73) at (3.5,-1.0)  {\cellTwo{$\AAbarE$}{$\PTIME^{\NP}$-complete}};
    \node [main node](74) at (-3.3,-2)  {\cellTwo{$\A\B$}{$\PTIME^{\NP}$-complete}};
    \node [main node](75) at (3.5,-2)  {\cellTwo{$\Abar\E$}{$\PTIME^{\NP}$-complete}};
    
    \node [main node](53) at (-4,-6) {\cellTwo{$\B$}{$\co\NP$-complete}};
    \node [main node](3) at (-4,-5) {\cellTwo{$\E$}{$\co\NP$-complete}};
    \node [main node](4) at (3.5,-5.5) {\cellTwo{$\HSprop$}{$\co\NP$-complete }};
    
    \node [main node](5) at (-4,4) {\cellThree{$\AAbarBBbarEbar$}{$\EXPSPACE$ }{$\Psp$-hard }};
    
    \node [main node](9) at (3,6) {\cellThree{$\BE$}{\textbf{nonELEMENTARY}}{$\EXPSPACE$-hard }};
    \node [main node](10) at (3,8) {\cellThree{full $\HS$}{\textbf{nonELEMENTARY}}{$\EXPSPACE$-hard}};
   
    \path
    (1) edge [swap] node {hard} (0) 
    (0) edge  [out=150,in=190] node {hard} (5.south)
    (4.west) edge [swap,near end] node {hard} (3.east)
    (4.west) edge [near end] node {hard} (53.east)
    (2.north east) edge [swap,out=370,in=170] node {upper-b.} (22.north west)
    (22.south west) edge [swap,out=190,in=-10] node {hard} (2.south east)
    (22.east) edge node {hard} (92.west)
    (9) edge [swap] node {hard} (10)
    (21.north) edge node {hard} (31.south)
    (1.west) edge [out=120,in=-45, near end] node {hard} (32.south)
    (74.west) edge [out=130, in=230, near end] node {hard} (72.west)
    (72.east) edge [out=310, in=50] node {upper-bound} (74.east)
    (75.west) edge [out=130, in=230, near end] node {hard} (73.west)
    (73.east) edge [out=310, in=50] node {upper-b.} (75.east)
    ;
    
    \draw [dashed,-,red] (-6.5,4) -- (6,4);
    \draw [dashed,-,red] (-6.5,-2.5) -- (6,-2.5);
    \draw [dashed,-,red] (-6.5,-4.5) -- (6,-4.5);
    \draw [dashed,-,red] (-6.5,-0.5) -- (6,-0.5);

    \draw [->,blue] (0,8) .. controls (0.1,3) .. node [right] {Chap.~\ref{chap:ICALP}} (0.8,3.3) ;
    \draw [->,blue] (0.5,2.2) .. controls (-0.4,3.5) and (-1.3,2.3) .. node [right, very near end] {Chap.~\ref{chap:TCS17}} (-1.8,3.4);
    \draw [->,blue] (0.5,2) .. controls (0.1,-0.8) .. node [left, near end] {Chap.~\ref{chap:IC17}} (-0.5,-1.4);
    \draw [->,blue] (0,-0.8) .. controls (0.7,-1.2) .. (1.2,-2.8);
\end{tikzpicture}
}

%% file: Chaps/Intro/LOMrelated.tex
\section{Related work}\label{sec:LOMrelated}
We would like to conclude the chapter by summarizing the content of other research papers that deal with $\HS$ MC.
As already mentioned, the MC problem for interval temporal logics has not been extensively studied in literature. Indeed the only three papers (apart from the ones by these authors) that study $\HS$ MC are all by A.\ Lomuscio and J.\ Michaliszyn~\cite{LM13,LM14,lm16}.

In~\cite{LM13,LM14,lm16}, they address the MC problem for some fragments of $\HS$ extended with epistemic modalities. Their semantic assumptions are different from the ones we make here, thus 
a systematic comparison of the two research lines is quite difficult. In both cases, formulas of $\HS$ are evaluated over traces/intervals of a Kripke structure; however, in~\cite{LM13,LM14} truth of proposition letters over an interval depends only on its endpoints. 

In~\cite{LM13}, the authors focus on the $\HS$ fragment $\B\E\D$ of Allen's relations \emph{started-by}, \emph{finished-by}, and \emph{contains} (since modality $\hsD$ is definable in terms of modalities $\hsB$ and $\hsE$, $\B\E\D$ is actually as expressive as $\BE$), extended with epistemic modalities. They consider a \emph{restricted} form of MC (\lq\lq local\rq\rq\ MC), which verifies a given specification against a single (finite) initial computation interval. Their goal is indeed to reason about a given computation of a multi-agent system, rather than on all its admissible computations.
They prove that the considered MC problem is $\PSPACE$-complete; furthermore, they show that the same problem restricted to the pure temporal fragment $\B\E\D$, that is, the one obtained by removing epistemic modalities, is in $\PTIME$ as, basically, modalities $\hsB$ and $\hsE$ allow one to access only sub-intervals of the initial one, whose number is quadratic in the length (number of states) of the initial interval.

In~\cite{LM14}, they show that the picture drastically changes with other $\HS$ fragments that allow one to access infinitely many intervals. In particular, they prove that the MC problem for the fragment $\ABbar\L$ of Allen's relations \emph{meets}, \emph{starts}, and \emph{before} (since modality $\hsL$ is definable in terms of modality $\hsA$, $\ABbar\L$ is actually as expressive as $\ABbar$) extended with epistemic modalities, is decidable in nonelementary time. Note that, thanks to modalities $\hsA$ and $\hsBt$, formulas of $\ABbar\L$ can possibly refer to infinitely many (future) intervals.

Finally, in~\cite{lm16}, Lomuscio and Michaliszyn show how to use regular expressions in order to specify the way in which intervals of a Kripke structure get labelled. Such an extension leads to a significant increase in 
expressiveness, as the labelling of an interval is no more determined by that of its endpoints only, 
but it depends on the ordered sequence of states the interval consists of. They also prove that there is no corresponding increase in computational complexity, as the bounds given in~\cite{LM13,LM14} still hold with the new semantic variant: MC for $\B\E\D$ is in $\PSPACE$, and it is nonelementarily decidable for $\ABbar\L$.
We will come back to this idea of using regular expressions to define interval labelling in Chapter~\ref{chap:Gand17}.

%% file: Chaps/ICALP_D/ICALPmain.tex
\chapter{The logic of sub-intervals $\D$}\label{chap:ICALP}
\begin{chapref}
The reference for this chapter is \cite{icalp17}.
\end{chapref}

\minitoc\mtcskip

\newcommand{\CL}{\mathsf{CL}}
\newcommand{\REQ}{\mathsf{REQ}}
\newcommand{\obsD}{\cO bs_D}
\newcommand{\reqD}{\cR eq_D}
\newcommand{\rank}{rank}
\newcommand{\row}{row}
\newcommand{\orow}{\overline{row}}
\newcommand{\Atoms}{\cA_{\varphi}}
\newcommand{\Rows}{\cR ows_{\varphi}}

\newcommand{\bbD}{\mathbb{S}}
\newcommand{\bbI}{\mathbb{I}}
\newcommand{\bbP}{\mathbb{P}}

\newcommand{\cA}{\mathcal{A}}
\newcommand{\cG}{\mathcal{G}}
\newcommand{\cH}{\mathcal{H}}
\newcommand{\cL}{\mathcal{L}}
\newcommand{\cO}{\mathcal{O}}
\newcommand{\cR}{\mathcal{R}}
\newcommand{\cV}{\mathpzc{V}}
\newcommand{\hA}{\hat{A}}
\newcommand{\hn}{\hat{n}}
\newcommand{\hm}{\hat{m}}

\newcommand{\om}{\overline{m}}

\newcommand{\bM}{\mathbf{M}}

\newcommand{\subint}{\sqsubset}
\newcommand{\subinteq}{\sqsubseteq}
\newcommand{\ssubint}{{\text{$\sqsubset$\llap{$\;\cdot\;$}}}}
\newcommand{\Dphi}{\mathrel{D_\varphi}}
\newcommand{\Dref}{\textsf{D$_\subinteq$}\xspace}
\newcommand{\Dirr}{\textsf{D$_\subint$}\xspace}
\newcommand{\Dstr}{\textsf{D$_\ssubint$}\xspace}
\newcommand{\Dsim}{\textsf{D$_\circ$}\xspace}

\newcommand{\genDphi}{\begin{tikzpicture}
\node(A)[inner sep=0pt]{$\scriptstyle\Dphi$};
\node[inner sep=0pt](B) at (0.45,0){};
\node[inner sep=0pt](C) at (-0.4,0){};
\draw[double](C) -- (A);
 \draw[double,->](A) --(B); 
\end{tikzpicture}}

\newcommand{\rownext}{\begin{tikzpicture}
\node(A)[inner sep=0pt]{$\scriptstyle row_{\varphi}$};
\node[inner sep=0pt](B) at (0.6,0){};
\node[inner sep=0pt](C) at (-0.5,0){};
\draw[double](C) -- (A);
 \draw[double,->](A) --(B); 
\end{tikzpicture}}

\newcommand{\genDphiS}{\begin{tikzpicture}
\node(A)[inner sep=0pt]{$\scriptstyle{\Dphi\ssubint}$};
\node[inner sep=0pt](B) at (0.7,0){};
\node[inner sep=0pt](C) at (-0.6,0){};
\draw[double](C) -- (A);
\draw[double,->](A) --(B); 
\end{tikzpicture}}

\newcommand{\rownextS}{\begin{tikzpicture}
\node(A)[inner sep=0pt]{$\scriptstyle row_{\varphi}\ssubint$};
\node[inner sep=0pt](B) at (0.8,0){};
\node[inner sep=0pt](C) at (-0.7,0){};
\draw[double](C) -- (A);
\draw[double,->](A) --(B); 
\end{tikzpicture}}

\newcommand{\hsDhom}{\D}

\input{Chaps/ICALP_D/intro}
\input{Chaps/ICALP_D/preliminaries}
\input{Chaps/ICALP_D/compass}
\input{Chaps/ICALP_D/decidability}
\input{Chaps/ICALP_D/D_modelchecking}
\input{Chaps/ICALP_D/Psp_hard}
\input{Chaps/ICALP_D/concl}

%% file: Chaps/ICALP_D/intro.tex
\lettrine[lines=3]{I}{n this chapter} we focus on the logic 
$\D$ (also known as \lq\lq the logic of sub-intervals\rq\rq) which features one modality only, corresponding to the Allen interval relation \emph{during}. Since any sub-interval is just an initial sub-interval of an ending one, or, equivalently, an ending sub-interval of an initial one, $\D$ is a (proper) fragment of $\BE$. 

The logic of sub-intervals comes into play in the study of \emph{temporal prepositions in natural language}~\cite{DBLP:journals/ai/Pratt-Hartmann05};
the connections between the temporal logic of (strict) sub-intervals and 
\emph{the logic of Minkowski space-time} have been explored by Shapirovsky and Shehtman~\cite{DBLP:journals/logcom/ShapirovskyS05}.
Finally, the temporal logic of reflexive sub-intervals has been studied for the
first time by van Benthem, who proved that, when interpreted over dense linear 
orderings, it is equivalent to the standard modal logic S4~\cite{DBLP:journals/jphil/Benthem91}.

From a computational point of view, $\D$\ is a real character! Its SAT problem is $\Psp$-complete over the class of dense linear orders~\cite{DBLP:journals/logcom/BresolinGMS10,DBLP:conf/aiml/Shapirovsky04} (whereas the problem is undecidable for $\BE$~\cite{DBLP:conf/asian/Lodaya00}), it becomes undecidable when the logic is interpreted over the classes of finite and discrete linear orders~\cite{DBLP:journals/fuin/MarcinkowskiM14}, and the situation is still unknown over the class of all linear orders. As for its expressiveness, unlike $\AAbar$---which is expressively complete with respect to the two-variable fragment of first-order logic for binary relational structures over various linearly-ordered domains~\cite{DBLP:journals/apal/BresolinGMS09,DBLP:journals/jsyml/Otto01}---\emph{three} variables are needed to encode $\D$\ in first-order logic (the two-variable property is a sufficient condition for decidability, but it is not a necessary one).

In this chapter we show that decidability of SAT for $\D$\ over the class of finite linear orders can be recovered \emph{under the homogeneity assumption}. This will allow us to show that also MC under homogeneity is decidable. We first prove that $\D$ SAT is in $\Psp$ by exploiting a suitable contraction method. Then we show that the proposed SAT algorithm can be transformed into a $\Psp$ MC procedure for $\D$\ formulas over finite Kripke structures; $\Psp$-hardness of both problems follows via a reduction from the language universality problem of nondeterministic finite-state automata. 

$\Psp$-completeness of $\D$\ MC strongly contrasts with the case of $\BE$, which we showed to be nonelementarily decidable and hard for $\EXPSPACE$.

\paragraph*{Organization of the chapter.}
\begin{itemize}
	\item In Section~\ref{sec:preliminaries} and \ref{sec:compass}, we provide some background knowledge;
in particular, we introduce the logic $\D$, interval models, and a spatial representation of interval models called \lq\lq \emph{compass structure}\rq\rq.
	\item In Section~\ref{sec:decidability}, we show the $\Psp$ membership of the SAT problem for $\D$\ over finite linear orders
(under homogeneity). 
This complexity result is proved via a contraction technique, applied to the mentioned compass structures, relying on a suitable finite-index equivalence relation.
	\item In Section~\ref{sec:mc}, we show that the MC problem for  $\D$\ over finite Kripke structures (again, under homogeneity) is in $\Psp$ as well. The proposed MC algorithm is basically a SAT procedure driven by the computation traces of the system model. 
	\item Finally, 
in Section \ref{sec:outlineSATMC}, we comment on the $\Psp$-hardness of both problems, referring to
Appendix~\ref{sec:MChard} and Appendix~\ref{sec:SAThard} for further details. 
\end{itemize}

%% file: Chaps/ICALP_D/preliminaries.tex
\section{Preliminaries}\label{sec:preliminaries}

To start with, we introduce some preliminary notions.
Let $\mathbb{S} = (\mathpzc{S}, <)$ be a  linear order. An
\emph{interval} over $\mathbb{S}$ is an ordered pair $[x, y]$,
where $x,y\in \mathpzc{S},\, x \leq y$, representing the set $\{z\in \mathpzc{S}\mid x\leq z\leq y\}$. We denote the set of all intervals over
$\mathbb{S}$ by $\mathbb{I(S)}$.
We consider three possible \emph{sub-interval relations}: 
\begin{enumerate}
    \item the \emph{reflexive} sub-interval relation (denoted as
    $\sqsubseteq$), defined by $[x, y] \sqsubseteq [x', y']$ if and only if
    $x' \leq x$ and $y\leq y'$,
    \item the \emph{proper (or
    irreflexive)} sub-interval relation (denoted as $\subint$), defined
    by $[x, y]\subint  [x', y']$ if and only if $[x, y] \sqsubseteq
    [x', y']$ and $[x, y] \neq [x', y']$, and 
    \item the \emph{strict}
    sub-interval relation (denoted as $\ssubint$), defined by 
    $[x, y] \ssubint [x', y'] $ if and only if $x' < x$ and $y<y'$.
\end{enumerate}

The three modal logics $\Dref$, $\Dirr$, and $\Dstr$ feature the same
language, consisting of a set $\AP$ of proposition
letters, the logical connectives $\neg$ and $\vee$, and the
modal operator $\hsD$. Formally, formulas are
defined by the grammar: \[\varphi ::= p \ |\ \neg\varphi\ |\ \varphi \vee \varphi\ |\ \hsD \varphi,\]
with $p\in\AP$.
The other connectives, as well as
the logical constants $\top$ (true) and $\bot$ (false), are defined as usual; moreover, the
dual universal modal operator $\hsDu\varphi$ is defined as $\neg\hsD\neg\varphi$. The length of a formula
$\varphi$, denoted as $|\varphi|$, is the number of subformulas of $\varphi$.

The semantics of~\Dstr, \Dirr, and \Dref only differ in the
interpretation of the $\hsD$ operator. For the sake of brevity, we use
$\circ \in \{\ssubint,\subint,\subinteq \}$ as a shorthand for
any of the three sub-interval relations. The semantics of a
sub-interval logic \Dsim is defined in terms of \emph{interval models}\footnote{Not to be confused with the previously introduced \emph{abstract interval models}.}
$\bM=(\mathbb{I(S)},\circ,\cV)$. The \emph{valuation
function} $\cV : \AP \to 2^{\mathbb{I(S)}}$ assigns to
every proposition letter $p$ the set of intervals $\cV(p)$ over
which $p$ holds. The \emph{satisfiability relation} $\models$ is
defined as:
\begin{itemize}
  \item for every proposition letter  $p \in \AP$, $\bM, [x,y]
        \models p$ if and only if $[x,y] \in \cV(p)$;

  \item $\bM, [x,y] \!\models\! \neg \psi$ if and only if $\bM, [x,y]
        \!\not\models\! \psi$ (i.e. it is not true that $\bM, [x,y]
        \!\models\! \psi$);

  \item $\bM, [x,y] \models \psi_1 \vee \psi_2$ if and only if $\bM, [x,y] \models \psi_1$ or $\bM, [x,y] \models \psi_2$;

  \item $\bM, [x,y] \models \hsD \psi$ if and only if there exists an interval $[x',y'] \in
        \bbI(\mathbb{S})$ such that $[x',y'] \circ [x,y]$ and
        $\bM, [x',y'] \models \psi$.
\end{itemize}
A \Dsim formula is \emph{satisfiable} if it holds over some interval of an interval model, and \emph{valid} if it holds over every interval of every interval model.

As we mentioned earlier on, it can be shown that the logic $\Dref$ turns out to be equivalent to the standard modal logic S4 \cite{DBLP:journals/jphil/Benthem91}. 
%
Here we restrict our attention to the finite SAT problem, that is, satisfiability over the class of finite linear orders. The problem has been shown to be \emph{undecidable} for $\Dirr$ and $\Dstr$~\cite{DBLP:journals/fuin/MarcinkowskiM14} and \emph{decidable} for $\Dref$~\cite{DBLP:conf/time/MontanariPS10}. In the following, we prove that decidability can be recovered for $\Dirr$ and $\Dstr$ by restricting to the class of \emph{homogeneous} interval models. We fully work out the case of $\Dirr$ (for the sake of simplicity, we will write $\D$\ for $\Dirr$), and then we briefly explain how to adapt the proofs to $\Dstr$. 

\begin{definition}\label{def:homogeneous_models}
An interval model
$\bM=(\mathbb{I(S)},\circ,\cV)$ is \emph{homogeneous} if,
for every interval $[x,y]  \in \mathbb{I(S)}$ and every proposition letter $p \in \AP$,
it holds that $[x,y] \in \cV(p)$ if and only if $[x',x'] \in \cV(p)$
for every $x\leq x'\leq y$.
\end{definition}
Hereafter we will assume the logic $\D$ to be interpreted over homogeneous interval models.

%% file: Chaps/ICALP_D/compass.tex
\section{A spatial representation of interval models}\label{sec:compass}

Let us now introduce some basic definitions and notation which will be
extensively used in the following, concluding the section with the notion of compass structure. Given a $\D$ formula $\varphi$,
we define the \emph{closure of $\varphi$}, denoted by
$\CL(\varphi)$, as the set of all subformulas $\psi$ of $\varphi$ and of their
negations $\neg\psi$ (we identify $\neg\neg \psi$ with $\psi$). 

\begin{definition}\label{def:d-atom}
Given a $\D$-formula $\varphi$, a $\varphi$-atom $A$ is a subset of
$\CL(\varphi)$ such that: 
\begin{itemize}
    \item  for every $\psi \in \CL(\varphi)$, $\psi \in A$ if and only if $\neg\psi \notin A$, and
    \item for every $\psi_1 \vee \psi_2 \in \CL(\varphi)$, $\psi_1 \vee \psi_2 \in A$ if and only if $\psi_1 \in A$ or $\psi_2 \in A$.
\end{itemize}
\end{definition}

The idea underlying atoms is to enforce a \lq\lq local\rq\rq{} (or Boolean) form of consistency among the formulas it contains, that is, a $\varphi$-atom $A$ is a \emph{maximal, locally consistent subset} of $\CL(\varphi)$. As an example, $\neg(\psi_1 \vee \psi_2) \in A$ if and only if $\neg\psi_1\in A$ and $\neg\psi_2\in A$. However, note that the definition does not set any constraint on $\hsD\psi$ formulas, hence the word ``local''.
We denote the set of all $\varphi$-atoms as $\cA_\varphi$; its cardinality is clearly bounded by $2^{|\varphi|}$ (by the first point of Definition~\ref{def:d-atom}). Atoms are connected by the following binary relation $\Dphi$.

\begin{definition}\label{def:Dphi-relation}
Let $\Dphi$ be a binary relation over $\cA_\varphi$ such that, for each pair of atoms $A, A' \in \cA_\varphi$, $A \Dphi A'$ holds if and only if both $\psi \in A'$ and $\hsDu\psi \in A'$ for each formula $\hsDu\psi \in A$.
\end{definition}

Let $A$ be a $\varphi$-atom. We denote by $\reqD(A)$ the set $\{\psi \in \CL(\varphi) \mid \hsD \psi \in A\}$ of \lq\lq \emph{temporal requests}\rq\rq{} of $A$. In particular, if $ \psi \notin \reqD(A)$, then $\hsDu \neg \psi \in A$ (by the definition of $\varphi$-atom). Moreover, we denote by $\REQ_{\varphi}$ the set of all  arguments of $\D$ formulas in $\CL(\varphi)$, namely, $\REQ_{\varphi}=\{ \psi \mid \hsD \psi\in \CL(\varphi)\}$. Finally, we denote by $\obsD(A)$ the set $\{\psi \in A \mid \psi \in \REQ_{\varphi}\}$ of \lq\lq \emph{observables}\rq\rq{} of $A$. 

It is easy to prove by induction the next proposition stating that, once the proposition letters of $A$ and its temporal requests are fixed, $A$ is determined.
\begin{proposition}\label{prop:unique}
    For any $\D$ formula $\varphi$,
    given a set $R \subseteq \REQ_{\varphi}$ and a set $P \subseteq \CL(\varphi)\cap \AP$, there exists a unique $\varphi$-atom $A$ that satisfies $\reqD(A)=R$ and $A\cap \AP = P$.
\end{proposition} 

We now provide a natural interpretation of $\D$\ over grid-like
structures, called \emph{compass structures}, by exploiting the existence
of a natural bijection between intervals $[x,y]$ and 
points $(x,y)$, with $x \leq y$, of an $\mathpzc{S}\times \mathpzc{S}$ grid, where $\mathbb{S} = (\mathpzc{S},<)$ is a finite linear order. Such an interpretation was originally proposed by Venema in~\cite{venema1990},
and it can also be given for $\HS$ and all its (other) fragments.

\begin{figure}[tb]
\centering
\begin{tikzpicture}[scale=1]

\draw[step=0.5cm,gray,very thin] (-2.9,-2.9) grid (2.9,2.9);
\fill[color=red, opacity=.5] (-2,1) -- (1,1)-- (-2,-2);
\draw (-2.9,-2.9) -- (2.9,2.9);
\fill[color=white] (-2.9,-2.9) -- (2.9,2.9) -- (2.9,-2.9);
\draw[dashed] (-2,1) -- (-2, -2);
\draw[dashed] (-2,1) -- (1, 1);

\node[shape=circle,draw=black,inner sep=2pt,fill=black, label={above :$(x_0,y_0)$}](A) at (0,1) {};

\node[shape=circle,draw=black,inner sep=2pt,fill=red, label={left :$(x_3,y_3)$}](C) at (-2,1) {};

\node[shape=circle,draw=black,inner sep=2pt,fill=blue, label={below :$(x_1,y_1)$}](B) at (-1,0.5) {};
\node[shape=circle,draw=black,inner sep=2pt,fill=green, label={below left:$(x_2,y_2)$}](D) at (-2,-0.5) {};

\pgftransformshift{\pgfpoint{4cm}{-0.5cm}}

\draw[very thick,|-|] (0.2,-0.3) -- (2,-0.3)node[pos=0.5, above](AI) {$[x_0,y_0]$};

\draw(AI.west) edge[->, bend left] (A);

\draw[very thick,|-|,red] (-3,1) -- (2,1)node[pos=0.5, above=0.001cm,black](CI) {$[x_3,y_3]$};

\draw(CI.north) edge[->, bend right=90, looseness=0.8] (C);

\draw[very thick,|-|,blue] (-1,-2.1) -- (1.5,-2.1)node[pos=0.7, above=0.001cm,black](BI) {$[x_1,y_1]$};

\draw(BI.north) edge[->] (B.east);

\draw[very thick,|-|,green] (-3,-1.4) -- (-0.5,-1.4)node[pos=0.5, above=0.001cm,black](DI) {$[x_2,y_2]$};

\draw(DI.west) edge[->] (D);

\end{tikzpicture}
\caption{Correspondence between intervals and points of the compass structure.}\label{fig:compassstructure}
\end{figure}
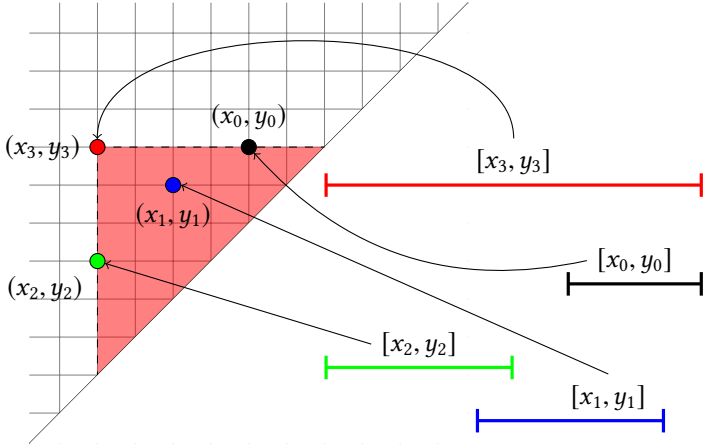

As an example, Figure~\ref{fig:compassstructure} shows four intervals
$[x_0,y_0],\ldots,[x_3,y_3]$, respectively represented by the points in the grid $(x_0,y_0),\ldots,(x_3,y_3)$, such that: 
\begin{itemize}
    \item $[x_0,y_0], [x_1,y_1], [x_2,y_2]\subint [x_3,y_3]$, 
    \item $[x_1,y_1] \ssubint [x_3,y_3]$, and 
    \item $[x_0,y_0], [x_2,y_2]\not\!\!\ssubint [x_3,y_3]$. 
\end{itemize}
The red 
region highlighted in Figure~\ref{fig:compassstructure} contains all and only the points $(x,y)$ such that $[x,y]\subint[x_3,y_3]$.
Allen interval relation \emph{contains} can thus be represented as a spatial relation between pairs of points. In the following, we make use of $\subint$ also for relating points, i.e., given two points $(x,y),(x',y')$ of the grid,
$(x',y')\subint (x,y)$ if and only if $(x',y')\neq (x,y)$ and  $x\leq x'\leq y'\leq y$.
Compass structures, repeatedly exploited in the following to establish the next complexity results, 
can be formally defined as follows.

\begin{definition}\label{def:compassstructure}
Given a finite linear order $\mathbb{S} = (\mathpzc{S}, <)$ and a $\D$ formula $\varphi$, a  \emph{compass}
$\varphi$-\emph{structure} is a pair $\cG=(\bbP_\bbD,\cL)$,
where $\bbP_\bbD$ is the set of points of the form $(x,y)$,
with $x,y \in \mathpzc{S}$ and $ x\leq y$, and $\cL$ is a function that
maps any point $(x,y)\in\bbP_\bbD$ to a $\varphi$-atom $\cL(x,y)$
in such a way that for every pair of points $(x,y)\neq(x',y')\in\bbP_\bbD$ ,
        if $x\leq x'\leq y'\leq y$ then $\cL(x,y) \Dphi \cL(x',y')$
        (\emph{temporal consistency}).
\end{definition}
Due to temporal consistency, the following important property holds in compass structures.
\begin{lemma}\label{lem:transitive_req}
Given a compass $\varphi$-structure $\cG=(\bbP_\bbD,\cL)$,
for all pairs of points $(x',y'), (x,y)\in \bbP_\bbD$,
if $(x',y')\subint (x,y)$,
then $\reqD(\cL(x',y')) \subseteq \reqD(\cL(x,y))$ and $\obsD(\cL(x',y')) \subseteq \reqD(\cL(x,y))$.
\end{lemma}
\begin{proof} 
By Definition \ref{def:compassstructure} we have  $\cL(x,y) \Dphi \cL(x',y')$.
Let us assume by contradiction that there exists  $\psi \in \reqD(\cL(x',y'))\setminus \reqD(\cL(x,y))$. By definition of $\reqD$ and by Definition  \ref{def:d-atom}, we have that 
$\psi \in \reqD(\cL(x',y'))$ 
implies   $\hsD\psi \in \cL(x',y')$, and
$\psi \notin \reqD(\cL(x,y))$ 
implies   $\neg\hsD\psi =\hsDu \neg \psi \in \cL(x,y)$. Since $\cL(x,y) \Dphi \cL(x',y')$,
then $\hsDu \neg \psi \in \cL(x',y')$ and thus we can conclude that
both $\hsDu \neg \psi$ and $\hsD \psi$ belong to $\cL(x',y')$ (contradiction).

$\obsD(\cL(x',y')) \subseteq \reqD(\cL(x,y))$ can analogously be proved by contradiction.
\end{proof}

\emph{Fulfilling} compass structures are defined as follows.
\begin{definition}\label{def:fulfillingcompass}
A compass $\varphi$-structure
$\cG=(\bbP_\bbD,\cL)$ is said to be \emph{fulfilling}
if, for every point $(x,y)\in\bbP_\bbD$ and each formula 
$\psi\in \reqD(\cL(x,y))$, 
there exists a point $(x',y')\subint (x,y)$ in $\bbP_\bbD$ such that 
$\psi \in \cL(x',y')$.
\end{definition}
Note that if $\cG$ is fulfilling, then $\reqD(\cL(x,x))=\emptyset$ for all points ``on the diagonal'' $(x,x)\in\bbP_\bbD$.
%
%
We say that a compass $\varphi$-structure $\cG=(\bbP_\bbD,\cL)$
\emph{features} a formula $\psi$ if there exists a point $(x,y)\in\bbP_\bbD$
such that $\psi \in \cL(x,y)$.
The following result holds.
\begin{proposition}\label{prop:compassstructure}
A $\D$ formula $\varphi$ is satisfiable if and only if there exists a fulfilling compass $\varphi$-structure that
features it.
\end{proposition}

In a fulfilling compass $\varphi$-structure  $\cG=(\bbP_\bbD,\cL)$, where $\mathpzc{S}=\{0,\ldots,t\}$, w.l.o.g., we will sometimes assume  $\varphi$ to be satisfied by the maximal interval $[0,t]$, that is, $\varphi\in\cL(0,t)$.

The notion of homogeneous models directly transfers to compass structures.
\begin{definition}\label{def:hom_compass}
A compass $\varphi$-structure $\cG=(\bbP_\bbD,\cL)$ is \emph{homogeneous}
if, for every point $(x,y)\in \bbP_\bbD$ and each proposition letter $p\in \AP$,
we have that $p\in \cL(x,y)$ if and only if $p\in \cL(x',x')$ for all $x\leq x'\leq y$.
\end{definition}

Proposition \ref{prop:compassstructure} can be tailored to homogeneous compass structures.
\begin{proposition}\label{prop:satiffcompass}
A $\hsDhom$ formula $\varphi$ is satisfiable if and only if there exists a fulfilling homogeneous compass $\varphi$-structure that
features it.
\end{proposition}

%% file: Chaps/ICALP_D/decidability.tex
\section{SAT of $\hsDhom$\ over finite linear orders}\label{sec:decidability}

In this section, we devise a SAT checking procedure for $\hsDhom$ formulas over finite linear orders, which will also allow us to easily derive a MC algorithm for $\hsDhom${} over finite Kripke structures.

To start with, we show that there is a ternary relation between $\varphi$-atoms, that we denote by $\genDphi$, such that if it holds among all atoms in consecutive positions of a compass $\varphi$-structure, then the structure is fulfilling. Hence, we may say that $\genDphi$ is the rule for labeling fulfilling compasses. 

Next, we introduce an equivalence relation $\sim$ between \emph{rows} of a compass $\varphi$-structure. Since it has finite index---exponentially bounded by $|\varphi|$---and it preserves fulfillment of compasses, it is intuitively possible to ``contract'' the structures when we can find two related rows. Moreover, any contraction done according to $\sim$ keeps the same atoms (only the number of their occurrences may vary), and thus if a compass features $\varphi$ before the contraction, then $\varphi$ is still featured after it. This fact is exploited to build a SAT algorithm for $\hsDhom$ formulas which makes use of \emph{polynomial working space}, because $(i)$~it only needs to keep track of two rows of a compass at a time, $(ii)$~all rows satisfy some nice properties that make it possible to succinctly encode them, and $(iii)$~compass contractions are implicitly performed by means of a reachability check in a suitable graph, whose nodes are the equivalence classes of $\sim$.

Let us now introduce the aforementioned ternary relation $\genDphi$ among atoms.
\begin{definition}\label{def:d_generator}
Given three $\varphi$-atoms $A_1, A_2$ and $A_3$, we say that 
$A_3$ is $\Dphi$-generated by $A_1, A_2$,
written $A_1A_2\genDphi A_3$, if:  
\begin{itemize}
    \item $A_3\cap\AP = A_1 \cap A_2 \cap \AP$ and 
    \item $\reqD(A_3)=\reqD(A_1)\cup \reqD(A_2) \cup \obsD(A_1 )\cup \obsD (A_2)$.
\end{itemize}
\end{definition}

%

It is immediate to show that $A_1A_2\genDphi A_3$ if and only if $A_2A_1\genDphi A_3$ (i.e., the order of the first two components in the ternary relation is irrelevant). 
Notice that the first point of the definition enforces the homogeneity assumption.

The next result, immediately following from Proposition~\ref{prop:unique}, proves that $\genDphi$ expresses a \emph{functional dependency} on $\varphi$-atoms.
\begin{lemma}\label{lem:functional}
Given two $\varphi$-atoms $A_1, A_2\in \Atoms$, there exists \emph{exactly one} $\varphi$-atom  $A_3\in \Atoms$  such that $A_1A_2\genDphi A_3$.
\end{lemma}

Definition~\ref{def:d_generator} and Lemma~\ref{lem:functional} can be exploited to label a homogeneous compass $\varphi$-structure $\cG$, namely, to determine the $\varphi$-atoms labeling all the points $(x,y)$ of $\cG$, starting from the ones on the diagonal.
The idea is the following: if two $\varphi$-atoms $A_1$ and $A_2$ label respectively the greatest proper prefix $[x,y-1]$, that is, the point $(x,y-1)$, and the greatest proper suffix $[x+1,y]$, that is, $(x+1,y)$, of the same interval $[x,y]$,  
then the atom $A_3$ labeling $[x,y]$ is unique, and it is precisely the one satisfying $A_1A_2\genDphi A_3$ (see Figure~\ref{fig:labelling}). The next lemma, proved in Appendix~\ref{proof:lem:compass_hom_gen}, claims that this is the general rule for labeling fulfilling homogeneous compasses.

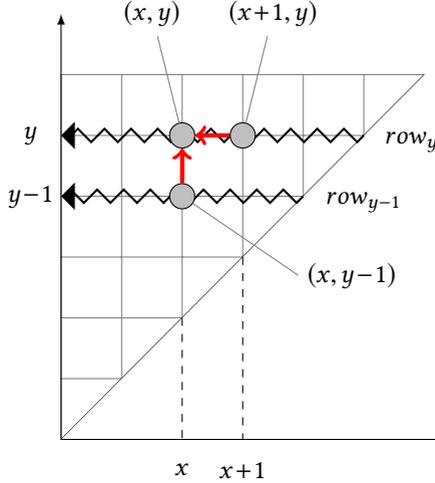
\begin{figure}[tb]
    \centering
    \input{Chaps/ICALP_D/phirows}
    \caption{Rule for labeling homogeneous fulfilling compass $\varphi$-structures}
    \label{fig:labelling}
\end{figure}

\begin{lemma}\label{lem:compass_hom_gen}
Let $\cG=(\bbP_\bbD,\cL)$. $\cG$ is a fulfilling homogeneous compass $\varphi$-structure if and only if, for every pair $x,y \in \mathpzc{S}$,  we have: 
\begin{itemize}
    \item $\cL(x,y-1)\cL(x+1, y)\genDphi \cL(x, y)$ if $x<y$, and 
    \item $\reqD(\cL(x,y))=\emptyset$ if $x=y$.
\end{itemize}
\end{lemma}
%


Now we introduce the concept of $\varphi$-row, which can be viewed as the ordered sequence of (the occurrences of) atoms labelling a row of a compass $\varphi$-structure. Given an atom $A \in \Atoms$, we call it \emph{reflexive} if 
$A \Dphi A$, and \emph{irreflexive} otherwise.
\begin{definition}\label{def:row}
A $\varphi$-row is a finite sequence of $\varphi$-atoms \[\row=A_0^{m_0}\cdots A_n^{m_n},\] where $A^m$ stands for $m$ repetitions of $A$, such that for each $0\leq i \leq n$, we have that $m_i>0$---if $m_i>1$, then $A_i$ is reflexive---and for each $0\leq j<i$, it holds that $A_i \Dphi A_j$,  $A_i \neq A_j$, and $(A_j\cap \AP)\supseteq (A_i\cap \AP)$. Moreover, $\reqD(A_0)=\emptyset$.
\end{definition}

The length of a $\varphi$-row $\row=A_0^{m_0} \cdots A_n^{m_n}$ is defined as $|\row|=\sum_{0\leq i \leq n} m_i$, and for each $0\leq j<|\row|$, the $j$-th element, denoted by $\row(j)$, is the $j$-th symbol in the word $A_0^{m_0} \cdots A_n^{m_n}$, e.g., $\row(0)=A_0$, $\row(m_0)=A_1$, \dots; finally
$\row(i, j)$, for  $0\leq i \leq j < |\row|$, represents the sub-word $\row(i)\row(i+1)\cdots \row(j)$.
We denote by $\Rows$ the set of all possible $\varphi$-rows. This set may be infinite.

The number of distinct atoms in any $\varphi$-row is bounded. 
Since for each $0\leq i \leq n$ and each $0\leq j<i$, $A_i \Dphi A_j$,
it holds that $\reqD(A_j) \subseteq \reqD(A_i)$.  
Therefore, two monotonic sequences for every 
$\varphi$-row can be considered, one increasing, i.e., $\emptyset=\reqD(A_0) \subseteq \reqD(A_1)\subseteq \ldots \subseteq \reqD(A_n)$,
and one decreasing, i.e., $(A_0 \cap \AP)\supseteq (A_1\cap \AP)\supseteq \ldots \supseteq (A_n\cap \AP)$. 
The number of distinct elements is bounded by $|\varphi|$ in the former sequence and by $|\varphi|+1$ in the latter (as $|\REQ_{\varphi}|\leq |\varphi|-1$ and $|\AP|\leq |\varphi|$--w.l.o.g., we can consider only the letters actually occurring in $\varphi$). Since, as already shown (Proposition~\ref{prop:unique}), a set of requests and a set of proposition letters uniquely determine a $\varphi$-atom, any $\varphi$-row may feature at most $2|\varphi|$ distinct atoms, namely, $n<2|\varphi|$.  

Given a homogeneous compass $\varphi$-structure $\cG= (\bbP_\bbD,\cL)$, for every  $y\in \mathpzc{S}$, we define $\row_y$ as the word of $\varphi$-atoms $\row_y=\cL(y,y)\cdots \cL(0, y)$, i.e., the sequence of atoms labeling points of $\cG$ with the same $y$-coordinate, starting from the one on the diagonal \emph{inwards}. See Figure~\ref{fig:labelling}. 

The next result holds; its proof can be found in Appendix~\ref{proof:lem:compas_implies_row}.
\begin{lemma}\label{lem:compas_implies_row}
Let $\cG= (\bbP_\bbD,\cL)$ be a fulfilling homogeneous compass $\varphi$-structure. For every $y\in \mathpzc{S}$, $\row_y$ is a $\varphi$-row.
\end{lemma}

We now define the \emph{successor} relation between pairs of $\varphi$-rows, denoted as $\rownext$, which is basically a component-wise application of $\genDphi$ over the elements of two $\varphi$-rows (remember that atoms on rows are collected right-to-left).
\begin{definition}\label{def:rownext}
Given two $\varphi$-rows $\row$ and $\row'$, we say that $\row'$ is a \emph{successor}
of $\row$, written as $\row \rownext \row'$, if $|\row'| = |\row| +1$, and for all $0\leq i < |\row|$,
$\row(i)\row'(i) \genDphi \row'(i+1)$. 
\end{definition}

The next lemma, whose proof is in Appendix~\ref{proof:lem:row_successor}, states that consecutive  rows in homogeneous fulfilling compass $\varphi$-structures respect the successor relation.
\begin{lemma}\label{lem:row_successor}
Let $\cG=(\bbP_\bbD,\cL)$, with $\reqD(\cL(x,x))=\emptyset$ for all $(x,x)\in\bbP_\bbD$. $\cG$ is a fulfilling homogeneous compass $\varphi$-structure
if and only if, for each $0\leq y < |\mathpzc{S}| -1$, $\row_y \rownext \row_{y+1}$.
\end{lemma}

\begin{figure}[tb]
    \centering
    \resizebox{\linewidth}{!}{\input{Chaps/ICALP_D/figRank}}
\caption{A graphical account of the proof of Lemma~\ref{lem:rank}}\label{fig:rank}
\end{figure}
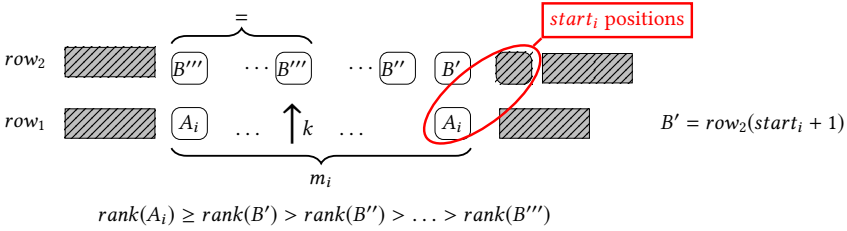

Given an atom $A \in \Atoms$, we define the \emph{rank of $A$}, written $\rank(A)$, as 
$|\REQ_\varphi| - |\reqD(A)|$.  Clearly, $\rank(A)< |\varphi|$. Whenever $A \Dphi A'$, for some $A' \in \Atoms$, $\reqD(A')\subseteq \reqD(A)$, and thus  $\rank(A)\leq \rank(A')$ and $|\reqD(A)\setminus \reqD(A')|\leq \rank(A')$.
We can see the $\rank$ of an atom as the ``number of degrees of freedom'' 
that it gives to 
the atoms  that stay ``above it''.
In particular, by definition, for every $\varphi$-row $\row=A_0^{m_0} \cdots A_n^{m_n}$, we have $\rank(A_0)\geq \ldots \geq  \rank(A_n)$.
The next lemmas use the notion of rank to provide an insight on how consecutive $\varphi$-rows are connected
(see Figure~\ref{fig:rank}).

\begin{lemma}\label{lem:rank0}
Let $\row_1$ and $\row_2$ be two $\varphi$-rows, with
$\row_1= A_0^{m_0}\cdots A_n^{m_n}$ and $\row_1\rownext \row_2$. 
For each $0\leq i\leq n$, let $st_i=\sum_{0\leq j <i} m_j$.
If, for some $st_i<j\leq st_i + m_i $, the atom $\row_2(j)$ is reflexive, then 
for each $j\leq j'\leq st_i + m_i$, $\row_2(j')=\row_2(j)$.
\end{lemma}
\begin{proof}
If $j=st_i+m_i$ there is nothing to prove. Thus we consider $j<st_i+m_i$.
Since $\row_2(j)$ is reflexive,   
$\obsD(\row_2(j))\subseteq \reqD(\row_2(j)) $. 
%
Since $\row_1 \rownext \row_2 $, we have that $\reqD(A_i),\obsD(A_i)\subseteq\reqD(\row_2(j))$ 
and $\reqD(\row_2(j+1))=\reqD(\row_2(j))\cup\obsD(\row_2(j))\cup\reqD(A_i)\cup\obsD(A_i)=\reqD(\row_2(j))$.
Moreover, from $\row_1 \rownext \row_2 $, we have $\row_2(j)\cap \AP = \row_2(j-1) \cap A_i\cap \AP$ and $\row_2(j+1)\cap \AP = \row_2(j) \cap A_i\cap \AP=\row_2(j-1) \cap A_i\cap \AP$. 
Thus, $\row_2(j+1) = \row_2(j)$, because the two atoms feature exactly the same requests and proposition letters (Proposition~\ref{prop:unique}).
Then,
since $A_i \ \row_2(j)\genDphi \row_2(j+1)$, 
by iterating the reasoning and exploiting Lemma~\ref{lem:functional}, we can conclude that
$\row_2(j)=\row_2(j')$ for each $j\leq j'\leq st_i + m_i$.
\end{proof}

\begin{lemma}\label{lem:rank}
Let $\row_1$ and $\row_2$ be two $\varphi$-rows, with
$\row_1= A_0^{m_0}\cdots A_n^{m_n}$ and $\row_1\rownext \row_2$. 
For each $0\leq i\leq n$, let $st_i=\sum_{0\leq j <i} m_j$.
If $m_i > rank(A_i)$, then there exists $st_i<k\leq st_i + m_i $
such that:
\begin{itemize}
\item $k\leq st_i+1+\rank(A_i)$;
\item $\row_2(k)$ is reflexive;
\item $\rank(\row_2(j))>\rank(\row_2(j+1))$ for each $st_i<j< k$; 
\item $\row_2(j)=\row_2(j+1)$ for each $k \leq j< st_i + m_i$;
\item if $m'$ is the exponent of the atom $\row_2(k)$ in $\row_2(st_i+1, st_i+m_i)$ then  $m'>  \rank(\row_2(k)) $.
\end{itemize}
\end{lemma}
\begin{proof}
If $m_i=1$, by hypothesis we have $ rank(A_i) = 0$. Hence, $rank(\row_2(st_i+1)) = 0$, because $\row_1\rownext \row_2$, and thus $\row_2(st_i+1)$ is (trivially) reflexive. All claims hold by choosing $k=st_i+1$. 

Let us then assume $m_i>1$. 
It can be easily shown that if we have two atoms $A$ and $A'$ such that $A\Dphi A'$ and $A'$ is irreflexive, then $rank(A)< rank(A')$.
Moreover, Lemma \ref{lem:rank0} proves that we cannot interleave reflexive atoms with irreflexive ones ``above'' the $A_i$'s (i.e., all irreflexive atoms must ``come before'' reflexive ones in the part of $row_2$ ``above'' the $A_i$'s).
Thus, in the worst case, the atoms $\row_2(st_i+1),\ldots ,\row_2(st_i+rank(A_i))$
may be irreflexive (as $rank(\row_2(st_i+1))>\ldots >rank(\row_2(st_i+rank(A_i)))$ and $rank(A_i)\geq rank(\row_2(st_i+1))$). Note that these irreflexive atoms may be the ``first''  $rank(A_i)$
atoms above the $A_i$'s only, and not the ``first'' $rank(A_i)+1$,
since any atom with rank equal to $0$ is reflexive. 
%
We conclude that $\row_2(st_i+rank(A_i)+1)$ must be reflexive. 
Thus, we can choose $k=st_i+rank(A_i)+1$. Since by hypothesis $m_i\geq rank(A_i)+1$, we get that
$st_i<k\leq st_i+m_i$. Finally, by Lemma \ref{lem:rank0}, $\row_2(j)=\row_2(j+1)$ for each $k \leq j< st_i + m_i$.

As for the last claim,  we have $rank(row_2(k))\leq rank(row_2(st_i+1)) - (k - st_i -1) \leq rank(A_i) - (k - st_i -1)$. 
Then, the exponent $m'$ of $row_2(k)$ in $\row_2(st_i+1, st_i+m_i)$ is such that
$m'\geq m_i - (rank(A_i) - rank(\row_2(k)))$,
that is, at least $m_i - (rank(A_i) - rank(\row_2(k)))$ atoms 
labelled by $\row_2(k)$ occur in the block $st_i+1,\ldots, st_i + m_i$  
of $\row_2$ (see Figure~\ref{fig:rank}). 
Since by hypothesis $m_i>rank(A_i)$, then $m_i-rank(A_i)>0$ and 
$rank(\row_2(k)) < m'$.
\end{proof}

Now we introduce an equivalence relation $\sim$ over $\Rows$ which is the key ingredient of the proofs showing that 
both SAT and MC for $\hsDhom$ formulas are decidable.
\begin{definition}\label{def:equivalence_class}
Given two $\varphi$-rows $\row_1=A_0^{m_0} \cdots A_n^{m_n}$ and $\row_2=
\hA_0^{\hm_0} \cdots \hA_{\hn}^{\hm_{\hn}}$, we say that they are \emph{equivalent}, written 
$\row_1 \sim \row_2$,
if 
\begin{itemize}
    \item $n=\hn$, and 
    \item for each $0\leq i \leq n$, $A_i =\hA_i$, and ${m_i=\hm_i}$ or
    both $m_i$ and $\hm_i$ are (strictly) greater than $\rank(A_i)$.
\end{itemize}
\end{definition}
Note that if two rows feature the same set of atoms, the lower the rank of an atom 
$A_i$, the lower the number of occurrences of $A_i$ both the rows have to feature in order to belong to the same equivalence class.
As an example, let $\row_1$ and $\row_2$ be two rows with $\row_1=A_0^{m_0}A_1^{m_1}$,
$\row_2=A_0^{\om_0}A_1^{\om_1}$, $\rank(A_0) = 4$, 
and $\rank(A_1)= 3$. If $m_1 = 4$ and $\om_1=5$, they are both greater than $\rank(A_1)$, and
thus they do not violate the condition for $\row_1 \sim \row_2$. On the other hand,
if $m_0 = 4$ and $\om_0=5$, we have that $m_0$  is less than or equal to $\rank(A_0)$. Thus, in this case, $\row_1 \not\sim \row_2$ due to the indexes of $A_0$. This happens because $\rank(A_0)$ is greater than $\rank(A_1)$. Two cases in which $\row_1 \sim \row_2$ 
are $m_0 = \om_0$ and $m_0, \om_0 \geq 5$. 

The relation $\sim$ has finite index, which is roughly bounded  by  the number of  all the possible $\varphi$-rows $\row=A_0^{m_0}\cdots A_n^{m_n}$, with exponents $m_i$ ranging from  $1$ to $ |\varphi|$. Since $(i)$~the number of possible atoms is $2^{|\varphi|}$, $(ii)$~the number of \emph{distinct} atoms in any $\varphi$-row is at most $2|\varphi|$, and $(iii)$~the number of possible functions
$f: \{1,\ldots , \ell\} \rightarrow \{ 1,\ldots ,|\varphi|\}$ is $|\varphi|^{\ell}$, we have that the number of distinct equivalence classes of $\sim$ is bounded by
\[
    \sum_{j=1}^{2|\varphi|} (2^{|\varphi|})^j\cdot |\varphi|^j\leq  2^{3|\varphi|^2},
\]
which is exponential in the length of the input formula 
$\varphi$. We denote the set of equivalence classes of $\sim$ over all the possible $\varphi$-rows by $\Rows^\sim$. 

Now we extend $\rownext$ to equivalence classes of $\sim$ in the following way. 
\begin{definition}\label{def:row_class_suc}
Given two $\varphi$-row
classes $[\row_1]_\sim$ and $[\row_2]_\sim$, we say that $[\row_2]_\sim$ is a successor of  $[\row_1]_\sim$, 
written $[\row_1]_\sim\rownext [\row_2]_\sim$, if there exist 
$\row_1' \!\in\! [\row_1]_\sim$ and $\row_2' \!\in\!  [\row_2]_\sim$ such that 
$\row_1' \rownext \row_2'$.
\end{definition}

The following fundamental result proves that if some $\row_1' \in [\row_1]_\sim$ has a successor in $[\row_2]_\sim$, 
then \emph{every $\varphi$-row} of $[\row_1]_\sim$ has a successor in $[\row_2]_\sim$ (hence the definition of $\rownext$ over the equivalence classes of $\sim$ is well-given). 

\begin{lemma}\label{lem:row_class_suc}
Given two $\varphi$-row
classes $[\row_1]_\sim$ and $[\row_2]_\sim$ such that 
$[\row_1]_\sim\rownext [\row_2]_\sim$, \emph{for every} $\row\!\in\! [\row_1]_\sim$ there exists 
$\row'\!\in\! [\row_2]_\sim$ such that $\row\rownext \row'$.
\end{lemma}

Before starting the proof we observe that,
since $[\row_1]_\sim
\rownext [\row_2]_\sim$, 
there exists $\row \in [\row_1]_\sim$
and $\orow\in [\row_2]_\sim$ such that $\row \rownext \orow$. Thus, if $|[\row_1]_\sim|=1$, 
the lemma trivially holds. 

In the following we suppose that $|[\row_1]_\sim|\geq 2$: then there is $\row' \neq \row$ such that 
$\row' \sim \row$. We let $\row = A_0^{m_0}\cdots A_n^{m_n}$; by definition we have
$\row' =A_0^{m_0'}\cdots A_n^{m_n'}$, where for every $0\leq i \leq n$,
$m_i'= m_i$ if  $m_i\leq \rank(A_i)$, and $m_i'> \rank(A_i)$ if $m_i > \rank(A_i)$.

Hereafter, we will say that \lq\lq $A_i$ exceeds its rank in $\row$\rq\rq{}
whenever $m_i> \rank(A_i)$.
Moreover, for the sake of the proof it is worth observing that any sub-word $\row(0, i)$ is a $\varphi$-row for every $0\leq i < |\row|$, and, 
since $\row\rownext \orow$, then $\row(0, i)\rownext \orow(0, i+1)$.
 
We now want to determine 
$\orow'$ such that 
$\row' \rownext \orow' $  and $\orow'\sim \orow$, proving the lemma.
To this aim, we define a list of $\varphi$-rows $\orow'_{-1}, \orow'_0,\ldots ,\orow'_n$ such that
$\orow'_{-1}=\orow(0)$ and,
for all $0\leq i\leq n$, $\orow'_i$ extends $\orow'_{i-1}$; we will show that $\orow'_n=\orow'$ (i.e., the row we want to determine).

The $\varphi$-rows $\orow'_i$, for $i\geq 0$, are formally defined as:
\begin{small}
\[
\orow'_i =
\begin{cases} 
   \orow'_{i-1} \cdot \orow(st_i+1, st_i+m_i) 		& \text{if } m_i=m'_i \\
   \orow'_{i-1} \cdot \orow(st_i+1, st_i+m'_i)		& \text{if } \rank(A_i)\!<\! m'_i\!<\! m_i \\
   \orow'_{i-1} \cdot \orow(st_i+1, st_i+m_i) \cdot (\orow(st_i+m_i))^{m'_i-m_i}	& \text{if } \rank(A_i)\!<\! m_i\!<\! m'_i 
  \end{cases}
\]
\end{small}
Intuitively, if $m_i=m'_i$, $\orow'_i$ is obtained just by appending the block $\orow(st_i+1, st_i+m_i)$ to $\orow'_{i-1}$. If $\rank(A_i)<m'_i<m_i$, we do almost the same thing, but we ``truncate'' the block after $m'_i$ positions, instead of $m_i$. Finally, if $\rank(A_i)<m_i<m'_i$, then $\orow'_i$ is obtained by appending $\orow(st_i+1, st_i+m_i)$ to $\orow'_{i-1}$, with $m'_i-m_i$ additional repetitions of $\orow(st_i+m_i)$ at the end, as we need to ``pump'' $\orow'_i$ in order for it to have a suitable length with respect to $\row'$, thus respecting $\rownext$.

We are now ready to prove Lemma~\ref{lem:row_class_suc}
\begin{proof}
We show that the following properties hold by induction on $0\leq i\leq n$, where 
$st_i=\sum_{0\leq j <i} m_j$ and $st'_i=\sum_{0\leq j <i} m'_j$ are respectively the starting positions of $A_i$ in $\row$ and $\row'$:
\begin{enumerate}
    \item[0.] $\orow'_i$ is a $\varphi$-row featuring atoms of $\orow(0, st_i+m_i)$;
	\item $\orow'_i\sim \orow(0, st_i+m_i)$;
	\item $\row'(0, st_i'+m_i'-1)\rownext \orow_i'$.
\end{enumerate}
In particular, for $i=n$, the properties allow us to show that $\orow'_n\sim \orow$ and $\row'\rownext \orow'_n$, proving the lemma.
Figure \ref{fig:generat} illustrates such properties.

\begin{sidewaysfigure}
    \resizebox{\linewidth}{!}{\includegraphics{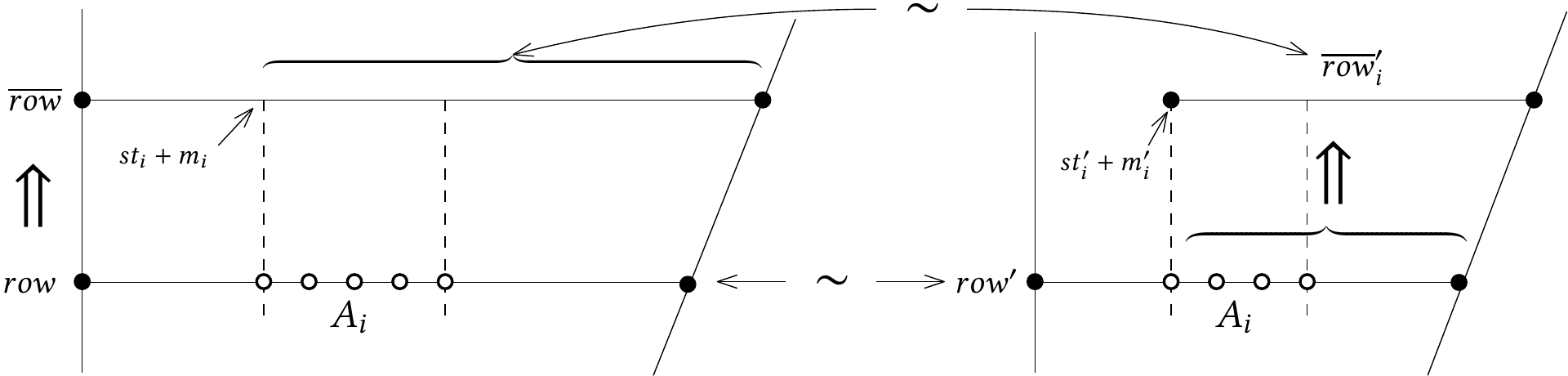}}
    \caption{Illustration of the properties of $\orow'_i$.}\label{fig:generat}
\end{sidewaysfigure}

The base case for $i=0$ is omitted, as it is just a simplification of the inductive step.
Let us consider $i\geq 1$. By the inductive hypothesis, we have $\orow'_{i-1}\sim \orow(0, st_{i-1}+m_{i-1})$ (hence in particular $\lst(\orow'_{i-1})=\orow(st_{i-1}+m_{i-1})$, where $\lst(\row)$ denotes the last atom of $\row$) and $\row'(0, st_{i-1}'+m_{i-1}'-1)\rownext \orow_{i-1}'$.
We consider three cases, corresponding to the cases of the definition of $\orow_i'$.
\begin{itemize}
	\item $m_i=m_i'$. Then $\orow'_i = \orow'_{i-1} \cdot \orow(st_i+1, st_i+m_i)$ is a $\varphi$-row. If $\lst(\orow'_{i-1})\neq (\orow(st_i+1))$, then it immediately follows that $\orow'_i\sim \orow(0, st_{i}+m_{i})$. 
	Conversely, if $\lst(\orow'_{i-1})=\orow(st_i+1)=\orow(st_{i-1}+m_{i-1})$, then $\orow(st_i+1)$ is reflexive. By Lemma~\ref{lem:rank0}, $\orow(st_i+1)=\ldots = \orow(st_i+m_i)$. It follows that $\orow'_i\sim \orow(0, st_{i}+m_{i})$, because we append to $\orow'_{i-1}$ exactly $\orow(st_i+1, st_i+m_i)$ (note that either $\orow(st_{i-1}+m_{i-1})$ already exceeded its rank in $\orow(0, st_{i-1}+m_{i-1})$ and so did $\lst(\orow_{i-1}')$ in $\orow_{i-1}'$, or the ranks of $\orow(st_{i-1}+m_{i-1})$ in $\orow(0, st_{i-1}+m_{i-1})$ and of $\lst(\orow_{i-1}')$ in $\orow_{i-1}'$ were equal).
	Moreover, in both cases, $\row'(0, st_i'+m_i'-1)\rownext \orow_i'$ by definition of $\genDphi$, as $\row'(st'_i)=\row(st_i)=A_i$ (recall that $\row\sim\row'$). 
	
	\item $\rank(A_i)<m'_i<m_i$. Then $\orow'_i = \orow'_{i-1} \cdot \orow(st_i+1, st_i+m'_i)$ is a $\varphi$-row. 
	First, we have $\row'(0, st_i'+m_i'-1)\rownext \orow_i'$ by definition of $\genDphi$, as $\row'(st'_i)=\row(st_i)=A_i$. 
	By Lemma~\ref{lem:rank}, being $\rank(A_i)<m_i$, there exists $st_i+1\leq k\leq st_i+m_i$ such that $\orow(k)$ exceeds its rank in $\orow(st_i+1, st_i+m_i)$, $\orow(k)=\ldots =\orow(st_i+m_i)$, and $k\leq st_i+1+\rank(A_i)$. We first note that, being $1+\rank(A_i)\leq m'_i$, we have $k\leq st_i+m'_i$, thus the atom $\orow(k)$ is present in the block $\orow(st_i+1, st_i+m'_i)$ of $\orow'_i$. 
	By Lemma~\ref{lem:rank}, being also $\rank(A_i)<m'_i$, we have that the atom $\orow(k)$ exceeds its rank in the block  $\orow(st_i+1, st_i+m'_i)$ of $\orow'_i$ as well. Thus $\orow'_i\sim \orow(0, st_i+m_i)$.
	
	\item $\rank(A_i)<m_i<m'_i$. Then $\orow'_i = \orow'_{i-1} \cdot \orow(st_i+1, st_i+m_i) \cdot (\orow(st_i+m_i))^{m'_i-m_i}$. By Lemma~\ref{lem:rank}, being $\rank(A_i)<m_i$, there exists $st_i+1\leq k\leq st_i+m_i$ such that $\orow(k)$ is reflexive, it exceeds its rank in $\orow(st_i+1, st_i+m_i)$, and $\orow(k)=\ldots =\orow(st_i+m_i)$.
	First we observe that $\orow'_i$ is a $\varphi$-row, being $\orow(st_i+m_i)$ reflexive. Then
	the atom $\orow(k)$ is (trivially) also present in the block $\orow(st_i+1, st_i+m_i)$ of $\orow'_i$. As a consequence $\orow'_i\sim \orow(0, st_i+m_i)$. Moreover, $A_i\, \orow(st_i+m_i) \genDphi \orow(st_i+m_i)$ (as $\orow(st_i+m_i)$ is reflexive), therefore $\row'(0, st_i'+m_i'-1)\rownext \orow_i'$.
	\qedhere
\end{itemize}
\end{proof}

The following result arranges the equivalence classes $\Rows^\sim$ in a graph $G_{\varphi \sim}$.
\begin{definition}\label{def:equivalencegraph}
Let $\varphi$ be a $\hsDhom$ formula. The \emph{$\varphi\!\sim$graph} of $\varphi$ is defined as $G_{\varphi \sim}=(\Rows^\sim,\rownext)$.
\end{definition}

The next theorem reduces the problem of SAT checking for a $\hsDhom$ formula $\varphi$ over finite linear orders (equivalent, by Proposition~\ref{prop:satiffcompass}, to deciding if there is a homogeneous fulfilling compass $\varphi$-structure that features $\varphi$) to a reachability problem in the $\varphi\!\sim$graph, allowing us to determine the computational complexity of the former problem.
\begin{theorem}\label{thm:path_iff_sat}
Given a $\hsDhom$ formula $\varphi$, there exists a homogeneous fulfilling compass $\varphi$-structure $\cG=(\bbP_\bbD, \cL)$ that features $\varphi$
if and only if there exists a path  in $G_{\varphi \sim}=(\Rows^\sim,\rownext)$
from some class $[\row]_\sim\in\Rows^\sim$ to some class $[\row']_\sim\in\Rows^\sim$ such that
\begin{enumerate}
    \item there exists $row_1\in[\row]_\sim$ with $|row_1|=1$, and
    \item there exist $row_2\in[\row']_\sim$ and $0\leq i<|row_2|$ such that $\varphi\in row_2(i)$.
\end{enumerate}
\end{theorem} 
\begin{proof}
Preliminarily we observe that, in (1.), if $|row_1|=1$, then $\{row_1\}=[\row]_\sim$; moreover, in (2.), if for $row_2\in[\row']_\sim$ and $0\leq i<|row_2|$ we have that $\varphi\in row_2(i)$, then for any $row_2'\in[\row']_\sim$, there is $0\leq i'<|row_2'|$ such that $\varphi\in row_2'(i')$.

($\Rightarrow$) 
Let us consider a homogeneous fulfilling compass $\varphi$-structure $\cG=(\bbP_\bbD, \cL)$ that features $\varphi$.
By Lemma~\ref{lem:compas_implies_row} and \ref{lem:row_successor}, 
$\cL(0,0) \rownext \allowbreak  \row_1 \rownext \cdots\linebreak \rownext \row_{\max(\mathpzc{S})}$. Thus
there exist two indexes $0\leq j\leq \max(\mathpzc{S})$ and $0\leq i<|row_j|$ for which $\varphi \in \row_j(i)$. 
By Definition~\ref{def:row_class_suc}, we get that $[\cL(0,0)]_\sim \rownext \allowbreak  [\row_1]_\sim \rownext \cdots \rownext [\row_{j}]_\sim$ is a path in $G_{\varphi \sim}$; it is immediate to check that it fulfills requirements (1.) and (2.).

($\Leftarrow$) Let us assume that there exists a path $[\row_0]_\sim 
\rownext \cdots \rownext [\row_{m}]_\sim$ in $G_{\varphi\sim}=\allowbreak (\Rows^\sim, \rownext)$
for which $|\row_0| = 1$ and there exists $i$ such that $\varphi\in \row_m(i)$. 
By applying repeatedly Lemma~\ref{lem:row_class_suc}
we get that there exists a sequence 
$\row'_0 \rownext \cdots \rownext \row'_{m}$ of $\varphi$-rows where $\row'_0 =\row_0$,
for every $0\leq j\leq m$, $\row'_j\in [\row_j]_\sim$,
and there exists $i'$ such that $\varphi\in \row_m'(i')$.
We observe that, by Definition~\ref{def:rownext},
$|\row_j'|=|\row_{j-1}'|+1$
 for $1\leq j \leq m$
and, since $|\row'_0|=1$, we have $|\row_{j}'|= j + 1$. 
Let us now define $\cG=(\bbP_\bbD, \cL)$ where $\mathpzc{S}= \{0,\ldots, m\}$ and
$\cL(x,y)=\row_y'(y-x)$ for every $0\leq x\leq y\leq m$.
By Lemma~\ref{lem:row_successor}, $\cG$ is a fulfilling homogeneous compass $\varphi$-structure.
Finally, since $\varphi\in \row_m'(i')$, $\cG$ features $\varphi$.
\end{proof}

The size of $G_{\varphi\sim}=(\Rows^\sim,\rownext)$ is bounded by $|\Rows^\sim|^2$, which is exponential 
in $|\varphi|$. However, 
it is possible to (non-deterministically) perform a reachability in $G_{\varphi \sim}$ by using space \emph{logarithmic} in $|\Rows^\sim|^2$.

\begin{algorithm}[b]
\begin{algorithmic}[1]
\caption{\texttt{SAT}$(\varphi)$}\label{NDAlgo}
    \State{$M\gets 2^{3|\varphi|^2}$, $step\gets 0$ and $\row\gets A$ for some atom $A\in \Atoms$ with  $\reqD(A)=\emptyset$}
    \If{there exists $0\leq i<|\row|$ such that $\varphi\in \row(i)$}
        \State{\textbf{return} ``satisfiable''}
    \EndIf
    \If{$step = M-1$}
        \State{\textbf{return} ``unsatisfiable''}
    \EndIf
    \State{Non-deterministically generate a $\varphi$-row $\row'$ and check that $\row \rownext \row'$}
    \State{$step \gets step +1$ and $\row \gets\row'$}
    \State{Go back to line 2}
\end{algorithmic}
\end{algorithm}

The \emph{non-deterministic} procedure \texttt{SAT}$(\varphi)$ in Algorithm~\ref{NDAlgo} exploits this fact in order to decide the satisfiability 
of a $\hsDhom$ formula $\varphi$, by using only a working space \emph{polynomial} in $|\varphi|$: it searches for a suitable path in $G_{\varphi \sim}$, namely  $[\row_0]_\sim \rownext \cdots \allowbreak \rownext [\row_{m}]_\sim$, where $\row_0 =A$ for $A\in \Atoms$ with  $\reqD(A)=\emptyset$, $m< M$, and $\varphi\in \row_m(i)$ for $0\leq i<|\row_m|$. At the $j$-th iteration of line 6., $\row_{j}$ is non-deterministically generated, and it is checked whether $\row_{j-1}\rownext\row_{j}$.
The procedure terminates after at most $M$ iterations, where $M$ is the maximum possible length of a simple path in $G_{\varphi\sim}$.

The working space used by the procedure is polynomial:
$M$
and $step$ (which ranges in $[0,M-1]$) can be encoded 
in binary with $\lceil \log_2 M \rceil +1=O(|\varphi|^2)$ bits.  
At each step, we need to keep track of two $\varphi$-rows at a time, the current one, $\row$, and its successor, $\row'$: each $\varphi$-row can be represented as a sequence of at most $2|\varphi|$ (distinct) atoms,
each one with an exponent that, by construction, cannot exceed $M$. 
Moreover, each $\varphi$-atom $A$ can be represented using exactly $|\varphi|$ bits (for each
$\psi \in \CL(\varphi)$, we set a bit to 1 if $\psi\in A$, and to 0 if $\neg\psi\in A$). Hence a $\varphi$-row can be encoded using $2|\varphi|\cdot(|\varphi|+\lceil \log_2 M \rceil +1)=O(|\varphi|^3)$ bits. 
Finally, the condition $\row \rownext \row'$ can be checked
by $O(|\varphi|^2)$ bits of space once we have guessed $\row'$. 
This analysis entails the following result (we recall that $\NPsp=\Psp$).
\begin{theorem}\label{thm:pspace}
The SAT problem for $\hsDhom$ formulas over finite linear orders is in $\Psp$.
\end{theorem}

We now outline which are the modifications to the previous concepts needed for proving the 
decidability of SAT for $\hsDhom${} with the strict relation $\ssubint$, in place of $\subint$.
It is sufficient to replace the definitions
of $\genDphi$,  $\varphi$-$\row$ and $\rownext$
with the following ones. 
For the sake of simplicity, we introduce a dummy atom $\boxdot$, for which we assume $\reqD(\boxdot)=\obsD(\boxdot)=\emptyset$.

\begin{definition}\label{def:d_generatorS}
Given four $\varphi$-atoms $A_1, A_3,A_4\in\Atoms$ and $A_2\in\Atoms\cup\{\boxdot\}$, we say that 
$A_4$ is \mbox{$\Dphi\ssubint$-generated} by $A_1,A_2,A_3$,
written $A_1,A_2, A_3\genDphiS A_4$, if:  
\begin{itemize}
    \item $A_4\cap\AP = A_1 \cap A_3 \cap \AP$ and 
    \item $\reqD(A_4)=\reqD(A_1)\cup \reqD(A_3) \cup \obsD(A_2 )$.
\end{itemize}
\end{definition}
The idea of this definition is that, if an interval $[x,y]$, with $x<y$, is labeled by $A_4$, and the three subintervals $[x,y-1]$, $[x+1,y-1]$, and $[x+1,y]$ by $A_1$, $A_2$, $A_3$, respectively, we want $A_1,A_2, A_3\genDphiS A_4$. In particular, if $x=y-1$, then $A_2=\boxdot$ (because $[x+1,y-1]$ is not a valid interval). Note that only $[x+1,y-1]\ssubint[x,y]$, hence we want $\obsD(A_2)\subseteq\reqD(A_4)$; moreover, since the requests of $A_1$ and $A_3$ are fulfilled by a strict subinterval of $[x,y]$, it must be $\reqD(A_1)\subseteq\reqD(A_4)$ and $\reqD(A_3)\subseteq\reqD(A_4)$. 

\begin{definition}\label{def:rowS}
A $\varphi$-$\ssubint$-row is a finite sequence of $\varphi$-atoms \[\row=A_0^{m_0}\cdots A_n^{m_n},\]
such that for each $0\leq i \leq n$, we have $m_i>0$, and for each $0\leq j<i$, it holds $\reqD(A_j) \subseteq \reqD (A_i)$, $A_i \neq A_j$, and $(A_j\cap \AP)\supseteq (A_i\cap \AP)$.
 Moreover, $\reqD(A_0)=\emptyset$.
\end{definition}

\begin{definition}\label{def:rownextS}
Given two $\varphi$-rows $\row$ and $\row'$, we say that $\row'$ is a successor
of $\row$, written as $\row\rownextS\row'$, if $|\row'| = |\row| +1$, and for all $0\leq i < |\row|$,
$\row(i)\row(i-1)\row'(i) \genDphiS \row'(i+1)$, where we assume 
$\row(i-1)=\boxdot$ if $i=0$. 
\end{definition}

We will come back to SAT later, showing the $\Psp$-completeness of such problem  (under both the \emph{strict} and the \emph{proper} semantic variants). In the next section, we will consider MC for $\hsDhom$.

%% file: Chaps/ICALP_D/phirows.tex
\begin{tikzpicture}[scale=0.8]

\draw[step=1cm,gray, thin] (-3,-3) grid (3,3);

\draw (-3,-3) -- (3,3);
\fill[color=white] (-3,-3) node (v1) {} -- (3.1,3.1) -- (3.1,-3);

\draw [-latex](-3,-3) -- (4,-3);
\draw [-latex](-3,-3) -- (-3,4);

\draw[thick,- triangle 90, decorate, decoration=zigzag] (2,2) -- (-3,2);
\draw[thick,- triangle 90, decorate, decoration=zigzag] (1,1) -- (-3,1);

\node[draw,circle,fill=lightgray] (v3) at (-1,2) {};
\node[draw,circle,fill=lightgray] (v2) at (-1,1) {};
\node[draw,circle,fill=lightgray] (v4) at (0,2) {};


\node (v6) at (-1.5,4) {$(x,y)$};
\node (v5) at (0.5,4) {$(x\!+\!1,y)$};
\node (v7) at (1.8,-0.3) {$(x,y\!-\! 1)$};

\draw [dashed](-1,-1) -- (-1,-3);
\draw [dashed](0,0) -- (0,-3);

\node at (-1,-3.5) {$x$};
\node at (0,-3.5) {$x\!+\!1$};
\node at (-3.5,2) {$y$};
\node at (-3.5,1) {$y\!-\! 1$};

\node at (2,0.9) {$row_{y - 1}$};
\node at (2.8,1.9) {$row_y$};

\draw [gray] (v5) edge (v4);
\draw [gray] (v6) edge (v3);
\draw [gray] (v7.west) edge (v2);

\draw [red,ultra thick,<-] (v3) edge (v2);
\draw [red,ultra thick,<-] (v3) edge (v4);
\end{tikzpicture}

%% file: Chaps/ICALP_D/figRank.tex
\begin{tikzpicture}

\node[inner sep=0pt, label={[label distance=0.3cm]180:$\row_1$}](R1A) {\phantom{$A$}};
\node[inner sep=0pt, right of=R1A](R1B) {\phantom{$A$}};
\node[fit=(R1A)(R1B), draw, fill=gray!50] (R1){};
 \fill[pattern=north east lines] (R1.north west) rectangle (R1.south east);

\node(AMI)[draw, rounded corners, right of=R1B,node distance=0.8cm]{$A_i$};
\node(AD)[right of=AMI,node distance=1.8cm]{$\ldots\quad${\huge $\uparrow$}$k\quad\ldots$};
\node(A1)[draw, rounded corners,right of=AD,node distance=2.5cm]{$A_i$};

\node(BJ)[draw, label={[label distance=-0.6cm]0:$B'$}, rounded corners,above of=A1,node distance=0.9cm]{$\phantom{A_i}$};
\node(BJ1)[draw , label={[label distance=-0.7cm]0:$B''$},rounded corners,left of=BJ,node distance=0.9cm]{$\phantom{A_i}$};
\node(BD)[left of=BJ1,node distance=0.6cm]{$\ldots$};
\node(BJ1K)[draw , label={[label distance=-0.7cm]0:$B'''$},rounded corners,left of=BD,node distance=1.1cm]{$\phantom{A_i}$};
\node(BD)[left of=BJ1K,node distance=0.6cm]{$\ldots$};
\node(BJ1KE)[draw , label={[label distance=-0.7cm]0:$B'''$},rounded corners,left of=BD,node distance=1.1cm]{$\phantom{A_i}$};
\draw[decorate, decoration={brace, raise=0.4cm, amplitude=0.25cm}, line width=0.3mm] (BJ1KE.west) -- (BJ1K.east)node[pos=0.5, above, yshift=0.6cm]{$=$};

\node[inner sep=0pt, label={[label distance=0.3cm]180:$ $}](R11A) [right of=A1] {\phantom{$A$}};
\node[inner sep=0pt, right of=R11A](R11B) {\phantom{$A$}};
\node[fit=(R11A)(R11B), draw, fill=gray!50] (R11){};
 \fill[pattern=north east lines] (R11.north west) rectangle (R11.south east);

\node(SB)[draw, rounded corners,node distance=0.8cm, fill=gray!50] [above of=R11A, node distance=0.9cm]{$\phantom{A_i}$};
 \fill[pattern=north east lines] (SB.north west) rectangle (SB.south east);

\node[inner sep=0pt, label={[label distance=0.3cm]180:$ $}](R21A) [right of=SB, node distance=0.7cm] {\phantom{$A$}};
\node[inner sep=0pt, right of=R21A](R21B) {\phantom{$A$}};

\node[inner sep=0pt, label={[label distance=0.3cm]180:$\row_2$}](R2A) [above of=R1A, node distance=1cm] {\phantom{$A$}};
\node[inner sep=0pt, right of=R2A](R2B) {\phantom{$A$}};
\node[fit=(R2A)(R2B), draw, fill=gray!50] (R2){};
 \fill[pattern=north east lines] (R2.north west) rectangle (R2.south east);
\node[fit=(R21A)(R21B), draw, fill=gray!50] (R21){};
 \fill[pattern=north east lines] (R21.north west) rectangle (R21.south east);

\draw[decorate, decoration={brace, mirror, raise=0.4cm, amplitude=0.25cm}, line width=0.3mm] (AMI.west) -- (A1.east)node[pos=0.5, below, yshift=-0.7cm]{$m_i$};

\path let
         \p1 = ($(SB.north east)-(A1.south west)+(1,0)$),
         \n1 = {veclen(\p1)}
         in
         (SB.south) -- (A1.north) 
         node[midway, sloped, draw, red, ellipse, line width=0.4mm,
          minimum width=\n1/1.29, rotate=18, minimum height=\n1/3.2, xshift=0.0cm, yshift=0.03cm](STARTS) {};
\node(STARTS_L)[above right of=STARTS, yshift=0.5cm, xshift=1.5cm, draw, line width=0.4mm, red]{$start_i$ positions};
\draw[red, line width=0.4mm] (STARTS) -- (STARTS_L.south west);
\pgftransformshift{\pgfpoint{4cm}{-1.5cm}}
\node{$\begin{array}{c}
rank(A_i)\geq rank(B')> rank(B'') > \ldots > rank(B''') 
\end{array}$};
\pgftransformshift{\pgfpoint{7cm}{1.5cm}}
\node{$\begin{array}{c}
B' = \row_2(start_i+1)\end{array}$};
\end{tikzpicture}

%% file: Chaps/ICALP_D/D_modelchecking.tex
\section{MC for $\hsDhom${} over Kripke structures} \label{sec:mc}

In this section we focus our attention on the MC problem for $\hsDhom$ formulas over Kripke structures, under homogeneity.
Let us start with the following (technical) definition, basically mapping traces of a Kripke structure into interval models.
\begin{definition}\label{def:trimodel}
The interval model
$\bM_\rho=(\mathbb{I(S)},\circ,\cV)$ \emph{induced by a trace $\rho$} of a finite Kripke structure $\Ku=\KuDef$
is the homogeneous interval model such that:
\begin{enumerate}
    \item $\mathpzc{S}=\{0,\ldots,|\rho|-1\}$, and
    \item for all $x\in \mathpzc{S}$ and $p\in\AP$, $[x,x]\in\cV(p)$ if and only if $p\in\mu(\rho(x+1))$.\footnote{We add 1 to the index $x$ of $\rho$ just because traces are 1-based (whereas linear orders and interval models are 0-based).}
\end{enumerate}
\end{definition}
Clearly we have $\Ku,\rho(i+1,j+1)\models \psi$ if and only if $\bM_\rho,[i,j]\models \psi$.

We now describe how, with a slight modification of the previous SAT procedure, it is possible to derive a MC algorithm for $\hsDhom$ formulas $\varphi$ over finite Kripke structures $\Ku$.
The idea is to consider some finite linear orders---not all the possible ones, unlike the case of SAT---precisely those corresponding to (some) initial traces of $\Ku$, checking whether $\neg\varphi$ holds over them: 
in such a case we have found a counterexample, and we can conclude that $\Ku\not\models\varphi$.
To ensure this kind of \lq\lq SAT driven by the traces of $\Ku$\rq\rq, we make a product between $\Ku$ and the previous graph $G_{\varphi\sim}$, getting what we call a \lq\lq\emph{$(\varphi\!\sim\!\Ku)$-graph}\rq\rq. In the following, we will also exploit the notion of \lq\lq compass structure \emph{induced by a trace $\rho$} of $\Ku$\rq\rq, which is a fulfilling homogeneous compass $\varphi$-structure built from $\rho$ and completely determined by it.

Given a finite Kripke structure $\Ku=\KuDef$ and a $\hsDhom$ formula $\varphi$, we consider the $(\varphi\!\sim\!\Ku)$-graph $G_{\varphi \sim\Ku}$, which is basically the product of $\Ku$ and $G_{\varphi \sim}=(\Rows^\sim,\rownext)$, and is formally defined as 
\[G_{\varphi \sim\Ku}=(\Gamma, \Xi),\] where:
\begin{itemize}
    \item $\Gamma$ is the maximal subset of $\States\times \Rows^\sim$ such that: if $(s,[row]_\sim)\in\Gamma$ then $\mu(s)=row(0)\cap\AP$;
    \item $\big((s_1,[row_1]_\sim), (s_2,[row_2]_\sim)\big)\in\Xi$ iff $(i)$~$\big((s_1,[row_1]_\sim), (s_2,[row_2]_\sim)\big)\in\Gamma^2$, $(ii)$~$(s_1,s_2)\in\Edges$, and $(iii)$ $[row_1]_\sim\rownext [row_2]_\sim$.
\end{itemize}
Note that the definition of $\Gamma$ is well-given, since for all $row'\in[row]_\sim$, $row'(0)=row(0)$. The size of $G_{\varphi\sim\Ku}$ is bounded by $(|\States|\cdot |\Rows^\sim|)^2$.

Given a generic trace $\rho$ of $\Ku$, we define the compass $\varphi$-structure \emph{induced by $\rho$} as the fulfilling homogeneous  compass $\varphi$-structure $\cG_{(\Ku,\rho)}=(\bbP_\bbD, \cL)$, where $\mathpzc{S}=\{0,\ldots, |\rho|-1\}$, and for $0\leq x<|\rho|$, $\cL(x,x)\cap\AP=\mu(\rho(x+1))$ and $\reqD(\cL(x,x))=\emptyset$. 
Note that, given $\rho$, $\cG_{(\Ku,\rho)}$ always exists and is unique: all $\varphi$-atoms $\cL(x,x)$ ``on the diagonal'' are determined by the labeling of $\rho(x+1)$ (and by the absence of requests). Moreover,
by Lemma~\ref{lem:row_successor}, all the other atoms $\cL(x,y)$, for $0\leq x<y<|\rho|$, are determined by the $\rownext$ relation between $\varphi$-rows. 

The following property can easily be proved by induction.
\begin{proposition}\label{prop:eqTrack}
Given a finite Kripke structure $\Ku$, a trace $\rho$ of $\Ku$, and a $\hsDhom$ formula $\varphi$, 
for all $0\leq x\leq y <|\rho|$ and for all subformulas $\psi$ of $\varphi$, we have 
$\Ku,\rho(x+1,y+1)\models \psi$ if and only if $\psi\in\cL(x,y)$ in $\cG_{(\Ku,\rho)}$.
\end{proposition}
%
%
We can now introduce Theorem~\ref{thm:path_iff_MC}, that can be regarded as a version of Theorem~\ref{thm:path_iff_sat} for MC.
Its proof is in Appendix~\ref{proof:thm:path_iff_MC}.
\begin{theorem}\label{thm:path_iff_MC}
Given a finite Kripke structure $\Ku=\KuDef$ and a $\hsDhom$ formula $\varphi$, 
there exists an initial trace $\rho$ of $\Ku$ such that $\Ku,\rho\models\varphi$
if and only if
there exists
a path  in $G_{\varphi \sim\Ku}=(\Gamma, \Xi)$
from some node $(\sinit,[\row]_\sim)\in\Gamma$ to some node $(s,[\row']_\sim)\in\Gamma$ such that:
\begin{enumerate}
    \item there exists $row_1\in[\row]_\sim$ with $|row_1|=1$, and
    \item there exists $row_2\in[\row']_\sim$ with $\varphi\in row_2(|row_2|-1)$.
\end{enumerate}
\end{theorem} 

\begin{algorithm}[t]
\caption{\texttt{Counterexample}$(\Ku,\varphi)$}\label{NDAlgoMC}
\begin{algorithmic}[1]
    \State{$M\gets |\States|\cdot 2^{3|\varphi|^2}$, $step\gets 0$ and $(s,\row)\gets(\sinit, A)$ for some atom $A\in \Atoms$ with  $\reqD(A)=\emptyset$ and $A\cap\AP=\mu(\sinit)$}
    \If{$\varphi\not\in \row(|\row|-1)$}
        \State{\textbf{return} ``yes''}
    \EndIf
    \If{$step = M-1$} 
        \State{\textbf{return} ``no''}
    \EndIf
    \State{Non-deterministically choose $s'$ such that $(s,s')\in\Edges$}
    \State{Non-deterministically generate a $\varphi$-row $\row'$ and check that $row'(0)\cap\AP=\mu(s')$ and $\row \rownext \row'$}
    \State{$step \gets step +1$ and $(s,\row) \gets(s',\row')$}
    \State{Go back to line 2}
\end{algorithmic}
\end{algorithm}

Now, analogously to the case of SAT,
we can perform a reachability in $G_{\varphi\sim\Ku}$, exploiting the previous theorem
to decide whether there is an initial trace $\rho$ of $\Ku$ such that $\Ku,\rho\models\neg\varphi$, for a $\hsDhom$-formula $\varphi$ (i.e., the complementary problem of MC $\Ku\models\varphi$); $\rho$ is called a \lq\lq counterexample\rq\rq{} to $\varphi$.
The \emph{non-deterministic} procedure \texttt{Counterexample}$(\Ku,\varphi)$ in Algorithm~\ref{NDAlgoMC} 
searches for a suitable path in $G_{\varphi \sim\Ku}$,  $(\sinit,[\row_0]_\sim) \stackrel{\Xi}{\rightarrow} \cdots \stackrel{\Xi}{\rightarrow} (s_m,[\row_{m}]_\sim)$, where $\row_0 =A\in \Atoms$ with  $\reqD(A)=\emptyset$, $A\cap\AP=\mu(\sinit)$, $m< M$, and $\neg\varphi\in \row_m(|\row_m|-1)$ (i.e., $\varphi\not\in \row_m(|\row_m|-1)$). At the $j$-th iteration of lines 6./7., $(s_{j-1},s_j)\in\Edges$ is selected, and $\row_{j}$ is non-deterministically generated checking that $row_j(0)\cap\AP=\mu(s_j)$ and $\row_{j-1}\rownext\row_{j}$.

Basically, the same observations about the working space of the procedure in Algorithm~\ref{NDAlgo} can be done also for this algorithm, except for the space used to encode in binary $M\leq |\States|\cdot 2^{3|\varphi|^2}$ and $step$, ranging in $[0,M-1]$, which is $O(\log|\States| + |\varphi|^2)$ bits. Moreover we need to store two states, $s$ and $s'$ of $\Ku$, that need $O(\log |\States|)$ bits to be represented.

We conclude the section by stating the main result.
\begin{theorem}\label{thm:pspaceMC}
The MC problem for $\hsDhom$ formulas over finite Kripke structures is in $\Psp$. Moreover, for constant-length formulas, it is in $\NLOGSP$.
\end{theorem} 
\begin{proof}
Membership is immediate by the previous space analysis, and the fact that the complexity classes $\NPsp=\Psp$ and $\NLOGSP$ are closed under complement.
\end{proof}
Finally, it is possible to adapt the procedure also for \emph{strict} $\hsDhom${} (using Definitions~\ref{def:d_generatorS}--\ref{def:rownextS}). 

In the next section we will 
comment on $\Psp$-hardness, and thus $\Psp$-completeness, of SAT and MC for $\hsDhom$ formulas.

%% file: Chaps/ICALP_D/Psp_hard.tex
\section{An outline of $\Psp$-completeness of SAT \allowbreak and MC for $\D$}\label{sec:outlineSATMC}
In Appendix~\ref{sec:MChard} we prove in detail that MC for $\D$ formulas is $\Psp$-hard: 
we provide a reduction from the $\Psp$-complete \mbox{\emph{problem of (non-)universality}} of the language of a non-deterministic finite state automaton ($\NFA$) $\Au$~\cite{holzer}.
Here we only give an account of the main ideas behind the reduction.

We build a Kripke structure $\Ku_\Au$ and a $\D$ formula $\Phi_\Au$ which, together, allow us to consider \emph{the} deterministic computation of a $\DFA$ $\Du$, equivalent to the original $\NFA$ $\Au$ (i.e., accepting the same language), over some word $w$ \emph{not} accepted by $\Au$ (if such $w$ exists). The computation is built \lq\lq on-the-fly\rq\rq{} (i.e., we do not construct the---possibly exponentially larger---equivalent $\DFA$ $\Du$), and it is \lq\lq captured\rq\rq{} by an initial trace of $\Ku_\Au$, which is a concatenation of suitable subtraces, each one encoding a state $\tilde{q}$ of $\Du$, where $\tilde{q}$ is a subset of  $\Au$-states, by listing the $\Au$-states that belong to $\tilde{q}$. Each $\Du$-state (subtrace) is copied two times along the initial trace: this is necessary to enforce a suitable \lq\lq orientation\rq\rq, something that $\D${} is unaware of, as such logic is completely \lq\lq symmetric\rq\rq.

By this construction and Theorem~\ref{thm:pspaceMC}, the following holds.
\begin{theorem}
The MC problem for $\hsDhom$ formulas over finite Kripke structures is $\Psp$-complete.
\end{theorem} 

Moreover,
\begin{theorem}
The MC problem for \emph{constant-length} $\hsDhom$ formulas over finite Kripke structures is $\NLOGSP$-complete.
\end{theorem} 
\begin{proof}
Membership is stated by Theorem~\ref{thm:pspaceMC}.
To prove the $\NLOGSP$-hardness, there exists a trivial reduction from the ($\NLOGSP$-complete) \emph{problem of \mbox{(non-)reachability}} of two nodes in a directed graph.
\end{proof}

We now briefly turn to $\D$ SAT,
outlining its $\Psp$-hardness over finite linear orders.
Our construction (thoroughly worked out in Appendix \ref{sec:SAThard}) mimics that of Section 3.2 and 3.3 of \cite{DBLP:journals/fuin/MarcinkowskiM14}, in which the authors show 
that it is possible to build a $\D$ formula $\Psi_{\Ku}$ that encodes a finite Kripke structure $\Ku$.
We instantiate it over $\Ku_\Au$, getting $\Psi_{\Ku_\Au}$, with the result that
any finite linear order satisfying $\Psi_{\Ku_\Au}$ represents an initial trace of $\Ku_\Au$.
In this way we get a reduction from the 
problem of non-universality of the language of $\Au$ to SAT for $\hsDhom$:
the language of $\Au$ is non-universal if and only if the formula $\Psi_{\Ku_\Au}\wedge\Phi_\Au$ is satisfiable.

By also recalling Theorem \ref{thm:pspace}, the next result is proved.
\begin{theorem}\label{cor:pspace_complete}
The SAT problem for $\hsDhom$ formulas over finite linear orders is $\Psp$-complete.
\end{theorem}

%% file: Chaps/ICALP_D/concl.tex
\section{Conclusions}
In this chapter we have shown $\Psp$-completeness of both SAT and MC for $\D$ formulas $\varphi$ under homogeneity.
This result is proved thanks to a finite-index equivalence relation $\sim$ between rows of a compass $\varphi$-structure, that preserves fulfillment of compasses, and allows us to ``contract'' these structures whenever we find two related rows. Moreover, if a compass features $\varphi$ before the contraction, then $\varphi$ is still featured after it. These facts are exploited to build a SAT and an MC algorithm for $\hsDhom$, which implicitly make compass contractions by a reachability check in suitable graphs, whose nodes are basically the equivalence classes of $\sim$.
Since $\sim$ has finite index, exponentially bounded by $|\varphi|$,
and the algorithms
only need to keep track of two rows of a compass at a time (encoding them succinctly), $\Psp$-membership of the problems follows.

In the next chapter, we will see that also the fragments $\AAbarBBbar$ and $\AAbarEEbar$ feature an MC problem complete for $\Psp$.
However, as the reader will see, completely different techniques and notions are used to prove the result.

%% file: Chaps/TCS17/TCS17main.tex
\chapter{$\HS$ fragments in $\Psp$ and $\EXPSPACE$}\label{chap:TCS17}
\begin{chapref}
{\footnotesize The references for this chapter 
are~\cite{BOZZELLI2018,ijcar16,csl15,MMP15,ictcs16}.}
\end{chapref}

\minitoc\mtcskip

\input{Chaps/TCS17/intro}

\input{Chaps/TCS17/AAbarEEbar}
\input{Chaps/TCS17/AAbarBBbarEbar}
\input{Chaps/TCS17/concl}

%% file: Chaps/TCS17/intro.tex
\lettrine[lines=3]{I}{n this chapter,} we prove that MC for the $\HS$ fragment $\AAbarBBbar$ (resp., $\AAbarEEbar$), that allows one to express properties of future and past intervals, interval prefixes (resp., suffixes), and right (resp., left) interval extensions,
is in $\PSPACE$ (under homogeneity).
Since MC for the $\HS$ fragment featuring only one modality for right (resp., left) interval extensions $\Bbar$ (resp., $\Ebar$) can be shown to be $\PSPACE$-hard (Appendix~\ref{sect:BbarHard}), $\PSPACE$-completeness immediately follows. 
Moreover, if we restrict $\HS$ to modalities either for interval prefixes $\B$ or for interval suffixes $\E$ only, MC turns out to be $\co\NP$-complete.

These results are proved by a small-model property based on the notion of \emph{induced trace}: given a trace $\rho$ in a finite Kripke structure and a formula $\varphi$ of $\AAbarBBbar$/$\AAbarEEbar$, it is always possible to build, by iteratively contracting $\rho$, another (\lq\lq induced\rq\rq ) trace, 
whose length is \emph{polynomially bounded} in the size of $\varphi$ and of the Kripke structure, which preserves the satisfiability of  $\varphi$  with respect to  $\rho$.
%

The lower bound for $\BE$ MC (recall Section \ref{sec:BEhard}) shows that there is no way to provide an MC algorithm for the extension of $\AAbarBBbar$ with $\E$ (resp., of $\AAbarEEbar$ with $\B$)  with a \lq\lq good\rq\rq{} computational complexity. 
The picture is not so clear for the extension of $\AAbarBBbar$ with $\Ebar$ (resp., $\AAbarEEbar$ with $\Bbar$):
membership of $\AAbarBBbarEbar$ (resp., $\AAbarEBbarEbar$) to $\EXPSPACE$  was already shown in~\cite{MMP15} and the $\PSPACE$-hardness of MC for $\Bbar$ (resp., $\Ebar$) currently gives the best (unmatching) complexity lower bound.
In this chapter we provide a much more understandable proof of membership to $\EXPSPACE$ of the MC problem for $\AAbarBBbarEbar$ (w.r.t. the one given in~\cite{MMP15}), which makes use of (a suitable extension of) the notion of induced trace, which we exploit in the proof of the small model property for $\AAbarBBbar$.

\paragraph*{Organization of the chapter.} 
\begin{itemize}
	\item In Section~\ref{sec:AAbarEEbar}, we first introduce the notion of induced trace and then we
    prove, via a contraction technique, a polynomial small-model property for $\AAbarBBbar$ and $\AAbarEEbar$ (Section~\ref{subsec:polyAAbarEEbar}), which allows us to devise a $\PSPACE$ MC algorithm for them (Section~\ref{subsec:MCpolyAAbarEEbar}). In addition we consider the one-modality fragments $\B$ and $\E$, and prove their $\co\NP$-completeness (Section~\ref{sec:TheFragmentE}).
	\item In Section~\ref{sec:AAbarBBbarEbar}, we focus on the fragment $\AAbarBBbarEbar$ and the symmetric fragment $\AAbarEBbarEbar$. We first define the equivalence relation of \emph{trace bisimilarity}, and then we introduce the    
    notion of \emph{prefix sampling}. With these tools, we prove an exponential small-model property for $\AAbarBBbarEbar$ (and $\AAbarEBbarEbar$), resulting into an easier proof of membership of the two fragments to $\EXPSPACE$.
\end{itemize}

%% file: Chaps/TCS17/AAbarEEbar.tex
\section{The fragments $\AAbarBBbar$ and $\AAbarEEbar$}\label{sec:AAbarEEbar}
In this section, we show that the MC problem for the fragments $\AAbarBBbar$ and $\AAbarEEbar$  is \PSPACE-complete.
Moreover, we prove that MC for the smaller fragments $\B$ and $\E$ is $\co\NP$-complete.

\subsection{Polynomial small-model property for $\AAbarBBbar$ and $\AAbarEEbar$}\label{subsec:polyAAbarEEbar}
We first prove membership to \PSPACE\ of the MC problem for the fragments $\AAbarBBbar$ and $\AAbarEEbar$ by showing that they feature a \emph{polynomial small-model property}, that is, we prove that if a trace $\rho$ of a finite Kripke structure $\Ku$ satisfies a given formula $\varphi$ of $\AAbarEEbar$ or $\AAbarBBbar$, then there exists a trace $\pi$, whose length is polynomial in the sizes of $\varphi$ and $\Ku$, starting from, and leading to, the same states as $\rho$, that satisfies $\varphi$. In the following, we focus on $\AAbarEEbar$, being the case of $\AAbarBBbar$ completely symmetric.

Let $\Ku= \KuDef$ be a finite Kripke structure.
We start by introducing the basic notions of \emph{induced trace} and \emph{well-formed trace}, that will be extensively used to prove the polynomial small-model property. 
Intuitively, we say that a trace $\pi\in\Trk_{\Ku}$ is induced by a trace $\rho\in\Trk_{\Ku}$ if it can be obtained from 
$\rho$ by suitably contracting it, that is, by concatenating some subtraces of $\rho$.
Well-formedness adds a condition on the suffixes of an induced trace.
For the sake of readability, hereafter we denote by $\rho^i$ the suffix $\rho(i,|\rho|)$ of $\rho$ (hence, $\rho^1$ is just $\rho$).

\begin{definition}[Induced and well-formed trace] \label{definition:inducedTrk}
Let $\Ku\!=\! \KuDef$ be a finite Kripke structure and let $\rho\in\Trk_\Ku$, with $|\rho| = n$. 
A trace $\pi\in\Trk_\Ku$ is \emph{induced} by $\rho$ if there exists an increasing sequence of 
$\rho$-positions $i_1<\ldots < i_k$, with $i_1=1$ and $i_k=n$, such that $\pi= \rho(i_1)\cdots \rho(i_k)$.
For $j = 1, \ldots, k$, the $\pi$-position $j$ and the $\rho$-position $i_j$ are called \emph{corresponding positions}.

An induced trace $\pi$ is \emph{well-formed} (with respect to $\rho$) if, for all $\pi$-positions $j$, with corresponding $\rho$-positions $i_j$, and all proposition letters $p\in \Prop$, it holds that
\[\Ku,\pi^{j} \models p \iff \Ku,\rho^{i_j} \models p.\]
\end{definition}

\begin{example}
As an example, let us consider Figure~\ref{fig:induced}. The trace $\pi = \rho(1)\rho(4)\rho(5)\rho(7)\rho(10)$ is induced by $\rho$, provided that both $\rho$ and $\pi$ are traces of a Kripke structure $\Ku$, and the positions $1, 2, 3, 4$, and $5$ of $\pi$ correspond to the positions $i_1=1$, $i_2=4$, $i_3=5$, $i_4=7$ and $i_5=10$ of $\rho$.

\begin{figure}[H]
    \centering
    \resizebox{\linewidth}{!}{\includegraphics{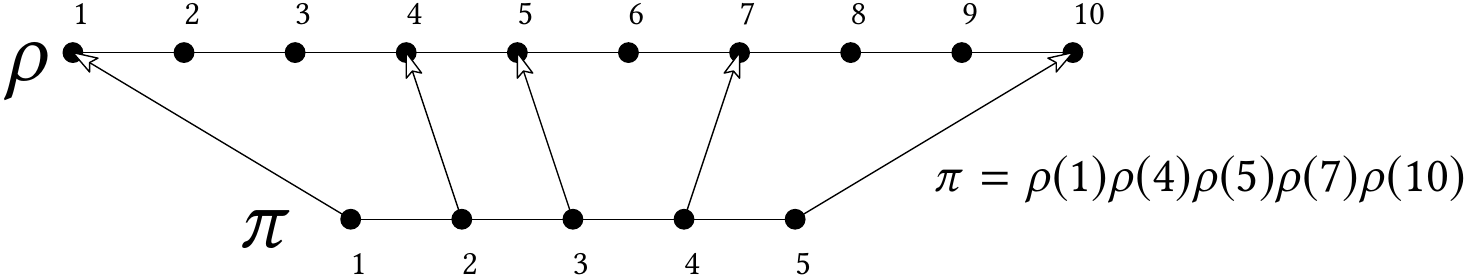}}
    \caption{A trace $\pi$ induced by $\rho$.
    }
    \label{fig:induced}
\end{figure}
\end{example}

Note that if $\pi$ is induced by $\rho$, then $\fst(\pi)=\fst(\rho)$, $\lst(\pi)=\lst(\rho)$, and $|\pi|\leq |\rho|$ ($|\pi| = |\rho|$ if and only if $\pi = \rho$). 

Well-formedness implies that the suffix of $\pi$ starting from position $j$ and that of $\rho$ starting from the corresponding position $i_j$ agree over all the proposition letters in $\Prop$, that is, they have the same \lq\lq satisfaction pattern\rq\rq{} of proposition letters. In particular, for all $p\in \Prop$, $\Ku,\pi \models p$ if and only if $\Ku,\rho \models p$. It can be easily checked that the \emph{well-formedness relation is transitive}.

The following proposition shows how it is possible to contract a trace while preserving the same satisfaction pattern of proposition letters with respect to suffixes.  Such a criterion represents a ``basic step'' in a contraction process that will allow us to prove the polynomial small-model property. 

\begin{proposition}\label{proposition:wellFormdness}
Let $\Ku = \KuDef$ be a finite Kripke structure. For any trace $\rho\in\Trk_\Ku$, there exists a
well-formed (with respect to $\rho$) trace $\pi\in\Trk_\Ku$ such that $|\pi| \leq |\States|\cdot (|\Prop|+1)$.
\end{proposition}
\begin{proof}
Let $\rho\in\Trk_\Ku$, with $|\rho| = n$. If $n\leq |\States|\cdot (|\Prop|+1)$, the thesis trivially holds.
Let us assume $n > |\States|\cdot (|\Prop|+1)$. We show that there exists $\pi\in\Trk_\Ku$, with $|\pi| < n$, which is well-formed with respect to $\rho$.

Since $n> |\States|\cdot (|\Prop|+1)$, there is some state $s\in \States$ occurring in $\rho$ at least $|\Prop|+2$ times.
Assume that for all $\rho$-positions $i$ and $j$, with $j>i$, if $\rho(i)=\rho(j)=s$, then there exists some $p\in\Prop$ such that $\Ku,\rho^j \models p$ and $\Ku,\rho^i \not\models p$. This assumption leads to a contradiction, as the suffixes of $\rho$ may feature at most $|\Prop|+1$ distinct satisfaction patterns of proposition letters (due to the homogeneity assumption in Definition~\ref{def:inducedmodel}), while there are at least $|\Prop|+2$ occurrences of $s$.
As a consequence, there are two $\rho$-positions $i$ and $j$, with $j>i$, such that $\rho(i)=\rho(j)=s$ and, for all $p\in\Prop$, $\Ku,\rho^j \models p$ if and only if $\Ku,\rho^i \models p$ (see Figure~\ref{fig:basicContr} for a graphical account).
It is easy to see that $\pi=\rho(1, i)\star\rho(j, n)\in\Trk_\Ku$ is well-formed with respect to $\rho$ and $|\pi| < n$.  If $|\pi| \leq |\States|\cdot (|\Prop|+1)$, the thesis is proved; otherwise, the same basic step can be iterated a finite number of times, and the thesis follows by transitivity of the well-formedness relation.
\end{proof}

\begin{figure}[tb]
\centering
    \scalebox{1.8}{
    \begin{tikzpicture}
    	\filldraw [gray] (0,0) circle (2pt)
    	(1.5,0) circle (2pt)
    	(2,0) circle (2pt)
    	(3.5,0) circle (2pt);
    	\filldraw [gray] (0,-0.5) circle (2pt)
    	(1.5,-0.5) circle (2pt)
    	(3,-0.5) circle (2pt);
    	\draw [red] (1.5,0) -- (2,0);
    	\draw [black]  (0,0) -- (1.5,0);
    	\draw [black] (2,0) -- (3.5,0);
    	\draw [black] (0,-0.5) -- (3,-0.5);
    	\draw [dashed, red] (1.5,0) -> (1.5,-0.5);
    	\draw [dashed, red] (2,0) -> (1.5,-0.5);
    	{\tiny
    		\node (a0) at (3.8,0) {$\rho$};	
    		\node (b0) at (4,-0.5) {$\pi\!=\!\rho(1,\! i)\!\star\! \rho^j$};	
    		\node (a1) at (1.4,0.2) {$\rho(i)$};
    		\node (a11) at (1.75,0.2) {$=$};
    		\node (a2) at (2.1,0.2) {$\rho(j)$};
    		\node (a3) at (1.65,0.5) {$\Prop(\rho,\! i)= \Prop(\rho,\! j)$};
    			
    	}
    \end{tikzpicture}}
    \caption{The contraction step of Proposition~\ref{proposition:wellFormdness} ($\Prop(\rho,k)=\{ p \in \Prop \mid \Ku,\rho^{k} \models p\}$).}\label{fig:basicContr}
\end{figure}

The next definition identifies some distinguished positions in a trace, called \emph{witness positions}. As we will see in the proof of Theorem~\ref{theorem:polynomialSizeModelProperty}, if we perform a contraction (see the proof of Proposition~\ref{proposition:wellFormdness}, and its graphical account in Figure~\ref{fig:basicContr}) between a pair of such positions, we get a trace which is equivalent to the original one with respect to satisfiability of the considered $\AAbarEEbar$ formula. In the following, we restrict ourselves to formulas in \emph{negation normal form} (\nnf), namely, formulas where negation is applied only to proposition letters. By using De Morgan's laws and the dual modalities $\hsEu$, $\hsEtu$, $\hsAu$, and $\hsAtu$ of $\hsE$, $\hsEt$, $\hsA$, and $\hsAt$,  respectively, we can trivially convert (in linear time) a formula into an equivalent one in \nnf, having at most double length.

\begin{definition}[Witness position]\label{definition:WitnessPositions} 
Let $\Ku = \KuDef$ be a finite Kripke structure, $\rho\in\Trk_\Ku$, and $\varphi$ be a formula of $\AAbarEEbar$. Let us denote by $E(\varphi,\rho)$ the set of subformulas of the form $\hsE\psi$ of $\varphi$ such that $\Ku, \rho\models\hsE\psi$.

The \emph{set $Wt(\varphi,\rho)$ of witness positions of $\rho$ for $\varphi$} is the \emph{minimal} set of $\rho$-positions satisfying the following constraint:
\begin{itemize}
    \item for each $\hsE\psi\in E(\varphi,\rho)$, the greatest $\rho$-position $i>1$ such that $\Ku,\rho^i\models\psi$ belongs to $Wt(\varphi,\rho)$.\footnote{Note that such a $\rho$-position exists by definition of $E(\varphi,\rho)$.}
\end{itemize}
\end{definition}
It is easy to see that the cardinalities of $E(\varphi,\rho)$ and of $Wt(\varphi,\rho)$ are at most $|\varphi|-1$.
We are now ready to prove the polynomial small-model property.

\begin{theorem}[Polynomial small-model property for $\AAbarEEbar$]\label{theorem:polynomialSizeModelProperty}
Let $\Ku= \KuDef$ be a finite Kripke structure, $\rho, \sigma \in \Trk_\Ku$, and $\varphi$ be an $\AAbarEEbar$ formula in \nnf{} such that $\Ku,\rho\star\sigma\models \varphi$. Then there exists $\pi$, induced by $\rho$, such that $\Ku,\pi\star\sigma\models \varphi$, and $|\pi|\leq |\States|\cdot (|\varphi|+1)^2$.
\end{theorem}
As a preliminary remark, we note that the theorem holds in particular if $|\sigma|=1$, and thus $\rho\star\sigma=\rho$ and $\pi\star\sigma=\pi$. In such a case, if $\Ku,\rho\models \varphi$, then $\Ku,\pi\models \varphi$, where $\pi$ is induced by $\rho$ and $|\pi|\leq |\States|\cdot (|\varphi|+1)^2$. The more general statement of Theorem~\ref{theorem:polynomialSizeModelProperty} is needed for technical reasons in the soundness/completeness proofs of the  algorithms for MC given in the following.

\begin{proof}
W.l.o.g., we restrict ourselves to the proposition letters occurring in  $\varphi$, thus having $|\Prop|\leq |\varphi|$.
Let $Wt(\varphi,\rho\star\sigma)$ be the set of witness positions of $\rho\star\sigma$ for $\varphi$, let $\{i_1,\ldots,i_k\}$ be the ordering of $Wt(\varphi,\rho\star\sigma)$ such that $i_1<\ldots <i_k$, and let $i_0=1$ and $i_{k+1}=|\rho\star\sigma|$. Hence, $1=i_0< i_1<\ldots <i_k \leq i_{k+1}=|\rho\star\sigma|$.

If $|\rho| \leq |\States|\cdot (|\varphi|+1)^2$, the thesis trivially holds.
Let us assume that $|\rho|> |\States|\cdot (|\varphi|+1)^2$. We show that there exists a trace $\pi$ induced by $\rho$, with $|\pi| < |\rho |$, such that $\Ku,\pi\star\sigma\models \varphi$.

W.l.o.g., we can assume that, for some $j\geq 0$, $i_0<i_1<\ldots <i_j$ are $\rho$-positions, while $i_{j+1}<\ldots <i_{k+1}$ are $(\rho\star\sigma)$-positions not in $\rho$. 
Then, either $(i)$ there exists $t\in [0,j-1]$ such that $i_{t+1}-i_t>|\States|\cdot(|\varphi|+1)$ or $(ii)$ $|\rho(i_j,|\rho|)|>|\States|\cdot(|\varphi|+1)$. By way of contradiction, suppose that neither $(i)$ nor $(ii)$ holds. We need to distinguish two cases. If $\rho\star\sigma=\rho$, then
$|\rho| = (i_{k+1}-i_0)+1\leq (k+1) \cdot |\States|\cdot (|\varphi|+1)+1$; otherwise ($|\rho| < |\rho\star\sigma|$), $|\rho| = (i_j - i_0) + |\rho(i_j,|\rho|)| \leq j \cdot |\States|\cdot (|\varphi|+1) + |\States|\cdot (|\varphi|+1) \leq (k+1) \cdot |\States|\cdot (|\varphi|+1)$. The contradiction follows since $(k+1) \cdot |\States|\cdot (|\varphi|+1)+1 \leq |\varphi| \cdot |\States|\cdot (|\varphi|+1)+1 \leq |\States|\cdot (|\varphi|+1)^2$.

Let 
$(\alpha,\beta)=(i_t,i_{t+1})$ in case $(i)$, and $(\alpha,\beta)=(i_j,|\rho|)$ in case $(ii)$ and let $\rho'=\rho(\alpha,\beta)$. In both cases, 
$|\rho'|> |\States|\cdot (|\varphi|+1)\geq |\States|\cdot (|\Prop|+1)$, being $|\Prop|\leq |\varphi|$.
By Proposition~\ref{proposition:wellFormdness}, there exists a trace $\pi'$ of $\Ku$, well-formed with respect to $\rho'$, such that $|\pi'|\leq |\States|\cdot(|\Prop|+1) < |\rho'|$. Let $\pi$ be the trace induced by $\rho$ obtained by replacing the subtrace $\rho'$ of $\rho$ by $\pi'$ (see Figure~\ref{fig:contr2} for a graphical account).
Since $|\pi|<|\rho|$, it remains to prove that  $\Ku,\pi\star\sigma\models \varphi$.

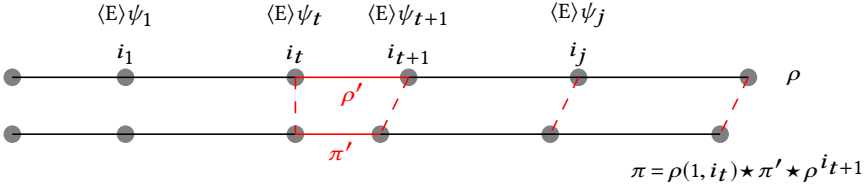
\begin{figure}[tb]
\centering
    \resizebox{\linewidth}{!}{\begin{tikzpicture}
				\filldraw [gray] (0,0) circle (2pt)
				(1.0,0) circle (2pt)
				(2.5,0) circle (2pt)
				(3.5,0) circle (2pt) 
				(5.0,0) circle (2pt)
				(6.5,0) circle (2pt);
				\filldraw [gray] (0,-0.5) circle (2pt)
				(1.0,-0.5) circle (2pt)
				(2.5,-0.5) circle (2pt)
				(3.25,-0.5) circle (2pt) 
				(4.75,-0.5) circle (2pt)
				(6.25,-0.5) circle (2pt);
				\draw [red] (2.5,0) -- (3.5,0);
				\draw [black]  (0,0) -- (2.5,0);
				\draw [black] (3.5,0) -- (6.5,0);
				\draw [black] (0,-0.5) -- (2.5,-0.5);
				\draw [red] (2.5,-0.5) -- (3.25,-0.5);
				\draw [black] (3.25,-0.5) -- (6.25,-0.5);
				\draw [dashed, red] (2.5,0) -> (2.5,-0.5);
					\draw [dashed, red] (3.5,0) -> (3.25,-0.5);
				\draw [dashed, red] (5,0) -> (4.75,-0.5);
				\draw [dashed, red] (6.5,0) -> (6.25,-0.5);
				{\tiny
					\node (a0) at (6.9,0) {$\rho$};	
					\node (b0) at (6.5,-0.8) {$\pi\!=\!\rho(1,\! i_t)\!\star\! \pi'\! \star\! \rho^{i_{t+1}}$};	
					\node (a1) at (1,0.2) {$i_1$};
					\node (a2) at (2.5,0.2) {$i_t$};
					\node (a3) at (3.5,0.2) {$i_{t+1}$};
					\node (a4) at (5,0.2) {$i_{j}$};
					\node (a1) at (1,0.55) {$\hsE\! \psi_1$};
					\node (a2) at (2.5,0.55) {$\hsE\! \psi_{t}$};
					\node (a3) at (3.5,0.55) {$\hsE\! \psi_{t+1}$};
					\node (a4) at (5,0.55) {$\hsE\! \psi_{j}$};
					\node [red] (a4) at (2.9,-0.65) {$\pi'$};
					\node [red] (a4) at (3.0,-0.15) {$\rho'$};
				}
				
			\end{tikzpicture}}
    \caption{Representation of the contraction step of Theorem~\ref{theorem:polynomialSizeModelProperty}---case $(i)$}\label{fig:contr2}
\end{figure}

Let us denote $\pi\star\sigma$ by $\overline{\pi}$ and $\rho\star\sigma$ by $\overline{\rho}$. Let $H:[1,|\overline{\pi}|] \rightarrow [1,|\overline{\rho}|]$ be the function mapping positions of $\overline{\pi}$ into positions of $\overline{\rho}$ in such a way that positions ``outside'' $\pi'$, i.e., outside the interval $[\alpha,\alpha+|\pi'|-1]$, are mapped into their original positions in $\overline{\rho}$ while those ``inside'' $\pi'$, i.e., in $[\alpha,\alpha+|\pi'|-1]$, are mapped into the corresponding positions in $\rho'$ (exploiting well-formedness of $\pi'$ w.r.t.\ $\rho'$):
\begin{equation*}
H(m) = \begin{cases}
	m& \text{if}\quad m<\alpha; \\
\alpha+\ell_{m-\alpha+1}-1 & \mbox{if}\quad \alpha\leq m<\alpha+|\pi'|; \\
m+(|\rho'| - |\pi'|) & \mbox{if}\quad m\geq \alpha+|\pi'|,
\end{cases}
\end{equation*}
where  $\ell_s$ is the $\rho'$-position corresponding to the $\pi'$-position $s$,
with $\ell_s\in\mathopen[1,|\rho'|\mathclose]$ and $s\in\mathopen[1,|\pi'|\mathclose]$.

It is easy to check that $H$ satisfies the following properties:
\begin{enumerate}
  \item $H$ is strictly monotonic, i.e., for all $j,j'\in [1,|\overline{\pi}|]$, $j<j'\iff H(j)<H(j')$;
  \item for all $j\in [1,|\overline{\pi}|]$, $\overline{\pi}(j)=\overline{\rho}(H(j))$;
  \item $H(1)= 1$ and $H(|\overline{\pi}|)=|\overline{\rho}|$;
  \item $Wt(\varphi,\overline{\rho})\subseteq \{H(j) \mid j \in [1,|\overline{\pi}|] \}$, i.e., all witness positions are preserved;
  \item for each $j\in [1,|\overline{\pi}|]$ and $p\in \Prop$, $\Ku,\overline{\pi}^{j}\models p\iff \Ku,\overline{\rho}^{H(j)}\models p$.
\end{enumerate}

The fact that
$\Ku,\overline{\pi}\models \varphi$ is an immediate consequence of the following claim, considering that  $H(1)=1$, $\Ku,\overline{\rho}\models \varphi$,  $\overline{\rho}^{1}=\overline{\rho}$, and $\overline{\pi}^{1}=\overline{\pi}$.

\begin{claim}  For all $j\in [1,|\overline{\pi}|]$, all subformulas $\psi$ of $\varphi$, and all $u\in\Trk_\Ku$, it holds that 
\[
\Ku,u\star \overline{\rho}^{H(j)}\models \psi \Longrightarrow \Ku,u\star\overline{\pi}^{j}\models \psi.
\]
\end{claim}
\begin{proof}
Assume that $\Ku,u\star\overline{\rho}^{H(j)}\models \psi $. Note that $u\star \overline{\rho}^{H(j)}$ is defined if and only if $u\star \overline{\pi}^{j}$ is defined. We prove by induction on the structure of $\psi$  that
$\Ku,u\star\overline{\pi}^{j}\models \psi$. Since $\varphi$ is in \nnf, only the following cases occur:
 \begin{itemize}
   \item $\psi= p$ or $\psi=\neg p$ for some $p\in \Prop$. By Property~5 of $H$, $\Ku, \overline{\pi}^{j}\models p$ if and only if $\Ku,\overline{\rho}^{H(j)}\models p$.
    Hence, $\Ku, u\star \overline{\pi}^{j}\models p$ if and only if $\Ku,u\star\overline{\rho}^{H(j)}\models p$, and the result holds.
    \item $\psi= \theta_1\wedge\theta_2$ or $\psi= \theta_1\vee\theta_2$, for some $\AAbarEEbar$ formulas $\theta_1$ and $\theta_2$: the result directly follows from the inductive hypothesis.
   \item $\psi = \hsEu\theta$. We need to show that for each proper suffix $\eta$ of $u\star\overline{\pi}^{j}$, $\Ku,\eta\models\theta$. We distinguish two cases:
 \begin{itemize}
   \item $\eta$ is \emph{not} a proper suffix of $\overline{\pi}^{j}$. Hence, $\eta$ is of the form $u^{h}\star \overline{\pi}^{j}$ for some $h\in [2,|u|]$. Since $\Ku,u\star\overline{\rho}^{H(j)}\models \hsEu\theta $, then
   $\Ku,u^{h}\star\overline{\rho}^{H(j)}\models \theta $.  By the inductive hypothesis, $\Ku,u^{h}\star\overline{\pi}^j\models \theta $.
   \item $\eta$ is a proper suffix of $\overline{\pi}^{j}$. Hence, $\eta = \overline{\pi}^{h}$ for some $h\in [j+1,|\overline{\pi}|]$.   By Property~1 of $H$, $H(h)>H(j)$, and since $\Ku,u\star\overline{\rho}^{H(j)}\models \hsEu\theta $, we have that $\Ku,\overline{\rho}^{H(h)}\models \theta $. By the inductive hypothesis,  $\Ku,\overline{\pi}^h\models \theta $.
 \end{itemize}
 Therefore, $\Ku,u\star \overline{\pi}^j\models \hsEu\theta $.
\item $\psi = \hsE\theta$. We need to show that there exists a proper suffix of $u\star\overline{\pi}^{j}$ satisfying $\theta$. 
Since $\Ku,u\star \overline{\rho}^{H(j)}\models \psi $, there exists a proper suffix $\eta'$ of  $u\star \overline{\rho}^{H(j)}$
 such that  $\Ku,\eta'\models \theta $. We distinguish two cases:
 \begin{itemize}
   \item $\eta'$ is \emph{not} a proper suffix of $\overline{\rho}^{H(j)}$. Hence, $\eta'$ is of the form $u^{h}\star \overline{\rho}^{H(j)}$ for some $h\in [2,|u|]$. By the inductive hypothesis, $\Ku,u^{h}\star\overline{\pi}^j\models \theta $. Hence, $\Ku,u\star\overline{\pi}^j\models \hsE\theta $.
   \item $\eta'$ is a proper suffix of $\overline{\rho}^{H(j)}$. Hence, $\eta' = \overline{\rho}^{i}$ for some $i\in [H(j)+1,|\overline{\rho}|]$, and $\Ku,\overline{\rho}^{i}\models \theta $.
   Let $i'$ be the greatest position of $\overline{\rho}$ such that $\Ku,\overline{\rho}^{i'}\models\theta$. Hence $i'\geq i$ and, by Definition~\ref{definition:WitnessPositions}, $i'\in Wt(\varphi,\overline{\rho})$. By Property 4 of $H$, $i'=H(h)$ for some $\overline{\pi}$-position $h$. Since $H(h)>H(j)$, it holds that $h>j$ (Property 1). By the inductive hypothesis, $\Ku,\overline{\pi}^h\models \theta $, and we obtain that $\Ku,u\star\overline{\pi}^j\models \hsE\theta $.
 \end{itemize}
  Therefore, in both the cases, $\Ku,u\star \overline{\pi}^j\models \hsE\theta $.
 \item $\psi=\hsEtu\theta$ or $\psi=\hsEt\theta$: the thesis holds as a direct consequence of the inductive hypothesis.
 \item $\psi = \hsAu\theta$, $\psi=\hsA\theta$, $\psi = \hsAtu\theta$, or $\psi=\hsAt\theta$. Since $u\star\overline{\pi}^j$ and $u\star\overline{\rho}^{H(j)}$ start at the same state and lead to the same state (by Properties 2 and 3 of $H$), the thesis trivially follows. This concludes the proof of the claim.\qedhere
\end{itemize}
\end{proof}
We have shown that $\Ku,\overline{\pi}\models \varphi$, with $|\pi| < |\rho |$. Now, if $|\pi|\leq |\States| \cdot (|\varphi|+1)^2$, the thesis is proved; otherwise, the above contraction step can be iterated a finite number of times, until the bound is reached, proving the thesis of Theorem~\ref{theorem:polynomialSizeModelProperty}.
  \end{proof}
  
\subsection{\PSPACE\ MC algorithm for $\AAbarEEbar$}\label{subsec:MCpolyAAbarEEbar}

By exploiting the polynomial small-model property stated by Theorem~\ref{theorem:polynomialSizeModelProperty}, it is easy to define a \PSPACE\ MC algorithm for $\AAbarEEbar$. The main MC procedure for $\AAbarEEbar$ formulas is \texttt{ModCheck}$(\Ku,\psi)$ (Algorithm~\ref{ModCheck2}). All the initial traces $\sigma$, obtained by visiting the unravelling of $\Ku$ from its initial state $\sinit$ up to depth $|\States|\cdot (2|\psi|+3)^2$, are checked with respect to $\psi$ by the function
$\texttt{Check}(\Ku, \psi,\sigma)$ (Algorithm~\ref{Chk2}) which decides whether $\Ku,\sigma\models \psi$. The $\texttt{Check}$ function is iteratively called until either some initial trace is found that does not satisfy $\psi$ or all bounded initial traces satisfy $\psi$ (and thus $\Ku\models\psi$). The invocation of $\texttt{Check}(\Ku, \psi,\sigma)$ (Algorithm~\ref{Chk2}) decides whether $\Ku,\sigma\models \psi$ or not by recursively calling itself on the subformulas  of $\psi$ either over $\sigma$ or over (bounded) traces obtained by unraveling $\Ku$ forward (starting from $\lst(\sigma)$) for occurrences of the modality  $\hsA$,  and backward  (starting from $\fst(\sigma)$) for occurrences of $\hsAt$ and
$\hsEt$.

\begin{algorithm}[p]
\begin{algorithmic}[1]
	\For{all initial traces $\sigma\in\Trk_\Ku$ such that $|\sigma|\leq |\States|\cdot (2|\psi|+3)^2$}
	    \If{$\texttt{Check}(\Ku,\psi,\sigma)=0$}
	        \Return{0: ``$\Ku,\sigma\not\models \psi$''}\Comment{Counterexample found}
	    \EndIf
	\EndFor
	\Return{1: ``$\Ku\models \psi$''}
\end{algorithmic}
\caption{\texttt{ModCheck}$(\Ku,\psi)$}\label{ModCheck2}
\end{algorithm}

\begin{algorithm}[p]
\begin{algorithmic}[1]
    \If{$\psi=p$, for $p\in\Prop$}
	    \If{$p\in \bigcap_{s\in \states(\sigma)}\mu(s)$}
	        \State{\textbf{return} 1 \textbf{else} \textbf{return} 0}
	    \EndIf
	\ElsIf{$\psi=\neg\varphi$}
        \Return{$1-\texttt{Check}(\Ku,\varphi,\sigma)$}
    \ElsIf{$\psi=\varphi_1\wedge\varphi_2$}
        \If{$\texttt{Check}(\Ku,\varphi_1,\sigma)=0$}
	        \Return{0}
	    \Else
	        \Return{$\texttt{Check}(\Ku,\varphi_2,\sigma)$}
	    \EndIf
	\ElsIf{$\psi=\hsA\varphi$}
	    \For{all $\pi\in\Trk_\Ku$ such that $\fst(\pi)=\lst(\sigma)$, and $|\pi|\leq |\States|\cdot (2|\varphi|+1)^2$}
	        \If{$\texttt{Check}(\Ku,\varphi,\pi)=1$}
	            \Return{1}
	        \EndIf
	    \EndFor
	    \Return{0}
	\ElsIf{$\psi=\hsE\varphi$}
	    \For{each proper suffix $\pi$ of $\sigma$}
	        \If{$\texttt{Check}(\Ku,\varphi,\pi)=1$}
	            \Return{1}
	        \EndIf
	    \EndFor
	    \Return{0}
	\ElsIf{$\psi=\hsEt\varphi$}
        \For{all $\pi\in\Trk_\Ku$ such that $\lst(\pi)=\fst(\sigma)$, and $2\leq |\pi|\leq |\States|\cdot (2|\varphi|+1)^2$}
            \If{$\texttt{Check}(\Ku,\varphi,\pi\star\sigma)=1$}
                \Return{1}
            \EndIf
        \EndFor
	    \Return{0}
	\EndIf
	\State{\dots}\Comment{$\psi=\hsAt\varphi$ is analogous to $\psi=\hsA\varphi$}
\end{algorithmic}
\caption{\texttt{Check}$(\Ku,\psi,\sigma)$}\label{Chk2}
\end{algorithm}

Note that the considered bound on the length of initial traces is actually  $|\States|\cdot (2|\psi|+3)^2\geq |\States|\cdot (|\nnf(\neg\psi)|+1)^2$ (line 1 of the \texttt{ModCheck} procedure).
The reason is that the correctness proof of the algorithm exploits the polynomial bound of Theorem~\ref{theorem:polynomialSizeModelProperty} over the formula $\neg\psi$, that has to be converted into \nnf .


The next results state soundness and completeness of  Algorithm~\ref{Chk2} and Algorithm~\ref{ModCheck2}, respectively. Their proofs can be found in Appendix~\ref{proof:lemmamdc} and \ref{proof:ThCorrComplMC}.

\begin{lemma}\label{lemmamdc}
Let $\psi$ be an $\AAbarEEbar$ formula, $\Ku$ be a finite Kripke structure, and $\sigma\in\Trk_\Ku$. Then, $\texttt{Check}(\Ku, \psi,\sigma)=1$ if and only if $\Ku,\sigma\models \psi$.
\end{lemma}

\begin{theorem}\label{ThCorrComplMC}
Let $\psi$ be an $\AAbarEEbar$ formula and $\Ku$ be a finite Kripke structure. Then, \texttt{ModCheck}$(\Ku,\psi)=1$ if and only if $\Ku\models \psi$.
\end{theorem}

The MC procedures require \emph{polynomial working space}, since:
\begin{itemize}
        \item \texttt{ModCheck} needs to store only a trace no longer than $|\States|\cdot (2|\psi|+3)^2$ (obviously, many traces are generated while visiting the unravelling of $\Ku$, but only one at a time needs to be stored);
        \item every recursive call to \texttt{Check} (possibly) needs space for a trace no longer than $|\States|\cdot
        (2|\varphi|+1)^2$, where $\varphi$ is a subformula of $\psi$ such that $|\varphi|\leq |\psi|-1$;
        \item at most one call to \texttt{ModCheck} and $|\psi|$ calls to \texttt{Check} can be simultaneously active.
\end{itemize}
Thus the maximum space needed by the given algorithms is $(|\psi|+1)\cdot O(\log |\States|)\cdot (|\States|\cdot (2|\psi|+3)^2)$ bits, where $O(\log |\States|)$ bits are needed to represent a state of $\Ku$. 

Theorem~\ref{ThCorrComplMC}, along with the above space analysis and the fact that
MC for the fragment $\Ebar$ is \PSPACE-hard (see Appendix~\ref{sect:BbarHard}), entail the following corollary.
\begin{corollary}
The MC problem for $\AAbarEEbar$ formulas over finite Kripke structures is \PSPACE-complete.
\end{corollary}
The same result, that is, \PSPACE-completeness, clearly holds also for any sub-fragment of $\AAbarEEbar$ which features modality $\Ebar$. 

\subsection{$\co\NP$ MC algorithm for $\B$ and $\E$}\label{sec:TheFragmentE}
We now conclude the section by showing that the MC problem for the  fragments $\B$ and $\E$ is in $\co\NP$, that is, they have the same complexity as the purely propositional fragment $\HSprop$ (proved to be hard for $\co\NP$ in~\cite{MMP15B}). We focus on $\E$, as the case  of  $\B$ is completely symmetric. As we will see, the MC algorithm heavily rests again on the polynomial small-model property. 

\begin{algorithm}[tp]
\begin{algorithmic}[1]
	\State{$\rho\gets \texttt{A\_trace}(\Ku,\sinit,|\psi|)$}\Comment{a trace of $\Ku$ from $\sinit$ of length $\leq |\States|\cdot (2|\psi|+3)^2$}
	\If{$\texttt{CheckE}(\Ku,\psi,\rho)=\bot$}
	    \Return{\textbf{Yes}: ``$\Ku,\rho\not\models \psi$''}\Comment{Counterexample found}
	\Else
	    \Return{\textbf{No}: ``$\Ku,\rho\models \psi$''}\Comment{Counterexample \emph{not} found}	
	\EndIf
\end{algorithmic}
\caption{\texttt{CounterExE}$(\Ku,\psi)$}\label{ModCheckE}
\end{algorithm}

\begin{algorithm}[tp]
\begin{algorithmic}[1]
\State{$T\gets \texttt{New\_Bool\_Table}(|\psi|,|\rho|)$}\Comment{creates new table of $|\psi|\cdot |\rho|$ Boolean entries}
\For{all subformulas $\varphi$ of $\psi$ by increasing length}
    \If{$\varphi=p$, for $p\in\Prop$}
        \State{$T[p,|\rho|]\gets p\in \mu(\lst(\rho))$}
        \For{$i=|\rho|-1,\ldots ,1$}
            \State{$T[p,i]\gets T[p,i+1]$ and $p\in \mu(\rho(i))$}
        \EndFor
	\ElsIf{$\varphi=\neg \varphi_1$}
	    \For{$i=|\rho|,\ldots ,1$}
            \State{$T[\varphi,i]\gets$ not $T[\varphi_1,i]$}
        \EndFor


    \ElsIf{$\varphi=\varphi_1\wedge\varphi_2$}
        \For{$i=|\rho|,\ldots ,1$}
            \State{$T[\varphi,i]\gets T[\varphi_1,i]$ and $T[\varphi_2,i]$}
        \EndFor
	\ElsIf{$\varphi=\hsE\varphi_1$}
	    \State{$T[\varphi,|\rho|]\gets\bot$}
	    \For{$i=|\rho|-1,\ldots ,1$}
            \State{$T[\varphi,i]\gets T[\varphi,i+1]$ or $T[\varphi_1,i+1]$}
	    \EndFor
	\EndIf
\EndFor	
\Return{$T[\psi,1]$}
\end{algorithmic}
\caption{\texttt{CheckE}$(\Ku,\psi,\rho)$}\label{ChkE}
\end{algorithm}

The algorithm is based on the \emph{non-deterministic} procedure
\texttt{CounterExE}$(\Ku,\psi)$  (Algorithm~\ref{ModCheckE}) which searches for counterexamples to the input $\E$ formula $\psi$ (that is, initial traces satisfying $\neg\psi$). If such a counterexample is found, clearly $\Ku\not\models \psi$. First, the procedure generates in a \emph{non-deterministic way} an initial trace $\rho$, whose length is at most $|\States|\cdot (2|\psi|+3)^2$, by means of $\texttt{A\_trace}(\Ku,\sinit,|\psi|)$. 
Then, the \emph{deterministic} function $\texttt{CheckE}(\Ku,\psi,\rho)$,  reported in Algorithm~\ref{ChkE}, evaluates $\psi$ over $\rho$. If \texttt{CheckE} returns $\bot$, a counterexample has been found and \texttt{CounterExE} returns \textbf{Yes} (thus the non-deterministic computation of the algorithm is successful). 
Otherwise, it returns \textbf{No} (the computation fails). 

As for the function \texttt{CheckE}, the following result holds. 
\begin{proposition} Let  $\psi$ be an $\E$ formula, $\Ku$ be a finite Kripke structure, and $\rho$ be a trace of $\Ku$. Then,
    \texttt{CheckE}$(\Ku,\psi,\rho)=\top $ if and only if $\Ku,\rho\models\psi$. 
\end{proposition}
\texttt{CheckE} exploits a Boolean
table $T$ with an entry for each pair consisting of a subformula of $\psi$ and the starting position of a suffix of $\rho$ (the size of $T$ is then $|\psi| \cdot |\rho|$). The function scans all the subformulas $\varphi$ of the input $\psi$ by increasing length, and it stores in the Boolean entry $T[\varphi,i]$, for $1\leq i\leq |\rho|$, whether $\Ku,\rho^i\models \varphi$ or not. 
Note that the result of the evaluation of $\psi$ over $\rho$ is stored in $T[\psi,1]$, as $\rho^1=\rho$.
Since subformulas of $\psi$ are considered by increasing length order, during an iteration starting at line 2,  when a subformula $\varphi$ of $\psi$ is being processed, it holds that $T[\xi,i]$ is defined for all other subformulas $\xi$ processed in some previous iteration, and $T[\xi,i]\!=\!\top$ iff $\Ku,\rho^i\!\models\! \xi$. Hence, at the end,
$T[\psi,1]\!=\!\top$ iff $\Ku,\rho \!\models\! \psi$.

We can now prove that the procedure \texttt{CounterExE} is sound and complete.
If \texttt{CounterExE}$(\Ku,\psi)$ has a successful computation, then there exists an initial trace $\rho$ such that $\texttt{CheckE}(\Ku,\psi,\rho)=\bot$. This means that $\Ku,\rho\not\models\psi$, and thus $\Ku\not\models\psi$.
Conversely, if $\Ku\not\models\psi$ then there exists an initial trace $\rho$ such that $\Ku,\rho\not\models \psi$.
By Theorem~\ref{theorem:polynomialSizeModelProperty}, there exists an initial trace $\pi$, whose length is bounded by $|\States|\cdot (|\psi'|+1)^2\leq |\States|\cdot (2|\psi|+3)^2$, such that $\Ku,\pi\models \psi'$, where $\psi'$ is the \nnf{} of $\neg\psi$.
Now, some non-deterministic instance of $\texttt{A\_trace}(\Ku,\sinit,|\psi|)$ generates exactly such $\pi$, being $|\pi|\leq |\States|\cdot (2|\psi|+3)^2$. Moreover, $\texttt{CheckE}(\Ku,\psi,\pi)=\bot$, and thus
\texttt{CounterExE}$(\Ku,\psi)$ has a successful computation.

\texttt{CounterExE}$(\Ku,\psi)$ is in $\NP$, as the generated trace(s) $\rho$ has (have) a length polynomial in $|\States|$ and $|\psi|$, and can thus be computed in polynomial time. Moreover, \texttt{CheckE} performs a polynomial number of steps, since all it has to do is filling in the table $T$, which features $|\psi|\cdot |\rho|$ entries.

\begin{corollary}\label{cor:E}
The MC problem for $\E$ formulas over finite Kripke structures is $\co\NP$-complete.
\end{corollary}
\begin{proof}
Since \texttt{CounterExE}$(\Ku,\psi)$ has a successful computation if and only if $\Ku\not\models \psi$, and such a procedure runs in (non-deterministic) polynomial time, the MC problem belongs to $\co\NP$. 
The $\co\NP$-hardness derives immediately from that of the purely propositional $\HS$ fragment $\HSprop$, as proved in~\cite{MMP15B}.
\end{proof}

In the next section we will see what happens if we add $\Bbar$ to $\AAbarEEbar$, and $\Ebar$ to $\AAbarBBbar$: we shall prove a different (this time, exponential) small-model property for $\AAbarEBbarEbar$ and $\AAbarBBbarEbar$, which will allow us to devise an $\EXPSPACE$ MC algorithm for them.

%% file: Chaps/TCS17/AAbarBBbarEbar.tex
\section{The fragments $\AAbarBBbarEbar$ and $\AAbarEBbarEbar$: ex\-po\-nen\-tial small-model property}\label{sec:AAbarBBbarEbar}

Before going into the details of the section, we would like to summarize the main ideas of~\cite{MMP15},
in which the authors gave the first exponential small-model for $\AAbarBBbarEbar$ and $\AAbarEBbarEbar$, different from the one we shall see in the following.

We recall from Chapter~\ref{chap:MCfullHShomo} that
being associated with the same $B$-descriptor is a sufficient condition for two traces to be indistinguishable with respect to the fulfillment of $\AAbarBBbarEbar$ formulas, provided that the B-nesting depth of the considered formula is less than or equal to the depth of the descriptor itself. 
Thus we may say that a $B_k$-descriptor provides a finite encoding for a possibly infinite set of traces (the traces associated with that descriptor). Unfortunately, the representation of $B_k$-descriptors as trees labelled over descriptor elements is highly redundant. For instance, given any pair of subtrees rooted in some children of the root of a descriptor, it is always the case that one of them is a subtree of the other:
the two subtrees are associated with two (different) prefixes of a trace and one of them is necessarily a prefix of the other. In practice, redundancy of the tree representation of $B_k$-descriptors prevents their direct use in MC algorithms, and makes it difficult to determine their intrinsic complexity.

In~\cite{MMP15}, the authors 
devise a more compact representation of $B_k$-de\-scrip\-tors. 
Their idea is selecting,
for each $B_k$-descriptor witnessed in $\Ku$, a trace called \emph{trace representative}, which is associated with such a descriptor, and whose length is (exponentially) bounded in both the size of $\States$ (the set of states of $\Ku$) and $k$. 
Clearly, they have to prove that 
such a finite bound exists. To the aim they consider suitable ordered sequences (possibly with repetitions) of descriptor elements of a $B_k$-descriptor. Let the \emph{descriptor sequence} for a trace be the ordered sequence of descriptor elements associated with its prefixes. In general, in a descriptor sequence, descriptor elements can be repeated. The authors introduce an equivalence relation that allows them to put together (contracting them) indistinguishable
occurrences of the same descriptor element in a descriptor sequence, that is, 
to detect those occurrences which are associated with prefixes of the trace
with the same $B_{k}$-descriptor. 
A trace representative for a $B_{k}$-descriptor should not feature indistinguishable occurrences of any descriptor elements: 
it can be obtained by iteratively applying such a \emph{contraction method}, which always leads to an exponential (in $|\States|$ and $k$) bounded-length trace indistinguishable from the original one. 
All this is done avoiding the expensive operation of explicitly constructing $B_k$-descriptors.

Thanks to trace representatives,
they finally devise an $\EXPSPACE$
\lq\lq representa\-tive-based\rq\rq{} MC algorithm for $\AAbarBBbarEbar$, which basically can restrict the verification of $\AAbarBBbarEbar$ formulas only over trace representatives, while retaining completeness.

In~\cite{MMP15}, the bound on the length of representatives is calculated in a very technical (and tricky) way.
We now give a much more understandable proof of  $\EXPSPACE$ membership of MC for $\AAbarBBbarEbar$, which makes use of the presented notion of induced trace (Definition~\ref{definition:inducedTrk}).

More precisely,
we prove that  $\AAbarBBbarEbar$ (and $\AAbarEBbarEbar$) features an \emph{exponential small-model property} guaranteeing that, for each $h\geq 0$ and trace $\rho$ of a finite Kripke structure $\Ku= \KuDef$,
it is always possible to find another trace of $\Ku$ \emph{induced by} $\rho$, whose length is at most $(|\States|+2)^{h+2}$,
 which is indistinguishable from $\rho$ with respect to the satisfiability of any $\AAbarBBbarEbar$ (resp., $\AAbarEBbarEbar$) formula $\varphi$ with B-nesting depth $\nestb(\varphi)$ (resp.,  E-nesting depth  $\neste(\varphi)$) at most $h$.

To prove such a property, we first introduce 
the notion of \emph{$h$-prefix bisimilarity} (resp., \emph{$h$-suffix bisimilarity}), 
which defines an equivalence relation over $\Trk_{\Ku}$    
ensuring that equivalent traces  satisfy the same   $\AAbarBBbarEbar$ (resp., $\AAbarEBbarEbar$) formulas with 
 B-nesting (resp., E-nesting) depth at most $h$.  

Then, 
we  show how to determine, for a given  trace $\rho$, a subset of positions of $\rho$  that allow us to build another trace $\rho'$, with length at most $(|\States|+2)^{h+2}$, such that $\rho$ and $\rho'$ are $h$-prefix bisimilar (resp., $h$-suffix bisimilar). We call such a set of $\rho$-positions \emph{prefix} (resp., \emph{suffix}) \emph{sampling} of $\rho$. Intuitively, they play a role which is analogous to that of witness positions (Definition~\ref{definition:WitnessPositions}) used in  the previous section.

%
%

Let $h\geq 0$; \emph{$h$-prefix bisimilarity} and \emph{$h$-suffix bisimilarity} between a pair of traces $\rho$ and $\rho'$ of a Kripke structure are defined as follows. 

\begin{definition}[Prefix-bisimilarity and Suffix-bisimilarity]
Let $h\geq 0$. Two traces  $\rho$ and $\rho'$ of a finite Kripke structure $\Ku$
are \emph{$h$-prefix bisimilar} if the following conditions inductively hold:
\begin{itemize}
  \item for $h=0$: $\fst(\rho)=\fst(\rho')$, $\lst(\rho)=\lst(\rho')$, and $\states(\rho)=\states(\rho')$;
  \item for $h>0$: $\rho$ and $\rho'$ are $0$-prefix bisimilar and for each proper prefix $\nu$ of $\rho$ (resp., proper prefix $\nu'$ of $\rho'$), there exists
  a proper prefix $\nu'$ of $\rho'$ (resp., proper prefix $\nu$ of $\rho$) such that $\nu$ and $\nu'$ are $(h-1)$-prefix bisimilar.
\end{itemize}

The notion of \emph{$h$-suffix bisimilarity} is defined in a symmetric way by considering suffixes of traces, instead of prefixes.
\end{definition}

As it will be established by Proposition~\ref{prop:fulfillmentPreservingSuffixPrefix} below, $h$-prefix (resp., $h$-suffix) bisimilarity is a sufficient condition for two traces  $\rho$ and $\rho'$ to be indistinguishable with respect to the satisfiability of $\AAbarBBbarEbar$ (resp., $\AAbarEBbarEbar$) formulas with B-nesting (resp., E-nesting) depth at most~$h$.

The following property can easily be shown.
\begin{property}
Given a finite Kripke structure $\Ku$, for all $h\geq 0$, $h$-prefix (resp., $h$-suffix) bisimilarity is an equivalence relation over $\Trk_\Ku$.
\end{property}

The following property states that $h$-suffix bisimilarity and $h$-prefix bisimilarity \lq\lq propagate downwards\rq\rq .
\begin{property}\label{property:bisimDown}
Let $h> 0$, and $\rho$ and $\rho'$ be two $h$-prefix (resp., $h$-suffix)  bisimilar traces of a finite Kripke structure $\Ku$. 
Then $\rho$ and $\rho'$ are also $(h-1)$-prefix  (resp., $(h-1)$-suffix) bisimilar.
\end{property}

$h$-prefix and $h$-suffix bisimilarity are preserved by left and right $\star$-concatena\-tion with the same trace. The property can be easily proved by induction on $h\geq 0$.

\begin{proposition}\label{prop:invarianceLeftRightPrefixSuffix} Let $h\geq 0$, and let $\rho$ and $\rho'$ be two $h$-prefix (resp., $h$-suffix) bisimilar traces of a finite Kripke structure $\Ku$. Then, for each trace $\rho''$ of $\Ku$, it holds that:
\begin{enumerate}
  \item $\rho''\star \rho$ and $\rho''\star \rho'$ are $h$-prefix (resp., $h$-suffix) bisimilar;
  \item $\rho\star \rho''$ and $\rho'\star \rho''$ are $h$-prefix (resp., $h$-suffix) bisimilar.
\end{enumerate}
\end{proposition}

By Proposition~\ref{prop:invarianceLeftRightPrefixSuffix} and a straightforward induction on the structural complexity of formulas, we obtain that $h$-prefix (resp., $h$-suffix) bisimilarity preserves the satisfiability of $\AAbarBBbarEbar$ (resp., $\AAbarEBbarEbar$) formulas with B-nesting (resp., E-nesting) depth at most~$h$. In other words, $h$-prefix (resp., $h$-suffix) bisimilarity is a sufficient condition for two traces to be indistinguishable with respect to any $\AAbarBBbarEbar$ (resp., $\AAbarEBbarEbar$) formula $\psi$ with $\nestb(\psi)\leq h$ (resp., $\neste(\psi)\leq h$).

\begin{proposition}\label{prop:fulfillmentPreservingSuffixPrefix} Let $h\geq 0$, and let $\rho$ and $\rho'$ be two $h$-prefix (resp., $h$-suffix)  bisimilar traces of a finite Kripke structure $\Ku$. 
For each $\AAbarBBbarEbar$ (resp., $\AAbarEBbarEbar$) formula $\psi$ with $\nestb(\psi)\leq h$ (resp., $\neste(\psi)\leq h$), it holds that 
\[\Ku,\rho\models\psi\iff \Ku,\rho'\models\psi.\]
\end{proposition}

%

In the remaining part of the section, we will focus on the fragment $\AAbarBBbarEbar$ (the case of $\AAbarEBbarEbar$ is completely symmetric). We show how to determine a subset of positions of a trace $\rho$ (a \emph{prefix sampling} of $\rho$), starting from which it is possible to build another trace $\rho'$, of bounded exponential length, which is indistinguishable from $\rho$ with respect to the satisfiability of $\AAbarBBbarEbar$ formulas up to a given B-nesting depth. 

We start by introducing the notions of   \emph{prefix-skeleton sampling}  and \emph{$h$-prefix sampling}, and prove some related properties.
In the following, we fix a finite Kripke structure $\Ku = \KuDef$, and,  given a set $I$ of natural numbers, by ``two consecutive elements of $I$'' we mean a pair of elements $i,j\in I$ such that $i<j$ and $I\cap [i,j]=\{i,j\}$.

\begin{definition}[Prefix-skeleton sampling]\label{def:skeleton}  Let $\rho$ be a trace of $\Ku$. Given two $\rho$-positions $i$ and $j$, with $i\leq j$, the \emph{prefix-skeleton sampling of $\rho$ in the interval} $[i,j]$ is the \emph{minimal} set $P$ of $\rho$-positions in the interval $[i,j]$ satisfying the next conditions:
\begin{itemize}
  \item $i,j\in P$;
  \item for each state $s\in \States$ occurring in $\rho(i+1,j-1)$, the minimum position $k\in [i+1,j-1]$ such that $\rho(k)=s$ belongs to $P$.
\end{itemize}
\end{definition}

\begin{example}
Figure~\ref{fig:pref} gives a graphical account of the prefix-skeleton sampling of a trace $\rho$ in $[i,j]$, where $\rho(i,j)=s_1^4s_2s_1s_3s_1\allowbreak s_2s_3s_1s_3$.

\begin{figure}[H]
    \centering
    \resizebox{\linewidth}{!}{\includegraphics{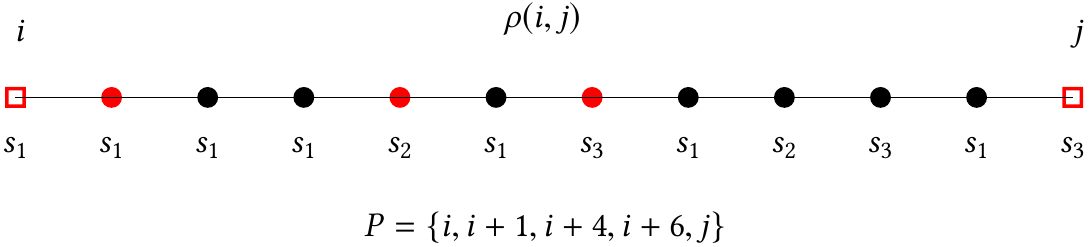}}
    \caption{The set $P$ is the prefix-skeleton sampling of $\rho(i,j)=s_1^4s_2s_1s_3s_1\allowbreak s_2s_3s_1s_3$.}
    \label{fig:pref}
\end{figure}
\end{example}

From Definition \ref{def:skeleton}, it immediately follows that the prefix-skeleton sampling $P$ of (any) trace $\rho$ in $[i,j]$ is such that $|P|\leq |\States|+2$, and if $i<j$, then $i+1\in P$.

\begin{definition}[$h$-prefix sampling] \label{def:hprefixsampling}
 Let $\rho$ be a trace of $\Ku$. For each $h\geq 1$, the \emph{$h$-prefix sampling of $\rho$} is the \emph{minimal} set $P_h$ of $\rho$-positions inductively satisfying the following conditions:
\begin{itemize}
  \item for $h=1$: $P_1$ is the prefix-skeleton sampling of $\rho$ in $[1,|\rho|]$;
  \item for $h>1$:
  $(i)$  $P_h\supseteq P_{h-1}$ and
  $(ii)$ for all pairs of consecutive positions $i,j\in P_{h-1}$,
  the prefix-skeleton sampling of $\rho$ in $[i,j]$ belongs to $P_h$.
\end{itemize}
\end{definition}

The following upper bound to the cardinality of prefix samplings easily follows from Definition \ref{def:hprefixsampling}.

\begin{property}\label{property:prefSamBound}
Let $h\geq 1$ and $\rho$ be a trace of  $\Ku$. The $h$-prefix sampling $P_h$ of $\rho$ is such that $|P_h|\leq (|\States|+2)^{h}$.
\end{property}

Lemma~\ref{lemma:prefixSamplingTwo} and Theorem ~\ref{theorem:singleExpTrackModel} below show how to derive, from any trace $\rho$ of the given finite Kripke structure $\Ku$, another trace $\rho'$, induced by $\rho$ and $h$-prefix  bisimilar to $\rho$, such that $|\rho'|\leq (|\States|+2)^{h+2}$. By Proposition~\ref{prop:fulfillmentPreservingSuffixPrefix}, $\rho'$ is indistinguishable from $\rho$ with respect to the satisfiability of any $\AAbarBBbarEbar$ formula $\psi$ with $\nestb(\psi)\leq h$.

In order to build $\rho'$, we first compute the $(h+1)$-prefix sampling $P_{h+1}$ of $\rho$. Then, for all the pairs of consecutive $\rho$-positions $i,j\in P_{h+1}$, we consider a trace induced by $\rho(i,j)$, with no repeated occurrences of any state, except at most the first and last ones (hence, it is not longer than $|\States|+2$). The trace $\rho'$ is just the ordered concatenation (by means of the $\star$-concatenation operator) of all these traces.
The aforementioned bound on $|\rho'|$ holds as, by Property~\ref{property:prefSamBound}, $|P_{h+1}|\leq (|\States|+2)^{h+1}$.
Lemma~\ref{lemma:prefixSamplingTwo} states that $\rho$ and $\rho'$ are indeed $h$-prefix bisimilar.
The proof of such lemma exploits the following technical result (proved in Appendix~\ref{proof:lemma:prefixSamplingOne}).

\begin{lemma}\label{lemma:prefixSamplingOne} Let $h\geq 1$, $\rho$ be a trace of $\Ku$, and let $i,j$ be two consecutive $\rho$-positions in the $h$-prefix sampling of $\rho$.
Then, for all $\rho$-positions $n,n'\in [i+1,j]$ such that $\rho(n)=\rho(n')$, it holds that $\rho(1,n)$ and $\rho(1,n')$ are $(h-1)$-prefix bisimilar.
\end{lemma}

\begin{lemma}\label{lemma:prefixSamplingTwo} Let $h\geq 1$, $\rho$ be a trace of $\Ku$, and $\rho'=\rho(i_1)\rho(i_2)\cdots \rho(i_k)$ be a trace induced by $\rho$, where $1= i_1<i_2 <\ldots < i_k=|\rho|$ and $P_{h+1} \subseteq\{i_1,\ldots,i_k\}$, with $P_{h+1}$ the \mbox{$(h+1)$-prefix} sampling of $\rho$.
Then for all $j\!\in\! [1,k]$, $\rho'(1,j)$ and $\rho(1,i_j)$ are \mbox{$h$-prefix} bisimilar.
\end{lemma}
Note that, as a straightforward consequence, $\rho$ and $\rho'$ are $h$-prefix bisimilar.
\begin{proof}
Let $Q=\{i_1,\ldots,i_k\}$ (hence $P_{h+1}\subseteq Q$) and
let $j\in [1,k]$. We prove by induction on $j$ that  $\rho'(1,j)$ and $\rho(1,i_j)$ are $h$-prefix bisimilar. As for the base case ($j=1$), the result holds, since $i_1=1$.

Let us assume that $j>1$. We first show that $\rho(1,i_j)$ and $\rho'(1,j)$ are $0$-prefix bisimilar. Clearly, $\rho(1)=\rho(i_1)=\rho'(1)$, $\rho(i_j)=\rho'(j)$, and $\states(\rho'(1,j))\subseteq\states(\rho(1,i_j))$. Now, if, by contradiction, there was a state $s$ such that $s\in\states(\rho(1,i_j))\setminus \states(\rho'(1,j))$, then for all $l\in Q$, with $l\leq i_j$, $\rho(l)\neq s$. However, the prefix-skeleton sampling $P_1$ of $\rho$ in $[1,|\rho|]$ is contained in $Q$, and the minimal $\rho$-position $l'$ such that $\rho(l')=s$ belongs to $P_1$. Since $s\in\states(\rho(1,i_j))$, we have $l'\leq i_j$ and we get a contradiction, implying that $\states(\rho'(1,j)) = \states(\rho(1,i_j))$.

It remains to prove that:
\begin{enumerate}
  \item for each proper prefix $\nu'$ of $\rho'(1,j)$, there exists a proper prefix $\nu$ of $\rho(1,i_j)$  such that $\nu$ and $\nu'$ are $(h-1)$-prefix bisimilar, and
  \item for each proper prefix $\nu$ of $\rho(1,i_j)$, there exists a proper prefix  $\nu'$ of $\rho'(1,j)$ such that $\nu$ and $\nu'$ are $(h-1)$-prefix bisimilar.
\end{enumerate}

As for (1.), let $\nu'$ be a proper prefix of $\rho'(1,j)$. Hence, there exists $m\in [1,j-1]$ such that
$\nu'=\rho'(1,m)$. By the inductive hypothesis, $\rho'(1,m)$ and $\rho(1,i_m)$ are $h$-prefix bisimilar, and thus $(h-1)$-prefix bisimilar as well by Property~\ref{property:bisimDown}.
Since $\rho(1,i_m)$ is a proper prefix of $\rho(1,i_j)$, by choosing $\nu=\rho(1,i_m)$, (1.) follows.

As for (2.), assume that $\nu$ is a proper prefix of $\rho(1,i_j)$. Therefore, there exists $n\in [1,i_j-1]$ such that $\nu= \rho(1,n)$. We distinguish two cases:
\begin{itemize}
  \item $n\in P_{h+1}$. Since $n<i_j$, there exists $m\in [1,j-1]$ such that $n=i_m$. By the inductive hypothesis, $\rho(1,n)$ and $\rho'(1,m)$ are $h$-prefix bisimilar, and thus $(h-1)$-prefix bisimilar as well by Property~\ref{property:bisimDown}.
  Since $\rho'(1,m)$ is a proper prefix of $\rho'(1,j)$, by choosing $\nu'=\rho'(1,m)$, (2.) follows.
  \item $n\notin P_{h+1}$. It follows that there exist two consecutive positions $i'$ and $j'$ in $P_{h+1}$, with $i'<j'$, such that $n\in [i'+1,j'-1]$. By definition of $(h+1)$-prefix sampling, there exist two consecutive positions $i''$ and $j''$ in the $h$-prefix sampling of $\rho$, with $i''<j''$, such that $i'$ and $j'$ are two consecutive positions in the prefix-skeleton sampling of $\rho$ in the interval $[i'',j'']$.

  First, we observe that $i'\neq i''$ (otherwise, $j'=i'+1$, which contradicts the fact that $[i'+1,j'-1]\neq \emptyset$, as $n\in [i'+1,j'-1]$). Thus, by definition of prefix-skeleton sampling applied to $\rho$ in $[i'',j'']$, and since $n\in [i'+1,j'-1]$, there must be $\ell\in [i''+1,i']$ such that $\rho(\ell)=\rho(n)$ and $\ell$ is in the prefix-skeleton sampling of $\rho$ in $[i'',j'']$. Hence, $\ell\in P_{h+1}$ by definition of $(h+1)$-prefix sampling. As a consequence, since $\ell<n<i_j$, there exists $m\in [1,j-1]$ such that $\ell=i_m$. By applying Lemma~\ref{lemma:prefixSamplingOne}, we deduce that $\rho(1,n)$ and $\rho(1,i_m)$ are $(h-1)$-prefix bisimilar. Moreover, by the inductive hypothesis, $\rho(1,i_m)$ and $\rho'(1,m)$ are $(h-1)$-prefix bisimilar. Thus, by choosing $\nu'=\rho'(1,m)$, $\nu'$ is a proper prefix of $\rho'(1,j)$ which is $(h-1)$-prefix bisimilar to $\nu=\rho(1,n)$.
\end{itemize}
This concludes the proof of Lemma~\ref{lemma:prefixSamplingTwo}.
\end{proof}

We are now ready to prove the exponential small-model property.
\begin{theorem}[Exponential small-model property for $\AAbarBBbarEbar$]\label{theorem:singleExpTrackModel}
Let $\rho$ be a trace of a finite Kripke structure $\Ku$ and let $h\geq 0$.
Then there exists a trace $\rho'$ induced by $\rho$, whose length is at most $(|\States|+2)^{h+2}$, such that for every $\AAbarBBbarEbar$ formula $\psi$  with $\nestb(\psi)\leq h$, it holds 
\[\Ku,\rho\models \psi\iff \Ku,\rho'\models \psi.\]
\end{theorem}
\begin{proof} Let $P_{h+1}$ be the $(h+1)$-prefix sampling of $\rho$. For all pairs of consecutive $\rho$-positions $i$ and $j$ in $P_{h+1}$,
there exists a trace induced by $\rho(i,j)$, whose length is at most $|\States|+2$, featuring no repeated occurrences of any internal state.
We now define $\rho'$ as the trace of $\Ku$ obtained by an ordered concatenation of all these induced traces by means of the $\star$-concatenation operator.
It is immediate to see that $\rho'=\rho(i_1)\rho(i_2)\cdots \rho(i_k)$, for some indexes $1= i_1<i_2 <\ldots < i_k=|\rho|$, where
$\{i_1,\ldots,i_k\}$ contains the $(h+1)$-prefix sampling $P_{h+1}$ of $\rho$. It holds that $|\rho'|\leq |P_{h+1}|\cdot (|\States|+2)$ and since, by Property~\ref{property:prefSamBound}, $|P_{h+1}|\leq (|\States|+2)^{h+1}$, we obtain that $|\rho'|\leq (|\States|+2)^{h+2}$. Moreover, by Lemma~\ref{lemma:prefixSamplingTwo}, $\rho$ and $\rho'$ are $h$-prefix bisimilar. By Proposition~\ref{prop:fulfillmentPreservingSuffixPrefix} the thesis follows.
\end{proof}

Theorem \ref{theorem:singleExpTrackModel} allows us to easily devise an \EXPSPACE\ MC algorithm for $\AAbarBBbarEbar$ formulas (and symmetrically for $\AAbarEBbarEbar$ formulas), which can be obtained from Algorithms \ref{ModCheck2} and \ref{Chk2} by adapting the bounds on the length of considered traces.

We observe that the polynomial small-model property for $\AAbarBBbar$/$\AAbarEEbar$ of the previous section depends \emph{on the specific formula $\varphi$} we are considering (as the input of the MC problem), whereas the exponential small-model property for $\AAbarBBbarEbar$/$\AAbarEBbarEbar$ states the existence of a shorter trace $\rho'$ equivalent to (a generic) $\rho$ with respect to \emph{all} formulas up to a given B/E-nesting depth $h$. As a matter of fact, the former relies on the witness positions, which are defined on $\varphi$; the latter on $h$-prefix/suffix bisimilarity and $h$-prefix/suffix samplings, that are independent of any formula (they are only based on $h$, i.e., the maximum B/E-nesting depth of formulas we want to consider).
Therefore, we can say that the latter small-model states a stronger property; however, this \emph{may} lead to a bound on the length of equivalent traces higher than necessary: 
we have proved the MC problem for $\AAbarBBbarEbar$/$\AAbarEBbarEbar$ to be in \EXPSPACE, but it is only known to be \PSPACE-hard (since $\Ebar$ or $\Bbar$ are enough for \PSPACE-hardness, as shown in Appendix~\ref{sect:BbarHard}). We do not know whether this  complexity gap is due to the small-model proving a loose bound (that might be strengthened by finding another characterization depending on the input formula as well), or to a weak complexity lower-bound (here, exploiting the other modalities $\A$, $\Abar$ and $\B$/$\E$, along with $\Bbar$ and $\Ebar$ jointly, may enable us to prove a stronger one), or to both at the same time. We will come back to $\AAbarBBbarEbar$/$\AAbarEBbarEbar$ in Chapter~\ref{chap:Gand17}, where we will relax homogeneity and will be able to decrease the complexity upper bound (yet still not matching the lower) also for the  homogeneous case.

%% file: Chaps/TCS17/concl.tex
\section{Conclusions}
In this chapter we have studied two well-behaved $\PSPACE$-complete fragments, $\AAbarBBbar$ and $\AAbarEEbar$, which
are quite promising from the point of view of applications.
MC for $\AAbarBBbarEbar$ and
$\AAbarEBbarEbar$ turns out to be in $\EXPSPACE$ and $\PSPACE$-hard.

Membership to $\PSPACE$ for the former two fragments and to $\EXPSPACE$ for the latter have been proved by means of
small-models which, in turn, rely on suitable (depending on the specific fragment) contraction techniques applied to the traces of a finite Kripke structure.
While the first result is novel, the second 
substantially simplifies the constructions and the complexity of the proofs
given for the same problem in~\cite{MMP15}.

%% file: Chaps/IC17/IC17main.tex
\chapter{$\HS$ fragments at the bottom of the polynomial hierarchy}\label{chap:IC17}
\begin{chapref}
The references for this chapter are \cite{BOZZELLI2018IC,gandalf16,kr16}.
\end{chapref}

\minitoc\mtcskip

\newcommand{\mods}{\mathsf{ModSubf}_\AAbar}

\newcommand{\TBSAT}{TB(SAT)}
\newcommand{\TBSATM}{TB(SAT)$_{1\times M}$}
\newcommand{\forw}{\textsc{forward}}
\newcommand{\back}{\textsc{backward}}

\input{Chaps/IC17/intro}

\input{Chaps/IC17/example}
\input{Chaps/IC17/classesCOMPLETO}

\input{Chaps/IC17/section01}
\input{Chaps/IC17/section02}
\input{Chaps/IC17/section03}
\input{Chaps/IC17/section04}
\input{Chaps/IC17/conclus}


%% file: Chaps/IC17/intro.tex
\lettrine[lines=3]{I}{n this chapter,} we study the MC problem for some of the sub-fragments of $\AAbarBBbar$ and of $\AAbarEEbar$---removing $\Bbar$ and $\Ebar$ which lead to $\Psp$-hardness---namely, $\A$, $\Abar$, $\AAbar$, $\AB$, $\AbarB$, $\AE$, $\AbarE$, $\AAbarB$, and  $\AAbarE$.
All these have a similar computational complexity, as their MC settles in one of the lowest levels of the \emph{polynomial-time hierarchy}, $\PTIME^{\NP}$, or even below. 
We recall that $\PTIME^{\NP}$ is the set of problems decided by a deterministic polynomial-time Turing machine, endowed with an oracle for the class $\NP$
which decides, in one computation step, whether an instance of a problem belonging to $\NP$ is positive or not. 
$\PTIME^{\NP}$ is also referred to as $\PTIME$ \emph{relative to} $\NP$.

In order to solve their MC, $\AAbarB$, $\AAbarE$, $\AB$, and $\AbarE$ require the $\PTIME$ Turing machine to perform $O(n^k)$
queries to the $\NP$ oracle, for an input of size $n$ and for some constant $k\geq 0$.
Conversely, $\A$, $\Abar$, $\AAbar$, $\AbarB$, and $\AE$ are \lq\lq easier\rq\rq{}
since, for them, $O(\log^2 n)$ queries are always enough:
the MC problem for the latter fragments witnesses
a \lq\lq non-standard\rq\rq{} complexity class in the polynomial-time hierarchy, called \emph{bounded-query} class and denoted as $\Thsq$, that will be presented in more detail below.

More formally, we first prove that MC for $\AB$, $\AbarE$, $\AAbarB$, and $\AAbarE$ is a $\PTIME^{\NP}$-complete problem.
To this end, we design a $\PTIME^{\NP}$ algorithm exploiting the polynomial small-model property of Section~\ref{sec:AAbarEEbar} and we prove a matching complexity lower bound.%
%
Next, we devise another MC algorithm for all the remaining fragments, $\A$, $\Abar$, $\AAbar$, $\AbarB$, and $\AE$, via a reduction \emph{to} the problem \TBSATM~\cite{schnoebelen2003} (a problem complete for the aforementioned bounded-query class), whose instances are complex circuits in which some of the gates feature $\NP$ oracles.
Finally, we prove a lower bound showing that at least $\log n$ queries are needed to solve MC; unfortunately, it does not match the upper bound, leaving open the question whether the problem can be solved by $o(\log^2 n)$ (i.e., strictly less than $O(\log^2 n)$) queries to an $\NP$ oracle, or a tighter lower bound can be proved (or both).
Figure~\ref{fig:lattice} depicts the scenario of complexity of MC for $\AAbarBBbar$ and all its sub-fragments.
\begin{sidewaysfigure}
    \centering
    \input{Chaps/IC17/lattice}
    \caption{The computational complexity of MC for the sub-fragments of $\AAbarBBbar$.}
    \label{fig:lattice}
\end{sidewaysfigure}

Before moving on, we want to intuitively explain why, for instance, $\AbarB$ is \lq\lq easier\rq\rq{} than $\AB$ (the same holds for the symmetric fragments $\AE$ and $\AbarE$); this is justified by the different expressiveness of the two fragments.
Let us consider an $\A\B$ formula of the form $\hsB \hsA \theta$. A trace $\rho$ satisfies $\hsB \hsA \theta$ if there exists a prefix $\tilde{\rho}$ of $\rho$ from which a branch, i.e., a trace starting from $\lst(\tilde{\rho})$, satisfying $\theta$ departs. 
Hence, $\A\B$ allows one to state specific properties of the branches departing from a state occurring in a given path. Such an ability will be exploited in Section~\ref{sec:ABhard} to prove the $\PTIME^{\NP}$-hardness of $\AB$.
This kind of properties cannot be expressed in the fragment $\AbarB$. Indeed, for any given trace $\rho$, modality $\hsAt$ only allows one to \lq\lq select\rq\rq{} traces leading to the first state of $\rho$, and modality $\hsB$ is of no help: if we consider any prefix $\tilde{\rho}$ of $\rho$, the set of traces leading to its first state is exactly the same as the set of those leading to the first state of $\rho$, as $\fst(\tilde{\rho}) = \fst(\rho)$. Therefore, pairing $\hsAt$ and $\hsB$ does not give any advantage in terms of expressiveness. 
%
Finally, since $\A$, $\Abar$, and $\AAbar$ are devoid of modalities for prefixes (and suffixes), they analogously belong to $\Thsq$.

\paragraph*{Organization of the chapter.}
\begin{itemize}
	\item In the next section, we start by recalling some complexity classes that come into play.
	\item In Section~\ref{sec:AAbarBalgo}, we describe a $\PTIME^{\NP}$ MC algorithm for $\AAbarB$ (and $\AAbarE$). 
	\item In Section~\ref{sect:AAbarAlg}, we provide a different MC algorithm for the fragments $\AAbar$, $\AbarB$, and $\AE$, whose computational complexity is lower, 
as it requires only $O(\log^2 n)$ queries to a $\NP$ oracle.
	\item In Section~\ref{sec:ABhard}, we prove a $\PTIME^{\NP}$ lower bound for $\AB$ (resp., $\AbarE$) MC. The bound immediately propagates to $\AAbarB$ (resp., $\AAbarE$), closing the complexity gap and proving that the MC problem for  $\AB$, $\AbarE$, $\AAbarB$, and $\AAbarE$ is $\PTIME^{\NP}$-complete. 
	\item In Section~\ref{sect:AHard}, we prove that MC for $\A$ and $\Abar$ formulas requires at least $\log n$ queries to a $\NP$ oracle: this bound propagates to $\AAbar$, $\AbarB$, and $\AE$.
\end{itemize}

%% file: Chaps/IC17/lattice.tex
\begin{tikzpicture}
\tikzset{
    every node/.style={
        draw
    },
    every edge/.style={
        draw, gray
    }
}

\tikzstyle{coNPstyle}=[draw,densely dashed,circle]
\tikzstyle{PNPlogstyle}=[draw,densely dotted,thick,rectangle, minimum width=1.2cm]
\tikzstyle{PNPstyle}=[draw,regular polygon,regular polygon sides=9,inner sep=0.5pt]
\tikzstyle{PSPstyle}=[draw,dash dot dot,thick,ellipse, minimum width=1.2cm]

\node [style=coNPstyle] (v2) at (-3.5,-3) {$\B$};
\node [style=PNPlogstyle] (v3) at (-0.5,-3) {$\A$};
\node [style=PNPlogstyle] (v4) at (2.5,-3) {$\Abar$};
\node [style=PSPstyle] (v5) at (5.5,-3) {$\Bbar$};

\node [style=PNPstyle](v6) at (-6.5,-1.5) {$\AB$};
\node [style=PNPlogstyle](v9) at (-3.5,-1.5) {$\AAbar$};
\node [style=PNPlogstyle] (v7) at (-0.5,-1.5) {$\AbarB$};
\node [style=PSPstyle] (v10) at (2.5,-1.5) {$\ABbar$};
\node [style=PSPstyle] (v11) at (5.5,-1.5) {$\mathsf{\overline{AB}}$};
\node [style=PSPstyle] (v8) at (8.5,-1.5) {$\mathsf{B\overline{B}}$};

\node [style=PNPstyle] (v12) at (-3.5,0) {$\AAbarB$};
\node [style=PSPstyle] (v13) at (-0.5,0) {$\ABBbar$};
\node [style=PSPstyle] (v14) at (2.5,0) {$\mathsf{A\overline{AB}}$};
\node [style=PSPstyle] (v15) at (5.5,0) {$\mathsf{\overline{A}B\overline{B}}$};

\node [style=PSPstyle] (v16) at (1,1.5) {$\AAbarBBbar$};
\node [style=coNPstyle] (v1) at (1,-4.5) {$\HSprop$};

\draw  (v1) edge (v2);
\draw  (v1) edge (v3);
\draw  (v1) edge (v4);
\draw  (v1) edge (v5);
\draw  (v2) edge (v6);
\draw  (v2) edge (v7);
\draw  (v2) edge (v8);
\draw  (v3) edge (v6);
\draw  (v3) edge (v9);

\draw  (v3) edge (v10);
\draw  (v4) edge (v9);
\draw  (v4) edge (v7);
\draw  (v4) edge (v11);
\draw  (v5) edge (v8);
\draw  (v5) edge (v11);
\draw  (v5) edge (v10);
\draw  (v6) edge (v12);
\draw  (v6) edge (v13);
\draw  (v9) edge (v12);
\draw  (v9) edge (v14);
\draw  (v7) edge (v12);
\draw  (v7) edge (v15);
\draw  (v10) edge (v13);
\draw  (v10) edge (v14);
\draw  (v11) edge (v14);
\draw  (v11) edge (v15);
\draw  (v12) edge (v16);
\draw  (v13) edge (v16);
\draw  (v14) edge (v16);
\draw  (v15) edge (v16);

\draw  (v8) edge (v15);
\draw  (v8) edge (v13);

\node [style=coNPstyle] at (-6.5,-5.5) {\scriptsize $\co\NP$};
\node [style=PNPlogstyle] at (-4.4,-5.5) {\scriptsize $\Thsq$};
\node [style=PNPstyle] at (-2.5,-5.5) {\scriptsize $\PTIME^{\NP}$};
\node [style=PSPstyle] at (-0.5,-5.5) {\scriptsize $\PSPACE$};

\end{tikzpicture}

%% file: Chaps/IC17/example.tex
Before going into the results, let us make the following example.

\begin{example}
We show how some meaningful properties, to be checked against
the Kripke structure
$\Ku_{Sched}$ of Figure~\ref{KSched} at page~\pageref{KSched},
can be expressed  by formulas of $\AbarE$ (a fragment that will be studied in detail next).

As in Example~\ref{example:Ksched}, in all formulas we force the validity of the considered property over all legal computation sub-intervals by using modality $\hsEu$ (all computation sub-intervals are suffixes of at least one initial trace).

Truth of the next statements can easily be checked (formulas express the same properties as those in Example~\ref{example:Ksched}, but are here adapted to $\AbarE$):
\begin{itemize}
    \item $\Ku_{Sched}\models\hsEu\big(\hsE^3\top \rightarrow (\chi(p_1,p_2) \vee \chi(p_1,p_3) \vee \chi(p_2,p_3))\big)$, \newline where $\chi(p,q)$ stands for $\hsE\hsAt p \wedge \hsE\hsAt q$;
    \item $\Ku_{Sched}\not\models\hsEu(\hsE^{10}\top \rightarrow \hsE\hsAt p_3)$;
    \item $\Ku_{Sched}\not\models\hsEu(\hsE^5 \rightarrow (\hsE\hsAt p_1 \wedge \hsE\hsAt p_2 \wedge \hsE\hsAt p_3))$.
\end{itemize}
The first formula states that in any suffix having length at least 4 of an initial trace, at least 2 proposition letters are witnessed. $\Ku_{Sched}$ satisfies the formula since a process cannot be executed twice in a row. 

The second formula states that in any suffix of an initial trace having length at least 11, process 3 is executed at least once in some internal states (\emph{non starvation}). $\Ku_{Sched}$ does not satisfy the formula since the scheduler can avoid executing a process ad libitum. 

The third formula requires that in any suffix of an initial trace having length at least 6, $p_1$, $p_2$ and $p_3$ are all witnessed.
The only way to satisfy this property is to constrain the scheduler to execute the three processes in a strictly periodic manner (\emph{strict alternation}), that is, $p_i p_j p_k p_i p_j p_k p_i p_j p_k\ldots$, $i,j,k \in \{1,2,3\}, i \neq j \neq k \neq i$, but this is not the case.
\end{example}

%% file: Chaps/IC17/classesCOMPLETO.tex
\section{Some complexity classes in the polynomial hierarchy}\label{sub:compl}

The \emph{polynomial-time hierarchy}, denoted by $\PH$, was introduced by Stockmeyer in~\cite{stockmeyer1976}, and is defined as \[\PH=\bigcup_{k\in\mathbb{N}}\Delta^p_k, \] 
where $\Delta_0^p=\Sigma_0^p=\Pi_0^p=\PTIME$
and, for all $k\geq 1$, 
\[
\Delta_k^p=\PTIME^{\Sigma_{k-1}^p},
\qquad 
\Sigma_k^p=\NP^{\Sigma_{k-1}^p},
\qquad 
\Pi_k^p=\co\Sigma_k^p.
\]
In particular, we have that $\Delta_1^p=\PTIME$, $\Sigma_1^p=\NP$, and $\Delta_2^p=\PTIME^{\NP}$. A well-known example of complete problem for $\Sigma^p_k$ (resp., $\Pi^p_k$) is to decide the truth of fully-quantified formulas of the form 
\[Q_1 x_1 Q_2 x_2\cdots Q_n x_n \phi(x_1,x_2,\ldots ,x_n),\] where $\phi(x_1,x_2,\ldots ,x_n)$ is a quantifier-free Boolean formula whose variables range in the set $\{x_1,x_2,\ldots ,x_n\}$, $Q_i\in\{\exists,\forall\}$, for all $2\leq i \leq n$, $Q_1=\exists$ (resp., $Q_1=\forall$), and there are $k-1$ quantifier alternations, that is, $k-1$ different indexes $j>1$ such that $Q_j\neq Q_{j-1}$.
On the contrary, $\Delta_k^p$ does not feature very popular complete problems. As an example, for each $k\geq 1$, a $\Delta_{k+1}^p$-complete problem is to decide whether, given a true quantified Boolean formula of the form 
\begin{equation*}
\exists x_1\cdots \exists x_r \forall x_{r+1} Q_{r+2} x_{r+2}\cdots Q_n x_n \phi(x_1,\ldots ,x_n),
\end{equation*}
with $k-1$ quantifier alternations, the lexicographically maximum truth assignment $\upsilon$ to the variables $(x_1,\ldots , x_r)$ such that 
\begin{equation*}
\forall x_{r+1} Q_{r+2} x_{r+2}\cdots Q_n x_n \phi(\upsilon (x_1),\ldots , \upsilon (x_r), x_{r+1}, \ldots , x_n)
\end{equation*}
is true assigns $1$ to $x_r$~\cite{gottlob1995}.

As a particular case, given a satisfiable Boolean formula $\phi(x_1,\ldots , x_n)$, the problem of deciding whether the lexicographically maximum truth assignment to $(x_1,\ldots , x_n)$ satisfying $\phi$ assigns $1$ to $x_n$ is complete for $\Delta_2^p=\PTIME^{\NP}$.
For other examples of $\PTIME^{\NP}$-complete problems (many of them are related to MC) we refer the reader to~\cite{batzold2009,LMS01,LMS02,Lmp10}.

Above $\NP$ and $\co\NP$, but below $\PTIME^{\NP}$, is the class $\Th$, introduced by Papadimitriou and Zachos in~\cite{Papadimitriou82}, which is the set of problems decided by a deterministic $\PTIME$ algorithm (Turing machine) which requires only $O(\log n)$ queries to an $\NP$ oracle (being $n$ the input size). Analogously, $\Thsq$ is the set of problems decided by a $\PTIME$ algorithm requiring $O(\log^2 n)$ queries to an $\NP$ oracle.\footnote{Here and in the following, we assume that the polynomial hierarchy $\PH$ is not collapsing, and that $\PTIME^{\NP}$, $\Th$, and $\Thsq$ are \emph{distinct}, as it is widely conjectured.} These complexity classes (and all others which set a bound on the number of allowed queries) are called \emph{bounded query classes}. Note that $\PTIME^{\NP}$, $\Th$ and $\Thsq$ are closed under complementation, as well as under $\LOGSPACE$ (many-one) reductions.

As for $\Th$, it has been proved (see~\cite{buss1991,wagner90}) that $\Th = \LOGSPACE^{\NP}$ $= \Thpar$, where $\Thpar$ is the class of problems decided by a deterministic $\PTIME$ algorithm which performs a single round (or a \emph{constant} number of rounds) of parallel queries to an $\NP$ oracle. By \emph{parallel queries}, we mean that each query is independent of the outcome of any other or, equivalently, that all queries have to be formulated before the oracle is consulted. Obviously, the constraint of parallelism is not necessarily fulfilled in the class $\PTIME^{\NP}$, where a query to the oracle may be \emph{adaptive}, that is, it  may depend on the results of previously performed queries.
An example of complete problem for $\Th$ is PARITY(SAT): given a set of Boolean formulas $\Gamma=\{\phi_1,\ldots , \phi_n\}$, the problem is to decide if the number of \emph{satisfiable} formulas in $\Gamma$ is odd or even~\cite{WAGNER87}.

As for $\Thsq$, it has been proved in~\cite{castro92} that $\Thsq=\Thparlogn$, (in $\Thparlogn$ a succession of $O(\log n)$ parallel query rounds are allowed). To the best of our knowledge, the first complete problems for this class were introduced in~\cite{schnoebelen2003}. Among these, a detailed account of the problem \TBSATM \ will be given in Section~\ref{sect:AAbarAlg}.

%% file: Chaps/IC17/section01.tex
\section{$\PTIME^{\NP}$ MC algorithm for $\AAbarB$ and $\AAbarE$}\label{sec:AAbarBalgo}
In this section, we present an MC algorithm for $\AAbarB$ formulas (see the procedure \texttt{MC} reported in Algorithm~\ref{MC}) with 
complexity in the class $\PTIME^{\NP}$. 
W.l.o.g., we restrict our attention to $\AAbarB$ formulas devoid of occurrences of conjunctions and universal modalities 
(definable, as usual, by means of disjunctions, negations, and existential modalities).

\begin{algorithm}[b]
\begin{algorithmic}[1]
	\For{all $\hsA\phi\in \mods(\psi)$}
		\State{\texttt{MC}$(\Ku,\phi,\textsc{forward})$}
	\EndFor
	\For{all $\hsAt\phi\in \mods(\psi)$}
		\State{\texttt{MC}$(\Ku,\phi,\textsc{backward})$}
	\EndFor
	
	\For{all $s\in \States$}
		\If{\textsc{direction} is \textsc{forward}}
			\State{$V_{\A}(\psi,s)\gets Success(\texttt{Oracle}(\Ku,\psi,s,\textsc{forward},V_{\A}\cup V_{\Abar}))$}
		\ElsIf{\textsc{direction} is \textsc{backward}}
			\State{$V_{\Abar}(\psi,s)\gets Success(\texttt{Oracle}(\Ku,\psi,s,\textsc{backward},V_{\A}\cup V_{\Abar}))$}
		\EndIf
	\EndFor
\end{algorithmic}
\caption{\texttt{MC}$(\Ku,\psi,\textsc{direction})$}\label{MC}
\end{algorithm}

The MC procedure \texttt{MC} for a formula $\psi$ against a Kripke structure $\Ku$ exploits two global vectors, $V_{\A}$ and $V_{\Abar}$, which can be seen as the tabular representations of two Boolean functions taking as arguments a subformula $\phi$ of $\psi$ and a state $s$ of $\Ku$. 
The function $V_{\A} (\phi,s)$ (resp., $V_{\Abar}(\phi,s)$) returns $\top$ if and only if there exists a 
trace $\rho \in \Trk_{\Ku}$ starting from the state $s$ (resp., leading to the state $s$) such that $\Ku, \rho \models \phi$. 
$\texttt{MC}$ is initially invoked with parameters $(\Ku,\neg\psi,\textsc{forward})$. During the execution, it instantiates the entries of
$V_{\A}$ and $V_{\Abar}$, which are exploited in order to answer the MC problem $\Ku\models \psi$. In the end, this is equivalent to checking whether $V_{\A}(\neg\psi,\sinit)=\bot$, where $\sinit$ is the initial state of $\Ku$.

Let us consider $\texttt{MC}$ in more detail. Along with the Kripke structure $\Ku$ and the formula $\psi$, $\texttt{MC}$ features a third parameter, $\textsc{direction}$, which can be assigned the value $\textsc{forward}$ (resp., $\textsc{backward}$), that is used in combination with the modality $\hsA$ (resp., $\hsAt$) for a forward (resp., backward) unravelling of $\Ku$. 
$\texttt{MC}$ is applied recursively (lines 1--4) on the nesting of modalities $\hsA$ and $\hsAt$ in the formula $\psi$ (in the base case, $\psi$ features no occurrences of $\hsA$ or $\hsAt$). In order to instantiate the Boolean vectors $V_{\A}$ and $V_{\Abar}$, an oracle is invoked (lines 5--9) for each state $s$ of the Kripke structure. Such an invocation is syntactically represented by $Success(\texttt{Oracle}(\Ku,\psi,s,\textsc{direction},V_{\A}\cup V_{\Abar}))$, and it returns $\top$ whenever there exists a computation of the non-deterministic algorithm $\texttt{Oracle}(\Ku,\psi,s,\textsc{direction},\allowbreak V_{\A}\cup V_{\Abar})$ returning $\top$, namely, whenever there is a suitable trace starting from, or leading to $s$ (depending on the value of the parameter $\textsc{direction}$), satisfying $\psi$.

We define now the set of $\AAbar$-\emph{modal subformulas} of $\psi$ ($\mods(\psi)$) used to ``direct'' the recursive calls of $\texttt{MC}$ (lines 1--4).
\begin{definition}\label{def:modsubf}
The set $\mods(\psi)$ of $\AAbar$-\emph{modal subformulas} of an $\AAbarB$ formula $\psi$ is the set of subformulas of $\psi$  either of the form $\hsA \psi'$ or of the form $\hsAt \psi'$, for some $\psi'$, which are \emph{not in the scope of any $\hsA$ or $\hsAt$ modality}.
\end{definition}
As an example, it holds that
\begin{itemize}
    \item $\mods(\hsA\hsAt q)=\{\hsA\hsAt q\}$, and
    \item $\mods\big(\big(\hsA p\, \wedge\, \hsA\hsAt q \big)\rightarrow \hsA p\big)\allowbreak =\{\hsA p,\hsA\hsAt q\}$.
\end{itemize}

$\texttt{MC}$ is recursively called on each formula $\phi$ such that $\hsA\phi$ or $\hsAt\phi$ belongs to the set $\mods(\psi)$ (lines 1--4). 
In this way, we can recursively gather in the Boolean vectors $V_{\A}$ and $V_{\Abar}$, by increasing nesting depth of the modalities $\hsA$ and $\hsAt$, the oracle answers for all the formulas $\psi'$ such that $\hsA\psi'$, or $\hsAt\psi'$, is a subformula (be it maximal or not) of $\psi$.

\begin{algorithm}[tp]
\begin{algorithmic}[1]
\State{$\tilde{\rho}\gets \texttt{A\_trace}(\Ku,s,|\States|\cdot(2|\psi|+1)^2,\textsc{direction})$}\Comment{a trace of $\Ku$ from/to $s$ having length $\leq |\States|\cdot (2|\psi|+1)^2$}
\For{all $\hsA \phi\in\mods(\psi)$}
    	\For{$i=1,\ldots ,|\tilde{\rho}|$}
    		\State{$T[\hsA\phi,i]\gets V_{\A}(\phi,\tilde{\rho}(i))$}
    	\EndFor
\EndFor
\For{all $\hsAt \phi\in\mods(\psi)$}
    	\For{$i=1,\ldots ,|\tilde{\rho}|$}
    		\State{$T[\hsAt\phi,i]\gets V_{\Abar}(\phi,\fst(\tilde{\rho}))$}
    	\EndFor
\EndFor

\For{all subformulas $\varphi$ of $\psi$, not contained in (or equal to) $\AAbar$-modal subformulas of $\psi$, by increasing length}
    \If{$\varphi=p$, for $p\in\Prop$}
        \State{$T[p,1]\gets p\in \Lab(\fst(\tilde{\rho}))$}
        \For{$i=2,\ldots ,|\tilde{\rho}|$}
            \State{$T[p,i]\gets T[p,i-1]$ and $p\in \Lab(\tilde{\rho}(i))$}
        \EndFor
 	\ElsIf{$\varphi=\neg \varphi_1$}
	    \For{$i=1,\ldots ,|\tilde{\rho}|$}
            \State{$T[\varphi,i]\gets$ not $T[\varphi_1,i]$}
        \EndFor        
    \ElsIf{$\varphi=\varphi_1\vee\varphi_2$}
        \For{$i=1,\ldots ,|\tilde{\rho}|$}
            \State{$T[\varphi,i]\gets T[\varphi_1,i]$ or $T[\varphi_2,i]$}
        \EndFor
	\ElsIf{$\varphi=\hsB\varphi_1$}
	    \State{$T[\varphi,1]\gets\bot$}
	    \For{$i=2,\ldots ,|\tilde{\rho}|$}
            \State{$T[\varphi,i]\gets T[\varphi,i-1]$ or $T[\varphi_1,i-1]$}
	    \EndFor
	\EndIf
\EndFor	
\Return{$T[\psi,|\tilde{\rho}|]$}
\end{algorithmic}
\caption{\texttt{Oracle}$(\Ku,\psi,s,\textsc{direction},V_{\A}\cup V_{\Abar})$}\label{Oracle}
\end{algorithm}

We now consider the \emph{non-deterministic polynomial time} procedure $\texttt{Oracle}(\Ku,\allowbreak \psi,s, \textsc{direction}, V_{\A}\cup V_{\Abar})$ (see Algorithm~\ref{Oracle}), which is used as the \lq\lq basic engine\rq\rq{} by the oracle in the aforementioned MC Algorithm~\ref{MC}. 
The idea underlying Algorithm~\ref{Oracle} is to first non-deterministically generate a trace $\tilde{\rho}$ by unravelling the Kripke structure $\Ku$
according to the parameter $\textsc{direction}$, and then to verify $\psi$ over $\tilde{\rho}$.
Such a procedure exploits the proved \emph{polynomial small-model property} for formulas of $\AAbarBBbar$ (Theorem~\ref{theorem:polynomialSizeModelProperty}), 
guaranteeing that, in order to check the satisfiability of an $\AAbarBBbar$ formula $\phi$, it is enough to consider traces whose length is bounded by $|\States|\cdot (2|\phi|+1)^2$. 

An execution of \texttt{Oracle}$(\Ku,\psi,s,\textsc{direction},V_{\A}\cup V_{\Abar})$ starts (line 1) by \emph{non-determin\-istically} generating a trace $\tilde{\rho}$ (having length at most $|\States|\cdot (2|\psi|+1)^2$), with $s$ as its first (resp., last) state if the \textsc{direction} parameter is \textsc{forward} (resp., \textsc{backward}). 
The trace is generated by visiting the unravelling of $\Ku$ (resp., of $\Ku$ with transposed edges). The remaining part of the algorithm \emph{deterministically} checks whether $\Ku,\tilde{\rho}\models\psi$ or not. 
Such a verification is performed in a bottom-up way: for all the subformulas $\phi$ of $\psi$ (starting from the minimal ones) and for all the prefixes $\tilde{\rho}(1,i)$ of $\tilde{\rho}$, with $1 \leq i \leq  |\tilde{\rho}|$ (starting from the shorter ones), the procedure establishes if $\Ku,\tilde{\rho}(1,i)\models\phi$, and this result is stored in the entry $T[\phi,i]$ of a Boolean table $T$. 
Note that if the considered subformula of $\psi$ is an element of $\mods(\psi)$, the algorithm does not need to perform any verification, since the result is already available in the Boolean vectors $V_{\A}$ and $V_{\Abar}$ (as a consequence of the previously completed calls to the procedure \texttt{Oracle}), and the table $T$ is updated accordingly (lines 2--7).
For the remaining subformulas, the entries of $T$ are computed, as we already said, in a bottom-up fashion (lines 8--22). The result of the overall verification is stored 
in $T[\psi,|\tilde{\rho}|]$ and returned (line 23).

Such an algorithm 
for checking formulas of $\AAbarB$ can trivially be adapted to check formulas of the symmetric fragment $\AAbarE$.

The next lemma states soundness and completeness of the procedure \texttt{Oracle} (its proof is in the
Appendix~\ref{proof:lemmaOracle}).

\begin{lemma}\label{lemmaOracle}
Let $\Ku=\KuDef$ be a finite Kripke structure, $\psi$ be an $\AAbarB$ formula, and $V_{\A}(\cdot,\cdot)$, $V_{\Abar}(\cdot,\cdot)$ be two Boolean arrays. We assume that 
\begin{enumerate}
	\item for each $\hsA \phi\in\mods(\psi)$ and $s'\in \States$, $V_{\A}(\phi,s')=\top$ if and only if there exists $\rho\in\Trk_{\Ku}$ such that $\fst(\rho)=s'$ and $\Ku,\rho\models \phi$, and
	\item for each $\hsAt \phi\in\mods(\psi)$ and $s'\in \States$, $V_{\Abar}(\phi,s')=\top$ if and only if there exists $\rho\in\Trk_{\Ku}$ such that $\lst(\rho)=s'$ and $\Ku,\rho\models \phi$.
\end{enumerate}
Then \texttt{Oracle}$(\Ku,\psi,s,\textsc{direction},V_{\A}\cup V_{\Abar})$ features a successful computation (returning $\top$) if and only if:
\begin{itemize}
	\item there exists $\rho\in\Trk_{\Ku}$ such that $\fst(\rho)=s$ and $\Ku,\rho\models \psi$, when \textsc{direction} is \textsc{forward};
	\item there exists $\rho\in\Trk_{\Ku}$ such that $\lst(\rho)=s$ and $\Ku,\rho\models \psi$, when \textsc{direction} is \textsc{backward}.
\end{itemize}
\end{lemma}

The following theorem states soundness and completeness of the MC procedure \texttt{MC} (Algorithm~\ref{MC}).
\begin{theorem}\label{MCsoundCompl}
Let $\Ku=\KuDef$ be a finite Kripke structure, $\psi$ be an $\AAbarB$ formula, and $V_{\A}(\cdot,\cdot)$ and $V_{\Abar}(\cdot,\cdot)$ be two Boolean arrays. If $\texttt{MC}(\Ku,\psi,\textsc{direction})$ is executed, then, for all $s\in \States$:
\begin{itemize}
	\item if \textsc{direction} is \textsc{forward}, $V_{\A}(\psi,s)=\top$ if and only if there exists $\rho\in\Trk_{\Ku}$ such that $\fst(\rho)=s$ and $\Ku,\rho\models \psi$;
	\item if \textsc{direction} is \textsc{backward}, $V_{\Abar}(\psi,s)=\top$ if and only if there exists $\rho\in\Trk_{\Ku}$ such that $\lst(\rho)=s$ and $\Ku,\rho\models\psi$.
\end{itemize}
\end{theorem}
\begin{proof}
The proof is by induction on the number $n$ of occurrences of $\hsA$ and $\hsAt$ modalities in $\psi$.

If $n=0$, since $\mods(\psi)=\emptyset$, (1.) and (2.) of Lemma~\ref{lemmaOracle} are satisfied and the thesis trivially holds.

Otherwise, $n>0$ and the formula $\psi$ contains at least an $\hsA$ or an $\hsAt$ modality, and thus $\mods(\psi)\allowbreak \neq \emptyset$. Since each recursive call to \texttt{MC} (either at line 2 or 4) is performed on a formula $\phi$ featuring a number of occurrences of $\hsA$ and $\hsAt$ which is strictly less than the number of their occurrences in $\psi$, we can apply the inductive hypothesis. As a consequence, when the control flow reaches line 5, it holds that:
\begin{enumerate}
	\item for each $\hsA \phi\in\mods(\psi)$ and $s'\in \States$, $V_{\A}(\phi,s')=\top$ if and only if there exists $\rho\in\Trk_{\Ku}$ such that $\fst(\rho)=s'$ and $\Ku,\rho\models \phi$;
	\item for each $\hsAt \phi\in\mods(\psi)$ and $s'\in \States$, $V_{\Abar}(\phi,s')=\top$ if and only if there exists $\rho\in\Trk_{\Ku}$ such that $\lst(\rho)=s'$ and $\Ku,\rho\models \phi$.
\end{enumerate}
This implies that (1.) and (2.) of Lemma~\ref{lemmaOracle} are fulfilled. Hence (assuming that \textsc{direction} is \textsc{forward}), it holds that, for $s\in \States$, $V_{\A}(\psi,s)=\top$ if and only if there exists $\rho\in\Trk_{\Ku}$ such that $\fst(\rho)=s$ and $\Ku,\rho\models \psi$. The case for  \textsc{direction} $=$  \textsc{backward} is symmetric, and thus omitted.
\end{proof}

As an immediate consequence, we have that the procedure  \texttt{MC} solves the MC problem for $\AAbarB$ formulas with an  algorithm belonging to the class $\PTIME^{\NP}$.

\begin{corollary}
Let $\Ku=\KuDef$ be a finite Kripke structure and $\psi$ be an $\AAbarB$ formula. If $\texttt{MC}(\Ku,\neg\psi,\textsc{forward})$ is executed, then $V_{\A}(\neg\psi,\sinit)=\bot$ if and only if $\Ku\models \psi$.
\end{corollary}

\begin{corollary}\label{coro:AAbarBalgo}
The MC problem for $\AAbarB$ formulas over finite Kripke structures is in $\PTIME^{\NP}$.
\end{corollary}
\begin{proof}
Given a finite Kripke structure $\Ku=\KuDef$ and an $\AAbarB$ formula $\psi$, the number of recursive calls performed by $\texttt{MC}(\Ku,\neg\psi,\textsc{forward})$ is at most $|\psi|$. Each one costs $O(|\psi|+|\States|\cdot (|\Ku|+|\psi|+|\psi|\cdot |\States|))$, where the first addend it is due to the search of $\psi$ for its modal subformulas (lines 1--4), and the second one to the preparation of the input for the oracle call, for each $s\in \States$ (lines 5--9). Therefore, its (deterministic) complexity is $O(|\psi|^2\cdot |\Ku|^2)$.

As for \texttt{Oracle}$(\Ku,\psi,s,\textsc{direction},V_{\A}\cup V_{\Abar})$, its (non-deterministic) complexity is $O(|\psi|^3\cdot |\Ku|)$, where $|\psi|$ is a bound to the number of subformulas and $O(|\psi|^2 \cdot |\Ku|)$ is the number of steps necessary to generate and check $\tilde{\rho}$.
\end{proof}

By a straightforward adaptation of the procedure \texttt{Oracle}, it is easy to prove that also the MC problem for the symmetric fragment $\AAbarE$ is in $\PTIME^{\NP}$.
As we will show in Section~\ref{sec:ABhard}, both problems are actually \emph{complete for} $\PTIME^{\NP}$.

%% file: Chaps/IC17/section02.tex
\section{$\Thsq$ MC algorithm for $\AAbar$, $\AbarB$, and $\AE$}\label{sect:AAbarAlg}

In this section, we first propose an MC algorithm for the fragment $\AAbar$ with complexity $\Thsq$, thus lower than the one described in the previous section for $\AAbarB$.
In fact, we do not directly devise an MC algorithm; we proceed instead \emph{via a reduction to} a
$\Thsq$-complete problem,  \TBSATM{} (a restriction of \TBSAT, see~\cite{schnoebelen2003}), whose instances are complex 
circuits where some of the gates are endowed with $\NP$ oracles.
%
%

\subsection{The problem \TBSATM{}}

\begin{figure}[tp]
    \centering
    \input{Chaps/IC17/block}
    \caption{General form of a block.}\label{blockFig}
\end{figure}
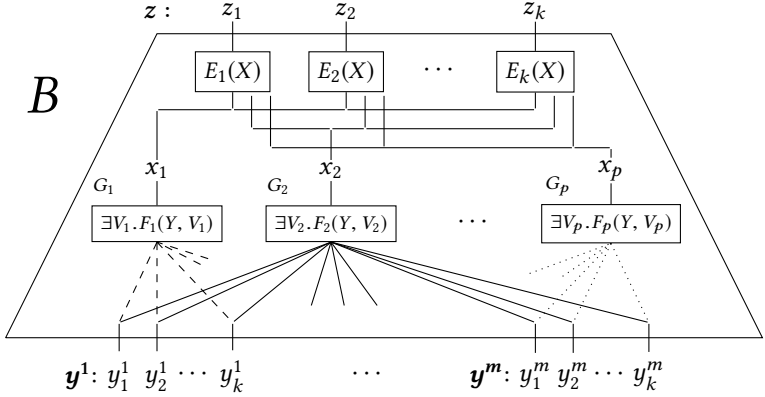

In order to introduce \TBSAT, we need to preliminarily describe its basic component, which is the \emph{block}. A block $B$ (see Figure~\ref{blockFig}) is a circuit
whose input lines are organized in $m$ bit vectors $\boldsymbol{y^1},\ldots,\boldsymbol{y^m}$, each of which has $k$ entries, namely $\boldsymbol{y^i}=(y^i_1, \ldots ,y^i_k)$. The values $m$ and $k$ are respectively called the \emph{degree} and the \emph{width} of $B$.
The input lines are connected to $p$ internal gates $G_1,\ldots,G_p$. Each gate $G_i$
features a Boolean formula $F_i(Y,V_i)$ associated with it, where $Y=\{y^j_s\mid j=1,\dots ,m,\; s=1,\dots , k\}$ and $V_i$ is a set of private variables of $F_i$, not occurring in any other $F_j$, with $j\neq i$, that is, $V_i\cap V_j=\emptyset$ for $j\neq i$.
The gate $G_i$ queries a SAT oracle in order to decide whether the associated Boolean formula is satisfiable.
The output of $G_i$ is denoted by $x_i$, and it evaluates to $\top$ if and only if $F_i(Y,V_i)$ is satisfiable. Finally, $k$ classic circuits (without oracles) $E_1,\ldots,E_k$ compute, from $X=\{x_1, \ldots,x_p\}$, their outputs $z_1,\ldots,z_k$, which are also the final $k$ outputs of the block $B$. 
The size of $B$ is defined as the total number of gates, plus the lengths of all the
associated Boolean formulas.
%
In the following, to make clear that a gate $G_i$ (respectively, input $y_i$, block output $z_i$, gate output $x_i$) is an element of a block $B$, we write  $B(G_i)$ (respectively, $B(y_i)$, $B(z_i)$, $B(x_i)$).

Given the $k\cdot m$ input bits, determining the output value of any $z_i$ is a $\Thpar$ problem: the $p$ queries associated with the oracle gates, which determine the outputs $x_j$'s, can be performed in parallel (they are independent of each other) and then the value of the block output $z_i$ can be calculated in deterministic polynomial time.

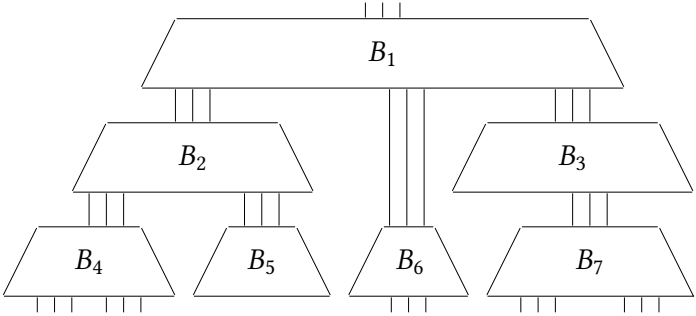
\begin{figure}[tp]
    \centering
    \resizebox{0.8\linewidth}{!}{\input{Chaps/IC17/treeBlock}}
        
    \caption{A tree of blocks ($B_5$ has degree $m=0$).}\label{tob}
\end{figure}%

Blocks of \emph{the same width} can be combined together to form a tree-structured complex circuit, called a \emph{tree of blocks}. See Figure~\ref{tob} for an example.
Every block in the tree-structure has a level: blocks which are leaves of the tree are at level 1;
a block $B_i$ whose inputs depend on (at least) a block $B_j$ at level $d-1$ and possibly on other blocks at levels less than $d$, is at level $d$. In Figure~\ref{tob}, 
$B_4$, $B_5$, $B_6$, $B_7$ are at level 1,
$B_2$ and $B_3$ at level 2, 
and $B_1$ at level 3. 
If the root of the tree-structure $T$ is at level $d$, the $k$ outputs of $T$ can be determined by $d$ rounds of parallel queries: all the queries relative to blocks placed at the same level $d'$ can be answered in parallel once all those at level $d'-1$ have been answered. 

\TBSAT{} is the problem of deciding whether a specific output $z_i$ of (the root of) a tree-structure of blocks $T$ is $\top$ or $\bot$, given the values for the inputs (of the leaf blocks) of $T$. As proved in~\cite{schnoebelen2003}, the problem \TBSAT{} is $\PTIME^{\NP}$-complete.

The problem \TBSATM{} is a constrained version of \TBSAT{}: any Boolean formula (SAT query) associated with a block $B$ of the tree-structure must have the following form: \[\exists\ell_1,\ldots ,\ell_m\in\{1,\ldots ,k\}\,\exists V'_i.F_i(y^1_{\ell_1},\ldots ,y^m_{\ell_m},\ell_1,\ldots ,\ell_m,V'_i),\] where $m$ and $k$ are respectively the degree and the width of $B$. This amounts to say that $F_i$ can use \emph{only one bit from each input vector} of $B$ (no matter which), hence ``$1\times M$''. 
The existential quantification over the indexes $\ell_1,\ldots ,\ell_m$ is an abuse of notation borrowed from~\cite{schnoebelen2003}: 
$\exists\ell_j\in\{1,\ldots ,k\}$ is just a shorthand for $k$ bits (belonging to the set of private variables) ``$\ell_j=1$'', \dots , ``$\ell_j=k$'', among which exactly one is $\top$. In the  formula above, $V'_i$ is $V_i$ deprived of such bits.

In~\cite{schnoebelen2003}, it is proved that \TBSATM{} is a $\Thsq$-complete problem. In particular, the proof of membership to $\Thsq$ exploits the \emph{squeeze} technique of~\cite{gottlob1995} applied to TREE(SAT) instances.  The particular form ``$1\times M$'' of the queries allows us to \lq\lq reshape\rq\rq{} the tree-structure of blocks, in such a way that the height becomes logarithmic in the number of blocks. Therefore, only $O(\log n)$ rounds of parallel queries are needed, allowing us to prove membership of the problem to $\Thparlogn=\Thsq$.

\subsection{Reduction of MC for $\AAbar$ to \TBSATM }
Let us show now how to reduce the MC problem for $\AAbar$ formulas to \TBSATM . 
As in the previous section, w.l.o.g., we assume that  only \emph{existential} modalities occur 
in the $\AAbar$ formula $\psi$ to be checked over a Kripke structure $\Ku=(\Prop,\States,\Edges,\Lab,s_1)$, 
with $\States=\{s_1,\ldots , s_{|\States|}\}$.\footnote{Here, for technical reasons, we assume an arbitrary order of the states of the Kripke structure, $s_1,\ldots , s_{|\States|}$, where $s_1$ is the initial state.}
We consider its negation $\neg\psi$ and build from it a tree-structure of blocks $T_{\Ku,\neg\psi}$.
Each block of $T_{\Ku,\neg\psi}$ has a type, \forw{} or \back, and it is associated with a subformula of $\neg\psi$. The root block, $B_{root}$, is always of type \forw\ and it is associated with $\neg\psi$. Each block $B$ has an output line $z_i$ for each state $s_i\in \States$, thus the \emph{width} of all blocks is $k=|\States|$.

Starting from $B_{root}$,  $T_{\Ku,\neg\psi}$ is built by recursive applications of the following \emph{basic step}, 
which are guided
by the $\AAbar$-modal subformulas (recall Definition~\ref{def:modsubf}):
if some (generic) block $B$ is associated with a formula $\varphi$, then
\begin{itemize}
	\item for every $\phi\in\mods(\varphi)$, where $\phi = \hsA \xi$, we create a \forw{} child $B'$ of $B$ associated with $\xi$, and
	\item for every $\phi'\in\mods(\varphi)$, where $\phi' = \hsAt \xi'$, we create a \back{} child $B''$ of $B$ associated with $\xi'$.
\end{itemize}
Then, the basic step is recursively applied to all the generated children of $B$, terminating when $\mods(\varphi)=\emptyset$.
Note that a block $B$ associated with a formula $\varphi$ has \emph{degree} $m=|\mods(\varphi)|$.

In such a way, we determine the 
tree-structure of blocks $T_{\Ku,\neg\psi}$. We now describe the internal structure of blocks. 

As a preliminary step, we suitably transform $\AAbar$ formulas $\varphi$ into Boolean ones by replacing all the occurrences of proposition letters and modal subformulas in $\varphi$ by Boolean variables, as described by the next definition.

\begin{definition}\label{psioverl}
Let $\Ku=(\Prop,\States,\Edges,\Lab,s_1)$ be a finite Kripke structure and let $\chi$ be
an $\AAbar$ formula. We define $\overline{\chi}(V_\Prop, V_{modSubf})$---where $V_\Prop=\{v_p\mid p\in\Prop\}$ and $V_{modSubf}=\{v_{\chi'}\mid\chi'\in\mods(\chi)\}$ are sets of Boolean variables---as the \emph{Boolean} formula obtained from $\chi$ by replacing
\begin{itemize}
	\item each (occurrence of a) $\AAbar$-modal subformula $\chi'\in \mods(\chi)$ by the variable $v_{\chi'}$,
	\item and \emph{subsequently} each (occurrence of a) proposition letter $p\in\Prop$ by the variable $v_p$.
\end{itemize}
\end{definition}

Given a trace $\rho\in\Trk_\Ku$ and an $\AAbar$ formula $\chi$,
it is easy to prove (by induction on the complexity of $\overline{\chi}(V_\Prop,V_{modSubf})$) that
if $\omega$ is an interpretation (truth assignment) of the variables of $V_\Prop\cup V_{modSubf}$ such that $\omega(v_p)=\top\iff \Ku,\rho\models p$, for all $p\in\Prop$, and $\omega(v_{\chi'})=\top\iff \Ku,\rho\models \chi'$, for all $\chi'\in\mods(\chi)$, then it holds that $\Ku,\rho\models \chi\iff\omega(\overline{\chi}(V_\Prop,V_{modSubf}))=\top$.

We are now ready to describe the internal structure of a block $B$ for a formula $\varphi$ in $T_{\Ku,\neg\psi}$. Let us assume that $B$ has type \forw{} and let us refer  to the block depicted in Figure~\ref{blockFig} for the description. The block features a gate $G_i$, with $1 \leq i \leq |\States|$, for each state of $\Ku$. Each output line $z_i$ of $B$ is directly linked to the output $x_i$ of the oracle gate $G_i$, avoiding the use of circuits $E_1,\ldots,E_k$.

Now, let $F_i(Y,V)$ be the Boolean formula for the gate $G_i$, with $1\leq i\leq |\States|$ (for the sake of simplicity, we write $V$ instead of $V_i$). The basic idea is that $F_i(Y,V)$ is satisfiable if and only if there is a trace having length at most $|\States|^2+2$, starting from the $i$-th state of $\States$, which satisfies $\overline{\varphi}(V_{\Prop},V_{modSubf})$, where $\varphi$
is the formula associated with the block $B$. To check the existence of such a witness trace,  we need a set of private variables $V_{trace}=\{v_1^1, \ldots, v_{|\States|}^1, v_1^2, \ldots, v_{|\States|}^2, \ldots, v_1^{|\States|^2+2}, \ldots, v_{|\States|}^{|\States|^2+2}\}$. In particular, 
the subset of variables $v_1^j, \ldots, v_{|\States|}^j$, with $1 \leq j \leq |\States|^2+2$, 
is used to \lq\lq encode\rq\rq{} the state in the $j$-th position of the trace. The encoding requires that exactly one variable $v_k^j$ of the subset is assigned to $\top$, for $1 \leq k \leq |\States|$, meaning that the $k$-th state of $\Ku$ occurs in the $j$-th position of the sequence.
Moreover, we use a set of private variables $V_{last}=\{v_1, v_2, \ldots, v_{|\States| }\}$ which are used to encode the last state of the witness trace (note that the length of the witness trace can be actually less than the bound  $|\States|^2+2$).

In detail, the Boolean formula $trace(V_{trace},V_{last},V_{\Prop})$, which ensures
that a truth assignment of the private variables
$V_{trace}$ properly encodes a trace $\rho$ of $\Ku$ of length $\ell$, for $1 \leq \ell \leq |\States|^2+2$, is as follows.

\begin{multline*}
trace(V_{trace},V_{last},V_{\Prop})=
\bigvee_{\ell=1}^{|\States|^2+2}
\Big[\bigwedge_{t=1}^{\ell} one_t(v_1^t, v_2^t, \ldots, v_{|\States|}^t) \wedge \\
\bigwedge_{t=1}^{\ell-1} edge_t(v_1^t, \ldots ,v_{|\States|}^t,v_1^{t+1}, \ldots ,v_{|\States|}^{t+1}) \wedge
\underbrace{\bigwedge_{t=1}^{|\States|} (v^\ell_t \leftrightarrow v_t)}_{(1)} \wedge\\
\underbrace{\bigwedge_{p\in\Prop}\!\! \Big(\big(v_p \!\rightarrow\! \bigwedge_{t=1}^\ell \bigwedge_{j=1}^{|\States|}(v_j^t\!\rightarrow\! VAL(s_j,p))\big) \!\wedge\! \big(\neg v_p \!\rightarrow\! \bigvee_{t=1}^\ell \bigwedge_{j=1}^{|\States|}(v_j^t\!\rightarrow\! \neg VAL(s_j,p))\big)\Big)}_{(2)}
\Big].
\end{multline*}

For any $1 \leq t \leq \ell$, being $\ell$ the length of 
the witness trace, the Boolean formula $one_t(v_1^t, v_2^t, \ldots, v_{|\States|}^t)$ \lq\lq checks\rq\rq{} that the variables  $v_1^t, v_2^t, \ldots, v_{|\States|}^t$ encode (exactly) one state for the $t$-th element of the trace:
\[one_t(v_1^t, v_2^t, \ldots, v_{|\States|}^t)=\Big(\bigvee_{j=1}^{|\States|}v_j^t\Big) \wedge\Big(\bigwedge_{j=1}^{|\States|}\bigwedge_{k=j+1}^{|\States|}\neg(v_j^t \wedge v_k^t)\Big).\]
Then, for any $1 \leq t \leq \ell -1$, the formula $edge_t(v_1^t, \ldots ,v_{|\States|}^t,v_1^{t+1}, \ldots ,v_{|\States|}^{t+1})$ checks that
if $s_k$ and $s_j$ are states which occur consecutively in the encoded trace ($v_k^t$ and $v_j^{t+1}$ are set to $\top$), then 
$(s_k,s_j)\in\Edges$:
\[edge_t(v_1^t, \ldots ,v_{|\States|}^t,v_1^{t+1}, \ldots ,v_{|\States|}^{t+1})=\smashoperator{\bigvee_{(s_k,s_j)\in\Edges}} (v_k^t \wedge v_j^{t+1}).\] 
Then, conjunct (1) ensures that the private variables in $V_{last}$ encode the last state of the witness trace, that is, the $\ell$-th state.
%
Finally, conjunct (2) enforces the homogeneity assumption, ensuring that a variable $v_p \in V_{\Prop}$ evaluates to $\top$ if and only if $p$ holds over all the states of the witness trace ($VAL(s_j,p)$ is just a shorthand for $\top$ if $p\in\Lab(s_j)$, and $\bot$ otherwise).


Now, taking the set of private variables $V= V_{last} \cup V_{trace} \cup V_{\Prop} \cup V_{modSubf}$, the Boolean formula $F_i(Y,V)$ for the gate $G_i$ is formally defined as:
\begin{multline*}
F_i(Y,V)=
v_i^1 \wedge 
\overline{\varphi}(V_{\Prop},V_{modSubf}) \wedge
trace(V_{trace},V_{last},V_{\Prop}) \wedge 
\\
\smashoperator{\bigwedge_{\hsA\xi\in\mods(\varphi)}}\big(v_{\hsA\xi}\leftrightarrow \bigvee_{j=1}^{|\States|} (v_j \wedge y_j^\xi)\big) \quad \wedge \quad \smashoperator{\bigwedge_{\hsAt\xi'\in\mods(\varphi)}}\big(v_{\hsAt\xi'}\leftrightarrow \bigvee_{j=1}^{|\States|} (v_j^1 \wedge y_j^{\xi'})\big)
\end{multline*}
 
The first conjunct of $F_i(Y,V)$ requires that the witness trace starts with the $i$-th state of $\Ku$. The fourth one requires that each private variable $v_{\hsA\xi}$, for $\hsA\xi \in\mods(\varphi)$ has exactly the same truth assignment as the $j$-th output, $y_j^\xi$, of the block for $\xi$
(which is a child of $B$)---provided that the final state of the trace is the $j$-th state of $\Ku$.
Since exactly one among the variables of $V_{last}=\{v_1,\ldots , v_{|\States|}\}$ is set to $\top$, it is guaranteed that at most one bit for every child-block is considered by $B$, thus fulfilling the restriction of \TBSATM. The last conjunct of $F_i(Y,V)$ forces the symmetric constraint for modal subformulas of the form $\hsAt\xi'$.

The formula $F_i(Y,V)$ for a gate $G_i$ in a \back{} block is very similar: we just need to replace the first conjunct $v_i^1$ by $v_i$.

The following proposition states the correctness of the encoding for traces.

\begin{proposition}\label{remk}
Given a trace $\rho\in\Trk_\Ku$, with $|\rho|\leq |\States|^2+2$, 
there exists a truth assignment $\omega$ to the variables in $V$ which satisfies the formula $trace(V_{trace},V_{last},V_{\Prop})$, and 
\begin{itemize}
\item for any $1 \leq t \leq |\rho|$ and $1 \leq j \leq |\States|$, $\rho(t)\!=\!s_j\!\iff\!\omega(v_j^t)\!=\!\top$ and $\omega(v_j^{|\rho|})=\omega(v_j)$;
\item for any $p\in\Prop$, $\omega(v_p)=\top\iff\Ku,\rho\models p.$
\end{itemize}
Conversely,
if a truth assignment $\omega$ to the variables in $V$ satisfies the $r$-th disjunct of $trace(V_{trace},V_{last},V_{\Prop})$,
then there exists a trace $\rho\in\Trk_\Ku$, with $|\rho|=r$, such that
\begin{itemize}
\item for any $1 \leq t \leq |\rho|$ and $1 \leq j \leq |\States|$, $\rho(t)=s_j\iff\omega(v_j^t)=\top$;
\item for any $p\in\Prop$, $\Ku,\rho\models p\iff\omega(v_p)=\top.$
\end{itemize}
\end{proposition}


The following theorem states the correctness of the construction of $T_{\Ku,\neg\psi}$ (the proof is given in Appendix~\ref{proof:th:cx}).

\begin{theorem}\label{th:cx} 
Let $\psi$  be an $\AAbar$ formula and  $\Ku=(\Prop,\States,\Edges,\Lab,s_1)$ be a finite Kripke structure.
For every block $B$ of $T_{\Ku,\neg\psi}$,
if $B$ is associated with an $\AAbar$ formula $\varphi$, then
\begin{itemize}
	\item if $B$ is a \forw{} block, for all $i\in\{1,\ldots,|\States|\}$, $B(z_i)=\top$ if and only if there exists a trace $\rho\in\Trk_\Ku$ such that $\fst(\rho)=s_i$ and $\Ku,\rho\models\varphi$;
	\item if $B$ is a \back{} block, for all $i\in\{1,\ldots,|\States|\}$, $B(z_i)=\top$ if and only if there exists a trace $\rho\in\Trk_\Ku$ such that $\lst(\rho)=s_i$ and $\Ku,\rho\models\varphi$.
\end{itemize}
\end{theorem}

The two next corollaries immediately follow.

\begin{corollary}\label{CorAAbar}
Let $\psi$ be an $\AAbar$ formula, $\Ku$
be a finite Kripke structure,
and $B_{root}$ be the root block of $T_{\Ku,\neg\psi}$. Then, it holds that
$B_{root}(z_1)=\bot$ if and only if $\Ku\models\psi$.
\end{corollary}

\begin{corollary}\label{th:AAbaralgo}
The MC problem for $\AAbar$ formulas over finite Kripke structures is in $\Thsq$.
\end{corollary}
\begin{proof}
The result follows from Corollary~\ref{CorAAbar} and the fact that the instance of \TBSATM{} generated from an $\AAbar$ formula $\psi$ and 
a Kripke structure $\Ku$ is polynomial in $|\psi|$ and $|\Ku|$.
\end{proof}

\subsection{Reduction of MC for $\AbarB$ (resp., $\AE$) to \TBSATM }

We conclude the section by showing that it is possible to adapt the above-described reduction to the fragment $\AbarB$ and the symmetric fragment $\AE$.
Let us focus on $\AbarB$ (the case for $\AE$ can be dealt with in an analogous way).

Having in mind that $\AbarB$ is a fragment of $\AAbarB$, by removing the case for modality $\hsA$ in Algorithm~\ref{Oracle}, we get a procedure for which Lemma~\ref{lemmaOracle} still holds. 
Since Algorithm~\ref{Oracle} is in $\NP$, there must exist a reduction to SAT, that is, given an instance $(\Ku,\psi,s,\textsc{direction},V_{\Abar})$ for \texttt{Oracle}, there exists a Boolean formula $\Psi_{(\Ku,\psi,s,\textsc{direction},V_{\Abar})}$, which depends on $(\Ku,\psi,s,$ $\textsc{direction}, V_{\Abar})$, that is satisfiable if and only if \texttt{Oracle}$(\Ku,\psi,s,\allowbreak \textsc{direction}, V_{\Abar})$ admits a successful computation on the given input. 
By Lemma~\ref{lemmaOracle}, this is the case if and only if:  
\begin{itemize}
	\item there exists $\rho\in\Trk_{\Ku}$ such that $\fst(\rho)=s$ and $\Ku,\rho\models \psi$, in the case \textsc{direction} is \textsc{forward};
	\item there exists $\rho\in\Trk_{\Ku}$ such that $\lst(\rho)=s$ and $\Ku,\rho\models \psi$, in the case \textsc{direction} is \textsc{backward},
\end{itemize}
provided that 
for each $\hsAt \phi\in\mods(\psi)$ and $s'\in \States$, $V_{\Abar}(\phi,s')=\top$ if and only if there exists $\rho'\in\Trk_{\Ku}$ such that $\lst(\rho')=s'$ and $\Ku,\rho'\models \phi$.

The idea is that such a formula $\Psi_{(\Ku,\psi,s,\textsc{direction},V_{\Abar})}$ can be used as a SAT query for an oracle gate $G_i$ in a generic block associated with the formula $\psi$. The role of the global Boolean vector $V_{\Abar}$ is instead played by the local communication among blocks in the tree-structure; 
$\Psi_{(\Ku,\psi,s,\textsc{direction},V_{\Abar})}$ 
basically has the same structure as the Boolean formula $F_i(Y,V)$, with some minor differences outlined below.

First of all, we observe that Algorithm~\ref{Oracle} works with traces whose length is bounded by $|\States|\cdot (2|\psi|+1)^2$ (instead of ${| \States |}^2+ 2$ as in $F_i(Y,V)$). A trace is then encoded exactly as in $F_i(Y,V)$ by using a set of private variables
$V_{trace}=\{v_j^t \mid j=1,\ldots ,|\States|$, $t=1,\ldots , |\States|\cdot (2|\psi|+1)^2 \}$;
$\Psi_{(\Ku,\psi,s,\textsc{direction},V_{\Abar})}$ has also to encode the Boolean table $T$ of \texttt{Oracle} with entries $T[\varphi,i]$, where $\varphi$ is a subformula of $\psi$, and $1 \leq i \leq |\States|\cdot (2|\psi|+1)^2$ is the length of a prefix of the considered trace. Therefore,
there is a variable $x_{\varphi,i}$ in  $\Psi_{(\Ku,\psi,s,\textsc{direction},V_{\Abar})}$ for each entry $T[\varphi,i]$, with the intuitive meaning that $x_{\varphi,i}$ is assigned $\top$ if and only if $T[\varphi,i]=\top$. Actually, in this encoding we do not need any entry for a modal subformula $\hsAt \xi$, whose truth value is conveyed by $y_j^\xi$, namely, the $j$-th input connected to the child block for the subformula $\xi$ (assuming that the starting state of the trace is the $j$-th state of $\Ku$).

It is worth noting that this construction is possible since all the prefixes of the trace $\rho$ encoded by the assignment to $V_{trace}$, 
and $\rho$ itself, share the same starting point, and thus agree on the truth value of any $\Abar$-modal subformula. The most relevant consequence of this property is that  the constraint of \TBSATM{} on the form of SAT queries is respected.

As for the construction of $T_{\mathpzc{K},\neg\psi}$, it is exactly as before where, in particular, the root block $B_{root}$ has type \forw, and all the other blocks have type \back. 

The following result can be stated. 
\begin{theorem}\label{th:AbarBalgo}
The MC problem for $\AbarB$ (resp., $\AE$) formulas over finite Kripke structures is in $\Thsq$.
\end{theorem}

The construction we have sketched cannot be adapted to the fragment $\AB$. This is due to the fact that the \emph{right} endpoints of the prefixes of a trace differ in general, and thus they do not necessarily agree on the truth value of $\A$-modal subformulas, hence the restriction of \TBSATM{} on the form of SAT queries cannot be respected.
In the next section, we will prove that MC for $\AB$ formulas is indeed \emph{inherently} more difficult than MC for $\AbarB$.
%

%% file: Chaps/IC17/block.tex
\begin{tikzpicture}
\draw (0,0) -- (10,0) -- (8,4) -- (2,4) -- (0,0);
\draw  [style={inner sep=1,outer sep=0}](0.5,3.2) node (x1) {{\huge $B$}};

\draw (2,1.5) node (f1) [draw] {\scriptsize $\exists V_1.F_1(Y,V_1)$}; \draw (1.3,2) node {{\scriptsize $G_1$}};
\draw (4.3,1.5) node (f2) [draw] {\scriptsize $\exists V_2.F_2(Y,V_2)$}; \draw (3.6,2) node {{\scriptsize $G_2$}};
\draw (6.2,1.5) node {$\cdots$};
\draw (8,1.5) node (fp) [draw] {\scriptsize $\exists V_p.F_p(Y,V_p)$}; \draw (7.3,2) node {{\scriptsize $G_p$}};

\draw  [style={inner sep=1,outer sep=0}](2,2.2) node (x1) {$x_1$};
\draw  [style={inner sep=1,outer sep=0}](4.3,2.2) node (x2) {$x_2$};
\draw  [style={inner sep=1,outer sep=0}](8,2.2) node (xp) {$x_p$};

\draw [-] (f1) -- (x1);
\draw [-] (f2) -- (x2);
\draw [-] (fp) -- (xp);

\draw (3,3.5) node [draw] (e1) {\small $E_1(X)$};
\draw (4.5,3.5) node [draw] (e2) {\small $E_2(X)$};
\draw (5.75,3.5) node {$\cdots$};
\draw (7,3.5) node [draw] (ek) {\small $E_k(X)$};

\draw  [style={inner sep=1,outer sep=0}](2,4.3) node {$\boldsymbol{z:}$};
\draw  [style={inner sep=1,outer sep=0}](3,4.3) node (z1) {$z_1$};
\draw  [style={inner sep=1,outer sep=0}](4.5,4.3) node (z2) {$z_2$};
\draw  [style={inner sep=1,outer sep=0}](7,4.3) node (zk) {$z_k$};

\draw  (e1) edge (z1);
\draw  (e2) edge (z2);
\draw  (ek) edge (zk);

\node at (1,-0.5) {$\boldsymbol{y^1}$:};
\node (v2) at (1.5,-0.5) {$y_1^1$};
\node (v4) at (2,-0.5) {$y_2^1$};
\node at (2.5,-0.5) {$\cdots$};
\node (v6) at (3,-0.5) {$y_k^1$};

\node [style={inner sep=0,outer sep=0}] (v1) at (1.5,0.2) {};
\node [style={inner sep=0,outer sep=0}] (v3) at (2,0.2) {};
\node  [style={inner sep=0,outer sep=0}](v5) at (3,0.2) {};

\node at (4.8,-0.5) {$\cdots$};

\node (v8) at (7,-0.5) {$y_1^m$};
\node (v10) at (7.5,-0.5) {$y_2^m$};
\node at (8,-0.5) {$\cdots$};
\node (v12) at (8.5,-0.5) {$y_k^m$};
\node at (6.4,-0.5) {$\boldsymbol{y^m}$:};

\node [style={inner sep=0,outer sep=0}] (v7) at (7,0.2) {};
\node [style={inner sep=0,outer sep=0}] (v9) at (7.5,0.2) {};
\node [style={inner sep=0,outer sep=0}] (v11) at (8.5,0.2) {};

\draw  (v1) edge (v2);
\draw  (v3) edge (v4);
\draw  (v5) edge (v6);
\draw  (v7) edge (v8);
\draw  (v9) edge (v10);
\draw  (v11) edge (v12);
\draw  (v1) edge (f2.south);
\draw  (v3) edge (f2.south);
\draw  (v5) edge (f2.south);
\draw  (v7) edge (f2.south);
\draw  (v9) edge (f2.south);
\draw  (v11) edge (f2.south);

\node [style={inner sep=0,outer sep=0}] (v13) at (2,3) {};
\node [style={inner sep=0,outer sep=0}] (v16) at (7,3) {};
\node [style={inner sep=0,outer sep=0}] (v19) at (3.5,2.5) {};
\node [style={inner sep=0,outer sep=0}] (v18) at (7.5,2.5) {};
\node [style={inner sep=0,outer sep=0}] (v17) at (8,2.5) {};
\node [style={inner sep=0,outer sep=0}] (v14) at (3,3) {};
\node [style={inner sep=0,outer sep=0}] (v15) at (4.5,3) {};
\draw  (x1) edge (v13);
\draw  (v13) edge (v14);
\draw  (v14) edge (v15);
\draw  (v15) edge (v16);
\draw  (v16) edge (ek);
\draw  (v15) edge (e2);
\draw  (e1) edge (v14);
\draw  (xp) edge (v17);
\draw  (v17) edge (v18);
\draw  (v18) edge (v19);
\node [style={inner sep=0,outer sep=0}] (v21) at (5,2.5) {};
\node [style={inner sep=0,outer sep=0}] (v20) at (3.5,3.2) {};
\node [style={inner sep=0,outer sep=0}] (v22) at (5,3.2) {};
\node [style={inner sep=0,outer sep=0}] (v23) at (7.5,3.2) {};
\draw  (v19) edge (v20);
\draw  (v21) edge (v22);
\draw  (v18) edge (v23);

\node [style={inner sep=0,outer sep=0}] (v25) at (3.25,2.75) {};
\node [style={inner sep=0,outer sep=0}] (v27) at (4.75,2.75) {};
\node [style={inner sep=0,outer sep=0}] (v29) at (7.25,2.75) {};
\node [style={inner sep=0,outer sep=0}] (v24) at (3.25,3.2) {};
\node [style={inner sep=0,outer sep=0}] (v28) at (4.75,3.2) {};
\node [style={inner sep=0,outer sep=0}] (v30) at (7.25,3.2) {};
\node [style={inner sep=0,outer sep=0}] (v26) at (4.3,2.75) {};
\draw  (v24) edge (v25);
\draw  (v25) edge (v26);
\draw  (v26) edge (v27);
\draw  (v27) edge (v28);
\draw  (v27) edge (v29);
\draw  (v29) edge (v30);
\draw  (v26) edge (x2);
\draw [dashed] (f1.south) edge (v1);
\draw [dashed] (f1.south) edge (v3);
\draw [dashed] (f1.south) edge (v5);
\draw [dotted] (fp.south) edge (v7);
\draw [dotted] (fp.south) edge (v9);
\draw [dotted] (fp.south) edge (v11);
\node (v32) at (2.7,0.9) {};
\node (v33) at (2.8,1) {};
\node (v31) at (7.2,0.7) {};
\node (v34) at (6.8,0.8) {};
\draw [dashed] (v32) edge (f1.south);
\draw [dashed] (v33) edge (f1.south);
\draw [dotted] (v34) edge (fp.south);
\draw [dotted] (fp.south) edge (v31);
\node (v35) at (4,0.3) {};
\node (v36) at (4.5,0.3) {};
\node (v37) at (5,0.3) {};
\draw (f2.south) edge (v35);
\draw (f2.south) edge (v36);
\draw (v37) edge (f2.south);
\end{tikzpicture}

%% file: Chaps/IC17/treeBlock.tex
\begin{tikzpicture}

\node [style={inner sep=1,outer sep=0}] (v1) at (0,0) {};
\node [style={inner sep=1,outer sep=0}] (v2) at (14,0) {};
\node [style={inner sep=1,outer sep=0}] (v4) at (1,2) {};
\node [style={inner sep=1,outer sep=0}] (v3) at (13,2) {};
\draw (v4) -- (v1) -- (v2) -- (v3) -- (v4);

\node [style={inner sep=1,outer sep=0}] (v11) at (-2,-3) {};
\node [style={inner sep=1,outer sep=0}] (v21) at (5,-3) {};
\node [style={inner sep=1,outer sep=0}] (v41) at (-1,-1) {};
\node [style={inner sep=1,outer sep=0}] (v31) at (4,-1) {};
\draw (v41) -- (v11) -- (v21) -- (v31) -- (v41);

\node [style={inner sep=1,outer sep=0}] (v12) at (-4,-6) {};
\node [style={inner sep=1,outer sep=0}] (v22) at (1,-6) {};
\node [style={inner sep=1,outer sep=0}] (v42) at (-3,-4) {};
\node [style={inner sep=1,outer sep=0}] (v32) at (0,-4) {};
\draw (v42) -- (v12) -- (v22) -- (v32) -- (v42);

\node [style={inner sep=1,outer sep=0}] (v13) at (1.5,-6) {};
\node [style={inner sep=1,outer sep=0}] (v23) at (5.5,-6) {};
\node [style={inner sep=1,outer sep=0}] (v43) at (2.5,-4) {};
\node [style={inner sep=1,outer sep=0}] (v33) at (4.5,-4) {};
\draw (v43) -- (v13) -- (v23) -- (v33) -- (v43);

\node [style={inner sep=1,outer sep=0}] (v14) at (6,-6) {};
\node [style={inner sep=1,outer sep=0}] (v24) at (9.5,-6) {};
\node [style={inner sep=1,outer sep=0}] (v44) at (7,-4) {};
\node [style={inner sep=1,outer sep=0}] (v34) at (8.5,-4) {};
\draw (v44) -- (v14) -- (v24) -- (v34) -- (v44);

\node [style={inner sep=1,outer sep=0}] (v15) at (9,-3) {};
\node [style={inner sep=1,outer sep=0}] (v25) at (16,-3) {};
\node [style={inner sep=1,outer sep=0}] (v45) at (10,-1) {};
\node [style={inner sep=1,outer sep=0}] (v35) at (15,-1) {};
\draw (v45) -- (v15) -- (v25) -- (v35) -- (v45);

\node [style={inner sep=1,outer sep=0}] (v16) at (10,-6) {};
\node [style={inner sep=1,outer sep=0}] (v26) at (16,-6) {};
\node [style={inner sep=1,outer sep=0}] (v46) at (11,-4) {};
\node [style={inner sep=1,outer sep=0}] (v36) at (15,-4) {};
\draw (v46) -- (v16) -- (16,-6) -- (v36) -- (v46);

\node [style={inner sep=1,outer sep=0}] (v29) at (1.5,-1) {};
\node [style={inner sep=1,outer sep=0}] (v37) at (2,-1) {};
\node [style={inner sep=1,outer sep=0}] (v27) at (1,-1) {};
\node [style={inner sep=1,outer sep=0}] (v5) at (-1.5,-4) {};
\node [style={inner sep=1,outer sep=0}] (v7) at (-1,-4) {};
\node [style={inner sep=1,outer sep=0}] (v9) at (-0.5,-4) {};
\node [style={inner sep=1,outer sep=0}] (v18) at (3.5,-4) {};
\node [style={inner sep=1,outer sep=0}] (v49) at (12.5,-4) {};
\node [style={inner sep=1,outer sep=0}] (v51) at (13,-4) {};
\node [style={inner sep=1,outer sep=0}] (v53) at (13.5,-4) {};
\node [style={inner sep=1,outer sep=0}] (v59) at (12,-1) {};
\node [style={inner sep=1,outer sep=0}] (v57) at (12.5,-1) {};
\node [style={inner sep=1,outer sep=0}] (v55) at (13,-1) {};
\node [style={inner sep=1,outer sep=0}] (v28) at (1,0) {};
\node [style={inner sep=1,outer sep=0}] (v30) at (1.5,0) {};
\node [style={inner sep=1,outer sep=0}] (v38) at (2,0) {};
\node [style={inner sep=1,outer sep=0}] (v60) at (12,0) {};
\node [style={inner sep=1,outer sep=0}] (v58) at (12.5,0) {};
\node [style={inner sep=1,outer sep=0}] (v56) at (13,0) {};

\node [style={inner sep=1,outer sep=0}] (v17) at (3,-3) {};
\node [style={inner sep=1,outer sep=0}] (v19) at (3.5,-3) {};
\node [style={inner sep=1,outer sep=0}] (v20) at (4,-3) {};
\node [style={inner sep=1,outer sep=0}] (v6) at (-1.5,-3) {};
\node [style={inner sep=1,outer sep=0}] (v8) at (-1,-3) {};
\node [style={inner sep=1,outer sep=0}] (v10) at (-0.5,-3) {};
\node [style={inner sep=1,outer sep=0}] (v50) at (12.5,-3) {};
\node [style={inner sep=1,outer sep=0}] (v52) at (13,-3) {};
\node [style={inner sep=1,outer sep=0}] (v54) at (13.5,-3) {};
\node [style={inner sep=1,outer sep=0}] (v61) at (3,-4) {};
\node [style={inner sep=1,outer sep=0}] (v62) at (4,-4) {};

\draw  (v5) edge (v6);
\draw  (v7) edge (v8);
\draw  (v9) edge (v10);
\draw  (v18) edge (v19);
\draw  (v27) edge (v28);
\draw  (v29) edge (v30);
\draw  (v37) edge (v38);

\draw  (v49) edge (v50);
\draw  (v51) edge (v52);
\draw  (v53) edge (v54);
\draw  (v55) edge (v56);
\draw  (v57) edge (v58);
\draw  (v59) edge (v60);
\draw  (v17) edge (v61);
\draw  (v20) edge (v62);

\node [style={inner sep=1,outer sep=0}] (v48) at (7.7,-4) {};
\node [style={inner sep=1,outer sep=0}] (v64) at (8.2,-4) {};
\node [style={inner sep=1,outer sep=0}] (v40) at (7.2,-4) {};
\node [style={inner sep=1,outer sep=0}] (v39) at (7.2,0) {};
\node [style={inner sep=1,outer sep=0}] (v47) at (7.7,0) {};
\node [style={inner sep=1,outer sep=0}] (v63) at (8.2,0) {};
\draw  (v39) edge (v40);
\draw  (v47) edge (v48);
\draw  (v63) edge (v64);

\node [style={inner sep=1,outer sep=0}] (v65) at (-3,-6) {};
\node [style={inner sep=1,outer sep=0}] (v67) at (-2.5,-6) {};
\node [style={inner sep=1,outer sep=0}] (v69) at (-2,-6) {};
\node [style={inner sep=1,outer sep=0}] (v71) at (-1,-6) {};
\node [style={inner sep=1,outer sep=0}] (v73) at (-0.5,-6) {};
\node [style={inner sep=1,outer sep=0}] (v75) at (0,-6) {};
\node [style={inner sep=1,outer sep=0}] (v66) at (-3,-6.5) {};
\node [style={inner sep=1,outer sep=0}] (v68) at (-2.5,-6.5) {};
\node [style={inner sep=1,outer sep=0}] (v70) at (-2,-6.5) {};
\node [style={inner sep=1,outer sep=0}] (v72) at (-1,-6.5) {};
\node [style={inner sep=1,outer sep=0}] (v74) at (-0.5,-6.5) {};
\node [style={inner sep=1,outer sep=0}] (v76) at (0,-6.5) {};
\node [style={inner sep=1,outer sep=0}] (v77) at (7.25,-6) {};
\node [style={inner sep=1,outer sep=0}] (v79) at (7.75,-6) {};
\node [style={inner sep=1,outer sep=0}] (v81) at (8.25,-6) {};
\node [style={inner sep=1,outer sep=0}] (v83) at (11,-6) {};
\node [style={inner sep=1,outer sep=0}] (v85) at (11.5,-6) {};
\node [style={inner sep=1,outer sep=0}] (v87) at (12,-6) {};
\node [style={inner sep=1,outer sep=0}] (v89) at (14,-6) {};
\node [style={inner sep=1,outer sep=0}] (v91) at (14.5,-6) {};
\node [style={inner sep=1,outer sep=0}] (v93) at (15,-6) {};
\node [style={inner sep=1,outer sep=0}] (v84) at (11,-6.5) {};
\node [style={inner sep=1,outer sep=0}] (v86) at (11.5,-6.5) {};
\node [style={inner sep=1,outer sep=0}] (v88) at (12,-6.5) {};
\node [style={inner sep=1,outer sep=0}] (v90) at (14,-6.5) {};
\node [style={inner sep=1,outer sep=0}] (v92) at (14.5,-6.5) {};
\node [style={inner sep=1,outer sep=0}] (v94) at (15,-6.5) {};
\node [style={inner sep=1,outer sep=0}] (v82) at (8.25,-6.5) {};
\node [style={inner sep=1,outer sep=0}] (v80) at (7.75,-6.5) {};
\node [style={inner sep=1,outer sep=0}] (v78) at (7.25,-6.5) {};
\draw  (v65) edge (v66);
\draw  (v67) edge (v68);
\draw  (v69) edge (v70);
\draw  (v71) edge (v72);
\draw  (v73) edge (v74);
\draw  (v75) edge (v76);
\draw  (v77) edge (v78);
\draw  (v79) edge (v80);
\draw  (v81) edge (v82);
\draw  (v83) edge (v84);
\draw  (v85) edge (v86);
\draw  (v87) edge (v88);
\draw  (v89) edge (v90);
\draw  (v91) edge (v92);
\draw  (v93) edge (v94);

\node [style={inner sep=1,outer sep=0}] (v97) at (7,2) {};
\node [style={inner sep=1,outer sep=0}] (v95) at (6.5,2) {};
\node [style={inner sep=1,outer sep=0}] (v99) at (7.5,2) {};
\node [style={inner sep=1,outer sep=0}] (v96) at (6.5,2.5) {};
\node [style={inner sep=1,outer sep=0}] (v98) at (7,2.5) {};
\node [style={inner sep=1,outer sep=0}] (v100) at (7.5,2.5) {};
\draw [style={inner sep=1,outer sep=0}]  (v95) edge (v96);
\draw [style={inner sep=1,outer sep=0}]  (v97) edge (v98);
\draw [style={inner sep=1,outer sep=0}]  (v99) edge (v100);

\node at (7,1) {\Huge $B_1$};
\node at (1.5,-2) {\Huge $B_2$};
\node at (12.5,-2) {\Huge $B_3$};
\node at (-1.5,-5) {\Huge $B_4$};
\node at (3.5,-5) {\Huge $B_5$};
\node at (7.8,-5) {\Huge $B_6$};
\node at (13,-5) {\Huge $B_7$};
\end{tikzpicture}

%% file: Chaps/IC17/section03.tex
\section{$\PTIME^{\NP}$-hardness of MC for $\AB$ and $\AbarE$}\label{sec:ABhard}

In this section, we show that the $\PTIME^{\NP}$-complete problem named \emph{Sequentially Nested SATisfiability} (SNSAT~\cite{LMS01}) can be reduced to MC for formulas of the fragment $\AB$ (and similarly $\AbarE$), thus proving $\PTIME^{\NP}$-hardness of the latter.
%
SNSAT 
is a logical problem featuring a series of nested satisfiability questions.

\begin{definition}\label{snsat}
An instance $\mathcal{I}$ of SNSAT consists of 
    a set of $n$ Boolean variables $X=\{x_1,\ldots ,x_n\}$ and
   a set of $n$ Boolean formulas 
\[\{F_1(Z_1), F_2(x_1,Z_2),\ldots , F_n(x_1,\ldots , x_{n-1}, Z_n)\}, \] where, for $i=1,\ldots , n$, $F_i(x_1, \ldots ,x_{i-1},Z_i)$ has variables in $\{x_1, \ldots ,x_{i-1}\}$ and in the set of private variables $Z_i=\{z_i^1,\ldots ,z_i^{j_i}\}$, 
%
%
that is, $Z_i\cap Z_t=\emptyset$, for $t \neq i$, and $X\cap Z_i=\emptyset$.
We denote by $|\mathcal{I}|$ the cardinality $|X| =n$.

Let $v_\mathcal{I}$ be a truth-assignment of the variables in $X$ defined as follows: 
\[
    v_\mathcal{I}(x_i)=\top \iff
    \begin{array}{l}
        F_i(v_\mathcal{I}(x_1), \ldots, v_\mathcal{I}(x_{i-1}), Z_i)\text{ is satisfiable} 
    \end{array}
\]
(by a suitable truth-assignment  to the private
variables $z_i^1,\ldots ,z_i^{j_i}\in Z_i$).  
 
SNSAT is the problem of deciding, given an instance $\mathcal{I}$ with $|\mathcal{I}|=n$, whether
$v_\mathcal{I}(x_n)=\top$ or not. In such a case, we say that $\mathcal{I}$ is a positive instance of SNSAT.
\end{definition}

Given an SNSAT instance $\mathcal{I}$, with $|\mathcal{I}|=n$, the truth-assignment $v_\mathcal{I}$ is unique and it can be easily computed by a $\PTIME^{\NP}$ algorithm as follows. A first query to a SAT oracle determines whether $v_\mathcal{I}(x_1)$ is $\top$ or $\bot$, since $v_\mathcal{I}(x_1)=\top$ if and only if $F_1(Z_1)$ is satisfiable. Then, we replace $x_1$ by the value $v_\mathcal{I}(x_1)$ in $F_2(x_1,Z_2)$ and another query to the SAT oracle is performed to determine whether $F_2(v_\mathcal{I}(x_1),Z_2)$ is satisfiable, yielding the value of $v_\mathcal{I}(x_2)$. This step is iterated other $n-2$ times, finally obtaining the value of $v_\mathcal{I}(x_n)$. 

Let $\mathcal{I}$ be an instance of SNSAT, with $|\mathcal{I}|=n$. We now show how to build a finite Kripke structure $\Ku_\mathcal{I}$ and an $\AB$ formula $\Phi_\mathcal{I}$ in polynomial time, such that $\mathcal{I}$ is a positive instance of SNSAT if and only if $\Ku_\mathcal{I}\models \Phi_\mathcal{I}$. Such a reduction is inspired by similar constructions from~\cite{LMS01}.

\begin{figure}[p]
  \centering
  \input{Chaps/IC17/gadget}
  \caption{The gadget for $x_i$.}\label{gadget}
\end{figure}
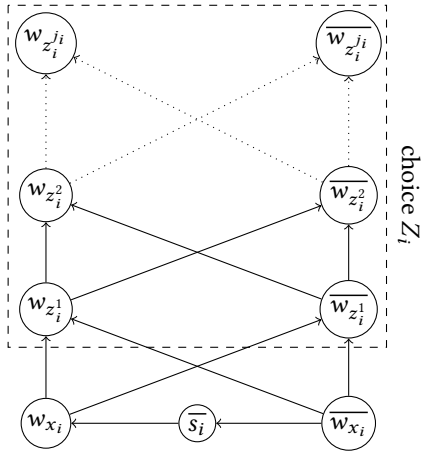

Let $Z=\bigcup_{i=1}^n Z_i$ and let $\tilde{R}=\{r_i \mid i=1,\ldots , n\}$ and $\tilde{R}_i=\tilde{R}\setminus \{r_i\}$ be $n+1$ sets of auxiliary variables.
The Kripke structure $\Ku_\mathcal{I}$ consists of a suitable composition of $n$ instances of a \emph{gadget} (an instance  for each variable $x_1,\ldots , x_n\in X$). The structure of the gadget for $x_i$, with $1\leq i\leq n$, is shown in Figure~\ref{gadget}, assuming that the labeling of states (nodes) is defined as follows:
\begin{itemize}
	\item $\Lab(w_{x_i})=X \cup Z \cup \{s,t\} \cup \tilde{R}_i$, and
    	$\Lab(\overline{w_{x_i}})=(X\setminus \{x_i\}) \cup Z \cup \{s,t\} \cup \tilde{R}_i \cup \{p_{\overline{x_i}}\}$;
	\item for $u_i=1,\ldots , j_i$, $\Lab(w_{z_i^{u_i}})=X \cup Z \cup \{s,t\} \cup \tilde{R}_i$, and\\
	    $\Lab(\overline{w_{z_i^{u_i}}})=X \cup (Z\setminus \{z_i^{u_i}\}) \cup \{s,t\} \cup \tilde{R}_i$;
	\item $\Lab(\overline{s_i})=X \cup Z \cup \{t\} \cup \tilde{R}_i$.
\end{itemize}

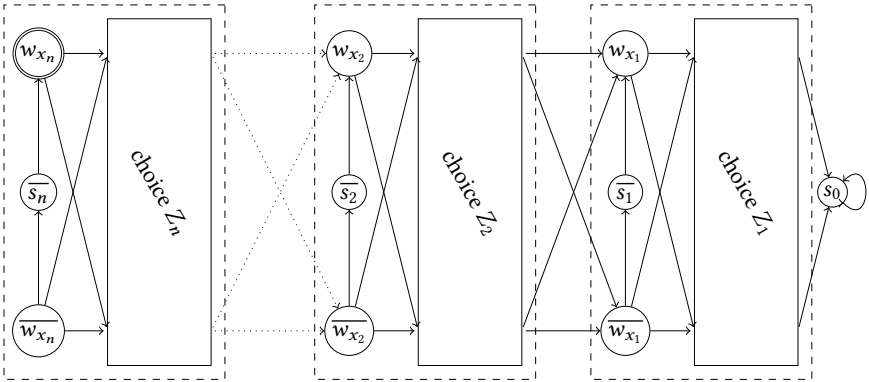
\begin{figure}[p]
  \rotatebox{-90}{\resizebox{!}{\linewidth}{\input{Chaps/IC17/fullKripke}}}
  \caption{Kripke structure $\Ku_\mathcal{I}$ associated with an SNSAT instance $\mathcal{I}$, with $|\mathcal{I}|=n$. Note that the states $\overline{s_n}$ and $\overline{w_{x_n}}$ are unreachable.}\label{fullKripke}
\end{figure}

The Kripke structure  $\Ku_\mathcal{I}$ is obtained by sequentializing (adding suitable edges) the $n$ instances of the gadget (in reverse order, from $x_n$ to $x_1$), adding a collector terminal state $s_0$, with labeling $\Lab(s_0)=X \cup Z \cup \{s\} \cup \tilde{R}$, and setting $w_{x_n}$ as the initial state. The overall construction is reported in Figure~\ref{fullKripke}. Formally, we have \[\Ku_\mathcal{I}=(\Prop, \States,\Edges,\Lab,w_{x_n}),\] 
where, in particular, $\Prop=X \cup Z \cup \{s,t\} \cup \tilde{R}\cup\{p_{\overline{x_i}}\mid i=1,\ldots ,n\}$ and \[\States=\{s_0\}\cup\bigcup_{i=1}^n\big(\{\overline{s_i},w_{x_i},\overline{w_{x_i}}\}\cup\{w_{z_i^{u_i}},\overline{w_{z_i^{u_i}}}\mid u_i=1,\ldots,j_i\}\big).\]
The Kripke structure $\Ku_\mathcal{I}$ features the following properties:
\begin{itemize}
	\item any trace satisfying $s$ does not pass through any $\overline{s_i}$, for $1\leq i\leq n$;
	\item any trace \emph{not} satisfying $t$ has $s_0$ as its last state;
	\item any trace \emph{not} satisfying $r_i$ passes through some state of the $i$-th gadget, for $1\leq i\leq n$;
	\item the only trace satisfying $p_{\overline{x_i}}$ is $\overline{w_{x_i}}$ (note that $|\overline{w_{x_i}}|=1$), for $1\leq i\leq n$.
\end{itemize}

A trace $\rho\in\Trk_{\Ku_\mathcal{I}}$ \emph{induces} a truth assignment of all the proposition letters, denoted by $\omega_\rho$, which is defined as $\omega_\rho(y)=\top \iff \Ku_\mathcal{I},\rho\models y$, for any letter $y$. 
In the following, we shall write $\omega_\rho(Z_i)$ for $\omega_\rho(z_i^1),\ldots , \omega_\rho(z_i^{j_i})$.
In particular, if $\rho$ starts from some state $w_{x_i}$ or $\overline{w_{x_i}}$, and satisfies $s\wedge \neg t$ (that is, it reaches the collector state $s_0$ without visiting any node $\overline{s_j}$, for $1\leq j \leq i$), $\omega_\rho$ fulfills the following conditions: 
for $1\leq m\leq i$,
\begin{itemize}
	\item if $w_{x_m}\!\in\!\states(\rho)$, then $\omega_\rho(x_m)\!=\!\top$, and if $\overline{w_{x_m}}\!\in\!\states(\rho)$, then $\omega_\rho(x_m)\!=\!\bot$;
	\item for $1\leq u_m\leq j_m$, if $w_{z_m^{u_m}}\in\states(\rho)$, then $\omega_\rho(z_m^{u_m})=\top$, and if $\overline{w_{z_m^{u_m}}}\in\states(\rho)$, then $\omega_\rho(z_m^{u_m})=\bot$.
\end{itemize}	
It easily follows that $\Ku_\mathcal{I},\rho\models F_m(x_1,\ldots , x_{m-1}, Z_m)$ if and only if 
$F_m(\omega_\rho(x_1), \ldots ,\allowbreak \omega_\rho(x_{m-1}),\omega_\rho(Z_m)) = \top$.
Finally, let $\mathcal{F}_\mathcal{I}=\{\psi_k \mid 0\leq k\leq n+1\}$ be the set of formulas defined as: 
\[\psi_0=\bot \mbox{ and, for  }k\geq 1,\]
\begin{equation*}
\psi_k = \hsA \underbrace{\left[\begin{array}{c}
(s\wedge \neg t) \wedge \displaystyle{\bigwedge_{i=1}^n} \Big((x_i\wedge \neg r_i)\rightarrow F_i(x_1, \ldots, x_{i-1}, Z_i)\Big) \\ 
\wedge \\ 
\hsBu \Big( \Big(\displaystyle{\bigvee_{i=1}^n} \hsA p_{\overline{x_i}}\Big)\rightarrow \hsA\big(\neg s \wedge \Length_2\wedge \hsA (\Length_2\wedge \neg\psi_{k-1})\big)\Big)
\end{array}\right]}_{\text{\normalsize $\varphi_k$}} ,
\end{equation*}
where $\Length_2=\hsB \top \wedge \hsBu\hsBu\bot$ (Example~\ref{example:length}) is satisfied only by traces of length 2.
The first conjunct of $\varphi_k$, i.e, $s\wedge \neg t$, forces the trace to reach the collector state $s_0$, without visiting any state $\overline{s_j}$. The second conjunct checks that if the trace assigns the truth value $\top$ to $x_m$ passing through $w_{x_m}$, with $1\leq m \leq n$, then $F_m(x_1,\ldots,x_{m-1},Z_m)$ is satisfied by $\omega_\rho$ (which amounts to say that the SAT problem connected with $Z_m$ has a positive answer, for the selected values of $x_1,\ldots,x_{m-1}$).
Conversely, the third conjunct ensures that if the trace assigns the truth value $\bot$ to some $x_m$ by passing through $\overline{w_{x_m}}$, then, intuitively, the SAT problem connected with $Z_m$ has no assignment satisfying $F_m(x_1, \ldots, x_{m-1}, Z_m)$. As a matter of fact, if $\rho$ satisfies $\varphi_k$, for some $k\geq 2$, and assigns $\bot$ to $x_m$, then there is a prefix $\tilde{\rho}$ of $\rho$ ending in $\overline{w_{x_m}}$. Since  $\bigvee_{i=1}^n\hsA p_{\overline{x_i}}$ is satisfied by $\tilde{\rho}$, $\hsA\big(\neg s \wedge \Length_2\wedge \hsA (\Length_2\wedge \neg\psi_{k-1})\big)$ must be satisfied as well. The only possibility is that the trace $\overline{s_m}\cdot w_{x_m}$ does not model $\psi_{k-1}$, as $\overline{w_{x_m}}\cdot \overline{s_m}$ has to model $\hsA (\Length_2\wedge \neg\psi_{k-1})$. However, since $\psi_{k-1}=\hsA\varphi_{k-1}$, this holds if and only if $\Ku, w_{x_m}\not \models\psi_{k-1}$.

The following theorem states the correctness of the construction. Its proof can be found in Appendix~\ref{proof:thcorr}.
\begin{theorem}\label{thcorr} 
Let $\mathcal{I}$ be an instance of SNSAT, with $|\mathcal{I}|=n$, and let $\Ku_\mathcal{I}$ and $\mathcal{F}_\mathcal{I}$
be defined as above. For all $0\leq k\leq n+1$ and all $r=1,\ldots , n$, it holds that:
	\begin{enumerate}
		\item if $k\geq r$, then $v_\mathcal{I}(x_r)=\top \iff \Ku_\mathcal{I},w_{x_r}\models \psi_k;$
		\item if $k\geq r+1$, then $v_\mathcal{I}(x_r)=\bot \iff \Ku_\mathcal{I},\overline{w_{x_r}}\models \psi_k.$
	\end{enumerate}
\end{theorem}

The correctness of the reduction from SNSAT to MC for $\AB$ follows.

\begin{corollary}\label{corol:c} Let $\mathcal{I}$ be an instance of SNSAT, with $|\mathcal{I}|=n$, and let $\Ku_\mathcal{I}$ and $\mathcal{F}_\mathcal{I}$
be defined as above. Then, 
$v_\mathcal{I}(x_n)=\top \iff \Ku_\mathcal{I}\models \hsBu\bot \rightarrow\psi_n$.
\end{corollary}
\begin{proof}
By Theorem~\ref{thcorr}, $v_\mathcal{I}(x_n)=\top \iff \Ku_\mathcal{I},w_{x_n}\models \psi_n$. If $v_\mathcal{I}(x_n)=\top$ then $\Ku_\mathcal{I},w_{x_n}\models \psi_n$ and, since $w_{x_n}$ is the only initial trace satisfying $\hsBu\bot$ (this formula is satisfied by traces having length equal to 1 only), $\Ku_\mathcal{I}\models \hsBu\bot \rightarrow\psi_n$. Conversely, if $\Ku_\mathcal{I}\models \hsBu\bot \rightarrow\psi_n$, then $\Ku_\mathcal{I},w_{x_n}\models \psi_n$, allowing us to conclude that $v_\mathcal{I}(x_n)=\top$.
\end{proof}

Eventually we can state the complexity of the problem.

\begin{corollary}\label{coro:ABhard}
	The MC problem for $\AB$ formulas over finite Kripke structures is $\PTIME^{\NP}$-hard (under polynomial-time reductions).
\end{corollary}
\begin{proof}
	The result follows from Corollary \ref{corol:c} considering that, for an instance of SNSAT $\mathcal{I}$, with $|\mathcal{I}|=n$, $\Ku_\mathcal{I}$ and $\psi_n\in \mathcal{F}_\mathcal{I}$ have a size polynomial in $n$ and in the length of the formulas of $\mathcal{I}$. 
\end{proof}

We can prove the same complexity lower bound for the symmetric fragment $\AbarE$, just by transposing the edges of $\Ku_\mathcal{I}$, and by replacing $\hsBu$ with $\hsEu$ and $\hsA$ with $\hsAt$ in the definition of $\psi_n$. This hardness result immediately propagates to the bigger fragments $\AAbarB$ and $\AAbarE$.

Finally,  we summarize the $\PTIME^{\NP}$-completeness results that can be obtained by putting together Corollary~\ref{coro:AAbarBalgo} in Section~\ref{sec:AAbarBalgo} and Corollary~\ref{coro:ABhard}.
\begin{corollary}
The MC problem for $\AB$, $\AbarE$, $\AAbarB$, and $\AAbarE$ formulas over finite Kripke structures is $\PTIME^{\NP}$-complete.
\end{corollary}

In the next section, we will prove a different complexity lower bound for the fragments $\A$, $\Abar$, $\AAbar$, $\AbarB$, and $\AE$, to which the present one does not apply.

%% file: Chaps/IC17/gadget.tex
\begin{tikzpicture}[every node/.style={circle, draw, inner sep=1pt}]

\node (v3) at (-2,0.5) {$w_{x_i}$};
\node (v2) at (0,0.5) {$\overline{s_i}$};
\node (v1) at (2,0.5) {$\overline{w_{x_i}}$};
\draw [->] (v1) edge (v2);
\draw [->] (v2) edge (v3);
\draw [dashed] (-2.5,6) rectangle (2.5,1.5);

\draw [use as bounding box,draw=none] (-2.5,6) rectangle (2.5,0.2);

\node (v4) at (-2,2) {$w_{z_i^1}$};
\node (v6) at (2,2) {$\overline{w_{z_i^1}}$};
\node (v5) at (-2,3.5) {$w_{z_i^2}$};
\node (v7) at (2,3.5) {$\overline{w_{z_i^2}}$};
\node (v8) at (-2,5.5) {$w_{z_i^{j_i}}$};
\node (v9) at (2,5.5) {$\overline{w_{z_i^{j_i}}}$};
\draw [->] (v4) edge (v5);
\draw [->] (v6) edge (v7);
\draw [->] (v4) edge (v7);
\draw [->] (v6) edge (v5);

\draw [->,dotted] (v5) edge (v8);
\draw [->,dotted] (v7) edge (v9);
\draw [->,dotted] (v7) edge (v8);
\draw [->,dotted] (v5) edge (v9);
\draw [->] (v3) edge (v4);
\draw [->] (v3) edge (v6);
\draw [->] (v1) edge (v4);
\draw [->] (v1) edge (v6);

\node [draw=none,rotate=-90] (ch) at (2.8,3.5) {choice $Z_i$};
\end{tikzpicture}

%% file: Chaps/IC17/fullKripke.tex
\begin{tikzpicture}[every node/.style={circle, draw, inner sep=1pt}]

\node [rotate=90] (v3) at (-2,0.5) {$w_{x_1}$};
\node [rotate=90] (v2) at (0,0.5) {$\overline{s_1}$};
\node [rotate=90] (v1) at (2,0.5) {$\overline{w_{x_1}}$};
\draw [->] (v1) edge (v2);
\draw [->] (v2) edge (v3);
\draw  (-2.5,3) rectangle (2.5,1.5) node[midway,draw=none,rotate=30] {choice $Z_1$};;

\node [draw=none] (v4) at (-2,1.5) {};
\node [draw=none] (v6) at (2,1.5) {};

\draw [->] (v3) edge (v4);
\draw [->] (v1) edge (v6);

\node [double,rotate=90]  (v13) at (-2,-8) {$w_{x_n}$};
\node [rotate=90] (v12) at (0,-8) {$\overline{s_n}$};
\node [rotate=90] (v11) at (2,-8) {$\overline{w_{x_n}}$};
\draw [->] (v11) edge (v12);
\draw [->] (v12) edge (v13);
\draw  (-2.5,-5.5) rectangle (2.5,-7) node[midway,draw=none,rotate=30] {choice $Z_n$};;

\node [draw=none] (v14) at (-2,-7) {};
\node [draw=none] (v16) at (2,-7) {};

\draw [->] (v13) edge (v14);
\draw [->] (v11) edge (v16);

\node [rotate=90] (v23) at (-2,-3.5) {$w_{x_2}$};
\node [rotate=90] (v22) at (0,-3.5) {$\overline{s_2}$};
\node [rotate=90] (v21) at (2,-3.5) {$\overline{w_{x_2}}$};
\draw [->] (v21) edge (v22);
\draw [->] (v22) edge (v23);
\draw  (-2.5,-1) rectangle (2.5,-2.5) node[midway,draw=none,rotate=30] {choice $Z_2$};;

\node [draw=none] (v24) at (-2,-2.5) {};
\node [draw=none] (v26) at (2,-2.5) {};

\draw [->] (v23) edge (v24);
\draw [->] (v21) edge (v26);

\node [draw=none] (v8) at (-2,-5.5) {};
\node [draw=none] (v18) at (2,-5.5) {};
\node [draw=none] (v29) at (2,-1) {};
\node [draw=none] (v19) at (-2,-1) {};
\draw [->, dotted] (v8) edge (v23);
\draw [->, dotted] (v18) edge (v21);
\draw [->] (v19) edge (v3);
\draw [->] (v29) edge (v1);

\node [rotate=90] (v9) at (0,3.5) {$s_0$};
\node [draw=none] (v30) at (-2,3) {};
\draw [->,looseness=5] (v9.south east) edge (v9.north east);

\node [draw=none] (v10) at (2,3) {};

\draw [->] (v19) edge (v1);
\draw [->] (v29) edge (v3);
\draw [->,dotted] (v18) edge (v23);
\draw [->,dotted] (v8) edge (v21);
\draw [->] (v30) edge (v9);
\draw [->] (v10) edge (v9);
\draw [->] (v13) edge (v16);
\draw [->] (v11) edge (v14);
\draw [->] (v23) edge (v26);
\draw [->] (v21) edge (v24);
\draw [->] (v3) edge (v6);
\draw [->] (v1) edge (v4);

\draw [dashed] (-2.7,3.2) rectangle (2.7,0);
\draw [dashed] (-2.7,-0.8) rectangle (2.7,-4);
\draw [dashed] (-2.7,-5.3) rectangle (2.7,-8.5);

\end{tikzpicture}

%% file: Chaps/IC17/section04.tex
\section{$\Th$-hardness of MC for $\A$ and $\Abar$}\label{sect:AHard}

In this section, we prove that MC for formulas of the fragment $\A$ (and of $\Abar$, respectively) over finite Kripke structures is $\Th$-hard, by reducing to it \emph{PARITY(SAT)}~\cite{WAGNER87}, a problem complete for $\Th$.
PARITY(SAT) is to decide, for a  set of Boolean formulas $\Gamma$,
if the number of \emph{satisfiable} formulas in $\Gamma$ is odd or even. Hardness of MC for $\A$ and  $\Abar$ immediately propagates to $\AAbar$, $\AE$, $\AAbar$, $\AbarB$.

Let $\Gamma$ be a set of $n$ Boolean formulas $\{\phi_i(x_1^i,\ldots , x_{m_i}^i)\mid 1 \leq i \leq n,\; m_i\in\mathbb{N}\}$. We provide a Kripke structure $\mathpzc{K}_{PAR}^{\Gamma}$ and an $\A$-formula $\Phi_{\Gamma}$ such that $\mathpzc{K}_{PAR}^{\Gamma}\models \Phi_{\Gamma}$ if and only if the number of satisfiable Boolean formulas in $\Gamma$ is \emph{odd}.
 
We start by defining a Boolean formula, $\text{parity}(F, Z)$, over two sets of Boolean variables, $F=\{f_1,\ldots, f_n\}$ and $Z=\{z_1,\ldots ,z_t\}$, with $t=3 \cdot (n-1)+1$. Such a formula allows one to decide the parity of the number of variables in $F$ that evaluate to $\top$. $Z$ is a set of auxiliary variables, whose truth values are \emph{functionally determined} by those assigned to the variables in $F$. Given a truth assignment, the number of variables in $F$ set to $\top$ is even if $\text{parity}(F, Z)$ evaluates to $\top$, and, in particular, its last variable $z_t$ evaluates to $\top$.
 The formula $\text{parity}(F, Z)$ is defined as follows:
\[\text{parity}(f_1,\ldots ,f_n,z_1,\ldots ,z_t)=z_t\wedge\text{par}_n(f_1,\ldots ,f_n,z_1,\ldots ,z_t), \]
where $t=3 \cdot (n-1)+1$ and, 
for $i \geq 1$, $\text{par}_{i}(f_1,f_2,\ldots, f_i,z_1, \ldots , z_{3(i-1)+1})$ is inductively defined as: \[\text{par}_1(f_1,z_1)=\neg f_1\leftrightarrow z_1,\] and, for all   $i\geq 2$,
\begin{multline*}
    \text{par}_{i}(f_1,f_2,\ldots, f_i,z_1, \ldots ,z_{\alpha+3})=\\
    \big(z_{\alpha+1}\leftrightarrow (f_i \wedge \neg z_{\alpha})\big) \wedge \big(z_{\alpha+2}\leftrightarrow (\neg f_i \wedge z_{\alpha})\big) \wedge \big(z_{\alpha+3}\leftrightarrow (z_{\alpha+2} \vee z_{\alpha+1})\big) \wedge\\
    \text{par}_{i-1}(f_1,f_2,\ldots, f_{i-1},z_1, \ldots ,z_{\alpha}),
\end{multline*}
with $\alpha= 3 \cdot (i-2)+1$.

Each assignment satisfying $\text{par}_i$ has to set $z_{\alpha}$ to the parity value for the set of Boolean variables $f_1,f_2, \ldots, f_{i-1}$. Such a value is then possibly changed according to the truth of $f_i$ and \lq\lq assigned\rq\rq{} to $z_{\alpha + 3}$. Note that the length of $\text{parity}(F,Z)$ is polynomial in $n$.

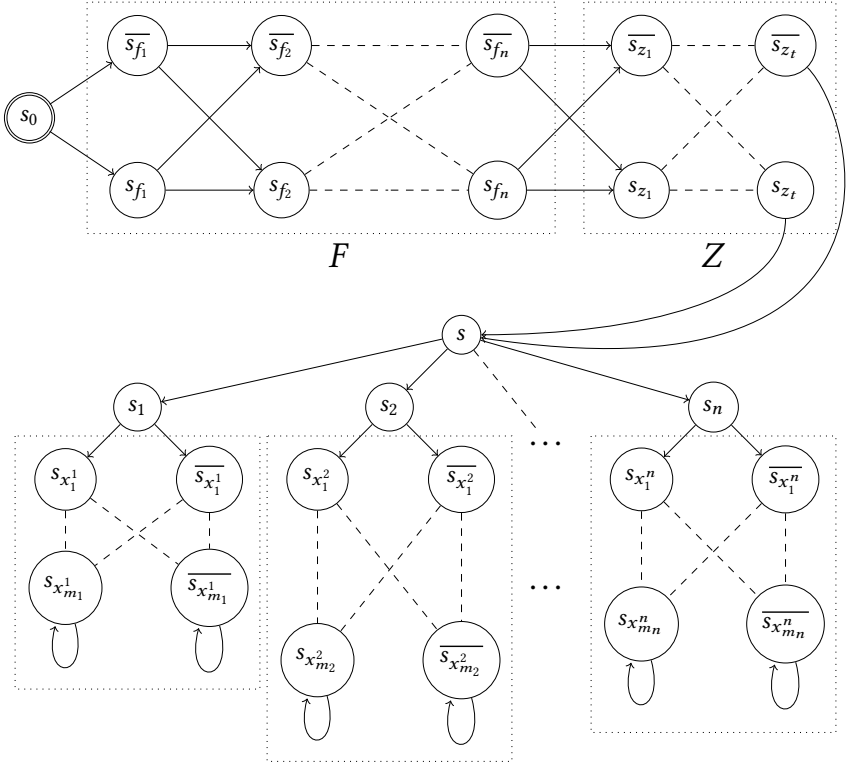
\begin{figure}[tb]
\centering
\resizebox{\linewidth}{!}{\input{Chaps/IC17/Kpar}}
\caption{The Kripke structure $\mathpzc{K}_{PAR}^{\Gamma}$.}\label{Kpar}
\end{figure}

We now show how to build the Kripke structure $\mathpzc{K}_{PAR}^{\Gamma}$ depicted in Figure~\ref{Kpar}, such that a subset of its traces encode all the possible truth assignments to the variables of $F\cup Z$ and to all the variables occurring in formulas of $\Gamma$.
%
%
$\mathpzc{K}_{PAR}^{\Gamma}$ features a pair of states for each Boolean variable in $F\cup Z$ as well as for all the variables of formulas in $\Gamma$ (one state for each truth value). Each path from the initial state $\sinit$ to the state $s$ represents a truth assignment to the variables in $F \cup Z$. Then, the structure branches into $n$ substructures, each one modeling 
the possible truth assignments to the variables of a formula in $\Gamma$. 

Formally, $\mathpzc{K}_{PAR}^{\Gamma}=\KuDef$, where
\begin{itemize}
    \item $\Prop=\{p,q\}\cup F \cup Z \cup \{aux_i\mid 1\leq i\leq n\}\cup \{x^i_{j_i}\mid 1\leq i\leq n,\; 1\leq j_i\leq m_i\}$,
    \item $\States=\{\sinit\}\cup\{s_{f_i}, \overline{s_{f_i}}\mid 1\leq i\leq n\}\cup\{s_z,\overline{s_z} \mid z\in Z\}\cup\{s\}\cup\{s_i\mid 1\leq i\leq n\}\cup $ \\
    $\{s_{x^i_{j_i}}, \overline{s_{x^i_{j_i}}}\mid 1\leq i\leq n,\; 1\leq j_i\leq m_i\}$,
    \item $\Edges\!=\!\{(\sinit,s_{f_1}),(\sinit,\overline{s_{f_1}})\}\cup \{(s_{f_i},s_{f_{i+1}}),(s_{f_i},\overline{s_{f_{i+1}}}),(\overline{s_{f_i}},s_{f_{i+1}}),(\overline{s_{f_i}},\overline{s_{f_{i+1}}})\mid 1\!\leq\! i\!<\! n\} \cup$\\
    $\{(s_{f_n},s_{z_1}),(\overline{s_{f_n}},s_{z_1}),(s_{f_n},\overline{s_{z_1}}),(\overline{s_{f_n}},\overline{s_{z_1}})\}\cup$\\
    $\{(s_{z_i},s_{z_{i+1}}),(s_{z_i},\overline{s_{z_{i+1}}}),(\overline{s_{z_i}},s_{z_{i+1}}),(\overline{s_{z_i}},\overline{s_{z_{i+1}}})| 1\leq i< t\} \cup$\\
    $\{(s_{z_t},s),(\overline{s_{z_t}},s)\}\cup \{(s,s_i),(s_i,s_{x^i_1}),(s_i,\overline{s_{x^i_1}})\mid 1\leq i\leq n\}\cup$ \\
    $\{(s_{x^i_{j_i}},s_{x^i_{j_i+1}}), (\overline{s_{x^i_{j_i}}},s_{x^i_{j_i+1}}),(\overline{s_{x^i_{j_i}}},s_{x^i_{j_i+1}}),(\overline{s_{x^i_{j_i}}},\overline{s_{x^i_{j_i+1}}}) | 1\leq i\leq n,\; 1\leq j_i< m_i\}\cup$\\
    $\{(s_{x^i_{m_i}},s_{x^i_{m_i}}),(\overline{s_{x^i_{m_i}}},\overline{s_{x^i_{m_i}}})|1\leq i\leq n\}$,
    \item and the labeling function $\Lab$ is defined as follows:
	\begin{itemize}
		\item $\Lab(\sinit)=\{p,q\}\cup F\cup Z$;
		\item for all $1\leq i\leq n$, $\Lab(s_{f_i})=\{p,q\}\cup F\cup Z$; $\Lab(\overline{s_{f_i}})=\{p,q\}\cup (F\setminus \{f_i\}) \cup Z$;
		\item for all $z\in Z$, $\Lab(s_z)=\{p,q\}\cup F\cup Z$; $\Lab(\overline{s_z})=\{p,q\}\cup F \cup (Z\setminus \{z\})$;
		\item $\Lab(s)=\{q\}\cup F\cup Z \cup \{aux_i\mid 1\leq i\leq n\}\cup \{x^i_{j_i}\mid 1\leq i\leq n,\; 1\leq j_i\leq m_i\}$;
		\item for all $1\leq i\leq n$, $\Lab(s_i)= \{aux_i\}\cup \{x^i_{j_i}\mid 1\leq j_i\leq m_i\}$;
		\item for all $1\leq i\leq n$, $1\leq k_i\leq m_i$, $\Lab(s_{x^i_{k_i}})= \{aux_i\}\cup \{x^i_{j_i}\mid 1\leq j_i\leq m_i\}$;  $\Lab(\overline{s_{x^i_{k_i}}})= \{aux_i\}\cup \{x^i_{j_i}\mid 1\leq j_i\leq m_i\}\setminus\{x^i_{k_i}\}$.
	\end{itemize}
\end{itemize}
According to the definition of $\mathpzc{K}_{PAR}^{\Gamma}$, it holds that:
\begin{enumerate}
\item each trace $\rho$ from $\sinit$ to $s$ encodes a truth assignment to the proposition letters in $F\cup Z$ (for all $y\in F\cup Z$, $y$ is $\top$ in $\rho$ if and only if $y\in\bigcap_{w\in\states(\rho)}\Lab(w)$). Conversely, for each truth assignment to the proposition letters in $F\cup Z$, there exists an initial trace $\rho$, reaching the state $s$, encoding such an assignment. Note that, among the initial traces, the ones leading to $s$ are exactly those satisfying $q \wedge \neg p$.
\item An initial trace leading to $s$ satisfies $\text{parity}(F, Z)$ if the induced assignment sets an even number of $f_i$'s to $\top$, and every $z\in Z$ to the truth value which is functionally implied by the values of the $f_i$'s.
\item A trace $\tilde{\rho}$ starting from $s$ and ending in a state $s$, $s_i$, $s_{x_j^i}$ or $\overline{s_{x_j^i}}$, with $1 \leq i \leq n$ and $1 \leq j \leq m_i$, encodes a truth assignment to the proposition letters $x_1, \ldots, x_{m_i}$ (if the trace 
ends in $s$ or $s_i$, all the variables are assigned to $\top$; if it ends in $s_{x_j^i}$ or $\overline{s_{x_j^i}}$, in particular all the variables 
$x^i_{j+1}, \ldots , x^i_{m_i}$ are assigned to $\top$, by homogeneity).
\item A Boolean formula $\phi_i(x_1^i,\ldots , x_{m_i}^i) \in \Gamma$ is satisfiable if and only if there exists a trace $\tilde{\rho}$ starting from $s$ and ending in a state $s$, $s_i$, $s_{x_j^i}$ or $\overline{s_{x_j^i}}$, for some $j=1,\ldots , m_i$, such that $\mathpzc{K}_{PAR}^{\Gamma},\tilde{\rho}\models \phi_i(x_1^i,\ldots , x_{m_i}^i)$. 
\end{enumerate}

Finally, let us consider the $\A$ formula 
\[\psi=q \wedge \neg p \wedge \text{parity}(F, Z) \wedge \bigwedge_{i=1}^n \big(f_i \leftrightarrow \hsA(aux_i \wedge \phi_i(x_1^i,\ldots , x_{m_i}^i))\big).\]
In view of the above observations, $\psi$ is satisfied by an initial trace $\overline{\rho}$ if (and only if) $(i)$ $\overline{\rho}$ leads to $s$, $(ii)$ $\overline{\rho}$ induces an assignment which sets an even number of $f_i$'s to $\top$ and all $z\in Z$ accordingly, and $(iii)$
for all $1 \leq i \leq n$, $f_i$ is $\top$ if and only if there exists a trace $\tilde{\rho}$ starting from $s$ and ending in a state $s$, $s_i$, $s_{x_j^i}$ or $\overline{s_{x_j^i}}$, such that $\mathpzc{K}_{PAR}^{\Gamma},\tilde{\rho}\models \phi_i(x_1^i,\ldots , x_{m_i}^i)$.
The length of $\psi$ is polynomial in the input size.

Let us now assume we are given an instance of PARITY(SAT) $\Gamma$ with an \emph{even} number of satisfiable Boolean formulas. Then, there exists an initial trace $\rho$ ending in $s$ such that, for all $i$, $s_{f_i}\in\states(\rho)$ if $\phi_i(x_1^i,\ldots , x_{m_i}^i)$ is satisfiable, and $\overline{s_{f_i}}\in\states(\rho)$ otherwise. Moreover, $\rho$ can be chosen in such a way that $\mathpzc{K}_{PAR}^{\Gamma},\rho\models  \text{parity}(F,Z)$. It immediately follows that, for all $i$, $\mathpzc{K}_{PAR}^{\Gamma},\rho\models f_i$ if and only if $\mathpzc{K}_{PAR}^{\Gamma},\rho\models \hsA(aux_i \wedge \phi_i(x_1^i,\ldots , x_{m_i}^i))$, concluding that $\mathpzc{K}_{PAR}^{\Gamma},\rho\models \psi$.

Conversely, let $\rho$ be  an initial trace such that $\mathpzc{K}_{PAR}^{\Gamma},\rho\models \psi$. It holds that 
$\rho$ ends in $s$ and sets an even number of $f_i$'s to $\top$. Furthermore, if $\mathpzc{K}_{PAR}^{\Gamma},\rho\models f_i$, then there exists $\tilde{\rho}$ starting from $s$ and ending in $s$, $s_i$, $s_{x_j^i}$ or $\overline{s_{x_j^i}}$, such that $\mathpzc{K}_{PAR}^{\Gamma},\tilde{\rho}\models\phi_i(x_1^i,\ldots , x_{m_i}^i)$, hence $\phi_i(x_1^i,\ldots , x_{m_i}^i)$ is satisfiable.
If $\mathpzc{K}_{PAR}^{\Gamma},\rho\models \neg f_i$, then there exists no $\tilde{\rho}$ starting from $s$ and ending in $s$, $s_i$, $s_{x_j^i}$ or $\overline{s_{x_j^i}}$, such that $\mathpzc{K}_{PAR}^{\Gamma},\tilde{\rho}\models\phi_i(x_1^i,\ldots , x_{m_i}^i)$. Thus $\phi_i(x_1^i,\ldots , x_{m_i}^i)$ is unsatisfiable. Hence, $\Gamma$ contains an even number of satisfiable formulas.

Therefore we have proved that the number of \emph{satisfiable} Boolean formulas of $\Gamma$ is \emph{even} if and only if there exists an initial trace $\rho$ such that $\mathpzc{K}_{PAR}^{\Gamma},\rho\models \psi$. 
This amounts to say that $\Gamma$ contains an \emph{odd} number of satisfiable Boolean formulas (the PARITY(SAT) problem) if and only if $\mathpzc{K}_{PAR}^{\Gamma}\models \Phi_{\Gamma}$, where $\Phi_{\Gamma}=\neg\psi$ (the MC problem).
The next theorem immediately follows.
\begin{theorem}
The MC problem for $\A$ formulas over finite Kripke structures is $\Th$-hard (under polynomial-time reductions).
\end{theorem}

A similar proof can be given for $\Abar$ (roughly speaking, we replace all the occurrences of $\hsA$ in $\Phi_{\Gamma}$ by $\hsAt$, and we stick the $n$ substructures of $\mathpzc{K}_{PAR}^{\Gamma}$---after transposing all their edges---on $\sinit$, instead of $s$).

Finally, we observe that the $\Th$-hardness of $\Abar$ and $\A$ immediately propagates to $\AAbar$, $\AbarB$ and $\AE$, yielding, together with Corollary~\ref{th:AAbaralgo} and Theorem~\ref{th:AbarBalgo}, the following result.
\begin{theorem}
The MC problem for $\A$, $\Abar$, $\AAbar$, $\AbarB$ and $\AE$ formulas over finite Kripke structures is in $\Thsq$ and it is hard for $\Th$.
\end{theorem}

It is still an open issue if MC for the above fragments can be solved by $o(\log^2 n)$ queries to the $\NP$ oracle, or it is possible to prove a stronger lower bound, or both (e.g., the problem may be complete for $\PTIME^{\NP[O(\log n \log\log n)]}$). As a matter of fact, any attempt to reduce \TBSATM{} \emph{to} MC for $\A$, $\AE$, or $\AAbar$ has failed, because in such \lq\lq reduction\rq\rq{} we need an $\HS$ formula of length $\Theta(n^{\log n})$, which clearly cannot be generated in polynomial time.

%% file: Chaps/IC17/Kpar.tex
\begin{tikzpicture}

\path [use as bounding box] (-6.5,1.1) rectangle (5.5,-9.4);

\node [draw,double,circle] (v1) at (-6,-0.5) {$\sinit$};
\node [draw,circle] (v2) at (-4.5,-1.5) {$s_{f_1}$};
\node [draw,circle] (v3) at (-4.5,0.5) {$\overline{s_{f_1}}$};
\node [draw,circle] (v4) at (-2.5,-1.5) {$s_{f_2}$};
\node [draw,circle] (v5) at (-2.5,0.5) {$\overline{s_{f_2}}$};
\node [draw,circle] (v6) at (0.5,-1.5) {$s_{f_n}$};
\node [draw,circle] (v8) at (0.5,0.5) {$\overline{s_{f_n}}$};
\node [draw,circle] (v7) at (2.5,-1.5) {$s_{z_1}$};
\node [draw,circle] (v9) at (2.5,0.5) {$\overline{s_{z_1}}$};
\node [draw,circle] (v10) at (4.5,-1.5) {$s_{z_t}$};
\node [draw,circle] (v12) at (4.5,0.5) {$\overline{s_{z_t}}$};
\node [draw,circle] (v11) at (0,-3.5) {$s$};

\draw [->] (v1) edge (v2);
\draw [->] (v1) edge (v3);
\draw [->] (v2) edge (v4);
\draw [->] (v2) edge (v5);
\draw [->] (v3) edge (v4);
\draw [->] (v3) edge (v5);
\draw [->] (v6) edge (v7);
\draw [->] (v8) edge (v9);
\draw [->] (v6) edge (v9);
\draw [->] (v8) edge (v7);
\draw [dotted] (-5.2,1.1) rectangle (1.3,-2.1);
\draw [dotted] (1.7,1.1) rectangle (5.2,-2.1);

\node [inner sep=0, outer sep=0] (v15) at (-1,0.5) {};
\node [inner sep=0, outer sep=0] (v14) at (-1,-0.5) {};
\node [inner sep=0, outer sep=0] (v13) at (-1,-1.5) {};
\draw [dashed] (v4) edge (v13);
\draw [dashed] (v13) edge (v6);
\draw [dashed] (v4) edge (v14);
\draw [dashed] (v14) edge (v6);
\draw [dashed] (v5) edge (v14);
\draw [dashed] (v14) edge (v8);
\draw [dashed] (v5) edge (v15);
\draw [dashed] (v15) edge (v8);
\draw [dashed] (v7) edge (v10);
\draw [dashed] (v7) edge (v12);
\draw [dashed] (v9) edge (v10);
\draw [dashed] (v9) edge (v12);
\node at (3.5,-2.4) {\Large $Z$};
\node at (-1.7,-2.4) {\Large $F$};

\node [draw,circle] (v16) at (-4.5,-4.5) {$s_1$};
\node [draw,circle] (v17) at (-1,-4.5) {$s_2$};
\node [draw,circle] (v18) at (3.5,-4.5) {$s_n$};
\node [draw,circle] (v19) at (-5.5,-5.5) {$s_{x^1_1}$};
\node [draw,circle] (v20) at (-3.5,-5.5) {$\overline{s_{x^1_1}}$};
\node [draw,circle] (v21) at (-2,-5.5) {$s_{x^2_1}$};
\node [draw,circle] (v22) at (0,-5.5) {$\overline{s_{x^2_1}}$};
\node [draw,circle] (v23) at (2.5,-5.5) {$s_{x^n_1}$};
\node [draw,circle] (v24) at (4.5,-5.5) {$\overline{s_{x^n_1}}$};
\node [draw,circle] (v25) at (-5.5,-7) {$s_{x^1_{m_1}}$};
\node [draw,circle] (v26) at (-3.5,-7) {$\overline{s_{x^1_{m_1}}}$};
\node [draw,circle] (v27) at (-2,-8) {$s_{x^2_{m_2}}$};
\node [draw,circle] (v28) at (0,-8) {$\overline{s_{x^2_{m_2}}}$};
\node [draw,circle] (v29) at (2.5,-7.5) {$s_{x^n_{m_n}}$};
\node [draw,circle] (v30) at (4.5,-7.5) {$\overline{s_{x^n_{m_n}}}$};
\draw [->] (v11) edge (v16);
\draw [->] (v11) edge (v17);
\draw [->] (v11) edge (v18);
\node (v31) at (1.2,-5) {\Large \dots};
\node at (1.2,-7) {\Large \dots};
\draw [->] (v16) edge (v19);
\draw [->] (v16) edge (v20);
\draw [->] (v17) edge (v21);
\draw [->] (v17) edge (v22);
\draw [->] (v18) edge (v23);
\draw [->] (v18) edge (v24);
\draw [->] (v25) edge[loop below] (v25);
\draw [->] (v26) edge[loop below] (v26);
\draw [->] (v27) edge[loop below] (v27);
\draw [->] (v28) edge[loop below] (v28);
\draw [->] (v29) edge[loop below] (v29);
\draw [->] (v30) edge[loop below] (v30);
\draw [dashed] (v19) edge (v25);
\draw [dashed] (v19) edge (v26);
\draw [dashed] (v20) edge (v25);
\draw [dashed] (v20) edge (v26);
\draw [dashed] (v21) edge (v27);
\draw [dashed] (v21) edge (v28);
\draw [dashed] (v22) edge (v27);
\draw [dashed] (v22) edge (v28);
\draw [dashed] (v23) edge (v29);
\draw [dashed] (v23) edge (v30);
\draw [dashed] (v24) edge (v29);
\draw [dashed] (v24) edge (v30);
\draw [dashed] (v11) edge (v31);
\draw [dotted] (-6.2,-4.9) rectangle (-2.8,-8.4);
\draw [dotted] (-2.7,-4.9) rectangle (0.7,-9.4);
\draw [dotted] (1.8,-4.9) rectangle (5.2,-9);

\draw [->](v12) .. controls (5.5,-0.5) and (6.5,-4.5) .. (v11);
\draw [->](v10) .. controls (4.5,-3) and (2.5,-3.5) .. (v11);
\end{tikzpicture}

%% file: Chaps/IC17/conclus.tex
\section{Conclusions}
In this chapter, we have proved that the fragments $\AB$, $\AbarE$, $\AAbarB$, and $\AAbarE$ are complete for $\PTIME^{\NP}$, thus joining other (point-based) temporal logics---e.g., $\CTL^+$, $\mathsf{F}\CTL$, and $\mathsf{E}\CTL^+$---whose MC problem is complete for that class~\cite{LMS01} as well. 
In addition, we have shown that MC for $\A$, $\Abar$, $\AAbar$, $\AbarB$, and $\AE$ has a lower complexity, placed in between $\Th$ and $\Thsq$. This result has been proved by reducing MC to \TBSATM, the problem of deciding the output value of a complex circuit, where some gates feature an $\NP$ oracle.

Both the MC algorithms we propose can be efficiently implemented in practice by means of a polynomial-time procedure which iteratively invokes a SAT-solver, whose
extreme efficiency can be \lq\lq imported\rq\rq{} in a straightforward way: the procedure just generates some suitable Boolean formulas, feeds the SAT-solver, and stores the results. 
The modular and repetitive structure of the required Boolean formulas allows us to efficiently generate them and also to exploit the \emph{warm-restart} feature (or \emph{incrementality}) of SAT-solvers to quickly solve formulas following a common structural pattern.

In the next chapter---as we anticipated in the introduction---we will (mostly) put aside complexity issues on $\HS$ MC, and, conversely, focus on expressiveness: we will compare the expressive power of three different semantic versions of $\HS$ among themselves, and to the standard point-based temporal logics $\CTL$, $\LTL$ and $\CTLStar$.

%% file: Chaps/TOCL17/TOCL17main.tex
\chapter[Interval vs.\ point logic MC: an expressiveness comparison]{Interval vs.\ point temporal logic MC: \\ an expressiveness comparison}\label{chap:TOCL17}
\begin{chapref}
The references for this chapter are~\cite{tocl18,fsttcs16}.
\end{chapref}

\minitoc\mtcskip

\newcommand{\stat}{\mathsf{st}}
\newcommand{\LinearPast}{\mathsf{ct}}
\newcommand{\LinearTime}{\mathsf{lin}}

\newcommand{\act}{{\textit{act}}}

\input{Chaps/TOCL17/introduction}
\input{Chaps/TOCL17/preliminaries}

\input{Chaps/TOCL17/expressiveness}

\input{Chaps/TOCL17/conclusion}

%% file: Chaps/TOCL17/introduction.tex
\lettrine[lines=3]{T}{he MC methodology} mainly considers two types of point-based temporal logics (PTLs) as the property specification language---\emph{linear} and \emph{branching}---which differ 
in the underlying model of time.
In linear PTLs, such as $\LTL$~\cite{Pnu77}, each moment in time has a unique possible future:
formulas are interpreted over (infinite) \emph{paths} of a Kripke structure, and thus they refer to a single computation of a system.
In branching PTLs,
such as $\CTL$ and $\CTLStar$~\cite{EH86}, each moment in time may evolve into several possible futures:
formulas are interpreted over \emph{states} of the Kripke structure, hence referring to all the possible system computations.

In the previous chapters we have assumed a \emph{state-based} semantics for $\HS$, which
induces a branching reference both in the future and in the past:
intervals/traces 
are \lq\lq forgetful\rq\rq{} of the history leading to their initial state, and  
the initial (resp., final) state of an interval may feature several predecessors (resp., successors).
A graphical account of the state-based semantics can be found in Figure~\ref{fig:ST}; a detailed explanation will be given in the following.

\input{Chaps/TOCL17/Imgs/semPicturesST.tex}

However, as we already said, $\HS$ MC has been simultaneously and independently studied also by Lomuscio and Michaliszyn in~\cite{LM13,LM14,lm16}:
there, the considered $\HS$ fragments are interpreted over the unwinding of a Kripke structure (\emph{computation-tree-based} semantics---see Figure~\ref{fig:CT}). 
This induces a linear reference in the past (the initial state of an interval may feature only one predecessor), but branching in the future (the final state features several successors). Moreover, the computation history is never forgotten and increases with time.

\input{Chaps/TOCL17/Imgs/semPicturesCT.tex}

In this chapter, we study the expressiveness of $\HS$, in the context of MC, in comparison with that of the standard PTLs $\LTL$, $\CTL$, and $\CTLStar$. The analysis is carried on \emph{enforcing the homogeneity assumption}.
We prove that $\HS$ endowed with the state-based semantics (hereafter denoted as $\HS_\stat$) is not comparable with $\LTL$, $\CTL$, and $\CTLStar$. On one hand, the result supports the intuition that $\HS_\stat$ gains some expressiveness by the ability of branching in the past. On the other hand, $\HS_\stat$ does not feature the possibility of forcing the verification of a property over an 
infinite path, thus implying that the formalisms are not comparable. With the aim of having a more ``effective'' comparison base, we consider two additional semantic variants of $\HS$, 
namely, the \emph{computation-tree-based semantic variant} (denoted as $\HS_\LinearPast$) and the \emph{trace-based} one ($\HS_\LinearTime$). 
\input{Chaps/TOCL17/Imgs/semPicturesLN.tex}

The state-based (see Figure~\ref{fig:ST}) and computation-tree-based (see Figure~\ref{fig:CT}) approaches rely on a \emph{branching}-time setting and differ in the nature of past. In the latter approach, past is \emph{linear}: each interval may have several possible futures, but only a unique past. Moreover, past is assumed to be \emph{finite} 
and \emph{cumulative}, that is, the story of the current situation increases with time, and is never forgotten. 
%
The trace-based approach relies on a \emph{linear}-time setting (see Figure~\ref{fig:LN}), where the infinite paths (computations) of the given Kripke structure are the main semantic entities. Branching is neither allowed in the past nor in the future.
Note that the linear-past (rather than branching) approach is more suited to the specification  of dynamic behaviors, because it considers states in a computation tree, while the branching-past approach considers machine states, where past is not very meaningful for the specification of behavioral constraints~\cite{LS95}.

The variant $\HS_\LinearPast$ is a natural candidate for an expressiveness comparison with the branching time logics  $\CTL$ and $\CTLStar$. The most interesting and technically involved result is the characterization of the expressive power of $\HS_\LinearPast$: $\HS_\LinearPast$ turns out to be expressively equivalent to finitary $\CTLStar$, that is, the variant of $\CTLStar$ with quantification over \emph{finite} paths. As for $\CTL$, a non comparability result can be stated.
%
The variant $\HS_\LinearTime$ is a natural candidate for an expressiveness comparison with $\LTL$: 
we prove that $\HS_\LinearTime$ and $\LTL$ are equivalent (this result holds true even for a very small fragment of $\HS_\LinearTime$), but the former is at least exponentially more succinct than the latter. 

We complete the picture with a comparison of the three semantic variants $\HS_\stat$, $\HS_\LinearPast$, and $\HS_\LinearTime$. We show that, as expected, $\HS_\LinearTime$ is not comparable with either of the branching versions, $\HS_\LinearPast$ and $\HS_\stat$. The interesting result is that, on the other hand, $\HS_\LinearPast$ is strictly included in $\HS_\stat$: this supports $\HS_\stat$ as a reasonable and adequate semantic choice.

\begin{figure}[b]
\centering
\input{Chaps/TOCL17/completepicture}
\vspace*{-0.6cm}
\caption{Overview of the expressiveness results.}\label{results}
\end{figure}
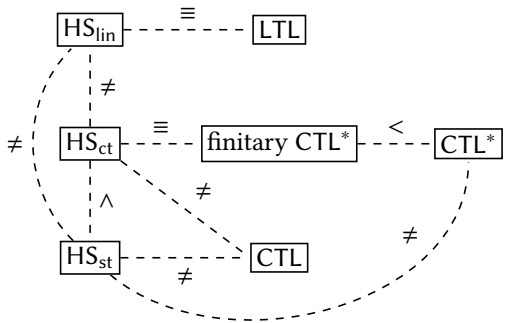

The complete picture of the expressiveness results is reported in Figure~\ref{results}
(the symbols $\neq$, $\equiv$, and $<$ denote incomparability, equivalence, and strict 
inclusion, respectively).

\paragraph*{Organization of the chapter.}
\begin{itemize}
	\item In the next section we start with some preliminaries; in particular, in Section~\ref{sect:PTL} we recall the well-known PTLs $\LTL$, $\CTL$ and $\CTLStar$; in Section~\ref{sect:3sem} we define the three semantic variants of $\HS$ ($\HS_\stat$, $\HS_\LinearPast$ and $\HS_\LinearTime$). In Section~\ref{subs:vendingMach} we provide a detailed example which gives an intuitive account of the three semantic variants and highlights their differences.
	\item In the next three sections we analyze and compare the expressiveness of these logics.  
In Section~\ref{sec:CharacterizezionOfLeniarTimeHS} we show the expressive equivalence of $\LTL$ and $\HS_\LinearTime$. Then, in Section~\ref{sec:characterizationHSLinearPast} we prove the equivalence of $\HS_\LinearPast$ and finitary $\CTLStar$. In Section~\ref{sect:allSems} we compare the expressiveness of 
$\HS_\stat$, $\HS_\LinearPast$ and $\HS_\LinearTime$.
\end{itemize}

%% file: Chaps/TOCL17/Imgs/semPicturesST.tex
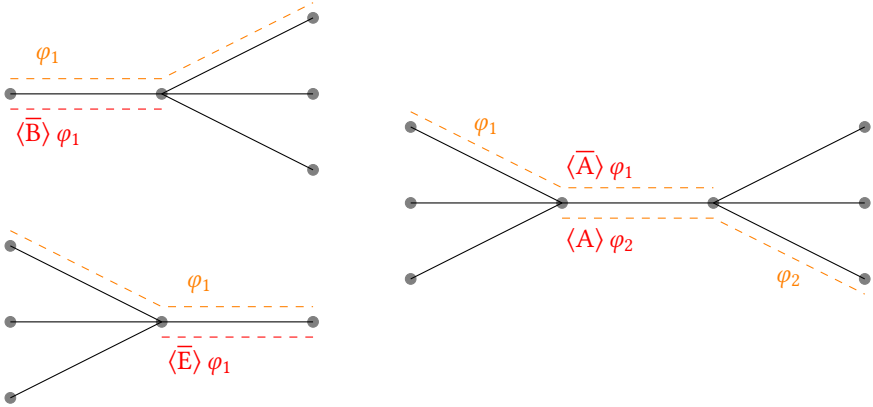
\begin{figure}[tp]
\centering
\begin{minipage}{0.36\linewidth}
\begin{tikzpicture}
				\filldraw [gray] (0,2) circle (2pt)
				(2,2) circle (2pt)
				(4,3) circle (2pt)
				(4,2) circle (2pt)
				(4,1) circle (2pt);
					\filldraw [gray] (4,-1) circle (2pt)
				(2,-1) circle (2pt)
				(0,0) circle (2pt)
				(0,-1) circle (2pt)
				(0,-2) circle (2pt);
				\draw [black]  (0,2) -- (4,2);
				\draw [black]  (0,-1) -- (4,-1);
				\draw [black] (2,2) -- (4,3);
				\draw [black] (2,2) -- (4,1);
				\draw [black] (0,0) -- (2,-1);
				\draw [black] (0,-2) -- (2,-1);
				\draw [dashed, orange] (0,2.2) -> (2,2.2) -> (4,3.2);
				\draw [dashed, red] (0,1.8) -> (2,1.8);
				\draw [dashed, orange] (0,0.2) -> (2,-0.8) -> (4,-0.8);
				\draw [dashed, red] (2,-1.2) -> (4,-1.2);
				
				\node [orange] at (0.5,2.5) {$\varphi_1$};	
				\node [red] at (0.5,1.5) {$\hsBt\varphi_1$};
			\node [orange] at (2.5,-0.5) {$\varphi_1$};	
				\node [red] at (2.5,-1.5) {$\hsEt\varphi_1$};
						
				
			\end{tikzpicture}
\end{minipage}	
\hfill 	
\begin{minipage}{0.54\linewidth}
			\begin{tikzpicture}
				%
					\filldraw [gray] (4,-1) circle (2pt)
				(2,-1) circle (2pt)
				(0,0) circle (2pt)
				(0,-1) circle (2pt)
				(0,-2) circle (2pt)
				(6,0) circle (2pt)
				(6,-1) circle (2pt)
				(6,-2) circle (2pt)	;
				\draw [black]  (0,-1) -- (6,-1);
				\draw [black] (4,-1) -- (6,0);
				\draw [black] (4,-1) -- (6,-2);
					\draw [black] (0,0) -- (2,-1);
				\draw [black] (0,-2) -- (2,-1);
				\draw [dashed, orange] (0,0.2) -> (2,-0.8) -> (4,-0.8);
				\draw [dashed, orange] (2,-1.2) -> (4,-1.2) -> (6,-2.2);
				
				\node [orange] at (1,0) {$\varphi_1$};	
				\node [red] at (2.5,-0.5) {$\hsAt\varphi_1$};
			\node [orange] at (5,-2) {$\varphi_2$};	
				\node [red] at (2.5,-1.5) {$\hsA \varphi_2$};
						
				
			\end{tikzpicture}
\end{minipage}
    \caption{State-based semantic variant $\HS_\stat$: past and future are branching}
    \label{fig:ST}
\end{figure}

%% file: Chaps/TOCL17/Imgs/semPicturesCT.tex
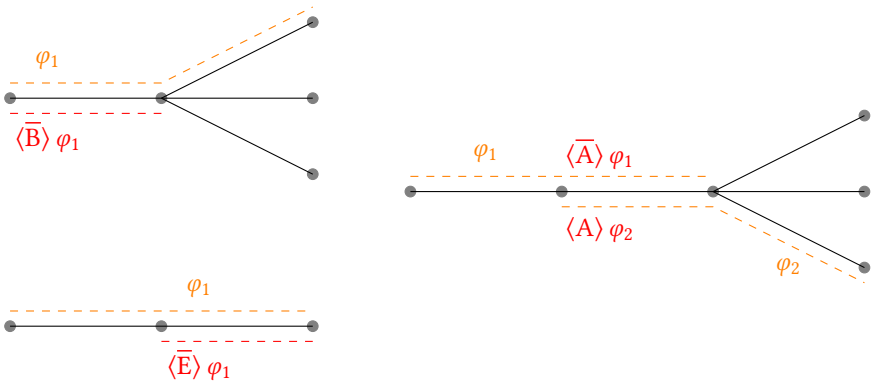
\begin{figure}[tp]
\centering
\begin{minipage}{0.36\linewidth}
\begin{tikzpicture}
				\filldraw [gray] (0,2) circle (2pt)
				(2,2) circle (2pt)
				(4,3) circle (2pt)
				(4,2) circle (2pt)
				(4,1) circle (2pt);
					\filldraw [gray] (4,-1) circle (2pt)
				(2,-1) circle (2pt)
				(0,-1) circle (2pt);
				%
				\draw [black]  (0,2) -- (4,2);
				\draw [black]  (0,-1) -- (4,-1);
				\draw [black] (2,2) -- (4,3);
				\draw [black] (2,2) -- (4,1);
				%
				\draw [dashed, orange] (0,2.2) -> (2,2.2) -> (4,3.2);
				\draw [dashed, red] (0,1.8) -> (2,1.8);
				\draw [dashed, orange] (0,-0.8) -> (2,-0.8) -> (4,-0.8);
				\draw [dashed, red] (2,-1.2) -> (4,-1.2);
				
				\node [orange] at (0.5,2.5) {$\varphi_1$};	
				\node [red] at (0.5,1.5) {$\hsBt\varphi_1$};
			\node [orange] at (2.5,-0.5) {$\varphi_1$};	
				\node [red] at (2.5,-1.5) {$\hsEt\varphi_1$};
						
				
			\end{tikzpicture}
\end{minipage}			
\hfill
\begin{minipage}{0.54\linewidth}
			\begin{tikzpicture}
				%
					\filldraw [gray] (4,-1) circle (2pt)
				(2,-1) circle (2pt)
				(0,-1) circle (2pt)
				(6,0) circle (2pt)
				(6,-1) circle (2pt)
				(6,-2) circle (2pt)	;
				\draw [black]  (0,-1) -- (6,-1);
				\draw [black] (4,-1) -- (6,0);
				\draw [black] (4,-1) -- (6,-2);
				%
				\draw [dashed, orange] (0,-0.8) -> (2,-0.8) -> (4,-0.8);
				\draw [dashed, orange] (2,-1.2) -> (4,-1.2) -> (6,-2.2);
				
				\node [orange] at (1,-0.5) {$\varphi_1$};	
				\node [red] at (2.5,-0.5) {$\hsAt\varphi_1$};
			\node [orange] at (5,-2) {$\varphi_2$};	
				\node [red] at (2.5,-1.5) {$\hsA \varphi_2$};
						
				
			\end{tikzpicture}
\end{minipage}
    \caption{Computation-tree-based semantic variant $\HS_\LinearPast$: future is branching, past is linear, finite and cumulative.}
    \label{fig:CT}
\end{figure}

%% file: Chaps/TOCL17/Imgs/semPicturesLN.tex
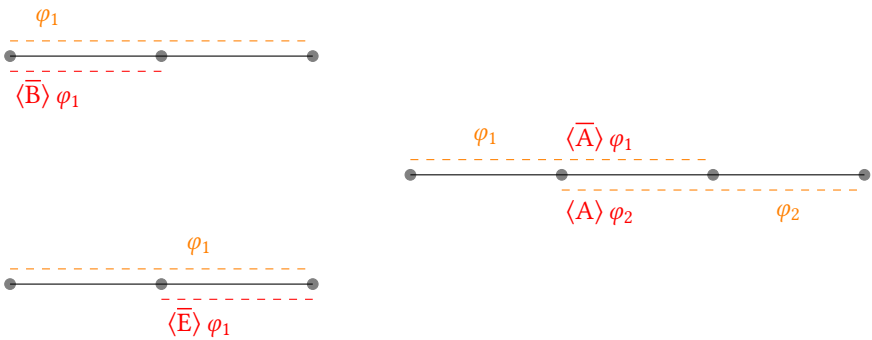
\begin{figure}[tp]
\centering
\begin{minipage}{0.36\linewidth}
\begin{tikzpicture}
				\filldraw [gray] (0,2) circle (2pt)
				(2,2) circle (2pt)
				(4,2) circle (2pt);
				%
					\filldraw [gray] (4,-1) circle (2pt)
				(2,-1) circle (2pt)
				(0,-1) circle (2pt);
				%
				\draw [black]  (0,2) -- (4,2);
				\draw [black]  (0,-1) -- (4,-1);
				%
				%
				\draw [dashed, orange] (0,2.2) -> (2,2.2) -> (4,2.2);
				\draw [dashed, red] (0,1.8) -> (2,1.8);
				\draw [dashed, orange] (0,-0.8) -> (2,-0.8) -> (4,-0.8);
				\draw [dashed, red] (2,-1.2) -> (4,-1.2);
				
				\node [orange] at (0.5,2.5) {$\varphi_1$};	
				\node [red] at (0.5,1.5) {$\hsBt\varphi_1$};
			\node [orange] at (2.5,-0.5) {$\varphi_1$};	
				\node [red] at (2.5,-1.5) {$\hsEt\varphi_1$};
						
				
			\end{tikzpicture}
\end{minipage}
\hfill
\begin{minipage}{0.54\linewidth}
        \begin{tikzpicture}
				%
					\filldraw [gray] (4,-1) circle (2pt)
				(2,-1) circle (2pt)
				(0,-1) circle (2pt)
				(6,-1) circle (2pt);
				%
				\draw [black]  (0,-1) -- (6,-1);
				%
				%
				\draw [dashed, orange] (0,-0.8) -> (2,-0.8) -> (4,-0.8);
				\draw [dashed, orange] (2,-1.2) -> (4,-1.2) -> (6,-1.2);
				
				\node [orange] at (1,-0.5) {$\varphi_1$};	
				\node [red] at (2.5,-0.5) {$\hsAt\varphi_1$};
			\node [orange] at (5,-1.5) {$\varphi_2$};	
				\node [red] at (2.5,-1.5) {$\hsA \varphi_2$};
						
				
			\end{tikzpicture}
\end{minipage}
    \caption{Trace-based semantic variant $\HS_\LinearTime$: neither past nor future are branching}
    \label{fig:LN}
\end{figure}

%% file: Chaps/TOCL17/completepicture.tex
\begin{tikzpicture}[-,>=stealth',shorten >=1pt,auto,semithick,main node/.style={rectangle,draw,inner sep=2pt}]  
\tikzstyle{gray node}=[fill=gray!30]
\node [main node](0) at (0,0) {$\HS_\LinearPast$};
\node [main node](1) at (0,1.5) {$\HS_\LinearTime$};
\node [main node](2) at (0,-1.5) {$\HS_\stat$};
\node [main node](3) at (2.5,0) {finitary $\CTLStar$};
\node [main node](4) at (2.5,1.5) {$\LTL$};
\node [main node](6) at (2.5,-1.5) {$\CTL$};
\node [main node](5) at (5,0) {$\CTLStar$};
\draw [dashed] (0.east) to node {$\equiv$} (3);
\draw [dashed] (1.east) to node {$\equiv$} (4);
\draw [dashed] (3.east) to node {$<$} (5);
\draw [dashed] (0.north) to node [swap] {$\neq$} (1);
\draw [dashed] (0.south) to node {\rotatebox{-90}{$<$}} (2);
\draw [dashed] (2.east) to node [swap] {$\neq$} (6.west);
\draw [dashed] (2) [out=-40,in=270] to node [near end] {$\neq$} (5);
\draw [dashed] (0.south east) to node {$\neq$} (6.west);
\draw [dashed] (2) [out=135,in=225] to node {$\neq$} (1);
\end{tikzpicture}

%% file: Chaps/TOCL17/preliminaries.tex
\section{Preliminaries}\label{sec:backgr}

In the following, let $\Sigma$ be an alphabet and $w$ be a non-empty finite or infinite word over $\Sigma$. We denote by $|w|$ the length of $w$ ($|w|=\infty$ if $w$ is infinite). For all  $i,j\in\Nat$ with $i\leq j$, $w(i)$ denotes the
$i$-th letter of $w$, while $w(i,j)$ denotes the finite subword of $w$ given by $w(i)\cdots w(j)$.
The set of all the finite words over $\Sigma$ is denoted by $\Sigma^*$, and $\Sigma^+$ represents $\Sigma^*\setminus\{\varepsilon\}$, where $\varepsilon$ is the empty word.

Clearly, a trace $\rho$ of a Kripke structure $\Ku=\KuDef$ can be considered as a finite word over $\States$, where $(\rho(i),\rho(i+1))\in\Edges$ for all $0\leq i<|\rho|-1$.
In addition to traces, we define an \emph{infinite path} $\pi$ of $\Ku$ as an infinite word over $\States$ such that $(\pi(i),\pi(i+1))\in \Edges$ for all $i\geq 0$.
Here, unlike the previous chapters, we assume traces and paths to be 0-based (and not 1-based) for notational convenience.

\subsection{Standard point-based temporal logics}\label{sect:PTL}

We start now by recalling the standard propositional temporal logics $\CTLStar$, $\CTL$  and $\LTL$~\cite{EH86,Pnu77}.
Given a set of proposition letters $\Prop$, the formulas $\varphi$ of
$\CTLStar$ are defined as follows:
\[
\varphi ::= \top \mid p \mid \neg \varphi \mid \varphi \wedge \varphi \mid \Next \varphi \mid \varphi \until \varphi \mid \EQ  \varphi ,
\]
where $p\in \Prop$, $\Next$ and $\until$ are the
``next'' and ``until'' temporal modalities,   and $\EQ$ is the
 existential path quantifier.%
\footnote{Hereafter, we denote by $\exists/\forall$ the existential/universal path quantifiers (instead of by the usual E/A), in order not to confuse them with the $\HS$ modalities $\mathsf{E}/\mathsf{A}$.}
 We also use the standard shorthands $\AQ\varphi=\neg\EQ\neg\varphi$ (``universal path quantifier''),
$\Eventually\varphi= \top \until \varphi$ (``eventually'' or ``in the future'') and its
dual $\Always \varphi=\neg \Eventually\neg\varphi$ (``always'' or ``globally'').
We denote by $|\varphi|$ the size of $\varphi$, that is, the number of its subformulas.
The logic $\CTL$ is the  fragment of $\CTLStar$ where each temporal modality is immediately preceded by a path quantifier, whereas  $\LTL$ corresponds to the path-quantifier-free  fragment of $\CTLStar$.

Given a Kripke
structure $\Ku=\KuDef$, an infinite path $\pi$ of
$\Ku$, and a position $i\geq 0$ along $\pi$, the
satisfaction relation $\Ku,\pi,i \models \varphi$ for
$\CTLStar$, written simply $\pi,i \models \varphi$ when $\Ku$ is clear from the context, is defined as follows (Boolean connectives are treated as usual):
\[ \begin{array}{ll}
\pi, i \models p  &  \Leftrightarrow  p \in \Lab(\pi(i)),\\
\pi, i \models \Next \varphi  & \Leftrightarrow   \pi, i+1 \models \varphi ,\\
\pi, i \models \varphi_1\until \varphi_2  &
  \Leftrightarrow  \text{for some $j\geq i$}: \pi, j
  \models \varphi_2,
  \text{ and }  \pi, k \models  \varphi_1 \text{ for all }i\leq k<j,\\
\pi, i \models \EQ \varphi  & \Leftrightarrow \text{for some infinite path } \pi'  \text{ starting from $\pi(i)$, }  \pi', 0 \models \varphi .
\end{array} \]
The MC problem is defined as follows: $\Ku$ is a model of $\varphi$, written $\Ku\models \varphi$, if for all initial infinite paths $\pi$ of $\Ku$, it holds that $\Ku,\pi, 0 \models\varphi$.

We also consider a variant of $\CTLStar$, called \emph{finitary} $\CTLStar$, where the path quantifier $\EQ$ of $\CTLStar$ is replaced by the finitary path
quantifier $\EQF$. In this setting, path quantification ranges over the traces starting from the current state.
The satisfaction relation $\rho, i \models \varphi$, where $\rho$ is a trace and $i$ is a position along $\rho$, is similar to that given for $\CTLStar$ with the only difference of finiteness of paths, and the fact that, for a formula $\Next\varphi$, we have  $\rho, i \models \Next\varphi$ if and only if $i+1<|\rho|$ and $\rho, i+1 \models \varphi$.    A Kripke structure $\Ku$  is a model of a finitary $\CTLStar$ formula if, for each initial trace $\rho$ of  $\Ku$,  it holds that $\Ku,\rho, 0 \models\varphi$.
  
The MC problem for both $\CTLStar$ and $\LTL$ is $\PSPACE$-complete~\cite{DBLP:conf/popl/EmersonL85,Sistla:1985}. It is not difficult to show that, as it happens with finitary $\LTL$~\cite{DBLP:conf/ijcai/GiacomoV13}, MC for finitary $\CTLStar$ is $\PSPACE$-complete as well.

\subsection{Three semantic variants of $\HS$ for MC}\label{sect:3sem}
In this section we formally define the three presented variants of $\HS$ semantics $\HS_\stat$ (state-based), $\HS_\LinearPast$ (computation-tree-based), and $\HS_\LinearTime$ (trace-based) for model checking $\HS$ formulas against Kripke structures.
For each variant, the related (finite) MC problem consists in deciding whether or not a finite Kripke structure is a model of an $\HS$ formula under such a semantic variant, as defined in the following.

\paragraph*{State-based variant $\HS_\stat$.} 
Let us recall the state-based variant,
which is basically the one we have been assuming so far,
where 
an abstract interval model  (Definition~\ref{def:AIM})
is naturally associated with a given Kripke structure $\Ku$ 
by considering the set of intervals as the set 
$\Trk_\Ku$
of traces of $\Ku$. 

\begin{definition}[Abstract interval model induced by a Kripke structure, Definition~\ref{def:inducedmodel}]
The abstract interval model induced by a finite Kripke structure $\Ku=\KuDef$ is
$\mathpzc{A}_\Ku=(\Prop,\mathbb{I},B_\mathbb{I},E_\mathbb{I},\sigma)$, where
\begin{itemize}
    \item $\mathbb{I}=\Trk_\Ku$,
    \item $B_\mathbb{I}=\{(\rho,\rho')\in\mathbb{I}\times\mathbb{I}\mid \rho'\in\Pref(\rho)\}$,
    \item $E_\mathbb{I}=\{(\rho,\rho')\in\mathbb{I}\times\mathbb{I}\mid \rho'\in\Suff(\rho)\}$, and
    \item $\sigma:\mathbb{I}\to 2^\Prop$ such that, for all $\rho\in\mathbb{I}$, 
        \begin{equation*}
            \sigma(\rho)=\bigcap_{s\in\states(\rho)}\Lab(s).
        \end{equation*}
\end{itemize}
\end{definition}

\begin{definition}[State-based $\HS$---$\HS_\stat$, Definition~\ref{def:satkripke} and~\ref{def:MCkripke}]
Let $\Ku$ be a finite Kripke structure and
$\psi$ be an $\HS$ formula.
A trace $\rho\in\Trk_\Ku$ satisfies $\psi$ under the state-based semantic variant,
denoted as $\Ku,\rho\models_\stat \psi$, if and only if $\mathpzc{A}_\Ku,\rho\models \psi$
(Definition~\ref{def:satisfaction}).

Moreover,
$\Ku$ is a model of $\psi$ under the state-based semantic variant, denoted as $\Ku\models_\stat \psi$, if and only if,
for all \emph{initial} traces $\rho\in\Trk_\Ku$, it holds that $\Ku,\rho\models_\stat \psi$.
\end{definition}

\paragraph*{Computation-tree-based semantic variant $\HS_\LinearPast$.}%
We now describe the com\-putation-tree-based semantic variant: to the aim, we consider the abstract interval model \emph{induced by the computation tree} of a Kripke structure (this will be formally defined in the following), and basically  proceed as in the previous case. 

We start by introducing the notion of \emph{$D$-tree structure}, namely, an infinite tree-shaped Kripke structure with branches over a set $D$ of directions.

\begin{definition}[$D$-tree structure] Given a set $D$ of directions,
a \emph{$D$-tree structure} over a set of proposition letters $\Prop$ is a Kripke structure $\Ku=\KuDef$
such that $\sinit\in D$, $\States$ is a prefix closed subset of $D^{+}$, and $\Edges$ is the set of pairs $(s,s')\in \States\times \States$ such that there exists $d\in D$
for which $s'=s\cdot d$ (note that $\Edges$ is completely specified by $\States$). The states of a $D$-tree structure are called \emph{nodes}.
\end{definition}

A Kripke structure $\Ku=\KuDef$ induces an $\States$-tree structure, called the \emph{computation tree of $\Ku$}, denoted by
$\mathpzc{C}(\Ku)$, which is obtained by unwinding $\Ku$ 
from the initial state (note that the directions are the set of states of $\Ku$).
Formally, $\mathpzc{C}(\Ku)= (\Prop,\Trk_\Ku^{0}, \Edges',\Lab',\sinit)$, where the set of nodes is the set of initial traces of
$\Ku$---hereafter denoted as $\Trk_\Ku^0$---and, for all $\rho,\rho'\in \Trk_\Ku^{0}$, $\Lab'(\rho)=\Lab(\lst(\rho))$ and $(\rho,\rho')\in \Edges'$ if and only if
$\rho'=\rho\cdot s$ for some $s\in \States$.
See Figure~\ref{fig:unr} for an example.

\input{Chaps/TOCL17/Imgs/unravelling.tex}

Notice that since each state in a computation tree has a unique predecessor (with the exception of the initial state), this $\HS$ variant enforces a linear reference in the past.

\begin{definition}[Computation-tree-based $\HS$---$\HS_\LinearPast$]
A finite Kripke structure $\Ku$ is a model of an $\HS$ formula $\psi$ under the computation-tree-based semantic variant, denoted as $\Ku\models_\LinearPast \psi$, if and only if
$\mathpzc{C}(\Ku)\models_\stat \psi$.
\end{definition}

\paragraph*{Trace-based semantic variant $\HS_\LinearTime$.}
Finally, we introduce the trace-based semantic variant, which exploits the interval structures induced by the infinite paths of a Kripke structure, as defined in the following.

We recall that,
given a strict partial ordering $\mathbb{S}=(\mathpzc{S},<)$, an \emph{interval} in $\mathbb{S}$ is an ordered pair
$[x,y]$, such that $x,y\in \mathpzc{S}$ and $x\leq y$, which represents the subset of $\mathpzc{S}$ given by all points $z\in \mathpzc{S}$ such
that $x\leq z\leq y$. We denote by $\mathbb{I}(\mathbb{S})$ the set of intervals in $\mathbb{S}$.

The following notion has already been presented in Section~\ref{sec:preliminaries}, but here we use (for technical convenience) a different notation.
\begin{definition}[Interval structure]
An \emph{interval structure (or interval model)} $\mathpzc{IS}$ over  $\Prop$ is a pair $\mathpzc{IS}=(\mathbb{S},\sigma)$ such that $\mathbb{S}=(\mathpzc{S},<)$ is a strict partial ordering
and $\sigma: \mathbb{I}(\mathbb{S}) \to 2^\Prop$ is a labeling function assigning a set of proposition letters
to each interval in $\mathbb{S}$.
\end{definition}

The next definition shows how to derive an abstract interval model from an interval structure,
allowing us to interpret $\HS$ over interval structures.

\begin{definition}[Abstract interval model induced by an interval structure] 
An interval structure 
$\mathpzc{IS}=(\mathbb{S},\sigma)$ over $\Prop$, where $\mathbb{S}=(\mathpzc{S},<)$ is a strict partial ordering, induces the abstract interval model
\[
\mathpzc{A}_{\mathpzc{IS}}=(\Prop,\mathbb{I}(\mathbb{S}), B_{\mathbb{I}(\mathbb{S})},E_{\mathbb{I}(\mathbb{S})},\sigma),
\]
where
 $[x,y] \, B_{\mathbb{I}(\mathbb{S})}\, [v,z]$ if and only if $x=v$ and $z<y$, and
 $[x,y] \, E_{\mathbb{I}(\mathbb{S})}\, [v,z]$ if and only if $y=z$ and $x<v$.
\end{definition}
Given an interval $I$ and an $\HS$ formula $\psi$, we write $\mathpzc{IS},I\models \psi$ to mean that $\mathpzc{A}_{\mathpzc{IS}},I\models \psi$.
We now interpret $\HS$ over the infinite paths of a Kripke structure by mapping them into interval structures.
\begin{definition}[Interval structure induced by an infinite path]\label{def:inducedPathIntervalStructure}
For a finite  Kripke structure $\Ku=\KuDef$ and an infinite path $\pi=\pi(0)\pi(1)\cdots$ of $\Ku$,
the \emph{interval structure induced by $\pi$} is
$\mathpzc{IS}_{\Ku,\pi}=((\Nat,<),\sigma)$, where
 for each interval $[i,j]$, we have  $\sigma([i,j])=\bigcap_{h=i}^{j}\Lab(\pi(h))$.
\end{definition}

\begin{definition}[Trace-based $\HS$---$\HS_\LinearTime$]
A finite Kripke structure $\Ku$ is a model of an $\HS$ formula $\psi$ under the trace-based semantic variant, denoted as $\Ku\models_\LinearTime \psi$, if and only if,
for each initial infinite path $\pi$  and for each initial interval $[0,i]$, it holds that  $\mathpzc{IS}_{\Ku,\pi},[0,i]\models \psi$. 
\end{definition}

%% file: Chaps/TOCL17/Imgs/unravelling.tex
\begin{figure}[tp]
    \centering
\begin{tikzpicture}[->,>=stealth,thick,shorten >=1pt,auto,node distance=1.6cm,every node/.style={ellipse,draw}]
			\node[double] (v0) {$s_0$};
			\node (v1) at (0,-1) {$s_0s_1$};
			\node (v01) at (-1,-2) {$s_0s_1s_0$};
			\node (v11) at (1,-2) {$s_0s_1s_1$};
			\node (v02) at (-1,-3) {$s_0s_1s_0s_1$};
			\node (v022) at (1,-3) {$s_0s_1s_1s_0$};
			\node (v12) at (3,-3) {$s_0s_1s_1s_1$};
			

			\node (v30) [draw=none, below =0.2cm of v02] {\Large $\cdots$};
			\node (v31) [draw=none, below left =0.21cm and 0.09cm of v12] {\Large $\cdots$};

			\draw (v0) to (v1);
			\draw (v1) to (v01); \draw (v1) to (v11);
			\draw (v01) to (v02);
			\draw (v11) to (v022);
			\draw (v11) to (v12);
		\end{tikzpicture}
	\vspace{-0.3cm}
    \caption{Computation tree $\mathpzc{C}(\mathpzc{K})$ of a finite Kripke structure $\Ku=\KuDef$ with $\States=\{s_0,s_1\}$ and $\Edges=\{(s_0,s_1),(s_1,s_1),(s_1,s_0)\}$.}
    \label{fig:unr}
\end{figure}

%% file: Chaps/TOCL17/expressiveness.tex
With this definition we conclude the presentation of the three semantic variants of $\HS$.
In the next sections  we will compare the expressiveness of the logics $\HS_{\stat}$, $\HS_{\LinearPast}$, $\HS_{\LinearTime}$, $\LTL$, $\CTL$, and $\CTLStar$ when interpreted over finite Kripke structures.

Given two logics $L_1$ and $L_2$, and two formulas $\varphi_1\in L_1$ and $\varphi_2\in L_2$, we say that   $\varphi_1$ in $L_1$ is \emph{equivalent} to  $\varphi_2$ in $L_2$ if, for every finite Kripke structure   $\Ku$, 
$\Ku$ is a model of $\varphi_1$ in $L_1$ if and only if $\Ku$ is a model of $\varphi_2$ in $L_2$. 
We say that \emph{$L_2$ is subsumed by $L_1$}, denoted as $L_1\geq L_2$, if for each formula $\varphi_2\in L_2$, there exists a formula $\varphi_1\in L_1$ such that $\varphi_1$ in $L_1$ is equivalent to $\varphi_2$
in $L_2$. Moreover, $L_1$ is \emph{as expressive as} $L_2$ (or, $L_1$  and $L_2$ have \emph{the same expressive power}), written $L_1\equiv L_2$, if both $L_1\geq L_2$ and $L_2\geq L_1$. We say that  $L_1$ is \emph{(strictly) more expressive than} $L_2$ if $L_1\geq L_2$ and $L_2\not\geq L_1$. Finally, $L_1$ and $L_2$ are \emph{expressively incomparable} if both $L_1\not\geq L_2$ and $L_2\not\geq L_1$.

\input{Chaps/TOCL17/vendingMach.tex}

\section{Equivalence between $\LTL$ and $\HS_\LinearTime$}\label{sec:CharacterizezionOfLeniarTimeHS}

In this section, we show that $\HS_\LinearTime$ is as expressive as $\LTL$ even for small syntactical fragments of $\HS_\LinearTime$. To this end, we exploit the well-known equivalence between $\LTL$ and the first-order fragment of monadic second-order logic over infinite words ($\FO$ for short). 

Recall that, given a countable set $\{x,y,z,\ldots\}$ of (position) variables, the $\FO$ formulas $\varphi$ over a set of proposition letters $\Prop=\{p,\ldots \}$
are defined as:
\[
\varphi ::= \top \mid  p(x) \mid x\leq y \mid x <y  \mid \neg\varphi \mid \varphi \wedge \varphi
 \mid \exists x. \varphi\; .
 \]

We interpret $\FO$ formulas $\varphi$ over infinite paths $\pi$ of Kripke structures $\Ku=\KuDef$. 
Given a variable \emph{valuation} $g$, assigning to each variable  a position $i\geq 0$,  the satisfaction relation
$(\pi,g)\models \varphi$ corresponds to the standard satisfaction relation $(\Lab(\pi),g)\models \varphi$, where $\Lab(\pi)$ is the infinite word over $2^{\Prop}$ given by $\Lab(\pi(0))\Lab(\pi(1))\cdots$. 
More precisely, 
$(\pi,g)\models \varphi$ is inductively defined as follows
 (we omit the standard rules for the Boolean connectives):
\[
\begin{array}{ll}
 (\pi,g)\models p(x) & \Leftrightarrow  p \in \Lab(\pi(g(x))),\\
 (\pi,g)\models x\ op\ y & \Leftrightarrow  g(x)\ op\ g(y), \mbox{ for } op \in \{<, \leq\},\\
   (\pi,g)\models \exists x. \varphi & \Leftrightarrow  (\pi,g[x\leftarrow i])\models \varphi \text{ for some } i\geq 0 ,
\end{array}
\]
where $g[x\leftarrow i](x)=i$ and $g[x\leftarrow i](y)=g(y)$ for $y\neq x$. Note that the satisfaction relation  depends only on the values assigned to the variables occurring free in the given formula $\varphi$.
We write $\pi\models \varphi$ to mean that $(\pi,g_0)\models \varphi$, where $g_0(x)=0$ for each variable $x$. An $\FO$ sentence is a formula with no free variables.  

The following is a well-known result (Kamp's theorem~\cite{Kamp}).

\begin{proposition}\label{theo:FromFOtoLTL} Given an $\FO$ sentence $\varphi$ over $\Prop$, one can construct an $\LTL$ formula $\psi$ such that, for all Kripke structures $\Ku$ over $\Prop$ and infinite paths $\pi$, it holds that
$\pi \models \varphi$ if and only if $\pi,0\models \psi$.
\end{proposition}

Given a $\HS_\LinearTime$ formula $\psi$, we now construct  an 
$\FO$ sentence $\psi_\FO$ such that, for all Kripke structures $\Ku$, 
$\Ku\models_\LinearTime \psi$ if and only if for each initial infinite path $\pi$ of $\Ku$, $\pi\models \psi_\FO$.

We start by defining a mapping $h$ assigning to each triple $(\varphi,x,y)$, consisting of a $\HS$ formula $\varphi$ and two distinct position variables $x,y$, an $\FO$ formula having as free variables $x$ and $y$.
The mapping $h$ returns the $\FO$ formula defining the semantics of the $\HS$ formula $\varphi$ interpreted over an interval bounded by the positions $x$ and $y$. 
 
 The function $h$ is homomorphic with respect to the Boolean connectives, and is defined for proposition letters and modal operators as follows (here $z$ is a fresh position variable):
\[ \begin{array}{ll}
h(p,x,y) & = \forall z.((z\geq x \wedge z\leq y) \rightarrow p(z)),\\
h(\hsE\psi,x,y) & = \exists z.(z> x \wedge z\leq y \wedge h(\psi,z,y)),\\
h(\hsB\psi,x,y) & = \exists z.(z\geq x  \wedge z< y \wedge h(\psi,x,z)),\\
h(\hsEt\psi,x,y) & = \exists z.(z< x \wedge h(\psi,z,y)),\\
h(\hsBt\psi,x,y) & = \exists z.(z> y  \wedge  h(\psi,x,z)).
\end{array} \]
It is worth noting that homogeneity plays a crucial role in the definition of $h(p,x,y)$ (without it, a binary predicate would be necessary to encode the truth of $p$ over the interval $[x,y]$).

Given a Kripke structure $\Ku$, an infinite path $\pi$, an interval of positions $[i,j]$, and an $\HS_\LinearTime$ formula $\psi$, by a straightforward induction on the structure of
$\psi$, we can show that 
$\mathpzc{IS}_{\Ku,\pi},[i,j]\models \psi$
if and only if $(\pi,g)\models h(\psi,x,y)$ for any valuation such that $g(x)= i$ and $g(y)=j$. 

Let us consider the
$\FO$ sentence $h(\psi)$ given by $\exists x ((\forall z.  z\geq x)\wedge \forall y.  h(\psi,x,y))$. Clearly $\Ku\models_\LinearTime \psi$ if and only if for each initial infinite path $\pi$ of $\Ku$, $\pi\models h(\psi)$.  By Proposition~\ref{theo:FromFOtoLTL}, it follows that one can construct an $\LTL$ formula $h'(\psi)$ such that $h'(\psi)$ in $\LTL$ is equivalent to $\psi$ in $\HS_\LinearTime$. 
Thus, we obtain the following expressiveness containment.
 
\begin{theorem}\label{theo:HSTracedBasedIsINLTL} $\LTL\geq \HS_\LinearTime$.
\end{theorem}

Now we show that also the converse containment holds, that is, $\LTL$ can be translated into $\HS_\LinearTime$.
Actually, it is worth noting that for such a purpose the fragment $\A\B$ of $\HS_\LinearTime$, featuring only modalities for $\hsA$ and $\hsB$, is expressive enough.

\begin{theorem}\label{theo:FromLTLTOSmallFragments} Given an $\LTL$ formula $\varphi$, one can construct in linear time an $\AB$  formula $\psi$ such that $\varphi$ in $\LTL$ is equivalent to $\psi$ in $\AB_\LinearTime$.
\end{theorem}
\begin{proof}
Let $f: \LTL \to \AB$ be the mapping, homomorphic with respect to the Boolean connectives, defined as follows:  
\[\begin{array}{ll}
f(p) & = p, \text{ for each } p\in\Prop,\\
f(\Next\psi) & = \hsA (\Length_{2}\wedge \hsA(\Length_1\wedge f(\psi))),\\
f(\psi_1\until\psi_2) & = \hsA \bigl(\hsA(\Length_{1}\wedge f(\psi_2))\wedge\hsBu(\hsA(\Length_{1}\wedge f(\psi_1))\bigr).
\end{array}\]

Given a Kripke structure $\Ku$, an infinite path $\pi$, a position $i\geq 0$, and an $\LTL$ formula $\psi$, by a straightforward induction on the structure of
$\psi$ we can show that $\pi,i\models \psi$ if and only if
$\mathpzc{IS}_{\Ku,\pi},[i,i]\models f(\psi)$.
Hence $\Ku\models \psi$ if and only if $\Ku\models_\LinearTime \Length_1\rightarrow f(\psi)$.
\end{proof}

The next corollary follows immediately from Theorem~\ref{theo:HSTracedBasedIsINLTL} and Theorem~\ref{theo:FromLTLTOSmallFragments}.
\begin{corollary}\label{cor:HSLinearCharacterization} $\HS_\LinearTime$ and $\LTL$ have the same expressive power.
\end{corollary}



While there is no difference in the expressive power between $\LTL$ and $\HS_\LinearTime$, things change if we consider succinctness. Whereas Theorem~\ref{theo:FromLTLTOSmallFragments} shows that it is possible to convert any $\LTL$ formula into an equivalent $\HS_\LinearTime$ one in linear time, the following theorem holds. 

\begin{theorem}\label{theo:succinctnessHSlin} $\HS_\LinearTime$ is at least exponentially more succinct than $\LTL$.
\end{theorem}
\begin{proof}
To prove the statement, it suffices to provide an $\HS_\LinearTime$ formula $\psi$ for which there exists no $\LTL$ equivalent formula whose size is polynomial in $|\psi|$.

To this end, we restrict our attention to the fragment $\B\E_\LinearTime$. Since modalities $\hsB$ and $\hsE$ only allow one to \lq\lq move\rq\rq{} from an interval to its subintervals, $\B\E_\LinearTime$ actually coincides with $\B\E_\stat$, whose MC is hard for $\EXPSPACE$ (Section~\ref{sec:BEhard}). Thus, in particular, it is possible to encode by means of a $\B\E_\LinearTime$ formula $\psi_{\text{cpt}}$ the (unique) computation of a deterministic Turing machine using $b(n)\in O(2^n)$ bits that, when executed on input $0^n$, for some natural number $n\geq 1$, counts in binary from $0$ to $2^{2^n}-1$, by repeatedly summing 1, and finally accepts. The length of $\psi_{\text{cpt}}$ is \emph{polynomial} in $n$, and the unique trace which satisfies it (that is, that encodes such a computation) has length $\ell(n)\geq b(n)\cdot 2^{2^n}$. 


Conversely, it is known that $\LTL$ features a \emph{single-exponential small-model property}~\cite{DBLP:books/cu/Demri2016}, 
stating that, for every \emph{satisfiable} $\LTL$ formula $\varphi$, there are $u,w\in S^*$ with $|u|\leq 2^{|\varphi|}$ and $|w|\leq |\varphi|\cdot 2^{|\varphi|}$, such that $u\cdot w^\omega ,0 \models \varphi$. 
This allows us to conclude (by a straightforward contradiction argument) that there is no polynomial-length (with respect to $|\psi_{\text{cpt}}|$, and thus to $n$) $\LTL$ formula that can encode the aforementioned computation. An \emph{exponential-length} $\LTL$ formula would be needed for such an encoding.
\end{proof}

Exactly the same argument can be used to show that $\HS_\LinearTime$ is at least exponentially more succinct than the extension of $\LTL$ with past modalities (denoted in the following as $\LTLP$)~\cite{DBLP:journals/igpl/LichtensteinP00}.

\section{A characterization of $\HS_\LinearPast$}\label{sec:characterizationHSLinearPast}

In this section, we will focus our attention on the computation-tree-based semantic variant $\HS_\LinearPast$, showing that  it is as expressive as \emph{finitary} $\CTLStar$. As a matter of fact, the result can be proved to hold already for the syntactical fragment $\A\B\E$ which does not feature inverse modalities. In addition, we show that $\HS_\LinearPast$ is subsumed by $\CTLStar$.

\subsection{From finitary $\CTLStar$ to $\HS_\LinearPast$}
We first show that finitary $\CTLStar$ is subsumed by $\HS_\LinearPast$. 
As a preliminary fundamental step, we prove that,  
when interpreted over finite words, the $\BE$ fragment of $\HS$ and $\LTL$ 
define the same class of finitary languages (Theorem \ref{theoremFromLTLtoBEoverFiniteWords}).

For an $\LTL$ formula $\varphi$ with proposition letters
over  an alphabet $\Sigma$ (in our case $\Sigma$ is  
 $2^{\Prop}$), let us denote by $\Lang_\act(\varphi)$ 
the set of non-empty finite words over $\Sigma$ satisfying $\varphi$ under the standard action-based semantics of $\LTL$, interpreted over finite words (see \cite{vardi1996automata}).
A similar notion can be given for $\BE$ formulas $\varphi$ with proposition letters in $\Sigma$ (under the homogeneity assumption).
Then, $\varphi$ denotes a language, written $\Lang_\act(\varphi)$, of non-empty finite words over $\Sigma$ inductively defined as:
\begin{itemize}
  \item $\Lang_\act(a)= a^{+} $, for each $a\in\Sigma$ (this definition reflects homogeneity);
  \item $\Lang_\act(\neg \varphi)= \Sigma^{+}\setminus \Lang_\act(\varphi)$;
  \item $\Lang_\act(\varphi_1\wedge \varphi_2)=\Lang_\act(\varphi_1)\cap  \Lang_\act(\varphi_2)$;
  \item $\Lang_\act(\hsB \varphi)= \{w\in \Sigma^{+}\mid \Pref(w)\cap \Lang_\act(\varphi)\neq \emptyset\}$;
    \item $\Lang_\act(\hsE \varphi)= \{w\in \Sigma^{+}\mid \Suff(w)\cap \Lang_\act(\varphi)\neq \emptyset\}$.
\end{itemize}

We prove that, under the action-based semantics, $\BE$ formulas and $\LTL$ formulas define the same class of finitary languages.

To prove that the finitary languages defined by $\LTL$ formulas are subsumed by 
those defined by $\BE$ formulas, we exploit an algebraic condition introduced by Wilke in~\cite{stacs/Wilke99}, called \emph{$\LTL$-closure}, which gives, for a class of finitary languages, a sufficient condition to guarantee the inclusion of the class of $\LTL$-definable languages.
The converse inclusion, that is, the class of finitary languages defined by the fragment $\BE$ is subsumed by that defined by $\LTL$, can be proved by a technique similar to that used in Section~\ref{sec:CharacterizezionOfLeniarTimeHS}, and is thus omitted.

We start by considering the former inclusion recalling from~\cite{stacs/Wilke99} a sufficient condition for a class of finitary languages to include the class of finitary languages which are $\LTL$-definable.

  \begin{definition}[\LTL-closure]\label{def:LTLClosure} A class $\mathcal{C}$  of languages of finite words over finite alphabets is \emph{$\LTL$-closed} if and only if the following conditions are satisfied, where $\Sigma$ and $\Delta$ are finite alphabets, $b\in \Sigma$ and $\Gamma=\Sigma\setminus\{b\}$:
  \begin{enumerate}
    \item $\mathcal{C}$ is closed under language complementation and intersection;
    \item if $\Lang \in \mathcal{C}$ with $\Lang\subseteq \Gamma^{+}$, then  $\Sigma^* b\Lang$, $\Sigma^* b(\Lang+\varepsilon)$,
    $ \Lang b \Sigma^* $, and $ (\Lang+\varepsilon) b \Sigma^* $ are in  $\mathcal{C}$;
    \item Let $U_0= \Gamma^{*}b$, $h_0:U_0 \rightarrow \Delta$, and $h:U_0^{+} \rightarrow \Delta^{+}$ be defined by
    $h(u_0u_1\cdots u_n)=h_0(u_0)\cdots h_0(u_n)$. Assume that for each $d\in \Delta$, the language $\Lang_d =\{u\in \Gamma^{+}\mid h_0(ub)= d\}$
    is in $\mathcal{C}$. Then for each language $\Lang\!\in\! \mathcal{C}$ such that $\Lang\!\subseteq\! \Delta^{+}$,
    the language $\Gamma^{*}b h^{-1}(\Lang)\Gamma^{*}$ is in $\mathcal{C}$.
  \end{enumerate}
  \end{definition}
  
\begin{figure}
    \centering
    \includegraphics[width=\linewidth]{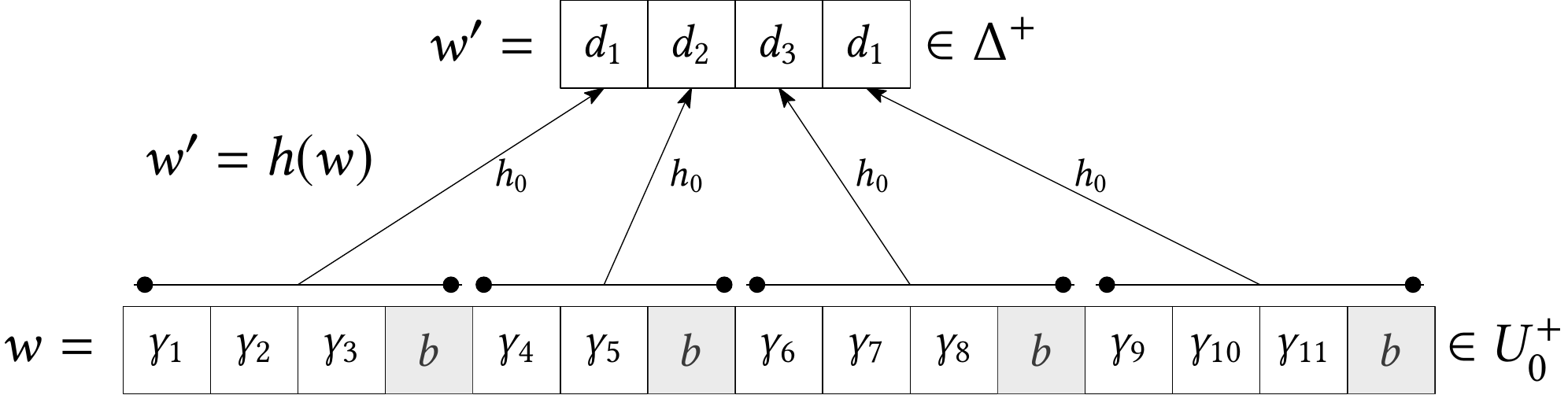}
    \caption{ 
            Visual description of condition 3 of Definition~\ref{def:LTLClosure} (\LTL-closure). 
    }
    \label{fig:LTLclos3}
\end{figure}
  
In Figure~\ref{fig:LTLclos3}, we graphically  depict (3.) of the definition of \LTL-closure. In the proposed example, we have:
              $(i)$~for all $i$, $d_i\in \Delta$ and $\gamma_i\in\Gamma$,
              $(ii)$~$w= (\gamma_1\gamma_2 \gamma_3 b) ( \gamma_4\gamma_5 b) ( \gamma_6\gamma_7 \gamma_8 b)( \gamma_9\gamma_{10} \gamma_{11} b)\in U_0^4$,
              $(iii)$~$w'=h(w)=h_0(\gamma_1\gamma_2 \gamma_3 b)h_0(\gamma_4\gamma_5 b)\allowbreak h_0(\gamma_6\gamma_7 \gamma_8 b)h_0(\gamma_9\gamma_{10} \gamma_{11} b)=d_1 d_2 d_3 d_1 \in\Delta^4$.
            For instance, $\gamma_1\gamma_2\gamma_3,\gamma_9\gamma_{10} \gamma_{11}\in L_{d_1}$ and $\gamma_4\gamma_5\in L_{d_2}$.

The following result holds~\cite{stacs/Wilke99}.
  \begin{theorem}\label{theo:CharacterizationFinitaryLTL} Any $\LTL$-closed class  $\mathcal{C}$ of finitary languages includes the class of  $\LTL$-definable finitary languages.
  \end{theorem}

Therefore, to prove that the finitary languages defined by  $\BE$ formulas subsume those defined by $\LTL$, as stated by Theorem~\ref{theoremFromLTLtoBEoverFiniteWords} below, it suffices to prove that \emph{the class of finitary languages definable by $\BE$ formulas is $\LTL$-closed}, and to apply Theorem~\ref{theo:CharacterizationFinitaryLTL}.
We observe that, by definition, the class of $\BE$-definable languages is obviously closed under language complementation and intersection ((1.) of Definition~\ref{def:LTLClosure}). The fulfillment of (2.) and (3.) of Definition~\ref{def:LTLClosure} is then proved by the two following Lemma~\ref{lemma1:CharacterizationFinitaryLTL} and~\ref{lemma2:CharacterizationFinitaryLTL}, respectively, whose proofs are in Appendix \ref{proof:lemma1:CharacterizationFinitaryLTL} and \ref{proof:lemma2:CharacterizationFinitaryLTL}.


 \begin{lemma}\label{lemma1:CharacterizationFinitaryLTL} Let $\Sigma$ be a finite alphabet,
  $b\in \Sigma$, $\Gamma=\Sigma\setminus\{b\}$, $\Lang\subseteq \Gamma^{+}$, and $\psi$ be a $\BE$ formula over $\Gamma$ such that $\Lang_\act(\psi)=\Lang$. Then,
  there are $\BE$ formulas defining (under the action-based semantics) the languages
   $b\Lang$, $\Sigma^* b\Lang$, $\Sigma^* b(\Lang+\varepsilon)$,
   $\Lang b$, $ \Lang b \Sigma^* $, $ (\Lang+\varepsilon) b \Sigma^* $, and $b \Lang b$.
  \end{lemma}

  \begin{lemma}\label{lemma2:CharacterizationFinitaryLTL}Let $\Sigma$ and $\Delta$ be finite alphabets,
  $b\in \Sigma$, $\Gamma=\Sigma\setminus\{b\}$, $U_0= \Gamma^{*}b$, $h_0:U_0 \rightarrow \Delta$ and $h:U_0^{+} \rightarrow \Delta^{+}$ be defined by
    $h(u_0u_1\cdots u_n)=h_0(u_0)\cdots h_0(u_n)$. Assume that, for each $d\in \Delta$, there is a $\BE$ formula capturing the language $\Lang_d =\{u\in \Gamma^{+}\mid h_0(ub)= d\}$.  Then, for each $\BE$ formula $\varphi$ over $\Delta$, one can construct a $\BE$ formula over $\Sigma$
    capturing
    the language $\Gamma^{*}b h^{-1}(\Lang_\act(\varphi))\Gamma^{*}$.
  \end{lemma}

Since, by Lemma~\ref{lemma1:CharacterizationFinitaryLTL} and~\ref{lemma2:CharacterizationFinitaryLTL}, the class of finitary languages definable by $\BE$ formulas is $\LTL$-closed, by Theorem~\ref{theo:CharacterizationFinitaryLTL} we finally get the following result.

\begin{theorem}\label{theoremFromLTLtoBEoverFiniteWords} Let $\varphi$ be an $\LTL$ formula over a finite alphabet $\Sigma$. Then there exists a $\BE$ formula
 $\varphi_\HS$ over $\Sigma$ such that $\Lang_\act(\varphi_\HS) =\Lang_\act(\varphi)$.
\end{theorem}

 The result of Theorem~\ref{theoremFromLTLtoBEoverFiniteWords} is used to prove that finitary $\CTLStar$ is subsumed by the fragment $\A\B\E$ under the state-based semantics.

  \begin{theorem}\label{theo:FromFInitaryCTLStarToFutureHSBranching} Let $\varphi$ be a finitary $\CTLStar$ 
  formula over $\Prop$. Then there is an $\A\B\E$ formula $\varphi_\HS$ over $\Prop$ such that, for all
  Kripke structures $\Ku$ over $\Prop$ and traces $\rho$, we have $\Ku,\rho,0\models \varphi$ if and only if  $\Ku,\rho\models_\stat \varphi_\HS$.
  \end{theorem}
  \begin{proof} The proof is by induction on the nesting depth of modality $\EQF$ in $\varphi$. In the base case, $\varphi$ is a finitary $\LTL$ formula over $\Prop$. Since what we need to deal with it is just the first part of the work we have to do for the inductive step, it is omitted and only the inductive step is detailed. 
  
  Let $H$ be the non-empty set of subformulas of $\varphi$ of the form $\EQF \psi$ which do not occur in the scope of the path quantifier $\EQF$, that is, the $\EQF \psi$ formulas which are maximal with respect to the nesting depth of modality $\EQF$. Then, $\varphi$ can be seen as an $\LTL$ formula over the extended set of proposition letters
  $\overline{\Prop}=\Prop\cup H$. Let $\Sigma = 2^{\overline{\Prop}}$ and $\overline{\varphi}$ be the $\LTL$ formula over $\Sigma$
  obtained from $\varphi$ by replacing each occurrence of $p \in \overline{\Prop}$ in $\varphi$ with the formula
  $\bigvee_{P\in \Sigma \; : \; p\in P} P$, according to the $\LTL$ action-based semantics. 
  
  Given a Kripke structure  $\Ku$ over $\Prop$ with labeling $\Lab$  and a trace $\rho$ of $\Ku$,
  we denote by $\rho_H$ the finite word over $2^{\overline{\Prop}}$ of length $|\rho|$ defined as \[\rho_H(i)= \Lab(\rho(i))\cup \{\EQF\psi\in H\mid\Ku,\rho,i\models \EQF\psi\},\] for all $i\in [0,|\rho|-1]$.
  One can easily prove by structural induction on $\overline{\varphi}$ that $\Ku,\rho,0\models \varphi$ if and only if $\rho_H\in \Lang_\act(\overline{\varphi})$.
  By Theorem~\ref{theoremFromLTLtoBEoverFiniteWords}, there exists a $\BE$ formula $\overline{\varphi}_\HS$ over $\Sigma$ such that
  $\Lang_\act(\overline{\varphi})= \Lang_\act(\overline{\varphi}_\HS)$.

Now, by the inductive hypothesis, for each formula   $\EQF \psi\in H$, there exists an $\A\B\E$ formula $\psi_\HS$ such that,
for all Kripke structures $\Ku$ and traces $\rho$ of $\Ku$, we have
$\Ku,\rho,0\models \psi$ if and only if $\Ku,\rho\models_\stat \psi_\HS$.
Since $\rho$ is arbitrary, 
$\Ku,\rho,i\models \EQF\psi$ if and only if $\Ku,\rho(i,i),0\models \EQF\psi$, if and only if $\Ku,\rho(i,i)\models_\stat \hsA\psi_\HS$,
for each $i\geq 0$.

Let $\varphi_\HS$ be the $\A\B\E$ formula over $\Prop$ obtained from the $\BE$ formula $\overline{\varphi}_\HS$ by replacing each occurrence of
$P\in \Sigma$ in $\overline{\varphi}_\HS$ with the formula
\[
\hsGu\Big(\Length_1 \longrightarrow  \smashoperator[r]{\bigwedge_{\EQF\psi\in H\cap P}} \hsA\psi_\HS \;\wedge  \smashoperator[r]{\bigwedge_{\EQF\psi\in H\setminus P}}\neg \hsA\psi_\HS \;\wedge \smashoperator[r]{\bigwedge_{p\in \Prop\cap P}} p\;\wedge  \smashoperator[r]{\bigwedge_{p\in \Prop\setminus P}} \neg p\Big).
\]

Since for all $i\geq 0$ and $\EQF\psi\in H$, we have $\Ku,\rho,i\models \EQF\psi$ if and only if  $\Ku,\rho(i,i)\models_\stat \hsA\psi_\HS$, it is possible to prove by a straightforward induction
on the structure of $\overline{\varphi}_\HS$ that, for any Kripke structure $\Ku$ and trace $\rho$ of $\Ku$, we have
$\Ku,\rho\models_\stat \varphi_\HS$ if and only if $\rho_H\in \Lang_\act(\overline{\varphi}_\HS)$.

   Therefore, since $\Ku,\rho,0\models \varphi$ if and only if $\rho_H\in \Lang_\act(\overline{\varphi})$ and
   $\Lang_\act(\overline{\varphi})= \Lang_\act(\overline{\varphi}_\HS)$, then $\Ku,\rho,0\models \varphi$ if and only if  $\Ku,\rho\models_\stat \varphi_\HS$, for any Kripke structure
 $\Ku$ and trace $\rho$ of $\Ku$. 
This concludes the proof of the theorem.
  \end{proof}

  Since the fragment $\A\B\E$ of $\HS$ does not feature any modalities unravelling a Kripke structure backward (namely, $\hsAt$ and $\hsEt$), the computation-tree-based semantics coincides with the state-based semantics (recall Figure~\ref{fig:ST} and \ref{fig:CT}), and thus the following corollary immediately follows from Theorem~\ref{theo:FromFInitaryCTLStarToFutureHSBranching}. 
  \begin{corollary}\label{cor:FromFInitaryCTLStarToFutureHSBranching} Finitary $\CTLStar$ is subsumed by both $\HS_\stat$ and $\HS_\LinearPast$.
  \end{corollary}

\subsection{From $\HS_\LinearPast$ to finitary $\CTLStar$}
We show now that $\HS_\LinearPast$ is subsumed by both $\CTLStar$ and its finitary variant.
To prove this result,  we first introduce a hybrid and linear-past extension of
$\CTLStar$, called \emph{hybrid} $\CTLStarLP$, and its finitary variant, called \emph{finitary hybrid} $\CTLStarLP$.

Besides standard modalities, hybrid logics make use of explicit variables and quantifiers that bind them~\cite{BS98}.
Variables and binders allow us to easily mark points in a path, which will be considered as starting and ending points of intervals, thus permitting a natural encoding of $\HS_\LinearPast$. Actually, we will show that the restricted use of variables and binders exploited in our encoding does not increase the expressive power of (finitary) $\CTLStar$ (as it happens for an unrestricted use), thus proving the desired result. We start defining \emph{hybrid} $\CTLStarLP$.

For a countable set $\{x,y,z,\ldots\}$ of (position) variables, the set of  formulas $\varphi$ of
hybrid $\CTLStarLP$  over $\Prop$ is defined as follows:
\[
\varphi ::= \top \mid p \mid x \mid \neg \varphi \mid \varphi \vee \varphi \mid \Downarrowx.\varphi \mid \Next \varphi\mid \varphi \until \varphi \mid \Next^{-} \varphi\mid \varphi \until^{-} \varphi\mid \EQ  \varphi,
\]
where $\Next^{-}$ (``previous'') and $\until^{-}$ (``since'') are the  past counterparts of the ``next'' and ``until'' modalities $\Next$ and $\until$, and $\Downarrowx$ is the \emph{downarrow binder operator}~\cite{BS98}, which binds $x$ to the current position along the given initial infinite path. We also use the standard shorthands $\Eventually^{-}\varphi:= \top \until^{-} \varphi$ (``eventually in the past'')  and its
dual  $\Always^{-} \varphi:=\neg \Eventually^{-}\neg\varphi$ (``always in the past''). As usual, a sentence is a formula with no free variables.

Let $\Ku$ be a Kripke structure and $\varphi$ be a hybrid $\CTLStarLP$ formula. For an \emph{initial}  infinite path $\pi$ of $\Ku$, a variable valuation $g$, that assigns to each  variable $x$  a position along $\pi$, and  $i\geq 0$, the satisfaction relation 
$\pi,g, i\models  \varphi$ 
is defined as follows (we omit the clauses for the Boolean connectives and for $\until$ and $\Next$):
\[ \begin{array}{ll}
\pi, g,i \models \Next^{-} \varphi  & \Leftrightarrow  i>0 \text{ and } \pi, g, i-1 \models \varphi ,\\
 \pi, g, i \models \varphi_1\until^{-} \varphi_2  &
  \Leftrightarrow  \text{for some $j\leq i$ we have } \pi, g, j
  \models \varphi_2,
  \text{ and }  \\
  & \phantom{\Leftrightarrow}\; \pi, g, k \models  \varphi_1 \text{ for all }j< k\leq i,\\
\pi, g, i \models \EQ \varphi  & \Leftrightarrow \text{for some initial infinite path } \pi'  \text{ such that } \\
& \phantom{\Leftrightarrow }\; \pi'(0,i)=\pi(0,i), \text{ we have } \pi', g, i \models \varphi,\\
\pi,g, i \models x  &  \Leftrightarrow  g(x)=i,\\
\pi,g, i \models \Downarrowx.\varphi  &  \Leftrightarrow  \pi,g[x\leftarrow i], i \models \varphi,
\end{array} \]
where $g[x\leftarrow i](x)=i$ and $g[x\leftarrow i](y)=g(y)$ for $y\neq x$.  A Kripke structure $\Ku$ is a model of a formula $\varphi$ if  $\pi,g_0, 0 \models \varphi$, for every initial infinite path $\pi$ of $\Ku$, with $g_0$ the variable valuation assigning $0$ to each variable. Note that the path quantification is \lq\lq memoryful\rq\rq ,
i.e., it  ranges over infinite paths that start at the root and visit  the current node of the computation tree. Clearly, the semantics for the syntactical fragment
$\CTLStar$ coincides with the standard one. 
If we disallow the use of variables and binder modalities, we obtain the  logic
$\CTLStarLP$, a well-known linear-past extension of $\CTLStar$ which is as expressive as $\CTLStar$ \cite{jcss/KupfermanPV12}.
We also consider the finitary variant of hybrid $\CTLStarLP$, where the path quantifier $\EQ$ is replaced with the finitary path quantifier $\EQF$. This logic corresponds to an extension of finitary $\CTLStar$ and its semantics is similar to that of hybrid $\CTLStarLP$ with the exception that path quantification ranges over the \emph{finite} paths (traces) that start at the root and visit the current node of the computation tree.

In the following, we will use the fragment of hybrid $\CTLStarLP$ consisting of \emph{well-formed} formulas,
namely, formulas $\varphi$ where:
 \begin{itemize}
   \item each subformula $\EQ \psi$   of $\varphi$ has at most one free variable (namely, not bound by the downarrow binder operator); 
	\item each  subformula $\EQ \psi(x)$ of $\varphi$ having $x$ as free variable occurs in $\varphi$ in the context $(\Eventually^{-} x) \wedge \EQ\psi(x)$.
\end{itemize}
Intuitively, the above conditions affirm that, for each state subformula  $\EQ \psi$, the unique free variable (if any) refers to ancestors of the current node in the computation tree.\footnote{The well-formedness constraint ensures that a formula captures only branching regular requirements. As an example,
the formula $\EQ\Eventually\Downarrowx.\Always^{-}(\neg \Next^{-} \top \rightarrow \AQ \Eventually(x\wedge p))$ is \emph{not} well-formed and requires that there is a level of the computation tree such that each node in the level satisfies $p$. This represents a non-regular context-free branching requirement (see, e.g., \cite{AlurCZ06}).} 
The notion of well-formed formula of finitary hybrid $\CTLStarLP$ is similar: the path quantifier $\EQ$ is replaced by its finitary version $\EQF$.

 We first show that
 $\HS_\LinearPast$ can be translated into the well-formed fragment of hybrid $\CTLStarLP$ (resp., well-formed fragment of finitary hybrid $\CTLStarLP$). Then we show that this fragment is subsumed by $\CTLStar$ (resp., finitary $\CTLStar$).

\begin{proposition}\label{prop:HSComputationToHybrid} Given a $\HS_\LinearPast$ formula $\varphi$, one can construct in linear-time an equivalent well-formed
sentence of hybrid $\CTLStarLP$ (resp., finitary hybrid $\CTLStarLP$).
\end{proposition}
\begin{proof}
We focus on the translation from $\HS_\LinearPast$ into the well-formed fragment of hybrid $\CTLStarLP$. The translation from $\HS_\LinearPast$ into the well-formed fragment
of finitary hybrid $\CTLStarLP$ is similar, and thus omitted.

Let $\varphi$ be a $\HS_\LinearPast$ formula. The desired hybrid $\CTLStarLP$ sentence is the formula
$\Downarrowx .\Always f(\varphi,x)$, where $f(\varphi,x)$ is a mapping which is homomorphic with respect to 
the Boolean connectives, and over proposition letters and modalities behaves in the following way:
\[ \begin{array}{ll}
f(p,x) & = \Always^{-}((\Eventually^{-}x ) \rightarrow p),\\
f(\hsB\psi,x) & = \Next^{-}\Eventually^{-}(f(\psi,x)\wedge \Eventually^{-}x) ,\\
f(\hsBt\psi,x) & = \EQ(\Next\Eventually f(\psi,x)) \wedge \Eventually^{-}x,\\
f(\hsE\psi,x) & = \Downarrowy.\Eventually^{-}\bigl(x \wedge \Next\Eventually \Downarrowx.\Eventually(y\wedge f(\psi,x))\bigr ) ,\\
f(\hsEt\psi,x) & = \Downarrowy.\Eventually^{-}\bigl((\Next\Eventually x) \wedge  \Downarrowx.\Eventually(y\wedge f(\psi,x))\bigr ),
\end{array} \]
where $y$ is a fresh variable.

Clearly $\Downarrowx .\Always f(\varphi,x)$ is well-formed. The formula $f(\varphi,x)$ intuitively states that $\varphi$ holds over an 
interval of the current path that starts at the position (associated with the variable) $x$ and ends at the current position.

More formally, let $\Ku$ be a Kripke structure, $[h,i]$ be an interval of positions, $g$ be a valuation assigning to 
the variable $x$ the position $h$, and $\pi$ be an initial infinite path.
By a straightforward induction on the structure of  $\varphi$, one can show that
 $\Ku,\pi,g,i\models f(\varphi,x)$ if and only if  $\mathpzc{C}(\Ku),\mathpzc{C}(\pi,h,i)\models_\stat \varphi$, where
$\mathpzc{C}(\pi,h,i)$ denotes the trace of the computation tree $\mathpzc{C}(\Ku)$ starting from $\pi(0,h)$ and leading to $\pi(0,i)$.
Hence, $\Ku$ is a model of $\Downarrowx .\Always f(\varphi,x)$ if, for each initial trace $\rho$ of $\mathpzc{C}(\Ku)$, we have that
$\mathpzc{C}(\Ku),\rho\models_\stat \varphi$.
\end{proof}

Let $\LTLP$ be the past extension of $\LTL$, obtained by adding the past modalities $\Next^{-}$ and $\until^{-}$.
By exploiting the well-known separation theorem for $\LTLP$ over finite and infinite words \cite{Gabbay87}, which states that any $\LTLP$ formula can be effectively converted into an equivalent Boolean combination of $\LTL$ formulas and pure past $\LTLP$ formulas, we can prove that, under the hypothesis of well-formedness, the extensions of $\CTLStar$ (resp., finitary $\CTLStar$) used to encode  $\HS_\LinearPast$ formulas do not increase the expressive power of $\CTLStar$ (resp., finitary $\CTLStar$). Such a result is the fundamental step to prove, together with Proposition~\ref{prop:HSComputationToHybrid},
that $\CTLStar$ subsumes $\HS_\LinearPast$. In addition, paired with Corollary~\ref{cor:FromFInitaryCTLStarToFutureHSBranching}, it will allow us to state the main result of the section, namely, that $\HS_\LinearPast$ and finitary $\CTLStar$ have the same expressiveness.

Let us now show that the well-formed fragment of hybrid $\CTLStarLP$ (resp., finitary hybrid $\CTLStarLP$) is not more expressive than $\CTLStar$ (resp., finitary $\CTLStar$). Once more, we focus on the well-formed fragment of hybrid $\CTLStarLP$ omitting the similar proof for the finitary variant.

We start with some additional definitions and auxiliary results.  A \emph{pure past} $\LTLP$ formula is an $\LTLP$ formula which does not contain occurrences of future temporal modalities. Given two formulas $\varphi$ and $\varphi'$ of hybrid $\CTLStarLP$, we say that
$\varphi$ and $\varphi'$ are \emph{congruent} if, for every  Kripke structure $\Ku$, initial infinite path $\pi$, valuation $g$, and current position $i$, it holds
$\Ku,\pi,g, i \models \varphi$ if and only if $\Ku,\pi,g, i \models \varphi'$ (note that congruence is a \emph{stronger} requirement than equivalence).

As usual, for a formula $\varphi$ of hybrid $\CTLStarLP$ with one free variable $x$, we write $\varphi(x)$. Moreover, since the satisfaction relation depends only on the variables occurring free in the given formula, for $\varphi(x)$ we  use the notation
$\Ku,\pi,i \models \varphi(x \leftarrow h)$ to mean that $\Ku,\pi,g,i \models \varphi$ for any valuation $g$ assigning $h$ to the unique free variable $x$.
For a formula $\varphi$ of hybrid $\CTLStarLP$, let $\EQSubf(\varphi)$ denote the set of subformulas of $\varphi$ of the form $\EQ \psi$ which do not occur in the scope of
the path quantifier $\EQ$.

Finally, for technical reasons, we introduce \emph{simple} hybrid $\CTLStarLP$ formulas.
\begin{definition} Given a variable $x$, a \emph{simple} hybrid $\CTLStarLP$ formula $\psi$ with respect to $x$  is a hybrid $\CTLStarLP$ formula satisfying the following syntactical constraints:
\begin{itemize}
    \item $x$ is the unique variable occurring in $\psi$; 
    \item $\psi$ does \emph{not} contain occurrences of the binder modalities and past temporal modalities;
    \item $\EQSubf(\psi)$ consists of $\CTLStar$ formulas.
\end{itemize}
\end{definition}

Intuitively, a \emph{simple} hybrid $\CTLStarLP$ (over $\Prop$) formula $\psi$ with respect to $x$  can be seen as a $\CTLStar$ formula over the set of proposition letters $\Prop\cup \{x\}$ such that $x$ does not occur in the scope of $\EQ$. The next lemma,
proved in Appendix \ref{proof:lemma:UsingSeparationHybridCTLPreliminary}, shows that $\psi$ can be further simplified whenever it is paired with the formula $\Eventually^{-} x$.

\begin{lemma}\label{lemma:UsingSeparationHybridCTLPreliminary} 
Let $\psi$ be  a \emph{simple} hybrid $\CTLStarLP$ formula  with respect to $x$. 
Then $(\Eventually^{-} x)\wedge \psi$ is congruent to a formula of the form $(\Eventually^{-} x)\wedge \xi$, where $\xi$ is a Boolean combination of the atomic formula $x$ and $\CTLStar$ formulas. 
\end{lemma}

The next lemma states an important technical property of  well formed formulas, which will be exploited in 
Theorem~\ref{Theo:HybridCTLInfiniteCase} to prove that the set of sentences of the well-formed fragment of hybrid $\CTLStarLP$ has the same expressiveness as $\CTLStar$. Intuitively, if the hybrid features of the language do not occur in the scope of existential path quantifiers, it is possible to remove the occurrences of the binder $\downarrow$ and to suitably separate past and future modalities. The result is obtained by exploiting the equivalence of $\FO$ and $\LTLP$ over infinite words and by applying the separation theorem for $\LTLP$ over infinite words~\cite{Gabbay87}, that we recall here for completeness. 

\begin{theorem}[$\LTLP$ separation over infinite words]\label{th:LTLpsep}
Any $\LTLP$ formula $\psi$ can be effectively transformed into a formula
\[
\psi'=   \bigvee_{i=1}^t(\psi_{p,i}\wedge \psi_{f,i}),
\]
for some $t\geq 1$, where $\psi_{p,i}$ is a pure past $\LTLP$ formula and $\psi_{f,i}$ is an $\LTL$ formula, such that
for all infinite words $w$ over $2^{\mathpzc{AP}}$ and $i\geq 0$, it holds that
$
 w,i\models \psi \text{ if and only if } w,i\models \psi'.
$
\end{theorem}

\begin{lemma}\label{lemma:UsingSeparationHybridCTL} Let $(\Eventually^{-} x)\wedge \EQ \varphi(x)$ (resp., $\EQ\varphi$) be a well-formed formula (resp., well-formed sentence) of hybrid $\CTLStarLP$ such that $\EQSubf(\varphi)$ consists of
$\CTLStar$ formulas. Then $(\Eventually^{-} x)\wedge \EQ \varphi(x)$ (resp., $\EQ\varphi$) is congruent to a well-formed formula of hybrid $\CTLStarLP$ which is a Boolean combination of $\CTLStar$ formulas and (formulas that
correspond to) \emph{pure past}  $\LTLP$ formulas over the set of proposition letters $\Prop\cup  \EQSubf(\varphi) \cup \{x\}$ (resp., $\Prop\cup  \EQSubf(\varphi)$).
\end{lemma}
\begin{proof} We focus on well-formed formulas of the form  $(\Eventually^{-} x)\wedge \EQ \varphi(x)$. The case of well-formed sentences of the form $\EQ\varphi$ is similar, and thus omitted.

Let $\overline{\Prop}= \Prop\cup  \EQSubf(\varphi) \cup \{x\}$. By hypothesis, $\EQSubf(\varphi)$ is a set of $\CTLStar$ formulas, that is, they are devoid of any hybrid feature.

Given a Kripke structure $\Ku=\KuDef$,  an initial infinite path $\pi$, and $h\geq 0$, we denote by $\pi_{\overline{\Prop},h}$ the infinite word over $2^{\overline{\Prop}}$, which, for every position $i\geq 0$, is defined as follows:
 \begin{itemize}
 \item $\pi_{\overline{\Prop},h}(i)\cap \Prop =\Lab(\pi(i))$;
 \item $\pi_{\overline{\Prop},h}(i)\cap \EQSubf(\varphi)= \{\psi\in \EQSubf(\varphi)\mid \Ku,\pi,i\models \psi\}$;
 \item $x\in \pi_{\overline{\Prop},h}(i)$ if and only if $i=h$.
 \end{itemize}

 By using a fresh position variable $\present$ to represent the current position, the formula $\varphi(x)$ can be easily converted into an $\FO$ formula
 $\varphi_\FO(\present)$
 over $ \overline{\Prop}$ having $\present$ as its unique free variable, such that for all Kripke structures $\Ku$, initial infinite paths $\pi$, and positions $i$ and $h$, we have:
 \begin{equation}\label{eq:UsingSeparationHybridCTL1}
  \Ku,\pi,i\models \varphi(x \leftarrow h) \text{ if and only if } \pi_{\overline{\Prop},h} \models \varphi_\FO(\present \leftarrow i).
\end{equation}
To this end, it suffices to map any proposition letter $\overline{p}\in \overline{\Prop}$ into a unary predicate $\overline{p}$,  and all the operators $\Next,\Next^{-},\until,\until^{-},\downarrow$ into  $\FO$ formulas expressing their semantics.

 By the equivalence of $\FO$ and $\LTLP$ and the separation theorem for $\LTLP$ over infinite words, starting from the $\FO$ formula $\varphi_\FO(\present)$, one can construct an $\LTLP$ formula $\varphi_\LTLP$ over $\overline{\Prop}$ of the form 
  \begin{equation}\label{eq:UsingSeparationHybridCTL2}
\varphi_\LTLP=   \bigvee_{i\in I}(\varphi_{p,i}\wedge \varphi_{f,i})
\end{equation}
such that $\varphi_{p,i}$ 
is a pure past $\LTLP$ formula,
$\varphi_{f,i}$ 
is an $\LTL$ formula, and for all infinite words $w$ over $2^{\overline{\Prop}}$ and $i\geq 0$, it holds that:
  \begin{equation}\label{eq:UsingSeparationHybridCTL3}
 w,i\models \varphi_\LTLP \text{ if and only if } w\models \varphi_\FO(\present \leftarrow i).
\end{equation}
 The $\LTLP$ formula $\varphi_\LTLP$ over $\overline{\Prop}$ corresponds to a hybrid $\CTLStarLP$ formula $\varphi_\LTLP(x)$ over $\Prop$ (note that the only hybrid feature is the possible occurrence of the variable $x$).
 By definition of the infinite words $\pi_{\overline{\Prop},h}$, one can easily show by structural induction that 
  for all Kripke structures $\Ku$, initial infinite paths $\pi$, and positions $i$ and $h$:
  \begin{equation}\label{eq:UsingSeparationHybridCTL4}
\pi_{\overline{\Prop},h},i \models  \varphi_\LTLP \text{ if and only if  }  \Ku,\pi,i\models \varphi_\LTLP(x \leftarrow h),
\end{equation}
the latter being a hybrid $\CTLStarLP$ formula.
Thus, by (\ref{eq:UsingSeparationHybridCTL1}), (\ref{eq:UsingSeparationHybridCTL3}), and (\ref{eq:UsingSeparationHybridCTL4}), we obtain that 
$\varphi(x)$ and $\varphi_\LTLP(x)$ are congruent.

Since in (\ref{eq:UsingSeparationHybridCTL2}),  for each $i\in I$, $\varphi_{p,i}$ is a pure past $\LTLP$ formula over $\overline{\Prop}$,  $\EQ \varphi_{p,i}(x)$ is trivially congruent to $\varphi_{p,i}(x)$. As a consequence, we have that
$(\Eventually^{-} x)\wedge \EQ \varphi(x)$ is congruent to
 $(\Eventually^{-} x)\wedge \bigvee_{i\in I}(\varphi_{p,i}(x)\wedge \EQ\varphi_{f,i}(x))$, which is congruent to $\bigvee_{i\in I}(\varphi_{p,i}(x)\wedge (\Eventually^{-} x)\wedge \EQ\varphi_{f,i}(x))$, which is in turn congruent to
 $\bigvee_{i\in I}(\varphi_{p,i}(x)\wedge \EQ((\Eventually^{-} x)\wedge \varphi_{f,i}(x)))$.

 Now, $\varphi_{f,i}(x)$ is a \emph{simple} hybrid $\CTLStarLP$ formula  with respect to $x$, and  $\EQ x$ (resp., $\EQ\neg x$) is trivially congruent to $x$ (resp., $\neg x$).
 By Lemma~\ref{lemma:UsingSeparationHybridCTLPreliminary} and some simple manipulation steps, we can prove the following sequence of equivalences:
 
 \begin{align*}
 \bigvee_{i\in I}\Big(\varphi_{p,i}(x)\wedge \EQ((\Eventually^{-} x)\wedge \varphi_{f,i}(x))\Big)&=
 \tag*{(Lemma~\ref{lemma:UsingSeparationHybridCTLPreliminary} and disjunctive normal form)}\\
   \bigvee_{i\in I}\Big(\varphi_{p,i}(x) \wedge\EQ \big((\Eventually^{-} x)\wedge\bigvee_{j\in J}(\tilde{x}_{i,j}\wedge \psi_{i,j})\big)\Big)& =\tag*{($\Eventually^{-} x$ is a pure past $\LTLP$ formula)}\\
   \bigvee_{i\in I}\Big(\varphi_{p,i}(x) \wedge(\Eventually^{-} x)\wedge\EQ \bigvee_{j\in J}(\tilde{x}_{i,j}\wedge \psi_{i,j})\Big)&=\tag*{(Distributive property of $\wedge$ over $\vee$)}\\
 (\Eventually^{-} x)\wedge \bigvee_{i\in I}\Big(\varphi_{p,i}(x) \wedge\EQ \bigvee_{j\in J}(\tilde{x}_{i,j}\wedge \psi_{i,j})\Big)&=\tag*{(Distributive property of $\EQ$ over $\vee$ and $\tilde{x}_{i,j}$ is a pure past $\LTLP$ formula)}\\
 (\Eventually^{-} x)\wedge \bigvee_{i\in I}\Big(\varphi_{p,i}(x) \wedge \bigvee_{j\in J}(\tilde{x}_{i,j}\wedge \EQ\psi_{i,j})\Big)&=\tag*{(Distributive property of $\wedge$ over $\vee$)}\\
 (\Eventually^{-} x)\wedge \bigvee_{i\in I}\bigvee_{j\in J}\Big(\varphi_{p,i}(x) \wedge \tilde{x}_{i,j}\wedge \EQ\psi_{i,j}\Big)&
 \end{align*}
where $\tilde{x}_{i,j}$ is either $x$, $\neg x$, or $\top$.
 
 Hence, 
 $(\Eventually^{-} x)\wedge \EQ \varphi(x)$ is congruent to a formula of the form
 \[(\Eventually^{-} x)\wedge \bigvee_{i\in I'}(\psi_{p,i}(x)\wedge \EQ\psi_{i}),\] for some $I'$, where $\psi_{p,i}(x)$ corresponds 
 to a  \emph{pure past}  $\LTLP$ formula over $\overline{\Prop} \,(=\Prop\cup  \EQSubf(\varphi) \cup \{x\})$
 and
 $\psi_i$ is a $\CTLStar$ formula.
\end{proof}

The following lemma generalizes the separation result given by Lemma~\ref{lemma:UsingSeparationHybridCTL} to any well-formed formula of the form $(\Eventually^{-} x)\wedge \EQ \varphi(x)$, that is, to formulas where $\varphi(x)$ is unconstrained. Its proof is by induction, and exploits Lemma~\ref{lemma:UsingSeparationHybridCTL} (Appendix~\ref{proof:cor:UsingSeparationHybridCTL}).

\begin{lemma}\label{cor:UsingSeparationHybridCTL} Let $(\Eventually^{-} x)\wedge \EQ \varphi(x)$ (resp., $\EQ\varphi$) be a well-formed formula (resp., well-formed sentence) of hybrid $\CTLStarLP$.  Then there exists a finite set $\mathpzc{H}$ of $\CTLStar$ formulas of the form $\EQ\psi$, such that $(\Eventually^{-} x)\wedge \EQ \varphi(x)$ (resp., $\EQ\varphi$) is congruent to a well-formed formula of hybrid $\CTLStarLP$ which is a Boolean combination of $\CTLStar$ formulas and (formulas that
correspond to) \emph{pure past}  $\LTLP$ formulas over the set of proposition letters $\Prop\cup  \mathpzc{H} \cup \{x\}$ (resp., $\Prop\cup  \mathpzc{H}$).
\end{lemma} 

We are now ready to prove that the well-formed sentences of hybrid $\CTLStarLP$ can be expressed in $\CTLStar$.

\begin{theorem}\label{Theo:HybridCTLInfiniteCase} 
The set of sentences of the well-formed fragment of hybrid $\CTLStarLP$ has the same expressiveness as $\CTLStar$.
\end{theorem}
\begin{proof}
Let $\varphi$ be a well-formed sentence of hybrid $\CTLStarLP$. To prove the thesis,  we construct a $\CTLStar$ formula which is equivalent to $\varphi$. 

Since $\varphi$ is equivalent to $\neg \EQ \neg \varphi$ and $\neg \EQ \neg \varphi$ is well-formed, by  Lemma~\ref{cor:UsingSeparationHybridCTL} one can convert  $\neg \EQ \neg \varphi$ into a congruent hybrid $\CTLStarLP$ formula which is a Boolean combination of $\CTLStar$ formulas and formulas $\theta$ which can be seen as pure past $\LTLP$ formulas over the set of proposition letters $\Prop\cup  \mathpzc{H}$, where $\mathpzc{H}$ is a set of $\CTLStar$ formulas of the form $\EQ\psi$. 

Since the past temporal modalities
in such  $\LTLP$ formulas $\theta$ refer to the initial position of the initial infinite paths, one can replace $\theta$ with an equivalent $\CTLStar$ formula $f(\theta)$, where the mapping $f$ is inductively defined as follows:
 \begin{itemize}
   \item $f(p)=p$ for all $p\in \Prop\cup  \mathpzc{H}$;
   \item $f$ is homomorphic with respect to the Boolean connectives;
   \item $f(\Next^{-}\theta)=\bot$ and $f(\theta_1\until^{-}\theta_2)=f(\theta_2)$.
 \end{itemize}
The resulting $\CTLStar$ formula is equivalent to $\neg \EQ \neg \varphi$, as required.
\end{proof}

By an easy adaptation of the proof of Theorem~\ref{Theo:HybridCTLInfiniteCase}, where one exploits the separation theorem for $\LTLP$ over finite words \cite{Gabbay87}, it is possible to characterize also the expressiveness of well-formed \emph{finitary} hybrid $\CTLStarLP$.

\begin{theorem}\label{theo:FinitaryHybridCTL} The set of sentences of the well-formed fragment of finitary hybrid $\CTLStarLP$ has the same expressiveness as finitary $\CTLStar$.
\end{theorem}


Together with Proposition~\ref{prop:HSComputationToHybrid}, Theorem~\ref{Theo:HybridCTLInfiniteCase} 
(resp., Theorem~\ref{theo:FinitaryHybridCTL})
allows us to conclude that $\CTLStar$ (resp., finitary $\CTLStar$) subsumes $\HS_\LinearPast$. 

Finally, by exploiting Corollary~\ref{cor:FromFInitaryCTLStarToFutureHSBranching}, we can state the main result of the section, namely, $\HS_\LinearPast$ and finitary $\CTLStar$ have the same expressiveness.

\begin{theorem}\label{Cor:CharacterizationHSCompTree} $\CTLStar\geq\HS_\LinearPast$. Moreover, $\HS_\LinearPast$ is as expressive as finitary~$\CTLStar$.
\end{theorem}

\section{Expressiveness comparison of  $\HS_\LinearTime$,  $\HS_\stat$, $\HS_\LinearPast$}\label{sect:allSems}

In this section, we compare the expressiveness of the three semantic variants of $\HS$, namely, $\HS_\LinearTime$,  $\HS_\stat$, and $\HS_\LinearPast$. 
The resulting picture was anticipated in Figure~\ref{results}. Here we give the proofs of the depicted results.

We start showing that $\HS_{\stat}$ is \emph{not} subsumed by $\HS_{\LinearPast}$. As a matter of fact, we show that   $\HS_{\stat}$ is sensitive to backward unwinding of finite Kripke structures, allowing us to sometimes discriminate finite Kripke structures with the same computation tree (these structures are always indistinguishable by $\HS_{\LinearPast}$). 

\begin{figure}[tp]
\centering
\begin{tikzpicture}[->,>=stealth,thick,shorten >=1pt,node distance=1.6cm]
\draw[draw=none, use as bounding box](-1.3,0.6) rectangle (9.3,-0.6);
   \node [right] at (-1.2,0) {$\Ku_1$:};
   \node [style={double,circle,draw}] (v0) {$\stackrel{s_0}{\phantom{p}}$};
   \node [style={circle,draw}, right of=v0] (v1) {$\stackrel{s_1}{p}$};
   \node [right] at (3.6,0) {$\Ku_2$:};
   \node [right of=v1] (z1) {$\phantom{p}$};
   \node [left of=v0] (z0) {$\phantom{p}$};
   \node [style={double,circle,draw}, right of=z1] (v2) {$\stackrel{s_0'}{\phantom{p}}$};
      \node [style={circle,draw}, right of=v2] (v3) {$\stackrel{s_1'}{p}$};
       \node [style={circle,draw}, right of=v3] (v4) {$\stackrel{s_2'}{p}$};
   \draw (v0) to [right] (v1);
    \draw (v1) to [loop right] (v1);
    \draw (v4) to [loop right] (v4);
    \draw (v2) to [right] (v3);
    \draw (v3) to [right] (v4);
\end{tikzpicture}
\caption{The finite Kripke structures $\Ku_1$ and $\Ku_2$.}\label{FigStateBasedExpressiveness}
\end{figure}
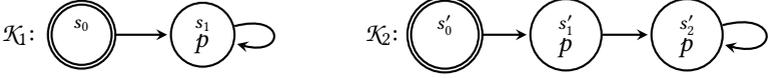

\begin{figure}[tp]
    \centering
\begin{tikzpicture}[->,>=stealth,thick,shorten >=1pt,node distance=1.6cm]

   \node [right] at (-1.2,0) {$\mathpzc{K}_1$:};
   \node [] (v0) {$\cdots$};
   \node [style={circle,draw}, right of=v0] (v1) {$\stackrel{s_1}{p}$};
   \node [style={double,circle,draw}, above left of=v1] (v1a) {$\stackrel{s_0}{\phantom{p}}$};
   \node [style={circle,draw}, right of=v1] (v11) {$\stackrel{s_1}{p}$};
   \node [style={double,circle,draw}, above left of=v11] (v110) {$\stackrel{s_0}{\phantom{p}}$};
   \node [ right of=v11] (v12) {$\cdots$};
   \draw (v0) to [right] (v1);
    \draw (v1) to (v11);
    \draw (v11) to (v12);
    \draw (v1a) to (v1);
    \draw (v110) to (v11);
\end{tikzpicture}

\bigskip

\begin{tikzpicture}[->,>=stealth,thick,shorten >=1pt,node distance=1.6cm]
   \node [right] at (-1.2,-1) {$\mathpzc{K}_2$:};
   \node [style={double,circle,draw}] (v2) {$\stackrel{s_0'}{\phantom{p}}$};
    \node [style={circle,draw}, right of=v2] (v3) {$\stackrel{s_1'}{p}$};
    \node [style={circle,draw}, below right of=v3] (v4) {$\stackrel{s_2'}{p}$};
    \node [left of=v4] (v4x) {$\cdots$};
    \node [style={circle,draw}, right=2.5cm of v4] (v40) {$\stackrel{s_2'}{p}$};
    \node [style={circle,draw}, above left of=v40] (v401) {$\stackrel{s_1'}{p}$};
    \node [style={double,circle,draw},left of=v401] (v401x) {$\stackrel{s_0'}{\phantom{p}}$};
    \node [right of=v40] (v41) {$\cdots$};
    \draw (v2) to (v3);
    \draw (v3) to (v4);
    \draw (v3) to (v4);
    \draw (v4) to (v40);
    \draw (v40) to (v41);
    \draw (v4x) to (v4);
    \draw (v401) to (v40);
    \draw (v401x) to (v401);
\end{tikzpicture}
\caption{Forward and backward unwinding of $\mathpzc{K}_1$ and $\mathpzc{K}_2$ of Figure~\ref{FigStateBasedExpressiveness}.}\label{FigFBUnwinding}
\end{figure}
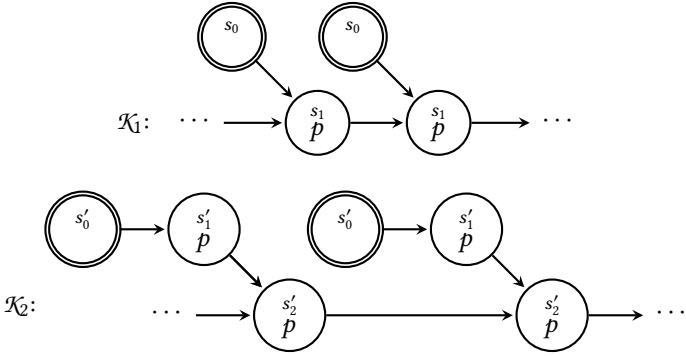

Let us consider, for instance, the two finite Kripke structures $\Ku_1$ and $\Ku_2$ of Figure~\ref{FigStateBasedExpressiveness}, whose forward and backward unwinding is shown in Figure~\ref{FigFBUnwinding}. Since $\Ku_1$ and $\Ku_2$ have the same computation tree, no $\HS$ formula $\varphi$ under the computation-tree-based semantics can distinguish $\Ku_1$ and $\Ku_2$, that is, $\Ku_1\models_\LinearPast\varphi$ if and only if $\Ku_2\models_\LinearPast\varphi$. On the other hand, the requirement ``\emph{each state reachable from the initial one where $p$ holds has a predecessor where $p$ holds as well}'' can be expressed, under the state-based semantics, by the $\HS$ formula
\[
\psi= \hsE(p\wedge \Length_1) \rightarrow \hsE(\Length_1 \wedge \hsAt(p\wedge \neg\Length_1)).
\]
It is easy to see that $\Ku_1\models_\stat\psi$: for any initial trace $\rho$ of $\Ku_1$,
we have $\Ku_1,\rho\models_\stat \hsE(p\wedge \Length_1)$ iff $\rho=s_0s_1^k$ for $k\geq 1$; the length-1 suffix $s_1$ is \emph{met-by} $s_1s_1$, and $\Ku_1,s_1s_1\models_\stat p\wedge \neg\Length_1$.
On the contrary, in $\Ku_2$ there is an initial trace, $s_0's_1'$, for which $\Ku_2,s_0's_1'\models_\stat\hsE(p\wedge \Length_1)$; however the only traces that meet the length-1 suffix $s_1'$ are $s_1'$ itself and $s_0's_1'$, but neither of them model $p\wedge \neg\Length_1$.
Therefore $\Ku_2\not\models_\stat\psi$. 
This allows us to prove the following proposition.

\begin{proposition}\label{Prop:StateBasedExpressiveness} 
$\HS_{\LinearPast} \not \geq \HS_\stat$.
\end{proposition}

Since, as stated by  Theorem~\ref{Cor:CharacterizationHSCompTree}, $\HS_{\LinearPast}$ and finitary $\CTLStar$ have the same expressiveness and finitary $\CTLStar$ is subsumed by $\HS_\stat$ (see Corollary~\ref{cor:FromFInitaryCTLStarToFutureHSBranching}), by Proposition~\ref{Prop:StateBasedExpressiveness} the next corollary immediately follows.

\begin{corollary} $\HS_\stat$ is more expressive than $\HS_{\LinearPast}$.
\end{corollary}

In the following, we focus on the comparison of $\HS_{\LinearTime}$ with $\HS_\stat$ and $\HS_{\LinearPast}$ showing that $\HS_{\LinearTime}$ is incomparable with both $\HS_\stat$ and $\HS_{\LinearPast}$. 

The fact that  $\HS_{\LinearTime}$ does not subsume either $\HS_\stat$ or $\HS_{\LinearPast}$ can be easily proved as follows. Consider the $\CTL$ formula $\forall \Always \exists \Eventually p$ asserting that from each state reachable from the initial one, it is possible to reach a state where $p$ holds. It is well-known that this formula is not $\LTL$-definable (see~\cite{Baier2008}, Theorem~6.21). Thus, by Corollary~\ref{cor:HSLinearCharacterization}, there is no equivalent $\HS_{\LinearTime}$ formula. On the other hand, the requirement $\forall \Always \exists \Eventually p$ can be trivially expressed under the state-based (resp., computation-tree-based) semantics by the $\HS$ formula $\hsBt\hsE p $, proving the following result.

\begin{proposition}\label{prop:nonBranchingExpressibilityOfLinearTime} $\HS_{\LinearTime} \not \geq \HS_\stat$ and $\HS_{\LinearTime} \not \geq \HS_\LinearPast$.
\end{proposition}

To prove the converse, namely, that $\HS_{\LinearTime}$ is not subsumed either by $\HS_\stat$ or by $\HS_{\LinearPast}$, 
we will show that the $\LTL$ formula $\Eventually p$ (equivalent to the $\CTL$ formula $\forall \Eventually p$)
cannot be expressed in either $\HS_{\LinearPast}$ or $\HS_{\stat}$ (Proposition~\ref{prop:nonBranchingExpressibilityOfEventually}). The proof is rather involved and requires a number of definitions and intermediate results; we consider only
the state-based semantics, as the case of the computation-tree-based one is very similar. 



Let us start by defining two families of Kripke structures $(\Ku_n)_{n\geq 1}$ and $(\mathpzc{M}_n)_{n\geq 1}$ over $\{p\}$ such that for all $n\geq 1$, the $\LTL$ formula $\Eventually p$
 distinguishes $\Ku_n$ and  $\mathpzc{M}_n$, and for every $\HS$ formula $\psi$ of size at most $n$, $\psi$ does \emph{not} distinguish
 $\Ku_n$ and  $\mathpzc{M}_n$ under the state-based semantics. 
 
For a given $n \geq 1$, the Kripke structures $\Ku_n$ and $\mathpzc{M}_n$ are depicted in
 Figure~\ref{FigEventuallyNOnBranchingExpressible}. Notice that the Kripke structure $\mathpzc{M}_n$  differs from  $\Ku_n$ only in that its initial state is $s_1$ instead of $s_0$. Formally, $\Ku_n=(\{p\},\States_n, \Edges_n,\Lab_n,s_0)$ and $\mathpzc{M}_n=(\{p\},\States_n, \Edges_n,\Lab_n,s_1)$, with
 $\States_n=\{s_0,s_1,\ldots, s_{2n},t\}$, $\Edges_n=\{(s_0,s_0),(s_0,s_1),\ldots, (s_{2n-1},s_{2n}),\allowbreak (s_{2n},t),(t,t)\}$,  $\Lab(s_i)=\emptyset$ for all $0\leq i\leq 2n$, and $\Lab(t)=\{p\}$. 
 
It is immediate to see that
%
 $\Ku_n\not\models \Eventually p$ and $\mathpzc{M}_n\models \Eventually p$.

\begin{figure}[t]
\centering
\begin{tikzpicture}[->,>=stealth,thick,shorten >=1pt,node distance=1.6cm]
   \node [right] at (-1.7,0) {$\Ku_n$:};
   \node [style={double,circle,draw}] (v0) {$\stackrel{s_0}{\phantom{p}}$};
   \node [style={circle,draw}, right of=v0] (v1) {$\stackrel{s_1}{\phantom{p}}$};
   \draw (v0) to [right] (v1);
       \draw (v0) to [loop left] (v0);
   \node [right of=v1] (v2) {$........$};
   \draw (v1) to [right] (v2);
   \node [style={circle,draw},right of=v2] (v2n) {$\stackrel{s_{2n}}{\phantom{p}}$};
   \draw (v2) to [right] (v2n);
   \node [style={circle,draw}, right of=v2n] (v) {$\stackrel{t}{p}$};
      \draw (v) to [loop right] (v);
   \draw (v2n) to [right] (v);
\end{tikzpicture}

\bigskip

\begin{tikzpicture}[->,>=stealth,thick,shorten >=1pt,node distance=1.6cm]
   \node [right] at (-1.7,0) {$\mathpzc{M}_n$:};
   \node [style={circle,draw}] (v0) {$\stackrel{s_0}{\phantom{p}}$};
   \node [style={double,circle,draw}, right of=v0] (v1) {$\stackrel{s_1}{\phantom{p}}$};
   \draw (v0) to [right] (v1);
       \draw (v0) to [loop left] (v0);
   \node [right of=v1] (v2) {$........$};
   \draw (v1) to [right] (v2);
   \node [style={circle,draw},right of=v2] (v2n) {$\stackrel{s_{2n}}{\phantom{p}}$};
   \draw (v2) to [right] (v2n);
   \node [style={circle,draw}, right of=v2n] (v) {$\stackrel{t}{p}$};
      \draw (v) to [loop right] (v);
   \draw (v2n) to [right] (v);
\end{tikzpicture}
\caption{The finite Kripke structures $\Ku_n$ and $\mathpzc{M}_n$ with $n\geq 1$.}\label{FigEventuallyNOnBranchingExpressible}
\end{figure}

On the contrary, with the next Lemma~\ref{lemma:MainnonBranchingExpressibilityOfEventually}, we are going to prove that  $\Ku_n\models_\stat \psi$ if and only if $\mathpzc{M}_n\models_\stat \psi$ for all balanced $\HS_\stat$ formulas $\psi$ of length at most $n$, with $n\geq 1$. 
An $\HS_\stat$ formula $\psi$ is \emph{balanced} if, for each subformula $\hsB \theta $ (resp., $\hsBt \theta $), $\theta$ has the form $\theta_1\wedge\theta_2$ with $|\theta_1| = |\theta_2|$. Proving the result for  balanced $\HS_\stat$ formulas allows us to state it for any $\HS_\stat$ formula, since it is possible to trivially 
convert an $\HS_\stat$ formula $\psi$ into a balanced one (by using conjunctions of $\top$) which is equivalent to $\psi$ under any of the considered $\HS$ semantic variants.

\begin{lemma}\label{lemma:MainnonBranchingExpressibilityOfEventually} For all $n\in\Nat^+$ and balanced $\HS_\stat$ formulas $\psi$, with $|\psi|\leq n$, 
it holds that $\Ku_n\models_\stat \psi$ if and only if $\mathpzc{M}_n\models_\stat \psi$.
\end{lemma}
The complete proof is in Appendix~\ref{proof:lemma:MainnonBranchingExpressibilityOfEventually}.

As an immediate consequence of Lemma~\ref{lemma:MainnonBranchingExpressibilityOfEventually} and of the fact that, for each $n\geq 1$,  $\Ku_n\not\models \Eventually p$ and $\mathpzc{M}_n\models \Eventually p$, we get the desired undefinability result.




\begin{proposition}\label{prop:nonBranchingExpressibilityOfEventually} 
The $\LTL$ formula $\Eventually p$ (equivalent to the $\CTL$ formula $\forall \Eventually p$) cannot be expressed in either $\HS_{\LinearPast}$ or $\HS_{\stat}$.
\end{proposition}

The next proposition follows from Corollary \ref{cor:HSLinearCharacterization} and Proposition~\ref{prop:nonBranchingExpressibilityOfEventually}.

\begin{proposition}\label{prop:oppositeDirection} $\HS_\stat \not \geq  \HS_{\LinearTime} $ and $\HS_\LinearPast \not \geq \HS_{\LinearTime}$.
\end{proposition}

Putting together Proposition~\ref{prop:nonBranchingExpressibilityOfLinearTime} and Proposition~\ref{prop:oppositeDirection}, we finally obtain the incomparability result.

\begin{theorem} 
$\HS_\LinearTime$ and $\HS_\stat$  are expressively incomparable, and so are $\HS_\LinearTime$ and $\HS_\LinearPast$.
\end{theorem}

The proved results also allow us to establish  the  expressiveness relations between  $\HS_\stat$, $\HS_\LinearPast$ and the standard branching temporal logics $\CTL$ and $\CTLStar$.


\begin{corollary}
The following expressiveness results hold:
\begin{enumerate}
\item $\HS_\stat$ and $\CTLStar$ are expressively incomparable;
\item $\HS_\stat$ and $\CTL$ are expressively incomparable;
\item $\HS_\LinearPast$ and finitary $\CTLStar$ are  less expressive than $\CTLStar$;
\item  $\HS_\LinearPast$ and $\CTL$ are expressively incomparable.
\end{enumerate}
\end{corollary}
\begin{proof}
\begin{enumerate}
\item By Proposition~\ref{prop:nonBranchingExpressibilityOfEventually} and the fact that  $\CTLStar$ is not sensitive to unwinding. 

\item Again, by Proposition~\ref{prop:nonBranchingExpressibilityOfEventually} and by the observation that $\CTL$ is not sensitive to unwinding.

\item By Theorem~\ref{Cor:CharacterizationHSCompTree},  $\HS_\LinearPast$ is subsumed by $\CTLStar$, and $\HS_\LinearPast$ and finitary $\CTLStar$ have the same expressiveness. 
Hence, by Proposition~\ref{prop:nonBranchingExpressibilityOfEventually}, the result follows.

\item Thanks to Proposition~\ref{prop:nonBranchingExpressibilityOfEventually}, it suffices to show that there exists a
$\HS_\LinearPast$ formula which cannot be expressed in $\CTL$. 

Let us consider the $\CTLStar$ formula
$\varphi= \EQ\bigl(((p_1 \until p_2) \vee (q_1 \until q_2)) \until r\bigr)$
over the set of propositions $\{p_1,p_2,q_1,q_2,r\}$. It is shown in~\cite{EH86} that $\varphi$
cannot be expressed in $\CTL$. Clearly, if
we replace the path quantifier $\EQ$ in $\varphi$ with the finitary path quantifier $\EQF$,  we obtain an equivalent formula of finitary $\CTLStar$.
Thus, since $\HS_\LinearPast$ and finitary $\CTLStar$ have the same expressiveness (Theorem~\ref{Cor:CharacterizationHSCompTree}),
the result follows.\qedhere
\end{enumerate}
\end{proof}





%% file: Chaps/TOCL17/vendingMach.tex
\subsection{An example: a vending machine}\label{subs:vendingMach}
In this section, we give an example highlighting the differences among the $\HS$ semantic variants $\HS_\stat$, $\HS_\LinearPast$, and $\HS_\LinearTime$.

\begin{figure}[tp]
\centering
\resizebox{\textwidth}{!}{
\begin{tikzpicture}[->,>=stealth',shorten >=1pt,auto,node distance=2.8cm,semithick]
  \tikzstyle{every state}=[inner sep=1pt, outer sep=0pt, minimum size = 40pt]

  \node[accepting,state] (A0)                    {$\stackrel{s_0}{p_\text{\$=0}}$};
  \node[state]         (B2) [below of=A0] {$\stackrel{s_2}{p_\text{\$=2}}$};
  \node[state]         (B1) [left of=B2] {$\stackrel{s_1}{p_\text{\$=1}}$};
  \node[state]         (B05) [right of=B2] {$\stackrel{s_3}{p_\text{\$=0.50}}$};
  \node[state]         (C1) [below of=B1] {$\stackrel{s_4}{p_\text{candy}}$};
  \node[state]         (C2) [below of=B2] {$\stackrel{s_5}{p_\text{hotdog}}$};
  \node[state]         (C05) [below of=B05] {$\stackrel{s_6}{p_\text{water}}$};
  \node[state]         (D) [below of=C2] {$\stackrel{s_7}{p_\text{change}}$};
  
  \node[state]         (M1) [right of=C05] {$\stackrel{s_8}{p_\text{maint}}$};
  \node[state]         (M2) [right of=B05] {$\stackrel{s_9}{p_\text{maint\_end}}$};
  
  \path (A0) edge  [swap]  node {ins\_\$2} (B2)
            edge  [bend right,swap]  node {ins\_\$1} (B1)
            edge  [bend left,swap]  node {ins\_\$0.50} (B05)
        (B2) edge  [near start]  node {sel} (C2)
             edge  [near end]  node {sel} (C1)
             edge   node {sel} (C05)
        (B1) edge   node {sel} (C1)
             edge  [very near start]  node {sel} (C05)
        (B05) edge  [near end]  node {sel} (C05)
        (C1) edge   [bend right,very near start]  node {dispensed} (D)
        (C2) edge  [swap]  node {dispensed} (D)
        (C05) edge  [bend left,swap]  node {dispensed} (D)
        (D) edge   [bend right,swap]  node {change\_given} (M1)
        (D) edge   [out=200,in=160,looseness=1.7]  node {change\_given} (A0)
        (M1) edge  [near end]  node {maint\_ongoing} (M2)
        (M2) edge [bend left] node {maint\_failed} (M1)
        (M2) edge   [bend right,swap]  node {maint\_success} (A0);
        
\draw (-3.6,1) [dashed] rectangle (3.6,-9.5);
\draw (4.5,-1.5) [dashed] rectangle (6.8,-7);
\node (P1) at (4.3,1) {$p_\text{operative}$};
\node (P2) at (7,-1.3) {$\neg p_\text{operative}$};
\end{tikzpicture}
}
\vspace{-1cm}
\caption{Kripke structure representing a vending machine.}\label{fig:vending}
\end{figure}
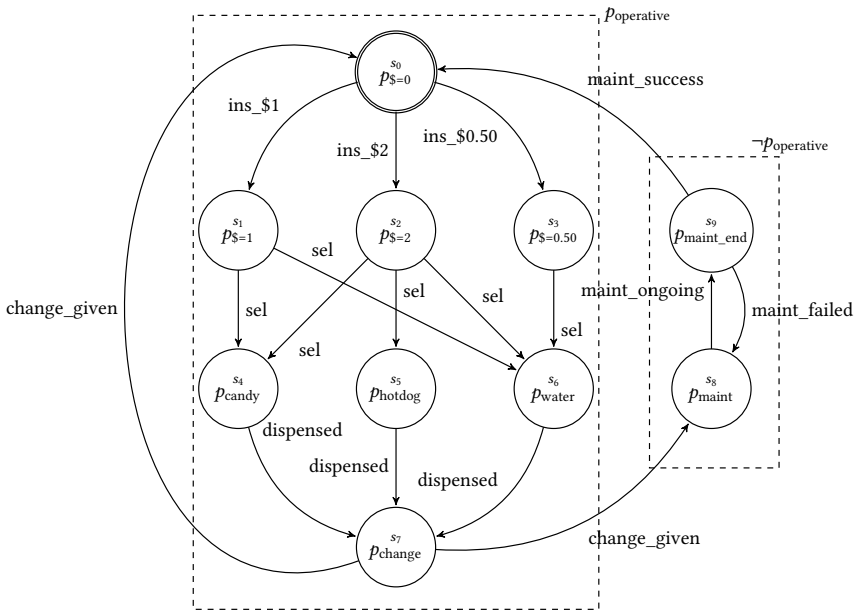

The Kripke structure of Figure~\ref{fig:vending} represents a \emph{vending machine}, which can dispense water, hot dogs, and candies.
In state $s_0$ (the initial one), no coin has been inserted into the machine (hence, the proposition letter $p_\text{\$=0}$ holds there).
Three edges, labelled by \lq\lq ins\_\$1\rq\rq, \lq\lq ins\_\$2\rq\rq, and \lq\lq ins\_\$0.50\rq\rq{}, connect $s_0$ to $s_1$, $s_2$, and $s_3$, respectively. 
Edge labels do not convey semantic value (they are neither part of the structure definition nor associated with proposition letters) and are simply used for an easy reference to edges. 
In $s_1$ (resp., $s_2$, $s_3$) the proposition letter $p_\text{\$=1}$ (resp., $p_\text{\$=2}$, $p_\text{\$=0.50}$) holds, representing the fact that 1 Dollar (resp., 2, 0.50 Dollars) has been inserted into the machine.
The cost of a bottle of water (resp., a candy, a hot dog) is \$0.50 (resp., \$1, \$2). A state $s_i$, for $i=1,2,3$, is connected to a state $s_j$, for $j=4,5,6$, only if the available credit allows one to buy the corresponding item.
Then, edges labelled by \lq\lq dispensed\rq\rq{} connect $s_4,s_5$, and $s_6$ to $s_7$. In $s_7$, the machine gives change, and  can nondeterministically move back to $s_0$ (ready for dispensing another item), or to $s_8$, where it begins an automatic maintenance activity ($p_\text{maint}$ holds there). Afterwards, state $s_9$ is reached, where maintenance ends. From there, if the maintenance activity fails (edge \lq\lq maint\_failed\rq\rq ), $s_8$ is reached again (another maintenance cycle is attempted); otherwise, maintenance concludes successfully (\lq\lq maint\_success\rq\rq ) and $s_0$ is reached. Since the machine is operating in states $s_0,\ldots,s_7$, and under maintenance in $s_8$ and $s_9$, $p_\text{operative}$ holds over the former, and it does not on the latter.

In the following, we will make use of the $\B$ formulas $\Length_{n}$, with $n\geq 1$, presented in Example~\ref{example:length}.

We now give some examples of properties we can formalize under all, or some, of the $\HS$ semantic variants $\HS_\stat$, $\HS_\LinearPast$, and $\HS_\LinearTime$.

\begin{itemize}
    \item In any run of length 50, during which the machine never enters maintenance mode, it dispenses at least a hotdog, a bottle of water and a candy. 
    \begin{multline*}
        \Ku\not\models (p_\text{operative} \wedge \Length_{50})\longrightarrow\\ \big((\hsB\hsE p_\text{hotdog}) \wedge  (\hsB\hsE p_\text{water}) \wedge (\hsB\hsE p_\text{candy}) \big)
    \end{multline*}
    Clearly this property is false, as the machine can possibly dispense only one or two kinds of items.
    We start by observing that the above formula is equivalent in all of the three semantic variants of $\HS$: since modalities $\hsB$ and $\hsE$ only allow one to \lq\lq move\rq\rq{} from an interval to its subintervals, $\B\E_\LinearTime$, $\B\E_\stat$, and $\B\E_\LinearPast$ coincide (for this reason, we have omitted the subscript from the symbol $\models$). 
    Homogeneity plays a fundamental role here: asking $p_\text{operative}$ to be true implies that such a letter is true along the whole trace (thus $s_8$ and $s_9$ are always avoided).
    
    It is worth observing that the same property can be expressed in $\LTL$, for instance as follows:
    \[
    \smashoperator{\bigwedge_{i\in\{0,\ldots,49\}}}\Next^i p_\text{operative} \wedge\smashoperator[r]{\bigvee_{i,j,k\in\{1,\ldots,48\}, i\neq j\neq k\neq i}} (\Next^i p_\text{hotdog}) \wedge (\Next^j p_\text{water}) \wedge (\Next^k p_\text{candy}).
    \]
    The length of this $\LTL$ formula is \emph{exponential} in the number of items (in this case, 3), whereas the length of the above $\HS$ one is only linear. As a matter of fact, we will prove (Theorem~\ref{theo:succinctnessHSlin}) that $\B\E$ is at least exponentially more succinct than $\LTL$.
    
    \item If the credit is \$0.50, then no hot dog or candy may be provided.
    \[
        \Ku\models (\hsE p_\text{\$=0.50})\longrightarrow \neg \hsA (\Length_{2} \wedge \hsE (p_\text{hotdog} \vee p_\text{candy})) 
    \]
    We observe that a trace satisfies $\hsE p_\text{\$=0.50}$ if and only if it ends in $s_3$.
    This property is satisfied under all of the three semantic variants, even though the nature of future differs among them (recall Figure~\ref{fig:ST}, \ref{fig:CT}, and \ref{fig:LN}). As we have already mentioned, a linear setting (rather than branching) is suitable for the specification  of dynamic behaviors, because it considers states \emph{of a computation}; conversely, a branching approach focuses on machine states (and thus on the structure of a system).
    
    In this case, only the state $s_6$ can be reached from $s_3$, regardless of the nature of future. For this reason, $\HS_\stat$, $\HS_\LinearPast$, and $\HS_\LinearTime$ behave in the same way. 
    
    \item Let us exemplify now a difference between $\HS_\stat$ (and $\HS_\LinearPast$) and $\HS_\LinearTime$.
    \[
        \begin{array}{l}
            \Ku\models_\stat \\
            \Ku\models_\LinearPast  \\
            \Ku\not\models_\LinearTime
        \end{array}
        (\hsE p_\text{maint\_end})\longrightarrow \hsA\hsE p_\text{operative}
    \]
    This is a structural property, requiring that when the machine enters state $s_9$ (where maintenance ends), it can become again operative reaching state $s_0$ ($s_9$ is not a lock state for the system). This is clearly true when future is branching and it is not when future is linear: $\HS_\LinearTime$ refers to system computations, and some of these may ultimately loop between $s_8$ and $s_9$.
    
    \item 
    Conversely, some properties make sense only if they are predicated over computations. This is the case, for instance, of fairness.
    \[
        \begin{array}{l}
            \Ku\models_\stat \\
            \Ku\models_\LinearPast  \\
            \Ku\not\models_\LinearTime
        \end{array}
        (\hsAu\hsA\hsE p_\text{maint})\longrightarrow \hsAu\hsA\hsE p_\text{operative}
    \]
    Assuming the trace-based semantics, the property requires that if a system computation enters infinitely often into maintenance mode, it will infinitely often enter operation mode.
    Again, this is not true, as some system computations may ultimately loop between $s_8$ and $s_9$ (hence, they are not fair). On the contrary, such a property is trivially true under $\HS_\stat$ or $\HS_\LinearPast$, as, for any initial trace $\rho$, it holds that $\Ku,\rho\models \hsA\hsE p_\text{operative}$.
    
    \item We conclude with a property showing the difference between linear and branching \emph{past}, that is, between $\HS_\stat$ and $\HS_\LinearTime$ (and $\HS_\LinearPast$).
    The requirement is the following: the machine may dispense water with any amount of (positive) credit.
    \[
        \begin{array}{l}
            \Ku\models_\stat \\
            \Ku\not\models_\LinearPast  \\
            \Ku\not\models_\LinearTime
        \end{array}
         (\hsE p_\text{water})\longrightarrow \hsE\big(p_\text{water}\wedge\smashoperator[r]{\bigwedge_{p\in\{p_\text{\$=2},p_\text{\$=1},p_\text{\$=0.50}\}}}\hsAt(\Length_{2} \wedge\hsB p)\big)
    \]
    Again, this one is a structural property, that cannot be expressed in $\HS_\LinearTime$ or $\HS_\LinearPast$, as these refer to a specific computation in the past. Conversely, it is true under $\HS_\stat$, since $s_6$ is backward reachable in one step by $s_1$, $s_2$, and $s_3$.
\end{itemize}

%% file: Chaps/TOCL17/conclusion.tex
\section{Conclusions}
In this chapter we have compared $\HS$ MC to MC for point-based temporal logics as far as it concerns expressiveness (and succinctness). To this end, we have taken into consideration three semantic variants  of $\HS$, namely, $\HS_\stat$, $\HS_\LinearPast$, and $\HS_\LinearTime$, under the homogeneity assumption. We have investigated their expressiveness and contrasted them with the point-based temporal logics $\LTL$, $\CTL$, finitary $\CTLStar$, and $\CTLStar$.

The resulting picture is as follows: $\HS_\LinearTime$ and $\HS_\LinearPast$ turn out to be as expressive as $\LTL$ and finitary $\CTLStar$, respectively. Moreover, $\HS_\LinearTime$ is at least exponentially more succinct than $\LTL$.
$\HS_\stat$ is expressively incomparable with $\HS_\LinearTime$/$\LTL$, $\CTL$, and $\CTLStar$, but it is strictly more expressive than $\HS_\LinearPast$/finitary $\CTLStar$.
We believe it possible to fill the expressiveness gap between $\HS_\LinearPast$ and $\CTLStar$ by considering abstract interval models, induced by Kripke structures, featuring worlds also for infinite paths/intervals, and extending the semantics of $\HS$ modalities accordingly. 

It is finally worth noting that the decidability of the MC problem for (full) $\HS_\LinearPast$ and $\HS_\LinearTime$ immediately follows, as a byproduct, from the proved results.

%% file: Chaps/Gandalf17RIVISTA/Gand17main.tex
\chapter[MC for $\HS$ and its fragments with regular expressions]{MC for $\HS$ and its fragments \\ extended with regular expressions}\label{chap:Gand17}
\begin{chapref}
The references for this chapter are \cite{sefm17,gandalf17,icRegex}.
\end{chapref}

\minitoc\mtcskip

\newcommand{\Trans}{\Edges}
\newcommand{\epist}{\text{epistemic}}
\renewcommand{\Instance}{\mathcal{I}}

\renewcommand{\Init}{\textit{Init}}

\newcommand{\GLab}{\textit{Lab}\,}

\renewcommand{\M}{{\mathcal{M}}}
\newcommand{\WS}{{\mathcal{W}}}

\newcommand{\FMC}{\mathsf{FMC}}

\newcommand{\acc}{\textit{acc}}

\newcommand{\subfB}{\mathsf{Subf}_{\hsB}}
\newcommand{\conf}{\mathsf{Conf}}
\newcommand{\SD}{\textsf{SD}}


\newcommand{\Summary}{\ensuremath{\mathcal{S}}}
\newcommand{\PrefS}{{\mathit{PS}}}
\renewcommand{\End}{\textit{end}}
\newcommand{\NU}{\textit{not\_unique}}
\newcommand{\Double}{\textit{double}}
\newcommand{\Single}{\textit{single}}
\newcommand{\Row}{\textit{row}}
\newcommand{\Column}{\textit{col}}
\newcommand{\CInc}{\textit{c,inc}}
\newcommand{\RInc}{\textit{r,inc}}
\newcommand{\Complete}{\textit{comp}}
\newcommand{\MC}{\textit{mc}}
\newcommand{\IC}{\textit{ic}}
\newcommand{\MT}{\textit{MT}}
\newcommand{\IMT}{\textit{IMT}}
\newcommand{\Coh}{\textit{coh}}
\newcommand{\Pos}{\textit{P}}
\newcommand{\Alt}{\Upsilon_w}
\newcommand{\AltN}{\Upsilon}

\input{Chaps/Gandalf17RIVISTA/intro}
\input{Chaps/Gandalf17RIVISTA/Preliminaries}

\input{Chaps/Gandalf17RIVISTA/ModelCheckingFullHS}

\input{Chaps/Gandalf17RIVISTA/TrackProperty}
\input{Chaps/Gandalf17RIVISTA/ModelCheckingAAbarBBbarEbar}

\input{Chaps/Gandalf17RIVISTA/AAbarEEbar}
\input{Chaps/Gandalf17RIVISTA/BBbar}

\input{Chaps/Gandalf17RIVISTA/AAbarElimin}
\input{Chaps/Gandalf17RIVISTA/PropHard}
\input{Chaps/Gandalf17RIVISTA/concl}


%% file: Chaps/Gandalf17RIVISTA/intro.tex
\lettrine[lines=3]{A}{\ good balancing} of \emph{expressiveness} and \emph{complexity} in the choice 
of the system 
model and of the specification language is a key factor for the effective 
exploitation of MC. 
Various improvements to both of them have been proposed in the literature. As 
for the latter we recall in particular the extensions of $\LTL$ with 
promptness, that make it 
possible to bound the delay with which a liveness request is fulfilled (see, 
e.g.,~\cite{DBLP:journals/fmsd/KupfermanPV09}). 
Another possible direction is
\emph{adding regular expressions}, that allow one to enrich the expressive 
power of existing logics. This has been investigated, for instance, in the 
cases of $\LTL$~\cite{Leucker2007} and $\CTL$~\cite{MATEESCU20112854}.
In this chapter, we study the MC problem for \emph{$\HS$ extended with regular 
expressions}, which are exploited as a means for \emph{relaxing the homogeneity 
assumption}, that otherwise limits the expressive power of $\HS$. 

As we have already said (Section~\ref{sec:LOMrelated}), in~\cite{lm16} 
Lomuscio and Michaliszyn propose to use regular expressions to define the 
labeling of proposition letters over intervals in terms of the component 
states---thus relaxing homogeneity, that can be trivially \lq\lq 
encoded\rq\rq{} by regular expressions, as shown later. In that work the 
authors prove decidability of MC with regular expressions for some very 
restricted fragments of epistemic $\HS$, giving some rough upper bounds to its 
computational complexity
(see the fourth column of Table~\ref{fig:overv}).
In this chapter, 
we define interval labeling via regular expressions in a way that 
can be shown to be equivalent to that of~\cite{lm16}, and
we give a detailed picture of decidability and complexity for $\HS$ with 
regular expressions, which was still missing. The results are summarized in the 
third column of Table~\ref{fig:overv}. 

\begin{table}
	\centering
	\caption{Complexity of MC for $\HS$ and its fragments ($^\dagger$local MC). 
	In red, the new results proved in this chapter.}\label{fig:overv}
	\resizebox{\textwidth}{!}{
		\input{Chaps/Gandalf17RIVISTA/overvTable}
	}
\end{table}

It is interesting to compare the complexity of MC for $\HS$ fragments extended with regular expressions with the same fragments under the homogeneity assumption. The rich spectrum of complexities for the less expressive fragments of $\HS$ under homogeneity (last four rows in the table) collapses to $\Psp$-completeness in the case of the corresponding fragments with regular expressions, witnessing that using regular expressions increases the expressive power of (syntactically) small fragments of $\HS$. Whether or not there exists an elementary algorithm for full $\HS$ remains an open issue, just like in the case of full $\HS$ under homogeneity. The main results of the chapter are summarized in the following account of the contents of the next sections.

\paragraph*{Organization of the chapter.}
\begin{itemize}

\item
In Section~\ref{sect:backgrRegex}, we start by recalling the notions of regular expression and finite-state automaton, and then give syntax and semantics of $\HS$ with regular expressions.

\item
In Section \ref{sect:fullHS}, we prove that MC for (full) $\HS$ extended with regular expressions (under the state-based semantics) is decidable,
by exploiting an automata-theoretic approach and the notion of $\Ku$-\NFA, a particular version of finite-state automaton. 
Moreover, the problem can be shown to be in $\PTIME$ when it is restricted to 
system models, assuming the formula to be of constant length. 

\item
Then, in Section~\ref{sec:AAbarBBbarEbarRegex},
we study the problems of MC for the two (syntactically) maximal (symmetric) 
fragments $\A\Abar\B\Bbar\Ebar$ and $\A\Abar\E\Bbar\Ebar$ with regular 
expressions, proving that both problems are $\LINAEXPTIME$-complete. 
$\LINAEXPTIME$ denotes the complexity class of problems decided by   
\emph{exponential-time bounded} alternating Turing machines making a 
\emph{polynomially 
bounded number of alternations}; such a class captures the exact complexity of 
some relevant problems~\cite{tcs15l,FR75}, such as, for instance, the 
first-order 
theory of real addition with order~\cite{FR75}.
First, we recall (Chapter~\ref{chap:TCS17}) that settling the exact complexity 
of these fragments under the homogeneity assumption is a difficult open 
question. Moreover,  considering that $\LINAEXPTIME \subseteq \AEXP = \EXPSPACE$ and 
that $\HS$ under homogeneity  is subsumed by $\HS$ with regular expressions (as 
we will see later), 
the results proved in this chapter improve  the complexity upper bounds for the 
fragments  
$\A\Abar\B\Bbar\Ebar$ and $\A\Abar\E\Bbar\Ebar$ given in 
Section~\ref{sec:AAbarBBbarEbar}. 
More in detail, we preliminarily establish an exponential small-model property 
for $\A\Abar\B\Bbar\Ebar$ (Section~\ref{sec:AAbarBBbarEbarTrackProperty}): for 
each interval (trace),  it is possible to find an interval of bounded 
exponential 
length that is indistinguishable with respect to the fulfillment of 
$\A\Abar\B\Bbar\Ebar$ formulas (respectively, $\A\Abar\E\Bbar\Ebar$ formulas).
Such a property allows us to devise an MC procedure belonging to the class $\LINAEXPTIME$ (Section~\ref{sec:UpperBound}). 
Finally, the 
matching lower bounds are obtained 
by polynomial-time reductions from the so-called \emph{alternating multi-tiling 
problem}, showing that they already hold for the fragments $\B\Ebar$ and 
$\E\Bbar$ of $\A\Abar\B\Bbar\Ebar$ and $\A\Abar\E\Bbar\Ebar$, respectively 
(Section~\ref{sec:LowerBound}). 

\item
Finally, in Section~\ref{sec:AABB}, we show that 
formulas of $\HS$ fragments featuring (any subset of) $\HS$ modalities for the Allen's relations \emph{meets, met-by, started-by}, and \emph{starts} ($\AAbarBBbar$) can be checked in polynomial working space (MC for all these is $\Psp$-complete). 
In particular, in Section~\ref{subsec:AAbarEEbar} we prove a small-model 
theorem for $\AAbarBBbar$ (and the symmetric $\AAbarEEbar$) with regular 
expressions, which is then 
exploited in Sections~\ref{sect:PspAlgo} and \ref{sect:genResult} to devise a 
$\Psp$ MC algorithm for $\AAbarBBbar$ (and $\AAbarEEbar$). Moreover, in 
Section~\ref{sect:genResult}, we prove that MC for the purely propositional 
fragment of $\HS$, denoted as $\HSprop$, is hard for $\Psp$, which is enough to 
conclude that MC for any sub-fragment of $\AAbarBBbar$ or $\AAbarEEbar$ is 
complete for $\Psp$.
Hence, relaxing the homogeneity assumption via regular expressions comes at no 
cost for $\AAbarBBbar$, $\AAbarEEbar$, $\B\Bbar$, $\E\Ebar$, $\Bbar$ and 
$\Ebar$---that remain in $\Psp$---while $\AAbarB$ and $\A\Abar\E$ and their 
sub-fragments
increase their complexity to $\Psp$ (see Table~\ref{fig:overv} once more). 
\end{itemize}

%% file: Chaps/Gandalf17RIVISTA/overvTable.tex
\begingroup
\renewcommand*{\arraystretch}{1.3}
\begin{tabular}{|@{\ }c@{\ }|@{\ }c@{\ }|@{\ }c@{\ }||@{\ }c@{\ }|}
\hline 
 & Homogeneity & Regular expressions & Endpoints~\cite{LM13,LM14,lm16}\\
\hline \hline 
\multirow{2}{*}{Full $\HS$, $\B\E$} & nonelementary  & \textcolor{red}{nonelementary}  & $\B\E$+$\epist^\dagger$: $\Psp$ \\
 & $\EXPSPACE$-hard & $\EXPSPACE$-hard & $\B\E^\dagger$: $\PTIME$\\
\hline 
\multirow{2}{*}{$\A\Abar\B\Bbar\Ebar,\A\Abar\E\Bbar\Ebar$} & $\in\EXPSPACE$ \textcolor{red}{$\in\LINAEXPTIME$}& nonelem.\  $\Psp$-hard &\\
 & $\Psp$-hard & \textcolor{red}{$\LINAEXPTIME$-complete}& \\
\hline 
\multirow{2}{*}{$\AAbar\Bbar\Ebar$} & \multirow{2}{*}{$\Psp$-complete} & \textcolor{red}{$\in\LINAEXPTIME$} & \\
 & & $\Psp$-hard & \\
\hline
$\AAbarBBbar,\B\Bbar,\Bbar,$ & \multirow{2}{*}{$\Psp$-complete} & \multirow{2}{*}{\textcolor{red}{$\Psp$-complete}} & \multirow{2}{*}{$\A\Bbar$+$\epist$: nonelementary}\\
$\AAbarEEbar,\E\Ebar,\Ebar$ & & & \\
\hline 
$\AAbar\B,\AAbar\E,\A\B,\Abar\E$ & $\PTIME^{\NP}\!$-complete & \textcolor{red}{$\Psp$-complete} & \\
\hline 
\multirow{2}{*}{$\A\Abar,\Abar\B,\A\E,\A,\Abar$} & $\in\Thsq$ & \multirow{2}{*}{\textcolor{red}{$\Psp$-complete}} & \\
 & $\Th$-hard &  & \\
\hline 
$\HSprop, \B,\E$ & $\co\NP$-complete & \textcolor{red}{$\Psp$-complete} & \\
\hline 
\end{tabular}

\endgroup

%% file: Chaps/Gandalf17RIVISTA/Preliminaries.tex
\section{Preliminaries}\label{sect:backgrRegex}
We start by recalling the notion of regular expressions over finite words. 
Since, as it will be clear later, we are interested in expressing requirements 
over finite words over $2^{\Prop}$, here we consider \emph{proposition-based} 
regular expressions (denoted as $\RE$s), where atomic expressions are 
propositional (Boolean) formulas over $\Prop$, instead of just letters over an 
alphabet. Formally, the set of $\RE$s $r$ over $\Prop$ is defined by the 
grammar:
\[
    r ::= \varepsilon\;\vert\; \phi\;\vert\; r\cup r\;\vert\; r\cdot r\;\vert\; r^{*},
\]
where $\phi$ is a propositional formula over $\Prop$. The length $|r|$ of an $\RE$ $r$ is the number of subexpressions of $r$.
 An $\RE$ $r$ denotes a language $\Lang(r)$ of finite words over $2^{\Prop}$ defined as:
\begin{itemize}
  \item $\Lang(\varepsilon)=\{\varepsilon\}$,
  \item $\Lang(\phi)=\{A\in 2^{\Prop}\mid A \text{ satisfies }\phi\}$,
  \item $\Lang(r_1\cup r_2)=\Lang(r_1)\cup \Lang(r_2)$,
  \item $\Lang(r_1\cdot r_2)=\Lang(r_1)\cdot \Lang(r_2)$, 
\item $\Lang(r^{*})=(\Lang(r))^{*}$.
\end{itemize}
By well-known results,  the class of $\RE$ over $\Prop$ captures the class of regular languages of finite words over $2^{\Prop}$.

\begin{example}\label{example:re}
An example of  $\RE$ is $r_1=\mathbf{(p\wedge s)} \cdot \mathbf{s}^* \cdot \mathbf{(p\wedge s)}$ that intuitively denotes the set of finite words where both $\mathbf{p}$ and $\mathbf{s}$ hold true on the endpoints, and $\mathbf{s}$ continuously holds in all internal symbols/sets of $2^{\Prop}$. The $\RE$ $r_2=\mathbf{(\neg p)}^*$ denotes the set of finite words such that $\mathbf{p}$ does not hold in any position.
\end{example}

We also recall the standard notion of
\emph{non-deterministic finite state automaton} ($\NFA$), which is a tuple 
$\Au =\tpl{\Sigma,Q,Q_0,\Delta,F}$, where $\Sigma$ is a finite alphabet, $Q$ is a finite set of states, $Q_0\subseteq Q$ is the set of initial states,
$\Delta: Q\times \Sigma \to  2^Q$ is the transition function (or, equivalently, 
$\Delta\subseteq Q\times \Sigma \times Q$), and $F\subseteq Q$ is the set of 
accepting states. 
An $\NFA$ $\Au$ is \emph{complete} if, for all $(q,\sigma)\in Q\times \Sigma$, $\Delta (q,\sigma)\neq \emptyset$.
Given a finite word $w$ over $\Sigma$, with $|w|=n$, and two states $q,q'\in 
Q$, a \emph{run} (or \emph{computation}) of $\Au$ from $q$ to $q'$ over $w$ is 
a finite sequence of states $q_1,\ldots,q_{n+1}$ such that $q_1=q$, 
$q_{n+1}=q'$, and $q_{i+1} \in \Delta(q_i,w(i))$ for all $i\in [1,n]$. The 
language $\Lang(\Au)$  \emph{accepted by} $\Au$ consists of the set of  finite 
words $w$ over $\Sigma$ such that there is a run over $w$ from some initial 
state to some accepting state.

A \emph{deterministic finite state automaton} ($\DFA$) is an $\NFA$ $\Du=\tpl{\Sigma,Q,Q_0,\Delta,F}$ such that $Q_0$ is a singleton, and for all $(q,c)\in Q\times \Sigma$, $\Delta(q,c)$ is a singleton. In the following, in the case of a $\DFA$, we will denote the transition function $\Delta$ as $\delta$.

\begin{remark}\label{remk:nfa}
By well-known results, given an $\RE$ $r$ over $\Prop$, one can construct, in a compositional way, an $\NFA$ $\Au_r$ with alphabet $2^{\Prop}$, whose number of states is at most $2|r|$, such that $\lang(\Au_r) = \lang(r)$. 
We call $\Au_r$ the \emph{canonical} $\NFA$ associated with $r$. 

Note that, though the number of edges of $\Au_r$ may be exponential in $|\Prop|$ (edges are labelled by assignments $A\in 2^\Prop$
satisfying propositional formulas $\phi$ of $r$), we can avoid explicitly 
storing edges, as they can be recovered in polynomial time from $r$. 

In  
Figure~\ref{fig:NFAex}, we depict the canonical $\NFA$ $\Au_{r_1}$ associated 
with the $\RE$ $r_1$ of  Example~\ref{example:re}.
We can avoid storing the edges of $\Au_{r_1}$ by remembering which propositional formulas of $r_1$ they are associated with.

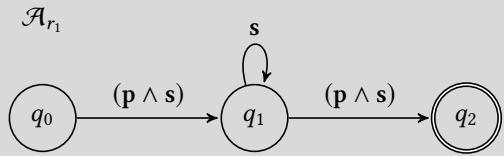
\begin{figure}[H]
\centering
\begin{tikzpicture}[->,>=stealth',shorten >=1pt,auto,node distance=2.8cm,
                    semithick]
  \node[state] 		(A) 			 {$q_0$};
  \node[state]         		(B) [right of=A] {$q_1$};
  \node[accepting,state]    (D) [right of=B] {$q_2$};
  \node[draw=none] 			(Z) [above of=A, yshift=-1.5cm] {$\mathcal{A}_{r_1}$};

  \path (A) edge              node {$\mathbf{(p\wedge s)}$} 	(B)
        (B) edge [loop above] node {$\mathbf{s}$} 			(B)
        (B) edge              node {$\mathbf{(p\wedge s)}$} 	(D);
\end{tikzpicture}
\caption{The canonical $\NFA$ $\Au_{r_1}$ associated with the $\RE$ $r_1$ of Example~\ref{example:re}.
}\label{fig:NFAex}
\end{figure}
\end{remark}

In the previous chapters, $\HS$ formulas are evaluated over intervals which correspond to the traces of a Kripke structure $\Ku$. The approach followed is subject to two restrictions: $(i)$ the set of  proposition letters of $\HS$ formulas and the set $\Prop$ of proposition letters of the  Kripke structure
coincide, and $(ii)$ a proposition letter holds over an interval if and only if it holds over all its sub-intervals (\emph{homogeneity assumption}). 
Here we adopt a more general and expressive approach.

Let $\mathpzc{P}_u$ be a finite set of \emph{abstract (uninterpreted) interval properties}.
An abstract interval property $p_u\in \mathpzc{P}_u$ denotes a regular language of finite words over $2^{\Prop}$.
More specifically,
every $p_u$ is a proposition-based regular expression over $\Prop$,
and it \emph{occurs as a proposition letter} in $\HS$ formulas. 
Thus, hereafter, an \HS\ formula $\varphi$ over $\Prop$ is an $\HS$ formula whose proposition letters (i.e., atomic formulas) are $\RE$s $r$ over $\Prop$.
For this reason, we define
the size (or length) $|\varphi |$ of $\varphi$ as the number of non-atomic subformulas of $\varphi$ plus $\sum_{r\in \SPEC} |r|$, where $\SPEC$ is the set of $\RE$s occurring in $\varphi$. 

We now define the semantics of an \HS\ formula $\varphi$ over $\Prop$ on a 
trace $\rho$ of a (finite) Kripke structure $\Ku=\KuDef$.
For the purpose we need the following definition.
\begin{definition}[Labelling sequence induced by a trace]
 A trace $\rho\in\Trk_\Ku$, with $|\rho|=n$, induces a finite word over $2^{\Prop}$, denoted as $\mu(\rho)$ and called \emph{labeling sequence}, defined as \[\mu(\rho(1))\cdots \mu(\rho(n)).\]
\end{definition}
Now, the satisfaction relation $\Ku,\rho\models \varphi$ can be defined inductively as follows (we omit the standard clauses for Boolean connectives):
\begin{itemize}
    \item $\Ku,\rho   \models r$ if and only if $ \mu(\rho) \in \Lang(r)$ for each $\RE$ $r$ over $\Prop$,
    \item $\Ku,\rho  \models \hsB\varphi$ if and only if there exists $\rho'\in\Pref(\rho)$ such that $\Ku,\rho' \models \varphi$,
    \item $\Ku,\rho  \models \hsE\varphi$ if and only if there exists $\rho'\in\Suff(\rho)$ such that $\Ku,\rho' \models \varphi$,
    \item $\Ku,\rho   \!\models\! \hsBt\varphi$ if and only if $\Ku,\rho'   \!\models\! \varphi$ for some trace $\rho'$ such that $\rho\!\in\!\Pref(\rho')$, 
    \item $\Ku,\rho   \!\models\! \hsEt\varphi$ if and only if $\Ku,\rho'   \!\models\! \varphi$ for some trace $\rho'$ such that $\rho\!\in\!\Suff(\rho')$. 
\end{itemize} 

As in the previous chapters, 
we say that $\Ku$ is a \emph{model} of $\varphi$, denoted as $\Ku\models 
\varphi$, if, for all \emph{initial} traces $\rho$ of $\Ku$, it holds that 
$\Ku,\rho\models \varphi$. The \emph{MC problem} for $\HS$ is the problem of 
checking, given a finite Kripke structure $\Ku$ and an $\HS$ formula $\varphi$, 
whether or not $\Ku\models \varphi$. Again, the problem is not trivially 
decidable since the set $\Trk_\Ku$ of traces of $\Ku$ is infinite.

With reference to Chapter~\ref{chap:TOCL17}, we recall that
the considered \emph{state-based semantics} provides a \emph{branching-time 
setting both in 
the past and in the future}. In particular, while modalities for $\hsB$ and 
$\hsE$ are linear-time (as they allow us to select prefixes and suffixes of the 
current trace only), modalities for $\hsA$ and $\hsBt$ (respectively, $\hsAt$ 
and $\hsEt$) are branching-time in the future (respectively, in the past) since 
they enable us to nondeterministically extend a trace in the future 
(respectively, in the past). 

As shown in Chapter~\ref{chap:TOCL17}, under the 
considered semantic version, the logics $\HS$ and $\CTLStar$ are expressively 
incomparable already under the homogeneity assumption. However, under such an 
assumption, the use of  the past branching-time modalities $\hsAt$ and $\hsEt$ 
is necessary for capturing requirements which cannot be expressed in 
$\CTLStar$. For instance, the constraint
``\emph{each state reachable from the initial one where $p$ holds has a 
predecessor where $p$ holds as well}'' cannot be expressed in \CTLStar, but can 
be easily stated in the fragment $\AbarE$ (see Section~\ref{sect:allSems}). 

Conversely, in the more expressive  setting based on regular expressions, the future branching-time modalities $\hsA$ and $\hsBt$ are already sufficient for capturing requirements which cannot be expressed in \CTLStar, such as the following branching-time bounded response property: ``\emph{for each state reachable from the initial one, where a request  $\emph{req}$ occurs, there is a computation starting from this state such that the request is followed by a response $\emph{res}$ after an  \emph{even number} of steps}''. This requirement can be expressed in the  fragment $\ABbar$ as follows: 
\[\hsAu(\text{req} \rightarrow \hsBt (\text{req}\cdot (\top\cdot \top)^{*}\cdot \text{res})).\] 
Note that it says nothing about the possible occurrence of a response 
$\emph{res}$ after an \emph{odd} number of steps.

Before moving on, we show that
in this setting it is easy to force homogeneity: we just have to impose that 
all 
regular expressions in the formula have the form $p\cdot p^*$, for some 
$p\in\Prop$.
Additionally, labelling of 
intervals by endpoints can be easily captured by regular expressions having the form:
\[\bigcup_{(i,j)\in I} (q_i\cdot \top^*\cdot q_j)\cup\bigcup_{i\in I'} q_{i},\]
for some suitable sets of indexes $I\subseteq \{1,\ldots,|S|\}^2$ and 
$I'\subseteq \{1,\ldots,|S|\}$, where $q_i\in\Prop$ is a letter labeling the 
state $s_i\in S$ of $\Ku$, only.

\begin{example}[Adapted from \cite{lm16}]
With this (toy) example we want to compare the effectiveness of regular expressions as rules for defining interval labelling, to homogeneity and to endpoint-based labelling.

In Figure~\ref{fig:printing}, a Kripke structure representing a printer is shown.
In $\sinit$ the printer starts printing a sheet; in $s_1$ the process is ongoing, and it ends in $s_2$. The printer then prints the subsequent sheet, by \lq\lq moving\rq\rq{} back to $\sinit$.

Imagine we want to label the process of printing a \emph{single sheet} by $p$ (i.e., only the trace $s_0s_1s_2$).
Under homogeneity, if $s_0s_1s_2$ is labeled by $p$, then $s_0$, $s_1$, $s_2$, $s_0s_1$, $s_1s_2$ must be all labeled by $p$ as well, against our idea.
In the endpoint-based approach, in order for $p$ to label $s_0s_1s_2$, $p$ must also label all traces $(s_0s_1s_2)^n$ for $n\in \Nat^+$, as the endpoints of all these are $s_0,s_2$. Thus \lq\lq several, consecutive sheets\rq\rq{} are labelled $p$.

Conversely, we just write the proposition-based $\RE$ $\mathbf{p_\text{st}} \cdot ( \mathbf{\neg p_\text{end}\wedge \neg p_\text{st}})^* \cdot \mathbf{p_\text{end}}$ to capture precisely the trace $s_0s_1s_2$.

\begin{figure}[H]
    \centering
    \begin{tikzpicture}[->,>=stealth,thick,shorten >=1pt,auto,node distance=2cm,every node/.style={circle,draw}]
		\node [style={double}](v0) {$\stackrel{s_0}{p,p_\text{st}}$};
		\node (v1) [right of=v0] {$\stackrel{s_1}{p}$};
		\node (v2) [right of=v1] {$\stackrel{s_2}{p,p_\text{end}}$};

		\draw (v0) to (v1);
		\draw (v1) to (v2);
		\draw (v2) to [bend right] (v0);
		\end{tikzpicture}
    \caption{Kripke structure representing a printer}
    \label{fig:printing}
\end{figure}
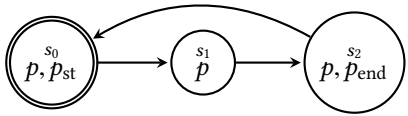
\end{example}

%% file: Chaps/Gandalf17RIVISTA/ModelCheckingFullHS.tex
\section{MC for full $\HS$}\label{sect:fullHS}

In this section, we develop an automata-theoretic approach to the MC problem 
for full $\HS$ with regular expressions. 

Given a finite Kripke structure $\Ku=\KuDef$ and an $\HS$ formula $\varphi$ over 
$\Prop$, we compositionally construct an \NFA\ over the set of states $S$ of 
$\Ku$ accepting the set of traces $\rho$ of $\Ku$ such that 
$\Ku,\rho\models\varphi$. The size of the resulting \NFA\ is nonelementary, but 
it is just \emph{linear in the size of $\Ku$}. To prove that the nonelementary 
blow-up does not depend on $\Ku$, we introduce a special subclass of  $\NFA$s, 
called  $\Ku$-\NFA, which intuitively represents the \lq\lq 
synchronization\rq\rq\ of an \NFA\ with the  Kripke structure $\Ku$. In this 
way, a $\Ku$-\NFA\ may only accept traces of $\Ku$. 

\begin{definition}[$\Ku$-\NFA]
A $\Ku$-\NFA\   is an \NFA\ $\Au=(S,Q,Q_0,\Delta,F)$ over $S$ satisfying the 
following conditions:
\begin{itemize}
  \item the set $Q$ of states has the form $M\times S$ ($M$ is called the 
  \emph{main component}, or the set of \emph{main states});
  \item $Q_0\cap F = \emptyset$, that is, the empty word $\varepsilon$ is not accepted;
  \item for all $(q,s)\in M\times S$ and $s'\in S$, we have 
  $\Delta((q,s),s')=\emptyset$ if $s' \neq s$,
and $\Delta((q,s),s)\subseteq M\times \Edges(s)$%
\footnote{We recall that $\Edges(s)$ is the set of successors of $s$ in $\Ku$.}.
\end{itemize}
\end{definition}
It is worth noticing that,
for all words $\rho\in S^{+}$, if there is a run of the $\Ku$-\NFA\ over $\rho$, then $\rho$ is a trace of  $\Ku$. In the following, we construct a $\Ku$-\NFA\ $\Au$ accepting the traces $\rho$ of $\Ku$ such that $\Ku,\rho\models\varphi$. 

In a standard automata-theoretic approach, an automaton accepting the set of models of $\varphi$ would be first defined, and then intersected with $\Ku$. In the following construction, the synchronization with $\Ku$ is instead implicitly associated with the construction of the  $\Ku$-\NFA\ itself.
Such a choice is motivated by the fact that proposition letters in the formula $\varphi$ (the base case in the construction) are regular expressions which have to be synchronized with 
the traces of $\Ku$. Such a synchronization  is then maintained along the whole process of
$\Ku$-\NFA{} construction. 

The recursive step for dealing with negation in 
$\varphi$ is noteworthy, since it is not just a pure complementation of the  $\Ku$-\NFA\ under construction. As a matter of fact, only the synchronized \NFA-component (for the regular expressions of $\varphi$) has to be complemented, whereas the synchronized $\Ku$-component does not. For this reason, the size of the final $\Ku$-\NFA{} is nonelementary, but linear in the size of $\Ku$.

%

In order to prove the main result of the section (stated in Theorem \ref{th-m}), we preliminarily 
describe the composition steps to build the required $\Ku$-\NFA. In particular, we give: $(i)$~in Proposition~\ref{prop:FromNFAtoK-NFA} the basic step to deal with propositions associated with regular expressions, $(ii)$~in Proposition \ref{prop:closureUnderPrefixSuffix} the closure of $\Ku$-\NFA s under language operations corresponding to \HS\ modalities, and $(iii)$~in Proposition~\ref{prop:booleanClosureNFA} the closure of $\Ku$-\NFA s  under Boolean operations.

In the following, let $\Ku=\KuDef$ be a finite Kripke structure over $\Prop$.

\begin{proposition}\label{prop:FromNFAtoK-NFA} Let $\Au$ be an \NFA\ over $2^{\Prop}$ with $n$ states. One can construct, in polynomial time, a $\Ku$-\NFA\  $\Au_{\Ku}$ with at most $n+1$ main states accepting the set of traces $\rho$ of $\Ku$ such that $\mu(\rho)\in \Lang(\Au)$.
\end{proposition}
\begin{proof}
Let $\Au = \tpl{2^{\Prop},Q,Q_0,\Delta,F}$. By using an additional state, we can assume $\varepsilon\notin \Lang(\Au)$, that is, $Q_0\cap F=\emptyset$. Then, $\Au_{\Ku}=\tpl{S,Q\times S,Q_0\times S,\Delta',F\times S}$, where for all $(q,s)\in Q\times S$ and $s'\in S$, it holds that $\Delta'((q,s),s')=\emptyset$ if $s'\neq s$, and $\Delta'((q,s),s) = \Delta(q,\mu(s))\times \Edges(s)$. Since $\Edges(s)\neq \emptyset$ for all $s \in S$, the thesis follows.
\end{proof}

We now define the operations on languages of  finite words over $S$ corresponding to the $\HS$ modalities $\hsB$, $\hsBt$, $\hsE$, and $\hsEt$.
Given a language $\Lang$ over $S$, we define the following languages of traces of $\Ku$:
%
%
%
\begin{itemize}
 \item   $\hsB_{\Ku}(\Lang) = \{ \rho\in\Trk_\Ku\mid  \exists\, \rho'\in \Lang\cap S^{+} \text{ and }\rho''\in S^{+} \text{ such that } \rho= \rho'\cdot \rho'' \}$;
 \item   $\hsBt_{\Ku}(\Lang) =\{ \rho\in \Trk_\Ku\mid \exists\, \rho'\in S^{+}\text{ such that }\rho\cdot \rho'\in\Lang\cap \Trk_\Ku \}$;
  \item   $\hsE_{\Ku}(\Lang) =\{ \rho\in\Trk_\Ku\mid \exists\, \rho''\in \Lang\cap S^{+} \text{ and } \rho'\in S^{+} \text{ such that } \rho= \rho'\cdot \rho''  \}$;
 \item   $\hsEt_{\Ku}(\Lang) = \{ \rho\in\Trk_\Ku\mid \exists\, \rho'\in S^{+}\text{ such that } \rho'\cdot \rho\in\Lang\cap \Trk_\Ku \}$.
\end{itemize}

We show that $\Ku$-$\NFA$s are closed under the above defined language operations $\hsB_{\Ku}(\cdot)$, $\hsE_{\Ku}(\cdot)$, $\hsBt_{\Ku}(\cdot)$ and $\hsEt_{\Ku}(\cdot)$.

\begin{proposition}\label{prop:closureUnderPrefixSuffix} Given a $\Ku$-$\NFA$ $\Au$ with $n$ main states, 
one can construct, in polynomial time, $\Ku$-$\NFA$s with $n+1$ main states accepting the languages
$\hsB_{\Ku}(\Lang(\Au))$, $\hsE_{\Ku}(\Lang(\Au))$, $\hsBt_{\Ku}(\Lang(\Au))$ and $\hsEt_{\Ku}(\Lang(\Au))$, 
respectively.
\end{proposition}
\begin{proof}
Let $\Au=\tpl{S, M\times S,Q_0,\Delta,F}$, where $M$ is the set of main states. 

\paragraph*{Language $\hsB_{\Ku}(\Lang(\Au))$.} Let $\Au_{\hsB}$ be the $\NFA$ over $S$ given by
\[\Au_{\hsB}=\tpl{S,(M\cup \{q_\acc\})\times S ,Q_0,\Delta',\{q_\acc\}\times S},\] where $q_\acc\notin M$ is a fresh main state, and for all $(q,s)\in (M\cup \{q_\acc\})\times S$ and $s'\in S$, we have $\Delta'((q,s),s')=\emptyset$, if $s'\neq s$, and 
%
\[
\Delta'((q,s),s) =  \left\{
    \begin{array}{ll}
    \Delta((q,s),s)
      &    \text{ if }   (q,s)\in (M\times S)\setminus F
      \\
    \Delta((q,s),s) \cup 
     (\{q_\acc\}\times \Edges(s))
      &    \text{ if }    (q,s)\in F
      \\
    \{q_\acc\}\times \Edges(s) &    \text{ if }    q = q_\acc.
    \end{array}
  \right.
\]
%
Given an input word $\rho$, from an initial state $(q_0,s)$ of $\Au$,   the automaton $\Au_{\hsB}$ simulates the behavior of $\Au$ from $(q_0,s)$ over $\rho$. When $\Au$ is in an accepting state $(q_f,s)$ and the current  input symbol is $s$, $\Au_{\hsB}$ can additionally choose
to  move to a state in  $\{q_\acc\}\times \Edges(s)$, which is accepting for $\Au_{\hsB}$ (a prefix of $\rho$ belongs to  $\Lang(\Au)$). From such states, $\Au_{\hsB}$ accepts if and only if the remaining part of the input is a trace of $\Ku$.
Since $\Au$ is a $\Ku$-\NFA, $\Au_{\hsB}$ is a $\Ku$-\NFA\  by construction. Moreover, a word $\rho$ over $S$ is accepted by $\Au_{\hsB}$  \emph{if and only if} $\rho$ is a trace of $\Ku$ having some proper prefix $\rho'$ in $\Lang(\Au)$ (note that $\rho'\neq \varepsilon$ since $\Au$ is a $\Ku$-\NFA). Thus, $\Lang(\Au_{\hsB})=\hsB_{\Ku}(\Lang(\Au))$.

\paragraph*{Language  $\hsBt_{\Ku}(\Lang(\Au))$.} Let $\Au_{\hsBt}$ be the $\NFA$ over $S$ given by 
\[\Au_{\hsBt}=\tpl{S,(M\cup \{q'_0\})\times S ,\{q'_0\}\times S,\Delta',F'},\] where $q'_0\notin M$ is a fresh main state and $\Delta'$ and $F'$ are defined as follows:
\begin{itemize}
  \item for all
$(q,s)\in (M\cup \{q'_0\})\times S$, $s'\in S$, we have $\Delta'((q,s),s')=\emptyset$, if $s'\neq s$, and 
\[
 \Delta'((q,s),s) = \left\{
    \begin{array}{ll}
   \displaystyle{\bigcup_{(q_0,s)\in Q_0}}\Delta((q_0,s),s)
      &    \text{ if } q=q'_0 \medskip
      \\
    \Delta((q,s),s)
      &    \text{ otherwise.}
    \end{array}
  \right.
\]
  \item The set $F'$ of accepting states is the set of states $(q,s)$ of $\Au$ such that there exists a run of $\Au$ from $(q,s)$ to some state in $F$ over some non-empty word.
\end{itemize}

Note that the set $F'$ can be computed in time polynomial in the size of $\Au$. Since   $\Au_{\hsBt}$ essentially simulates $\Au$, and $\{q'_0\}\times S$ and $F'$ are disjoint,
by construction it easily follows that $\Au_{\hsBt}$ is a $\Ku$-\NFA. Moreover, $\Au_{\hsBt}$ accepts a word $\rho$ \emph{if and only if} $\rho$ is a non-empty proper prefix of some word accepted by $\Au$.
Thus, since $\Au$ is a $\Ku$-\NFA, we obtain that $\Lang(\Au_{\hsBt})=\hsBt_{\Ku}(\Lang(\Au))$.


The constructions for $\hsE_{\Ku}(\Lang(\Au))$ and $\hsEt_{\Ku}(\Lang(\Au))$---which are symmetric to the ones for $\hsB_{\Ku}(\Lang(\Au))$ and $\hsBt_{\Ku}(\Lang(\Au))$---can be found in \ref{sec:prop:closureUnderPrefixSuffix}.
\end{proof}

Now we show that $\Ku$-$\NFA$s are closed under Boolean operations.

\begin{proposition}\label{prop:booleanClosureNFA} 
Given two $\Ku$-$\NFA$s  $\Au$ and $\Au'$   with $n$ and $n'$ main states, respectively, 
one can construct:
\begin{itemize}
  \item a  $\Ku$-$\NFA$ with $n+n'$ main states accepting $\Lang(\Au)\cup \Lang(\Au')$ in time~\mbox{$O(n+n')$};
   \item a $\Ku$-$\NFA$ with $2^{n+1}+1$ main states accepting   $\Trk_\Ku\setminus \Lang(\Au)$ in time~$2^{O(n)}$.
\end{itemize}
\end{proposition}
\begin{proof}     
The construction for union is standard and thus omitted. The construction for 
complementation follows.

%
%
%
%
%
Let $\Au=\tpl{ S,M\times S,Q_0,\Delta,F}$.
%
First, we need a preliminary construction. Let us consider the \NFA\ 
\[\Au''= \tpl{S,(M \cup \{q_\acc\}) \times S ,Q_{0},\Delta'',\{q_\acc\}\times S},\]
where $q_\acc\notin M$ is a fresh main state, and for all
$(q,s)\in (M \cup \{q_\acc\}) \times S$ and $s'\in s$, we have $\Delta''((q,s),s')=\emptyset$, if $s'\neq s$, and  %
\[
\Delta''((q,s),s)= \left\{
    \begin{array}{ll}
     \Delta((q,s),s)\cup (\{q_\acc\}\times S)
      &    \text{ if }   q\in M \text{ and } \Delta((q,s),s)\cap F\neq \emptyset
      \\
      \Delta((q,s),s)
      &    \text{ if }   q\in M \text{ and } \Delta((q,s),s)\cap F = \emptyset
      \\
      \emptyset
      &   \text{ if }   q = q_\acc.
    \end{array}
  \right.
\]
Note that  $\Lang(\Au'')=\Lang(\Au)$, but $\Au''$ is actually \emph{not} a  $\Ku$-$\NFA$.

Next, we show that it is possible to  construct in time $2^{O(n)}$ a \emph{weak} $\Ku$-$\NFA$ $\Au_c$ with $2^{n+1}$ main states   accepting   $(\Trk_\Ku\setminus \Lang(\Au''))\cup \{\varepsilon\}$, where a \emph{weak} $\Ku$-$\NFA$ is just a $\Ku$-$\NFA$ without the requirement that the empty word $\varepsilon$ is not accepted. Thus, since a weak $\Ku$-$\NFA$ can be easily converted into an equivalent $\Ku$-$\NFA$ by using an additional main state, and $\Lang(\Au'')=\Lang(\Au)$, the result follows.  The weak $\Ku$-$\NFA$ $\Au_c$ is given by
$\Au_c=\tpl{S,2^{\tilde{M}}\times S ,Q_{0,c},\Delta_c,F_c}$, where $\tilde{M}=M\cup \{q_\acc\}$, and $Q_{0,c}$,  $F_c$ and $\Delta_c$ are defined as follows:
\begin{itemize}
  \item $Q_{0,c}=\{(P,s)\in 2^{M}\times S \mid P=\{q\in M\mid (q,s)\in Q_0\}\}$;
  \item $F_c=\{(P,s)\in 2^{M}\times S \}$;
  \item for all
$(P,s)\in 2^{\tilde{M}}\times S$ and $s'\in S$, we have $\Delta_c((P,s),s')=\emptyset$, if $s'\neq s$, and 
\[\Delta_c((P,s),s)=
 \bigcup_{s'\in \Edges(s)} \Big\{ (\{q'\in \tilde{M} \mid (q',s')\in \bigcup_{p\in P}\Delta''(p,s)\},s') \Big\}.
\]
\end{itemize}
By construction, $\Au_c$ is a weak $\Ku$-$\NFA$ not accepting words in $S^{+}\setminus \Trk_\Ku$. Since $Q_{0,c}\subseteq F_c$, we have $\varepsilon\in \Lang(\Au_c)$.
Let $\rho\in \Trk_\Ku$ with $|\rho|=k$. To conclude the proof, we have to show that $\rho\in \Lang(\Au'')$ if and only if $\rho \notin \Lang(\Au_c)$.

Assuming that $\rho\in \Lang(\Au'')$, we prove by contradiction that $\rho \notin \Lang(\Au_c)$. Let us assume that there is a run of $\Au_c$ over $\rho$ having the form $(P_0,s_0)\cdots (P_{k},s_k)$ such that
$(P_0,s_0)\in Q_{0,c}$ and $(P_k,s_k)\in F_c$ implying that $q_\acc\notin P_k$. By construction, $P_0 = \{q\in M\mid (q,s_0)\in Q_0\}$,   and  for all $i\in [0,k-1]$, $s_i = \rho(i)$
and $P_{i+1} = \{p\in \tilde{M} \mid (p,s_{i+1})\in \Delta''(q,s_i) \text{ for some } q\in P_i  \}$. Since $\rho\in \Lang(\Au'')$, there is $s\in S$, $(q_0,s_0)\in Q_0$ and an accepting run of
$\Au''$ over $\rho$ having the form $(q_0,s_0)\cdots  (q_{k-1},s_{k-1}) (q_k,s)$ where $q_k = q_\acc$.  By definition of the transition function $\Delta ''$ of  $\Au''$, we can also assume that
$s=s_k$. It follows that
$q_i\in P_i $ for all $i\in [0,k]$, which is a contradiction since $q_\acc\notin P_k$. Therefore $\rho \notin \Lang(\Au_c)$.

As for the converse direction, let us assume that $\rho \notin \Lang(\Au_c)$.  We have to show that $\rho\in \Lang(\Au'')$.  By construction, there exists some run of
$\Au_c$ over $\rho$ starting from an initial state (recall that $\Edges(s)\neq \emptyset$  for all $s\in S$). Moreover, each of these runs has the form $(P_0,s_0)\cdots (P_{k},s_k)$
such that  $P_0 = \{q\in M\mid (q,s_0)\in Q_0\}$, $q_\acc \in P_k$, and    for all $i\in [0,k-1]$, $s_i = \rho(i)$
and $P_{i+1} = \{p\in \tilde{M} \mid (p,s_{i+1})\in \Delta(q,s_i) \text{ for some } q\in P_i  \}$. It easily follows that there is an accepting run of
$\Au''$ over $\rho$ from some initial state in $P_0\times \{s_0\}$, thus proving the thesis. 
\end{proof}

An MC algorithm for full $\HS$ can be built as follows.
Let $\varphi$ be an $\HS$ formula. First of all, we convert $\varphi$ into an equivalent formula, called \emph{existential form of $\varphi$}, that makes use of negations, disjunctions, and the existential modalities
$\hsB$, $\hsBt$, $\hsE$, and $\hsEt$, only. 
For all $h\geq 1$, let $\HS_h$ denote the syntactical $\HS$ fragment consisting only of formulas $\varphi$ such that the \emph{nesting depth of negation in the existential form
of $\varphi$ is at most $h$}. Moreover $\neg\HS_h$ is the set of formulas $\varphi$ such that $\neg\varphi\in \HS_h$.
  Now, given an $\HS$ formula $\varphi$ (in existential form), checking whether  $\Ku\not\models \varphi$ reduces to checking the existence of
an initial trace $\rho$ of $\Ku$ such that $\Ku,\rho\models \neg\varphi$. By exploiting Proposition~\ref{prop:FromNFAtoK-NFA}, \ref{prop:closureUnderPrefixSuffix} and~\ref{prop:booleanClosureNFA}, we can build  in a compositional way (driven by the structure of $\neg\varphi$) a $\Ku$-\NFA\ $\Au$ accepting the set of initial traces $\rho$ such that $\Ku,\rho\models \neg\varphi$ and check $\Au$ for emptiness.

For all $h,n\geq 0$, let $\Tower(h,n)$ denote a tower of exponentials of height $h$ and argument $n$, that is, $\Tower(0,n)=n$ and $\Tower(h+1,n)=2^{\Tower(h,n)}$. Moreover, let \mbox{$h$-$\EXPTIME$} denote the class of languages decided by deterministic Turing machines whose number of computation steps is bounded by functions of $n$ in $O(\Tower(h,n^c))$, for some constant $c\geq 1$. Note that \mbox{$0$-$\EXPTIME$} is $\PTIME$.

The next theorem states the main result of the section.
  \begin{theorem}\label{th-m} There exists a constant $c$ such that, given a finite Kripke structure $\Ku$ and  an $\HS$ formula $\varphi$, one can construct 
  a $\Ku$-\NFA\ with $O(|\Ku| \cdot \Tower(h,|\varphi|^{c}))$ states  accepting the set of traces $\rho$ of $\Ku$
  such that $\Ku,\rho\models  \varphi$, where $h$ is the nesting depth of negation in the existential form of $\varphi$. 
  
  Moreover,  for each $h\geq 0$, the MC problem for $\neg\HS_h$
is in \mbox{$h$-$\EXPTIME$}. Additionally, for a constant-length formula,
the MC problem is in $\PTIME$.
  \end{theorem} 
  
On one hand, this algorithm can be adapted to $\HS$ under homogeneity in a straightforward way, as an alternative to the one presented in 
Section~\ref{sec:decidProof}. On the other, the complexity lower bound proved in Section~\ref{sec:BEhard} for MC of $\HS$ formulas under  homogeneity immediately propagates to the complexity of MC for $\HS$ extended with regular expressions.

\begin{theorem}\label{theorem:lowerBoundBERegex} The MC problem for $\HS$ formulas extended with regular expressions over finite Kripke structures is $\EXPSPACE$-hard (under polynomial-time reductions).
\end{theorem}
  
We now focus on the complexity of $\HS$ fragments with regular expressions.

%% file: Chaps/Gandalf17RIVISTA/TrackProperty.tex
\section{The fragments $\AAbarBBbarEbar$ and $\AAbarEBbarEbar$}
\label{sec:AAbarBBbarEbarRegex}
In this section, we focus on the syntactically maximal fragments $\AAbarBBbarEbar$ and $\AAbarEBbarEbar$ showing that they feature  a lower computational complexity with respect to the general case of full $\HS$.
 
First of all, in Section~\ref{sec:AAbarBBbarEbarTrackProperty}  we prove that they feature an exponential small-model property, stating that
if $\rho$ is a trace of a finite Kripke structure $\Ku$ and  $\psi$ is an $\AAbarBBbarEbar$ formula, then there is a trace $\rho'$ such that $\Ku,\rho\models \psi$ if and only if $\Ku,\rho'\models \psi$, and $|\rho'|$ is exponential in the B-nesting depth of  $\psi$.

In Section~\ref{sec:UpperBound}, we exploit this small-model property to design a MC algorithm for $\AAbarBBbarEbar$  belonging to the complexity class $\LINAEXPTIME$, namely, the class of problems decidable by singly exponential-time bounded alternating Turing machines  with a polynomial-bounded number of alternations.

Finally, in Section~\ref{sec:LowerBound}, we show that MC for $\AAbarBBbarEbar$ (actually the smaller fragment $\BEbar$ suffices)
is hard for $\LINAEXPTIME$, hence proving completeness for that class.

\subsection{Exponential small-model property for $\AAbarBBbarEbar$}\label{sec:AAbarBBbarEbarTrackProperty}
Here we prove the \emph{exponential small-model property} for $\AAbarBBbarEbar$, which will be used as the basic step to prove that the MC problem for $\AAbarBBbarEbar$  belongs to $\LINAEXPTIME$.
The case of $\AAbarEBbarEbar$ is, as usual, completely symmetric and thus omitted.

Let us consider a finite Kripke structure $\Ku=\KuDef$ and a finite set $\SPEC=\{r_1,\ldots,r_H\}$ of (proposition-based)
regular expressions over $\Prop$.
The small-model guarantees that for each $h\geq 0$ and trace $\rho$ of $\Ku$,
it is possible to build another trace $\rho'$ of $\Ku$, of bounded exponential length, which is indistinguishable from $\rho$ with respect to the fulfilment of any $\AAbarBBbarEbar$ formula $\varphi$ having atomic formulas in $\SPEC$ and B-nesting depth at most $h$. 

In order to prove the result, 
we resort again to the notion of
\emph{$h$-prefix bisimilarity}  between a pair of traces $\rho$ and $\rho'$ of $\Ku$,
already introduced in Section~\ref{sec:AAbarBBbarEbar}, but adapted here to account for regular expressions. 
As proved by Proposition \ref{prop:fulfillmentPreservingPrefix} below, $h$-prefix bisimilarity   is a sufficient condition for two traces  $\rho$ and $\rho'$ to be indistinguishable with respect to the fulfillment of any $\AAbarBBbarEbar$ formula $\varphi$ over $\SPEC$ with $\nestb(\varphi)\leq h$.

Then we recall (and adapt) also the \emph{$h$-prefix sampling} of a trace: for a given trace $\rho$,  the \emph{$h$-prefix sampling} of $\rho$ is a subset of $\rho$-positions that allows us to build another trace $\rho'$ having single exponential length (both in $h$ and $|\SPEC|$, where $|\SPEC|$ is defined  as  $\sum_{r\in\SPEC}|r|$) such that $\rho$ and $\rho'$ are $h$-prefix bisimilar.

Even though we start from adaptations of already presented notions,
as it will be clear in the next section the resulting algorithm for $\AAbarBBbarEbar$ MC with regular expressions is completely novel w.r.t.\ that for the same fragment under homogeneity (Section~\ref{sec:AAbarBBbarEbar}).

For a regular expression $r_\ell$ in $\SPEC$, with $\ell\!\in\! [1,H]$, let $\Au_\ell\!=\!\tpl{2^{\Prop},Q_\ell,Q_\ell^{0},\Delta_\ell,F_\ell}$ be the \emph{canonical (complete)} \NFA\ accepting $\Lang(r_\ell)$ (recall that
$|Q_\ell|\leq 2|r_\ell|$, Remark~\ref{remk:nfa}).
W.l.o.g., we assume that the sets of states of these automata are pairwise disjoint.  

Prefix bisimilarity now uses the notion of \emph{summary}  of a trace $\rho$ of $\Ku$, namely, a tuple ``recording'' the initial and final states of $\rho$, and, for each automaton $\Au_\ell$, with $\ell\in [1,H]$, the pairs of states $q,q'\in Q_\ell$  such that some run of $\Au_\ell$ over $\mu(\rho)$ takes from $q$ to $q'$.

\begin{definition}[Summary of a trace]
Let $\rho$ be a trace of $\Ku$ with $|\rho|=n$. The summary $\Summary(\rho)$ of $\rho$ (w.r.t.\ $\SPEC$) is the triple
$(\rho(1),\Pi,\rho(n))$, where 
\begin{multline*}
\Pi=\{(q,q')\mid q,q'\in Q_\ell \text{ for some } \ell\in [1,H], \\ \text{and there is a run of $\Au_\ell$ from $q$ to $q'$ over $\Lab(\rho)$}\}.
\end{multline*}
\end{definition}
Note that the number of summaries is at most $|S|^2\cdot 2^{(2|\SPEC|)^2}$. The following result can be easily proved.

\begin{proposition}\label{prop:Summaries} Let $h\geq 0$, and $\rho$ and $\rho'$ be two traces of  $\Ku$ such that $\Summary(\rho)=\Summary(\rho')$. Then, for all regular expressions $r\in \SPEC$ and traces $\rho_L$ and $\rho_R$ of $\Ku$ such that $\rho_L \star \rho$ and  $\rho \star \rho_R$ are defined, the next properties hold:
\begin{enumerate}
    \item $\mu(\rho)\in\Lang(r)$ if and only if $\mu(\rho')\in\Lang(r)$;
    \item $\Summary(\rho_L\star \rho)=\Summary(\rho_L\star \rho')$;
    \item $\Summary(\rho\star \rho_R)=\Summary(\rho'\star \rho_R)$.
\end{enumerate}
\end{proposition}

We now introduce the \lq\lq new version\rq\rq{} of \emph{prefix bisimilarity}  between a pair of traces $\rho$ and $\rho'$ of $\Ku$.
\begin{definition}[Prefix bisimilarity]
Let $h\geq 0$. Two traces $\rho$ and $\rho'$  of $\Ku$  are \emph{$h$-prefix bisimilar} (w.r.t.\ $\SPEC$) if the next conditions inductively hold:
\begin{itemize}
  \item for $h=0$: $\Summary(\rho)=\Summary(\rho')$;
  \item for $h>0$: $\Summary(\rho)=\Summary(\rho')$  and for each proper prefix $\nu$ of $\rho$ (resp., proper prefix $\nu'$ of $\rho'$), there exists
  a proper prefix $\nu'$ of $\rho'$ (resp., proper prefix $\nu$ of $\rho$) such that $\nu$ and $\nu'$ are $(h-1)$-prefix bisimilar.
\end{itemize}
\end{definition}
\begin{property}
For all $h\geq 0$, $h$-prefix   bisimilarity is an equivalence relation over $\Trk_\Ku$.
\end{property}

The $h$-prefix  bisimilarity of two traces $\rho$ and $\rho'$ is preserved by left (resp., right) \mbox{$\star$-concatenation} with another trace of $\Ku$.
\begin{proposition}\label{prop:invarianceLeftRightPrefix} Let $h\geq 0$, and $\rho$ and $\rho'$ be two $h$-prefix  bisimilar traces of  $\Ku$. Then, for all traces $\rho_L$ and $\rho_R$ of $\Ku$ such that $\rho_L \star \rho$ and  $\rho \star \rho_R$ are defined, the following properties hold:
\begin{enumerate}
    \item $\rho_L\star \rho$ and $\rho_L\star \rho'$ are $h$-prefix   bisimilar; 
    \item $\rho\star \rho_R$ and $\rho'\star \rho_R$ are $h$-prefix  bisimilar.
\end{enumerate}
\end{proposition}
The proof can be found in Appendix \ref{proof:prop:invarianceLeftRightPrefix}. 
By exploiting Proposition~\ref{prop:Summaries} and~\ref{prop:invarianceLeftRightPrefix},
we can prove that $h$-prefix  bisimilarity preserves the satisfiability of $\AAbarBBbarEbar$ formulas over $\SPEC$ having B-nesting depth at most $h$.

\begin{proposition}\label{prop:fulfillmentPreservingPrefix} Let $h\geq 0$, and $\rho$ and $\rho'$ be two $h$-prefix bisimilar traces of $\Ku$. Then, for each $\AAbarBBbarEbar$
formula $\psi$ over $\SPEC$ with $\nestb(\psi)\leq h$, we have that
\[\Ku,\rho\models\psi\iff \Ku,\rho'\models\psi.\]
\end{proposition}
The proof is in Appendix~\ref{proof:prop:fulfillmentPreservingPrefix}.

In the following, by analogy with Section~\ref{sec:AAbarBBbarEbar}, we show how a trace $\rho$, whose length exceeds a suitable exponential bound---precisely, $(|S|\cdot 2^{(2|\SPEC|)^2})^{h+2}$---can be contracted preserving $h$-prefix bisimilarity and, consequently, satisfiability of formulas $\varphi$ with $\nestb(\varphi) \leq h$. The basic contraction step of $\rho$ is performed by choosing the subset of $\rho$-positions called $h$-\emph{prefix sampling} ($\PrefS_h$). A contraction can be performed whenever there are two positions $\ell < \ell'$ satisfying $\Summary(\rho(1,\ell))=\Summary(\rho(1,\ell'))$ in between two consecutive positions in the linear ordering of $\PrefS_h$. We will prove that  by taking the contraction $\rho'=\rho(1,\ell)\cdot \rho(\ell'+1,|\rho|)$, we obtain a trace of $\Ku$ which is $h$-prefix bisimilar to $\rho$. The basic contraction step can then be iterated over $\rho'$ until the length bound is reached.

The notion of $h$-prefix sampling is, again, inductively defined using the definition of \emph{prefix-skeleton sampling} (now  accounting for trace summaries).
%
%
We recall that, for a set $I$ of natural numbers, by ``two consecutive elements of $I$'' we mean a pair of elements $i,j\in I$ such that $i<j$ and $I\cap [i,j]=\{i,j\}$.

\begin{definition}[Prefix-skeleton sampling]\label{def:skeletonRegex}  Let $\rho$ be a trace of $\Ku$. Given two $\rho$-positions $i$ and $j$, with $i\leq j$, the \emph{prefix-skeleton sampling of $\rho$ in the interval $
[i,j]$} is the \emph{minimal} set $\Pos$ of $\rho$-positions in the interval $[i,j]$ satisfying the conditions:
\begin{itemize}
    \item $i,j\in \Pos\,$;
    \item  for each $k\in [i+1,j-1]$, the minimum position $k'\in [i+1,j-1]$ such that $\Summary(\rho(1,k'))=\Summary(\rho(1,k))$ belongs to $\Pos$.
\end{itemize}
\end{definition}

\begin{example}
An example of prefix-skeleton sampling $\Pos$ of a trace $\rho$ in the interval $[i,j]$ is shown in  Figure~\ref{fig:prefsk}. Assuming that $\Summary(\rho(1,u))=\Summary_1$ for $u\in\{i+1,i+2,i+3,i+5,i+7,i+10\}$, $\Summary(\rho(1,u'))=\Summary_2$ for $u'\in\{i+4,i+8\}$, and $\Summary(\rho(1,u''))=\Summary_3$ for $u''\in\{i+6,i+9\}$, we have that $\Pos= \{i, i+1, i+4,i+6,j\}$.

\begin{figure}[H]
    \centering
    \includegraphics[width=\linewidth]{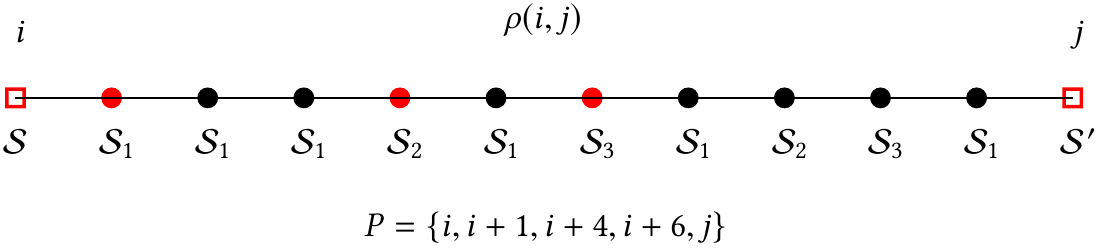}
    \caption{Example of prefix-skeleton sampling $\Pos$ of a trace $\rho$ in the interval $[i,j]$. }
    \label{fig:prefsk}
\end{figure}
\end{example}

Note that, as an immediate consequence of Definition~\ref{def:skeletonRegex}, the prefix-skeleton sampling $\Pos$  of (any) trace $\rho$ in $[i,j]$ is such that $|\Pos\,|\leq (|S|\cdot 2^{(2|\SPEC|)^2})+2$. 

\begin{definition}[$h$-prefix sampling and $h$-sampling word] Let $h\geq 0$.
The \emph{$h$-prefix sampling of a trace $\rho$} of $\Ku$ is the \emph{minimal} set $\PrefS_h$ of $\rho$-positions inductively satisfying the following conditions:
\begin{itemize}
  \item for $h=0$: $\PrefS_0=\{1,|\rho|\}$;
  \item for $h> 0$: 
  $(i)$~$\PrefS_h\supseteq\PrefS_{h-1}$ and
  $(ii)$~for all pairs of consecutive positions $i,j\in\PrefS_{h-1}$,
  the prefix-skeleton sampling of $\rho$ in the interval $[i,j]$ belongs to $\PrefS_h$.
\end{itemize}

Let $i_1<\ldots<i_N$ be the ordered sequence of positions in $\PrefS_h$ (note that $i_1=1$ and $i_N=|\rho|$). The \emph{$h$-sampling word of $\rho$} is the sequence of summaries 
$\Summary(\rho(1,i_1))\cdots \Summary(\rho(1,i_N))$.
\end{definition}

 We can state the following upper bound to the cardinality of prefix samplings.

\begin{property}\label{property:prefSamBoundRegex}
Let $h\geq 0$. The $h$-prefix sampling $\PrefS_h$ of a (any) trace $\rho$ of $\Ku$ is such that $|\PrefS_h|\leq (|S|\cdot 2^{(2|\SPEC|)^2})^{h+1}$.
\end{property}

As stated by the following lemma, for two traces, the property of having the same $h$-sampling word is a sufficient condition to guarantee that they are $h$-prefix bisimilar.
The proof is in Appendix~\ref{proof:lemma:prefixSamplingOneRegex}.
\begin{lemma}\label{lemma:prefixSamplingOneRegex} For $h\geq 0$, any two traces
$\rho$ and  $\rho'$ of $\Ku$ having the same
$h$-sampling word are $h$-prefix bisimilar.
\end{lemma}

The sufficient condition of Lemma~\ref{lemma:prefixSamplingOneRegex} allows us to finally state the exponential small-model property for $\AAbarBBbarEbar$. In the proof of Theorem~\ref{theorem:singleExpTrackModelRegex} below, it is shown  how to derive from any trace $\rho$ of $\Ku$, an $h$-prefix  bisimilar trace $\rho'$ \emph{induced by} $\rho$ 
(Definition~\ref{definition:inducedTrk}) 
such that $|\rho'|\leq (|S|\cdot 2^{(2|\SPEC|)^2})^{h+2}$. By Proposition~\ref{prop:fulfillmentPreservingPrefix}, $\rho'$ is indistinguishable from $\rho$ w.r.t.\ the fulfillment of any $\AAbarBBbarEbar$ formula $\varphi$ over the set of atomic formulas in $\SPEC$ such that $\nestb(\varphi)\leq h$. 

\begin{theorem}[Exponential small-model property for $\AAbarBBbarEbar$]\label{theorem:singleExpTrackModelRegex}
Let $\rho$ be a trace of $\Ku$ and $h\geq 0$.
Then there exists a trace $\rho'$ induced by $\rho$, whose length is at most $(|S|\cdot 2^{(2|\SPEC|)^2})^{h+2}$, which is $h$-prefix bisimilar to $\rho$. In particular,  for every $\AAbarBBbarEbar$ formula $\psi$ with atomic formulas in $\SPEC$ and $\nestb(\psi)\leq h$, it holds that \[\Ku,\rho\models \psi\iff \Ku,\rho'\models \psi.\]
\end{theorem}
\begin{proof} 
We show that if $|\rho|>(|S|\cdot 2^{(2|\SPEC|)^2})^{h+2}$, then there exists a trace $\rho'$ induced by $\rho$ such that $|\rho'|<|\rho|$,
 and $\rho$ and $\rho'$ have the same $h$-sampling word. 
 
 Assume that $|\rho|>(|S|\cdot 2^{(2|\SPEC|)^2})^{h+2}$.
 Let $\PrefS_{h}: 1=i_1<\ldots <i_N=|\rho|$ be the $h$-prefix sampling of $\rho$. By Property~\ref{property:prefSamBoundRegex}, $|\PrefS_{h}|\leq (|S|\cdot 2^{(2|\SPEC|)^2})^{h+1}$.
  Since  the number of distinct summaries (w.r.t.\ $\SPEC$) associated with the prefixes of $\rho$ is at most $|S|\cdot 2^{(2|\SPEC|)^2}$, there must be two consecutive positions $i_j$ and
  $i_{j+1}$ in $\PrefS_h$ such that for some $\ell,\ell'\in [i_j+1,i_{j+1}-1]$ with $\ell<\ell'$, $\Summary(\rho(1,\ell))=\Summary(\rho(1,\ell'))$. It easily follows that
  the sequence $\rho'$ given by $\rho'=\rho(1,\ell)\cdot \rho(\ell'+1,|\rho|)$ is a trace induced by $\rho$ such that $|\rho'|<|\rho|$, and $\rho$ and $\rho'$ have the same
  $h$-sampling word.
  Now, by Lemma~\ref{lemma:prefixSamplingOneRegex}, $\rho$ and $\rho'$ are $h$-prefix bisimilar and by applying Proposition~\ref{prop:fulfillmentPreservingPrefix} we get that $\Ku,\rho\models \psi\iff \Ku,\rho'\models \psi$.
  Now, if $|\rho'|\leq (|S|\cdot 2^{(2|\SPEC|)^2})^{h+2}$, the thesis follows, otherwise a sequence of contraction steps as shown above can be performed, until the length of the contracted trace fulfills the requirement.
\end{proof}

%% file: Chaps/Gandalf17RIVISTA/ModelCheckingAAbarBBbarEbar.tex
\subsection{$\LINAEXPTIME$ MC algorithm for $\AAbarBBbarEbar$}\label{sec:UpperBound}

In this section, taking advantage of the  exponential small-model property proved previously, we design a MC algorithm for $\AAbarBBbarEbar$ formulas with regular expressions belonging to the complexity class $\LINAEXPTIME$.
We recall that  $\LINAEXPTIME$ is the class of problems solvable by  singly exponential-time bounded alternating Turing machines (ATMs, for short) performing at most a polynomial-bounded number of alternations. More formally, an ATM $\M$ (we refer to~\cite{CKS81} for standard syntax and semantics of ATMs)  is \emph{singly exponential-time bounded} if there exists an integer constant $c\geq 1$ such that, for each input $\alpha$, any computation starting on  $\alpha$
 halts after at most $2^{|\alpha|^{c}}$ steps. The ATM $\M$ makes a \emph{polynomial-bounded number of alternations} if there exists an integer constant $c'\geq 1$ such that, for all inputs $\alpha$ and computations $\pi$ starting from $\alpha$, the number of alternations of existential and universal configurations along $\pi$ is at most $|\alpha|^{c'}$.

In the sequel, we restrict ourselves w.l.o.g.\ to
$\AAbarBBbarEbar$  formulas in
\emph{negation normal form} (\nnf). 
 For $\varphi$ in \nnf, the \emph{dual}
 of  $\varphi$, denoted as $\widetilde{\varphi}$, is defined as  the \nnf\ of $\neg\varphi$.
The complexity measure of an $\AAbarBBbarEbar$ formula $\varphi$ that we will consider is
the standard  \emph{alternation depth}, denoted by $\AltN(\varphi)$, between the existential $\hsX$  and universal
modalities $\hsXu$ (and vice versa) occurring in the \nnf{} of $\varphi$, \emph{for $X\in \{\overline{B},\overline{E}\}$}. Note that the definition does not consider the modalities associated with the Allen's relations in $\{A,\overline{A},B\}$. 
Moreover, let $\FMC$ be the set of pairs $(\Ku,\varphi)$ consisting of a finite Kripke structure $\Ku$ and an $\AAbarBBbarEbar$ formula $\varphi$ such that $\Ku\models\varphi$ (i.e., $\Ku$ is a model of $\varphi$). 

The following theorem states the complexity upper bound of MC for $\AAbarBBbarEbar$ formulas with regular expressions.


\begin{theorem}\label{Theorem:UpperBoundAAbrBBarEbar} One can construct a singly exponential-time bounded ATM accepting $\FMC$ whose number of alternations on an input $(\Ku,\varphi)$ is at most $\AltN(\varphi)+2$.
\end{theorem}

To prove the assertion  of Theorem~\ref{Theorem:UpperBoundAAbrBBarEbar}, we define a procedure in the remaining part of the section. Such a procedure can be easily translated into an ATM (the translation is omitted).

We start with some auxiliary notation. Let us fix a finite Kripke structure $\Ku$ with set of states $\States$ and an $\AAbarBBbarEbar$ formula $\varphi$ in \nnf.
Let $h=\nestb(\varphi)$, and
$\SPEC$ be the set of regular expressions occurring in $\varphi$.
A \emph{certificate} of $(\Ku,\varphi)$ is a trace $\rho$ of $\Ku$ whose length is less than $(|S|\cdot 2^{(2|\SPEC|)^2})^{h+2}$ (the bound for the exponential small-model property of Theorem~\ref{theorem:singleExpTrackModelRegex}).
A \emph{$\overline{B}$-witness} (resp., \emph{$\overline{E}$-witness}) of a certificate $\rho$ for $(\Ku,\varphi)$ is a certificate $\rho'$ of  $(\Ku,\varphi)$ such that $\rho'$ is $h$-prefix bisimilar to a trace having the form $\rho\star \rho''$ (resp., $\rho''\star \rho$) for some
\emph{certificate} $\rho''$ of $(\Ku,\varphi)$ with $|\rho''|>1$. Finally, by $\SD(\varphi)$ we denote the set consisting of the subformulas $\psi$ of $\varphi$ and  the \emph{duals} $\widetilde{\psi}$.

The results stated in Section~\ref{sec:AAbarBBbarEbarTrackProperty} are used to prove the properties of certificates, which are listed in the following Proposition~\ref{prop:EbarBbarWitness}, and then exploited in the MC algorithm.
\begin{proposition}\label{prop:EbarBbarWitness} Let $\Ku$ be a finite Kripke structure,  $\varphi$ be an $\AAbarBBbarEbar$ formula  in \nnf, and $\rho$ be a certificate for $(\Ku,\varphi)$. The next properties hold:
\begin{enumerate}
  \item for each $\hsX\psi\in\SD(\varphi)$, with $X\in \{\overline{B}, \overline{E}\}$, it holds $\Ku,\rho\models \hsX\psi$ if and only if there exists an $X$-witness $\rho'$ of $\rho$
  for $(\Ku,\varphi)$ such that $\Ku,\rho'\models \psi$;
  \item for each trace having the form $\rho\star\rho'$ (resp., $\rho'\star\rho$) such that $\rho'$ is a certificate for $(\Ku,\varphi)$, one can construct in time singly exponential in the size of $(\Ku,\varphi)$,
  a certificate $\rho''$ which is $h$-prefix bisimilar to $\rho\star\rho'$ (resp., $\rho'\star\rho$), with $h=\nestb(\varphi)$.
\end{enumerate}
\end{proposition}
\begin{proof} 
$(1.)$ Let $\hsX\psi\in\SD(\varphi)$  with $X\in \{\overline{B}, \overline{E}\}$, $h=\nestb(\varphi)$, and $\rho$ be a certificate for $(\Ku,\varphi)$. Let us assume that $X= \overline{E}$  (the  case for $X= \overline{B}$ is similar). 

First we assume that there exists an  $\overline{E}$-witness $\rho'$ of $\rho$ for $(\Ku,\varphi)$ such that $\Ku,\rho'\models \psi$. Hence $\rho'$ is $h$-prefix bisimilar to a trace having the form $\rho''\star \rho$, with $|\rho''|>1$. Since $\hsEt\psi\in\SD(\varphi)$, it holds that
$\nestb(\hsEt\psi)\leq h$. By Proposition~\ref{prop:fulfillmentPreservingPrefix} we have that  $\Ku, \rho''\star \rho \models \psi$ and, then, $\Ku,\rho\models \hsEt\psi$.

To prove the converse implication, we assume that $\Ku,\rho\models \hsEt\psi$. Then, there exists a trace having the form $\rho''\star \rho$ with $|\rho''|>1$ such that
$\Ku,\rho''\star\rho\models \psi$. By Theorem~\ref{theorem:singleExpTrackModelRegex} there exists a certificate $\nu$ for $(\Ku,\varphi)$ which is $h$-prefix bisimilar
to $\rho''$. By Proposition~\ref{prop:invarianceLeftRightPrefix}, $\nu\star \rho$ is $h$-prefix bisimilar to  $\rho''\star\rho$. By applying Proposition~\ref{prop:fulfillmentPreservingPrefix} we deduce that $\Ku,\nu\star\rho\models \psi$. By applying again Theorem~\ref{theorem:singleExpTrackModelRegex}, there exists
a certificate $\rho'$ for $(\Ku,\varphi)$ which is $h$-prefix bisimilar to  $\nu\star \rho$ such that $\Ku,\rho'\models \psi$. Thus, since $\rho'$ is an
$\overline{E}$-witness of $\rho$ for $(\Ku,\varphi)$, $(1.)$ follows. 

$(2.)$ From the trace $\rho\star\rho'$ (resp., $\rho'\star\rho$), where both $\rho$ and $\rho'$ are certificates for $(\Ku,\varphi)$, we first compute the $h$-prefix sampling of $\rho\star\rho'$ (resp., $\rho'\star\rho$), where $h=\nestb(\varphi)$. Then, proceeding as in the proof of Theorem~\ref{theorem:singleExpTrackModelRegex}, we extract from
$\rho\star\rho'$ (resp., $\rho'\star\rho$) a trace which is $h$-prefix bisimilar to $\rho\star\rho'$ (resp., $\rho'\star\rho$). Since $|\rho|$ and $|\rho'|$ are singly exponential in the sizes of $(\Ku,\varphi)$, $(2.)$ follows.
\end{proof}

Let $\AAbar(\varphi)$ be the set of formulas in $\SD(\varphi)$ having the form $\hsX\psi'$ or $\hsXu\psi'$, with $X\in \{A,\overline{A}\}$.
An $\AAbar$-labeling $\GLab$ for $(\Ku,\varphi)$ is a mapping associating with each state $s$ of $\Ku$ a maximally consistent set of subformulas of $\AAbar(\varphi)$. More precisely, for all $s \in \States$, $\GLab(s)$ is such that for all $\psi,\widetilde{\psi}\in \AAbar(\varphi)$, $\GLab(s)\cap \{\psi,\widetilde{\psi}\}$ is a singleton.
 We say that $\GLab$ is \emph{valid} if, for all states $s \in \States$ and $\psi\in \GLab(s)$, we have $\Ku,s\models \psi$ (we consider $s$ as a length-1 trace). 
Note that
if $\GLab$ is valid, then 
\begin{itemize}
    \item for each trace $\rho$ of $\Ku$ and $\hsA\psi'\in \AAbar(\varphi)$ (resp.,  $\hsAt\psi'\in \AAbar(\varphi)$), it holds that $\Ku,\rho\models \hsA\psi'$ (resp., $\Ku,\rho\models \hsAt\psi'$) if and only if $\hsA\psi'\in \GLab(\lst(\rho))$ (resp., $\hsAt\psi'\in \GLab(\fst(\rho))$). 
    \item Analogously, for each trace $\rho$ of $\Ku$ and $\hsAu\psi'\in \AAbar(\varphi)$ (resp.,  $\hsAtu\psi'\in \AAbar(\varphi)$), it holds that $\Ku,\rho\models \hsAu\psi'$ (resp., $\Ku,\rho\models \hsAtu\psi'$) if and only if $\hsAu\psi'\in \GLab(\lst(\rho))$ (resp., $\hsAtu\psi'\in \GLab(\fst(\rho))$).
\end{itemize}

Finally, a \emph{well-formed set for $(\Ku,\varphi)$} is a finite set $\WS$ consisting of pairs $(\psi,\rho)$ such that $\psi\in\SD(\varphi)$ and $\rho$ is a certificate of $(\Ku,\varphi)$.   
$\WS$ is said to be \emph{universal}
if each formula occurring in $\WS$ has the form $\hsXu\psi$, with $X\in \{\overline{B},\overline{E}\}$.  The \emph{dual} $\widetilde{\WS}$ of $\WS$ is the well-formed set  obtained by replacing each pair $(\psi,\rho)\in \WS$ with
   $(\widetilde{\psi},\rho)$.  A well-formed set $\WS$ is \emph{valid} if, for each $(\psi,\rho)\in \WS$, it holds that
  $\Ku,\rho\models \psi$.
 
\begin{algorithm}[tp]
\caption{$\texttt{check}(\Ku,\varphi)$ }\label{fig-proc-check}
\begin{algorithmic}[1]
\State{existentially choose an $\AAbar$-labeling  $\GLab$ for  $(\Ku,\varphi)$}
\For{each state $s$ and $\psi\in\GLab(s)$}
    \Case{ $\psi=\hsA \psi'$ (resp., $\psi=\hsAt \psi'$)}          
    \State{existentially choose a certificate $\rho$ with $\fst(\rho)=s$ (resp., $\lst(\rho)=s$)} 
    \State{$\texttt{checkTrue}_{\tpl{\Ku,\varphi,\GLab}}(\{(\psi',\rho)\})$}
    \case{ $\psi=\hsAu \psi'$ (resp., $\psi=\hsAtu \psi'$)}  
        \State{universally choose a certificate $\rho$ with $\fst(\rho)=s$ (resp., $\lst(\rho)=s$)} \State{$\texttt{checkTrue}_{\tpl{\Ku,\varphi,\GLab}}(\{(\psi',\rho)\})$}
    \EndCase
\EndFor
\State{universally choose a certificate $\rho$ for $(\Ku,\varphi)$ with $\fst(\rho)=s_0$}\Comment{$s_0$ is the initial state of $\Ku$}
 \State{$\texttt{checkTrue}_{\tpl{\Ku,\varphi,\GLab}}(\{(\varphi,\rho)\})$}
\end{algorithmic}
\end{algorithm}

We can now introduce the procedure $\texttt{check}$, reported in Algorithm~\ref{fig-proc-check}, that defines the ATM required to prove the assertion of Theorem~\ref{Theorem:UpperBoundAAbrBBarEbar}.
The procedure $\texttt{check}$ takes a pair $(\Ku,\varphi)$ as input, being $\Ku$ a finite Kripke structure and $\varphi$ an $\AAbarBBbarEbar$ formula in \nnf, and: (1)~it guesses
  an $\AAbar$-labeling  $\GLab$ for $(\Ku,\varphi)$ (line 1); (2)~it checks that $\GLab$ is valid (lines 2--9); (3)~for every certificate $\rho$ starting from the initial state, it verifies that $\Ku,\rho\models \varphi$ (lines 10--11). To perform steps (2) and (3), it exploits the
  auxiliary ATM procedure $\texttt{checkTrue}$, reported in Algorithm~\ref{fig-proc-checkTRUE}. 
  
\begin{algorithm}[p]
\caption{$\texttt{checkTrue}_{\tpl{\Ku,\varphi,\GLab}}(\WS)$}\label{fig-proc-checkTRUE}
\begin{algorithmic}[1]
\While{$\WS$ is \emph{not} universal}
    \State{deterministically select $(\psi,\rho)\in \WS$ s.t.\ $\psi$ is not of the form  $\hsEtu\psi'$ and $\hsBtu\psi'$}
    \State{$\WS\leftarrow \WS\setminus \{(\psi,\rho)\}$}
    \Case{ $\psi=r$  with $r\in \RE$}
        \If{$\rho\notin \Lang(r)$}
            \State{reject the input}
        \EndIf
    \case{ $\psi=\neg r$ with $r\in \RE$}
        \If{$\rho\in \Lang(r)$}
            \State{reject the input}
        \EndIf
    \case{ $\psi=\hsA \psi'$ or $\psi=\hsAu \psi'$}
        \If{$\psi\notin\GLab(\lst(\rho))$}
            \State{reject the input}
        \EndIf
    \case{ $\psi=\hsAt \psi'$ or $\psi=\hsAtu \psi'$}
        \If{$\psi\notin\GLab(\fst(\rho))$}
            \State{reject the input}
        \EndIf
    \case{ $\psi=\psi_1\vee \psi_2$}
        \State{existentially choose $i=1,2$}
        \State{$\WS\leftarrow \WS\cup \{(\psi_i,\rho)\}$}
        
        
    \case{ $\psi=\psi_1\wedge \psi_2$}
        \State{$\WS\leftarrow \WS\cup \{(\psi_1,\rho),(\psi_2,\rho)\}$}
    \case{ $\psi=\hsB\psi'$}
        \State{existentially choose $\rho'\in \Pref(\rho)$}
        \State{$\WS\leftarrow \WS\cup \{(\psi',\rho')\}$}
    \case{ $\psi=\hsBu\psi'$}
        \State{$\WS\leftarrow \WS\cup \{(\psi',\rho')\mid \rho'\in\Pref(\rho)\}$}
    \case{ $\psi=\hsX\psi'$ with $X\in \{\overline{E},\overline{B}\}$}
        \State{existentially choose an $X$-\emph{witness} $\rho'$ of $\rho$
        for $(\Ku,\varphi)$}
        \State{$\mathcal{\WS}\leftarrow \mathcal{\WS}\cup \{(\psi',\rho')\}$}
    \EndCase
\EndWhile
\If{$\mathcal{\WS}=\emptyset$}
    \State{accept the input}
\Else
    \State{universally choose $(\psi,\rho)\in \widetilde{\WS}$}
    \State{$\texttt{checkFalse}_{\tpl{\Ku,\varphi,\GLab}}(\{(\psi,\rho)\})$}
\EndIf
\end{algorithmic}
\end{algorithm}  
  
  The procedure $\texttt{checkTrue}$ takes as input a well-formed set  $\WS$ for $(\Ku,\varphi)$ and, assuming that the current $\AAbar$-labeling $\GLab$ for $(\Ku,\varphi)$ is valid,  checks whether $\WS$ is valid. For each pair $(\psi,\rho) \in \WS$ such that $\psi$ does not have the form $\hsXu\psi'$, with $X\in \{\overline{B},\overline{E}\}$, $\texttt{checkTrue}$ directly checks  whether or not $\Ku,\rho\models \psi$ (lines 4--29). 

  In order to allow for a deterministic choice of the current element of the iteration (line 2), we assume that the set $\WS$ is implemented as an ordered data structure.
  At each iteration of the while loop in $\texttt{checkTrue}$, 
  the current pair  $(\psi,\rho)\in \WS$ is processed
   according to the semantics of $\HS$, exploiting the guessed $\AAbar$-labeling  $\GLab$ for modalities 
   $\hsA$, $\hsAt$, $\hsAu$ and $\hsAtu$ (lines 10--15), and $\hsEt$-witnesses and $\hsBt$-witnesses (guaranteed by Proposition~\ref{prop:EbarBbarWitness}) for $\hsEt$ and $\hsBt$ (lines 26--28).
   The processing is either deterministic or based on an existential choice,
   and the currently processed pair $(\psi,\rho)$  is either removed from $\WS$, or replaced with pairs $(\psi',\rho')$ such that $\psi'$ is a strict subformula of $\psi$ (this is the case of Boolean connectives and modalities $\hsB$, $\hsBu$, $\hsEt$ and $\hsBt$, at lines 16--28).
   
   At the end of the while loop, the resulting well formed set $\WS$ is either empty or universal. In the former case, the procedure accepts (lines 30--31). In the latter, 
   %
   %
   there is a switch in the current operation mode (line 33). For each element $(\psi,\rho)$ in the dual of $\WS$---note that the root modality  of $\psi$ is either $\hsEt$ or $\hsBt$---the auxiliary ATM procedure $\texttt{checkFalse}$ (reported in Appendix~\ref{sec:chkFalse}) is invoked, which accepts the input $\{(\psi,\rho)\}$ if and only if $\Ku,\rho \not\models \psi$. The procedure $\texttt{checkFalse}$ is the \lq\lq dual\rq\rq{} of $\texttt{checkTrue}$, as it is simply obtained from $\texttt{checkTrue}$ by switching \emph{accept} and \emph{reject},  by switching existential and universal choices, and by converting the last call to $\texttt{checkFalse}$ into $\texttt{checkTrue}$. 
  Thus $\texttt{checkFalse}$  accepts an input $\WS$ if and only if $\WS$ is \emph{not} valid.
  
  Notice that the number of alternations of the ATM $\texttt{check}$  between existential  and universal choices is clearly the number of switches between the calls to the procedures $\texttt{checkTrue}$ and $\texttt{checkFalse}$, plus 2, i.e. $\AltN(\varphi)+2$.

The correctness of the procedure $\texttt{check}$ and its complexity bound is stated by the following proposition, that immediately implies Theorem~\ref{Theorem:UpperBoundAAbrBBarEbar}.   

\begin{proposition}\label{prop:correctnessATMcheck}The ATM $\texttt{check}$ is a singly exponential-time bounded ATM accepting $\FMC$,  whose number of alternations on input $(\Ku,\varphi)$ is at most $\AltN(\varphi)+2$.
\end{proposition}

The proof of Proposition~\ref{prop:correctnessATMcheck} is given in details in Appendix~\ref{APP:correctnessATMcheck}: it exploits the exponential small-model property for $\AAbarBBbarEbar$ (Theorem~\ref{theorem:singleExpTrackModelRegex}) which allows us to consider only certificates, that are singly exponential in the size of the input $(\Ku,\varphi)$, instead of traces of arbitrary length. 

Clearly, this algorithm improves on the complexity of the one presented in Section~\ref{sec:AAbarBBbarEbar} for $\AAbarBBbarEbar$ under homogeneity.
This concludes the section.

\subsection{$\LINAEXPTIME$-hardness of MC for $\BEbar$}\label{sec:LowerBound}
In this section we prove that the MC problem for the fragment $\BEbar$, extended with regular expressions, is $\LINAEXPTIME$-hard (implying the $\LINAEXPTIME$-hardness of $\AAbarBBbarEbar$). The result is obtained by a polynomial-time reduction from
a variant of the domino-tiling problem for grids with exponential-length rows and columns, 
 called \emph{alternating multi-tiling problem}.
  
An instance of this problem is a tuple $\Instance=\tpl{n,D,D_0,H,V, M,D_\acc}$, where: $n$ is a positive \emph{even} natural number encoded in unary; $D$ is a  non-empty finite set of \emph{domino types}; $D_0\subseteq D$ is a set of \emph{initial domino types}; $H\subseteq D\times D$ and $V\subseteq D\times D$  are the \emph{horizontal} and \emph{vertical matching relations}, resp.;
 $M\subseteq D\times D$ is the \emph{multi-tiling matching relation};
 $D_\acc\subseteq D$ is a set of \emph{accepting domino types}.

\begin{figure}[p]
\centering
    \includegraphics[width=0.9\textwidth]{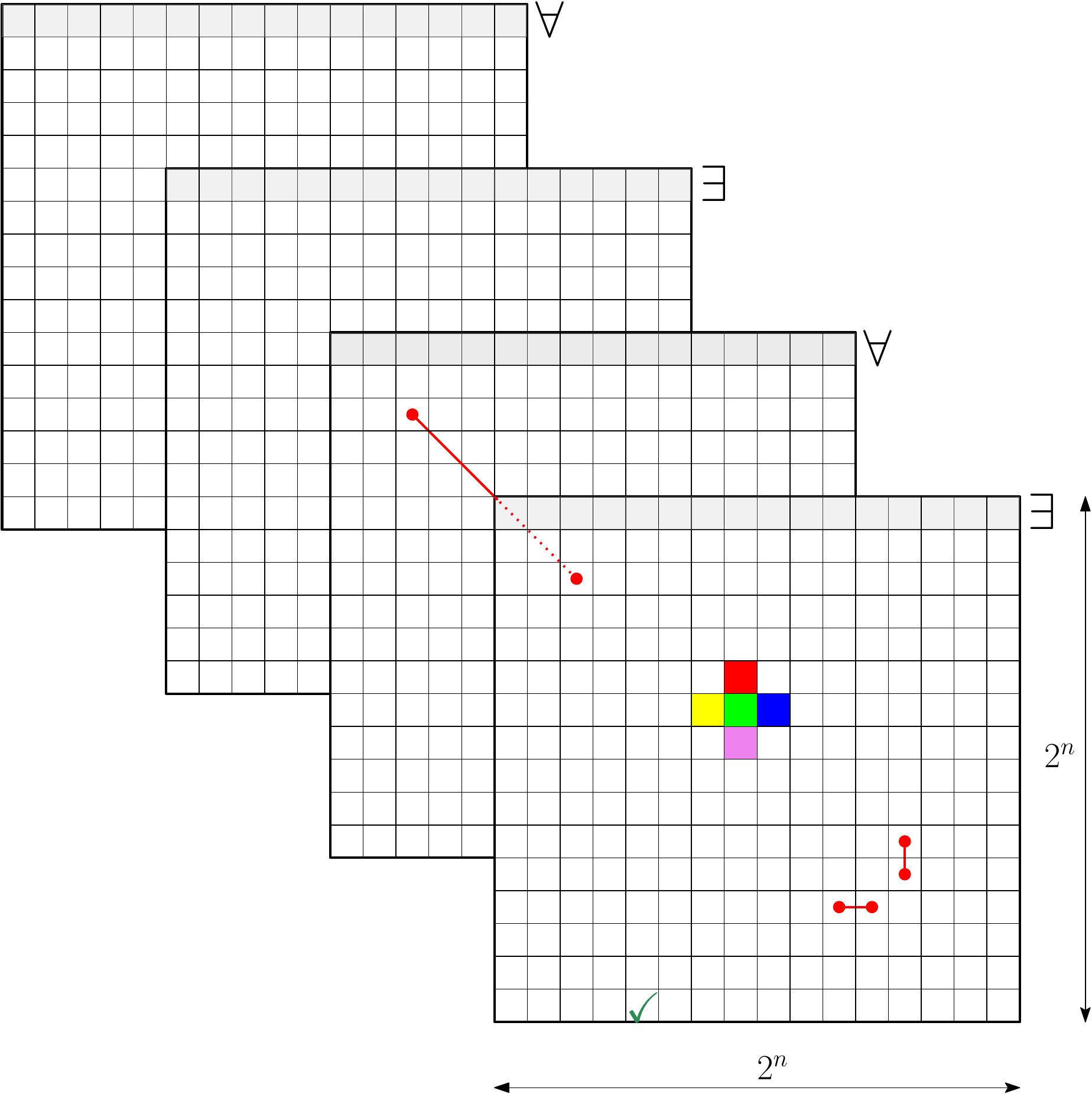}
\caption{The alternating multi-tiling problem (for $n=4$).
The red lines represent the row-adjacency,  column-adjacency, and multi-cell requirements. The green tick denotes the acceptance requirement.
The quantifiers $\forall/\exists$ associated with  the first rows of each tiling mean that the content of these rows has to be universally (resp., existentially) selected, if they belong to an odd (resp., even) tiling.}\label{fig:altmultitil}
\end{figure}

A \emph{tiling of $\Instance$} is a map assigning a domino type to each cell of a $2^{n} \times 2^{n}$ squared grid 
coherently with
the horizontal and vertical matching relations. Formally, a tiling of $\Instance$  is a map   $f:[0,2^{n}-1]\times [0,2^{n}-1] \rightarrow D$ such that:
\begin{itemize}
  \item for all $i,j\in [0,2^{n}-1]\times [0,2^{n}-1]$ with $j<2^{n}-1$, $(f(i,j),f(i,j+1))\in H$ (\emph{row-adjacency requirement});
  \item  for all $i,j\in [0,2^{n}-1]\times [0,2^{n}-1]$ with $i<2^{n}-1$, $(f(i,j),f(i+1,j))\in V$ (\emph{column-adjacency requirement}).
\end{itemize}
The \emph{initial condition} $\Init(f)= f(0,0)f(0,1)\cdots f(0,2^{n}-1)$ of the tiling $f$ is  the content of the first row of $f$.
%
A \emph{multi-tiling of $\Instance$} is a tuple $\tpl{f_1,\ldots,f_n}$ of $n$ tilings which are coherent w.r.t.\ the multi-tiling matching relation $M$, that is, such that:
\begin{itemize}
  \item for all  $i,j\in [0,2^{n}-1]\times [0,2^{n}-1]$ and $\ell\in[1,n-1]$,  $(f_\ell(i,j),f_{\ell+1}(i,j))\in M$ (\emph{multi-cell requirement}), and
 \item $f_n(2^{n}-1,j)\in D_\acc$ for some $j\in [0,2^{n}-1]$ (\emph{acceptance requirement}).
\end{itemize}
The \emph{alternating multi-tiling problem} for an instance $\Instance$ is checking whether
\[\forall w_1\in (D_0)^{2^{n}},\exists w_2 \in (D_0)^{2^{n}},\ldots,\forall w_{n-1}\in (D_0)^{2^{n}}, \exists w_n\in (D_0)^{2^{n}}\] 
such that there exists a multi-tiling $\tpl{f_1,\ldots,f_n}$, where for all $i\in [1,n]$, $\Init(f_i)=w_i$.
See Figure~\ref{fig:altmultitil} for a visual representation of the alternating multi-tiling problem.

The following complexity result holds (the proof is in Appendix~\ref{proof:theo:ComplexityAlternatingMT}).
\begin{theorem}\label{theo:ComplexityAlternatingMT} The alternating multi-tiling problem is $\LINAEXPTIME\!$-complete
\end{theorem}

The fact that the MC problem for the fragment $\BEbar$ with regular expressions is $\LINAEXPTIME$-hard is an immediate corollary of the following result.

\begin{theorem}\label{theo:MainLowerBoundResult}
One can construct, in time polynomial in the size of $\Instance$, a finite Kripke structure $\Ku_\Instance$ and a $\BEbar$ formula
$\varphi_\Instance$ over the set of proposition letters \[\Prop = D \cup (\{r,c\}\times \{0,1\}) \cup \{\bot,\End\},\]
such that  
$\Ku_\Instance\models \varphi_\Instance$ if and only if $\Instance$ is a \emph{positive} instance of the alternating multi-tiling problem.
\end{theorem}


The rest of this section is devoted to the construction of the Kripke structure $\Ku_\Instance$ and the $\BEbar$ formula
$\varphi_\Instance$ proving Theorem~\ref{theo:MainLowerBoundResult}. Let $\Prop$ be as above. The Kripke structure $\Ku_\Instance$ is given by
$\Ku_\Instance=\KuDef$, where $\States=\Prop$, $\sinit=\End$, $\Lab$ is the identity mapping (we identify
a singleton set $\{p\}$ with $p$), and $\Trans = \{(s,s') \mid s\in \Prop\setminus \{\End\}, s'\in\Prop\}$. Note that the initial state $\End$ has no successors,\footnote{This violates Definition~\ref{def:kripkestructure}, but we define the state $\End$ to have no successors only for technical convenience.}
and that a trace of $\Ku_\Instance$ can be identified with its induced labeling sequence.

The construction of the $\BEbar$ formula
$\varphi_\Instance$ is based on a suitable encoding of multi-tilings which is described in the following. The symbols
$\{r\}\times   \{0,1\}$ and  $\{c\} \times \{0,1\}$ in $\Prop$
are used to encode the values of two $n$-bits counters numbering the $2^{n}$ rows and  columns, respectively, of a  tiling.

For a multi-tiling $F=\tpl{f_1,\ldots,f_n}$ and for all $i,j\in [0,2^{n}-1]$, the $(i,j)$-{th} \emph{multi-cell} $\tpl{f_1(i,j),\ldots,f_n(i,j)}$ of $F$ is encoded by
the word $C$ of length $3n$ over $\Prop$, called \emph{multi-cell code}, given by
\[C= d_1 \cdots d_n (r, b_1)\cdots(r, b_n)(c,b'_1)\cdots(c, b'_n),\]
where $b_1 \cdots b_n$ and $b'_1\cdots b'_n$ are the binary encodings of the row number $i$ and column number $j$, resp., and for all $\ell\in[1,n]$, $d_\ell= f_\ell(i,j)$ (i.e., the content of the $(i,j)$-th cell of component $f_\ell$).
The \emph{content} of  $C$ is $d_1\cdots d_n$. 
%
Since $F$ is a multi-tiling,  the following well-formedness requirement must  be satisfied by the encoding $C$: for all $\ell\in [1,n-1]$, $(d_\ell,d_{\ell+1})\in M$. We call such words \emph{well-formed multi-cell codes}.

\begin{definition}[Multi-tiling codes]\label{Def:multiTilingCodes} 
A \emph{multi-tiling code} is a finite word $w$ over $\Prop$  obtained by concatenating well-formed multi-cell codes in such a way  that the following conditions hold:
\begin{itemize}
  \item for all $i,j\in [0,2^{n}-1]$, there is a multi-cell code in $w$  with row number $i$ and column number $j$ \emph{(completeness requirement)};
  \item for all multi-cell codes $C$ and $C'$ occurring in $w$, if $C$ and $C'$ have the same row number and column number, then $C$ and $C'$ have the same content \emph{(uniqueness requirement)};
   \item for all multi-cell codes $C$ and $C'$ in $w$ having the same row number
   (resp., column number), column numbers  (resp., row numbers) $j$ and $j+1$, resp., and contents $d_1\cdots d_n$ and $d'_1\cdots d'_n$, resp., it holds that $(d_\ell,d'_\ell)\in H$ (resp. $(d_\ell,d'_\ell)\in V$) for all $\ell\in [1,n]$ \emph{(row-adjacency requirement)} (resp., \emph{(column-adjacency requirement)});
  \item there is a multi-cell code in $w$ with row number $2^{n}-1$   whose content is in $D^{n-1}\cdot d_\acc$ for some $d_\acc\in D_\acc$ \emph{(acceptance requirement)}.
\end{itemize}
\end{definition}

Finally, we encode the initial conditions of the components of a multi-tiling.
An \emph{initial cell code} encodes a cell of the first row of a tiling
%
and is a word $w$ of length $n+1$ having the form $w =d (c,b_1) \cdots (c,b_n)$, where $d\in D_0$ and $b_1,\ldots,b_n\in \{0,1\}$. We say that $d$ is the \emph{content} of $w$ and the integer in $[0,2^{n}-1]$ encoded by $b_1\cdots b_n$ is the \emph{column number} of $w$.

\begin{definition}[Multi-initialization codes]\label{Def:InitializationCodes} An \emph{initialization code} is a finite word $w$ over $\Prop$ which is the concatenation of initial cell codes such that:
\begin{itemize}
  \item for all $i \in [0,2^{n}-1]$, there is an initial cell code in $w$  with column number $i$; 
  \item for all initial cell codes $C$ and $C'$ occurring in $w$, if $C$ and $C'$ have the same  column number, then $C$ and $C'$ have the same content.
\end{itemize}
A \emph{multi-initialization code} is a finite word over $\Prop$ having the form $\bot\cdot w_n\cdots \bot \cdot w_1\cdot \End$
such that for all $\ell\in [1,n]$, $w_\ell$ is an initialization code.
\end{definition}

\begin{definition}[Initialized multi-tiling codes]\label{Def:IMTCodes} An \emph{initialized multi-tiling code} is a finite word over $\Prop$ having the form
$\bot \cdot w \cdot \bot\cdot w_n\cdots \bot \cdot w_1\cdot \End$
such that $w$ is a multi-tiling code, $\bot\cdot w_n\cdots \bot \cdot w_1\cdot \End$ is a multi-initialization code, and the following requirement holds:
\begin{itemize}
  \item for each multi-cell code in $w$ having row number $0$, column number $i$, and content $d_1\cdots d_n$, and for all $\ell\in [1,n]$, there is
  an initial cell code in $w_\ell$ having column number $i$ and content $d_\ell$  \emph{(initialization coherence requirement)}.
\end{itemize}
\end{definition}

Before proving Theorem~\ref{theo:MainLowerBoundResult},
we sketch the idea for the construction of the $\B\Ebar$ formula $\varphi_\Instance$ which guarantees that $\Ku_\Instance\models \varphi_\Instance$ if and only if $\Instance$ is a positive instance of the alternating multi-tiling problem. 

We preliminarily  observe that  since the initial state of $\Ku_\Instance$ has no successors, the only initial trace of $\Ku_\Instance$ is the trace $end$ having length 1. To guess a trace corresponding to an initialized multi-tiling code, $\Ku_\Instance$ is unraveled backward starting from $end$, exploiting the modality $\Ebar$. The structure of the formula $\varphi_\Instance$ is 
\[
\varphi_\Instance= \hsEtu(\varphi_1 \rightarrow \hsEt(\varphi_2\wedge(\ldots (\hsEtu(\varphi_{n-1} \rightarrow \hsEt(\varphi_n\wedge \hsEt \varphi_\IMT)))\ldots))).
\]
It features $n+1$ unravelling steps starting from the initial trace $end$. The first $n$ steps are used to guess a sequence of $n$ initialization codes. Intuitively, each formula $\varphi_i$ is used to constrain 
the $i$-th unravelling to be an initialization code, in such a way that at depth $n$ in the formula a multi-initialization code is under evaluation. 
The last unravelling step (the innermost in the formula) is used to guess the multi-tiling code. The innermost formula $\varphi_\IMT$ is evaluated over a trace corresponding to an initialized multi-tiling code, and checks its structure: 
multi-cell codes are \lq\lq captured\rq\rq{} by regular expressions (encoding in particular their row and column numbers and contents).
The completeness, uniqueness, row- and column-adjacency requirements for the multi-tiling of Definition~\ref{Def:multiTilingCodes} are enforced by the combined use of the $\hsEtu$ modality and regular expressions.

\begin{figure}[tp]
    \centering
    \includegraphics[width=\textwidth]{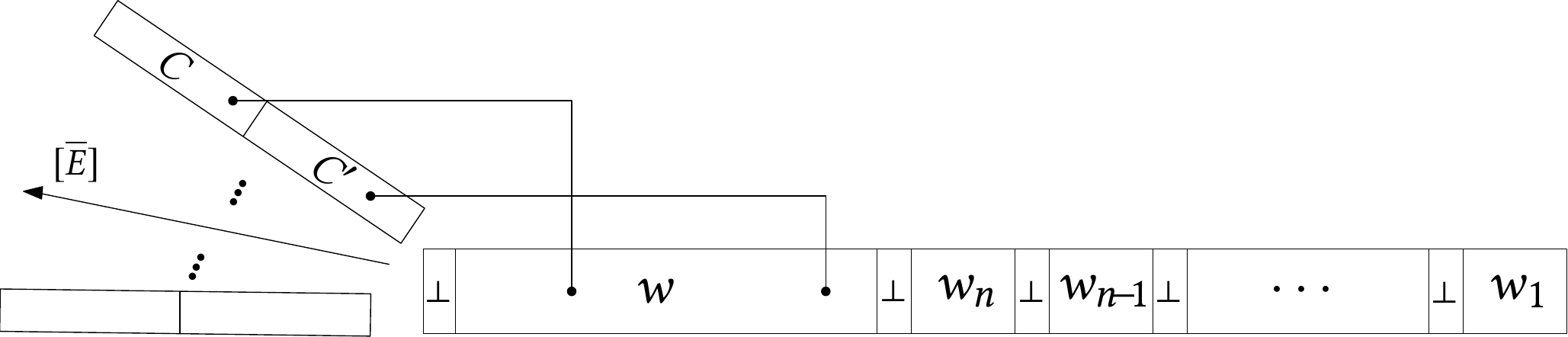}
    \caption{Checking constraints between pairs of multi-cell codes $C$ and $C'$ in an initialized multi-tiling code.}
    \label{fig:multicod}
\end{figure}

The intuition of such a technique is graphically depicted in Figure~\ref{fig:multicod}, where $w$ is a multi-tiling code.
Since the problem is to check constraints between pairs of multi-cell codes occurring in arbitrary positions of $w$, we use the following trick. A copy of two multi-cell codes $C$ and $C'$ (see Figure~\ref{fig:multicod}) are generated next to each other, as backward extensions of the initialized multi-tiling code, by means of modality $\hsEtu$. We then check that both $C$ and $C'$ occur in (arbitrary positions of) $w$, and, if this is the case, the required constraint is checked against the generated copies $C$ and $C'$, taking advantage of their adjacency. 
%
The initialization coherence requirement of Definition~\ref{Def:IMTCodes} is guaranteed in an analogous way, by comparing initial cell codes and multi-cell codes.

Note that the first $n-1$ occurrences of alternations between universal and existential modalities  $\hsEtu$ and $\hsEt$ correspond to the alternations of universal and existential quantifications in the definition of alternating multi-tiling problem. 

The correctness of the construction of $\varphi_\Instance$ is stated by the next proposition.
\begin{proposition}\label{prop:FormulasForMultiTilingCodes}  One can build, in  time polynomial in the size of $\Instance$,
$n+1$ $\BEbar$ formulas $\varphi_\IMT,\varphi_1,\ldots,\varphi_n$ with $\AltN(\varphi_\IMT)=\AltN(\varphi_1) =\ldots =\AltN(\varphi_n)= 0$, fulfilling the following conditions:
\begin{itemize}
  \item for all finite words $\rho$ over $\Prop$ having the form $\rho =\rho' \cdot \bot \cdot w_n  \cdots  \bot \cdot w_1\cdot \End$ such that $\rho'\neq \varepsilon$ and
  $\bot \cdot w_n  \cdots  \bot \cdot w_1\cdot \End$ is a multi-initialization code, it holds that $\Ku_\Instance,\rho\models\varphi_\IMT$ if and only if
  $\rho$ is an initialized multi-tiling code;
  \item for all $\ell\in [1,n]$ and words $\rho$ having the form $\rho=\rho'\cdot \bot \cdot w_{\ell-1} \cdots \bot \cdot w_1\cdot \End$ such that
 $\rho'\neq \varepsilon$ and $w_j\in (\Prop\setminus \{\bot\})^{*}$ for all $j\in [1,\ell-1]$, it holds that $\Ku_\Instance,\rho\models\varphi_\ell$ if and only if
$\rho'$ has the form  $\rho'=\bot\cdot w_\ell$, where $w_\ell$ is an initialization code.
\end{itemize}
\end{proposition}
\begin{proof}
Since each state of the Kripke structure $\Ku_\Instance$ is labeled by exactly one proposition letter of $\Prop$, in the proof we exploit the standard regular expressions, where atomic expressions are single letters of $\Prop$. Evidently, a standard regular expression can be converted into a proposition-based one, where each $p\in\Prop$ is replaced by the formula $p\wedge \bigwedge_{p'\in\Prop\setminus \{p\}}\neg p'$. 

Let us focus on the construction of the $\BEbar$ formula $\varphi_\IMT$ (as $\varphi_1,\ldots,\varphi_n$ are simpler).
First, we define a $\BEbar$ formula $\varphi_\MT$ ensuring the following property:
\begin{itemize}
  \item for all finite words $\rho$ over $\Prop$ having the form  $\rho =\rho' \cdot \bot \cdot w_n \cdots \bot \cdot w_1\cdot \End$ such that $\rho'\neq \varepsilon$ and
  $\bot \cdot w_n \cdots \bot \cdot w_1\cdot \End$ is a multi-initialization code, it holds $\Ku_\Instance,\rho\models\varphi_\MT$ if and only if
  $\rho'=\bot\cdot w$ for some multi-tiling code $w$.
\end{itemize}

In order to build $\varphi_\MT$, we need some auxiliary formulas and regular expressions.
\begin{itemize}
\item A regular expression $r_{\MC} = D^{n}\cdot (\{r\}\times \{0,1\})^{n}\cdot  (\{c\}\times \{0,1\})^{n}$ capturing the multi-cell codes.
\item    A $\B$ formula $\psi_{\Complete}$ requiring that for each word $C\cdot   \bot \cdot C_1 \cdots C_N\cdot \bot$ such that
  $C, C_1,\ldots,C_N$ are multi-cell codes,
   there is $i\in [1,N]$ such that $C$ and $C_i$ have the same row number and column number.
\begin{multline*}
    \psi_{\Complete} \! = \! \hsB \Bigl((r_{\MC}\cdot \bot \cdot (r_{\MC})^{+}) \wedge \!\!\!
    \bigwedge_{i\in [1,n]}\bigvee_{b\in \{0,1\}}\!\!\!\!\!(\Prop^{n+i-1}\cdot (r,b) \cdot \Prop^{+}\cdot  (r,b)\cdot \Prop^{2n-i}) \wedge
    \\
     \bigwedge_{i\in [1,n]}\bigvee_{b\in \{0,1\}}(\Prop^{2n+i-1}\cdot (c,b) \cdot \Prop^{+}\cdot   (c,b)\cdot \Prop^{n-i})\Bigr)
\end{multline*}

 \item A propositional formula $\psi_=$ requiring that for each word having as a proper prefix $C\cdot C'$ such that $C$ and $C'$ are multi-cell codes,
 $C$ and $C'$ have the same row number and column number.
\begin{multline*}
    \psi_{=}  =  \bigwedge_{i\in [1,n]}\bigvee_{b\in \{0,1\}}(\Prop^{n+i-1}\cdot (r,b) \cdot \Prop^{3n-1}\cdot (r,b)\cdot \Prop^{+})\wedge
    \\
    \bigwedge_{i\in [1,n]}\bigvee_{b\in \{0,1\}}(\Prop^{2n+i-1}\cdot (c,b) \cdot \Prop^{3n-1}\cdot (c,b)\cdot \Prop^{+})
\end{multline*}
 \item A propositional formula $\psi_{\RInc}$ (resp., $\psi_{\CInc}$) requiring that for each word having as a proper prefix $C\cdot C'$ such that $C$ and $C'$ are multi-cell codes,
 $C$ and $C'$ have the same column number (resp., the same row number), and there is $h\in [0,2^{n}-2]$ such that $C$ and $C'$ have row numbers (resp., column numbers) $h$ and $h+1$, resp.. We consider the formula $\psi_{\RInc}$ (the definition of $\psi_{\CInc}$ is similar).
\begin{multline*}
    \psi_{\RInc}  = \bigwedge_{i\in [1,n]}\bigvee_{b\in \{0,1\}}(\Prop^{2n+i-1}\cdot (c,b) \cdot \Prop^{3n-1}\cdot (c,b)\cdot \Prop^{+})\wedge
    \\
    \bigvee_{i\in [1,n]}\Bigl(\bigwedge_{j\in [1,i-1]}(\Prop^{n+j-1}\cdot (r,1) \cdot \Prop^{3n-1}\cdot (r,0)\cdot \Prop^{+})\wedge
    \\
     (\Prop^{n+i-1}\cdot (r,0) \cdot \Prop^{3n-1}\cdot (r,1)\cdot \Prop^{+})\wedge
    \\
    \bigwedge_{j\in [i+1,n]} \bigvee_{b\in \{0,1\}}(\Prop^{n+j-1}\cdot (r,b) \cdot \Prop^{3n-1}\cdot (r,b)\cdot \Prop^{+})\Bigr)
\end{multline*}
 \item A $\B$ formula $\psi_{\Double}$ requiring that for each word $C\cdot C'\cdot \bot \cdot C_1 \cdots C_N\cdot \bot$ such that
  $C,C',C_1,\ldots,C_N$ are multi-cell codes, there are $i,j\in [1,N]$ such that $C=C_i$ and $C'=C_j$;
  $\psi_{\Double} = \theta \wedge \theta'$, where $\theta$ (resp., $\theta'$) requires that
  there is $i\in [1,N]$ such that $C_i=C$ (resp., $C_i=C'$). We consider $\theta'$ (the definition of $\theta$ is similar).
\begin{multline*}
    \theta'  = \hsB \Bigl((r_{\MC}\cdot r_{\MC} \cdot \bot \cdot (r_{\MC})^{+})\wedge
    \bigwedge_{i\in [1,n]}\bigvee_{d\in D}(\Prop^{3n+i-1}\cdot d \cdot \Prop^{+}\cdot  d\cdot \Prop^{3n-i})\wedge
    \\
    \bigwedge_{i\in [1,n]}\bigvee_{b\in \{0,1\}}(\Prop^{4n+i-1}\cdot (r,b) \cdot \Prop^{+}\cdot  (r,b)\cdot \Prop^{2n-i})\wedge
    \\
    \bigwedge_{i\in [1,n]}\bigvee_{b\in \{0,1\}}(\Prop^{5n+i-1}\cdot (c,b) \cdot \Prop^{+}\cdot   (c,b)\cdot \Prop^{n-i})\Bigr)
\end{multline*}
  \item A $\B$ formula $\psi_{\NU}$ requiring that for each word $C\cdot C'\cdot \bot \cdot C_1 \cdots C_N\cdot \bot$ such that
  $C,C',C_1,\ldots,C_N$ are multi-cell codes, the next properties hold:
  \begin{itemize}
    \item $C$ and $C'$ have the same row and column numbers, but different content;
    \item there are $i,j\in [1,N]$ such that $C=C_i$ and $C'=C_j$.
  \end{itemize}
  The construction of $\psi_{\NU}$ is based on the formulas $\psi_{\Double}$ and $\psi_=$:
\begin{multline*}
    \psi_{\NU}  = \psi_{\Double} \wedge  \psi_= \wedge \bigvee_{i\in [1,n]}\bigvee_{d,d'\in D:d\neq d'}(\Prop^{i-1}\cdot d \cdot \Prop^{3n-1}\cdot d'\cdot \Prop^{+}).
\end{multline*}
\item A $\B$ formula $\psi_{\Row}$ (resp., $\psi_{\Column}$) requiring that for each word $C\cdot C'\cdots C_1 \cdots C_N\cdot \bot$ such that
  $C,C',C_1,\ldots,C_N$ are multi-cell codes, the next condition holds.
  \begin{itemize}
    \item Let us denote by $d_1\cdots d_n$ the content of $C$ and by $d'_1\cdots d'_n$ the content of $C'$. Whenever $(1)$~there are $i,j\in [1,N]$ such that $C=C_i$ and $C'=C_j$, and $(2)$~$C$ and $C'$ have the same row number and column numbers $h$ and $h+1$, resp. (resp., $C$ and $C'$ have the same column number and row numbers $h$ and $h+1$, resp.) for some $h\in [0,2^{n}-2]$,
         then it holds that $(d_\ell,d'_\ell)\in H$ (resp., $(d_\ell,d'_\ell)\in V$), for all
        $\ell\in [1,N]$.
  \end{itemize}
We focus on $\psi_{\Row}$ (the definition of $\psi_{\Column}$ is similar): 
\begin{multline*}
    \psi_{\Row}  = (\psi_{\Double} \wedge  \psi_{\CInc})  \longrightarrow  \bigwedge_{i\in [1,n]}\bigvee_{(d,d')\in H}(\Prop^{i-1}\cdot d \cdot \Prop^{3n-1}\cdot d'\cdot \Prop^{+}).
\end{multline*}
\end{itemize}

Finally, the $\BEbar$ formula  $\varphi_\MT$ is defined as follows:
\begin{multline*}
\neg (\Prop^{*}\cdot \bot \cdot \Prop^{*})^{n+2} \wedge
 \hsB \Bigl(\underbrace{(\bot\cdot (r_{\MC})^{+} \cdot \bot)\wedge\neg \bigvee_{(d,d')\in D^{2}\setminus M}(\Prop^{+}\cdot d\cdot d'\cdot \Prop^{+})}_{\text{Concatenation of well-formed multi-cell codes}}\wedge
\\
 \underbrace{\hsEtu((r_{\MC}\cdot \bot\cdot (r_{\MC})^{+} \cdot \bot) \longrightarrow \psi_\Complete)}_{\text{Completeness requirement of Definition \ref{Def:multiTilingCodes}}}\wedge
 \\
\underbrace{\hsEtu((r_{\MC}\cdot r_{\MC}\cdot \bot\cdot (r_{\MC})^{+} \cdot \bot) \longrightarrow \neg\psi_{\NU})}_{\text{Uniqueness requirement of Definition \ref{Def:multiTilingCodes}}}\wedge
\\
\underbrace{\hsEtu((r_{\MC}\cdot r_{\MC}\cdot \bot\cdot (r_{\MC})^{+} \cdot \bot) \longrightarrow (\psi_{\Row}\wedge \psi_{\Column}))}_{\text{Row-adjacency and column-adjacency requirements of Definition \ref{Def:multiTilingCodes}}}\wedge
\\
\underbrace{\bigvee_{d_\acc\in D_\acc}(\Prop^{+}\cdot d_\acc \cdot (r,1)^{n}\cdot \Prop^{+}) }_{\text{Acceptance requirement of Definition \ref{Def:multiTilingCodes}}}\Bigr).
\end{multline*}

The $\BEbar$ formula $\varphi_\IMT$ is given by
$\varphi_\MT\wedge \varphi_\Coh$,
where $\varphi_\Coh$ ensures the initialization coherence requirement of Definition~\ref{Def:IMTCodes}. In order to define
$\varphi_\Coh$, we need some auxiliary formulas and regular expressions.
\begin{itemize}
\item A regular expression $r_{\IC} = D_0 \cdot   (\{c\}\times \{0,1\})^{n}$ capturing the  initial cell codes.
\item    A $\B$ formula $\psi_{\Single}$ requiring that for each word $C\cdot   \bot \cdot C_1 \cdots C_N\cdot \bot$ such that
  $C, C_1,\ldots,C_N$ are multi-cell codes,
   there is $i\in [1,N]$ such that $C=C_i$ and  the row number of $C$ is $0$. The definition of $\psi_\Single$ is similar to $\psi_\Double$.
 \item A $\B$ formula $\psi_{\Coh}$ requiring that for each word $C\cdot   \bot \cdot C_1 \cdots C_N\cdot \bot \cdot w_n \cdots \bot \cdot w_1 \cdot \End$ such that   $C, C_1,\ldots,C_N$ are multi-cell codes and $\bot \cdot w_n \cdots \bot \cdot w_1 \cdot \End$ is a multi-initialization code, the following constraint holds:
     if   there is $i\in [1,N]$ such that $C=C_i$, the row number of $C$ is $0$ and the content of $C$ is $d_1\cdots d_n$, then for all $\ell\in [1,n]$, there exists an initial code in $w_\ell$ having the same column number as $C$ and content $d_\ell$.
\[
\psi_\Coh  =   \bigl( \hsB( [(\Prop\setminus \{\bot\})^{+} \cdot \bot \cdot (\Prop\setminus \{\bot\})^{+} \cdot \bot] \wedge \psi_\Single)  \bigr) \longrightarrow \displaystyle{\bigwedge_{\ell\in [1,n]}\psi_\ell};
\]

\begin{multline*}
\psi_\ell =   \hsB\Bigl(\bigl[(\Prop\setminus \{\bot\})^{+}\cdot (\bot \cdot (\Prop\setminus \{\bot\})^{+})^{n-\ell+1}\cdot \bot \cdot r_\IC^{+}  \bigr] \wedge
\\
\bigwedge_{i\in [1,n]}\bigvee_{b\in \{0,1\}}(\Prop^{2n+i-1}\cdot (c,b) \cdot \Prop^{+}\cdot (c,b)\cdot \Prop^{n-i})\wedge
\\
 \bigvee_{d\in D}(\Prop^{\ell-1}\cdot d \cdot \Prop^{+}\cdot d\cdot \Prop^{n})\Bigr).
\end{multline*}
\end{itemize}
The  $\BEbar$ formula $\varphi_\Coh$ is then 
$\varphi_\Coh = \hsEtu\bigl([r_\MC\cdot (\bot \cdot (\Prop\setminus \{\bot\})^{+})^{n+1}  ]  \rightarrow \psi_\Coh\bigr)$.

This concludes the proof of Proposition~\ref{prop:FormulasForMultiTilingCodes}.
\end{proof}

Recall that
$\varphi_\Instance\!=\! \hsEtu(\varphi_1 \rightarrow \hsEt(\varphi_2\wedge(\ldots (\hsEtu(\varphi_{n-1} \rightarrow \hsEt(\varphi_n\wedge \hsEt \varphi_\IMT)))\ldots))),$
where $\varphi_\IMT,\varphi_1,\ldots,\varphi_n$ are the formulas defined in Proposition~\ref{prop:FormulasForMultiTilingCodes}.
Since the initial state of $\Ku_\Instance$ has no successors and the only initial trace has length 1 and corresponds to the proposition letter $\End$, by Definitions~\ref{Def:multiTilingCodes}--\ref{Def:IMTCodes}, we have that
$\Ku_\Instance\models \varphi_\Instance$ if and only if $\Instance$ is a positive instance of the alternating multi-tiling problem, proving  Theorem~\ref{theo:MainLowerBoundResult}. This result, combined with Theorem~\ref{Theorem:UpperBoundAAbrBBarEbar}, implies the following.

\begin{corollary}
The MC problem for $\AAbarBBbarEbar$ (and $\AAbarEBbarEbar$) formulas extended with regular expressions over finite Kripke structures is \mbox{$\LINAEXPTIME$-complete}.
\end{corollary}

%% file: Chaps/Gandalf17RIVISTA/AAbarEEbar.tex
\section{The fragments $\AAbarBBbar$ and $\AAbarEEbar$}\label{sec:AABB}
In this section we show that the two symmetric fragments $\AAbarBBbar$ and $\AAbarEEbar$,
extended with regular expressions, 
feature a better complexity, showing MC for them to be in $\Psp$.  To this end we first prove, in Section~\ref{subsec:AAbarEEbar}, that they feature an \emph{exponential small-model property}, that is, if a trace $\rho$ of a finite Kripke structure $\Ku$ satisfies a formula $\varphi$ of $\AAbarBBbar$/$\AAbarEEbar$, then there is always a trace $\pi$, whose length is exponential 
in the sizes of $\varphi$ and $\Ku$, starting from and leading to the same states as $\rho$, that satisfies $\varphi$.  
Therefore, without loss of generality, one can limit the verification of traces of $\Ku$ to those having at most exponential length. It is worth recalling that, in Section~\ref{sec:AAbarEEbar}, we proved a \emph{polynomial small-model property} in the sizes of the  $\AAbarBBbar$/$\AAbarEEbar$ formula $\varphi$ and the Kripke structure $\Ku$ \emph{under the homogeneity assumption}.

Then, in Section~\ref{sect:PspAlgo} and~\ref{sect:genResult} we provide a $\Psp$ MC algorithm which exploits the exponential small-model property. Such an algorithm is completely different from the one presented in Section~\ref{sec:AAbarEEbar} for 
the MC problem of the same fragments under the homogeneity assumption, which can
exploit the aforementioned polynomial small-model property. As a matter of fact, unlike that of Section~\ref{sec:AAbarEEbar}, this algorithm cannot store even a single---possibly exponential-length---trace, being bound to use polynomial working space. For this reason it visits the (exponential-length) traces of the input Kripke structure $\Ku$ by means of a \emph{binary reachability} technique that allows it to use logarithmic space in the length of traces, hence guaranteeing the $\Psp$ complexity upper bound. The surprising fact is that both the algorithm of Section~\ref{sec:AAbarEEbar} and the one presented here use polynomial working space (independently of the different size small-models), thus showing that relaxing the homogeneity assumption comes at no additional computational cost for the fragments $\AAbarBBbar$ and $\AAbarEEbar$.

Finally, in Section~\ref{sect:genResult} we prove the $\Psp$-completeness of MC for $\AAbarBBbar$ and $\AAbarEEbar$ with regular expressions.

\subsection{Exponential small-model property for $\AAbarBBbar$ and $\AAbarEEbar$}
\label{subsec:AAbarEEbar}
In this section we prove the exponential small-model property for the fragments $\AAbarBBbar$ and $\AAbarEEbar$ 
(actually, we focus only on $\AAbarBBbar$ being the case for $\AAbarEEbar$ symmetric). Most results are just adaptations---with the aim of accounting for regular expressions---of those already presented in Section~\ref{sec:AAbarEEbar}.

Given a $\DFA$ $\Du=(\Sigma,Q,q_0,\delta,F)$, we denote by $\Du(w)$ (resp., $\Du_q(w)$) the state reached by the computation of $\Du$ from $q_0$ (resp., $q\in Q$) over the word $w\in\Sigma^*$.

We now consider \emph{well-formedness} of induced traces (recall Definition~\ref{definition:inducedTrk}) w.r.t.\ a set of $\DFA$s:%
\footnote{Another variant of well-formedness of induced traces, independent of $\DFA$s, was already given in Definition~\ref{definition:inducedTrk}.}
a well formed trace $\pi$ induced by $\rho$ preserves the states of the computations of the  $\DFA$s reached by reading prefixes of $\rho$ and $\pi$ bounded by corresponding positions.
Hereafter, for $i\in [1,|\rho|]$, $\rho^i$ denotes the prefix $\rho(1,i)$.

\begin{definition}[Well-formed trace w.r.t.\ a set of $\DFA$s]
Let $\Ku=\KuDef$ be a finite Kripke structure, $\rho\in\Trk_\Ku$ be a trace, and
$\Du^s=(2^\Prop,Q^s,q_0^s,\delta^s,F^s)$ for $s=1,\ldots ,k$, be $\DFA$s.
  A trace $\pi\in\Trk_\Ku$ \emph{induced by} $\rho$ is 
\emph{$(q^1_{\ell_1}, \ldots , q^k_{\ell_k})$-well-formed w.r.t.\ }$\rho$, with $q^s_{\ell_s}\in Q^s$  for all $s=1,\ldots ,k$, if and only if:
\begin{itemize}
     \item for all $\pi$-positions $j$, with corresponding $\rho$-positions $i_j$, and all $s=1,\ldots ,k$, it holds that $\Du^s_{q^s_{\ell_s}}(\mu(\pi^{j}))= \Du^s_{q^s_{\ell_s}}(\mu(\rho^{i_j}))$.
\end{itemize}
\end{definition}
It is easy to see that, 
for $q^s_{\ell_s}\in Q^s$, $s=1,\ldots ,k$, the $(q^1_{\ell_1}, \ldots , q^k_{\ell_k})$-well-formed\-ness relation is \emph{transitive}.

It is possible to show that a trace, whose length exceeds a suitable
exponential threshold, induces a shorter, well-formed trace. 
Such a contraction pattern (Proposition~\ref{proposition:wellFormdnessRegex})  represents a ``basic step'' in a contraction process which will allow us to prove the exponential small-model property for $\AAbarBBbar$.

Let us consider 
an $\AAbarBBbar$ formula $\varphi$ and let $r_1,\ldots ,r_k$ be the $\RE$'s over $\Prop$ in $\varphi$. Let 
 $\Du^1,\ldots, \Du^k$ be the $\DFA$s such that $\mathcal{L}(\Du^t)=\mathcal{L}(r_t)$, for $t=1,\ldots, k$, where $|Q^t|\leq 2^{2|r_t|}$ (see Remark~\ref{remk:nfa}). We denote $Q^1\times \ldots \times Q^k$ by $Q(\varphi)$, and $\Du^1,\ldots, \Du^{k}$ by $\Du(\varphi)$.

\begin{proposition}\label{proposition:wellFormdnessRegex} 
Let $\Ku=\KuDef$ be a finite Kripke structure, $\varphi$ be an $\AAbarBBbar$ formula  with $\RE$'s $r_1,\ldots ,r_k$ over $\Prop$, $\rho\in\Trk_\Ku$ be
a trace, and $(q^1,\ldots , q^k)\in Q(\varphi)$. There exists a trace $\pi\in\Trk_\Ku$, which is $(q^1,\ldots , q^k)$-well-formed w.r.t.\ $\rho$, such that $|\pi| \leq |\States|\cdot 2^{2\sum^k_{\ell=1}|r_\ell|}$.
\end{proposition}
The proof, which is an adaptation of that of Proposition~\ref{proposition:wellFormdness}, can be found in Appendix~\ref{proof:proposition:wellFormdnessRegex}.

The next step is to determine some conditions for contracting traces while preserving the equivalence w.r.t.\ the satisfiability of the considered $\AAbarBBbar$ formula. 
 In the following, we restrict ourselves again to formulas in \nnf.

For a trace $\rho$ and a formula $\varphi$ of $\AAbarBBbar$ (in \nnf), we fix a set of distinguished $\rho$-positions, called \emph{witness positions} (recall Definition~\ref{definition:WitnessPositions}), each one corresponding to the minimum prefix of $\rho$ which satisfies a formula $\psi$ occurring in $\varphi$ as a subformula of the form $\hsB\psi$ (provided that $\hsB\psi$ is satisfied by $\rho$). Such set is denoted by $Wt(\varphi,\rho)$, and we have $|Wt(\varphi,\rho)|\leq |\varphi|-1$.
We can show that, when a contraction is performed
 in between a pair of \emph{consecutive} witness positions (thus no witness position is ever removed), we get a trace induced by $\rho$ (according to Definition~\ref{definition:inducedTrk}) equivalent with respect to the satisfiability of $\varphi$. 

We are now ready to state the exponential small-model property for $\AAbarBBbar$.
\begin{theorem}[Exponential small-model for $\AAbarBBbar$]\label{theorem:expSizeModelPropertyAAbarBBbarRegex}
Let $\Ku=(\Prop,\States,\allowbreak \Edges,\Lab,\sinit)$ be a finite Kripke structure, $\sigma,\rho \in \Trk_\mathpzc{K}$, and $\varphi$ be an $\AAbarBBbar$ formula in \nnf{}, with $\RE$'s  $r_1,\ldots ,r_u$ over $\Prop$, such that $\Ku,\sigma\star\rho\models \varphi$. There exists $\pi\in \Trk_\mathpzc{K}$, induced by $\rho$, such that $\Ku,\sigma\star\pi\models \varphi$ and $|\pi|\leq |\States|\cdot (|\varphi|+1) \cdot 2^{2\sum_{\ell=1}^u |r_\ell|}$.
\end{theorem}
The theorem holds in particular if $|\sigma|\!=\!1$, and thus $\sigma\star\rho\!=\!\rho$ and $\sigma\star\pi\!=\!\pi$. In this case, if $\Ku,\rho\models \varphi$, then $\Ku,\pi\models \varphi$, where $\pi$ is induced by $\rho$ and $|\pi|\leq |\States|\cdot (|\varphi|+1) \cdot 2^{2\sum_{\ell=1}^u |r_\ell|}$. 
The proof, which is an adaptation of the one of Theorem~\ref{theorem:polynomialSizeModelProperty}, can be found in Appendix~\ref{proof:theorem:expSizeModelPropertyAAbarBBbarRegex}.

The exponential small-model property allows us to devise a trivial \emph{exponential working space} algorithm for $\AAbarBBbar$ (and $\AAbarEEbar$)---as already anticipated, we will actually present a polynomial space 
one in the next sections---which basically unravels the Kripke structure and 
checks all subformulas of the input formula. At every step it can consider traces not longer than $O(|\States|\cdot |\varphi| \cdot 2^{2\sum_{\ell=1}^u |r_\ell|})$. 
Conversely, the following example shows that the exponential small-model is strict, that is, there exist a formula and a Kripke structure such that the shortest trace satisfying the formula has exponential length in the size of the formula itself. This is the case even for purely propositional formulas (of the $\HS$ fragment $\HSprop$).
\begin{example} 
Let $pr_i$ be the $i$-th smallest prime number. It is well-known that $pr_i\in O(i\log i)$. Let $w^{\otimes k}$ denote the string obtained by concatenating $k$ times $w$.
Let us fix some $n\in\mathbb{N}$, and 
let $\Ku=(\{p\},\{s\}, \Edges,\mu,s)$ be the trivial Kripke structure having only one state with a self-loop, where $\Edges=\{(s,s)\}$, and $\mu(s)=\{p\}$.

The shortest trace satisfying \[\psi=\bigwedge_{i=1}^n (p^{\otimes (pr_i)})^*\] is $\rho=s^{\otimes (pr_1\cdots pr_n)}$, since its length is the least common multiple of $pr_1,\ldots, pr_n$, which is, indeed, $pr_1\cdots pr_n$.
It is immediate to check that the length of $\psi$ is $O(n\cdot pr_n)=O(n^2\log n)$. On the other hand, the length of $\rho$ is $pr_1\cdots pr_n \geq 2^n$.
\end{example}

In the following, we will exploit the exponential small-model property of the two symmetrical fragments $\AAbarBBbar$ and $\AAbarEEbar$ to prove the $\Psp$-completeness of their MC problems.
First, in Section~\ref{sect:PspAlgo}, we will provide a $\Psp$ MC algorithm for $\B\Bbar$ (resp., $\E\Ebar$). Then, in Section~\ref{sect:genResult}, we will show that the \emph{meets} and \emph{met-by} modalities $\A$ and $\Abar$ can be suitably encoded by regular expressions without increasing the complexity of $\B\Bbar$ (resp., $\E\Ebar$).

%% file: Chaps/Gandalf17RIVISTA/BBbar.tex
\subsection{$\Psp$ MC algorithm for $\B\Bbar$}\label{sect:PspAlgo}
In this section, to start with, we describe a $\Psp$ MC algorithm for $\B\Bbar$ formulas, extended with regular expressions.  W.l.o.g., we assume that the processed formulas do not contain occurrences of the universal modalities $\hsBu$ and $\hsBtu$. Moreover, for a formula $\psi$, we denote by $\subfB(\psi)=\{\varphi\mid \hsB\varphi \text{ is a subformula of } \psi\}$. In such an algorithm, $\Phi$ represents the overall formula to be checked, while the parametric formula $\psi$ ranges over its subformulas. 

Due to the result of the previous section, the algorithm can consider only traces having length bounded by the exponential small-model property. Note that
an algorithm required to work in polynomial space cannot explicitly
store the $\DFA$s for the regular expressions occurring in $\Phi$
(their states are \emph{exponentially} many in the length of the associated regular expressions). 
For this reason, while checking a formula against a trace, the algorithm just stores the \emph{current states} of the computations of the $\DFA$s associated with the regular expressions in $\Phi$, from the respective initial states 
(in the following such states are denoted---with a little abuse of notation---again by $\Du(\Phi)$, and called the \emph{\lq\lq current configuration\rq\rq{} of the $\DFA$s}) and calculates on-the-fly the successor states in the $\DFA$s, once they have read some state of $\Ku$ used to extend the considered trace 
(this can be done by 
exploiting a \emph{succinct} encoding of the $\NFA$s for the regular expressions of $\Phi$, see again Remark~\ref{remk:nfa} in Section~\ref{sect:backgrRegex}).

A call to the recursive procedure \texttt{Check}$(\Ku,\psi,s,G,\Du(\Phi))$ (Algorithm~\ref{Check}) verifies the satisfiability of a subformula $\psi$ of $\Phi$ w.r.t.\ any trace $\rho$ fulfilling the next conditions:
\begin{enumerate}
    \item $G \subseteq \subfB(\psi)$ is the set of formulas that hold true on at least a prefix of $\rho$;
    \item after reading $\Lab(\rho(1,|\rho|-1))$ the current configuration of the $\DFA$s for the regular expressions of $\Phi$ is $\Du(\Phi)$; 
    \item the last state of $\rho$ is $s$.
\end{enumerate}

\begin{algorithm}[tp]
    \caption{\texttt{Check}$(\Ku,\psi,s,G,\Du(\Phi))$}\label{Check}
    \begin{algorithmic}[1]
        \If{$\psi=r$}\Comment{$r$ is a regular expression}
            \If{the current state of the $\DFA$ for $r$ in \texttt{advance}$(\Du(\Phi),\mu(s))$ is final}
                \Return{$\top$}
            \Else
                \Return{$\bot$}
            \EndIf
        \ElsIf{$\psi=\neg\psi'$}
            \Return{\textbf{not} \texttt{Check}$(\Ku,\psi',s,G,\Du(\Phi))$}
        \ElsIf{$\psi=\psi_1\wedge \psi_2$}
            \Return{\texttt{Check}$(\Ku,\psi_1,s,G\cap \subfB(\psi_1),\Du(\Phi))$ \textbf{and} \texttt{Check}$(\Ku,\psi_2,s,G\cap \subfB(\psi_2),\Du(\Phi))$}
        \ElsIf{$\psi=\hsB\psi'$}
            \If{$\psi'\in G$}
                \Return{$\top$}
            \Else
                \Return{$\bot$}
            \EndIf
        \ElsIf{$\psi=\hsBt\psi'$}
            \For{each $b\in\{1,\ldots ,|\States|\cdot (2|\psi'|+1)\cdot 2^{2\sum_{\ell=1}^u |r_\ell|} -1 \}$ and each $(G',\Du(\Phi)',s')\in\conf(\Ku,\psi)$}\Comment{$r_1,\ldots ,r_u$ are the regular expressions of $\psi'$}
                \If{\texttt{Reach}$(\Ku,\psi',(G,\Du(\Phi),s),(G',\Du(\Phi)',s'),b)$ \textbf{and}
                    \texttt{Check}$(\Ku,\psi',\allowbreak s',G',\allowbreak \Du(\Phi)')$}
                    \Return{$\top$}
                \EndIf
            \EndFor
            \Return{$\bot$}
        \EndIf
    \end{algorithmic}
\end{algorithm}

Intuitively, since the algorithm cannot store the already checked portion of a
trace (whose length could be exponential), the relevant information is \emph{summarized} in a triple $(G,\Du(\Phi),s)$. 
Hereafter, the set of all possible summarizing triples $(\overline{G},\overline{\Du(\Phi)},\overline{s})$, where $\overline{G}\subseteq\subfB(\psi)$, $\overline{\Du(\Phi)}$ is any current configuration of the $\DFA$s for the regular expressions of $\Phi$, and $\overline{s}$ is a state of $\Ku$, is denoted 
by $\conf(\Ku,\psi)$.

Let us consider in detail the body of the procedure.
First, \texttt{advance}($\Du(\Phi),\Lab(s)$), invoked at line 2, updates the current configuration of the $\DFA$s after reading the symbol $\Lab(s)$. 
If $\psi$ is a regular expression $r$ (lines 1--5), we just check whether the (computation of the) $\DFA$ associated with $r$ is in a final state (i.e., the summarized trace is accepted).
Boolean connectives are easily dealt with recursively (lines 6--9).
If $\psi$ has the form $\hsB\psi'$ (lines 10--14), then $\psi'$ has to hold over a proper prefix of the summarized trace, i.e. $\psi'$ must belong to $G$.

The only involved case is $\psi=\hsBt\psi'$ (lines 15--19): we have to unravel the Kripke structure $\Ku$ to find an \emph{extension} $\rho'$ of $\rho$, summarized by the triple $(G',\Du(\Phi)',s')$, satisfying $\psi'$. The idea is 
checking whether there exists a summarized 
trace $(G',\Du(\Phi)',s')$, suitably extending $(G,\Du(\Phi),s)$, namely, such that:
\begin{enumerate}
\item $\Du(\Phi)'$ and $s'$ are \emph{synchronously} reachable from $\Du(\Phi)$ and $s$, respectively; 
\item $G'\supseteq G$ contains any formula of $\subfB(\psi')$ satisfied by a prefix of the extension; 
\item the extension
$(G',\Du(\Phi)',s')$ satisfies $\psi'$.
\end{enumerate}

In order to check $(1.)$, i.e., synchronous reachability,
we can exploit the exponential small-model property and 
consider only the unravelling of $\Ku$ starting from $s$ having depth at most $|\States|\cdot (2|\psi'|+1)\cdot 2^{2\sum_{\ell=1}^u |r_\ell|} -1$\footnote{
The factor 2 of $|\psi'|$ is added since the exponential small-model for $\AAbarBBbar$ requires a formula to be in \nnf .
}.
The verification of $(1.)$ and $(2.)$ is performed by the procedure
\texttt{Reach} (Algorithm~\ref{Reach}), which accepts as input two summarized traces and a bound $b$ on the depth of the unravelling of $\Ku$.
The proposed reachability algorithm is reminiscent of  the binary reachability technique of Savitch's theorem~\cite{Garey79}. 

\begin{algorithm}[tp]
    \caption{\texttt{Reach}($\Ku,\psi,(G_1,\Du(\Phi)_1,s_1),(G_2,\Du(\Phi)_2,s_2),b$)}\label{Reach}
    \begin{algorithmic}[1]
        \If{$b=1$}
            \Return{\texttt{Compatible}($\Ku,\psi,(G_1,\Du(\Phi)_1,s_1),(G_2,\Du(\Phi)_2,s_2)$)}
        \Else\Comment{$b\geq 2$}
            \State{$b'\gets \lfloor b/2\rfloor$}
            \For{each $(G_3,\Du(\Phi)_3,s_3)\in\conf(\Ku,\psi)$}
                \If{\texttt{Reach}($\Ku,\psi,(G_1,\Du(\Phi)_1,s_1),(G_3,\Du(\Phi)_3,s_3),b'$) \textbf{and} \texttt{Reach}($\Ku,\psi,\allowbreak (G_3,\Du(\Phi)_3,s_3),\allowbreak (G_2,\Du(\Phi)_2,s_2),b-b'$)}
                    \Return{$\top$}
                \EndIf
            \EndFor
            \Return{$\bot$}
        \EndIf
    \end{algorithmic}
\end{algorithm}

The procedure \texttt{Reach} proceeds recursively (lines 3--8) by halving at each step the value $b$ of the length bound, until it gets called over two states $s_1$ and $s_2$ which are
adjacent in a trace. At each halving step, an intermediate summarizing triple is generated to be associated with the split point.
At the base of recursion (for $b=1$, lines 1--2), the auxiliary procedure \texttt{Compatible} (Algorithm~\ref{Compatible}) is invoked. 

\begin{algorithm}[tp]
    \caption{\texttt{Compatible}($\Ku,\psi,(G_1,\Du(\Phi)_1,s_1), (G_2,\Du(\Phi)_2, s_2)$)}\label{Compatible}
    \begin{algorithmic}[1]        
        \If{$(s_1,s_2)\in \Edges$ \textbf{and} \texttt{advance}$(\Du(\Phi)_1,\mu(s_1))=\Du(\Phi)_2$ \textbf{and} $G_1\subseteq G_2$}
            \For{each $\varphi\in (G_2\setminus G_1)$}
                \State{$G\gets G_1\cap\subfB(\varphi)$}
                \If{\texttt{Check}$(\Ku,\varphi,s_1,G,\Du(\Phi)_1)=\bot$}
                    \Return{$\bot$}
                \EndIf
            \EndFor
            \For{each $\varphi\in (\subfB(\psi)\setminus G_2)$}
                \State{$G\gets G_1\cap\subfB(\varphi)$}
                \If{\texttt{Check}$(\Ku,\varphi,s_1,G,\Du(\Phi)_1)=\top$}
                    \Return{$\bot$}
                \EndIf
            \EndFor
            \Return{$\top$}
        \Else
            \Return{$\bot$}
        \EndIf
    \end{algorithmic}
\end{algorithm}

At line 1, \texttt{Compatible} checks whether there is an edge between  $s_1$ and $s_2$ ($(s_1,s_2)\in \Edges$), and if, at the considered step,
the current configuration of the $\DFA$s $\Du(\Phi)_1$ is transformed into the configuration $\Du(\Phi)_2$   (i.e., $s_2$ and  $\Du(\Phi)_2$ are synchronously reachable from $s_1$ and  $\Du(\Phi)_1$). 
At lines 2--9, \texttt{Compatible} checks that each formula $\varphi$ in $(G_2\setminus G_1)$, where $G_2\supseteq G_1$, is satisfied by 
a trace summarized by $(G_1,\Du(\Phi)_1,s_1)$ (lines 2--5). Intuitively, $(G_1,\Du(\Phi)_1,s_1)$ summarizes the maximal prefix of 
$(G_2,\Du(\Phi)_2,s_2)$, and thus a subformula satisfied by a 
prefix of a trace summarized by $(G_2,\Du(\Phi)_2,s_2)$ either belongs to $G_1$ or it is satisfied by the trace summarized by $(G_1,\Du(\Phi)_1,s_1)$.
Moreover, (lines 6--9) \texttt{Compatible} checks that $G_2$ is maximal (i.e., no subformula that must be in $G_2$ has been forgotten).
%
Note that by exploiting this binary reachability technique,
the recursion depth of \texttt{Reach} is logarithmic in the length of the trace to be visited, hence it can use only polynomial space.

Theorem~\ref{corrComplCheckBBbar}, proved in Appendix~\ref{proof:corrComplCheckBBbar}, establishes the soundness of \texttt{Check}.

\begin{theorem}\label{corrComplCheckBBbar}
Let $\Phi$ be a $\B\Bbar$ formula, $\psi$ be a subformula of $\Phi$, and $\rho\in\Trk_{\Ku}$ be a trace with $s=\lst(\rho)$. Let $G$ be the subset of formulas in $\subfB(\psi)$ that hold on some proper prefix of $\rho$.
Let $\Du(\Phi)$ be the current configuration of the $\DFA$s associated with the regular expressions in $\Phi$ after reading $\mu(\rho(1,|\rho|-1))$.
%
Then $\texttt{Check}(\Ku,\psi,s,G,\Du(\Phi))=\top \!\iff\! \Ku,\rho\models \psi$.
\end{theorem}

\begin{algorithm}[tp]
    \caption{\texttt{CheckAux}($\Ku,\Phi$)}\label{CheckAux}
    \begin{algorithmic}[1]        
        \State{$\texttt{create}(\Du(\Phi)_0)$}\Comment{Creates the (succinct) $\NFA$s and the initial states of the $\DFA$s for all the $\RE$s in $\Phi$}
        \If{\texttt{Check}$(\Ku,\neg\Phi,\sinit,\emptyset,\Du(\Phi)_0)$ \textbf{or}                 \texttt{Check}$(\Ku,\hsBt\neg\Phi,\sinit,\emptyset,\Du(\Phi)_0)$}
            \Return{$\bot$}
        \Else
            \Return{$\top$}  
        \EndIf
    \end{algorithmic}
\end{algorithm}

Algorithm~\ref{CheckAux} reports the main MC procedure \texttt{CheckAux}($\Ku,\Phi$) for $\B\Bbar$.  It starts constructing the $\NFA$s and the initial states of the $\DFA$s for the regular expressions of $\Phi$ (line 1). Then \texttt{CheckAux} invokes the procedure \texttt{Check} two times (line 2): the former to check the special case of the trace $\sinit$ consisting of the initial state of $\Ku$ only, and the latter for all right-extensions of $\sinit$ (i.e., the initial traces having length at least 2). 
Notice that the $\NFA$s and $\DFA$s for the regular expressions of $\hsBt\neg\Phi$, $\neg\Phi$ and $\Phi$ are the same (i.e. $\Du(\Phi)_0=\Du(\hsBt\neg\Phi)_0=\Du(\neg\Phi)_0$), allowing us to simultaneously apply the result of Theorem~\ref{corrComplCheckBBbar} for both the invocations of \texttt{Check} at line 2.

The next theorem, proved in Appendix~\ref{proof:corrCheckAux}, establishes the soundness and completeness of \texttt{CheckAux}.
\begin{theorem}\label{corrCheckAux}
    Let $\Ku=\KuDef$ be a finite Kripke structure, and $\Phi$ be a $\B\Bbar$ formula. Then, \texttt{CheckAux}$(\Ku,\Phi)=\top\iff\Ku\models\Phi$.
\end{theorem}

The next corollary states the upper bound to the complexity of MC for $\B\Bbar$.
\begin{corollary}
The MC problem for $\B\Bbar$ formulas extended with regular expressions over finite Kripke structures is in $\Psp$.
\end{corollary}
\begin{proof}
    The procedure \texttt{CheckAux} decides the problem using \emph{polynomial working space} basically due to two facts. 
    The first one is the number of simultaneously active recursive calls of \texttt{Check}, which is $O(|\Phi|)$.
    The second is the space (in bits) used for any call of \texttt{Check}, that is,
        \begin{multline*}
            O\Big(|\Phi| + |\States| + \sum_{\ell=1}^u |r_\ell| + \underbrace{\log (|\States|\cdot |\Phi|\cdot 2^{2\sum_{\ell=1}^u |r_\ell|})}_{(1)} +\\
            \underbrace{(|\Phi| + |\States| + \sum_{\ell=1}^u |r_\ell|)}_{(2)}\cdot\underbrace{\log (|\States|\cdot |\Phi|\cdot 2^{2\sum_{\ell=1}^u |r_\ell|})}_{(3)}\Big),
        \end{multline*}
         
        In particular, $(1)$~$O(\log (|\States|\cdot |\Phi|\cdot 2^{2\sum_{\ell=1}^u |r_\ell|}))$ bits are used for the bound $b$ on the trace length, $(3)$~for \emph{each subformula} $\hsBt\psi'$ of $\Phi$ at most $O(\log (|\States|\cdot |\Phi|\cdot 2^{2\sum_{\ell=1}^u |r_\ell|}))$ recursive calls of \texttt{Reach} may be simultaneously active (the recursion depth of \texttt{Reach} is logarithmic in $b$), and $(2)$~each call of \texttt{Reach} requires $O(|\Phi| + |\States| + \sum_{\ell=1}^u |r_\ell|)$ bits.\qedhere
%
\end{proof}

Finally, since a Kripke structure can be unravelled against the direction of its edges,
and a language $\lang$ is regular if and only if its reversed version $\lang^{\text{Rev}}=\{w(|w|)\cdots w(1)\mid w\in\lang\}$ is, the proposed algorithm can be easily modified to deal with the symmetric fragment $\E\Ebar$, proving that also the MC problem for $\E\Ebar$ is in $\Psp$.

%% file: Chaps/Gandalf17RIVISTA/AAbarElimin.tex
\subsection{$\Psp$-completeness of MC for $\AAbarBBbar$} \label{sect:genResult}

We now show that the algorithm \texttt{CheckAux} can be used as a basic engine to design a $\Psp$ MC algorithm for the bigger fragment $\AAbarBBbar$. 

The idea is that, being proposition letters (related with) regular expressions, modalities $\hsA$ and $\hsAt$ do not augment the expressiveness of the fragment $\B\Bbar$. In particular,
we will show that the occurrences of modalities  $\hsA$ and $\hsAt$ in an $\AAbarBBbar$ formula can suitably be ``absorbed'' and replaced by fresh proposition letters.

We recall that
$\Ku,\rho\models \hsA \psi$ if and only if there exists a trace $\tilde{\rho}\in\Trk_{\Ku}$ such that $\lst(\rho)=\fst(\tilde{\rho})$ and $\Ku,\tilde{\rho}\models \psi$. An immediate consequence is that, for any $\rho'\in\Trk_{\Ku}$ with $\lst(\rho)=\lst(\rho')$, $\Ku,\rho\models \hsA \psi\iff \Ku,\rho'\models \hsA \psi$ and similarly for the symmetrical modality $\hsAt$ with respect to the first state of the trace. In general, if two traces have the same final state (resp., first state), either both of them satisfy a formula $\hsA \psi$ (resp., $\hsA \psi$), or none of them does.

As a consequence, for a formula $\hsA\psi$ (resp., $\hsAt\psi$), we can determine the subset $S_{\hsA\psi}$ (resp., $S_{\hsAt\psi}$) of the set of states $\States$ of the Kripke structure
such that, for any $\rho\in\Trk_{\Ku}$, $\Ku,\rho\models\hsA\psi$ (resp., 
$\Ku,\rho\models\hsAt\psi$) if and only if $\lst(\rho)\in S_{\hsA\psi}$
(resp., $\fst(\rho)\in S_{\hsAt\psi}$).

Now, for a formula $\hsA\psi$ (resp., $\hsAt\psi$), we provide a regular expression $r_{\hsA\psi}$ (resp., $r_{\hsAt\psi}$) characterizing the set of traces which model the formula.
To this end we identify the states in $\States$ by a set of  fresh proposition letters $\{q_s\mid s\in \States\}$ and we 
replace the Kripke structure  $\Ku=\KuDef$ by
$\Ku'=\tpl{\Prop',\States,\Edges,\Lab',\sinit}$, with $\Prop'=\Prop\cup \{q_s\mid s\in \States\}$ and $\Lab'(s) = \{q_s\}\cup \Lab(s)$ for any $s \in \States$.
The regular expressions $r_{\hsA\psi}$ and $r_{\hsAt\psi}$ are  
\[r_{\hsA\psi}=\top^*\cdot\Big( \bigcup_{s\in S_{\hsA\psi}} q_s\Big)  \quad \text{ and } \quad r_{\hsAt\psi}=\Big( \bigcup_{s\in S_{\hsAt\psi}}q_s \Big)\cdot \top^*.\]
By definition $\Ku,\rho\models r_{\hsA \psi}$ if and only if $\lst(\rho)\in S_{\hsA\psi}$, if and only if $\Ku,\rho\models \hsA \psi$.

We can now sketch the procedure for \lq\lq reducing\rq\rq{} the MC problem for $\AAbarBBbar$ to the MC problem for $\B\Bbar$: we iteratively rewrite a formula $\Phi$ of $\AAbarBBbar$ until it gets converted to an (equivalent) formula of $\B\Bbar$.
At each step, we select an occurrence of a subformula of $\Phi$, either having the form $\hsA\psi$ or $\hsAt\psi$, devoid of any  occurrence of modalities $\hsA$ and $\hsAt$ in $\psi$. For such an occurrence, say $\hsA\psi$, we have to compute the set $S_{\hsA\psi}$. For that purpose we can run a variant \texttt{CheckAux'}($\Ku,\Psi,s$) of the MC procedure \texttt{CheckAux}($\Ku,\Psi$), which invokes \texttt{Check} at line 2 on the additional parameter (state) $s$, instead of $\sinit$. 
For each $s \in \States$, we invoke \texttt{CheckAux'}($\Ku,\neg\psi,s$), deciding that $s\in S_{\hsA\psi}$ if and only if the procedure returns $\bot$.
Then we \emph{replace} $\hsA\psi$ in $\Phi$ with the regular expression $r_{\hsA\psi}$, obtaining  a formula $\Phi'$. To deal with subformulas of the form $\hsAt\psi$, we have to introduce a slight variant of the procedure \texttt{Check}, which finds traces leading to (and not starting from) a given state.
Now, if the resulting formula $\Phi'$ is in $\B\Bbar$, the rewriting process ends; otherwise, we can perform another rewriting step over $\Phi'$.
 
Considering that the sets  $S_{\hsA\psi}$, $S_{\hsAt\psi}$ and the regular expressions $r_{\hsA\psi}$ and $r_{\hsAt\psi}$ have a size linear in $|\States|$, we can conclude with the following result.

\begin{theorem}\label{th:ABBA}
The MC problem for $\AAbarBBbar$ formulas extended with regular expressions over finite Kripke structures is in $\Psp$.
\end{theorem}

By symmetry we can show that the MC problem for $\A\Abar\E\Ebar$ is also in $\Psp$.

%% file: Chaps/Gandalf17RIVISTA/PropHard.tex
The $\Psp$-hardness of MC for $\B\Bbar$ and 
$\AAbarBBbar$ directly follows from that of the smallest fragment $\HSprop$ (the purely propositional fragment of $\HS$), which is stated by Theorem~\ref{th:hard}: in Appendix~\ref{sec:th:hard}, we prove that $\HSprop$ is hard for $\Psp$ by a reduction from the $\Psp$-complete \emph{universality problem for regular expressions}~\cite{Garey79}, namely, the problem of deciding, for a (standard) regular expression $r$ with $\lang(r) \subseteq \Sigma^*$ and $|\Sigma|\geq 2$, whether $\lang(r)=\Sigma^*$ or not.

\begin{theorem}\label{th:hard}
The MC problem for $\HSprop$ formulas extended with regular expressions over finite Kripke structures is $\Psp$-hard (under polynomial-time reductions).
\end{theorem}

By Theorem~\ref{th:ABBA} and Theorem~\ref{th:hard} we obtain the following complexity result.
\begin{theorem}\label{th:glob}
The MC problem for formulas of any (proper or improper) sub-fragment of $\AAbarBBbar$ (and $\AAbarEEbar$) extended with regular expressions over finite Kripke structures is $\Psp$-complete.
\end{theorem}

%% file: Chaps/Gandalf17RIVISTA/concl.tex
\section{Conclusions}
In this chapter we have studied the MC problem for $\HS$ extended with regular expressions  used to define interval labelling. The approach, stemming from~\cite{lm16}, generalizes both the one of the previous chapters, in which we enforce the homogeneity principle, and of~\cite{LM13,LM14} where labeling is endpoint-based. 
In the general case, MC for (full) $\HS$ turns out to be nonelementarily decidable---the proof exploits an automata-theoretic approach based on the notion of $\Ku$-\NFA---%
but, for a constant-length formula,
it is in $\PTIME$.
Moreover, the problem is $\EXPSPACE$-hard (the hardness follows from that of $\BE$ under homogeneity).

We have also investigated the MC problem for two maximal fragments of $\HS$, namely $\AAbarBBbar\Ebar$ and $\AAbar\E\Bbar\Ebar$ with 
regular expressions, and we have showed that it is 
$\LINAEXPTIME$-complete.
The complexity upper bound has been proved by providing an alternating algorithm which performs an exponential number of computation steps, but only polynomially many alternations (in the length of the formula to be checked). 
Conversely, the lower bound has been shown by a reduction from the $\LINAEXPTIME$-complete alternating multi-tiling problem.
In this way, we have also improved the known complexity result for the same fragments under the homogeneity assumption.

Finally, we have proved that the $\HS$ fragments $\AAbarBBbar$ and $\AAbarEEbar$, and all  their sub-fragments, are $\Psp$-complete.
The bedrock is a small-model property that allows us to restrict the verification of formulas of $\AAbarBBbar$/$\AAbarEEbar$ to traces having at most exponential length. 
Conversely, the matching complexity lower bound has been proved by a reduction from the $\Psp$-complete universality problem for regular expressions.

%% file: Chaps/Timelines/TimelinesMain.tex
\chapter{Interval-based system models: timelines}\label{chap:Timelines}
\begin{chapref}
The references for this chapter are~\cite{kr18,gand18,ictcs18}.
\end{chapref}

\minitoc\mtcskip

\newcommand{\Instruct}{\mathsf{Inst}}
\newcommand{\InstructLab}{\mathsf{InstLab}}
\newcommand{\InstL}{\mathsf{L}}
\newcommand{\Inc}{\mathsf{Inc}}
\newcommand{\Dec}{\mathsf{Dec}}
\newcommand{\halt}{{\textit{halt}}}
\newcommand{\init}{{\textit{init}}}
\newcommand{\main}{{\textit{main}}}
\newcommand{\Checking}{{\textit{check}}}
\newcommand{\inc}{{\mathsf{inc}}}
\newcommand{\dec}{{\mathsf{dec}}}

\newcommand{\zero}{{\mathsf{zero}}}
\newcommand{\instr}{{\textit{op}}}
\newcommand{\gainy}{{\textit{gainy}}}
\newcommand{\From}{{\textit{from}}}
\newcommand{\To}{{\textit{to}}}
\newcommand{\cont}{{\textit{sec}}}
\newcommand{\Tag}{{\textit{Tag}}}
\newcommand{\Succ}{{\textit{succ}}}
\newcommand{\op}{{\textit{op}}}

\newcommand{\dummy}{{\textit{dummy}}}

\newcommand{\Con}{\mathsf{con}}

\newcommand{\start}{\mathsf{s}}
\newcommand{\startTime}{\mathsf{s}}
\newcommand{\Ending}{\mathsf{e}}

\newcommand{\der}[1]{\ensuremath{\;\;{\mathop{{ %
            \longrightarrow}}\limits^{{#1}}}\!}\;\;} %

\newcommand{\derG}[1]{\ensuremath{\;\;{\mathop{{ %
            \longrightarrow}}\limits^{{#1}}}\!}_\gainy\;\;} %

\newcommand{\TA}{\text{\sffamily TA}}
\newcommand{\MTL}{\text{\sffamily MTL}}
\newcommand{\TPTL}{\text{\sffamily TPTL}}
\newcommand{\MITL}{\text{\sffamily MITL}}
\newcommand{\MITLR}{\text{\sffamily MITL}_{(0,\infty)}}

\newcommand{\RealP}{{\mathbb{R}_+}}
\newcommand{\RatP}{{\mathbb{Q}_+}}
\newcommand{\val}{{\mathit{val}}}
\newcommand{\Main}{{\mathit{Main}}}
\newcommand{\EqTime}{{\mathit{EqTime}}}
\newcommand{\Past}{{\mathit{past}}}
\newcommand{\code}{{\mathit{code}}}
\newcommand{\Res}{\textit{Res}}
\newcommand{\TLang}{{\mathcal{L}_T}}
\newcommand{\StrictUntil}{\textsf{U}}

\newcommand{\Deriv}{{\mathit{Deriv}}}
\newcommand{\Intv}{{\mathit{Intv}}}
\newcommand{\IntvR}{{\mathit{Intv}_{(0,\infty)}}}

\input{Chaps/Timelines/intro.tex}
\input{Chaps/Timelines/prelim.tex}
\input{Chaps/Timelines/undecidability.tex}
\input{Chaps/Timelines/DecidFutTPsimple.tex}
\input{Chaps/Timelines/PSPACEhard.tex}
\input{Chaps/Timelines/TriggerlessNPcomplete.tex}
\input{Chaps/Timelines/MCTimelines.tex}
\input{Chaps/Timelines/conclusions.tex}

%% file: Chaps/Timelines/intro.tex
\lettrine[lines=3]{I}{n all the previous chapters,} we have assumed finite Kripke structures as system models.
On the positive side, these labelled state-transition graphs are simple, in that no \lq\lq feasibility check\rq\rq\ needs to
be performed over them and the described system (the latter exists by definition of the structure itself); moreover they are commonly employed for several industrial purposes, being many modeling languages translated into them, before the MC process (running phase) starts.
On the negative side, they are \lq\lq inherently point-based\rq\rq , as they make explicit how a system evolves \emph{state-by-state}
(i.e., how from a state it can move to another one, according to the transition function), and describe which are the atomic properties (proposition letters) that hold true at every state.

In this chapter, we study a possible replacement of Kripke structures by a more expressive model, which allows us to describe systems in terms of their \emph{interval-based} behaviour and properties.
We first identified as natural candidates \emph{interval graphs}~\cite{intvgraphs}, on which there has been a good deal of work, also from the algorithmic point of view; however, we immediately realized that they are too basic to capture meaningful properties of systems.
We then turned to a different kind of structures, 
called \emph{timelines}, which have been fruitfully exploited in temporal planning for quite a long time.
For this reason, we want now to start a short digression on timeline-based planning, and come back to MC later, explaining why the former is a sort of \lq\lq necessary condition\rq\rq\ for the latter, which can then be solved straightforwardly once the former has, under certain conditions.

\emph{Timeline-based planning} (TP for short) can be viewed as an alternative to the classic action-based approach to planning. Action-based planning aims at determining a sequence of actions that, given the initial state of the world 
and a goal, transforms, step by step, the state of the world until we reach a state satisfying the goal.  
TP focuses on what has to happen in order to satisfy the goal instead of what an agent has to do, and thus it can be considered as a more declarative approach with respect to action-based planning: it models the planning domain as a set of independent (but interacting) components, each one consisting of a number of \emph{state variables}. The evolution of the values of state variables over time is described by means of a set of \emph{timelines} (in turn these are sequences of time intervals called \emph{tokens}), and it is governed by a set of transition functions, one for each state variable, and a set of synchronization rules, that constrain the temporal relations among (values of) state variables. This standard lexicon of TP---that some readers, from our experience, may find misleading---will be formally defined and become clearer in the next sections. In the meanwhile, looking (ahead) at Figure~\ref{fig:timelineEx} may give an intuition.

TP has been successfully exploited in a number of application domains, for instance, in space missions, constraint solving, activity scheduling (see, e.g.,~\cite{barreiro2012europa,CestaCFOP07,aspen2010,FrankJ03,JonssonMMRS00,Muscettola94}), but a systematic study of its expressiveness and complexity has been undertaken only very recently. The temporal domain is commonly assumed to be \emph{discrete}.
In~\cite{GiganteMCO16}, Gigante et al.\ showed that TP with bounded temporal relations and token durations, and no temporal horizon, is $\EXPSPACE$-complete and  expressive enough to capture action-based temporal planning. Later, in~\cite{GiganteMCO17}, they proved that $\EXPSPACE$-completeness still holds for TP with unbounded interval relations, and that the problem becomes $\NEXPTIME$-complete if an upper bound to the temporal horizon is added. 

In the following sections we will study 
TP over a \emph{dense temporal domain} 
(without having recourse to any form of artificial discretization, which is quite a common trick).
The reason why we assume this different version of time domain is to avoid discreteness in system descriptions,
which can then be abstracted at a higher level, enabling us to neglect details which are unnecessary, and paving the way for a really interval-based MC:
in this respect TP can be regarded, as we said, as a sort of necessary condition for MC, the former playing the role of a \lq\lq feasibility check\rq\rq\ of the system description (which is not immediately feasible by definition---as opposed to Kripke structures---given the presence of synchronization rules). 
Moreover, if both the system model and the specifications (temporal formulas) can be translated into a common formalism (in our case, as we will show, \emph{timed automata}), \lq\lq adding\rq\rq\ MC on top of TP is just a matter of technical aspects. This is why 
we shall now focus mainly on TP, and come back to MC at the end of the chapter.

In the next sections, we will study suitable restrictions on the TP problem that allow us to recover its decidability: as a matter of fact, the first result we establish is a negative one, 
namely, that TP over dense time, in its general formulation, is \emph{undecidable}. Then we will also show how to obtain better computational complexities, which are appropriate to the practical exploitation of timeline-based TP and MC,  by suitably constraining the logical structure of synchronization rules. 

In the general case, a synchronization rule allows a universal quantification over the tokens of a timeline (such a quantification is called \emph{trigger}).
When a token is \lq\lq selected\rq\rq{} by a trigger, the rule allows us to compare 
tokens of the timelines both preceeding (past) and following (future) the trigger token. 
The first restriction we consider consists in limiting the comparison to tokens in the future with respect to the trigger (\emph{future semantics} of trigger rules). 
The second imposes that the name of a non-trigger token appears exactly once in the constraints set by the rule (\emph{simple} trigger rules):
this syntactical restriction avoids comparisons of multiple token time-events with a non-trigger reference time-event.  
Better complexity results can be obtained by restricting also the type of \emph{intervals} used in rules in order to compare tokens.

We now describe the organization of this chapter, outlining in particular which are the complexity results implied by the aforementioned restrictions of TP.

\paragraph{Organization of the chapter.}
\begin{itemize}
    \item In Section~\ref{sec:preliminTimelines} we start by introducing the TP framework, providing some background knowledge on it. 
    \item In Section~\ref{sec:undecidability} we prove that TP is \emph{undecidable} in the general case, by a reduction from the \emph{halting problem for Minsky $2$-counter machines}. The section is concluded commenting on \emph{non-primitive recursive-hardness} of TP under the future semantics of trigger rules (this is formally proved in Appendix~\ref{sec:NPRHardness}).    
    \item In Section~\ref{sec:DecisionProcedures}, we establish that future TP with simple trigger rules is \emph{decidable} (in non-primitive recursive time), and show membership in $\EXPSPACE$ (respectively, $\Psp$) under the restriction to \emph{non-singular intervals} (respectively, intervals of the forms $\mathopen[0,a\mathclose]$ and $\mathopen[b,+\infty\mathclose[\,$).
    \item Matching complexity lower bounds for the last two restrictions are given in Section~\ref{sec:pspace}. 
    \item In Section~\ref{sec:NPtriggerless} we outline an \NP\ planning algorithm for TP with \emph{trigger-less rules only} (which disallow the universal quantification/trigger and have a purely existential form) stemming from the results of the previous sections. With a trivial hardness proof, we also show TP with trigger-less rules to be $\NP$-complete.
    \item In Section~\ref{sec:modelcheckingTimelines}, we finally tackle the MC problem for systems described as timelines, where property specifications are given in terms of formulas of \emph{Metric Interval Temporal Logic} (\MITL), a timed logic which extends \LTL. The reason why here we drop $\HS$ and employ the latter is the following: enriching system models claims for extensions of the property specification language; in our case, timelines would naturally require a timed extension of $\HS$ which, however, has not been studied yet (in the literature, only \emph{metric} extensions of $\HS$ have been proposed over the natural numbers~\cite{DBLP:journals/sosym/BresolinMGMS13}).
    Thus the well-known and thoroughly studied \MITL\ comes to the rescue as a sort of \lq\lq approximation\rq\rq\ of $\HS$; moreover, it allows us to link the world of interval-based models/timelines with that of \emph{timed automata}, another famous formalism used in planning as well as in system verification, that will be heavily used for many results we are about to prove. 
\end{itemize}

%% file: Chaps/Timelines/prelim.tex
\section{The TP Problem}\label{sec:preliminTimelines}

As in the previous chapters, let $\Nat$ be the set of natural numbers and $\RealP$ be the set of non-negative real numbers; moreover, $\Intv$ denotes the set of intervals of $\RealP$ whose endpoints are in $\Nat\cup\{\infty\}$, and $\Intv_{(0,\infty)}$ is the set of non-singular%
\footnote{An interval of the form $[a,a]$, for $a\in\Nat$, is called \emph{singular}.} 
intervals $I\in \Intv$ such that
  either $I$ is unbounded, or $I$
  is left-closed with left endpoint $0$. The latter intervals $I$ can be replaced by expressions of the form $\sim n$, for some $n\in\Nat$
  and $\sim\,\in\{<,\leq,>,\geq\}$.
%


We now introduce notation and basic notions of
the TP framework as presented in~\cite{MayerOU16,GiganteMCO16}.
In TP, domain knowledge is encoded by a set of state variables, whose behaviour over time is described by transition functions and synchronization rules.

\begin{definition}[State variable]
  \label{def:statevar}
  A \emph{state variable} $x$ is a triple \[x= (V_x,T_x,D_x),\] where 
  \begin{itemize}
      \item $V_x$ is the \emph{finite domain} of the state variable $x$,
      \item $T_x:V_x\to 2^{V_x}$ is the \emph{value transition function}, which maps
        each $v\in V_x$ to the (possibly empty) set of successor values, and
      \item $D_x:V_x\to \Intv$ is the \emph{constraint (or duration) function} that maps each $v\in V_x$
        to an interval of $\Intv$.
  \end{itemize}   
\end{definition}

A \emph{token} for a variable $x$ is a pair $(v,d)$ consisting of a value $v\in V_x$ and a duration $d\in \RealP$
such that $d\in D_x(v)$. Intuitively, a token for $x$ represents an interval of time where the state variable $x$ takes value $v$.
In order to clarify the variable to which a token refers, we shall often denote $(v,d)$ as $(x,v,d)$.

\begin{figure}
    \centering
    \includegraphics[scale=0.7]{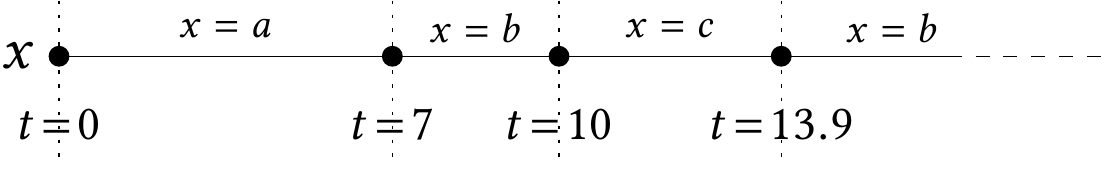}
    \caption{An example of timeline $(a,7)(b,3)(c,3.9)\cdots$ for the state variable $x= (V_x,T_x,D_x)$, where $V_x=\{a,b,c,\ldots\}$, $b\in T_x(a)$, $c\in T_x(b)$, $b\in T_x(c)$\dots\ and $D_x(a)=[5,8]$, $D_x(b)=[1,4]$, $D_x(c)=[2,\infty[$\dots}
    \label{fig:timelineEx}
\end{figure}

The behavior of the state variable $x$ is specified by means of a \emph{timeline}, which is a non-empty sequence of tokens
$\pi = (v_0,d_0)\cdots  (v_n,d_n)$  consistent with the value transition function $T_x$, namely, such that
$v_{i+1}\in T_x(v_i)$ for all $0\leq i<n$. The \emph{start time} $\start(\pi,i)$ and the \emph{end time} $\Ending(\pi,i)$ of the $i$-th token of the timeline
$\pi$  are defined respectively as follows: 
\[\start(\pi,i)=0 \text{ if } i=0, \qquad \start(\pi,i)=\sum_{h=0}^{i-1} d_h \text{ otherwise},\]
and
\[\Ending(\pi,i)=\sum_{h=0}^{i} d_h.\]
See Figure~\ref{fig:timelineEx} for an example.

Given a finite set $SV$ of state variables, a \emph{multi-timeline} of $SV$ is a mapping $\Pi$ assigning to
each state variable $x\in SV$ a timeline for $x$.

Multi-timelines of $SV$ can be constrained by a set of \emph{synchronization
rules}, which relate tokens, possibly belonging to different timelines, through
temporal constraints on the start/end times of tokens (time-point constraints) and on the difference
between start/end times of tokens (interval constraints). The synchronization rules exploit
an alphabet $\Sigma=\{o,o_0,o_1,o_2,\ldots\}$ of token names to refer to the tokens along a multi-timeline, and are based on the notions of
\emph{atom} and \emph{existential statement}.

\begin{definition}[Atom]
  \label{def:timelines:atom}
  An \emph{atom} $\rho$ is either a clause of the form \mbox{$o_1\leq^{e_1,e_2}_{I} o_2$}
  (\emph{interval atom}), or of the forms $o_1\leq^{e_1}_{I} n$ or  $n\leq^{e_1}_{I}
  o_1$ (\emph{time-point atom}), where $o_1,o_2\in\Sigma$, $I\in\Intv$, $n\in\Nat$, and $e_1,e_2\in\{\start,\Ending\}$.
\end{definition}

An atom $\rho$ is evaluated with respect to a \emph{$\Sigma$-assignment $\lambda_\Pi$} for a given multi-timeline $\Pi$,
which is a mapping assigning to each token name $o\in \Sigma$ a pair $\lambda_\Pi(o)=(\pi,i)$ such that $\pi$ is a timeline of $\Pi$ and $0\leq i<|\pi|$ is a position along $\pi$ (intuitively,
$(\pi,i)$ represents the token of $\Pi$ referenced by the name $o$).

An interval atom $o_1\leq^{e_1,e_2}_{I} o_2$  \emph{is satisfied by  $\lambda_\Pi$} if $e_2(\lambda_\Pi(o_2))-e_1(\lambda_\Pi(o_1))\in I$.
A point atom $o\leq^{e}_{I} n$  (respectively, $n\leq^{e}_{I}o$)   \emph{is satisfied by  $\lambda_\Pi$} if $n-e(\lambda_\Pi(o))\in I$ (respectively, $e(\lambda_\Pi(o))-n\in I$).

\begin{definition}[Existential statement]
 An \emph{existential statement} $\mathcal{E}$ for a finite set $SV$ of state variables is a statement of the form
\[
\mathcal{E}=  \exists o_1[x_1=v_1]\cdots \exists o_n[x_n=v_n].\mathcal{C},
\]
  where $\mathcal{C}$ 
  is a conjunction of atoms,
  $o_i\!\in\!\Sigma$, $x_i\!\in\! SV$, $v_i\!\in\! V_{x_i}$, for
  $1\leq i\leq n$. 

The elements $o_i[x_i=v_i]$ are called
  \emph{quantifiers}. A token name used in $\mathcal{C}$, but not occurring in any
  quantifier, is said to be \emph{free}. 
\end{definition}

  Given a $\Sigma$-assignment $\lambda_\Pi$ for a multi-timeline $\Pi$ of $SV$,
  we say that \emph{$\lambda_\Pi$ is consistent with the existential statement $\mathcal{E}$} if, for each quantifier $o_i[x_i=v_i]$, we have
   $\lambda_\Pi(o_i)=(\pi,h)$, where $\pi=\Pi(x_i)$ and the $h$-th token of $\pi$ has value $v_i$. A multi-timeline $\Pi$ of $SV$ \emph{satisfies} $\mathcal{E}$
   if there exists a $\Sigma$-assignment $\lambda_\Pi$ for $\Pi$ consistent with $\mathcal{E}$ such that each atom in $\mathcal{C}$ is satisfied by
   $\lambda_\Pi$.

We can now introduce synchronization rules, which constrain tokens, possibly belonging to different timelines.
\begin{definition}[Synchronization rule]
  A \emph{synchronization rule} $\mathcal{R}$ for a finite set $SV$ of state variables is a rule of one of the forms
  \[
  o_0[x_0=v_0] \to \mathcal{E}_1\lor \mathcal{E}_2\lor \ldots \lor \mathcal{E}_k, \qquad
          \true \to \mathcal{E}_1\lor \mathcal{E}_2\lor \ldots \lor \mathcal{E}_k,
  \]
  where $o_0\in\Sigma$, $x_0\in SV$, $v_0\in V_{x_0}$, and $\mathcal{E}_1, \ldots, \mathcal{E}_k$
  are \emph{existential statements}. 
  In rules of the first
  form (which are called \emph{trigger rules}), the quantifier $o_0[x_0=v_0]$ is called \emph{trigger}; we require that only $o_0$ may appear free in $\mathcal{E}_i$, for all $1\leq i\leq n$. In rules of the second form (\emph{trigger-less rules}), we require
  that no token name appears free.
  \newline
  A trigger rule $\mathcal{R}$ is \emph{simple} if, for each existential statement $\mathcal{E}$ of $\mathcal{R}$ and each token name $o$ distinct from the trigger, there is at most one \emph{interval atom}
  of $\mathcal{E}$ where $o$ occurs.
\end{definition}

Intuitively, the  trigger $o_0[x_0=v_0]$ acts as a universal quantifier, which
states that \emph{for all} the tokens of the timeline for
$x_0$, where $x_0$ takes the
value $v_0$, at least one of the existential statements $\mathcal{E}_i$ must be satisfied. 
As an example, \[o_0[x_0=v_0]\to \exists o_1[x_1=v_1].o_0\leq^{\mathsf{e},\mathsf{s}}_{[2,\infty[} o_1\] states that after \emph{every} token for $x_0$ with value $v_0$ there exists a token for $x_1$ with value $v_1$ \emph{starting} at least 2 time instants after the \emph{end} of the former.
Trigger-less rules simply assert the
satisfaction of some existential statement. The intuitive meaning of \emph{simple} trigger rules is that they disallow simultaneous comparisons of multiple time-events
 (start/end times of tokens) with a non-trigger reference time-event. 
 
 The semantics of synchronization rules is formally defined as follows.
\begin{definition}[Semantics of synchronization rules]\label{def:semanticsRules}
Let $\Pi$ be a multi-timeline of a set $SV$ of state variables.

Given a \emph{trigger-less rule} $\mathcal{R}$ of $SV$, \emph{$\Pi$ satisfies $\mathcal{R}$} if $\Pi$ satisfies some existential statement of $\mathcal{R}$.

Given a \emph{trigger rule} $\mathcal{R}$ of $SV$ with trigger $o_0[x_0=v_0]$, \emph{$\Pi$ satisfies   $\mathcal{R}$} if, for every position $i$ of the
 timeline $\pi=\Pi(x_0)$ for $x_0$ such that $\pi(i)=(v_0,d)$, there exists an existential statement $\mathcal{E}$ of $\mathcal{R}$  and a $\Sigma$-assignment
 $\lambda_\Pi$ for $\Pi$ consistent with $\mathcal{E}$ such that $\lambda_\Pi(o_0)= (\pi,i)$ and $\lambda_\Pi$ satisfies all the atoms of $\mathcal{E}$.
\end{definition}

In the paper, we shall also focus on a stronger notion of satisfaction of trigger rules, called \emph{satisfaction under the future semantics}: it requires that all non-trigger tokens selected by some quantifier
do not start \emph{strictly before} the start time of the trigger token.

\begin{definition}[Future semantics of trigger rules]\label{def:futurerules}
  A multi-timeline $\Pi$ of $SV$ satisfies a trigger rule \[\mathcal{R}= o_0[x_0=v_0] \to \mathcal{E}_1\vee \mathcal{E}_2\vee
  \ldots \vee \mathcal{E}_k\]   \emph{under the future semantics} if $\Pi$ satisfies the trigger rule obtained from
  $\mathcal{R}$ by replacing each existential statement \[\mathcal{E}_i=\exists o_1[x_1=v_1]\cdots \exists o_n[x_n=v_n].\mathcal{C}\]
  by \[\mathcal{E}_i'=\exists o_1[x_1=v_1]\cdots \exists o_n[x_n=v_n].\Big(\mathcal{C}\wedge  \bigwedge_{i=1}^{n} o_0\leq^{\start,\start}_{[0,+\infty[} o_i\Big).\]
\end{definition}

A \emph{TP domain} $P=(SV,R)$ is specified by a finite set $SV$ of state variables and
a finite set $R$ of synchronization rules for $SV$ modeling their admissible behaviors.
Trigger-less rules can be used to express initial, as well as intermediate conditions
and the goals of the problem, while trigger rules are much more powerful and useful, for instance, to specify invariants and response requirements. 

A \emph{plan for $P=(SV,R)$} is a  multi-timeline of $SV$ satisfying all the rules in $R$. A \emph{future plan for $P$} is defined in a similar way, but we require satisfaction under the future semantics of \emph{all} trigger rules.

In the next sections we will study the following decision problems:
\begin{description}
  \item[TP problem] Given a TP domain $P=(SV,R)$, is there a plan for $P$?
  \item[Future TP problem] Given a TP domain $P\!=\!(SV\!,R)$, is there a \emph{future} plan for $P$?
\end{description}
%

Table~\ref{tab:complex} summarizes all the decidability and complexity results described in the following about the mentioned problems:
we will consider mixes of restrictions on TP involving trigger rules with future semantics, simple trigger rules, and intervals in atoms (of trigger rules) which are non-singular or in $\Intv_{(0,\infty)}$.

\begin{table}[t]
    \centering    
    \caption{Decidability and complexity of restrictions of the TP problem.}
    \label{tab:complex}
    \resizebox{\linewidth}{!}{
    \begin{tabular}{r|c|c}
    	& TP problem & Future TP problem \\ 
    	\hline 
    	Unrestricted & Undecidable & (Decidable?) Non-primitive recursive-hard \\ 
    	\hline 
    	Simple trigger rules & Undecidable & Decidable (non-primitive recursive) \\ 
    	\hline 
    	Simple trigger rules, & \multirow{2}{*}{?} & \multirow{2}{*}{$\EXPSPACE$-complete} \\ 
    	non-singular intervals & & \\
    	\hline 
    	Simple trigger rules, & \multirow{2}{*}{?} & \multirow{2}{*}{$\Psp$-complete} \\ 
    	intervals in $\Intv_{(0,\infty)}$ & & \\
    	\hline 
    	Trigger-less rules only & $\NP$-complete & // \\ 
    \end{tabular} 
    }
\end{table}

%% file: Chaps/Timelines/undecidability.tex
\section{TP over dense temporal domains is an undecidable problem}\label{sec:undecidability}

In this section, we
start by settling an important negative result, namely, we
show that the TP problem, in its full generality, is undecidable over dense temporal domains, even when a single state variable is involved.
Undecidability is proved via a reduction from the halting problem for \emph{Minsky $2$-counter machines}~\cite{Minsky67}. The proof somehow resembles the one for the satisfiability problem of Metric Temporal Logic (which will be formally introduced later, in Section~\ref{sec:DecisionProcedures}) with both past and future temporal modalities, interpreted on dense time~\cite{AlurH93}.

As a preliminary step, we give a short account of Minsky 2-counter machines. A Minsky 2-counter machine (or just \emph{counter machine} for short) is a tuple $M = \tpl{\Instruct,\ell_\init,\ell_\halt}$ consisting of a finite set $\Instruct$ of labeled instructions $\ell: \imath$, where $\ell$ is a label  and $\imath$ is an instruction for either
\begin{itemize}
  \item \emph{increment} of counter $h$: $c_h:= c_h+1$; \texttt{goto} $\ell_r$, or
  \item  \emph{decrement} of counter $h$: \texttt{if} $c_h\!>\!0$ \texttt{then} $c_h:= c_h-1$; \texttt{goto} $\ell_s$ \texttt{else goto} $\ell_t$,
\end{itemize}
where $h \in \{1, 2\}$,  $\ell_s\neq \ell_t$, and $\ell_r$ (respectively, $\ell_s, \ell_t$) is either a label of an instruction in $\Instruct$ or the halting label $\ell_\halt$. Moreover, $\ell_\init\in\Instruct$ is the label of a designated (\lq\lq initial\rq\rq) instruction.

An \emph{$M$-configuration} is a triple of the form $C=(\ell, n_1, n_2)$, where $\ell$ is the label of an instruction (intuitively, which is the one to be executed next), and $n_1,n_2\in\Nat$ are the current values of the two counters $c_1$ and $c_2$, respectively. 

$M$ induces a transition relation, denoted by $\stackrel{M}{\longrightarrow}$, over pairs of $M$-configurations: 
\begin{itemize}
    \item for an instruction with label $\ell$ incrementing $c_1$, we have $(\ell, n_1, n_2)\stackrel{M}{\longrightarrow} (\ell_r, n_1+1, n_2)$, and
    \item for an instruction decrementing $c_1$, we have $(\ell, n_1, n_2)\stackrel{M}{\longrightarrow} (\ell_s, n_1-1, n_2)$ if $n_1>0$, and $(\ell, 0, n_2)\stackrel{M}{\longrightarrow} (\ell_t, 0, n_2)$ otherwise. 
\end{itemize}
The analogous for instructions changing the value of $c_2$.

An \emph{$M$-computation} is a \emph{finite} sequence $C_1,\ldots ,C_k$ of $M$-configurations such that $C_i \stackrel{M}{\longrightarrow} C_{i+1}$ for all $1\leq i<k$.
$M$ \emph{halts} if there exists an $M$-computation starting at $(\ell_\init, 0, 0)$ and leading to 
$(\ell_{\halt}, n_1, n_2)$, for some $n_1,n_2\in\Nat$. 
Given a counter machine $M$,
the \emph{halting problem for $M$} is to decide whether $M$ halts, and it was proved to be \emph{undecidable} by Minsky~\cite{Minsky67}.

The rest of the section is devoted to showing the following result.
\begin{theorem}\label{theorem:undecidability}
The TP problem over dense temporal domains is undecidable (even when a single state variable is involved).
\end{theorem}
\begin{proof}
We prove the thesis by a reduction from the halting problem for Minsky $2$-counter machines. 
Let us introduce the following notational conventions:
\begin{itemize}
  \item for increment instructions $\ell : c_h:= c_h+1$; \texttt{goto} $\ell_r$, we define $c(\ell)= c_h$ and $\Succ(\ell)= \ell_r$;
  \item  for decrement instructions $\ell:$ \texttt{if} $c_h>0$ \texttt{then} $c_h:= c_h-1$; \texttt{goto}  $\ell_r$ \texttt{else goto} $\ell_s$,
  we define $c(\ell)= c_h$, $\dec(\ell)= \ell_r$, and $\zero(\ell)= \ell_s$.
\end{itemize}
Moreover, let $\InstructLab$ be the set of instruction labels, including $\ell_\halt$, and let
$\Inc$ (resp., $\Dec$) be the set of labels for increment (resp., decrement) instructions.
We consider a counter machine $M = \tpl{\Instruct,\ell_\init,\ell_\halt}$ assuming without loss of generality that
no instruction of $M$ leads to $\ell_\init$, and that $\ell_\init$ is the label of an increment instruction.
To prove the thesis, we build in polynomial time a state variable $x_M=(V,T,D)$ and a finite set $R_M$ of synchronization rules
over $x_M$ such that $M$ halts if and only if there is a timeline for $x_M$ which satisfies all the rules in $R_M$, that is, a plan for $P=(\{x_M\},R_M)$.

\paragraph*{Encoding of $M$-computations.}

First, we define a suitable encoding of a computation of $M$ as a timeline for $x_M$.
For such an encoding we exploit the
finite set of symbols $V= V_{\main}\cup V_{\Checking}$ corresponding to the finite domain of the state variable $x_M$.
The sets of \emph{main} values $V_{\main}$ and \emph{check} values $V_{\Checking}$
are defined as
\begin{multline*}
V_{\main} = \bigcup_{\ell\in \Inc\cup\{\ell_\halt\}}\; \smashoperator[r]{\bigcup_{h\in\{1,2\}}}\; \Big(\{\ell\}\cup \{(\ell,c_h)\}\Big)\cup \\
 \bigcup_{\ell\in \Dec}\;\bigcup_{\ell' \in \{\zero(\ell),\dec(\ell)\}}\;\bigcup_{h\in\{1,2\}}\Big( \{(\ell,\ell')\}\cup   \{(\ell,\ell',c_h)\}\cup \{(\ell,\ell',(c_h,\#))\} \Big)
\end{multline*}
and
\[
V_{\Checking} = \bigcup_{\ell\in \InstructLab}\;\bigcup_{i,h\in\{1,2\}}\;\smashoperator[r]{\bigcup_{\op_i \in \{\inc_i,\dec_i,\zero_i\}}}\; \Big( \{(\ell,\op_i)\}\cup   \{(\ell,\op_i,c_h)\}\cup \{(\ell,\op_i,(c_h,\#))\} \Big).
\]
 
For each $h\in\{1,2\}$, we denote by $V_{c_h}$ the set of $V$-values $v$
having 
the 
form $v=(\ell,c)$, 
$v=(\ell,\ell',c)$, or $v=(\ell,\op,c)$, where $c\in\{c_h,(c_h,\#)\}$:  if $c=c_h$, we say that $v$ is an \emph{unmarked} $V_{c_h}$-value; otherwise ($c=(c_h,\#)$), $v$ is a \emph{marked} $V_{c_h}$-value.

An $M$-configuration is encoded by a finite word over $V$ consisting of the concatenation of a $\Checking$-code and a $\main$-code.
The $\main$-code $w_\main$  for a $M$-configuration $(\ell,n_1,n_2)$, where the instruction label  $\ell\in \Inc\cup\{\ell_\halt\}$, $n_1\geq 0$, and $n_2\geq 0$, has the form:
\[
w_\main = \ell \cdot  \underbrace{(\ell,c_1)\cdots(\ell,c_1)}_{n_1 \text{ times }} \cdot \underbrace{(\ell,c_2)\cdots(\ell,c_2)}_{n_2 \text{ times }}.
\]

In the case of a \emph{decrement} instruction label $\ell\in \Dec$  such that $c(\ell)=c_1$, the $\main$-code $w'_\main$ has one of the following two forms, depending on whether the value of $c_1$ in the encoded configuration is equal to, or greater than zero.
 \[
(\ell,\zero(\ell)) \cdot  \underbrace{(\ell,\zero(\ell),c_2)\cdots(\ell,\zero(\ell),c_2)}_{n_2 \text{ times }},
\]
\begin{multline*}
 (\ell,\dec(\ell)) \cdot (\ell,\dec(\ell),(c_1,\#))\cdot \\  \underbrace{(\ell,\dec(\ell), c_1)\cdots(\ell,\dec(\ell), c_1)}_{n_1 \text{ times }}
 \cdot \underbrace{(\ell,\dec(\ell), c_2)\cdots(\ell, \dec(\ell), c_2)}_{n_2 \text{ times }}.
\end{multline*}
In the first case, $w'_\main$ encodes the configuration $(\ell,0,n_2)$ and in the second case the configuration $(\ell,n_1+1,n_2)$. Note that, in the second case, there is exactly one occurrence of a \emph{marked} $V_{c_1}$-value which intuitively \lq\lq marks\rq\rq{} the unit of the counter which will be removed by the decrement. Analogously, the $\main$-code  for a  \emph{decrement} instruction label $\ell$ with $c(\ell)=c_2$ has two forms symmetric with respect to the previous cases.
%
%

The $\Checking$-code is used to trace both an $M$-configuration $C$ and the type of instruction associated with the configuration $C_p$ preceding $C$ in the considered computation. 
The type of instruction is given by the symbols $\inc_i$, $\dec_i$, and $\zero_i$, with $i\in\{1,2\}$: 
$\inc_i$ (resp.,  $\dec_i$, $\zero_i$) means that $C_p$ is associated with an instruction incrementing the counter $c_i$
(resp., decrementing $c_i$ with $c_i$ greater than $0$ in $C_p$,  decrementing $c_i$ with $c_i$ equal to $0$ in $C_p$).

The $\Checking$-code  for an instruction label $\ell\in\InstructLab$ and an $\inc_1$-operation has the following form
\[
 (\ell,\inc_1)  \cdot (\ell,\inc_1,(c_1,\#))\cdot \underbrace{(\ell,\inc_1,c_1)\cdots(\ell,\inc_1,c_1)}_{n_1 \text{ times }}
 \cdot 
 \underbrace{(\ell,\inc_1,c_2)\cdots(\ell,\inc_1,c_2)}_{n_2 \text{ times }},
\]
and encodes the configuration $(\ell,n_1+1,n_2)$. Note that there is exactly one occurrence of a \emph{marked} $V_{c_1}$-value which intuitively represents the unit added to the counter by the increment operation.

The $\Checking$-code  for an instruction label $\ell\in \InstructLab$ and an operation $\op_1\in\{\dec_1,\zero_1\}$ for the counter $c_1$ has 
the form
\[
 (\ell,\op_1) \cdot   \underbrace{(\ell,\op_1,c_1)\cdots(\ell,\op_1,c_1)}_{n_1 \text{ times }}
 \cdot \underbrace{(\ell,\op_1,c_2)\cdots(\ell,\op_1,c_2)}_{n_2 \text{ times }},
\]
where we require that $n_1=0$ if $\op_1=\zero_1$.  The $\Checking$-code  for a label $\ell\in \InstructLab$ and an operation associated with the counter $c_2$
is defined in a similar way.

A \emph{configuration}-code is a word $w= w_{\Checking}\cdot w_\main $ such that  $w_{\Checking}$ is a $\Checking$-code, $w_\main$ is
a $\main$-code, and  $w_{\Checking}$ and $w_\main$ are associated with the same instruction label.
The configuration code is \emph{well-formed} if $ w_{\Checking}$ and $w_\main $ encode the same configuration.

Figure~\ref{fig:counters} depicts the encoding of a configuration-code for the instruction $\ell_{i+1}$. The check-code for the instruction $\ell_{i+1}$ is associated with an increment of the counter $c_1$ (the type of instruction $\ell_{i}$).

\begin{figure}
    \centering
    \includegraphics[width=\textwidth]{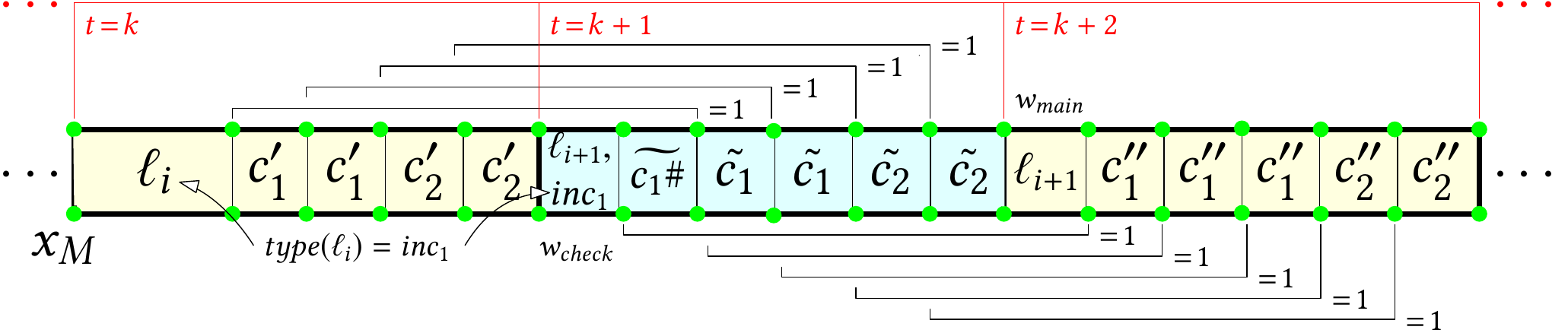}
    \caption{
A fragment of a computation code with a configuration code for 
an instruction $\ell_{i+1}$. Main-codes are highlighted in yellow and check-codes in cyan. Each square can also be seen as a token of a timeline for $x_M$ (tokens are decorated with their start time and their temporal constraints).
In the figure, for $h\in\{1,2\}$, the symbols $c_h'$, $\tilde{c_h}$, $\widetilde{c_h\#}$, and $c_h''$, stand respectively for $(\ell_i ,c_h)$, $(\ell_{i+1},inc_1,c_h)$, $(\ell_{i+1},inc_1,(c_h,\#))$, and $(\ell_{i+1} ,c_h)$.
    }
    \label{fig:counters}
\end{figure}
A \emph{computation}-code is a 
sequence of configuration-codes $\pi= w_{\Checking}^{1}\cdot w_\main^{1}\cdots\allowbreak  w_{\Checking}^{n}\cdot w_\main^{n}$ such that, for all $1\leq j<n$, the following holds (we assume $\ell_i$ to be the instruction label associated with the configuration code $w_{\Checking}^{i}\cdot w_\main^{i}$):
\begin{itemize}
  \item $\ell_j\neq \ell_\halt$;
  \item if $\ell_j\in\Inc$ with $c(\ell_j)=c_h$, then $\ell_{j+1}=\Succ(\ell_j)$ and  $w_{\Checking}^{j+1}$ is associated with the operation
  $\inc_h$;
  \item if $\ell_j\in\Dec$ with $c(\ell_j)=c_h$, and the first symbol of $w_\main^{j}$ is $(\ell_j,\zero(\ell_j))$ (resp., $(\ell_j,\dec(\ell_j))$),
  then $\ell_{j+1}=\zero(\ell_j)$ (resp., $\ell_{j+1}=\dec(\ell_j)$) and  $w_{\Checking}^{j+1}$ is associated with the operation
  $\zero_h$ (resp., $\dec_h$).
\end{itemize}
The computation-code $\pi$ is \emph{well-formed} if, additionally, each configuration-code in $\pi$ is \emph{well-formed} and, for all $1\leq j<n$, the following holds  (we assume  $(\ell_i,n_1^{i},n_2^{i})$
to be the configuration encoded by $w_{\Checking}^{i}\cdot w_\main^{i}$):
\begin{itemize}
  \item if $\ell_j\in\Inc$, with $c(\ell_j)=c_h$,  then $n_h^{j+1}= n_h^{j}+1$ and $n_{3-h}^{j+1}= n_{3-h}^{j}$;
  \item if $\ell_j\in\Dec$, with $c(\ell_j)=c_h$, then $n_{3-h}^{j+1}= n_{3-h}^{j}$. Moreover, if  $w_{\Checking}^{j+1}$ is associated with 
  $\dec_h$, then $n_h^{j+1}= n_h^{j}-1$.
\end{itemize}
Clearly, a well-formed computation code  $\pi$ encodes a computation of the Minsky 2-counter machine. 

A computation-code $\pi$ is \emph{initial} if it starts with the  prefix $(\ell_\init,\zero_1)\cdot  \ell_\init$, and it is \emph{halting} if it leads to a configuration-code associated with the halting label $\ell_\halt$. The counter machine $M$ halts if and only if there is an initial and halting well-formed computation-code.

\paragraph*{Definition of $x_M$ and $R_M$.}
Let us show now how to reduce the problem of checking the existence of an initial and halting well-formed computation-code to a TP problem for the state variable $x_M$. 

The idea is to define a timeline where the sequence of values of its tokens is a well-formed computation-code. The durations of tokens are suitably exploited to guarantee well-formedness of computation-codes. We refer the reader again to Figure~\ref{fig:counters} for an intuition. 
Each symbol of the computation-code is associated with a token having a positive duration. The overall duration of the sequence of tokens corresponding to a check-code or a main-code amounts exactly to one time unit. To allow for the encoding of 
arbitrarily large values of counters in one time unit, the duration of such tokens is not fixed (taking advantage of the dense temporal domain). In two adjacent check/main-codes, the time elapsed between the start times of corresponding elements in the representation of the value of a counter (see elements in Figure~\ref{fig:counters} connected by horizontal lines) amounts exactly to one time unit. Such a constraint allows us to compare the values of counters in adjacent codes, either checking for equality, or simulating (by using marked symbols) increment and decrement operations. Note that there is a single \emph{marked} token $c_1$ in the check-code---that represents the unit added to $c_1$ by the instruction $\ell_i$---which does not correspond to any of the $c_1$'s of the preceding main-code.

We now formally define a state variable $x_M$ and a set $R_M$ of synchronization rules for $x_M$ such that the untimed part of any timeline (i.e., neglecting tokens' durations)
for $x_M$ satisfying the rules in $R_M$ is (represents) an initial and halting well-formed computation-code. Thus, $M$ halts if and only if there exists a timeline for $x_M$ satisfying the rules in $R_M$.

As for $x_M$, we let $x_M= (V,T,D)$ where, for each $v\in V$, 
$D(v)=\mathopen]0,1\mathclose]$. This sets the \emph{strict time monotonicity} constraint, namely, the duration of a token along a timeline is always greater than zero and less than or equal to 1. 

The value transition function $T$ of $x_M$ ensures the following requirement.
\begin{claim}\label{ref:claim}
The untimed part of any timeline for $x_M$ whose first token has value $(\ell_\init,\zero_1)$
 is a prefix of some initial computation-code. Moreover, $(\ell_\init,\zero_1)\notin T(v)$ for all $v\in V$.
\end{claim}
It is a straightforward task to define $T$ in  such a way that the previous requirement is fulfilled (for details, see Appendix~\ref{sec:undecidabilityTrans}).
 
Finally, the synchronization rules in $R_M$ ensure the following requirements.
 \begin{itemize}
   \item \emph{Initialization:} every timeline starts with two tokens, the first one having value
   $(\ell_\init,\zero_1)$, and the second having value $\ell_\init$. By Claim~\ref{ref:claim} and the fact that no instruction
   of $M$ leads to $\ell_\init$, it suffices to require that a timeline has a token with value $(\ell_\init,\zero_1)$ and a token with value
   $\ell_\init$.
   This is ensured by the following two trigger-less rules:
   \[
   \true \rightarrow \exists \, o[x_M=(\ell_\init,\zero_1)].\, \true
   \] and \[
   \true \rightarrow \exists \, o[x_M= \ell_\init].\, \true
   .\]
   \item \emph{Halting:} every timeline leads to a configuration-code associated with the halting label. By the rules for initialization and Claim~\ref{ref:claim}, it suffices to require that a timeline has a token with value $\ell_\halt$. This is ensured by the following trigger-less rule:
   \[
   \true \rightarrow \exists \, o[x_M=\ell_\halt].\, \true
   .\]
   \item \emph{1-Time distance between consecutive control values:} a \emph{control $V$-value} corresponds to the first symbol of a $\main$-code or a $\Checking$-code, i.e., it is an element in $V\setminus (V_{c_1}\cup V_{c_2})$. We require  that the difference of the start times of two consecutive tokens along a timeline having a control $V$-value is exactly $1$. Formally, 
    for each pair $tk$ and $tk'$ of tokens along a timeline such that $tk$ and $tk'$ have a control $V$-value, $tk$ precedes $tk'$, and there is no token between $tk$ and $tk'$ having a control $V$-value, it holds that $\startTime(tk')-\startTime(tk)=1$ (we write this with a little abuse of notation). By Claim~\ref{ref:claim}, strict time monotonicity, and the halting requirement, it suffices to ensure that each token $tk$ having a control $V$-value distinct from $\ell_\halt$ is eventually followed by a token $tk'$ such that $tk'$ has a control $V$-value and $\startTime(tk')-\startTime(tk)=1$. To this aim, for each  $v\in V_\Con\setminus \{\ell_\halt\}$, being $V_\Con$ the set of control $V$-values,  we write the following trigger rule:
   \[
   o[x_M=v] \rightarrow \displaystyle{\bigvee_{u\in V_\Con}} \exists\, o'[x_M= u].\, o\leq^{\start,\start}_{[1,1]} o'. 
   \]
   
   \item \emph{Well-formedness of configuration-codes:} we need to guarantee that for each configuration-code $w_\Checking\cdot w_\main$ occurring along a timeline and 
    each counter $c_h$, the value of $c_h$ along the $\main$-code $w_\main$ and the $\Checking$-code $w_\Checking$ coincide.
    By Claim~\ref{ref:claim}, strict time monotonicity, initialization, and 1-Time distance between consecutive control values,  it suffices to ensure that $(i)$~each token $tk$  with
    a $V_{c_h}$-value in $V_\Checking$ is eventually followed by a token $tk'$ with a $V_{c_h}$-value such that  $\startTime(tk')-\startTime(tk)=1$, and vice versa $(ii)$~each token $tk$  with
    a $V_{c_h}$-value in $V_\main$ is eventually preceded by a token $tk'$ with a $V_{c_h}$-value such that  $\startTime(tk)-\startTime(tk')=1$. As for the former requirement, for each   $v\in V_{c_h}\cap V_\Checking$, we write the rule:
   \[
   o[x_M=v] \rightarrow \displaystyle{\bigvee_{u\in V_{c_h}}} \exists\, o'[x_M= u].\, o\leq^{\start,\start}_{[1,1]} o'.
   \]
   
   For the latter, for each   $v\in V_{c_h}\cap V_\main$, we have the rule:
   \[
   o[x_M=v] \rightarrow \displaystyle{\bigvee_{u\in V_{c_h}}} \exists\, o'[x_M= u].\, o'\leq^{\start,\start}_{[1,1]} o.
   \]
   
   \item \emph{Increment and decrement:} we need to guarantee that the increment and decrement instructions are correctly simulated.
    By Claim~\ref{ref:claim} and the previously defined synchronization rules, we can assume that the untimed part $\pi$ of a timeline is an initial and halting
    computation-code such that all  configuration-codes occurring in $\pi$ are well-formed.

    Let $w_\main \cdot w_\Checking$ be a subword occurring in $\pi$ such that
    $w_\main$ (resp., $w_\Checking$) is a $\main$-code (resp., $\Checking$-code). Let $\ell_\main$ (resp., $\ell_\Checking$) be the instruction label
    associated with $w_\main$ (resp., $w_\Checking$) and for $i=1,2$, let $n_i^{\main}$ (resp., $n_i^{\Checking}$) be the value of counter $c_i$ encoded by
     $w_\main$ (resp., $w_\Checking$). Let $c_h=c(\ell_\main)$. By construction $\ell_\main\neq \ell_\halt$, end either $\ell_\main\in\Inc$   and
     $\ell_\Checking =\Succ(\ell_\main)$, or $\ell_\main\in \Dec$  and $\ell_\Checking\in \{\zero(\ell_\main),\dec(\ell_\main)\}$. Moreover, if $\ell_\main\in \Dec$
     and $\ell_\Checking = \zero(\ell_\main)$, then $n_h^{\Checking}= n_h^{\main}=0$.
     Thus, it remains to ensure the following two requirements:
     \begin{itemize}
       \item[(*)] if $\ell_\main\in\Inc$, then $n_h^{\Checking}= n_h^{\main}+1$ and $n_{3-h}^{\Checking}= n_{3-h}^{\main}$;
       \item[(**)] if $\ell_\main\in\Dec$, then $n_{3-h}^{\Checking}= n_{3-h}^{\main}$, and whenever
       $\ell_\Checking = \dec(\ell_\main)$, then $n_h^{\Checking}= n_h^{\main}-1$.
     \end{itemize}

First we observe that, if $\ell_\main\in\Inc$, our encoding ensures that all $V_{c_{3-h}}$-values in $w_\main$ and in $w_\Checking$ are unmarked,
all $V_{c_{h}}$-values in  $w_\main$ are unmarked, and there is exactly one marked $V_{c_{h}}$-value  in  $w_\Checking$.
If instead $\ell_\main\in\Dec$, our encoding ensures that all $V_{c_{3-h}}$-values in $w_\main$ and in $w_\Checking$ are unmarked,
all $V_{c_{h}}$-values in  $w_\Checking$ are unmarked, and in case $\ell_\Checking= \dec(\ell_\main)$, then there is exactly one marked $V_{c_{h}}$-value  in  $w_\main$.
Thus, by strict time monotonicity and 1-Time distance between consecutive control values, it follows that requirements~(*) and~(**) are captured by the following
rules, where $U_{c_i}$ denotes the set of \emph{unmarked} $V_{c_i}$-values, for $i=1,2$, and $V_\init$ (resp., $V_\halt$) is the set of $V$-values associated with the label $\ell_\init$ (resp., $\ell_\halt$). For each   $v\in (U_{c_i}\cap V_\main)\setminus V_\halt$, we have the rule:
   \[
   o[x_M=v] \rightarrow \displaystyle{\bigvee_{u\in U_{c_i}}} \exists\, o'[x_M= u].\, o\leq^{\start,\start}_{[1,1]} o'. 
   \]
   
   For each $v\in (U_{c_i}\cap V_\Checking)\setminus V_\init$, we have the rule:
   \[
   o[x_M=v] \rightarrow \displaystyle{\bigvee_{u\in U_{c_i}}} \exists\, o'[x_M= u].\, o'\leq^{\start,\start}_{[1,1]} o .
   \]
 \end{itemize}
This concludes the proof of the theorem.
\end{proof}

It is worth observing that all the above trigger rules are \emph{simple}, hence \emph{undecidability of the TP problem holds also under the restriction to simple trigger rules}.

In order to ensure the well-formedness of configuration-codes and the increment/decrement requirements, a one-to-one correspondence between (suitable) pairs of tokens in main- and check-codes is enforced thanks to the above trigger rules. Whereas most of such rules are (already) satisfied under the future semantics (as the extra conjoined atoms added by Definition~\ref{def:futurerules} would be \lq\lq subsumed\rq\rq\ by already-existing ones), some rules are not (the second ones of the well-formedness and increment/decrement requirements are unsatisfiable under the future semantics). 
As a result, intuitively, having only rules under the future semantics,
we can only force the presence, for every token with value $c_h$ (for $h=1,2$), of another token with value $c_h$ starting exactly one time instant later, in the following main-/check-code. However, we cannot prevent extra \lq\lq spurious\rq\rq\  tokens to appear moving from a code to the following one.
This is the reason why, with only rules under the future semantics, we lose the ability of encoding computations of (exact) Minsky machines. Only \emph{gainy counter machines}~\cite{DemriL09}---a variant of Minsky machines whose counters may \lq\lq erroneously\rq\rq\ increase---can be captured, thus proving, as a consequence, 
\emph{non-primitive recursive-hardness} of the future TP problem (the halting problem for gainy counter machines is known to be non-primitive recursive~\cite{DemriL09}). 

\begin{theorem}\label{theorem:NPRHardness}
The future TP problem, even with \emph{one state variable}, is non-primitive recursive-hard also under one of the following two assumptions: \emph{either} $(1)$ the trigger rules are simple,
\emph{or} $(2)$ the intervals are in $\Intv_{(0,\infty)}$%
\footnote{We refer to intervals in rules' atoms and in the constraint functions of state variables.}.
\end{theorem}

Since this result is just an adaptation of the previous one (apart from some technicalities), we report its proof in Appendix~\ref{sec:NPRHardness}. 

In the next section, we will show that future TP with simple trigger rules is indeed decidable in non-primitive recursive time. 

%% file: Chaps/Timelines/DecidFutTPsimple.tex
\section{Decidability of future TP with simple trigger rules}\label{sec:DecisionProcedures}

In this section, we show that the decidability of the TP problem can be recovered assuming that the trigger rules are \emph{simple} and \emph{interpreted under the future semantics}. Moreover, under the additional assumption that intervals in trigger rules are non-singular (respectively, are in $\Intv_{(0,\infty)}$), the problem is
in $\EXPSPACE$ (respectively, in $\Psp$). 
The decidability status of \emph{future TP with arbitrary trigger rules remains an open problem}.

The rest of this section is organized as follows: in Section~\ref{sec:TimedAutomata}, we recall Timed Automata (\TA)~\cite{ALUR1994183}  and Metric Temporal logic (\MTL)~\cite{Koymans90}. In Section~\ref{sec:Reduction}, we reduce the future TP problem
with simple trigger rules to the \emph{existential MC problem} for \TA s against \MTL\ over \emph{finite timed words}. The latter problem is known to be decidable~\cite{OuaknineW07}.

\subsection{Timed automata and the logic \MTL}\label{sec:TimedAutomata}

We start by recalling the notion of timed automaton (\TA)~\cite{ALUR1994183} and the logic \MTL~\cite{Koymans90}.

Let $\Sigma$ be a finite alphabet. A  \emph{timed word} $w$ over  $\Sigma$ is
a \emph{finite}  word $w=(a_0,\tau_0)\cdots (a_n,\tau_n)$ over $\Sigma\times \RealP$ 
($\tau_i$ is called a \emph{timestamp}, and intuitively represents the time at which the \lq\lq event\rq\rq\ $a_i$ occurs) such that  $\tau_{i}\leq \tau_{i+1}$ for all $0\leq i<n$ (\emph{monotonicity} requirement).
The timed word $w$ is also denoted by $(\sigma,\tau)$, where $\sigma$ is the finite (untimed) word $a_0 \cdots a_n$
and $\tau$ is the sequence of timestamps $\tau_0, \ldots, \tau_n$.
A \emph{timed language}  over $\Sigma$ is a set of timed words over $\Sigma$.

\paragraph{Timed Automata (\TA).} Let $C$ be a finite set of clocks. A clock valuation $\val:C\to \RealP$ for $C$ is a function
 assigning a non-negative real value to each clock in $C$.
Given a value $t\in\RealP$ and a set $\Res\subseteq C$ (that we call \emph{reset set}), $(\val+ t)$ and $\val[\Res]$ denote the valuations for $C$ defined respectively as follows: for all $c\in C$,
 $(\val +t)(c) = \val(c)+t$, and $\val[\Res](c)=0$ if $c\in \Res$ and $\val[\Res](c)=\val(c)$ otherwise.

 A \emph{clock constraint} $\theta$ over $C$ is a Boolean combination of atomic formulas of the form
$c \in I$ or $c-c'\in I$, where $c,c'\in C$ and
$I\in\Intv$.
Given a clock valuation $\val$ and a clock constraint $\theta$, $\val$ is said to satisfy $\theta$, written
$\val\models \theta$, if 
$\theta$ evaluates to true after replacing each occurrence of a clock $c$ in $\theta$ by $\val(c)$, and interpreting Boolean connectives and membership to intervals in the standard way. 
%
We denote by $\Phi(C)$ the set of all possible clock constraints over $C$.

\begin{definition}[Timed automaton \TA]
 A  \TA\ over  $\Sigma$ is a tuple
$\Au=\tpl{\Sigma, Q,q_0,C,\Delta,F}$, where $Q$ is a finite
set of (control) states, $q_0\in Q$ is the initial
state, $C$ is a finite set of clocks,
$F\subseteq Q$ is the set of accepting states, and $\Delta \subseteq Q\times \Sigma \times \Phi(C) \times 2^{C} \times Q $ is the transition relation.

The \emph{maximal constant of $\Au$} is the greatest integer occurring as an endpoint of some interval in the clock constraints of the transitions of $\Au$.
\end{definition}

Intuitively, in a \TA\  $\Au$, while transitions are instantaneous, time can elapse in a control
state. The clocks  progress at the same speed  and can
be reset independently of each other when a transition is executed, in such a way that each clock
keeps track of the time elapsed since the last reset. Moreover, clock constraints
are used as guards of transitions to restrict the behavior of the
automaton.

A configuration of $\Au$ is a pair $(q,\val)$, where $q\in Q$ and $\val$ is a clock valuation for $C$.
A run $r$ of $\Au$ on a timed word $w=(a_0,\tau_0)\cdots (a_n,\tau_n)$ over $\Sigma$
is a sequence  of configurations
 $r=(q_0,\val_0)\cdots (q_{n+1},\val_{n+1})$ starting at the initial configuration $(q_0,\val_0)$,
where $\val_0(c)=0$ for all $c\in C$ (\emph{initiation requirement}), and 
\begin{itemize}
\item for all $0\leq i\leq n$ we have (\emph{consecution requirement}): 
  $(i)$~\mbox{$(q_{i},a_i,\theta,\Res,q_{i+1})\!\in\!\Delta$} for some $\theta\in\Phi(C)$ and reset set $\Res$, $(ii)$~$(\val_{i} +\tau_i-\tau_{i-1})\models \theta$ and $(iii)$~$\val_{i+1}= (\val_{i} +\tau_i-\tau_{i-1})[\Res]$ (we let $\tau_{-1}=0$).
\end{itemize}
The intuitive behavior of the \TA\ $\Au$ is the following.
Assume that $\Au$ is on state $q\in Q$ after reading the symbol $(a',\tau_i)$ at time $\tau_i$ and, 
at that time, the clock valuation is $\val$. On reading 
$(a,\tau_{i+1})$, $\Au$ chooses a transition of the form $\delta=(q,a,\theta,\Res, q')\in
\Delta$ such that the constraint $\theta$
is fulfilled by $(\val+t)$, with
$t=\tau_{i+1}-\tau_{i}$. 
The control then changes from $q$ to $q'$ and $\val$ is updated 
in such a way as to record the amount of time elapsed $t$ in the clock valuation, and to reset the clocks in $\Res$,
namely, $\val$ is updated to $(\val +t)[\Res]$.

A run $r$ is \emph{accepting} if $q_{n+1}\in F$.
The \emph{timed language} $\TLang(\Au)$ of $\Au$ is the set of  timed words $w$ over $\Sigma$
such that there is an accepting run of $\Au$ on $w$.

As shown in~\cite{ALUR1994183}, given two \TA s $\mathcal{A}_1$, with $s_1$ states and $k_1$ clocks, and $\mathcal{A}_2$, with $s_2$ states and $k_2$ clocks, the union (resp., intersection) automaton $\mathcal{A}_\vee$ (resp., $\mathcal{A}_\wedge$) such that $\TLang(\mathcal{A}_\vee)=\TLang(\mathcal{A}_1)\cup\TLang(\mathcal{A}_2)$ (resp., $\TLang(\mathcal{A}_\wedge)=\TLang(\mathcal{A}_1)\cap\TLang(\mathcal{A}_2)$) 
can be effectively calculated, and 
has $s_1+s_2$ states (resp., $s_1\cdot s_2$ states) and $k_1+k_2$ clocks (resp., $k_1+k_2$ clocks).

\paragraph{The logic \MTL.}Let us now recall the framework of Metric Temporal Logic (\MTL)~\cite{Koymans90},  a well-known  timed linear-time temporal logic which extends standard \LTL\ with time
constraints on the until modality.

Given a finite set $\Prop$ of proposition letters, the set of \MTL\ formulas $\varphi$ over $\Prop$ is defined by the following grammar:
\[
\varphi::= \top \mid
p \mid
\varphi \vee \varphi \mid
\neg \varphi      \mid
\varphi \StrictUntil_I\varphi,
\]
where $p\in \Prop$, $I\in\Intv$, and
$\StrictUntil_I$  is the \emph{strict timed until} \MTL\ modality. 

\MTL\ formulas over $\Prop$ are interpreted over  timed words over $2^{\Prop}$.
Given an \MTL\ formula $\varphi$, a  timed word $w=(\sigma,\tau)$ over $2^{\Prop}$, and a position $0\leq  i< |w|$, the satisfaction relation
$(w,i)\models\varphi$---meaning that $\varphi$ holds at position $i$ of $w$---is  defined as follows (we omit the clauses for Boolean connectives):
\begin{itemize}
\item $(w,i)\models p \iff p\in\sigma(i)$,
\item $(w,i)\models \varphi_1 \StrictUntil_I\varphi_2
              \iff $ there exists $j>i$ such that $(w,j)\models \varphi_2$, $\tau_j-\tau_i\in I$, and $(w,k)\models \varphi_1$ for all $i<k<j$.
\end{itemize}
A \emph{model of $\varphi$} is a  timed word $w$ over $2^{\Prop}$ such that $(w,0)\models \varphi$. The \emph{timed language} $\TLang(\varphi)$ of $\varphi$ is the set of  models of $\varphi$.

The \emph{existential MC problem for \TA s against \MTL} is the problem of checking, for a given \TA\ $\Au$ over $2^{\Prop}$ and an \MTL\ formula $\varphi$ over $\Prop$, whether
$\TLang(\Au)\cap \TLang(\varphi)\neq \emptyset$.

In \MTL, we use standard shortcuts such as: $\Eventually_I \varphi$ for $\varphi \vee (\true 
\StrictUntil_I \varphi)$ (\emph{timed eventually} or \emph{timed future}), and $\Always_I \varphi$ for $\neg \Eventually_I  \neg\varphi$ (\emph{timed always} or \emph{timed globally}).
 
 We also consider two fragments of \MTL, namely, \MITL\ (Metric Interval Temporal Logic)
and $\MITLR$~\cite{Alur:1996}: \MITL\ is obtained by allowing only non-singular intervals of $\Intv$ at the subscript of $\StrictUntil$, while $\MITLR$ is the fragment of \MITL\ obtained by allowing only intervals
in $\IntvR$. 

The \emph{maximal constant} of an \MTL\ formula $\varphi$ is the greatest integer occurring as an endpoint of some interval of (the occurrences of) the $\StrictUntil_I$ modality in $\varphi$.

\subsection{Reduction to existential MC for \TA s against \MTL}\label{sec:Reduction}
We now solve the future TP problem with simple trigger rules by means of an exponential-time reduction to the existential MC problem for \TA s against \MTL. 

In the following, we fix an instance $P=(SV,R)$ of the problem where the trigger rules in $R$ are simple. The \emph{maximal constant} of $P$, denoted by $K_P$, is the greatest integer occurring in the atoms of the rules in $R$ and in the constraint  functions of the state variables in $SV$.

The proposed reduction consists of three steps:
\begin{enumerate}
  \item first, we define an encoding of the multi-timelines of $SV$ by means of timed words over $2^{\Prop}$ for a suitable finite set $\Prop$ of proposition letters,
  and show how to construct a \TA\ $\Au_{SV}$ over $2^{\Prop}$ accepting such encodings;
  \item next, we build an \MTL\ formula $\varphi_{\forall}$ over $\Prop$ such that for each multi-timeline $\Pi$ of $SV$ and encoding $w_\Pi$ of $\Pi$, $w_\Pi$ is a model of $\varphi_\forall$
  if and only if $\Pi$ satisfies all the trigger rules in $R$ under the future semantics;
  \item finally, we construct a \TA\ $\Au_{\exists}$ over $2^{\Prop}$ such that for each multi-timeline $\Pi$ of $SV$ and encoding $w_\Pi$ of $\Pi$, $w_\Pi$ is accepted by $\Au_{\exists}$
  if and only if $\Pi$ satisfies all the trigger-less rules in $R$.
\end{enumerate}
Hence, there is a future plan for $P=(SV,R)$  iff  $\TLang(\Au_{SV})\cap \TLang(\Au_\exists )\cap \TLang(\varphi_\forall)\neq \emptyset$. 

For each $x\in SV$, we let $x=\tpl{V_x,T_x,D_x}$.
Given an interval $I\in\Intv$ and a natural number $n\in \Nat$, let $n+I$ (respectively, $n-I$) denote the set of non-negative real numbers
$\tau\in\RealP$ such that $\tau-n\in I$ (respectively, $n-\tau \in I$). Note that $n+I$ (respectively, $n-I$) is a (possibly empty) interval in $\Intv$ whose endpoints can be trivially calculated.

For an atom $\rho$ in $R$ involving a time constant (time-point atom), let $I(\rho)$ be the interval in $\Intv$ defined as follows:
\begin{itemize}
  \item if $\rho$ has the form  $o\leq^{e}_{I} n$ (resp., $n\leq^{e}_{I} o$), then $I(\rho)= n-I$ (resp., $I(\rho)= n+I$).
\end{itemize}
We finally define $\Intv_R$ as the set of intervals $J\in\Intv$ such that $J=I(\rho)$ for some time-point atom $\rho$ occurring in a trigger rule of  $R$.

\paragraph{Encodings of multi-timelines of $SV$.} We assume that for distinct state variables $x$ and $x'$, the sets $V_x$ and  $V_{x'}$
are disjoint. We exploit the following set $\Prop$ of proposition letters to encode multi-timelines of $SV$:
\[
\Prop = \bigcup_{x\in SV}\Main_x \cup \Deriv ,
\]
\[
\Main_x = ((\{\Beg_x \}\cup V_x) \times V_x)   \cup   (V_x \times \{\End_x\}),
\]
\[
\Deriv = \Intv_R \cup \{p_>\} \cup \bigcup_{x\in SV}\bigcup_{v\in V_x}\{\Past_v^{\start},\Past_v^{\Ending}\}.
\]
Intuitively, we use the propositions in $\Main_x$ to encode a token along a timeline for $x$. The propositions in $\Deriv$, as explained below, represent
enrichments of the encoding, used for translating simple trigger rules in \MTL\ formulas under the future semantics.
 The tags $\Beg_x$ and $\End_x$ in $\Main_x$ are used to mark the start and the end of a timeline  for $x$. 
 
\begin{figure}[tb]
    \centering
    \includegraphics[width=0.85\linewidth]{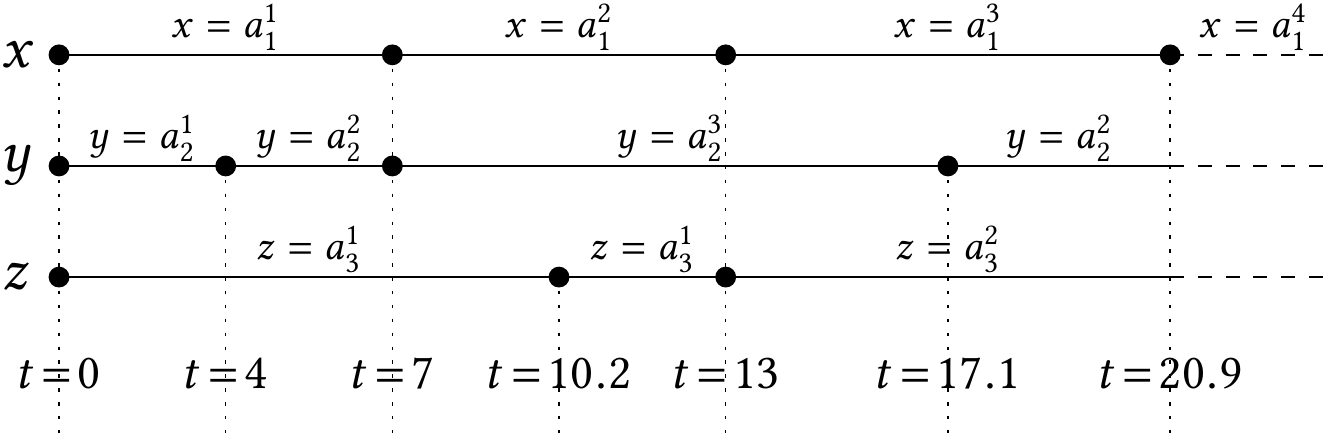}
    \caption{An example of multi-timeline of $SV=\{x,y,z\}$, where in particular $V_x=\{a_1^i\mid 1\leq i\leq 4\}$, $V_y=\{a_2^i\mid 1\leq i\leq 3\}$ and $V_z=\{a_3^i\mid 1\leq i\leq 2\}$. \\
    The encoding of the timeline for $x$ depicted in the figure (we show only values in $\Main_x$) is $\big(\{(\Beg_x,a_1^1)\},0\big)\big(\{(a_1^1,a_1^2)\},7\big)\big(\{(a_1^2,a_1^3)\},13\big)\big(\{(a_1^3,a_1^4)\},20.9\big)\cdots$ \\
    The encoding of the multi-timeline of $SV$ depicted in the figure (we show only values in $\Main_x\cup\Main_y\cup\Main_z$) is $\big(\{(\Beg_x,a_1^1),(\Beg_y,a_2^1),(\Beg_z,a_3^1)\},0\big)\allowbreak
    \big(\{(a_2^1,a_2^2)\},4\big)\allowbreak
    \big(\{(a_1^1,a_1^2),(a_2^2,a_2^3)\},7\big)
    \big(\{(a^1_3,a^1_3)\},10.2\big)
    \big(\{(a_1^2,a_1^3),(a^1_3,a^2_3)\},13\big)
    \big(\{(a^3_2,a^2_2)\},17.1\big)\cdots$}
    \label{fig:ktimelines}
\end{figure}

 A token $tk$ with value $v$ along a timeline for
 $x$ is encoded by two events: the \emph{start-event} (occurring at the start time of $tk$) and
 the \emph{end-event} (occurring at the end time of $tk$). The start-event of $tk$ is specified by a main proposition of the form
 $(v_p,v)$, where either $v_p=\Beg_x$ ($tk$ is the first token of the timeline) or $v_p$ is the value of the token for $x$
preceding $tk$. The end-event of $tk$ is instead specified by a main proposition of the form
 $(v,v_s)$, where either $v_s=\End_x$ ($tk$ is the last token of the timeline) or $v_s$ is the value of the token for $x$
following $tk$. 
See Figure~\ref{fig:ktimelines} for an example.

Now we explain the meaning of the proposition letters in $\Deriv$.
The elements in  $\Intv_R$ reflect the semantics of
the time-point atoms in the trigger rules of $R$: for each $I\in \Intv_R$, $I$ holds at the current position if the current timestamp $\tau$ satisfies
$\tau\in I$.  The tag $p_>$ keeps track of whether the current timestamp is strictly greater than the previous one.
Finally,  the propositions in $\bigcup_{x\in SV}\bigcup_{v\in V_x}\{\Past_v^{\start},\Past_v^{\Ending}\}$ keep track of past token events occurring at timestamps \emph{coinciding}
with the current timestamp. 

We start by defining the encoding of timelines for $x\in SV$.
An \emph{encoding of a timeline for $x$} is
 a timed word $w$ over $2^{\Main_x \cup \Deriv}$ of the form
 \[
 w =(\{(\Beg_x,v_0)\}\cup S_0,\tau_0)(\{(v_0,v_1)\}\cup S_1,\tau_1)\cdots (\{(v_n,\End_x)\}\cup S_{n+1},\tau_{n+1})
 \]
 where, for all $0\leq i\leq n+1$, $S_i\subseteq \Deriv$, and 
 \begin{itemize}
  \item $v_{i+1}\in T_x(v_i)$ for $i<n$;
 \item $\tau_0=0$ and $\tau_{i+1}-\tau_i \in D_x(v_i)$ for $i\leq n$;
  \item   $S_i\cap \Intv_R$ is the set of intervals $I\in\Intv_R$
  such that $\tau_i\in I$;
  \item $p_>\in S_i$ iff either $i=0$ or $\tau_i>\tau_{i-1}$;
  \item for all $v\in V_x$, $\Past^{\start}_v\in S_i$ (resp., $\Past^{\Ending}_v\in S_i$) iff there is $0\leq h<i$ such that $\tau_h=\tau_i$ and $v= v_h$ (resp., $\tau_h=\tau_i$, $v=v_{h-1}$ and $h>0$).
\end{itemize}
Note that the length of $w$ is at least $2$. The timed word $w$ encodes the timeline for $x$ of length $n+1$
given by $\pi=(v_0,\tau_1) (v_1,\tau_2-\tau_1)\cdots (v_n,\tau_{n+1}-\tau_n)$. Note that in the encoding, $\tau_i$ and $\tau_{i+1}$ represent the start time and the end time
of the $i$-th token of the timeline $\pi$ ($0\leq i\leq n$).
See the caption of Figure~\ref{fig:ktimelines} for an example of encoding.

Next, we define the encoding of a multi-timeline of $SV$.  For a set $P\subseteq \Prop$ and $x\in SV$, let $P[x]= P\setminus \bigcup_{y\in SV\setminus \{x\}}\Main_y$.
An \emph{encoding of a multi-timeline of $SV$} is
 a timed word $w$ over $2^{\Prop}$ of the form
$
 w =(P_0,\tau_0)\cdots (P_n,\tau_n)
$
 such that  the following conditions hold:
 \begin{itemize}
  \item  for all $x\in SV$, the timed word obtained from $(P_0[x],\tau_0)\cdots (P_n[x],\tau_n)$ by removing
  the pairs $(P_i[x],\tau_i)$ such that $P_i[x]\cap \Main_x=\emptyset$ is an encoding of a timeline for $x$;
    \item $P_0[x]\cap \Main_x\neq \emptyset$ for all $x\in SV$ (initialization).
\end{itemize}
See again Figure~\ref{fig:ktimelines} for an example of encoding of a multi-timeline.

We now construct a \TA\ $\Au_{SV}$ over $2^{\Prop}$ accepting the encodings of the multi-timelines of $SV$,
as shown in the proof of the next proposition. 

 \begin{proposition}\label{prop:AtutomataForMultiTimeline} One can construct in exponential time a \TA\ $\Au_{SV}$ over $2^{\Prop}$, with $2^{O(\sum_{x\in SV}|V_x|)}$ states, $|SV|+2$ clocks, and
 maximal constant $O(K_P)$, such that $\TLang(\Au_{SV})$ is the set of encodings of the multi-timelines of $SV$.
 \end{proposition}
 \begin{proof}
Let us fix an ordering $SV=\{x_1,\ldots,x_N\}$ of the state variables. Let $\mathcal{H}= \Deriv\setminus (\Intv_R \cup \{p_>\})$ and $V'_i = V_{x_i}\cup \{\Beg_{x_i},\End_{x_i}\}$ for all $1\leq i\leq N$.

The  \TA\ $\Au_{SV}=\tpl{2^{\Prop},Q,q_0,C,\Delta,F}$ is defined as follows.
\begin{itemize}
\item The set of states is given
by $Q= V'_1\times \ldots \times V'_N \times 2^{\mathcal{H}}$. Intuitively, for a state $(v_1,\ldots,v_N, H)$, the $i$-th component $v_i$ keeps track of the value of the last (start-event for a) token for $x_i$ read so far
  if $v_i \notin \{\Beg_{x_i},\End_{x_i}\}$. If $v_i =\Beg_{x_i}$ (resp., $v_i =\End_{x_i}$), then no start-event for a token for $x_i$ has been read so far (resp., no start-event for a token for $x_i$ can be read). Moreover, the last component $H$ of the state
  keeps track of past token events occurring at a timestamp coinciding with the last timestamp.
\item The initial state $q_0$ is $(\Beg_{x_1},\ldots,\Beg_{x_N},\emptyset)$.
\item The set $F$ of accepting states is the set of all states
  $(\End_{x_1},\ldots,\End_{x_N},\allowbreak H)$ for any $H\subseteq \mathcal{H}$.

\item   The set of clocks $C$ is given by $C=\{c_1,\ldots,c_N,c_>,c_{glob}\}$. We have a clock $c_i$ for each state variable $x_i$,  which is used to check that the duration
  of a token for $x_i$ with value $v$ is in $D_{x_i}(v)$. Moreover, $c_>$ is a clock which is always reset and is used to capture the meaning of proposition $p_>$,
   whereas $c_{glob}$ is a clock that measures the current (global) time and is never reset.

\item   
The relation $\Delta$ consists of the transitions 
\[((v_1,\ldots,v_N,H),\; P,\; \theta_1\wedge \ldots \wedge \theta_N \wedge \theta_> \wedge \theta_{glob},\; \Res,\; (v'_1,\ldots,v'_N,H'))\]
such that:
\begin{itemize}
   \item if $(v_1,\ldots,v_N,H)=q_0$, then $P\cap \Main_{x}\neq \emptyset$ for all $x\in SV$ (this ensures initialization);
     \item for all $1\leq i\leq N$, the following holds:
     \begin{itemize}
       \item \emph{either}  $P\cap \Main_{x_i}=\emptyset$, $v'_i=v_i$, $\theta_i=\true$, and $c_i\notin \Res$ (intuitively, no event associated with $x_i$ occurs in this case),
       \item \emph{or}   $P\cap \Main_{x_i}=(v_i,v'_i)$ (hence, $v_i\neq \End_{x_i}$),
       $v'_i\in T_{x_i}(v_i)$ if both $v_i\in V_{x_i}$ and $v'_i\in V_{x_i}$; $c_i\in \Res$ and $\theta_i= c_i\in D_{x_i}(v_i)$
            (resp., $\theta_i= c_i\in [0,0]$) if $v_i\neq \Beg_{x_i}$ (resp., if $v_i=\Beg_{x_i}$);
     \end{itemize}
      \item $c_{glob}\notin \Res$ and \[\theta_{glob}= \bigwedge_{I\in P\cap \Intv_R } c_{glob}\in I \wedge \bigwedge_{I\in \Intv_R\setminus P } (c_{glob}\in \overrightarrow{I} \vee c_{glob}\in \overleftarrow{I}),\] where, for each $I\in \Intv_R\setminus P$, $\overrightarrow{I}$ and $\overleftarrow{I}$ are (possibly empty) maximal intervals in $\RealP$ disjoint from $I$ (e.g., if $I=\mathopen[3,5\mathclose[$, then $\overleftarrow{I}=\mathopen[0,3\mathclose[$ and  $\overrightarrow{I}=\mathopen[5,+\infty\mathclose[$). Note that $\overrightarrow{I}, \overleftarrow{I}\in \Intv$. Recall that, 
      for each $I\in \Intv_R$, $I$ must be in $P$ if and only if the current time (given by $c_{glob}$) is in $I$;
     \item $c_>\in \Res$; moreover, if $(v_1,\ldots,v_N,H)=q_0$, then  $p_>\in P$   and $\theta_> = \true$, otherwise, \emph{either} $p_>\in P$ and $\theta_>=c_>\in \mathopen]0,+\infty\mathclose[$, \emph{or} $p_>\notin P$ and $\theta_>=c_>\in [0,0]$;
     \item $P\cap \mathcal{H}=\emptyset$ if $p_>\in P$; otherwise $P\cap \mathcal{H}=H$;
     \item for all $x\in SV$ and $v\in V_{x}$, $\Past^{\start}_v\in H'$ iff \emph{either} $P\cap \Main_{x_i}$ is of the form $(v',v)$, \emph{or}
     $p_>\notin P$ and $\Past^{\start}_v\in H$;
     \item for all $x\in SV$ and $v\in V_{x}$, $\Past^{\Ending}_v\in H'$ iff \emph{either} $P\cap \Main_{x_i}$ is of the form $(v,v')$, \emph{or}
     $p_>\notin P$ and $\Past^{\Ending}_v\in H$.
\end{itemize}
\end{itemize}
 This concludes the proof. 
 \end{proof}

 \paragraph{Encodings of simple trigger rules by \MTL\ formulas.} 
 We now construct an \MTL\ formula $\varphi_{\forall}$ over $\Prop$ capturing the simple trigger rules in $R$, 
  under the future semantics.

 \begin{proposition}\label{prop:MTLTriggerRules} One can construct in linear time an \MTL\ formula $\varphi_{\forall}$, with maximal constant $O(K_P)$,
  such that for each multi-timeline $\Pi$ of $SV$ and encoding $w_\Pi$ of $\Pi$, $w_\Pi$ is a model of $\varphi_\forall$
  iff $\Pi$ satisfies all the simple trigger rules in $R$ under the future semantics.

  The formula $\varphi_\forall$ is an \MITL\ formula (resp., $\MITLR$ formula) if the intervals in the trigger rules are non-singular (resp., belong to $\IntvR$). 
  
  The formula $\varphi_\forall$ has $O(|R|\!\cdot\! N_A \!\cdot\! N_{\mathcal{E}}\!\cdot\! \left(|\Intv_R| + (\sum_{x\in SV}|V_x|)^2\right))$ distinct subformulas, with $N_A$ the maximum number of atoms in a trigger rule of $R$, and $N_{\mathcal{E}}$ the maximum number of existential statements in a trigger rule of $R$.
 \end{proposition}
\begin{proof} We first introduce some auxiliary propositional (Boolean) formulas over $\Prop$. Let $x\in SV$ and $v\in V_x$. We denote by
$\psi(\start,v)$ and $\psi(\Ending,v)$ the two propositional formulas over $\Main_x$ defined as follows:
\[
\psi(\start,v)= (\Beg_x,v)\vee \displaystyle{\bigvee_{u\in V_x}}(u,v),
\]
\[
\psi(\Ending,v)= (v,\End_x)\vee \displaystyle{\bigvee_{u\in V_x}}(v,u).
\]
Intuitively, $\psi(\start,v)$ (resp., $\psi(\Ending,v)$) states that a start-event (resp., end-event) for a token for $x$ with value $v$ occurs at the current time.
We also use the formula \[\psi_{\neg x}= \neg \bigvee_{m\in \Main_x} m\] asserting that no event for a token for $x$ occurs at the current time.
Additionally, given an \MTL\ formula $\theta$, we define the \MTL\ formula \[\EqTime(\theta) = \theta \vee [\neg p_> \StrictUntil_{\geq 0}(\neg p_> \wedge \theta)]\] which is satisfied
by an encoding of a multi-timeline of $SV$ at the current time if $\theta$ eventually holds at a position whose timestamp coincides with the current timestamp.

The \MTL\ formula $\varphi_{\forall}$ has a conjunct   
 $\varphi_{\mathcal{R}}$ for each trigger rule $\mathcal{R}\in R$. 
 Let $\mathcal{R}$ be a trigger rule of the form
 $o_t[x_t =v_t] \to \mathcal{E}_1\vee \mathcal{E}_2\vee \ldots \vee \mathcal{E}_k$. 
 Then $\varphi_{\mathcal{R}}$ is given by 
 \[
\varphi_{\mathcal{R}}= \Always_{\geq 0} \big(\psi(\start,v_t) \rightarrow \displaystyle{\bigvee_{i=1}^{k}}\Phi_{\mathcal{E}_i}\big),
 \]
where $\Phi_{\mathcal{E}_i}$, with $1\leq i\leq k$, ensures the fulfillment of the existential statement $\mathcal{E}_i$
of $\mathcal{R}$ under the future semantics. 

Let $\mathcal{E}\in \{\mathcal{E}_1,\ldots,\mathcal{E}_k\}$, $O$ be the set of token names existentially quantified
in $\mathcal{E}$, $\mathbf{A}$ be  the set of \emph{interval} atoms in $\mathcal{E}$ and, for each $o\in O$, $val(o)$  be the value of the token referenced by $o$ in the associated quantifier. 
In the construction of $\Phi_{\mathcal{E}}$, we crucially exploit the assumption that  $\mathcal{R}$ is simple: 
for each token name $o\in O$, there is at most one atom in $\mathbf{A}$ where $o$ occurs.

For each token name $o\in \{o_t\}\cup O$, 
we denote by $\Intv_o^{\start}$ (resp., $\Intv_o^{\Ending}$) the set of intervals $J\in\Intv$ such that $J=I(\rho)$ for some time-point atom $\rho$ occurring in  $\mathcal{E}$, which imposes a time constraint on the start time (resp., end time) of the token referenced by $o$. Note that $\Intv_o^{\start},\Intv_o^{\Ending}\subseteq \Prop$, and we exploit the propositional formulas $\xi^{\start}_o = \bigwedge_{I\in \Intv^{\start}_o}I$ and $\xi^{\Ending}_o = \bigwedge_{I\in \Intv^{\Ending}_o}I$  to ensure the fulfillment of the time constraints imposed by the
time-point atoms associated with the token $o$.  

The \MTL\ formula $\Phi_{\mathcal{E}}$ is thus given by:
\[
\Phi_{\mathcal{E}}=\xi^{\start}_{o_t} \wedge [\psi_{\neg x_t}\StrictUntil_{\geq 0}(\psi(\Ending,v_t)\wedge \xi^{\Ending}_{o_t})]\wedge \displaystyle{\bigwedge_{\rho\in \mathbf{A}}} \chi_\rho,
\]
where, for each atom $\rho\in \mathbf{A}$, the formula $\chi_\rho$ captures the future semantics of $\rho$. 

The construction of $\chi_\rho$ depends on the form of $\rho$. We distinguishes four cases.
\begin{enumerate}
   \item $\rho = o \leq_I^{e_1,e_2} o_t$  and $o\neq o_t$. We assume $0\in I$ (the other case being simpler). First, assume that $e_2=\start$. Under the future semantics,
  $\rho$ holds iff the start time of the trigger token $o_t$ coincides with the $e_1$-time of token $o$. Hence, in this case ($e_2=\start$), $\chi_\rho$ is given by:
  \[
  \chi_\rho = \xi_o^{e_1}\wedge  \bigl(\Past_{val(o)}^{e_1}\vee \EqTime(\psi(e_1,val(o)))\bigr).
  \]
  If instead $e_2 = \Ending$, then $\chi_\rho$ is defined as follows:
 \begin{multline*}
  \chi_\rho =  \big[\psi_{\neg x_t}\StrictUntil_{\geq 0}\{\xi_o^{e_1}\wedge \psi(e_1,val(o))\wedge \psi_{\neg x_t} \wedge (\psi_{\neg x_t}\StrictUntil_I \psi(\Ending,v_t))\}\big]   \vee 
  \\
  \big[(\psi(e_1,val(o))\vee \Past_{val(o)}^{e_1}) \wedge \xi_o^{e_1} \wedge\\ \big(\EqTime(\psi(\Ending,v_t)) \vee (\psi_{\neg x_t} \wedge (\psi_{\neg x_t}\StrictUntil_I \psi(\Ending,v_t)))\big)\big]   \vee 
  \\
       \big[\psi_{\neg x_t}\StrictUntil_{\geq 0}\{\psi(\Ending,v_t)  \wedge \EqTime(\psi(e_1,val(o))\wedge\xi_o^{e_1})\}\big].
\end{multline*}
 The first disjunct (in square brackets) considers the case where the $e_1$-event of token $o$ occurs strictly between the start-event and the end-event of the trigger token $o_t$ (along the encoding of a multi-timeline of $SV$).
 The second considers the case where the  $e_1$-event of token $o$ precedes the start-event of the trigger token: thus, under the future semantics, it holds that
 the $e_1$-time of token $o$ coincides with the start time of the trigger token. Finally, the third disjunct considers the case where the $e_1$-event of token $o$ follows the
 end-event of the trigger (hence, the related timestamps must coincide).
 \item $\rho = o_t \leq_I^{e_1,e_2} o$ and $o\neq o_t$. We assume $e_1=\Ending$  and $0\in I$ (the other cases being simpler). Then,
  \begin{multline*}
  \chi_\rho = \big[\psi_{\neg x_t}\StrictUntil_{\geq 0}(\psi(\Ending,u_t)\wedge \Eventually_I (\psi(e_2, val(o))\wedge \xi_o^{e_2}) ) \big]\vee\\
  \big[\psi_{\neg x_t}\StrictUntil_{\geq 0}(\psi(\Ending,u_t)\wedge \Past_{val(o)}^{e_2} \wedge \xi_o^{e_2})\big],
  \end{multline*}
  where the second disjunct captures the situation where the $e_2$-time  of $o$ coincides with the end time of the trigger token $o_t$, but the $e_2$-event of $o$ occurs before the end-event of the trigger token.
  \item $\rho = o_t \leq_I^{e_1,e_2} o_t$. This case is straightforward and we omit the details.
  \item $\rho = o_1 \leq_I^{e_1,e_2} o_2$, with $o_1\neq o_t$ and $o_2 \neq o_t$. We assume $o_1\neq o_2$  and $0\in I$ (the other cases are simpler).  Then,
\begin{multline*}
  \chi_\rho =  \big[\Past_{val(o_1)}^{e_1} \wedge \xi_o^{e_1} \wedge \Eventually_I (\psi(e_2,val(o_2)) \wedge \xi_o^{e_2}) \big]   \vee \\    \big[\Eventually_{\geq 0}\{\psi(e_1,val(o_1)) \wedge \xi_o^{e_1} \wedge \Eventually_I (\psi(e_2,val(o_2)) \wedge \xi_o^{e_2})\} \big]   \vee \\
    \big[\Past_{val(o_1)}^{e_1} \wedge \xi_o^{e_1} \wedge \Past_{val(o_2)}^{e_2} \wedge \xi_o^{e_2}\big] \vee \\
    \big[\Past_{val(o_2)}^{e_2} \wedge \xi_o^{e_2} \wedge \EqTime(\psi(e_1,val(o_1)) \wedge\xi_o^{e_1})\big]   \vee \\
  \big[\Eventually_{\geq 0}\{\psi(e_2,val(o_2)) \wedge \xi_o^{e_2} \wedge \EqTime(\psi(e_1,val(o_1)) \wedge\xi_o^{e_1})\}\big].
\end{multline*}
The first two disjuncts handle the cases where (under the future semantics) the $e_1$-event of token $o_1$ precedes the $e_2$-event of token $o_2$, while the last three disjuncts consider the dual situation.
In the latter three cases, the $e_1$-time of token $o_1$ and the $e_2$-time of token $o_2$ are equal.
\end{enumerate}
Note that the \MTL\ formula $\varphi_\forall$ is an \MITL\ formula (resp., $\MITLR$ formula) if the intervals in the trigger rules are non-singular (resp., belong to $\IntvR$). 
 \end{proof}

\paragraph{Encoding of trigger-less rules by a \TA.} 
We now deal with trigger-less rules.
We start by noting that an existential statement $\mathcal{E}$ in a trigger-less rule requires
the existence of an \emph{a priori bounded number} of temporal events satisfying mutual temporal relations (namely, in the worst case, the start time and end time of all tokens associated with some quantifier of $\mathcal{E}$). Thus we can construct a \TA\ for $\mathcal{E}$
which guesses such a chain of events and then checks the temporal relations by means of suitable clock constraints and clock resets.
Finally, by the closure of \TA s under language union~\cite{ALUR1994183}, we can build a \TA\ for the whole trigger-less rule. Additionally, exploiting also the closure of \TA s under intersection, we construct a \TA\ accepting (encodings of) multi-timelines satisfying all trigger-less rules.%

 \begin{proposition}\label{prop:TATriggerLessRules} One can construct in exponential time a \TA\ $\Au_{_\exists}$ over $2^{\Prop}$  such that, for each multi-timeline $\Pi$ of $SV$ and encoding $w_\Pi$ of $\Pi$, $w_\Pi$ is accepted by $\Au_{\exists}$
  iff $\Pi$ satisfies all the  trigger-less  rules in $R$. 
  
  $\Au_{_\exists}$ has  $2^{O(N_q)}$ states, $O(N_q)$ clocks and maximal constant $O(K_P)$, where
  $N_q$  is the overall number of quantifiers   in the trigger-less  rules of $R$.
 \end{proposition}
 
 We recall that, in the encoding of multi-timelines of $SV$, we assume that, for distinct state variables $x,x'\in SV$, the domains $V_x$ and  $V_{x'}$
are disjoint.

 \begin{proof} Let $\mathcal{E}$ be an existential statement for $SV$ such that no token name appears free in $\mathcal{E}$. We first show how to construct
  a \TA\ $\Au_{\mathcal{E}}$ over $2^{\Prop}$  such that for each multi-timeline $\Pi$ of $SV$ and encoding $w_\Pi$ of $\Pi$, $w_\Pi$ is accepted by $\Au_{\mathcal{E}}$
  iff $\Pi$ satisfies $\mathcal{E}$. Then,  we exploit the well-known effective closure of \TA\ under language union and language intersection to prove the proposition.

Let $O$ be the set of token names existentially quantified
in the existential statement $\mathcal{E}$ and,  for each $o\in O$, let $val(o)$ be  the value of the token referenced by
$o$ in the associated quantifier. For each token name $o\in  O$, we denote by $\Intv_o^{\start}$ (resp., $\Intv_o^{\Ending}$) the set of intervals $J\in\Intv$ such that $J=I(\rho)$ for some time-point atom $\rho$ occurring in  $\mathcal{E}$ which imposes a time constraint on the start time (resp., end time) of the token referenced by $o$.

We first outline the construction of  $\Au_{\mathcal{E}}$. We associate two clocks with  each token name $o\in O$,
namely $c_o^{\start}$ and $c_o^{\Ending}$ which, intuitively, are reset when the token chosen for $o$ starts and ends, respectively.
The clocks $c_o^{\start}$ and $c_o^{\Ending}$ are non-deterministically reset when a start-event for $val(o)$ and the related end-event occur along an encoding of a multi-timeline.
The automaton $\Au_{\mathcal{E}}$ ensures that the clocks $c_o^{\start}$ and $c_o^{\Ending}$ are reset exactly once.
$\Au_{\mathcal{E}}$ moves to an accepting state only if all the clocks $c_o^{\start}$ and $c_o^{\Ending}$ for each $o\in O$ have been reset and the time constraints that encode the interval atoms in $\mathcal{E}$ are fulfilled.
To deal with time-point atoms, we also exploit, like in the previous proofs, a global clock $c_{glob}$ which measures the current time and is never reset: whenever the clock $c_o^{\start}$ (resp., $c_o^{\Ending}$) is reset, we require that the clock constraint
  $\bigwedge_{I\in \Intv_o^{\start} } c_{glob}\in I$ (resp., $\bigwedge_{I\in \Intv_o^{\Ending} } c_{glob}\in I$) is fulfilled.

The  \TA\ $\Au_{\mathcal{E}}=\tpl{2^{\Prop},Q,q_0,C,\Delta,F}$ is formally defined as follows.
\begin{itemize}
    \item The set $C$ of clocks is  $\{c_{glob}\}\cup \bigcup_{o\in O}\{c_o^{\start},c_o^{\Ending}\}$.
    \item The set of states is $2^{C\setminus \{c_{glob}\}}$. Intuitively, a state keeps track of the clocks in $C\setminus \{c_{glob}\}$ which have been reset so far.
    \item The initial  state $q_0$ is  $\emptyset$.
    \item The set of final states $F$ is given by the singleton $\{C\setminus \{c_{glob}\}\}$. In such a state all clocks different from $c_{glob}$ have been reset.
    \item The transition relation
$\Delta$ consists of the transitions $(C_1,P,\theta\wedge \theta_{glob},\Res,C_2)$ such that \emph{either} $(i)$~$C_1 = C\setminus \{c_{glob}\}$, $C_2=C_1$, $\Res=\emptyset$,  $\theta = \true$, and
$\theta_{glob} =\true$ (intuitively $\Au_{\mathcal{E}}$ loops unconditionally in its final state),
\emph{or} $(ii)$~$C_1 \subset C\setminus \{c_{glob}\}$, $C_2 \supseteq C_1$ ($\Au_{\mathcal{E}}$ has not reached its final state yet), and the following conditions hold:
\begin{itemize}
    \item for each $c_o^{\start}\in C_2\setminus C_1$, there is a main proposition  in $P$ of the form $(v',val(o))$  for some $v'$.
    \item  for each $o\in O$, $c_o^{\Ending}\in C_2\setminus C_1$ if and only if $c_o^{\start}\in C_1$ and $(val(o),v')\in P$ for some $v'$.
     \item if $C_2\subset C\setminus \{c_{glob}\}$ (in this case $\Au_{\mathcal{E}}$ is not transitioning to its final state), then $\theta= \true$.
     
     Conversely, if $C_2= C\setminus \{c_{glob}\}$ (here $\Au_{\mathcal{E}}$ moves to the final state), then $\theta = \bigwedge_{\rho\in \mathbf{A}}\code(\rho)$,
     where  $\mathbf{A}$  is the set of \emph{interval} atoms of $\mathcal{E}$ and for each interval atom $\rho\in \mathbf{A}$ of the form $o_1\leq^{e_1,e_2}_{I} o_2$, the clock constraint $\code(\rho)$ is defined as follows:
     \begin{itemize}
       \item if $c_{o_2}^{e_2}\notin C_1$ and $c_{o_1}^{e_1}\notin C_1$, then $\code(\rho)= c_{o_2}^{e_2}-c_{o_1}^{e_1} \in I$ (in this case, both $c_{o_2}^{e_2}$ and $c_{o_1}^{e_1}$ are reset simultaneously by the transition to the final state $C_2$, meaning that $o_2$'s $e_2$-event and $o_1$'s $e_1$-event have the same timestamp; hence it must be that $c_{o_2}^{e_2}-c_{o_1}^{e_1} = 0 \in I$ for the atom to be satisfied);
       \item  if $c_{o_2}^{e_2}\in C_1$ and $c_{o_1}^{e_1}\in C_1$, then $\code(\rho)= c_{o_1}^{e_1}-c_{o_2}^{e_2} \in I$;
       \item  if $c_{o_2}^{e_2}\in C_1$ and $c_{o_1}^{e_1}\notin C_1$, then $\code(\rho)= c_{o_2}^{e_2}\in [0,0]\wedge c_{o_2}^{e_2}\in I$ ($o_2$'s $e_2$-event and $o_1$'s $e_1$-event must have the same timestamp; as before, it must be that $0\in I$);
       \item  if $c_{o_2}^{e_2}\notin C_1$ and $c_{o_1}^{e_1}\in C_1$, then $\code(\rho)= c_{o_1}^{e_1}\in I$.
     \end{itemize}
     \item $\theta_{glob} = \displaystyle{\bigwedge_{c_o^{e}\in C_2\setminus C_1 } \,\,\bigwedge_{I\in \Intv_o^{e} }} c_{glob}\in I$.
     \item $\Res = C_2 \setminus C_1$.
     \end{itemize}
\end{itemize}
  Note that $\Au_{\mathcal{E}}$ has $2^{O(m)}$ states, $O(m)$ clocks and maximal constant $O(K)$, where
  $m$ is the number of quantifiers in  $\mathcal{E}$
  and $K$ is the maximal constant in $\mathcal{E}$.

 Given a trigger-less rule $\mathcal{R}=\true \to \mathcal{E}_1\lor \mathcal{E}_2\lor \ldots \lor \mathcal{E}_k$, we construct the \TA\  $\Au_{\mathcal{R}}$ resulting from the union of the automata
 $\Au_{\mathcal{E}_1},\ldots,\Au_{\mathcal{E}_k}$. Then the \TA\ $\Au_\exists$ is obtained as intersection of the automata $\Au_{\mathcal{R}}$, for all $\mathcal{R}\in R$ being trigger-less rules.
 By~\cite{ALUR1994183},  $\Au_\exists$ has  $2^{O(N_q)}$ states, $O(N_q)$ clocks, and maximal constant $O(K_P)$, where
  $ N_q$ is the overall number of quantifiers in the  trigger-less rules of  $R$. 
%
 \end{proof} 

\paragraph{Conclusion of the construction.}
By applying Proposition~\ref{prop:AtutomataForMultiTimeline}, \ref{prop:MTLTriggerRules}, \ref{prop:TATriggerLessRules} and well-known results about \TA s and \MTL\ over finite timed words~\cite{ALUR1994183,OuaknineW07},
we obtain the main result of this section.

\begin{theorem}\label{theorem:UpperBounds}
The future TP problem with simple trigger rules is decidable (with non-primitive recursive complexity).
Moreover, if the intervals in the atoms of the trigger rules are non-singular
(resp., belong to $\Intv_{(0,\infty)}$), then the problem is in $\EXPSPACE$ (resp., in $\Psp$).
\end{theorem}
\begin{proof} 
Let us consider an instance $P=(SV,R)$ of the problem with maximal constant $K_P$.
Let $N_v = \sum_{x\in SV}|V_x|$, $N_q$ be the overall number of quantifiers in the trigger-less  rules of $R$, 
$N_A$ the maximum number of atoms in a trigger rule of $R$, and $N_{\mathcal{E}}$ the maximum number of existential statements in a trigger rule of $R$.

By Proposition~\ref{prop:AtutomataForMultiTimeline}, \ref{prop:MTLTriggerRules}, \ref{prop:TATriggerLessRules} and the effective closure
of \TA s under language intersection~\cite{ALUR1994183}, we can build:
\begin{itemize}
    \item a \TA\ $\Au_P$---namely, the intersection of $\Au_{SV}$ from Proposition~\ref{prop:AtutomataForMultiTimeline} and $\Au_{_\exists}$ from Proposition~\ref{prop:TATriggerLessRules}---having $2^{O(N_q+N_v)}$ states, $O(N_q+|SV|)$ clocks, and maximal constant $O(K_P)$,
    \item and an \MTL\ formula $\varphi_\forall$ with $O(|R|\cdot N_A \cdot N_{\mathcal{E}}\cdot (|\Intv_R| + N_v^2))$ distinct subformulas and maximal constant $O(K_P)$,
\end{itemize}
    such that there is a future plan for $P$ iff
$\TLang(\Au_P)\cap\TLang(\varphi_\forall)\neq \emptyset$. By~\cite{OuaknineW07}, checking non-emptiness of  $\TLang(\Au_P)\cap\TLang(\varphi_\forall)$ is decidable. Thus the first part of the theorem holds. 

For the
second part, assume that  the intervals in 
the trigger rules are non-singular
(resp., belong to $\Intv_{(0,\infty)}$). By Proposition~\ref{prop:MTLTriggerRules}, $\varphi_\forall$ is an \MITL\ (resp., $\MITLR$) formula. By~\cite{Alur:1996}, one can build a \TA\ $\Au_\forall$ accepting $\TLang(\varphi_\forall)$ having 
\begin{itemize}
    \item $2^{O(K_P\cdot |R|\cdot N_A \cdot N_{\mathcal{E}}\cdot (|\Intv_R| + N_v^2))} $ states, $O(K_P\cdot |R|\cdot N_A \cdot N_{\mathcal{E}}\cdot (|\Intv_R| + N_v^2))$ clocks
    \item (resp., $2^{O(|R|\cdot N_A \cdot N_{\mathcal{E}}\cdot (|\Intv_R| + N_v^2))}$ states, $O(|R|\cdot N_A \cdot N_{\mathcal{E}}\cdot (|\Intv_R| + N_v^2))$ clocks),
\end{itemize}
and maximal constant $O(K_P)$.

Non-emptiness of a \TA\ $\Au$ can be solved by an $\NPsp=\Psp$ search algorithm over the \emph{region automaton} of $\Au$,%
\footnote{The \emph{region automaton} of $\mathcal{A}$ features states of the form $(q,r)$, 
where $q$ is a state of $\mathcal{A}$ and $r$ a \emph{region}: every region specifies, for each clock $c$ of $\mathcal{A}$, whether its value is integer or not (and, if it is, its value up to $K_c$, the maximum constant to which $c$ is compared), and the ordering of the fractional parts of the clocks.}
 which uses work space \emph{logarithmic} in the number of control 
states of $\Au$ and \emph{polynomial}
in the number of clocks and in the length of the encoding of the maximal constant of $\Au$~\cite{ALUR1994183}.
  Thus, since $\Au_P$, $\Au_\forall$, and the intersection $\Au_\wedge$ of $\Au_P$ and $\Au_\forall$ can be constructed on the fly---that is, by looking at their transition relations $\Delta$, one can determine, given a state $q$, a successor $q'$ and the connecting transition, along with the associated constraints and clocks to reset---and the search in the region automaton of
$\Au_\wedge$ can be done without explicitly constructing $\Au_\wedge$, the result follows.
\end{proof} 

In the next section, we consider future TP with simple trigger rules and non-singular intervals in the atoms of trigger rules (resp., intervals in $\Intv_{(0,\infty)}$), and prove a matching complexity \emph{lower bound}: $\EXPSPACE$-completeness (resp., $\Psp$-completeness) of the problem follows.

%% file: Chaps/Timelines/PSPACEhard.tex
\section[Future TP, simple trigger rules, non-singular intervals: hardness]{Future TP with simple trigger rules and non-singular intervals: hardness\label{sec:pspace}}

In this section, we first consider the future TP problem with simple trigger rules and non-singular intervals, and prove that it is $\EXPSPACE$-hard by a polynomial-time reduction from the \emph{domino-tiling problem for grids with rows of single exponential length} (it has already been presented in Section~\ref{sec:BEhard}), which is known to be  $\EXPSPACE$-complete~\cite{harel92}. Since the reduction is standard, we refer the reader to Appendix~\ref{sec:EXPSPHardFutTP} for the details of the construction.

\begin{theorem}\label{theorem:EXPSPlowerBound} 
The future TP problem, even with \emph{one state variable}, with simple trigger rules and non-singular intervals is $\EXPSPACE$-hard (under polynomial-time reductions).
\end{theorem}
By putting together Theorem~\ref{theorem:UpperBounds}, $\EXPSPACE$-completeness follows.

We now focus on the case 
with intervals in $\Intv_{(0,\infty)}$, proving that the problem is $\Psp$-hard (and thus $\Psp$-complete by Theorem~\ref{theorem:UpperBounds}) by reducing periodic SAT to it in polynomial time.

The problem \emph{periodic SAT} is defined as follows~\cite{Pap94}.
We are given a Boolean formula $\varphi$ in \emph{conjunctive normal form},
defined over two sets of variables, $\Gamma=\{x_1,\ldots, x_n\}$ and $\Gamma^{+1}=\{x_1^{+1},\ldots , x_n^{+1}\}$, namely,
\[
    \varphi = \bigwedge_{t=1}^m \Big(\smashoperator[r]{\bigvee_{x\in (\Gamma \cup \Gamma^{+1})\cap L^+_t}}\quad x  \vee \smashoperator[r]{\bigvee_{x\in (\Gamma \cup \Gamma^{+1})\cap L^-_t}}  \neg x\;\Big),
\]
where $m$ is the number of conjuncts of $\varphi$ and, for $1\leq t\leq m$, $L^+_t$ (resp., $L^-_t$) is the set of variables occurring non-negated (resp., negated) in the $t$-th conjunct of $\varphi$.
Moreover, the formula $\varphi^j$, for $j\in\Nat\setminus\{0\}$, is defined as $\varphi$ in which we replace each variable
$x_i\in \Gamma$ by a fresh one $x_i^j$, and $x_i^{+1}\in \Gamma^{+1}$ by $x_i^{j+1}$.
Periodic SAT is then the problem of deciding the satisfiability of the (infinite-length) formula 
\[\Phi= \bigwedge_{j\in\Nat\setminus\{0\}} \varphi^j,\] that is, deciding the existence 
of a truth assignment of (infinitely many) variables $x_i^j$, for $i=1,\ldots, n,\, j\in\Nat\setminus\{0\}$, satisfying $\Phi$.

Periodic SAT is $\Psp$-complete~\cite{Pap94}; in particular membership to such a class is proved by showing that one can equivalently check the satisfiability of the (finite-length) formula $\Phi_f= \bigwedge_{j= 1}^{2^{2n}+1} \varphi^j$. Intuitively, $2^{2n}$ is the number of possible truth assignments to variables of $\Gamma\cup \Gamma^{+1}$, thus, after $2^{2n}+1$ copies of $\varphi$, we can find a repeated assignment: from that point, we can just loop through the previous assignments. 

We now 
reduce periodic SAT to our problem.
Hardness also holds when only a single state variable is involved, and also restricting to intervals of the form $[0,a]$.

\begin{theorem}\label{theorem:PSPlowerBound} 
The future TP problem, even with \emph{one state variable}, with simple trigger rules and intervals $[0,a]$, $a\in\Nat\setminus\{0\}$, is $\Psp$-hard  (under polynomial-time reductions).
\end{theorem}
\begin{proof}
Let us define the state variable $y=(V,T,D)$, where 
\begin{itemize}
    \item $V=\{\$,\tilde{\$},stop\}\cup \{x_i^\top,x_i^\bot, \tilde{x_i}^\top, \tilde{x_i}^\bot \mid i=1,\ldots, n\}$,
    \item $T(\$)=\{x_1^\top,x_1^\bot\}$, $T(\tilde{\$})=\{\tilde{x_1}^\top, \tilde{x_1}^\bot\}$ and $T(stop)=\{stop\}$,
    \item for $i=1,\ldots, n-1$, $T(x_i^\top)=T(x_i^\bot)=\{x_{i+1}^\top,x_{i+1}^\bot\}$,
    \item for $i=1,\ldots, n-1$, $T(\tilde{x_i}^\top)=T(\tilde{x_i}^\bot)=\{\tilde{x_{i+1}}^\top,\tilde{x_{i+1}}^\bot\}$,
    \item $T(x_n^\top)=T(x_n^\bot)=\{\tilde{\$},stop\}$,
    \item $T(\tilde{x_n}^\top)=T(\tilde{x_n}^\bot)=\{\$,stop\}$, and
    \item for all $v\in\ V$, $D(v)=[2,+\infty[$.
\end{itemize}
Intuitively, we represent an assignment of variables $x_i^j$ by means of a timeline for $y$:
after every occurrence of the symbol $\$$, $n$ tokens are present, one for each $x_i$, and the value $x_i^\top$ (resp., $x_i^\bot$) represents a positive (resp., negative) assignment of $x_i^j$, for some \emph{odd} $j\geq 1$. Then, there is an occurrence of $\tilde{\$}$, after which $n$ more tokens occur, again one for each $x_i$, and the value $\tilde{x_i}^\top$ (resp., $\tilde{x_i}^\bot$) represents a positive (resp., negative) assignment of $x_i^j$, for some \emph{even} $j\geq 2$.
See Figure~\ref{fig:phij} for an example.
\begin{figure}[t]
    \centering
    \includegraphics[width=\textwidth]{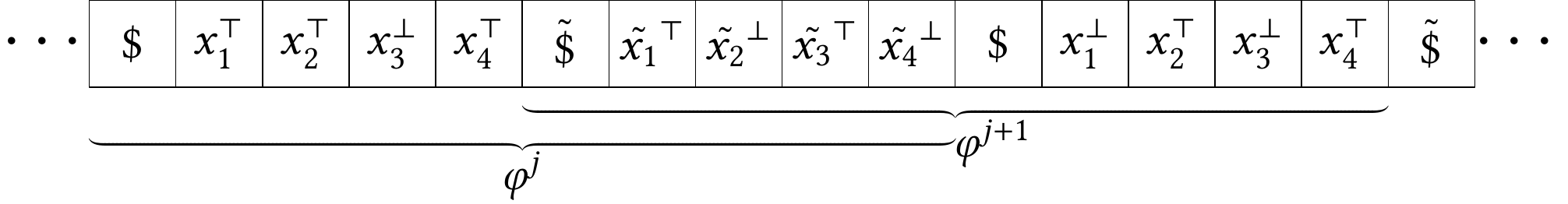}
    \caption{Let
    the formula $\varphi$ be defined over two sets of variables, $\Gamma=\{x_1,x_2,x_3,x_4\}$ and $\Gamma^{+1}=\{x_1^{+1},x_2^{+1},x_3^{+1},x_4^{+1}\}$. 
    The $j$-th copy (we assume $j$ is odd) of $\varphi$, i.e., $\varphi^j$, is satisfied by the assignment $x_1^j\mapsto \top$, $x_2^j\mapsto \top$, $x_3^j\mapsto \bot$, $x_4^j\mapsto \top$, $x_1^{j+1}\mapsto \top$, $x_2^{j+1}\mapsto \bot$, $x_3^{j+1}\mapsto \top$, $x_4^{j+1}\mapsto \bot$. The analogous for $\varphi^{j+1}$. }
    \label{fig:phij}
\end{figure}

We start with the next simple trigger rules, one for each $v\in V$: 
\[o[y=v]\to o\leq^{\mathsf{s},\mathsf{e}}_{[0,2]} o.\]
Paired with the constraint function $D$, they enforce
all tokens' durations to be \emph{exactly} 2: intuitively, since we exclude singular intervals, requiring, for instance, that a token $o'$ starts  $t$ instants of time after the end of $o$, with $t\in [\ell,\ell+1]$ and even $\ell\in\Nat$, boils down to $o'$ starting \emph{exactly} $\ell$ instants after the end of $o$. We also observe that, given the constant token duration, the density of the time domain does not play any role in this proof.

We now add the next rules:
\begin{itemize}
\item
    $\true \to \exists o[y=\$]. o\geq^{\mathsf{s}}_{[0,1]} 0$;
\item
    $\true \to \exists o[y=\tilde{\$}]. o\geq^{\mathsf{s}}_{[0,1]} (2^{2n}+1)\cdot 2(n+1)$;
\item
    $\true \to \exists o[y=stop]. o\geq^{\mathsf{s}}_{[0,1]} (2^{2n}+2)\cdot 2(n+1)$.
\end{itemize}
They respectively impose that $(i)$~a token with value $\$$ starts exactly at $t=0$ (recall that the duration of every token is 2);
$(ii)$~there exists a token with value $\tilde{\$}$ starting at $t=(2^{2n}+1)\cdot 2(n+1)$; 
$(iii)$~a token with value $stop$ starts at $t=(2^{2n}+2)\cdot 2(n+1)$. 
We are forcing the timeline to encode truth assignments for variables $x_1^1,\ldots, x_n^1,\ldots, x_1^{2^{2n}+2} ,\ldots , x_n^{2^{2n}+2}$: as a matter of fact, we will decide satisfiability of the finite formula $\Phi_f= \bigwedge_{j= 1}^{2^{2n}+1} \varphi^j$, which is equivalent to $\Phi$.


We now consider the next rules, that enforce the satisfaction of each $\varphi^j$ or,
equivalently, of $\varphi$ over the assignments of $(x_1^j,\ldots, x_n^j, x_1^{j+1},\ldots, x_n^{j+1})$.

For the $t$-th conjunct of $\varphi$, 
we define the future simple rule:
\begin{multline*}
    o[y=\tilde{\$}]\to \\
    \Big(\smashoperator{\bigvee_{x_i\in \Gamma\cap L^+_t}} \exists o'[y=\tilde{x_i}^\top]. o\leq ^{\mathsf{e},\mathsf{s}}_{[0,4n]} o' \Big) \vee 
        \Big(\smashoperator{\bigvee_{x_i^{+1}\in \Gamma^{+1}\cap L^+_t}} \exists o'[y=x_i^\top]. o\leq ^{\mathsf{e},\mathsf{s}}_{[0,4n]} o' \Big) \vee \\
        \Big(\smashoperator{\bigvee_{x_i\in \Gamma\cap L^-_t}} \exists o'[y=\tilde{x_i}^\bot]. o\leq ^{\mathsf{e},\mathsf{s}}_{[0,4n]} o' \Big) \vee 
        \Big(\smashoperator{\bigvee_{x_i^{+1}\in \Gamma^{+1}\cap L^-_t}} \exists o'[y=x_i^\bot]. o\leq ^{\mathsf{e},\mathsf{s}}_{[0,4n]} o' \Big) \vee \\
        \exists o''[y=stop]. o\leq ^{\mathsf{e},\mathsf{s}}_{[0,2n]} o''.
\end{multline*}
Basically, this rule (the rule where the trigger has value $\$$ being analogous) states that, after every occurrence of $\tilde{\$}$, a token $o'$, making true at least a (positive or negative) literal in the conjunct, must occur by $4n$ time instants (i.e., before the following occurrence of $\tilde{\$}$).
The disjunct $\exists o''[y=stop]. o\leq ^{\mathsf{e},\mathsf{s}}_{[0,2n]} o'' $ is present just to avoid evaluating $\varphi$ on the $n$ tokens before (the first occurrence of) $stop$.

The variable $y$ and all synchronization rules can be generated in time polynomial in $|\varphi|$ (in particular, all interval bounds and time constants of time-point atoms have a value, encoded in binary, in $O(2^{2n})$).
\end{proof}

By Theorem~\ref{theorem:UpperBounds} and Theorem~\ref{theorem:PSPlowerBound}, $\Psp$-completeness of 
future TP with simple trigger rules and intervals in $\Intv_{(0,\infty)}$ follows.

In the next section we focus on a different restriction of the TP problem, which will allow us to devise a $\NP$ planning algorithm for it.

%% file: Chaps/Timelines/TriggerlessNPcomplete.tex
\section{TP with trigger-less rules only is \NP-complete}\label{sec:NPtriggerless}

In this section we describe a TP algorithm,
for planning domains where only trigger-less rules are allowed,
which requires a polynomial number of (non-deterministic) computation steps.
We recall that trigger-less rules are useful, for instance, to express initial, intermediate conditions and reachability goals.

We want to start with the following example, with which we highlight that there is no \emph{polynomial-size} plan for some problem instances/domains. Thus, an explicit enumeration of all tokens of a multi-timeline \emph{does not} represent a suitable polynomial-size certificate.

\begin{example}
Let us consider the following planning domain.
We denote by $p(i)$ the $i$-th prime number, assuming $p(1)=1$, $p(2)=2$, $p(3)=3$, $p(4)=5$,\dots .
We define, for $i=1,\ldots,n$, the state variables $x_i=(\{v_i\},\{(v_i,v_i)\},D_{x_i})$ with $D_{x_i}(v_i)=\mathopen[p(i),p(i)\mathclose]$.
The following rule
\[
\true\to \exists o_1[x_1=v_1] \cdots \exists o_n[x_n=v_n].\bigwedge_{i=1}^{n-1} o_i\leq_{[0,0]}^{\mathsf{e},\mathsf{e}} o_{i+1}
\]
is asking for the existence of a \lq\lq synchronization point\rq\rq , where $n$ tokens (one for each variable) have their ends aligned.
Due to the allowed token durations, the first such time point is $\prod_{i=1}^{n} p(i)\geq 2^{n-1}$.
Hence, in any plan, the timeline for $x_1$ features at least $2^{n-1}$ tokens: no explicit polynomial-time enumeration of such tokens is possible.
\end{example}
As a consequence, there exists no trivial guess-and-check \NP\ algorithm.
Conversely, one can easily prove the following result.
\begin{theorem}
The TP problem with trigger-less rules only is \NP-hard, even with one state variable (under polynomial-time reductions).
\end{theorem}
\begin{proof}
There is a trivial reduction from the problem of the existence of a Hamiltonian path in a directed graph.

Given a directed graph $G=(V,E)$, with $|V|=n$, 
we define the state variable $x=(V,E,D_x)$, where $D_x(v)=[1,1]$ for each $v\in V$.
We add the following trigger-less rules, one for each $v\in V$:
\[
\true \to \exists o[x=v]. o\geq ^{\mathsf s}_{[0,n-1]} 0 .
\]
The rule for $v\in V$ requires that there is a token $(x,v,1)$ along the timeline for $x$, which starts no later than $n-1$.
It is easy to check that $G$ contains a Hamiltonian path if and only if there exists a plan for the defined planning domain.
\end{proof}

We now present the aforementioned non-deterministic polynomial-time algorithm, proving that timeline-based planning with trigger-less rules is in \NP.

We preliminarily have to derive a \emph{finite horizon} (namely, the end time of the last token) for the plans of a (any) instance of TP with trigger-less rules. That is, 
if an instance $P=(SV,R)$ admits a plan, then $P$ also has a plan whose horizon is no greater than a given bound. Analogously, we have to calculate a \emph{bound to the maximum number of tokens} in a plan.
Both can be obtained from 
the constructions of the \TA s described in the proof of 
Theorem~\ref{theorem:UpperBounds}:
since only trigger-less rules are now allowed,
we disregard the construction of the \MTL\ formula $\varphi_\forall$,
and restrict our attention to the \TA\  
$\Au_P$ (i.e., the intersection between $\Au_{SV}$ for the state variables in $SV$ from Proposition~\ref{prop:AtutomataForMultiTimeline} and $\Au_{_\exists}$ for the trigger-less rules in $R$ from Proposition~\ref{prop:TATriggerLessRules}), which has $\alpha_s=2^{O(N_q+\sum_{x\in SV} |V_x|)}$ states, $\alpha_c=O(N_q+|SV|)$ clocks and maximum constant $\alpha_K=O(K_p)$, where
$N_q$  is the overall number of quantifiers   in the trigger-less  rules of $R$, 
and accepts all and only the encodings $w_\Pi$ of multi-timelines $\Pi$ of $SV$ satisfying all the trigger-less rules in $R$.

The language emptiness checking algorithm for \TA s executed over $\Au_P$ visits the (untimed) region automaton for $\Au_P$~\cite{ALUR1994183},
which features $\alpha=\alpha_s\cdot O(\alpha_c!\cdot 2^{\alpha_c}\cdot 2^{2 N_q^2}\cdot (2\alpha_K +2)^{\alpha_c})$
states%
\footnote{The factor $2^{2 N_q^2}$ is present due to diagonal clock constraints in $\Au_P$.}%
, trying to find a path, from the initial state to a final state, whose length can clearly be bounded by the number of states.
We observe that each edge/transition of the region automaton in such a path corresponds, in the worst case, to the start point of a token for each timeline for the variables in $SV$ (i.e., assuming that all these tokens start simultaneously).
This yields a bound on the number of tokens, which is $\alpha \cdot |SV|$.
We can also derive a bound on the horizon of the plan, which is $\alpha \cdot |SV| \cdot (\alpha_K+1)$, as every transition taken in $\Au_P$ may let at most $\alpha_K+1$ time units pass, as $\alpha_K$ accounts in particular for the maximum constant to which a (any) clock is compared.\footnote{Clearly, and unbounded quantity of time units may pass, but after $\alpha_K+1$ the last region of the region automaton will certainly have been reached.}

Having this pair of bounds, 
we are now ready to describe the two main phases of the algorithm, corresponding to the following pair of observations.
On the one hand, $(i)$~each trigger-less rule requires, as we said,
the existence of an \emph{a priori bounded number} of temporal events satisfying mutual temporal relations (namely, in the worst case, the start time and end time of all tokens associated with the quantifiers of one of its existential statements).
On the other hand, $(ii)$~timelines for different state variables evolve independently of each other.
In order to deal with $(i)$, we non-deterministically position such temporal events along timelines; as for $(ii)$, we enforce a correct evolution of each timeline between pairs of \lq\lq positioned\rq\rq\ events, completely independently of the other timelines.



\paragraph{Non-deterministic token positioning.}
The algorithm starts by non-determin\-istically selecting, for every trigger-less rule in $R$, a disjunct---and deleting all the others. Then, for every (left) quantifier $o_i[x_i=v_i]$, it generates the integer part of both the start and the end time of the token for $x_i$ to which $o_i$ is mapped. We call such time instants, respectively, $\mathsf{s}_{int}(o_i)$ and $\mathsf{e}_{int}(o_i)$.\footnote{We can assume w.l.o.g.\ that all quantifiers refer to distinct tokens. As a matter of fact, the algorithm can non-deterministically choose to make two (or more) quantifiers $o_i[x_i=v_i]$ and $o_j[x_i=v_i]$ over the same variable and value \lq\lq collapse\rq\rq\ to the same token just by rewriting all occurrences of $o_j$ as $o_i$ in the atoms of the rules.} We observe that all start/end time $\mathsf{s}_{int}(o_i)$ and $\mathsf{e}_{int}(o_i)$, being less or equal to $\alpha \cdot |SV| \cdot (\alpha_K+1)$ (the finite horizon bound), have an integer part that can be encoded with polynomially many bits (and thus can be generated in polynomial time). 

Let us now consider the fractional parts of the start/end time of the tokens associated with quantifiers. We denote them by $\mathsf{s}_{frac}(o_i)$ and $\mathsf{e}_{frac}(o_i)$. The algorithm non-deterministically generates an \emph{order} of all such fractional parts. In particular we have to specify, for every token start/end time, whether it is integer ($\mathsf{s}_{frac}(o_i)=0$, $\mathsf{e}_{frac}(o_i)=0$) or not ($\mathsf{s}_{frac}(o_i)>0$, $\mathsf{e}_{frac}(o_i)>0$).
Every such possibility can be generated in polynomial time.

Some trivial tests should now be performed, namely that, for all $o_i$, $\mathsf{s}_{int}(o_i)\leq \mathsf{e}_{int}(o_i)$, each token is assigned an end time equal or greater than its start time, and no two tokens for the same variable are overlapping.

It is routine to check that, if we change the start/end time of (some of the) tokens associated with quantifiers, 
but we leave unchanged $(i)$~all the integer parts, $(ii)$~zeroness/non-zeroness of fractional parts, and $(iii)$~the fractional parts' order,
then the satisfaction of the (atoms in the) trigger-less rules does not change. This is due to all the constants being integers.%
\footnote{We may observe that, by leaving unchanged all the integer parts and the fractional parts' order, the region of the region graph of the timed automaton does not change.} Therefore we can now check whether all rules are satisfied.

\paragraph{Enforcing legal token durations and timeline evolutions.}

We now continue by checking that: $(i)$ all tokens associated with a quantifier have a legal duration, and that $(ii)$ there exists a legal timeline evolution between pairs of adjacent such tokens over the same variable (here \emph{adjacent} means that there is no other token associated with a quantifier in between). 
We will enforce all these requirements as constraints of a \emph{linear problem}, which can be solved in deterministic polynomial time (e.g., using the ellipsoid algorithm).
When needed, we use \emph{strict inequalities}, which are not allowed in linear programs. We shall show later how to convert these into non-strict ones.

We start by associating non-negative variables $\alpha_{o_i,s}, \alpha_{o_i,e}$ with the fractional parts of the start/end times $\mathsf{s}_{frac}(o_i)$, $\mathsf{e}_{frac}(o_i)$ of every token for a quantifier $o_i[x_i=v_i]$.
First, we add the linear constraints
\begin{equation*}
    0\leq \alpha_{o_i,s}<1,\quad 0\leq \alpha_{o_i,e}<1.
\end{equation*}
Then, we also need to enforce that the values of $\alpha_{o_i,s}, \alpha_{o_i,e}$ respect the decided order of the fractional parts: for example,
\begin{equation*}
    0=\alpha_{o_i,s}=\alpha_{o_j,s}<\alpha_{o_k,s}<\ldots <\alpha_{o_j,e}<\alpha_{o_i,e}=\alpha_{o_k,e}<\ldots
\end{equation*}
To enforce requirement $(i)$, we set, for all $o_i[x_i=v_i]$,
\begin{equation*}
    a\leq (\mathsf{e}_{int}(o_i)+\alpha_{o_i,e})-(\mathsf{s}_{int}(o_i)+\alpha_{o_i,s})\leq b
\end{equation*}
where $D_{x_i}(v_i)=\mathopen[a,b\mathclose]$. Clearly, strict ($<$) inequalities must be used for a left/right open interval.

To enforce requirement $(ii)$, namely that there exists a legal timeline evolution between each pair of adjacent tokens for the same state variable, say $o_i[x_i=v_i]$ and $o_j[x_i=v_j]$, we proceed as follows (for a correct evolution between $t=0$ and the first token, analogous considerations can be made).

Let us consider each state variable $x_i=(V_i,T_i,D_i)$ as a directed graph $G=(V_i,T_i)$ where 
$D_i$ is a function associating with each vertex $v\in V_i$ a duration range.
We have to decide whether or not there exist
\begin{itemize}
    \item a path in $G$, possibly with repeated vertices and edges, $v_0 \cdot v_1 \cdots v_{n-1}\cdot v_n$, where $v_0\in T_i(v_i)$ and $v_n$ with $v_j\in T_i(v_n)$ are non-deterministically generated, and
    \item a list of non-negative real values $d_0,\ldots,d_n$,
such that 
\[\sum_{t=0}^n d_t = (\mathsf{s}_{int}(o_j)+\alpha_{o_j,s}) - (\mathsf{e}_{int}(o_i)+\alpha_{o_i,e}),\]
and for all $s=0,\ldots, n$, $d_s\in D_i(v_s)$.
\end{itemize}


We guess 
a set of integers $\{\alpha'_{u,v}\mid (u,v)\in T_i\}$.
Intuitively, $\alpha'_{u,v}$ is the number of times the solution path
traverses $(u,v)$. Since every time an edge is traversed a new token starts, each $\alpha'_{u,v}$ is bounded by the number of tokens, i.e., by $\alpha \cdot |SV|$. Hence the binary encoding of $\alpha'_{u,v}$ can be generated in polynomial time.

We then perform the following deterministic steps.
\begin{enumerate}
\item We consider the subset $E'$ of edges of $G$, $E'=\{(u,v)\in T_i\mid \alpha'_{u,v}>0\}$. We check whether $E'$ induces a strongly (undirected) connected subgraph of $G$.
\item We check whether 
    \begin{itemize}
        \item $\sum_{(u,v)\in E'} \alpha'_{u,v}=\sum_{(v,w)\in E'} \alpha'_{v,w}$, for all $v \in V_i\setminus\{v_0,v_n\}$;
        \item $\sum_{(u,v_0)\in E'} \alpha'_{u,v_0}=\sum_{(v_0,w)\in E'} \alpha'_{v_0,w}-1$;
        \item $\sum_{(u,v_n)\in E'} \alpha'_{u,v_n}=\sum_{(v_n,w)\in E'} \alpha'_{v_n,w}+1$.
    \end{itemize}

\item For all $v \in V_i\setminus\{v_0\}$, we define $y_v=\sum_{(u,v)\in E'} \alpha'_{u,v}$ ($y_v$ is the number of times the solution path gets into $v$). Moreover, 
$y_{v_0} = \sum_{(v_0,u)\in E'} \alpha'_{v_0,u}$.
\item We define the real non-negative variables $z_v$, for every $v \in V_i$ ($z_v$ is the total waiting time of the path on the node $v$), subject to the following constraints:
\[a\cdot y_v \leq z_v \leq b\cdot y_v,\]
where $D_i(v)=[a,b]$ (an analogous constraint should be written for open intervals). Finally we set:
\[\sum_{v \in V_i} z_v = (\mathsf{s}_{int}(o_j)+\alpha_{o_j,s}) - (\mathsf{e}_{int}(o_i)+\alpha_{o_i,e}).\]
\end{enumerate}

Steps (1.) and (2.) together check that the values $\alpha'_{u,v}$ for the arcs specify
a directed Eulerian path from $v_0$ to $v_n$ in a multigraph. Indeed,
the following theorem holds.
\begin{theorem}\cite{Jung}
Let $G'=(V',E')$ be a directed multigraph ($E'$ is a multiset).
$G'$ has a (directed) Eulerian path from $v_0$ to $v_n$ if and only if:
\begin{itemize}
    \item the undirected version of $G'$ is connected, and
    \item $|\{(u,v)\in E'\}| =| \{(v,w)\in E'\}|$, for all $v \in V'\setminus\{v_0,v_n\}$;
    \item $|\{(u,v_0)\in E'\}|=|\{(v_0,w)\in E'\}|-1$;
    \item $|\{(u,v_n)\in E'\}|=|\{(v_n,w)\in E'\}|+1$.
\end{itemize}
\end{theorem}

Steps (3.) and (4.) evaluate the waiting times of the path in some
vertex $v$ with duration interval $\mathopen[a,b\mathclose]$.
If the solution path visits the vertex $y_v$ times, then every single
visit must take at least $a$ and at most $b$ units of time.
Hence the overall visitation time is in between $a\cdot y_v$ and
$b\cdot y_v$.
Vice versa, if the total visitation time is in between $a\cdot y_v$ and
 $b\cdot y_v$, then it can be slit into $y_v$ intervals, each one falling into $\mathopen[a,b\mathclose]$. 

The algorithm concludes by solving the linear program given by the variables $\alpha_{o_i,s}$ and $\alpha_{o_i,e}$ for each quantifier $o_i[x_i=v_i]$, and for each pair of adjacent tokens in the same timeline for $x_i$, for each $v\in V_i$, the variables $z_v$ subject to their constraints.

Finally, in order to conform to linear programming, we have to replace all strict inequalities with non-strict ones.
It is straightforward to observe that all constraints involving strict inequalities we have written so far are of
(or can easily be converted into) the following forms: 
$\xi s<\eta q+k$ or $\xi s>\eta q+k$, where $s$ and $q$ are variables, and $\xi$, $\eta$, $k$ are constants.
We replace them, respectively, by $\xi s-\eta q-k+\beta_t\leq 0$ and $\xi s-\eta q-k-\beta_t\geq 0$, where $\beta_t$ is an additional fresh non-negative variable, which is \emph{local} to a single constraint. 
We observe that the original inequality and the new one are equivalent if and only if $\beta_t$ is a small enough \emph{positive} number.
Moreover, we add another non-negative variable, say $r$, which is subject to a constraint $r\leq \beta_t$, for each of the introduced variables $\beta_t$ (i.e., $r$ is less than or equal to the minimum of all $\beta_t$'s). Finally, we maximize the value of $r$ when solving the linear program. We have that $\max r>0$ if and only if there is an admissible solution where the values of all $\beta_t$'s are positive (and thus the original strict inequalities hold true). 

This ends the description of the planning algorithm. We can thus conclude the section with the main result.
\begin{theorem}
The TP problem with trigger-less rules only is \NP-complete.
\end{theorem}

In the next section, using the results on the variants of the TP problem, we shall study MC over timelines.

%% file: Chaps/Timelines/MCTimelines.tex
\section{MC for \MITL\ over timelines}\label{sec:modelcheckingTimelines}

In this section we deal with the ultimate goal of the present chapter:
as mentioned, we want to model check systems specified in terms of timelines. More precisely, 
a system is described as a set of state variables along with a set of synchronization rules over them (a TP domain) $P=(SV,R)$.
As mentioned in the introduction, the property specification language we will be assuming is the logic \MITL.

We first recall the encoding of multi-timelines already adopted in Section~\ref{sec:Reduction}, 
over which we interpret \MITL,
that exploits 
the set $\Prop$ of proposition letters
\[
\Prop = \bigcup_{x\in SV}\Main_x \cup \Deriv ,
\] 
where in particular
\[
\Main_x = ((\{\Beg_x \}\cup V_x) \times V_x)   \cup   (V_x \times \{\End_x\}).
\]
The tags $\Beg_x$ and $\End_x$ mark the beginning and the end of a timeline for $x$, and a pair $(v,v')\in \Main_x$ represents a 
transition of the value taken by $x$ from $v$ to $v'$ (a token for $x$ with value $v'$ follows a token with value $v$).
The already introduced formula 
\[
\psi(\start,v)= (\Beg_x,v)\vee \displaystyle{\bigvee_{u\in V_x}}(u,v),
\]
states that a start-event for a token for $x$ with value $v$ occurs at the current time.
Finally, $\EqTime(\theta) = \theta \vee [\neg p_> \StrictUntil_{\geq 0}(\neg p_> \wedge \theta)]$, where $p_>\in\Deriv$, is satisfied
by an encoding of a multi-timeline at the current time if $\theta$ eventually holds at a position whose timestamp coincides with the current one.

Before formalizing the \emph{MC problem for \MITL\ formulas over timelines}, we want to start with an easy example of a system whose components are described by timelines, over which we check some properties encoded by \MITL\ formulas.

\begin{figure}[p]
    \centering
    \resizebox{!}{0.9\textheight}{\rotatebox{90}{\includegraphics{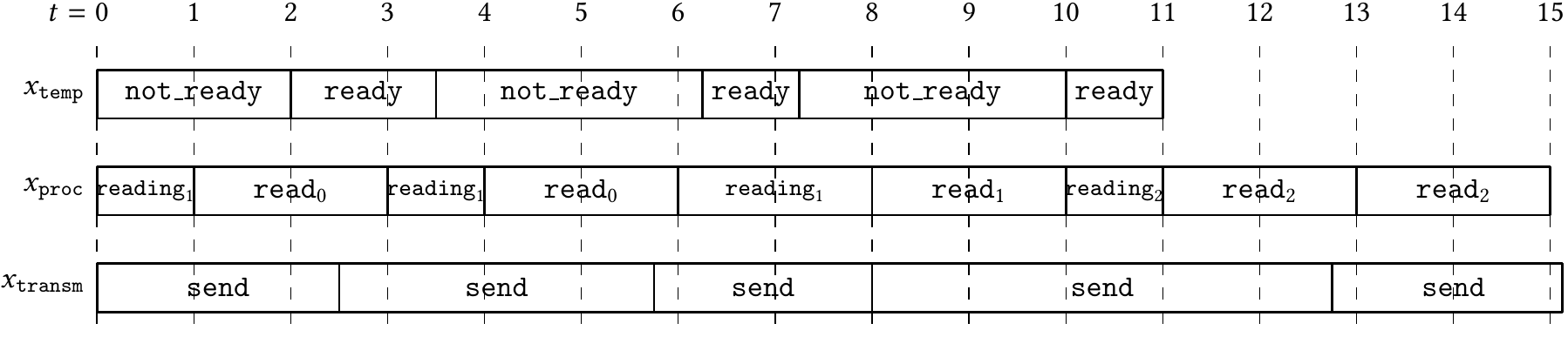}}}
    \caption{Example of computation for the defined system}
    \label{fig:explan}
\end{figure}

\begin{example}
Let us consider the following system, consisting of a \emph{temperature sensor}, a \emph{processing unit}, and a \emph{data transmission unit}.
These components are modelled by three state variables, $x_\mathtt{temp}=(V_\mathtt{temp},T_\mathtt{temp},D_\mathtt{temp})$, $x_\mathtt{proc}=(V_\mathtt{proc},T_\mathtt{proc},D_\mathtt{proc})$ and $x_\mathtt{transm}=(V_\mathtt{transm},T_\mathtt{transm},D_\mathtt{transm})$, where
\begin{itemize}
    \item $V_\mathtt{temp}=\{\mathtt{ready}, \mathtt{not\_ready}\}$, \\
    $T_\mathtt{temp}(\mathtt{ready})=\{\mathtt{not\_ready}\}$, $T_\mathtt{temp}(\mathtt{not\_ready})=\{\mathtt{ready}\}$, \\ $D_\mathtt{temp}(\mathtt{ready})=[1,2]$, $D_\mathtt{temp}(\mathtt{not\_ready})=[2,3]$;
    \item $V_\mathtt{proc}=\{\mathtt{reading}_1,\mathtt{reading}_2,\mathtt{read}_0,\mathtt{read}_1,\mathtt{read}_2\}$, \\
    $T_\mathtt{proc}(\mathtt{reading}_1)=\{\mathtt{read}_0,\mathtt{read}_1\}$,
    $T_\mathtt{proc}(\mathtt{reading}_2)=\{\mathtt{read}_1,\allowbreak \mathtt{read}_2\}$,
    $T_\mathtt{proc}(\mathtt{read}_0)=\{\mathtt{reading}_1\}$, $T_\mathtt{proc}(\mathtt{read}_1)=\{\mathtt{reading}_2\}$,
    $T_\mathtt{proc}(\mathtt{read}_2)=\{\mathtt{read}_2\}$, \\
    $D_\mathtt{proc}(\mathtt{reading}_1)=D_\mathtt{proc}(\mathtt{reading}_2)=[1,2]$, $D_\mathtt{proc}(\mathtt{read}_0)=D_\mathtt{proc}(\mathtt{read}_1)=D_\mathtt{proc}(\mathtt{read}_2)=[2,3]$;
    \item $V_\mathtt{transm}=\{\mathtt{send}\}$,\\
    $T_\mathtt{transm}(\mathtt{send})=\{\mathtt{send}\}$,\\
    $D_\mathtt{transm}(\mathtt{send})=[2,5]$.
\end{itemize}

The temperature sensor alternates between the two states $\mathtt{ready}$ and $\mathtt{not\_ready}$. In the former, it senses the temperature of the environment and \emph{possibly} sends the temperature value to the processing unit.
The purpose of the processing unit is receiving \emph{two} temperature samples from the sensor, and sending the average value to the data transmission unit. While in state $\mathtt{read}_i$, for $i=0,1,2$, it has read $i$ samples. While in $\mathtt{reading}_j$, for $j=1,2$, it is attempting to read the $j$-th sample. A reading is possible only if the sensor and the processing unit are synchronized: a time interval (token) with value $\mathtt{reading}_j$ has to \emph{contain} (Allen relation, see Table~\ref{allen}) a token $\mathtt{ready}$. 
Analogously, the processing unit can send data to the transmitter only if a token with value $\mathtt{send}$ contains one with value $\mathtt{read}_2$.

The sensor starts in state $\mathtt{not\_ready}$. This is specified by the trigger-less rule $\true\to\exists o[x_\mathtt{temp} = \mathtt{not\_ready}]. o\leq^\start_{[0,0]} 0$. The processing unit starts in state $\mathtt{reading}_1$: $\true\to\exists o[x_\mathtt{proc} = \mathtt{reading}_1]. o\leq^\start_{[0,0]} 0$.
(Recall that trigger-less rules may also contain singular intervals at no extra computational cost.)
The goal of the system is encoded by the rule
$\true\to \exists o_1 [x_\mathtt{proc} = \mathtt{read}_2] \exists o_2 [x_\mathtt{transm} = \mathtt{send}]. (o_2 \leq^{\start,\start}_{\mathopen[0,+\infty\mathclose[} o_1 \wedge o_1 \leq^{\Ending,\Ending}_{\mathopen[0,+\infty\mathclose[} o_2)$.

Let us now encode the fact that the sensor and the processing unit must be synchronized for the latter to receive a temperature sample.
We assume the next simple trigger rule to be \emph{interpreted under the future semantics}.
\begin{multline}\label{eq:sync}
o[x_\mathtt{proc} = \mathtt{reading}_1] \to (\exists o_1[x_\mathtt{proc} = \mathtt{read}_0]. o \leq_{[0,1]}^{\Ending,\start} o_1) \vee\\ (\exists o_2[x_\mathtt{proc} = \mathtt{read}_1] \exists o_3[x_\mathtt{temp} = \mathtt{ready}]. o \leq_{[0,1]}^{\Ending,\start} o_2\wedge o_3\leq_{\mathopen[0,+\infty\mathclose[}^{\Ending,\Ending} o).
\end{multline}
Let us observe that, due to the future semantics, the token (referenced by the name) $o$ starts no later than $o_3$: this with the interval atom \mbox{$o_3\leq_{\mathopen[0,+\infty\mathclose[}^{\Ending,\Ending} o$} force the \emph{contains} Allen relation between $o$ and $o_3$.
An analogous rule can be written for the second temperature sample (where $x_\mathtt{proc}=\mathtt{reading}_2$).

In Figure~\ref{fig:explan} we show an example of plan/computation for the system described by $P=(\{x_\mathtt{temp},x_\mathtt{proc},x_\mathtt{transm}\},R)$.

Let us now specify some properties in \MITL\ (more precisely, $\MITLR$) to check on the system model. The idea is that such properties must hold true over \emph{all possible computations (plans)} of the described system, in order for the MC problem to be satisfied.

\begin{itemize}
    \item $\Always_{<2}\;\neg \psi(\start,\mathtt{ready})$. This property holds true in any system computation, as the sensor does not ever get ready by 2 seconds;
    \item $\Eventually_{\leq 8}\; \psi(\start,\mathtt{read}_1)$. This property is not true in all computations (but it is, e.g., in the one of Figure~\ref{fig:explan}), because the sensor and the processing unit may synchronize for the first time after 8 seconds;
    \item $\Eventually_{\geq 0}\big(\psi(\start,\mathtt{ready}) \wedge (\true\until_{>0}\;\psi(\start,\mathtt{ready}))\big)$. This property holds true in any system computation, since the system guarantees, after some time, to eventually send the data via the transmitter. In order for this to happen, the sensor must become ready (at least) twice.
    \item $\Always_{\geq 0} \big(\psi(\start,\mathtt{read}_1) \to \Eventually_{\leq 3}\; \psi(\start,\mathtt{read}_2)\big)$. 
    This property is not true in all computations as the processing unit, after reading the first sample, may not be able to read the second one by 3 time units (e.g., when the transmitter and the processing unit do not synchronize as soon as possible).
    \item {\small $\Always_{\geq 0} \big(\psi(\start,\mathtt{reading}_1) \!\wedge\! (\EqTime(\psi(\start,\mathtt{ready}))\!\vee\! \Past_\mathtt{ready}^{\start}) \!\to\! \Eventually_{\leq 2}\, \psi(\start,\mathtt{read}_1) \big)$}. 
    We recall that the proposition letter $\Past_\mathtt{ready}^{\start}$ is true at the time it is interpreted if there is a past token for $x_\mathtt{temp}$ with value $\mathtt{ready}$ starting at the same timestamp. 
    The formula considers a situation where a token with value $\mathtt{reading}_1$ starts together with a token $\mathtt{ready}$. 
    The property expressed by the formula is not true in general, as either 
    $(i)$~the token $\mathtt{reading}_1$ may not contain the token $\mathtt{ready}$, hence $x_\mathtt{proc}$ will not move to the state $\mathtt{read}_1$ by 2 time units, or
    $(ii)$~the token $\mathtt{reading}_1$ is followed by a token $\mathtt{read}_0$. As for the latter case,
    the system description rule (\ref{eq:sync}) states that if there is a transition from state $\mathtt{reading}_1$ to $\mathtt{read}_1$, then the processing unit and the sensors must have synchronized. However, the converse implication need not hold: the two component may fail to communicate anyway (the processing unit remaining in $\mathtt{read}_0$).
\end{itemize}
\end{example}

Let us now formally define the MC problem for \MITL\ formulas over timelines.
As shown in the proof of Theorem~\ref{theorem:UpperBounds}, given a system model $P_\text{sys}=(SV,R)$, it is possible to build a \TA\ $\mathcal{A}_\text{sys}$ that
accepts all and only the encodings $w_\Pi$ of multi-timelines $\Pi$ of $SV$ satisfying all the rules in $R$.

\begin{definition}[Model checking]
Given a system model $P_\text{sys}=(SV,R)$ and a \MITL\ formula $\varphi$ over $\Prop$, the MC problem for \MITL\ formulas over timelines is to decide whether or not $\TLang(\mathcal{A}_\text{sys})\subseteq \TLang(\varphi)$.
\end{definition}

We recall that~\cite{Alur:1996} 
given a \MITL\ (resp., $\MITLR$) formula $\psi$, where $N$ is the number of distinct subformulas of $\psi$, and $K$ the largest integer constant appearing in $\psi$, we can build a \TA\ $\mathcal{A}_\psi$ accepting the models of $\psi$, with $O(2^{N\cdot K})$ (resp., $O(2^{N})$) states, $O(N\cdot K)$ (resp., $O(N)$) clocks, and maximum constant $O(K)$. Deciding its emptiness requires
space \emph{logarithmic} in the number of states of $\mathcal{A}_\psi$ and \emph{polynomial}
in the number of clocks and in the length of the encoding of $K$, hence
exponential (resp., polynomial) space.

To decide if $\TLang(\mathcal{A}_\text{sys})\subseteq \TLang(\varphi)$, we check whether $\TLang(\mathcal{A}_\text{sys})\cap \TLang(\mathcal{A}_{\neg \varphi})= \emptyset$ by making the intersection $\mathcal{A}_\wedge$ of 
$\mathcal{A}_\text{sys}$ and $\mathcal{A}_{\neg \varphi}$, and checking for emptiness of its timed language. 
The size of $\mathcal{A}_\wedge$ is polynomial in those of $\mathcal{A}_\text{sys}$ and $ \mathcal{A}_{\neg \varphi}$.
Moreover $\mathcal{A}_\text{sys}$, $\mathcal{A}_{\neg \varphi}$ and $\mathcal{A}_\wedge$ can be built on the fly, and the emptiness test can be done without explicitly constructing them as well.
The next result follows by these observations and by Theorem~\ref{theorem:UpperBounds}.

\begin{theorem}
The MC problem for \MITL\ formulas over timelines, with simple future trigger rules and non-singular intervals, is in \EXPSPACE.

The MC problem for $\MITLR$ formulas over timelines, with simple future trigger rules and intervals in $\Intv_{(0,\infty)}$, is in \Psp.
\end{theorem}

Clearly, \EXPSPACE- and \Psp-completeness of the above MC problems follow by the underlying future TP problems.

%% file: Chaps/Timelines/conclusions.tex
\section{Conclusions}
In this chapter we started by considering the
timeline-based planning problem (TP) over dense temporal domains.
Timelines have been fruitfully used in temporal planning for quite a long time to describe planning domains.
Having recourse to dense time is important for expressiveness: in this way one can express interval-based properties of planning domains, can abstract from unnecessary (or even \lq\lq forced\rq\rq) details often artificially added due to the necessity of discretizing time, and can suitably represent actions with duration, accomplishments and temporally extended goals.
Since TP turns out to be undecidable in its general form, 
we identified and studied \lq\lq intermediate\rq\rq\ decidable cases of the problem, which enforce 
forms of synchronization rules having lower expressive power than that of 
general ones.

The final purpose of the chapter is however exploiting timelines as system descriptions/models, 
which are checked against properties specified in the logic \MITL\ (which replaces $\HS$, in that timed extensions of the latter are not available in literature).
TP is a sort of \lq\lq necessary condition\rq\rq\ for timeline-based MC: 
TP boils down to a feasibility check of the system description; moreover MC can easily be solved once TP has, as both timelines and the property specification language \MITL\ can be translated into  timed automata~\cite{ALUR1994183}, which have been studied for a long time and are at the basis of well-known model checkers (e.g., Uppaal~\cite{UPP}).

\paragraph*{Acknowledgment} 
I would like to sincerely thank Gerhard Woeginger (RWTH Aachen University) for his fundamental contribution to the \lq\lq enforcing legal token durations and timeline evolutions\rq\rq\ phase of the $\NP$ planning algorithm for TP with trigger-less rules.

%% file: Chaps/Concl/conclMain.tex
\chapter{Conclusions}\label{chap:concl}


\lettrine[lines=3]{T}{he main topic} of this thesis is the MC problem for the interval temporal logic $\HS$.
The first results we have established regard $\HS$ under the homogeneous semantics, interpreted over finite Kripke structures: on one hand, we have shown the decidability of the problem 
in nonelementary time, and on the other, its $\EXPSPACE$-hardness.
Then, we have studied many $\HS$ fragments under the homogeneity assumption, proposing, for each of these,
ad-hoc MC algorithms that rest on concepts and techniques different from one another (e.g., small model properties, trace contractions, Boolean circuits with oracles, finite state automata,\ldots). The complexity of the MC problem for them ranges from $\EXPSPACE$ down to $\PSPACE$ and to some of the lowest levels of the polynomial time hierarchy. 
Still under homogeneity, we have studied the expressive power of three semantic variants of $\HS$: we have compared such variants among themselves, and to the standard point-based logics $\LTL$, $\CTL$ and $\CTLStar$.

Homogeneity readily enables us to interpret $\HS$ formulas over Kripke structures, which are inherently point-based models; however, it limits the expressive power of $\HS$. The first way of relaxing such an assumption, while still having Kripke structures as system models, is by adding regular expressions into $\HS$ formulas, in place of vanilla proposition letters. This gives us a means of selecting some traces fulfilling specific propositional patterns, and avoiding all the others. By an automata-theoretic approach, we have shown that MC for full $\HS$ extended with regular expressions retains the same complexity upper and lower bounds as those of the homogeneous case.
As for the analyzed $\HS$ fragments, some of them, when extended with regular expressions, feature an increased computational complexity, while others do not.

Finally, in the last chapter of the thesis
we have replaced Kripke structures by timelines, a more expressive type of model with dense temporal domain, which can capture the interval-based behaviour and properties of systems.
In order to finally deal with timeline-based MC, we have first considered
the timeline-based planning problem,  
identifying suitable restrictions on it, in order to overcome the undecidability of its general formulation.
As for the language for specifying properties of timelines in MC, we have adopted the logic \MITL, which represents a sort of \lq\lq compromise solution\rq\rq , because a timed extension of $\HS$ over dense domains is not available from the literature.

In this thesis we have completely worked out several problems; however, we now list some issues that remain still open from all previous chapters. 
\begin{itemize}
    \item The first noticeable one is the precise complexity of MC for full $\HS$ (and $\BE$)---with or without regular expressions. We conjecture the problem to be \emph{elementarily decidable} (and, maybe, $\EXPSPACE$-complete). The machinery presented in Chapter~\ref{chap:ICALP} for $\D$ does not generalize to $\BE$, and neither do the results of Section~\ref{sec:AAbarBBbarEbar} on $\AAbarBBbarEbar$/$\AAbarEBbarEbar$. 
    \item MC for $\AAbarBBbarEbar$ and $\AAbarEBbarEbar$ under homogeneity is in $\LINAEXPTIME$, but only known to be $\PSPACE$-hard; hardness derives from that of $\Bbar$/$\Ebar$. We recall that, conversely, MC for $\AAbarBBbarEbar$/$\AAbarEBbarEbar$ with regular expressions is complete for $\LINAEXPTIME$ (Section~\ref{sec:AAbarBBbarEbarRegex}). It is not clear if removing regular expression from such fragments lowers the complexity from $\LINAEXPTIME$ to other classes \lq\lq above\rq\rq\ $\PSPACE$ (or to $\PSPACE$ itself).
    \item MC for $\A$, $\Abar$, $\AAbar$, $\AbarB$, and $\AE$ under homogeneity has complexity in between $\Th$ and $\Thsq$. We do not know if the problem can be solved by less than $O(\log^2 n)$ queries to the $\NP$ oracle, or a tighter lower bound can be proved, or both (e.g., $\Theta(\log n \log\log n)$ queries may be needed).
    \item It is not known whether the future timeline-based planning problem 
    with unrestricted trigger rules is decidable, but we conjecture that this 
    is \emph{not} the case. In Section~\ref{sec:DecisionProcedures} 
    decidability has been shown only for \emph{simple} trigger rules, as they 
    allow for translation into \MTL/\MITL\ formulas.
\end{itemize}

Now we would like to conclude the thesis by outlining possible future research directions and themes.
\begin{itemize}
    \item An interesting problem to study is MC for $\HS$ over \emph{visibly pushdown systems} (VPS) that,
    given their infinite state (configuration) space, allow us to represent \emph{infinite state systems}, hence enabling us to deal, for instance, with \emph{recursive programs}.
    In this case, $\HS$ may be extended with a \emph{binding} and an \emph{unbinding} operator---the former for restricting the valuation of its argument to the sub-intervals of the current context-interval (each interval being a computation trace of the VPS), and the latter for removing the current binding. Such logic may then be compared, as for its expressiveness, to the context-free linear-time $\LTL$ extensions  $\mathsf{CARET}$~\cite{AlurEM04}, $\mathsf{NWTL}$~\cite{AlurABEIL07}, and $\mathsf{CARET}$ with the \emph{within} modality~\cite{AlurABEIL07}.
    \item The current semantic definition of $\HS$---either with or without regular expressions---does not allow us to express a meaningful property of intervals, the \emph{average}.
    Adding proposition letters in $\HS$ such as $p_{\geq 0.5}$, having the intuitive meaning of \lq\lq $p$ holds true in at least half of the states of the interval being considered\rq\rq , would be fairly natural and give the possibility of expressing a peculiar interval-based property.
    \item A \emph{timed} extension of $\HS$ \emph{over dense domains} would be naturally required by timelines, where time is a fundamental dimension. However, in the literature, only \emph{metric} extensions of $\HS$ have been proposed over the natural numbers~\cite{DBLP:journals/sosym/BresolinMGMS13}. Defining such an extension, and perhaps 
    linking it with known results regarding other timed logics and/or timed automata, is an interesting research theme.
    \item The \emph{synthesis} problem for $\HS$, a natural evolution of MC, useful, for example, for the development of digital sequential circuits, has already been studied in~\cite{DBLP:journals/corr/MontanariS14}, where the authors consider $\AAbarBBbar$ and $\AAbar\Bbar$ (with an equivalence relation). Synthesis for such fragments is either \emph{decidable and non-primitive recursive-hard}, or \emph{undecidable} (depending on the underlying linear orders).
    Thus, suitable restrictions should be identified on the fragments, on the semantics, and/or on the models, in order for the results to be useful for practical purposes.
\end{itemize}

%% file: Chaps/Appendices/appendixICALP_D.tex
\chapter{Proofs and complements of Chapter~\ref{chap:ICALP}}
\minitoc\mtcskip

\section{Proof of Lemma~\ref{lem:compass_hom_gen}}\label{proof:lem:compass_hom_gen}

\begin{lemma*}[\ref{lem:compass_hom_gen}]
Let $\cG=(\bbP_\bbD,\cL)$. $\cG$ is a fulfilling homogeneous compass $\varphi$-structure if and only if, for every pair $x,y \in \mathpzc{S}$,  we have: 
\begin{itemize}
    \item $\cL(x,y-1)\cL(x+1, y)\genDphi \cL(x, y)$ if $x<y$, and 
    \item $\reqD(\cL(x,y))=\emptyset$ if $x=y$.
\end{itemize}
\end{lemma*}

\begin{proof}
$(\Rightarrow)$
Let us consider $x,y \in \mathpzc{S}$.
First we note that, since $\cG$ is fulfilling, it must be $\reqD(\cL(x,y))=\emptyset$ whenever $x=y$.
Otherwise, if $x<y$, we consider the labelings $\cL(x,y-1)$ and $\cL(x+1, y)$. By the homogeneity property
of Definition~\ref{def:hom_compass}, $\cL(x,y)\cap\AP = \cL(x,y-1) \cap \cL(x+1,y)\cap \AP$: the first condition 
of Definition~\ref{def:d_generator} holds. 
Moreover,
since $\cG$ is fulfilling, for every $\psi \in \reqD(\cL(x,y))$ we have that either $\psi\in \cL(x,y-1)$, or   $\psi\in \cL(x+1,y)$, or $\psi \in \cL(x',y')$
for some $x<x'\leq y'<y$. In the first two cases $\psi \in \obsD(\cL(x,y-1)) \cup \obsD(\cL(x+1,y))$. As for the last case,  by Lemma~\ref{lem:transitive_req}, $\obsD(\cL(x',y'))\subseteq \reqD(\cL(x,y-1))$ and $\obsD(\cL(x',y'))\subseteq \reqD(\cL(x+1,y))$,
hence
$\psi \in \reqD(\cL(x,y-1))$ and  $\psi \in \reqD(\cL(x+1,y))$.
We can conclude that $\reqD(\cL(x,y))\subseteq\obsD(\cL(x,y-1)) \cup \obsD(\cL(x+1,y)) \cup  \reqD(\cL(x,y-1)) \cup \reqD(\cL(x+1,y))$. The converse inclusion ($\supseteq$) follows by Lemma~\ref{lem:transitive_req}, hence the second condition of Definition~\ref{def:d_generator} holds. We conclude that $\cL(x,y-1)\cL(x+1, y)\genDphi \cL(x, y)$.

$(\Leftarrow)$ Let us consider $\cG=(\bbP_\bbD,\cL)$ such that, for every pair $x,y \in \mathpzc{S},\, x\leq y$, we have $\cL(x,y-1)\cL(x+1, y)\genDphi \cL(x, y)$ if $x<y$, and $\reqD(\cL(x,y))=\emptyset$ if $x=y$. We have to prove that $\cG$ is a homogeneous fulfilling compass $\varphi$-structure. 

First, we prove consistency w.r.t.\ the relation $\Dphi$.
Let us show that, for all pairs of 
points $(x,y)$ and $(x',y')$ with $(x',y')\subint(x,y)$,
we have $\cL(x,y)\Dphi\cL(x',y')$. 
The proof is by induction on $\Delta =(x'-x) + (y - y')\geq 1$.
If $\Delta=1$, either $(x',y')=(x+1,y)$
or $(x',y')=(x,y-1)$. Let us consider $(x',y')=(x+1,y)$
(the other case is symmetric). Since 
$\cL(x,y-1)\cL(x+1, y)\genDphi \cL(x, y)$,
we easily get that $\cL(x,y)\Dphi \cL(x+1, y)$.
If $\Delta \geq 2$, 
since $(x',y')\subint(x,y)$, then
$(x',y'+1)\subint(x,y)$
or $(x'-1,y')\subint(x,y)$. We only consider
$(x'-1,y')\subint(x,y)$, being the other case symmetric. By the inductive hypothesis, 
$\cL(x,y)\Dphi \cL(x'-1,y')$.
Since $\cL(x'-1,y'-1)\cL(x', y')\genDphi \cL(x'-1, y')$,
we have $\cL(x'-1, y')\Dphi \cL(x',y')$.
 Let us observe that $\Dphi$ is a transitive relation,
and thus $\cL(x, y)\Dphi \cL(x',y')$.

We now show that $\cG$ is fulfilling.
We prove that for every point 
$(x,y)\in\bbP_\bbD$ and for every 
$\psi\in \reqD(\cL(x,y))$, there exists $(x',y')\in\bbP_\bbD,\,(x',y')\subint(x,y)$ such that
$\psi\in\cL(x',y')$. The proof is by induction on 
$y-x\geq 0$. If $x=y$, we have $\reqD(\cL(x,y))=\emptyset$, hence the thesis holds vacuously. 
If $y-x\geq 1$, since 
$\cL(x,y-1)\cL(x+1, y)\genDphi \cL(x, y)$,
we have $\reqD(\cL(x, y))= \reqD(\cL(x,y-1))\cup\reqD(\cL(x+1, y))\cup \obsD(\cL(x,y-1))\cup\obsD(\cL(x+1, y))$. If $\psi \in \obsD(\cL(x,y-1))\cup\obsD(\cL(x+1, y))$, the thesis is verified. If $\psi \in \reqD(\cL(x+1, y))$ (the case $\psi \in \reqD(\cL(x, y-1))$ is symmetric and thus omitted), by the inductive hypothesis, 
$\psi \in \cL(x'',y'')$ for some
$(x'',y'') \subint (x+1, y) \subint (x,y)$. 

It remains to prove that $\cG$ 
is homogeneous. 
We have to show that, for every $(x,y)\in\bbP_\bbD$ and every $p\in \AP$, $p\in \cL(x,y)$
if and only if for every point $(x',x')$, with $x\leq x' \leq y$, we have $p\in \cL(x',x')$.
The proof is by induction on the length of the interval $(x,y)$. If $x=y$
the property trivially holds. Let us consider now $y-x>0$ (inductive step).
By the inductive hypothesis, since $(x+1, y)$ and $(x,y-1)$ are shorter than $(x,y)$, 
we have $p\in \cL(x+1,y)$  (resp., $p\in \cL(x, y-1)$)
if and only if, for every $(x',x')$ with $x+1 \leq x' \leq y$, (resp., $x\leq x'\leq y-1$), $p\in \cL(x',x')$.
Thus $p \in \cL(x+1,y) \cap \cL(x, y-1)$ if and only if
for every $(x',x')$ with $x\leq x'\leq y$, $p\in \cL(x',x')$.
Since $\cL(x+1,y)\cL(x,y-1)\genDphi \cL(x, y)$, we have
$\cL(x, y)\cap \AP = \cL(x+1,y) \cap \cL(x, y-1) \cap \AP$.
Therefore $p\in \cL(x,y)$ if and only if for every $(x',x')$, with $x\leq x'\leq y$, we have $p\in \cL(x',x')$.
 \end{proof}

\section{Proof of Lemma~\ref{lem:compas_implies_row}}\label{proof:lem:compas_implies_row}

\begin{lemma*}[\ref{lem:compas_implies_row}]
Let $\cG= (\bbP_\bbD,\cL)$ be a fulfilling homogeneous compass $\varphi$-structure. For every $y\in \mathpzc{S}$, $\row_y$ is a $\varphi$-row.
\end{lemma*}

\begin{proof}
Let  $\row_y=\cL(y,y)^{m_0}\cL(y- m_0, y )^{m_1}\cdots\cL(y - \sum_{0\leq i< n} m_i, y)^{m_n}$ 
where, for every $0\leq j\leq n$, $\cL(y - \sum_{0\leq i< j} m_j, y)^{m_j}$ is a maximal substring of identical atoms (note that any $\row_y$ can be represented w.l.o.g.\  in this way, for $m_i> 0$). 
Since $(y,y) \subint  \ldots \subint (0,y)$, by Lemma~\ref{lem:transitive_req}, $\reqD(\cL(y,y))\subseteq \reqD( \cL(y- m_0, y ))\subseteq\ldots \subseteq \reqD(\cL(y - \sum_{0\leq i< n} m_i, y))$.
Moreover, by homogeneity, $(\cL(y,y)\cap \AP) \supseteq (\cL(y- m_0, y )\cap \AP)\supseteq \ldots \supseteq(\cL(y - \sum_{0\leq i< n} m_i, y)\cap \AP )$. 
By maximality, $\cL(y - \sum_{0\leq i< j} m_i, y)\neq \cL(y - \sum_{0\leq i< j-1} m_i, y)$ for every $0<j\leq n$, and thus, since $\varphi$-atoms are 
uniquely determined by a pair $R\subseteq \REQ_{\varphi}$ and $P\subseteq \AP$ which are monotonically arranged, we can conclude that $\cL(y - \sum_{0\leq i< j} m_i, y)\neq \cL(y - \sum_{0\leq i< j'} m_i, y)$ for every $j'< j$. Now we prove that $m_j=1$ if    $\cL(y - \sum_{0\leq i< j} m_i, y)$
is irreflexive. By contradiction let us suppose that $m_j>1$; then $\cL(y - \sum_{0\leq i< j} m_i, y)=
\cL(y - (\sum_{0\leq i< j} m_i)-1, y)$. Since $\cL(y - (\sum_{0\leq i< j} m_i)-1, y)\Dphi \cL(y - \sum_{0\leq i< j} m_i, y)
$, then $\cL(y - \sum_{0\leq i< j} m_i, y)$ is reflexive (contradiction).
Finally we have $\reqD(\cL(y,y))=\emptyset$, as $\cG$ is fulfilling.
\end{proof}

\section{Proof of Lemma~\ref{lem:row_successor}}\label{proof:lem:row_successor}

\begin{lemma*}[\ref{lem:row_successor}]
Let $\cG=(\bbP_\bbD,\cL)$, with $\reqD(\cL(x,x))=\emptyset$ for all $(x,x)\in\bbP_\bbD$. $\cG$ is a fulfilling homogeneous compass $\varphi$-structure
if and only if, for each $0\leq y < |\mathpzc{S}| -1$, $\row_y \rownext \row_{y+1}$.
\end{lemma*}

\begin{proof}
$(\Rightarrow$) By Lemma~\ref{lem:compas_implies_row}, the rows $\row_0,\ldots,\row_{|\mathpzc{S}|-1}$ of $\cG$ are $\varphi$-rows.
By Lemma~\ref{lem:compass_hom_gen}, for every $0\leq x\leq y$, 
$\cL(x,y)\cL(x+1,y+1)\genDphi \cL(x, y+1)$. Since $\cL(x,y)= \row_y(y-x)$,
$\cL(x+1, y+1)= \row_{y+1}((y+1)- (x +1))$, and $\cL(x,y+1)= \row_{y+1}((y+1)- x )$,
we can conclude that $\row_y \rownext \row_{y+1}$.

$(\Leftarrow)$ Since for each $0\leq y < |\mathpzc{S}| -1$, $\row_y \rownext \row_{y+1}$, we have that for all $0\leq i \leq y$,
$\row_y(i)\row_{y+1}(i) \genDphi \row_{y+1}(i+1)$, namely, $\cL(y-i,y)\cL(y+1-i,y+1)\genDphi\cL(y-i,y+1)$. Let $x=y-i$, for $0\leq x\leq y$. We get $\cL(x,y)\cL(x+1,y+1)\genDphi\cL(x,y+1)$. By Lemma~\ref{lem:compass_hom_gen}, $\cG$ is a fulfilling homogeneous compass $\varphi$-structure.
\end{proof}

\section{Proof of Theorem~\ref{thm:path_iff_MC}}\label{proof:thm:path_iff_MC}

\begin{theorem*}[\ref{thm:path_iff_MC}]
Given a finite Kripke structure $\Ku=\KuDef$ and a $\hsDhom$ formula $\varphi$, 
there exists an initial trace $\rho$ of $\Ku$ such that $\Ku,\rho\models\varphi$
if and only if
there exists
a path  in $G_{\varphi \sim\Ku}=(\Gamma, \Xi)$
from some node $(\sinit,[\row]_\sim)\in\Gamma$ to some node $(s,[\row']_\sim)\in\Gamma$ such that:
\begin{enumerate}
    \item there exists $row_1\in[\row]_\sim$ with $|row_1|=1$, and
    \item there exists $row_2\in[\row']_\sim$ with $\varphi\in row_2(|row_2|-1)$.
\end{enumerate}
\end{theorem*}

\begin{proof}
Preliminarily we observe that, in $(1.)$, if $|row_1|=1$, then $\{row_1\}=[\row]_\sim$; moreover, in $(2.)$, if for $row_2\in[\row']_\sim$ we have $\varphi\in row_2[|row_2|-1]$, then for any $row_2'\in[\row']_\sim$ we have $\varphi\in row_2'[|row_2'|-1]$.

($\Rightarrow$) 
Let us consider an initial trace $\rho$ such that $\Ku,\rho\models\varphi$, hence, by Proposition~\ref{prop:eqTrack}, 
$\varphi\in\cL(0,|\rho|-1)$ in the fulfilling homogeneous compass $\varphi$-structure induced by $\rho$, $\cG_{(\Ku,\rho)}=(\bbP_\bbD, \cL)$.
By Lemmata~\ref{lem:compas_implies_row} and \ref{lem:row_successor}, 
$\cL(0,0) \rownext \allowbreak  \row_1 \rownext \cdots \rownext \row_{|\rho|-1}$, and $\varphi \in \row_{|\rho|-1}[|\rho|-1]$. 
By definition of $(\varphi\!\sim\!\Ku)$-graph, 
$(\rho(0),[\cL(0,0)]_\sim) \stackrel{\Xi}{\rightarrow} (\rho(1),[\row_1]_\sim) \stackrel{\Xi}{\rightarrow} \cdots \stackrel{\Xi}{\rightarrow} (\rho(|\rho|-1),[\row_{|\rho|-1}]_\sim)$ is a path in $\cG_{(\Ku,\rho)}$ (since $\row_y[0]\cap\AP\!=\!\Lab(\rho(y))$ for all $0\!\leq\! y\!<\!|\rho|$) satisfying $(1.),(2.)$.

($\Leftarrow$) Let us assume there  is a path $(\sinit,[\row_0]_\sim) \stackrel{\Xi}{\rightarrow} (s_1,[\row_1]_\sim) \stackrel{\Xi}{\rightarrow} \cdots \stackrel{\Xi}{\rightarrow} (s_m,[\row_m]_\sim)$
in the $(\varphi\!\sim\!\Ku)$-graph $G_{\varphi \sim\Ku}=(\Gamma, \Xi)$, satisfying $(1.)$ and $(2.)$.
Hence, by definition of $(\varphi\!\sim\!\Ku)$-graph, $\rho\!=\!s_0s_1\cdots s_m$ is an (initial) trace,  $[\row_0]_\sim \rownext \allowbreak \cdots \rownext [\row_{m}]_\sim$, and $\Lab(s_y)=row_y[0]\cap\AP$ for all $0\leq y\leq m$.
%
Applying repeatedly Lemma~\ref{lem:row_class_suc}
we get that there is a sequence 
$\row'_0 \rownext \cdots \rownext \row'_{m}$ of $\varphi$-rows where $\row'_0 =\row_0$,
for every $0\leq j\leq m$, $\row'_j\in [\row_j]_\sim$,
and $\varphi\in \row_m'[|\row_m'|-1]$.
We observe that, by Definition~\ref{def:rownext},
$|\row_j'|=|\row_{j-1}'|+1$
 for $1\leq j \leq m$
and, since $|\row'_0|=1$, we have $|\row_{j}'|= j + 1$. 
Let us now define $\cG=(\bbP_\bbD, \cL)$ where $\mathpzc{S}= \{0,\ldots, m\}$ and
$\cL(x,y)=\row_y'[y-x]$ for every $0\leq x\leq y\leq m$.
Note that $\reqD(\cL(y,y))=\emptyset$ for every $0\leq y\leq m$ (by definition of $\varphi$-row).
By Lemma~\ref{lem:row_successor}, $\cG$ is a fulfilling homogeneous compass $\varphi$-structure.
Since $\Lab(s_y)=row_y[0]\cap\AP(=\cL(y,y)\cap\AP)$ for all $0\leq y\leq m$, then $\cG$ is precisely the compass $\varphi$-structure induced by $\rho$.
Finally, since $\varphi\in \row_m'[m]=\cL(0,m)$, by Proposition~\ref{prop:eqTrack} we can conclude that $\Ku,\rho\models\varphi$.
\end{proof}

\section{$\Psp$-Hardness of MC for $\hsDhom${} over finite Kri\-pke structures}\label{sec:MChard}
In this section we prove the $\Psp$-hardness of MC for $\hsDhom${} over finite Kripke structures
by a reduction from the $\Psp$-complete \mbox{\emph{problem of (non-)universality}} of the language of a non-deterministic finite state automaton ($\NFA$)~\cite{holzer}.
We refer to Section~\ref{sect:backgrRegex} for notation and definitions on $\NFA$s. 

Let us consider an $\NFA$ 
$\Au =(\Sigma,Q,q_1,\Delta,F)$, where, for convenience, $q_1\in Q$ is the only initial state.
The problem of deciding whether $\Lang(\Au)\neq\emptyset$ can be solved by \emph{logarithmic working space} by means of a non deterministic reachability from the initial state of $\Au$ to an accepting state. On the other hand, deciding if $\Lang(\Au)\neq\Sigma^*$ (namely, deciding if $\Lang(\Au)$ is \emph{non-universal}, i.e., there is some word $w\in\Sigma^*$ such that $w\notin\Lang(\Au)$) is more difficult, and it is $\Psp$-complete~\cite{holzer}. This is due to the fact that the shortest word \emph{not accepted} by an $\NFA$ can have length exponential in the number of states of the $\NFA$.

It is well-known that, by a subset construction, we can build from $\Au$ a $\DFA$ $\Du=(\Sigma,\tilde{Q},\tilde{q_1},\tilde{\delta},\tilde{F})$ such that $\Lang(\Du)=\Lang(\Au)$ and $\tilde{Q}= 2^{Q}$, $\tilde{q_1}=\{q_1\}$. We say that $\Du$ is \emph{equivalent} to $\Au$.
In order to decide if a word $w$ is accepted by $\Au$, it is then possible to build \lq\lq on the fly\rq\rq{} \emph{the} computation of $\Du$ over $w$ from $\tilde{q_1}$ (hereafter we will omit \lq\lq from $\tilde{q_1}$\rq\rq).

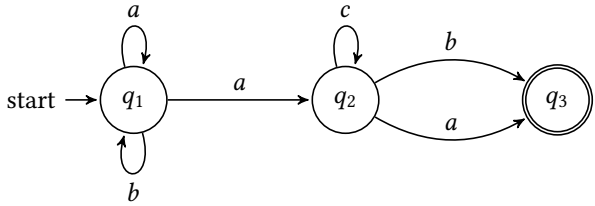
\begin{figure}[t]
    \centering
    \begin{tikzpicture}[->,>=stealth',shorten >=1pt,auto,node distance=2.8cm, semithick]

  \node[initial,state] (A)              {$q_1$};
  \node[state]         (B) [right of=A] {$q_2$};
  \node[state,accepting](C)[right of=B] {$q_3$};

  \path (A) edge              node {$a$} (B)
        (B) edge [bend right] node {$a$} (C)
        (B) edge [loop above] node {$c$} (B)
        (B) edge [bend left] node {$b$} (C)
        (A) edge [loop above] node {$a$} (A)
        (A) edge [loop below] node {$b$} (A);
    \end{tikzpicture}
    \caption{An example of $\NFA$, where $q_1$ is the initial state, and $q_3$ the only final state.}\label{fig:Nfa}
\end{figure}
For example, let us consider the $\NFA$ $\Au$ of Figure \ref{fig:Nfa}.
The computation of $\Du$ (its equivalent $\DFA$) over $aab$ is:
\begin{equation}
    (Q_1=\{q_1\}) \stackrel{a}{\to} 
    (Q_2=\{q_1,q_2\}) \stackrel{a}{\to} 
    (Q_3=\{q_1,q_2,q_3\}) \stackrel{b}{\to} (Q_4=\{q_1,q_3\}). 
\label{eq:1}\end{equation}
The word $aab$ is accepted since there exists a final state of $\Au$, $q_3\in Q_4$ ($Q_4$ is the state of $\Du$ reached by the computation). Note that $Q_1$ is the initial state of $\Du$.
The computation of $\Du$ over $aac$ is:
\begin{equation}
    (Q_1=\{q_1\}) \stackrel{a}{\to} 
    (Q_2=\{q_1,q_2\}) \stackrel{a}{\to} 
    (Q_3=\{q_1,q_2,q_3\}) \stackrel{c}{\to} (Q_4'=\{q_2\}), 
\end{equation}
hence $aac$ is \emph{not} accepted since $Q_4'$ does not contain final states of $\Au$.

Note that, as a rule, in the computation of $\Du$ over $w$, with $|w|=n$,
\[
    Q_1 \stackrel{w(0)}{\to}
    Q_2 \stackrel{w(1)}{\to}
    \cdots \stackrel{w(n-2)}{\to}
    Q_{n} \stackrel{w(n-1)}{\to} Q_{n+1},
\]
we have that the $\Au$-state $q$ belongs to $Q_i$, for $0\leq i\leq n$, \emph{if and only if} there exists some computation of $\Au$ over $w$, 
\[  q_1 \stackrel{w(0)}{\to}
    q_2 \stackrel{w(1)}{\to}
    \cdots \stackrel{w(n-2)}{\to}
    q_{n} \stackrel{w(n-1)}{\to} q_{n+1},
\]
where $q=q_i$.
Moreover, if some $q\in Q_{i}$ then all $q'\in\Delta(q,w(i-1))$ must be in $Q_{i+1}$. Conversely, if some $q'\in Q_{i+1}$ then it has to exist some $q\in Q_i$ such that $q'\in\Delta(q,w(i-1))$.

We are ready to reduce the $\Psp$-complete problem of (non-)universality of the language of an $\NFA$ to the MC problem for $\hsDhom${} over finite Kripke structures, proving that the latter is $\Psp$-hard.

Given an $\NFA$ $\Au =(\Sigma,Q,q_1,\Delta,F)$, we build the Kripke structure $\Ku_\Au=\KuDef$, where:
\begin{itemize}
    \item $\States=\{q_i^\top,q_i^\bot,q_i'^\top,q_i'^\bot\mid i=1,\ldots, |Q|\}\cup\{x_1,x_2,v_1,v_2,v'_1,v'_2,\widehat{q_1}^\top,\widehat{q_2}^\bot,\ldots,\allowbreak \widehat{q_{|Q|}}^\bot\}\cup\Sigma$;
    \item $\sinit=v_1'$;
    \item $\AP=\{q_i,q_i'\mid i=1,\ldots, |Q|\}\cup\Sigma\cup\{e_1,e_2,f_1,f_2\}$;  
    \item $\Lab({q_i}^\top)\!=\!\Lab({q_i'}^\top)\!=\!\AP$, $\Lab({q_i}^\bot)\!=\!\AP\setminus\{q_i\}$, $\Lab({q'_i}^\bot)\!=\!\AP\setminus\{q_i'\}$, for $1\leq i\leq |Q|$;\\
    $\Lab(a)=\AP\setminus(\Sigma\setminus\{a\})$ for $a\in\Sigma$;\\
    $\Lab(x_1)=\AP\setminus\{e_1\}$, $\Lab(x_2)=\AP\setminus\{e_2\}$, $\Lab(v_1)=\Lab(v_1')=\AP\setminus\{f_1\}$,
    $\Lab(v_2)=\Lab(v_2')=\AP\setminus\{f_2\}$;\\ $\Lab(\widehat{q_1}^\top)=\AP$,
    $\Lab(\widehat{q_2}^\bot)=\AP\setminus\{q_2\}$,\dots , $\Lab(\widehat{q_{|Q|}}^\bot)=\AP\setminus\{q_{|Q|}\}$.
\end{itemize}

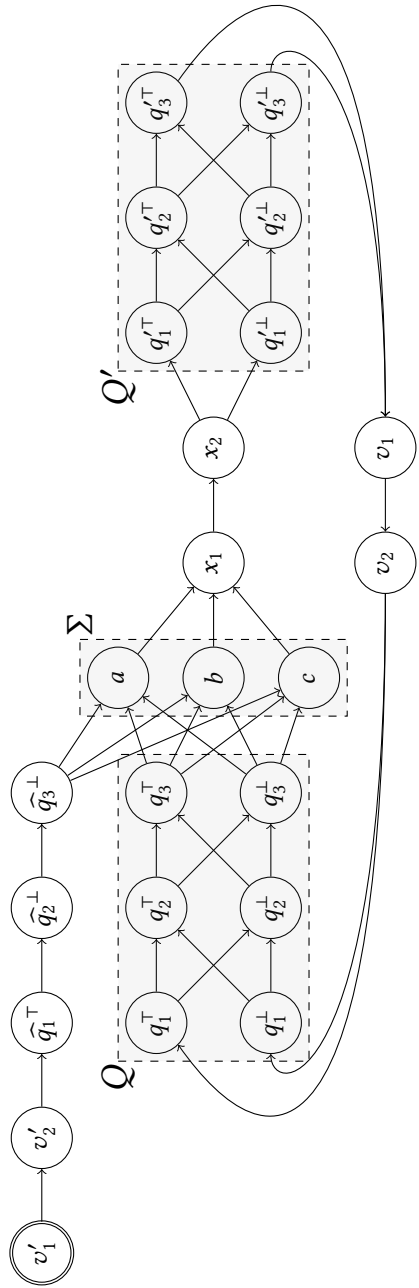
\begin{sidewaysfigure}
    \centering
    \resizebox{0.96\textwidth}{!}{\input{Chaps/ICALP_D/HardKripke}}
    \caption{The Kripke structure $\Ku_\Au$ built for an $\NFA$ with set of states $Q=\{q_1,q_2,q_3\}$  and alphabet $\Sigma=\{a,b,c\}$}
    \label{fig:HardKu}
\end{sidewaysfigure}

The edges $\Edges$ of $\Ku_\Au$ can easily be deduced from Figure~\ref{fig:HardKu}, which is an example of Kripke structure built for an $\NFA$ with set of states $Q=\{q_1,q_2,q_3\}$, and alphabet $\Sigma=\{a,b,c\}$.

The idea is that the computation of $\Du$ on a word, say $aab$, which we have already seen in (\ref{eq:1}), should be represented by the following initial trace of $\Ku_\Au$:
\begin{equation}\begin{split}
v_1'v_2'\underbrace{(\widehat{q_1}^\top\widehat{q_2}^\bot\widehat{q_3}^\bot)}_{Q_1} a x_1x_2 \underbrace{({q'_1}^\top {q'_2}^\top {q'_3}^\bot)}_{Q'_2}\cdots\\
v_1v_2 \underbrace{({q_1}^\top {q_2}^\top {q_3}^\bot)}_{Q_2} a x_1x_2 \underbrace{({q'_1}^\top {q'_2}^\top {q'_3}^\top)}_{Q'_3} \cdots\\
v_1v_2 \underbrace{({q_1}^\top {q_2}^\top {q_3}^\top)}_{Q_3} b x_1x_2 \underbrace{({q'_1}^\top {q'_2}^\bot {q'_3}^\top)}_{Q'_4} \cdots\\
v_1v_2 \underbrace{({q_1}^\top {q_2}^\bot {q_3}^\top)}_{Q_4}.
\end{split}
\label{eq:extrack}
\end{equation}
 
The states $v_1,v'_1,v_2,v'_2,x_1,x_2$ are there only for technical reasons (explained later). 
A triple of states $({q_1}^* {q_2}^* {q_3}^*)$ denoted by $Q_i$, where $^*$ stands for $\top$ or $\bot$,  represents a state of $\Du$, reached at the $(i-1)$-th step of the computation before reading $w(i-1)$: we have ${q_j}^\top$ if $q_j\in Q_i$, and ${q_j}^\bot$ if $q_j\not\in Q_i$.
Moreover the subtraces denoted by $Q_i$ and $Q'_i$ must be copies (i.e., ${q_j}^\top\in Q_i$ iff ${q'_j}^\top\in Q'_i$).
In between $Q_i$ and $Q'_{i+1}$ in the trace we have $w(i-1)\in\Sigma$.
The states $\widehat{q_1}^\top$, $\widehat{q_2}^\bot$ and $\widehat{q_3}^\bot$ of $\Ku_\Au$ are just \lq\lq copies\rq\rq{} of ${q_1}^\top$, ${q_2}^\bot$ and ${q_3}^\bot$ respectively, added to ensure that the first state of the $\DFA$ $\Du$ is $Q_1=\{q_1\}$ (represented by $\widehat{q_1}^\top \widehat{q_2}^\bot \widehat{q_3}^\bot$). 
Finally note that there is an intuitive match between subtraces and proposition letters satisfied. For example,
\begin{multline*}
    \Ku_\Au, v_1v_2 ({q_1}^\top {q_2}^\top {q_3}^\top) b x_1x_2 ({q'_1}^\top {q'_2}^\bot {q'_3}^\top)\models\\
    (q_1\wedge q_2\wedge q_3)\wedge (q'_1\wedge\neg q'_2\wedge q'_3)\wedge (\neg a \wedge b \wedge \neg c).
\end{multline*}

Let us now come to the formula $\Phi_\Au$, built from $\Au$. We assume the \emph{strict} semantic variant of $\hsDhom${}.
Preliminarily, we define the following formulas, which exploit the auxiliary states $v_1,v'_1,v_2,v'_2,x_1,x_2$ in order to ``select'' some suitable traces:
\[
    \varphi_{trans}=\neg f_1\wedge \neg f_2 \wedge \hsDu(f_1\wedge f_2) \wedge \hsD\top ,
\]
\[
    \varphi_{copy}= \neg e_2\wedge \neg e_1 \wedge \hsDu(e_1\wedge e_2) \wedge \hsD\top .
\]
%
We can prove that:
\begin{itemize}
    \item $\Ku_\Au,\rho\models\varphi_{trans}$ if and only if $\rho=\tilde{v_2}\cdots v_1$ and $v_1,v_2$ do not occur as internal states of $\rho$ (where $\tilde{v_2}$ can be either $v_2$ or $v'_2$);
    \item $\Ku_\Au,\rho\models\varphi_{copy}$ if and only if $\rho=x_2\cdots x_1$ and $x_1,x_2$ do not occur as internal states of $\rho$.
\end{itemize}

Moreover the following formulas have an intuitive meaning (in particular $\Length_{\geq 3}$ is satisfied by a trace $\rho$ if and only if $|\rho|\geq 3$):
\[
    \varphi_{reject}= \bigwedge_{q_i\in F}\neg q_i,
\]
\[
    \Length_{\geq 3}=\hsD\top .
\]

The formula $\Phi_\Au$ is defined as follows (for the sake of brevity, for $q_i,q_j\in Q$ and $c\in\Sigma$, we denote $q_j\in\Delta(q_i,c)$ as $(q_i,c,q_j)\in\Delta$).
\begin{multline*}
    \Phi_\Au=
     \underbrace{\hsDu\Big(\varphi_{trans}\rightarrow \Big(\big(\smashoperator{\bigwedge_{(q_i,a,q'_j)\in\Delta}}((q_i\wedge a)\rightarrow q'_j)\big) \wedge \big(\smashoperator{\bigwedge_{q'_i\in Q}} (q'_i\rightarrow \smashoperator{\bigvee_{(q_j,a,q'_i)\in\Delta}} (q_j\wedge a))\big) \Big)\Big)}_{(1)}\wedge\\
     \underbrace{\hsDu\big( \varphi_{copy} \rightarrow\!\! \smashoperator{\bigwedge_{q_i\in Q}} (q_i\leftrightarrow q'_i) \big)}_{(2)}\!\wedge\! \underbrace{\Big((e_1\!\wedge\! \Length_{\geq 3}\wedge \varphi_{reject})\!\vee\! \hsD\!\Big(\varphi_{copy}\!\wedge\!\varphi_{reject}\Big)\Big)}_{(3)}
\end{multline*}

Let us now prove the following lemma.
\begin{lemma}\label{lemma:nuniv}
$\Lang(\Au)\neq \Sigma^*$ if and only if there exists an initial trace $\rho$ of $\Ku_\Au$ such that $\Ku_\Au,\rho\models \Phi_\Au$.
\end{lemma}
\begin{proof}
$(\Rightarrow)$ If $\Lang(\Au)\neq \Sigma^*$, then there is $w\notin\Lang(\Au)$. Therefore the computation of $\Du$ over $w$ is \emph{not} accepting. Let us consider the initial trace $\rho$ of $\Ku_\Au$ encoding such a computation as explained before, see (\ref{eq:extrack}). We distinguish two cases:
\begin{itemize}
    \item $w=\varepsilon$: then we consider $\rho=v'_1v'_2\widehat{q_1}^\top\widehat{q_2}^\bot\cdots \widehat{q_{|Q|}}^\bot $. No strict subtrace satisfies $\varphi_{trans}$ or $\varphi_{copy}$, hence conjuncts (1) and (2) are trivially satisfied. Moreover, since $\varepsilon\notin\Lang(\Au)$, $q_1\notin F$, thus $\rho$ models also $e_1\wedge \Length_{\geq 3}\wedge \varphi_{reject}$.
    \item $w\neq \varepsilon$; then we consider the initial trace $\rho$ of $\Ku_\Au$ encoding the computation over $w$, w.l.o.g.\ extended with some $c\in\Sigma$ (any symbol is fine), and finally $x_1x_2$: its generic form is $\rho=v'_1v'_2\widehat{q_1}^\top\widehat{q_2}^\bot\cdots \widehat{q_{|Q|}}^\bot w(0)(x_1x_2{q'_1}^*{q'_2}^*\cdots {q'_{|Q|}}^*\cdot\allowbreak v_1v_2{q_1}^*{q_2}^*\cdots {q_{|Q|}}^*c)^+x_1x_2$, where $^*$ is $^\bot$ or $^\top$, and $^+$ denotes a positive number of occurrences of the string in brackets. Every strict subtrace satisfying $\varphi_{trans}$ models the right part of the implication in conjunct (1), which enforces the consistency conditions of a computation. 
    Every strict subtrace satisfying $\varphi_{copy}$ features ${q'_i}^\top$ if it features ${q_i}^\top$, and ${q'_i}^\bot$ if it features ${q_i}^\bot$, hence it satisfies $\bigwedge_{q_i\in Q} (q_i\leftrightarrow q'_i)$. Finally, the last part of $\rho$, $x_2{q'_1}^*{q'_2}^*\cdots {q'_{|Q|}}^*v_1v_2{q_1}^*{q_2}^*\cdots {q_{|Q|}}^*c x_1$,
    models $\varphi_{copy}$, and, since $w$ is \emph{not} accepted, it also fulfills $\varphi_{reject}$.
\end{itemize}
Therefore, in both cases, there exists an initial trace $\rho$ such that $\Ku_\Au,\rho\models \Phi_\Au$.

$(\Leftarrow)$ Let us assume there exists an initial trace $\rho$ of $\Ku_\Au$ such that $\Ku_\Au,\rho\models \Phi_\Au$. We distinguish some cases, according to the structure of $\rho$.
\begin{enumerate}
    \item $\rho=v'_1(v'_2)^?$ ($^?$ denotes 0 or 1 occurrences of the string in brackets).\\ 
    This trace does not model (3), thus it cannot be the trace we are looking for.
    \item $\rho\!=\!v'_1v'_2\widehat{q_1}^\top\widehat{q_2}^\bot \cdots \widehat{q_j}^\bot$ for $j\geq 1$, or $\rho\!=\!v'_1v'_2\widehat{q_1}^\top\widehat{q_2}^\bot \cdots \widehat{q_{|Q|}}^\bot c$ for some $c\!\in\!\Sigma$.\\
    No subtrace satisfies $\varphi_{copy}$, thus, by the conjunct (3), $\rho$ models $\varphi_{reject}$. Hence $q_1\notin F$, and $\varepsilon$ is rejected by $\Au$.
    \item $\rho=v'_1v'_2\widehat{q_1}^\top\widehat{q_2}^\bot \cdots \widehat{q_{|Q|}}^\bot c x_1(x_2)^?$\\
    This trace does not model the conjunct (3).
    \item $\rho=v'_1v'_2\widehat{q_1}^\top\widehat{q_2}^\bot \cdots \widehat{q_{|Q|}}^\bot c x_1x_2{q'_1}^*{q'_2}^*\cdots {q'_j}^*$  for $j\geq 1$\\
    This trace does not model the conjunct (3).
    \item $\rho=v'_1v'_2\widehat{q_1}^\top\widehat{q_2}^\bot \cdots \widehat{q_{|Q|}}^\bot c x_1x_2{q'_1}^*{q'_2}^*\cdots {q'_{|Q|}}^*v_1(v_2)^?$\\
    This trace does not model the conjunct (3).
    \item $\rho=v'_1v'_2\widehat{q_1}^\top\widehat{q_2}^\bot \cdots \widehat{q_{|Q|}}^\bot c x_1x_2{q'_1}^*{q'_2}^*\cdots {q'_{|Q|}}^*v_1v_2{q_1}^*{q_2}^*\cdots {q_j}^*$\\
    This trace does not model the conjunct (3).
    \item $\rho=v'_1v'_2\widehat{q_1}^\top\widehat{q_2}^\bot \cdots \widehat{q_{|Q|}}^\bot c x_1x_2{q'_1}^*{q'_2}^*\cdots {q'_{|Q|}}^*v_1v_2{q_1}^*{q_2}^*\cdots {q_{|Q|}}^*c$\\
    This trace does not model the conjunct (3).
    \item $\rho=v'_1v'_2\widehat{q_1}^\top\widehat{q_2}^\bot \cdots \widehat{q_{|Q|}}^\bot c x_1x_2{q'_1}^*{q'_2}^*\cdots {q'_{|Q|}}^*v_1v_2{q_1}^*{q_2}^*\cdots {q_{|Q|}}^*cx_1$\\
    This trace does not model the conjunct (3).
    \item\label{case:A} $\rho=v'_1v'_2\widehat{q_1}^\top\widehat{q_2}^\bot \cdots \widehat{q_{|Q|}}^\bot c (x_1x_2{q'_1}^*{q'_2}^*\cdots {q'_{|Q|}}^*v_1v_2{q_1}^*{q_2}^*\cdots {q_{|Q|}}^*c)^+x_1x_2$\\
    We have that, since $\rho$ models the conjunct (1), \emph{all} adjacent pairs of occurrences of ${q_1}^*{q_2}^*\cdots {q_{|Q|}}^*\rightsquigarrow{q'_1}^*{q'_2}^*\cdots {q'_{|Q|}}^*$ consistently model a transition of $\Du$; moreover \emph{all} adjacent pairs of occurrences of ${q'_1}^*{q'_2}^*\cdots {q'_{|Q|}}^*\rightsquigarrow {q_1}^*{q_2}^*\cdots {q_{|Q|}}^*$ are ``copies''. Put all together, a legal computation of $\Du$ over some string $w$ is encoded. 
    Finally, by the conjunct (3), a strict subtrace $x_2{q'_1}^*{q'_2}^*\cdots {q'_{|Q|}}^*v_1v_2{q_1}^*{q_2}^*\cdots {q_{|Q|}}^*c x_1$ models $\varphi_{reject}$. Thus either $w$ (if such subtrace is the last one) or one of its prefixes (if it is not the last one) is rejected by $\Du$.
    \item\label{case:B} $\rho=v'_1v'_2\widehat{q_1}^\top\widehat{q_2}^\bot \cdots \widehat{q_{|Q|}}^\bot c (x_1x_2{q'_1}^*{q'_2}^*\cdots {q'_{|Q|}}^*v_1v_2{q_1}^*{q_2}^*\cdots {q_{|Q|}}^*c)^+x_1x_2\cdot\allowbreak \underline{{q'_1}^*{q'_2}^*\cdots {q'_j}^*}$\\
    In this case and in the following ones, we underline the final part of $\rho$ which may be ``garbage'', namely, it may encode an illegal suffix of a computation, just because it is not forced to ``behave correctly'' by $\Phi_\Au$.
    However, since conjunct (3) is satisfied, a strict subtrace $x_2{q'_1}^*{q'_2}^*\!\cdots {q'_{|Q|}}^*v_1v_2{q_1}^*{q_2}^*\!\cdots {q_{|Q|}}^*c x_1$ models $\varphi_{reject}$ (and this is not part of the garbage). Thus, as before, some word $w$ or one of its prefixes is rejected by $\Du$.
    \item $\rho=v'_1v'_2\widehat{q_1}^\top\widehat{q_2}^\bot \cdots \widehat{q_{|Q|}}^\bot c (x_1x_2{q'_1}^*{q'_2}^*\cdots {q'_{|Q|}}^*v_1v_2{q_1}^*{q_2}^*\cdots {q_{|Q|}}^*c)^+x_1x_2\cdot\allowbreak\underline{{q'_1}^*{q'_2}^*\cdots {q'_{|Q|}}^*v_1}$\\
    Like the previous case.
    \item\label{case:C} 
    $
    \rho=v'_1v'_2\widehat{q_1}^\top\widehat{q_2}^\bot \cdots \widehat{q_{|Q|}}^\bot c (x_1x_2{q'_1}^*{q'_2}^*\cdots {q'_{|Q|}}^*v_1v_2{q_1}^*{q_2}^*\cdots {q_{|Q|}}^*c)^+x_1x_2\cdot \allowbreak
    {q'_1}^*{q'_2}^*\cdots {q'_{|Q|}}^*v_1v_2
    $\\
    Like case~\ref{case:A}, but a prefix of $w$ is necessarily rejected, such that $\rho$ encodes the computation of $\Du$ over $w$.
    \item 
        $
        \rho=v'_1v'_2\widehat{q_1}^\top\widehat{q_2}^\bot \cdots \widehat{q_{|Q|}}^\bot c (x_1x_2{q'_1}^*{q'_2}^*\cdots {q'_{|Q|}}^*v_1v_2{q_1}^*{q_2}^*\cdots {q_{|Q|}}^*c)^+x_1x_2\cdot \allowbreak 
        {q'_1}^*{q'_2}^*\cdots {q'_{|Q|}}^*v_1v_2\underline{{q_1}^*{q_2}^*\cdots {q_j}^*},
        $
        \\
        $
        \rho=v'_1v'_2\widehat{q_1}^\top\widehat{q_2}^\bot \cdots \widehat{q_{|Q|}}^\bot c (x_1x_2{q'_1}^*{q'_2}^*\cdots {q'_{|Q|}}^*v_1v_2{q_1}^*{q_2}^*\cdots {q_{|Q|}}^*c)^+x_1x_2\cdot \allowbreak
        {q'_1}^*{q'_2}^*\cdots {q'_{|Q|}}^*v_1v_2\underline{{q_1}^*{q_2}^*\cdots {q_{|Q|}}^*c},
        $
        \\
        $
        \rho=v'_1v'_2\widehat{q_1}^\top\widehat{q_2}^\bot \cdots \widehat{q_{|Q|}}^\bot c (x_1x_2{q'_1}^*{q'_2}^*\cdots {q'_{|Q|}}^*v_1v_2{q_1}^*{q_2}^*\cdots {q_{|Q|}}^*c)^+x_1x_2\cdot \allowbreak
        {q'_1}^*{q'_2}^*\cdots {q'_{|Q|}}^*v_1v_2\underline{{q_1}^*{q_2}^*\cdots {q_{|Q|}}^*c x_1}.
        $ 
    
   All like case~\ref{case:C}, just with the addition of final garbage, which is not considered, since it is not part of a strict subtrace satisfying $\varphi_{copy}$.
\end{enumerate}
Thus, in all possible (legal) cases, some string is rejected by $\Du$ (and by $\Au$).
\end{proof}

It follows that $\Lang(\Au)= \Sigma^*$ if and only if $\Ku_\Au\models \neg\Phi_\Au$. Since also the \emph{problem of universality} of the language of an $\NFA$ is $\Psp$-complete (because $\Psp$ is closed under complement), and both $\Ku_\Au$ and $\Phi_\Au$ can be generated in polynomial time, 
we have proved the following.
\begin{theorem}
The MC problem for $\hsDhom$ formulas over finite Kripke structures is $\Psp$-hard (under polynomial-time reductions).
\end{theorem}
By slightly modifying $\Phi_\Au$, we can adapt the proof to the \emph{proper} variant of $\hsDhom${}.

\section{$\Psp$-Hardness of SAT for $\hsDhom${} over finite linear orders}\label{sec:SAThard}
In this section we outline a $\Psp$-hardness proof for the SAT problem 
for $\hsDhom$ formulas over finite linear orders.

The construction mimics that of Sections 3.2 and 3.3 of \cite{DBLP:journals/fuin/MarcinkowskiM14}, in which the authors show 
that it is possible to build a formula $\Psi$ of $\D$ which encodes accepting computations of an $\NFA$. More precisely the set of letters of $\Psi$ equals the union of the alphabet of the $\NFA$ and the set of its states (plus some auxiliary letters, to enforce the \lq\lq orientation\rq\rq{} in the linear order, something that $\D${} is unaware of), and $\Psi$ is satisfied by all and only the models such that the point-intervals are labeled with an accepting computation of the $\NFA$ over the word written in its point-intervals.

The idea is then to exploit $\Psi$ to encode the Kripke structure of the previous section, thus getting a reduction from the 
problem of non-universality of the language of an $\NFA$ to the SAT problem for $\hsDhom$.
As a matter of fact, a Kripke structure can be regarded as a trivial $\NFA$ over a unary alphabet, say $\{a\}$, such that all the states are final, as we are interested only in the structure of traces (i.e., any word/trace is accepted under the only constraint that it exists in the structure).

By an easy adaptation of the results of Sections 3.2 and 3.3 of \cite{DBLP:journals/fuin/MarcinkowskiM14} we get the following.
\begin{proposition}\label{prop:MM}
    Given a Kripke structure $\Ku=\KuDef$ devoid of self-loops, there exists a $\hsDhom$ formula $\Psi_{\Ku}$ whose set of proposition letters is $\AP\cup\States\cup Aux$---being $Aux$ a set of auxiliary letters---such that any finite linear order satisfying $\Psi_{\Ku}$ represents an initial trace of $\Ku$.
    Moreover $\Psi_{\Ku}$ is polynomial in the size of $\Ku$.
\end{proposition}
Every linear order satisfying $\Psi_{\Ku}$ features states of $\Ku$ labeling point-intervals (exactly one state for each point). 
Moreover we can easily force, for each occurrence of some state $s$ of $\Ku$ along the order, the set of letters $\Lab(s)$ to hold on the same position (point).
The structure $\Ku$ in Proposition~\ref{prop:MM} must not feature self-loops for a technical reason: by fulfilling this requirement, there is no way for a state of $\Ku$ to \lq\lq span\rq\rq{} (by homogeneity) more than one point in a linear order satisfying $\Psi_{\Ku}$. 
We observe that in~\cite{DBLP:journals/fuin/MarcinkowskiM14} the authors do not assume homogeneity; however homogeneity does not cause problems in our construction, as, intuitively, all the significant properties stated by $\Psi_{\Ku}$ are related to point-intervals. 

Let us observe that the Kripke structure of the previous section does not contain self-loops.
By Lemma~\ref{lemma:nuniv}, the language of an $\NFA$ $\Au$ is non-universal if and only if there exists an initial trace $\rho$ such that $\Ku_\Au,\rho\models \Phi_\Au$ (the Kripke structure and formula built from $\Au$ in the previous section), if and only if (by Proposition~\ref{prop:MM} applied to $\Ku_\Au$) the formula $\Psi_{\Ku_\Au}\wedge\Phi_\Au$ is satisfiable. We have proved the following result.
\begin{theorem}
The SAT problem for $\hsDhom$ formulas over finite linear orders is $\Psp$-hard.
\end{theorem}

%% file: Chaps/ICALP_D/HardKripke.tex
\begin{tikzpicture}[every node/.style={draw,circle,inner sep=1pt,minimum size=0.8cm}]

\draw[draw=none, use as bounding box]  (-7,5.5) rectangle (9.7,0);

\node (v1) at (-3.5,3.5) {$q_1^\top$};
\node (v4) at (-3.5,2) {$q_1^\bot$};
\node (v2) at (-2,3.5) {$q_2^\top$};
\node (v5) at (-2,2) {$q_2^\bot$};
\node (v3) at (-0.5,3.5) {$q_3^\top$};
\node (v6) at (-0.5,2) {$q_3^\bot$};
\node (v8) at (1,2.75) {$b$};
\node (v7) at (1,4) {$a$};
\node (v9) at (1,1.5) {$c$};
\node (v10) at (2.5,2.75) {$x_1$};
\node (v11) at (4,2.75) {$x_2$};
\node (v12) at (5.5,3.5) {$q_1'^\top$};
\node (v13) at (5.5,2) {$q_1'^\bot$};
\node (v16) at (7,2) {$q_2'^\bot$};
\node (v14) at (7,3.5) {$q_2'^\top$};
\node (v15) at (8.5,3.5) {$q_3'^\top$};
\node (v17) at (8.5,2) {$q_3'^\bot$};
\node (v23) at (4,0.5) {$v_1$};
\node (v24) at (2.5,0.5) {$v_2$};

\draw [->] (v1) edge (v2);
\draw [->] (v2) edge (v3);
\draw [->] (v4) edge (v5);
\draw [->] (v5) edge (v6);
\draw [->] (v4) edge (v2);
\draw [->] (v1) edge (v5);
\draw [->] (v5) edge (v3);
\draw [->] (v2) edge (v6);
\draw [->] (v3) edge (v7);
\draw [->] (v3) edge (v8);
\draw [->] (v3) edge (v9);
\draw [->] (v6) edge (v7);

\draw [->] (v6) edge (v8);
\draw [->] (v6) edge (v9);
\draw [->] (v7) edge (v10);
\draw [->] (v8) edge (v10);
\draw [->] (v9) edge (v10);
\draw [->] (v10) edge (v11);
\draw [->] (v11) edge (v12);
\draw [->] (v11) edge (v13);
\draw [->] (v12) edge (v14);
\draw [->] (v14) edge (v15);
\draw [->] (v13) edge (v16);
\draw [->] (v16) edge (v17);
\draw [->] (v13) edge (v14);
\draw [->] (v12) edge (v16);
\draw [->] (v14) edge (v17);
\draw [->] (v16) edge (v15);

\node (v22) at (-0.5,5) {$\widehat{q_3}^\bot$};
\node (v21) at (-2,5) {$\widehat{q_2}^\bot$};
\node (v20) at (-3.5,5) {$\widehat{q_1}^\top$};
\node (v19) at (-5,5) {$v_2'$};
\node [double] (v18) at (-6.5,5) {$v_1'$};
\draw [->] (v18) edge (v19);
\draw [->] (v19) edge (v20);
\draw [->] (v20) edge (v21);
\draw [->] (v21) edge (v22);
\draw [->] (v22) edge (v7);
\draw [->] (v22) edge (v8);
\draw [->] (v22) edge (v9);

\draw [->](v15.south east) .. controls (10.5,2) and (10.5,0.5) .. (v23.east);
\draw [->](v17.east) .. controls (9.5,2) and (9.5,0.5) .. (v23.east);
\draw [->] (v23) edge (v24);
\draw [->](v24.west) .. controls (-5.5,0.5) and (-5.5,2) .. (v1.south west);
\draw [->](v24.west) .. controls (-4.5,0.5) and (-4.5,2) .. (v4.west);

\draw [dashed,fill=gray,fill opacity=0.07] (-4,4) rectangle (0,1.5);
\draw [dashed,fill=gray,fill opacity=0.07]  (5,4) rectangle (9,1.5);
\draw [dashed,fill=gray,fill opacity=0.07]  (1.5,4.5) rectangle (0.5,1);

\node [draw=none] at (-4.2,4) {\Large $Q$};
\node [draw=none] at (1.7,4.5) {\Large $\Sigma$};
\node [draw=none] at (4.8,4) {\Large $Q'$};

\end{tikzpicture}

%% file: Chaps/Appendices/appendixTCS17.tex
\chapter{Proofs and complements of Chapter~\ref{chap:TCS17}}
\minitoc\mtcskip

\section{Proof of Lemma~\ref{lemmamdc}}\label{proof:lemmamdc}

\begin{lemma*}[\ref{lemmamdc}]
Let $\psi$ be an $\AAbarEEbar$ formula, $\Ku$ be a finite Kripke structure, and $\sigma\in\Trk_\Ku$. Then, $\texttt{Check}(\Ku, \psi,\sigma)=1$ if and only if $\Ku,\sigma\models \psi$.
\end{lemma*}

\begin{proof}
The proof is by induction on the structure of $\psi$.
 The base case where $\psi=p$, for some $p\in\Prop$, directly follows from the definition (line 2 of Algorithm~\ref{Chk2}).
  The cases in which $\psi=\neg\varphi$ and $\psi=\varphi_1\wedge\varphi_2$ are also trivial and thus omitted. We focus on the remaining cases.
\begin{itemize}
    \item $\psi=\hsA\varphi$. If $\Ku,\sigma\models \psi$, then there exists a trace $\rho\in \Trk_\Ku$ such that $\lst(\sigma)=\fst(\rho)$ and $\Ku,\rho\models \varphi$.
    By Theorem~\ref{theorem:polynomialSizeModelProperty}, there exists a trace $\pi\in\Trk_\Ku$, with $|\pi|\leq |\States|\cdot(|\varphi'|+1)^2$ and $\fst(\pi)=\fst(\rho)(=\lst(\sigma))$, such that $\Ku,\pi\models\varphi'$, where $\varphi'$ is the \nnf{} of $\varphi$. Thus, being $|\pi|\leq |\States|\cdot(2|\varphi|+1)^2$, such trace $\pi$ is considered in the for-loop at line 12. By the inductive hypothesis, $\texttt{Check}(\Ku,\varphi,\pi)=1$ and thus \texttt{Check}$(\Ku,\psi,\sigma)=1$.

    Vice versa, if \texttt{Check}$(\Ku,\psi,\sigma)=1$, then there exists a trace $\pi\in \Trk_\Ku$, with $\lst(\sigma)=\fst(\pi)$, such that \texttt{Check}$(\Ku,\varphi,\pi)=1$. By the inductive hypothesis, $\Ku,\pi\models \varphi$, hence $\Ku,\sigma\models \psi$.

    \item $\psi=\hsAt\varphi$ is analogous to the previous case.
    \item $\psi=\hsE\varphi$. If $\Ku,\sigma\models \psi$, there exists a trace $\pi\in\Suff(\sigma)$ such that $\Ku,\pi\models \varphi$. By the inductive hypothesis, \texttt{Check}$(\Ku,\varphi,\pi)=1$. Since all the proper suffixes of $\sigma$ are checked (line 17), \texttt{Check}$(\Ku,\psi,\sigma)=1$.

    Vice versa, if \texttt{Check}$(\Ku,\psi,\sigma)=1$, then for some $\pi\in\Suff(\sigma)$, it holds that \texttt{Check}$(\Ku,\varphi,\pi)=1$. By the inductive hypothesis $\Ku,\pi\models \varphi$ implying that $\Ku,\sigma\models \psi$.
    \item $\psi=\hsEt\varphi$. If $\Ku,\sigma\models \psi$, then there exists a trace $\rho\in \Trk_\Ku$, with $|\rho|\geq 2$, such that $\Ku,\rho\star\sigma\models \varphi$.
    By Theorem~\ref{theorem:polynomialSizeModelProperty}, there exists a trace $\pi\in\Trk_\Ku$ induced by $\rho$, with $|\pi|\leq |\States|\cdot(|\varphi'|+1)^2$, such that $\Ku,\pi\star\sigma\models\varphi'$, where $\varphi'$ is the \nnf{} of $\varphi$. Such trace $\pi$ is considered in the for-loop at line 22, since $|\pi|\leq |\States|\cdot(2|\varphi|+1)^2$ and $|\pi|\geq 2$ as it is induced by $\rho$. By the inductive hypothesis, \texttt{Check}$(\Ku,\varphi,\pi\star\sigma)=1$ implying that  \texttt{Check}$(\Ku,\psi,\sigma)=1$.

    Vice versa, if \texttt{Check}$(\Ku,\psi,\sigma)=1$, then there exists a trace $\pi \in \Trk_\Ku$, with $|\pi|\geq 2$, such that \texttt{Check}$(\Ku,\varphi,\pi\star\sigma)=1$. By the inductive hypothesis, $\Ku,\pi\star\sigma\models \varphi$, hence $\Ku,\sigma\models \psi$.\qedhere
\end{itemize}
\end{proof}

\section{Proof of Lemma~\ref{ThCorrComplMC}}\label{proof:ThCorrComplMC}

\begin{theorem*}[\ref{ThCorrComplMC}]
Let $\psi$ be an $\AAbarEEbar$ formula and $\Ku$ be a finite Kripke structure. Then, \texttt{ModCheck}$(\Ku,\psi)=1$ if and only if $\Ku\models \psi$.
\end{theorem*}

\begin{proof}
	($\Leftarrow$)
	If $\Ku\models \psi$ then, for all initial traces $\rho\in\Trk_\Ku$, we have that $\Ku,\rho\models \psi$.
	By Lemma~\ref{lemmamdc},
it follows that $\texttt{Check}(\Ku,\psi,\rho)=1$. Now, since the for-loop at line 1 considers a subset of the initial traces, it holds that \texttt{ModCheck}$(\Ku,\psi)=1$.

    ($\Rightarrow$)
	If \texttt{ModCheck}$(\Ku,\psi)=1$, then, for any initial trace $\rho$ considered by the for-loop at line 1, that is, with $|\rho|\leq |\States|\cdot (2|\psi|+3)^2$, it holds that $\texttt{Check}(\Ku,\psi,\rho)=1$.
	Let us assume by contradiction that $\Ku\not\models\psi$, that is, there exists an initial trace $\rho'\in\Trk_\Ku$ such that $\Ku,\rho'\models\neg\psi$, or, equivalently, $\Ku,\rho'\models\overline{\psi}$, where $\overline{\psi}$ is the \nnf{} of $\neg\psi$. Thus, by Theorem~\ref{theorem:polynomialSizeModelProperty}, there exists an initial trace $\sigma$, with $|\sigma|\leq |\States|\cdot (|\overline{\psi}|+1)^2\leq |\States|\cdot (2|\psi|+3)^2$, such that $\Ku,\sigma\models \overline{\psi}$, namely, $\Ku,\sigma\not\models\psi$. By Lemma~\ref{lemmamdc}, it holds that $\texttt{Check}(\Ku,\psi,\sigma)=0$, leading to a contradiction and proving that $\Ku\models \psi$.
\end{proof}

\section{$\Psp$-hardness of MC for $\Bbar$ and $\Ebar$}\label{sect:BbarHard}
In this section, we prove that the MC problem for formulas of $\Bbar$ and of $\Ebar$, over finite Kripke structures, is $\Psp$-hard by means of a reduction from 
the QBF problem, that is, the problem of determining the truth of a \emph{fully-quantified} Boolean formula in \emph{prenex normal form}, which is known to be $\Psp$-complete (see, for example, \cite{Sip12}). The proof for  $\Bbar$ can easily be modified to show the $\Psp$-hardness of the symmetric fragment $\Ebar$.

Let $\psi=Q_n x_n Q_{n-1} x_{n-1}$ $\cdots Q_1 x_1 \phi(x_n,x_{n-1},\ldots ,x_1)$ be a quantified Boolean formula where, for $i=1,\ldots ,n,$ $Q_i\in \{\exists, \forall\}$ and $\phi(x_n,x_{n-1},\ldots ,x_1)$ is a quantifier-free Boolean formula over the set of variables $Var = \{x_n,\ldots ,x_1\}$. We define a Kripke structure $\Ku_{QBF}^{Var}$, whose initial traces represent all the possible assignments to the variables of $Var$. For each variable $x \in Var$, $\Ku_{QBF}^{Var}$ features a pair of states $s_x^{\top}$ and $s_x^{\bot}$, that represent a $\top$ and $\bot$ truth assignment to $x$, respectively. An example of $\Ku_{QBF}^{Var}$, with $Var=\{x,y,z\}$, is given in Figure~\ref{Kqbf}.

Formally, let $\Ku_{QBF}^{Var}=\KuDef$, where:
\begin{itemize}
    \item $\Prop= Var \cup \{t\} \cup \{ \tilde{x_i} \mid 1\leq i\leq n\}$;
    \item $\States= \{s_{x_i}^\ell \mid 1\leq i\leq n,\; \ell \in \{\bot,\top\}\} \cup \{s_0,sink\}$;
    \item if $n=0$, $\Edges=\{(s_0,sink),(sink,sink)\}$;\newline
            if $n>0$,  
            $\Edges = \{(s_0,s_{x_n}^{\top}),(s_0,s_{x_n}^{\bot})\}\cup 
            \{(s_{x_i}^\ell,s_{x_{i-1}}^m) \mid \ell \, , m \in \{\bot,\top\},\: 2 \leq i \leq n\} \cup 
            \{(s_{x_1}^{\top},sink),(s_{x_1}^{\bot},sink),(sink,sink) \}$. 
    \item $\Lab(s_0) = Var \cup \{t\}\cup \{ \tilde{x_i} \mid x_i \in Var\}$; \newline
            for all $1\leq i\leq n$, $\Lab(s_{x_i}^\top) = Var \cup \{\tilde{x_j} \mid 1\leq j\leq i\}$
            and $\Lab(s_{x_i}^\bot) = (Var \setminus \{x_i\}) \cup \{\tilde{x_j} \mid 1\leq j\leq i\}$; \newline
            $\Lab(sink) = Var$.
\end{itemize}

\begin{figure}[tb]
\centering
\begin{tikzpicture}[->,>=stealth',shorten >=1pt,node distance=2.2cm,semithick,every node/.style={circle,draw,inner sep=-2pt,outer sep=0},every loop/.style={max distance=8mm}]  
    \node (0) [double] {$\begin{array}{c} s_0 \\ x,y,z \\ t,\tilde{x},\tilde{y},\tilde{z} \end{array}$};
    \node (1a) [above right of= 0] {$\begin{array}{c} s_x^{\top} \\ x,y,z \\ \tilde{x},\tilde{y},\tilde{z} \end{array}$};
    \node (1b) [below right of= 0] {$\begin{array}{c} s_x^{\bot} \\ y,z \\ \tilde{x},\tilde{y},\tilde{z} \end{array}$};
    
    \node (2a) [right of=1a] {$\begin{array}{c} s_y^{\top} \\ x,y,z \\ \tilde{y},\tilde{z} \end{array}$};
    \node (2b) [right of=1b] {$\begin{array}{c} s_y^{\bot} \\ x,z \\ \tilde{y},\tilde{z} \end{array}$};

    \node (3a) [right of=2a] {$\begin{array}{c} s_z^{\top} \\ x,y,z \\ \tilde{z} \end{array}$};
    \node (3b) [right of=2b] {$\begin{array}{c} s_z^{\bot} \\ x,y \\ \tilde{z} \end{array}$};
    
    \node (pit) [below right of=3a] {$\begin{array}{c} sink \\ x,y,z \\  \end{array}$};
  
  \path
    (0) edge (1a)
        edge (1b)
        
    (1a) edge (2a)
        edge (2b)
    (1b) edge (2a)
        edge (2b)

    (2a) edge (3a)
        edge (3b)
    (2b) edge (3a)
        edge (3b)

    (3a) edge (pit)
    (3b) edge (pit)
    (pit) edge [loop above] (pit)
    ;
\end{tikzpicture}
\caption{The finite Kripke structure $\Ku_{QBF}^{x,y,z}$.}\label{Kqbf}
\end{figure}

The QBF formula $\psi$ is reduced to the $\Bbar$ formula $\xi=t\rightarrow \xi_n$, where
\begin{equation*}
\xi_i=
\begin{cases}
\phi(x_n,x_{n-1},\ldots ,x_1) & \text{ if } i=0\\
\hsBt\big(\tilde{x_i} \wedge \xi_{i-1}\big) & \text{ if } i>0 \text{ and }  Q_i=\exists\\
\hsBtu\big(\tilde{x_i} \rightarrow \xi_{i-1}\big) & \text{ if } i>0 \text{ and }  Q_i=\forall
\end{cases}.
\end{equation*}
Note that both $\Ku_{QBF}^{Var}$ and $\xi$ can be built in polynomial time. 
In Theorem~\ref{th:ABbarHard}, we shall show the correctness of the reduction, i.e., that $\psi$
is true iff $\Ku_{QBF}^{Var}\models\xi$.
As a preliminary step, we introduce some technical definitions and an auxiliary lemma.

Given a Kripke structure $\Ku=\KuDef$ and a $\Bbar$ formula $\chi$, we denote by $p\ell(\chi)$ the set of proposition letters occurring in $\chi$ and by $\Ku_{\,|p\ell(\chi)}$ the 
structure obtained from $\Ku$ by restricting the labelling of each state to $p\ell(\chi)$, namely, the 
Kripke 
structure $(\overline{\Prop},\States, \Edges,\overline{\Lab},s_0)$, where $\overline{\Prop}=\Prop\cap p\ell(\chi)$ and $\overline{\Lab}(s)=\Lab(s)\cap p\ell(\chi)$, for all $s\in \States$.

Moreover, for $v\in \States$, we denote by $reach(\Ku,v)$ the subgraph of $\Ku$ induced by the states reachable from $v$, namely, the 
Kripke 
structure $(\Prop,\States',\Edges',\Lab',v)$, where $\States'=\{s\in \States \mid \text{ there exists } \rho\in \Trk_\Ku \text{ with } \fst(\rho)=v \text{ and } \lst(\rho)=s\}$, $\Edges'=\Edges \cap (\States'\times \States')$, and $\Lab'(s)=\Lab(s)$, for all $s\in \States'$. 

As usual, we say that two Kripke structures $\Ku=\KuDef$ and $\Ku'=(\Prop',\States', \Edges',\Lab',\sinit')$ are \emph{isomorphic} (written $\Ku\sim \Ku'$) if and only if there is a \emph{bijection} $f:\States\to \States'$ such that 
\begin{itemize}
    \item $f(s_0)=s_0'$;
    \item for all $u,v\in \States$, $(u,v)\in \Edges \iff (f(u),f(v))\in\Edges'$;
    \item for all $v\in \States$, $\Lab(v)=\Lab'(f(v))$.
\end{itemize}

Finally, if  $\mathpzc{A}_\Ku=(\Prop,\mathbb{I},A_\mathbb{I},B_\mathbb{I},E_\mathbb{I},\sigma)$ is the abstract interval model induced by a Kripke structure  $\Ku$ and $\rho\in\Trk_{\Ku}$,
 we denote $\sigma(\rho)$ by $\mathpzc{L}(\Ku,\rho)$.

Let $\Ku$ and $\Ku'$ be two Kripke structures. The following lemma, which is an immediate consequence of Lemma~1 of \cite{MMP15B}, states that, for any $\Bbar$ formula $\psi$, if the same set of proposition letters, restricted to $p\ell(\psi)$, holds over two traces $\rho \in \Trk_{\Ku}$ and $\rho' \in \Trk_{\Ku'}$, and the subgraphs consisting of the states reachable from, respectively, $\lst(\rho)$ and $\lst(\rho')$ are isomorphic, then $\rho$ and $\rho'$ are equivalent with respect to $\psi$.

\begin{lemma}\label{lemmaBbar}
Given a $\Bbar$ formula $\psi$, two finite Kripke structures \[\Ku=\KuDef \text{ and } \Ku'=(\Prop',\States', \Edges',\Lab',\sinit'),\] and two traces $\rho\in\Trk_\Ku,\, \rho'\in \Trk_{\Ku'}$ such that  \[\mathpzc{L}(\Ku_{\,|p\ell(\psi)},\rho)=\mathpzc{L}(\Ku'_{\,|p\ell(\psi)},\rho')\] and 
\[reach(\Ku_{\,|p\ell(\psi)},\lst(\rho))\sim reach(\Ku'_{\,|p\ell(\psi)},\lst(\rho')),\]
it holds that $\Ku,\rho\models\psi\iff \Ku',\rho'\models\psi$.
\end{lemma}

\begin{theorem}\label{th:ABbarHard}
The MC problem for $\Bbar$ formulas over finite Kripke structures is $\PSPACE$-hard (under polynomial-time reductions).
\end{theorem}
\begin{proof}
Let $\psi=Q_n x_n$ $Q_{n-1} x_{n-1} \cdots Q_1 x_1 \phi(x_n,x_{n-1},\ldots ,x_1)$.
We prove by induction on the number $n$ of variables of $\psi$ that $\psi$ is true 
if and only if $\Ku_{QBF}^{x_n,\cdots , x_1}\models\xi$. In the following, $\phi(x_n,x_{n-1},\ldots , x_1)\{x_i/\upsilon\}$, with $\upsilon\in\{\top ,\bot\}$, denotes the formula obtained from $\phi(x_n,x_{n-1},\ldots ,x_1)$ by replacing all 
occurrences of $x_i$ by $\upsilon$. 
Notice that $\Ku_{QBF}^{x_n,x_{n-1},\cdots , x_1}$ and $\Ku_{QBF}^{x_{n-1},\cdots , x_1}$ are isomorphic if restricted to the states 
$s_{x_{n-1}}^\top$, $s_{x_{n-1}}^\bot$, $\cdots$, $s_{x_1}^\top$, $s_{x_1}^\bot$, $sink$ (i.e., the initial parts of both Kripke structures are eliminated), and the labelling of states is suitably restricted. Moreover, notice that only the trace $s_0$ satisfies the proposition letter $t$ in $\xi$.

Base case ($n=0$). In this case, $\psi=\phi(\emptyset)$ is a Boolean formula devoid of variables. 
If $\psi$ is true, then in particular $\Ku_{QBF}^{\emptyset},s_0\models\phi(\emptyset)$ and thus $\Ku_{QBF}^{\emptyset}\models s\rightarrow \phi(\emptyset)(=\xi)$. Conversely, if $\Ku_{QBF}^{\emptyset}\models s\rightarrow \phi(\emptyset)$, then $\Ku_{QBF}^{\emptyset},s_0\models\phi(\emptyset)$, and since $\phi(\emptyset)$ has no variables, it must be true.

Case $n\geq 1$.
We first prove that if $\psi\!=\!Q_n x_n Q_{n-1} x_{n-1}\!\cdots Q_1 x_1 \phi(x_n,x_{n-1},\ldots\! , x_1)$ is true, then $\Ku_{QBF}^{x_n,\cdots , x_1}\models\xi$. 

If
$Q_n=\exists$, one possibility is that $Q_{n-1} x_{n-1} \cdots Q_1 x_1 \phi'(x_{n-1},\ldots , x_1)$ is true, where $\phi'(x_{n-1},\ldots , x_1)=\phi(x_n,x_{n-1},\ldots , x_1)\{x_n/ \top\}$. By the inductive hypothesis, it holds that $\Ku_{QBF}^{x_{n-1},\cdots , x_1}\models\xi'$, where $\xi'=t\rightarrow\xi_{n-1}'$ and $\xi_{n-1}'=\xi_{n-1}\{x_n/\top\}$. Thus $\Ku_{QBF}^{x_{n-1},\cdots , x_1},s_0'\models \xi_{n-1}'$ ($s_0'$ is the initial state of the structure $\Ku_{QBF}^{x_{n-1},\cdots , x_1}$). By Lemma~\ref{lemmaBbar}, $\Ku_{QBF}^{x_n,\cdots , x_1},s_0s_{x_n}^\top\models \xi_{n-1}'$. Since every right extension of $s_0s_{x_n^\top}$ models $x_n$, $\Ku_{QBF}^{x_n,\cdots , x_1},s_0s_{x_n}^\top\models \xi_{n-1}$, and thus $\Ku_{QBF}^{x_n,\cdots , x_1},s_0\models \hsBt(\tilde{x_n}\wedge\xi_{n-1})(=\xi_n)$. To conclude, $\Ku_{QBF}^{x_n,\cdots , x_1}\models t\rightarrow\xi_n(=\xi)$. The only other possible case is that $Q_{n-1} x_{n-1} \cdots Q_1 x_1 \phi'(x_{n-1},\ldots , x_1)$ is true, with $\phi'(x_{n-1},\ldots , x_1)=\phi(x_n,x_{n-1},\ldots ,\allowbreak x_1)\{x_n/ \bot\}$. As before, it follows that $\Ku_{QBF}^{x_n,\cdots , x_1},s_0s_{x_n}^\bot\models \xi_{n-1}\{x_n/\bot\}$ and thus $\Ku_{QBF}^{x_n,\cdots , x_1},s_0\models \hsBt(\tilde{x_n}\wedge\xi_{n-1})$.

Now, let 
$Q_n=\forall$. Both formulas $Q_{n-1} x_{n-1} \cdots$ $Q_1 x_1 \phi(x_n,x_{n-1},\ldots , x_1)\{x_n/ \top\}$ and  $Q_{n-1} x_{n-1} \cdots Q_1 x_1 \phi(x_n,x_{n-1},\ldots , x_1)\{x_n/ \bot\}$ are true. By reasoning as in the existential case, we have $\Ku_{QBF}^{x_n,\cdots , x_1},s_0s_{x_n}^\top\models \xi_{n-1}$ and $\Ku_{QBF}^{x_n,\cdots , x_1},s_0s_{x_n}^\bot\models \xi_{n-1}$. Thus, $\Ku_{QBF}^{x_n,\cdots , x_1},s_0\models \hsBtu(\tilde{x_n}\rightarrow \xi_{n-1})$ and $\Ku_{QBF}^{x_n,\cdots , x_1}\models s\rightarrow \hsBtu(\tilde{x_n}\rightarrow \xi_{n-1})$.

We now prove that if $\Ku_{QBF}^{x_n,\cdots , x_1}\models\xi$, then $\psi$
is true.

If $Q_n=\exists$, then $\Ku_{QBF}^{x_n,\cdots , x_1},s_0\models \hsBt(\tilde{x_n}\wedge \xi_{n-1})$. Thus  $\Ku_{QBF}^{x_n,\cdots , x_1},s_0s_{x_n}^\top\models\xi_{n-1}$ or $\Ku_{QBF}^{x_n,\cdots , x_1},s_0s_{x_n}^\bot\models\xi_{n-1}$. In the former case, $\Ku_{QBF}^{x_n,\cdots , x_1},s_0s_{x_n}^\top\models \xi_{n-1}\{x_n/\top\}$ (since every right extension of $s_0s_{x_n}^\top$ models $x_n$). 
By Lemma~\ref{lemmaBbar}, $\Ku_{QBF}^{x_{n-1},\cdots , x_1},s_0'\models\xi_{n-1}\{x_n/\top\}$, and $\Ku_{QBF}^{x_{n-1},\cdots , x_1}\models t\rightarrow \xi_{n-1}\{x_n/\top\}$. By the inductive hypothesis, $Q_{n-1} x_{n-1} \cdots Q_1 x_1 \phi(x_n,x_{n-1},\ldots , x_1)\{x_n/\top\}$ is true, thus the formula $\psi=\exists x_n Q_{n-1} x_{n-1} \cdots Q_1 x_1 \phi(x_n,x_{n-1},\ldots , x_1)$ is true. In the latter case, we symmetrically have that $Q_{n-1} x_{n-1} .. Q_1 x_1 \phi(x_n,x_{n-1},\ldots , x_1)\{x_n/\bot\}$ is true, thus the formula $\psi=\exists x_n Q_{n-1} x_{n-1} \cdots Q_1 x_1 \phi(x_n,x_{n-1},\ldots , x_1)$ is true.

If $Q_n=\forall$, then $\Ku_{QBF}^{x_n,\cdots , x_1}, s_0 \models \hsBtu(\tilde{x_n}\rightarrow \xi_{n-1})$. So both $\Ku_{QBF}^{x_n,\cdots , x_1},$ $s_0s_{x_n}^\top\models  \xi_{n-1}$ and $\Ku_{QBF}^{x_n,\cdots , x_1},s_0s_{x_n}^\bot\models  \xi_{n-1}$.
By reasoning as in the existential case, both  
$Q_{n-1} x_{n-1} \cdots Q_1 x_1 \phi(x_n,x_{n-1},\ldots , x_1)\{x_n/\top\}$ and
$Q_{n-1} x_{n-1} \!\cdots Q_1 x_1 \phi(x_n,x_{n-1},\ldots ,\allowbreak x_1)\{x_n/\bot\}$ are true, thus $\psi=\forall x_n Q_{n-1} x_{n-1} \cdots  Q_1 x_1 \phi(x_n,x_{n-1},\ldots ,$ $x_1)$ is true.
\end{proof}

\section{Proof of Lemma~\ref{lemma:prefixSamplingOne}}\label{proof:lemma:prefixSamplingOne}

\begin{lemma*}[\ref{lemma:prefixSamplingOne}] Let $h\geq 1$, $\rho$ be a trace of $\Ku$, and let $i,j$ be two consecutive $\rho$-positions in the $h$-prefix sampling of $\rho$.
Then, for all $\rho$-positions $n,n'\in [i+1,j]$ such that $\rho(n)=\rho(n')$, it holds that $\rho(1,n)$ and $\rho(1,n')$ are $(h-1)$-prefix bisimilar.
\end{lemma*}

\begin{proof} The proof is by induction on $h\geq 1$.

Base case: $h=1$. The $1$-prefix sampling of $\rho$ is the prefix-skeleton sampling of $\rho$ in $[1,|\rho|]$. Hence, being $i$ and $j$ consecutive positions in this sampling, for each position $k\in [i,j-1]$, there is $\ell\leq i$ such that $\rho(\ell)=\rho(k)$. Since $\rho(n)=\rho(n')$, it holds that $\states(\rho(1,n))=\states(\rho(1,n'))$, and thus $\rho(1,n)$ and $\rho(1,n')$ are $0$-prefix bisimilar.

Inductive step: $h>1$. By definition of $h$-prefix sampling, there are two consecutive positions $i',j'$ in the $(h-1)$-prefix
  sampling of $\rho$ such that    $i,j$ are consecutive positions of the prefix-skeleton sampling of $\rho(i',j')$.

  If $i=i'$, then $j=i+1$, and thus, being $n,n'\in [i+1,j]$, we get that $n=n'$, and the result trivially holds.

  Let $i\neq i'$, thus $i>i'$.
     As in the base case, we easily deduce that $\rho(1,n)$ and $\rho(1,n')$ are $0$-prefix bisimilar. It remains to show that
for each proper prefix $\nu$ of $\rho(1,n)$ (resp., $\nu'$ of $\rho(1,n')$), there is a proper prefix $\nu'$ of $\rho(1,n')$ (resp., $\nu$ of $\rho(1,n)$) such that $\nu$ and $\nu'$ are $(h-2)$-prefix bisimilar. Let us consider a proper prefix $\nu$ of $\rho(1,n)$ (the proof for the other direction is symmetric). It holds that $\nu = \rho(1,m)$, for some $m<n$. We distinguish two cases:
\begin{itemize}
  \item $m\leq i$. Hence, $\rho(1,m)$ is a proper prefix of $\rho(1,n')$ and the result follows.
  \item $m>i$. Since $i$ and $j$ are consecutive positions of the prefix-skeleton sampling of $\rho(i',j')$, $i>i'$,
   and $m\in [i+1,j-1]$ (hence $m<j'$), there is $m'\in [i'+1,i]$ such that $\rho(m')=\rho(m)$ and $m'$ is in the prefix-skeleton sampling of $\rho(i',j')$. Let $\nu'=\rho(1,m')$. Clearly, $\nu'$ is a proper prefix of $\rho(1,n')$ (as $n'\geq i+1$). Moreover, since
       $m,m'\in [i'+1,j']$ and $i',j'$ are consecutive positions in the $(h-1)$-prefix sampling of $\rho$, by the inductive   hypothesis, $\nu=\rho(1,m)$ and $\nu'=\rho(1,m')$ are $(h-2)$-prefix bisimilar.
 \end{itemize}
This concludes the proof of Lemma~\ref{lemma:prefixSamplingOne}.
\end{proof}

%% file: Chaps/Appendices/appendixIC17.tex
\chapter{Proofs of Chapter~\ref{chap:IC17}}
\minitoc\mtcskip

\section{Proof of Lemma~\ref{lemmaOracle}}\label{proof:lemmaOracle}

\begin{lemma*}[\ref{lemmaOracle}]
Let $\Ku=\KuDef$ be a finite Kripke structure, $\psi$ be an $\AAbarB$ formula, and $V_{\A}(\cdot,\cdot)$, $V_{\Abar}(\cdot,\cdot)$ be two Boolean arrays. We assume that 
\begin{enumerate}
	\item for each $\hsA \phi\in\mods(\psi)$ and $s'\in \States$, $V_{\A}(\phi,s')=\top$ if and only if there exists $\rho\in\Trk_{\Ku}$ such that $\fst(\rho)=s'$ and $\Ku,\rho\models \phi$, and
	\item for each $\hsAt \phi\in\mods(\psi)$ and $s'\in \States$, $V_{\Abar}(\phi,s')=\top$ if and only if there exists $\rho\in\Trk_{\Ku}$ such that $\lst(\rho)=s'$ and $\Ku,\rho\models \phi$.
\end{enumerate}
Then \texttt{Oracle}$(\Ku,\psi,s,\textsc{direction},V_{\A}\cup V_{\Abar})$ features a successful computation (returning $\top$) if and only if:
\begin{itemize}
	\item there exists $\rho\in\Trk_{\Ku}$ such that $\fst(\rho)=s$ and $\Ku,\rho\models \psi$, when \textsc{direction} is \textsc{forward};
	\item there exists $\rho\in\Trk_{\Ku}$ such that $\lst(\rho)=s$ and $\Ku,\rho\models \psi$, when \textsc{direction} is \textsc{backward}.
\end{itemize}
\end{lemma*}

\begin{proof}
 It is easy to check that if $\tilde{\rho}$ is the trace non-deterministically generated by \texttt{A\_trace} at line 1  then, for $i=1,\ldots, |\tilde{\rho}|$, it holds that $\Ku,\tilde{\rho}(1,i)\models\phi$ if and only if $T[\phi,i]=\top$, either by hypothesis, when  $\phi$ occurs in  $\mods(\psi)$ (lines 2--7), or by construction, when $\phi$ does not occur in $\mods(\psi)$ (lines 8--22).

Let us now assume that the value of the parameter \textsc{direction} is \textsc{forward} (the proof for the other direction is analogous).

\begin{itemize}
	\item[$(\Rightarrow)$] If \texttt{Oracle}$(\Ku,\psi,s,\textsc{forward},V_{\A}\cup V_{\Abar})$ features a successful computation, it means that there exists a trace $\tilde{\rho} \in\Trk_{\Ku}$ (generated at line 1) such that $\fst(\tilde{\rho})=s$ and $T[\psi,|\tilde{\rho}|]=\top$ implying that $\Ku,\tilde{\rho}\models\psi$.

	\item[$(\Leftarrow)$] If there exists $\rho\in\Trk_{\Ku}$ such that $\fst(\rho)=s$ and $\Ku,\rho\models \psi$, by Theorem~\ref{theorem:polynomialSizeModelProperty} there exists $\tilde{\rho}\in\Trk_{\Ku}$ such that $\Ku,\tilde{\rho}\models \psi$, $\fst(\tilde{\rho})=\fst(\rho)$, and $|\tilde{\rho}|\leq
	|\States|\cdot (2|\psi|+1)^2$.
	It follows that in some non-deterministic instance of \texttt{Oracle}$(\Ku,\psi,s,\textsc{forward},V_{\A}\cup V_{\Abar})$, $\texttt{A\_trace}(\Ku,s,|\States|\cdot(2|\psi|+1)^2,\textsc{forward})$ returns such $\tilde{\rho}$ (at line 1). Finally, we have that $T[\psi,|\tilde{\rho}|]=\top$ as $\Ku,\tilde{\rho}\models \psi$, and hence the considered instance of \texttt{Oracle}$(\Ku,\psi,s,\allowbreak \textsc{forward},V_{\A}\cup V_{\Abar})$ is successful.\qedhere
\end{itemize}
\end{proof}

\section{Proof of Theorem~\ref{th:cx}}\label{proof:th:cx}

\begin{theorem*}[\ref{th:cx}]
Let $\psi$  be an $\AAbar$ formula and  $\Ku=(\Prop,\States,\Edges,\Lab,s_1)$ be a finite Kripke structure.
For every block $B$ of $T_{\Ku,\neg\psi}$,
if $B$ is associated with an $\AAbar$ formula $\varphi$, then
\begin{itemize}
	\item if $B$ is a \forw{} block, for all $i\in\{1,\ldots,|\States|\}$, $B(z_i)=\top$ if and only if there exists a trace $\rho\in\Trk_\Ku$ such that $\fst(\rho)=s_i$ and $\Ku,\rho\models\varphi$;
	\item if $B$ is a \back{} block, for all $i\in\{1,\ldots,|\States|\}$, $B(z_i)=\top$ if and only if there exists a trace $\rho\in\Trk_\Ku$ such that $\lst(\rho)=s_i$ and $\Ku,\rho\models\varphi$.
\end{itemize}
\end{theorem*}

\begin{proof}
The proof is by induction on the level $L \geq 1$ of the block $B$. The proof of the base case, i.e., for $L=1$, is just a simpler version of the inductive step and it is therefore omitted.

Assume that $B$ is a \forw{} block  at level $L\geq 2$ associated with a formula $\varphi$ (the \back{} case is symmetric).

We first prove the implication $(\Leftarrow$). We have to show that if there exists a trace $\rho\in\Trk_\mathpzc{K}$ such that $\fst(\rho)=s_i$ (for some $i\in\{1,\ldots,|\States|\}$) and $\mathpzc{K},\rho\models\varphi$, then $B(z_i)=\top$ that is, there exists a truth assignment $\omega$ to the variables in $V$ satisfying the formula $F_i(Y,V)$ of
$G_i$. In \cite{MMP15}, it is proved that if $\varphi$ is an $\AAbar$ formula and $\mathpzc{K},\rho\models\varphi$ (as in this case), there exists a trace $\rho'\in\Trk_\mathpzc{K}$, with $|\rho'|\leq |\States|^2+2$, such that $\fst(\rho)=\fst(\rho')=s_i$, $\lst(\rho)=\lst(\rho')$, and $\mathpzc{K},\rho'\models\varphi$. 
Thus, by Proposition~\ref{remk}, there exists a truth assignment $\omega$ to the variables in $V$, that satisfies $trace(V_{trace},V_{last},V_{\mathpzc{AP}})$, such that for all $1 \leq r \leq |\rho'|$ and $1 \leq j \leq |\States|$, $\rho(r)=s_j\iff\omega(v_j^r)=\top$ and $\omega(v_j^{|\rho|})=\omega(v_j)$, and for all $p\in\mathpzc{AP}$, $\omega(v_p)=\top\iff\mathpzc{K},\rho'\models p$ $(\star)$. 

Since $L\geq 2$, it holds that $\mods(\varphi)\neq \emptyset$. Let us consider a \forw{} child $B'$ of $B$ (if any), at a level lower than $L$, associated with some formula $\xi$ such that $\hsA\xi\in\mods(\varphi)$. By the inductive hypothesis, for all $j$, $B'(z_j)=\top$ if and only if there exists a trace $\overline{\rho}\in\Trk_\mathpzc{K}$ such that $\fst(\overline{\rho})=s_j$ and $\mathpzc{K},\overline{\rho}\models\xi$. Thus, $\mathpzc{K},\rho'\models\hsA\xi$ if and only if there exists $\tilde{\rho}\in\Trk_\mathpzc{K}$, with $\fst(\tilde{\rho})=\lst(\rho')=s_j$, for some $j$, and $\mathpzc{K},\tilde{\rho}\models\xi$ if and only if $B'(z_j)(=y_j^\xi)=\top$. Thus if $\mathpzc{K},\rho'\models\hsA\xi$, then $y_j^\xi=\top$, and $\omega(v_j)\wedge y_j^\xi=\top$. Now, to satisfy $F_i(Y,V)$, the truth assignment $\omega$ has to be such that $\omega(v_{\hsA\xi})=\top$. If $\mathpzc{K},\rho'\not\models\hsA\xi$, then $y_j^\xi=\bot$, thus $\bigvee_{u=1}^{|\States|} (\omega(v_u) \wedge y_u^{\xi})$ is false, and $\omega$ must be such that $\omega(v_{\hsA\xi})=\bot$. To conclude, $\mathpzc{K},\rho'\models\hsA\xi$ if and only if $\omega(v_{\hsA\xi})=\top$ $(\star\star)$. The symmetric reasoning can be applied to \back{} children of $B$.
Since $\mathpzc{K},\rho'\models\varphi$, by $(\star)$ and $(\star\star)$, we have $\omega(\overline{\varphi}(V_{\mathpzc{AP}},V_{modSubf}))=\top$.

We prove now the implication $(\Rightarrow$). If $B(z_i)=\top$, then there exists a truth assignment $\omega$ of $V$ satisfying $F_i(Y,\! V)$. In particular, $\omega$ satisfies $trace(V_{trace},\! V_{last},\! V_{\mathpzc{AP}})$ and $v_i^1$, thus, by Proposition~\ref{remk}, there exists a trace $\rho\in\Trk_\mathpzc{K}$ such that $\fst(\rho)=s_i$, $\lst(\rho)=s_j$, for some $j$, and $\mathpzc{K},\rho\models p\iff \omega(v_p)=\top$, for any $p\in\mathpzc{AP}$. 
By the inductive hypothesis, for all the formulas $\hsA\xi\in\mods(\varphi)$, $\mathpzc{K},\rho\models\hsA\xi$ if and only if $\omega(v_{\hsA\xi})=\top$, and symmetrically, for all $\hsAt\xi'\in\mods(\varphi)$, $\mathpzc{K},\rho\models\hsAt\xi'$ if and only if $\omega(v_{\hsAt\xi'})=\top$. 
Since $\omega(\overline{\varphi}(V_{\mathpzc{AP}},V_{modSubf}))=\top$, then we have $\mathpzc{K},\rho\models \varphi$.
\end{proof}

\section{Proof of Theorem~\ref{thcorr}}\label{proof:thcorr}

\begin{theorem*}[\ref{thcorr}]
Let $\mathcal{I}$ be an instance of SNSAT, with $|\mathcal{I}|=n$, and let $\Ku_\mathcal{I}$ and $\mathcal{F}_\mathcal{I}$
be defined as above. For all $0\leq k\leq n+1$ and all $r=1,\ldots , n$, it holds that:
	\begin{enumerate}
		\item if $k\geq r$, then $v_\mathcal{I}(x_r)=\top \iff \Ku_\mathcal{I},w_{x_r}\models \psi_k;$
		\item if $k\geq r+1$, then $v_\mathcal{I}(x_r)=\bot \iff \Ku_\mathcal{I},\overline{w_{x_r}}\models \psi_k.$
	\end{enumerate}
\end{theorem*}

\begin{proof}
The proof is by induction on $k\geq 0$. If $k=0$, the thesis trivially holds.
Therefore, let us assume that $k\geq 1$. We first prove the $(\Leftarrow)$ implication for both (1.) and (2.).
\begin{itemize}
\item[(1.)] Assume that $k\geq r$ and $\Ku_\mathcal{I},w_{x_r}\models \psi_k$. Thus, there exists $\rho\in\Trk_{\Ku_\mathcal{I}}$ such that $\rho=w_{x_r}\cdots s_0$ does not pass through any $\overline{s_m}$, for $1\leq m\leq r$, and $\Ku_\mathcal{I},\rho\models \varphi_k$. We show by induction on $1\leq m\leq r$ that $\omega_\rho(x_m)=v_\mathcal{I}(x_m)$. 
	\begin{itemize}
		\item Let us consider first the case where $\rho$ passes through $w_{x_m}$, implying that $\omega_\rho(x_m)=\top$; thus $\Ku_\mathcal{I},\rho\!\models\! x_m\wedge \neg r_m$ and $\Ku_\mathcal{I},\rho\!\models\! F_m(x_1,\ldots ,x_{m-1},\allowbreak Z_m)$. If $m=1$ (base case), since $F_1$ is satisfiable, then $v_\mathcal{I}(x_1)=\top$. If $m\geq 2$ (inductive step), by the inductive hypothesis $\omega_\rho(x_1)=v_\mathcal{I}(x_1)$, \dots , $\omega_\rho(x_{m-1})=v_\mathcal{I}(x_{m-1})$. Since $\Ku_\mathcal{I},\rho\models F_m(x_1,\ldots ,x_{m-1},Z_m)$ or, equivalently, $F_m(\omega_\rho(x_{1}),\ldots ,\allowbreak \omega_\rho(x_{m-1}), \omega_\rho(Z_m))=\top$, it holds that $F_m(v_\mathcal{I}(x_{1}),\ldots , v_\mathcal{I}(x_{m-1}), \allowbreak  \omega_\rho(Z_m))=\top$ and, by definition of $v_\mathcal{I}$, we have $v_\mathcal{I}(x_m)=\top$.
		
		\item Conversely, let us consider the case where $\rho$ passes through $\overline{w_{x_m}}$, implying that $\omega_\rho(x_m)=\bot$ and $m<r$, as we are assuming $\fst(\rho)=w_{x_r}$. In this case, the prefix $w_{x_r}\cdots \overline{w_{x_m}}$ of $\rho$ satisfies both $\bigvee_{i=1}^n \hsA p_{\overline{x_i}}$ and $\hsA\big(\neg s \wedge \Length_2\wedge \hsA (\Length_2\wedge \neg\psi_{k-1})\big)$. Therefore, $\Ku_\mathcal{I},\overline{w_{x_m}}\cdot \overline{s_m}\models \hsA (\Length_2\wedge \neg\psi_{k-1})$ and $\Ku_\mathcal{I}, \overline{s_m}\cdot w_{x_m}\not\models \psi_{k-1}$, with $\psi_{k-1}=\hsA\varphi_{k-1}$. Hence $\Ku_\mathcal{I}, w_{x_m}\not\models \psi_{k-1}$. Since $1\leq m<r$, we have $1\leq m<r\leq k$, thus $k'=k-1\geq m\geq 1$. By the inductive hypothesis (on $k'=k-1$), we get that $v_\mathcal{I}(x_m)=\bot$.
	\end{itemize}
Therefore $v_\mathcal{I}(x_r)=\omega_\rho(x_r)$ and, since $w_{x_r}\in\states(\rho)$, we have $\omega_\rho(x_r)=\top$ and then $v_\mathcal{I}(x_r)=\top$ proving the thesis.
	
\item[(2.)] Assume that $k\geq r+1$ and $\Ku_\mathcal{I},\overline{w_{x_r}}\models \psi_k$. The proof follows the same steps as the previous case and it is thus only sketched: there exists $\rho\in\Trk_{\Ku_\mathcal{I}}$ such that $\rho=\overline{w_{x_r}}\cdots s_0$ does not pass through any $\overline{s_m}$, for $1\leq m\leq r$, and $\Ku_\mathcal{I},\rho\models \varphi_k$. The only difference is that the prefix $\overline{w_{x_r}}$ satisfies $\bigvee_{i=1}^n \hsA p_{\overline{x_i}}$, thus, as before, we get $\Ku_\mathcal{I}, w_{x_r}\not\models \psi_{k-1}$. Now, $k'=k-1\geq r\geq 1$ and, by the inductive hypothesis (on $k'=k-1$),  $v_\mathcal{I}(x_r)=\bot$.
\end{itemize}

We prove now the converse implication $(\Rightarrow)$ for both (1.) and (2.).
\begin{itemize}
\item[(1.)] Assume that $k\geq r$ and $v_\mathcal{I}(x_r)=\top$. Let us consider the trace $\rho\in\Trk_{\Ku_\mathcal{I}}$, $\rho=w_{x_r}\cdots s_0$ never passing through any $\overline{s_m}$, for $1\leq m\leq r$, such that $w_{x_m}\in\states(\rho)$ if $v_\mathcal{I}(x_m)=\top$, and $\overline{w_{x_m}}\in\states(\rho)$ if $v_\mathcal{I}(x_m)=\bot$, for $1\leq m\leq r$. Such a choice of 
$\rho$ ensures that $v_\mathcal{I}(x_m)=\omega_\rho(x_m)$. In addition, the choice of $\rho$ has to induce also the proper truth-assignment of private variables,
that is, if $v_\mathcal{I}(x_m)=\top$, then for $1\leq u_m\leq j_m$, $w_{z_m^{u_m}}\in\states(\rho)$ if $F_m(v_\mathcal{I}(x_1),\ldots , v_\mathcal{I}(x_{m-1}),Z_m)$ is satisfied for $z_m^{u_m}= \top$, and $\overline{w_{z_m^{u_m}}}\in\states(\rho)$ otherwise. Note that such a choice of $\rho$ is always possible.
We have to show that $\Ku_\mathcal{I},\rho\models \varphi_k$, hence $\Ku_\mathcal{I},w_{x_r}\models \psi_k$.
\begin{itemize}
	\item For all $1\leq m\leq r$ such that $v_\mathcal{I}(x_m)=\top$, it holds that $F_m(v_\mathcal{I}(x_1),\ldots ,\allowbreak v_\mathcal{I}(x_{m-1}), Z_m)$ is satisfiable. Hence, by our choice of $\rho$, $F_m(\omega_\rho(x_1),\ldots ,\allowbreak \omega_\rho(x_{m-1}),\omega_\rho(Z_m))\!=\!\top$, or, equivalently, $\Ku_\mathcal{I},\rho\!\models\! F_m(x_1, \ldots, x_{m-1}, Z_m)$. Thus, $\Ku_\mathcal{I},\rho\models \bigwedge_{i=1}^n \Big((x_i\wedge \neg r_i)\rightarrow F_i(x_1, \ldots, x_{i-1}, Z_i)\Big)$. 

	\item Conversely, for all $1\leq m< r$ such that $v_\mathcal{I}(x_m)=\bot$ ($m\neq r$ as, by hypothesis, $v_\mathcal{I}(x_r)=\top$), it holds that $\overline{w_{x_m}}\in\states(\rho)$. Since $m<r$, we have $k\geq r>m$ and $k-1\geq m\geq 1$. By the inductive hypothesis, $\Ku_\mathcal{I},w_{x_m}\not\models \psi_{k-1}$. It follows that $\Ku_\mathcal{I},\overline{s_m}\cdot w_{x_m}\models \neg\psi_{k-1}\wedge \Length_2$, $\Ku_\mathcal{I},\overline{w_{x_m}}\cdot\overline{s_m}\models \neg s\wedge \Length_2\wedge\hsA(\neg\psi_{k-1}\wedge \Length_2)$ and $\Ku_\mathcal{I},\overline{w_{x_m}}\models\hsA( \neg s\wedge\Length_2\wedge\hsA(\neg\psi_{k-1}\wedge \Length_2))$. Hence, $\Ku_\mathcal{I},\rho\models \hsBu((\bigvee_{i=1}^n \hsA p_{\overline{x_i}})\rightarrow\hsA( \neg s\wedge\Length_2\wedge\hsA(\neg\psi_{k-1}\wedge \Length_2)))$.
\end{itemize} Combining the two cases, we can conclude that $\Ku_\mathcal{I},\rho\models \varphi_k$.
\item[(2.)] Assume that $k\geq r+1$ and $v_\mathcal{I}(x_r)=\bot$. The proof is as before and it is only sketched. In this case, we choose a trace $\rho=\overline{w_{x_r}}\cdots s_0$. Since $k'=k-1\geq r$, by the inductive hypothesis, $\Ku_\mathcal{I},w_{x_r}\not\models \psi_{k-1}$, and we can prove that $\Ku_\mathcal{I},\overline{w_{x_r}}\models\hsA( \neg s\wedge\Length_2\wedge\hsA(\neg\psi_{k-1}\wedge \Length_2))$.\qedhere
\end{itemize}
\end{proof}

%% file: Chaps/Appendices/appendixTOCL17.tex
\chapter{Proofs of Chapter~\ref{chap:TOCL17}}
\minitoc\mtcskip

\section{Proof of Lemma~\ref{lemma1:CharacterizationFinitaryLTL}}\label{proof:lemma1:CharacterizationFinitaryLTL}

\begin{lemma*}[\ref{lemma1:CharacterizationFinitaryLTL}] Let $\Sigma$ be a finite alphabet,
  $b\in \Sigma$, $\Gamma=\Sigma\setminus\{b\}$, $\Lang\subseteq \Gamma^{+}$, and $\psi$ be a $\BE$ formula over $\Gamma$ such that $\Lang_\act(\psi)=\Lang$. Then,
  there are $\BE$ formulas defining (under the action-based semantics) the languages
   $b\Lang$, $\Sigma^* b\Lang$, $\Sigma^* b(\Lang+\varepsilon)$,
   $\Lang b$, $ \Lang b \Sigma^* $, $ (\Lang+\varepsilon) b \Sigma^* $, and $b \Lang b$.
  \end{lemma*}
  
  \begin{proof}
We focus on the cases for the languages $b\Lang$, $\Sigma^* b\Lang$, $\Sigma^* b$, and  $b \Lang b$ (for the other languages, the proof is similar: $\Sigma^*b(\Lang + \varepsilon)=\Sigma^* b\Lang+ \Sigma^* b$, $\Lang b$ is symmetric to $b\Lang$, $\Lang b\Sigma^*$ to $\Sigma^* b\Lang$, and $ (\Lang+\varepsilon) b \Sigma^* $ to $\Sigma^*b(\Lang + \varepsilon)$). 

Let $\psi$ be a $\BE$ formula over $\Gamma$ such that $\Lang_\act(\psi)=\Lang$. 

\paragraph{Language $b\Lang$.} The $\BE$ formula defining the language $b\Lang$ is:
\begin{equation}
\label{eq-bL}
(\neg \Length_1 \wedge \hsB b \wedge \hsEu(\neg b \wedge \hsBu\neg b)) \wedge h_b(\psi),
\end{equation}
where the  formula  $h_b(\psi)$ is inductively defined on the structure of $\psi$ in the following way. The mapping $h_b$ is homomorphic with respect to the Boolean connectives, while for the atomic actions in $\Gamma$ and the modalities $\hsE$ and $\hsB$, it is defined  as follows:
 \begin{itemize}
   \item for all $a\in \Gamma$, $h_b(a)= a\vee (\hsB b \wedge \hsE a \wedge \hsEu a)$;
   \item $h_b(\hsB\theta)= (\hsB h_b(\theta) \wedge \neg\hsB b) \vee \hsB(h_b(\theta)\wedge \hsB b)$;
   \item $h_b(\hsE\theta)= (\hsE h_b(\theta) \wedge \neg\hsB b) \vee (\hsB b \wedge \hsE \hsE h_b(\theta))$.
 \end{itemize}
The first conjunct of (\ref{eq-bL}) ensures that a word $u'$ in the defined language has length at least $2$ and it has the form $bu$ without any occurrence of $b$ in $u$. The second conjunct $h_b(\psi)$ ensures that $u$ belongs to the language defined by $\psi$. For atomic actions and temporal modalities, $h_b(\psi)$ is a disjunction of two possible choices; the appropriate one is forced at top level by the first conjunct of (\ref{eq-bL}), that constrains one and only one $b$ to occur in the word in the first position.

By a straightforward structural induction on $\psi$, it can be shown that the following fact holds.

\begin{claim}\label{cl:claim1} Let $u\in \Gamma^{+}$, $u'= bu$, and $|u|= n+1$. Then, for all $i,j \in [0,n]$ with $i\leq j$, $u(i,j) \in \Lang_\act(\psi)$
if and only if $u'(\hat{i},j+1)\in \Lang_\act(h_b(\psi))$, where  $\hat{i}= i$ if $i=0$, and $\hat{i}=i+1$ otherwise.
\end{claim}

 By Claim~\ref{cl:claim1}, for each $u\in \Gamma^{+}$, $u\in \Lang_\act(\psi)$ if and only if $bu\in \Lang_\act(h_b(\psi))$. Therefore, (\ref{eq-bL}) captures the language $b\Lang_\act(\psi)$. 

\paragraph{Languages $\Sigma^* b\Lang$ and $\Sigma^* b$.} Following the proof given for the case of the language $b\Lang$, with $\Lang\subseteq \Gamma^{+}$, one can construct a
 $\BE$ formula  $\varphi$ defining the language $b\Lang$. Hence, the $\BE$ formula $\varphi\vee \hsE\varphi$ defines $\Sigma^* b\Lang$. The $\BE$ formula defining $\Sigma^* b$ is $b \vee \hsE b$. 

\paragraph{Language $b\Lang b$.} By the proof given for the language $b\Lang$, with $\Lang\subseteq \Gamma^{+}$, one can build a $\BE$ formula  $\varphi$ defining the language $b\Lang$. The $\BE$ formula defining the language $b\Lang b$ is the formula:
\begin{equation}
\label{eq-bLb}
(\neg \Length_1 \wedge \neg \Length_2 \wedge \hsB b \wedge \hsE b \wedge \hsEu\hsBu\neg b) \wedge k_b(\varphi)
\end{equation}
where the formula  $k_b(\varphi)$ is inductively defined on the structure of $\varphi$  in the following way. 
The mapping $k_b$ is homomorphic with respect to the Boolean connectives, while for the atomic actions in $\Sigma$ and the modalities $\hsE$ and $\hsB$, it is defined  as follows:
 \begin{itemize}
   \item for all $a\in \Gamma$, $k_b(a)= a\vee (\hsE b \wedge \hsB a \wedge \hsBu a)$;
   \item $k_b(b)=b$;
   \item $k_b(\hsB\theta)= (\hsB k_b(\theta) \wedge \neg\hsE b) \vee (\hsE b \wedge \hsB \hsB k_b(\theta))$.
      \item $k_b(\hsE\theta)= (\hsE k_b(\theta)\wedge \neg\hsE b) \vee \hsE(k_b(\theta)\wedge \hsE b)$.
 \end{itemize}

The first conjunct of (\ref{eq-bLb}) ensures that a word $u'$ in the defined language has length at least $3$ and it has the form $bub$ without any occurrence of $b$ in $u$. The second conjunct $k_b(\varphi)$ ensures that $bu$ belongs to the language defined by $\varphi$. Similarly to the case of the language $b\Lang$, for atomic actions (different from $b$) and temporal modalities, $k_b(\psi)$ is a disjunction of two possible choices; the appropriate one is forced at top level by the first conjunct of  (\ref{eq-bLb}), that constrains one and only one $b$ to occur in the word in the last position.

By a straightforward structural induction on $\varphi$, it can be shown that the following fact holds.

\begin{claim}\label{cl:claim2} Let $u\in \Gamma^{+}$ and $|bu|= n+1$. Then, for all $i,j \in [0,n]$ with $i\leq j$, $bu(i,j) \in \Lang_\act(\varphi)$
if and only if $bub(i,\hat{j})\in \Lang_\act(k_b(\varphi))$ where  $\hat{j}= j$ if $j<n$, and $\hat{j}=n+1$ otherwise.
\end{claim}

 By Claim~\ref{cl:claim2}, for each $u\in \Gamma^{+}$, $bu\in \Lang_\act(\varphi)$ if and only if $bub\in \Lang_\act(k_b(\varphi))$ implying that the formula of (\ref{eq-bLb})
 defines the language $\Lang_\act(\varphi)b$. 
%
\end{proof}

\section{Proof of Lemma~\ref{lemma2:CharacterizationFinitaryLTL}}\label{proof:lemma2:CharacterizationFinitaryLTL}

\begin{lemma*}[\ref{lemma2:CharacterizationFinitaryLTL}] Let $\Sigma$ and $\Delta$ be finite alphabets,
  $b\in \Sigma$, $\Gamma=\Sigma\setminus\{b\}$, $U_0= \Gamma^{*}b$, $h_0:U_0 \rightarrow \Delta$ and $h:U_0^{+} \rightarrow \Delta^{+}$ be defined by
    $h(u_0u_1\cdots u_n)=h_0(u_0)\cdots h_0(u_n)$. Assume that, for each $d\in \Delta$, there is a $\BE$ formula capturing the language $\Lang_d =\{u\in \Gamma^{+}\mid h_0(ub)= d\}$.  Then, for each $\BE$ formula $\varphi$ over $\Delta$, one can construct a $\BE$ formula over $\Sigma$
    capturing
    the language $\Gamma^{*}b h^{-1}(\Lang_\act(\varphi))\Gamma^{*}$.
  \end{lemma*}
  
  \begin{proof} By hypothesis and Lemma~\ref{lemma1:CharacterizationFinitaryLTL}, for each $d\in \Delta$ there exists a
  $\BE$ formula $\theta_d$ over $\Sigma$ defining the language $b \Lang_d b$, where $\Lang_d =\{u\in \Gamma^{+}\mid h_0(ub)= d\}$.
  Hence there exists a
  $\BE$ formula $\hat{\theta}_d$ over $\Sigma$ capturing the language $b \hat{\Lang}_d b$, where $\hat{\Lang}_d =\{u\in \Gamma^{*}\mid h_0(ub)= d\}$ (note that
 $\Lang_d= \hat{\Lang}_d\setminus\{\varepsilon\}$).

 Let $\varphi$ be a  $\BE$ formula over $\Delta$. By structural induction over $\varphi$, we construct a $\BE$ formula $\varphi^{+}$ over $\Sigma$
 such that $\Lang_\act(\varphi^{+})= \Gamma^{*}b h^{-1}(\Lang_\act(\varphi))\Gamma^{*}$.
The formula $\varphi^{+}$ is defined as follows:
\begin{itemize}
  \item $\varphi= d$ with $d\in \Delta$. We have that $\Lang_\act(d)=d^{+}$ and $\Gamma^{*}b h^{-1}(\Lang_\act(d))\Gamma^{*}$ is the set of finite words in
  $\Gamma^{*}b \Sigma^{*} b\Gamma^{*}$ such that each subword $u(i,j)$ of $u$ which is in $b\Gamma^{*}b$ is in $b\hat{\Lang}_d b$ as well. Using the formula  $\psi_b:= \neg \Length_1 \wedge \hsB b \wedge \hsE b \wedge \hsEu\hsBu\neg b$ to define the language $b\Gamma^{*}b$, $\varphi^{+}$ is defined as follows:
  \[
  \varphi^{+} =(\hsG\psi_b) \wedge \hsGu(\psi_b \rightarrow \hat{\theta}_d).
  \]
  \item $\varphi = \neg \theta$. We have that
  \begin{multline*}
 \Gamma^{*}b h^{-1}(\Lang_\act(\varphi))\Gamma^{*} = 
 \Gamma^{*}b h^{-1}(\Delta^+ \setminus \Lang_\act(\theta))\Gamma^{*} = \\
 \Gamma^{*}b h^{-1}(\Delta^+)\Gamma^{*} \cap \overline{\Gamma^{*}b  h^{-1}(\Lang_\act(\theta))\Gamma^{*}}, 
  \end{multline*}
  where $\Gamma^{*}b h^{-1}(\Delta^+)\Gamma^{*}$ restricts the set of \lq\lq candidate\rq\rq{} models to the well-formed ones.
  
  Thus, taking $\psi_b$ as defined in the previous case, $\varphi^{+}$ is given by: 
  \[\varphi^{+}=(\hsG\psi_b) \wedge \hsGu(\psi_b \rightarrow \bigvee_{d\in \Delta}\hat{\theta}_d) \wedge \neg \theta^{+},\]
  where, by the inductive hypothesis, $\Lang_\act(\theta^{+})=\Gamma^{*}b  h^{-1}(\Lang_\act(\theta))\Gamma^{*}$. 
  \item $\varphi = \theta\wedge \psi$. We simply have $\varphi^{+}= \theta^{+}\wedge \psi^{+}$.
  \item $\varphi =\hsB \theta$. First, we note that $\Gamma^{*}b h^{-1}(\Lang_\act(\hsB \theta))\Gamma^{*}$ is the set of finite words in the language  $\Gamma^{*}b h^{-1}(\Lang_\act(\theta))h^{-1}(\Delta^+)\Gamma^{*}$, which is included in the
  language  $\Gamma^{*}bh^{-1}(\Delta^+)\Gamma^{*}$ defined by the formula
  $ \hsGu(\psi_b \rightarrow \bigvee_{d\in \Delta}\hat{\theta}_d).$ Note also that,
  by the inductive hypothesis, $\Gamma^{*}b h^{-1}(\Lang_\act(\theta))$ is included in the language of $\theta^{+}$. 
  Thus, $\varphi^{+}$ is given by (for $\xi=(\hsE b) \wedge  \hsB(\theta^{+}\wedge \hsE b)$):
 \[
 \varphi^{+} =  \hsGu(\psi_b \rightarrow \bigvee_{d\in \Delta}\hat{\theta}_d) \wedge (\xi\vee \hsB\xi).
 \]
 %
%
  \item $\varphi =\hsE \theta$. $\Gamma^{*}b h^{-1}(\Lang_\act(\hsE \theta))\Gamma^{*}$ is the set  $\Gamma^{*}b h^{-1}(\Delta^+) h^{-1}(\Lang_\act(\theta))\Gamma^{*}$ included in the
  language  $\Gamma^{*}bh^{-1}(\Delta^+)\Gamma^{*}$, symmetrically to the previous case.
  %
  %
  Thus, $\varphi^{+}$ is given by (for $\xi'=(\hsB b) \wedge  \hsE(\theta^{+}\wedge \hsB b)$):
 \[
 \varphi^{+} =  \hsGu(\psi_b \rightarrow \bigvee_{d\in \Delta}\hat{\theta}_d) \wedge (\xi'\vee \hsE\xi'). \qedhere
\]
 \end{itemize}
%
  \end{proof}

\section{Proof of Lemma~\ref{lemma:UsingSeparationHybridCTLPreliminary}}\label{proof:lemma:UsingSeparationHybridCTLPreliminary}
\begin{lemma*}[\ref{lemma:UsingSeparationHybridCTLPreliminary}] 
Let $\psi$ be  a \emph{simple} hybrid $\CTLStarLP$ formula  with respect to $x$. 
Then $(\Eventually^{-} x)\wedge \psi$ is congruent to a formula of the form $(\Eventually^{-} x)\wedge \xi$, where $\xi$ is a Boolean combination of the atomic formula $x$ and $\CTLStar$ formulas. 
\end{lemma*}

\begin{proof} Let  $\psi$ be  a \emph{simple} hybrid $\CTLStarLP$ formula  with respect to $x$. From a syntactic point of view, $\psi$ is not, in general, a $\CTLStar$ formula due to the occurrences of the free variable $x$. We show that these occurrences can be separated whenever $\psi$ is paired with $\Eventually^{-} x$, obtaining a Boolean combination of the atomic formula $x$ and $\CTLStar$ formulas. 

The base case where $\psi=x$, $\psi=p\in\Prop$, or $\psi=\EQ\psi'$, is obvious.

As for the inductive step, let $\psi$ be a Boolean combination of simple hybrid $\CTLStarLP$ formulas $\theta$, where $\theta$ is either $p\in\Prop$, the variable $x$, a $\CTLStar$ formula, or a simple hybrid $\CTLStarLP$ formula (with respect to $x$) of the forms $\Next\theta_1$ or 
$\theta_1\until \theta_2$. Thus, we just need to consider the cases where $\theta =\Next \theta_1$ or $\theta =\theta_1\until \theta_2$.

Let us consider the case $\theta =\Next \theta_1$. Since there are no past temporal modalities in $\theta_1$, 
$\Next \theta_1$ forces the free occurrence of $x$ in $\psi$ to be interpreted in a (strictly) future position. 
However, $\psi$ is conjunct with the formula $\Eventually^{-} x$, which turns out to be false when $x$ is associated with a (strictly) future position.  
Let us denote by $\widehat{\theta}$  the $\CTLStar$ formula obtained from $\theta$ by replacing each occurrence of $x$ in $\psi$ with $\bot$ (false).
Now, when $x$ is mapped to a (strictly) future position, $\Eventually^{-} x$ is false and, when $x$ is mapped to a present/past position, $\Eventually^{-} x$ is  true, and $\theta$ and $\widehat{\theta}$ are congruent.
As a consequence it is clear that $(\Eventually^{-} x)\wedge \theta$ is congruent to $(\Eventually^{-} x)\wedge  \widehat{\theta}$.

Let us consider the case for $\theta =\theta_1\until \theta_2$. Using the same arguments of the previous case, we have that
 $(\Eventually^{-} x)\wedge \theta$ is congruent to $(\Eventually^{-} x)\wedge (\theta_2 \vee (\theta_1\wedge\Next (\widehat{\theta_1\until \theta_2})))$. By distributivity of $\wedge$ over $\vee$, we get $((\Eventually^{-} x)\wedge\theta_2) \vee ((\Eventually^{-} x)\wedge\theta_1\wedge\Next (\widehat{\theta_1\until \theta_2})))$. The thesis follows by applying the inductive hypothesis to $(\Eventually^{-} x)\wedge\theta_2$ and to $(\Eventually^{-} x)\wedge\theta_1$, and by factorizing $\Eventually^{-} x$ (note that $\widehat{\theta_1\until \theta_2}$ is a $\CTLStar$ formula).
\end{proof}

\section{Proof of Lemma \ref{cor:UsingSeparationHybridCTL}}\label{proof:cor:UsingSeparationHybridCTL}

\begin{lemma*}[\ref{cor:UsingSeparationHybridCTL}] Let $(\Eventually^{-} x)\wedge \EQ \varphi(x)$ (resp., $\EQ\varphi$) be a well-formed formula (resp., well-formed sentence) of hybrid $\CTLStarLP$.  Then there exists a finite set $\mathpzc{H}$ of $\CTLStar$ formulas of the form $\EQ\psi$, such that $(\Eventually^{-} x)\wedge \EQ \varphi(x)$ (resp., $\EQ\varphi$) is congruent to a well-formed formula of hybrid $\CTLStarLP$ which is a Boolean combination of $\CTLStar$ formulas and (formulas that
correspond to) \emph{pure past}  $\LTLP$ formulas over the set of proposition letters $\Prop\cup  \mathpzc{H} \cup \{x\}$ (resp., $\Prop\cup  \mathpzc{H}$).
\end{lemma*} 

\begin{proof}
As in the case of Lemma~\ref{lemma:UsingSeparationHybridCTL}, we focus on well-formed formulas of the form  $(\Eventually^{-} x)\wedge \EQ \varphi(x)$ (the case of well-formed sentences of the form $\EQ\varphi$ is similar).
The proof is by induction on the nesting depth of the path quantifier $\exists$ in $\varphi(x)$. 

In the base case we have $\EQSubf(\varphi)=\emptyset$: we apply Lemma~\ref{lemma:UsingSeparationHybridCTL}, and the result follows by taking  $\mathpzc{H}=\emptyset$. 

As for the inductive step, 
we let $\EQ \psi\in \EQSubf(\varphi)$. Since $(\Eventually^{-} x)\wedge \EQ \varphi(x)$ is well-formed, either $\psi$ is a sentence, or $\psi$ has a unique free variable $y$ and $\EQ \psi(y)$ occurs in $\varphi(x)$ in the context $(\Eventually^{-} y)\wedge \EQ \psi(y)$. Assume that the latter case holds (the former is similar). 
By definition of well-formed formula, $y$ is not free in $\varphi(x)$, and $(\Eventually^{-} y)\wedge \EQ \psi(y)$ must occur in the scope of some occurrence of  $\Downarrowy$.
By the inductive hypothesis, the thesis holds for $(\Eventually^{-} y)\wedge \EQ \psi(y)$. Hence there exists 
a finite set $\mathpzc{H}'$ of $\CTLStar$ formulas of the form $\EQ\theta$ such that $(\Eventually^{-} y)\wedge \EQ \psi(y)$ is congruent to a well-formed formula of hybrid $\CTLStarLP$, say $\xi(y)$, which is a Boolean combination of $\CTLStar$ formulas and formulas that correspond to  \emph{pure past}  $\LTLP$ formulas over the set of proposition letters
$\Prop\cup  \mathpzc{H'} \cup \{y\}$. 

By replacing each occurrence of $(\Eventually^{-} y)\wedge \EQ \psi(y)$ in $\varphi(x)$ with $\xi(y)$, and repeating the procedure for all the formulas in   $\EQSubf(\varphi)$, we obtain a well-formed formula of hybrid $\CTLStarLP$ of the form $(\Eventually^{-} x)\wedge \EQ \theta(x)$ which is congruent to $(\Eventually^{-} x)\wedge \EQ \varphi(x)$ (note that the congruence relation is closed under substitution) and such that $\EQSubf(\theta)$ consists of $\CTLStar$ formulas. At this point we can apply Lemma~\ref{lemma:UsingSeparationHybridCTL} proving the assertion.
\end{proof}

\section{Proof of Lemma~\ref{lemma:MainnonBranchingExpressibilityOfEventually}}\label{proof:lemma:MainnonBranchingExpressibilityOfEventually}

To prove the lemma, we need some technical definitions. 
Let $\rho$ be a trace of $\Ku_n$ (note that $\Ku_n$ and $\mathpzc{M}_n$ feature the same traces).
By construction, $\rho$ has the form $\rho'\cdot \rho''$, where $\rho'$ is a (possibly empty) trace visiting only states where $p$ does not hold, and $\rho''$
is a (possibly empty) trace visiting only the state $t$, where $p$ holds. We say that $\rho'$ (resp., $\rho''$) is the $\emptyset$-part (resp., $p$-part) of $\rho$.

Let $N_\emptyset(\rho)$, $N_p(\rho)$, and $D_p(\rho)$ be the natural numbers defined as follows:
\begin{itemize}
  \item $N_\emptyset(\rho) = |\rho'|$ (the length of the $\emptyset$-part of $\rho$);
  \item $N_p(\rho) = |\rho''|$ (the length of the $p$-part of $\rho$);
  \item $D_p(\rho)=0$ if $N_p(\rho)>0$ (i.e., $\lst(\rho)=t$); otherwise, $D_p(\rho)$ is the length of the minimal trace starting from  $\lst(\rho)$ and leading to
  $s_{2n}$. Notice that $D_p(\rho)$ is well defined and $0\leq D_p(\rho)\leq 2n+1$.
\end{itemize}

By construction, the following property holds.

\begin{proposition}\label{remark:HcompatibilityOne} For all traces $\rho$ and $\rho'$ of $\Ku_n$, if $D_p(\rho)= D_p(\rho')$, then $\lst(\rho)=\lst(\rho')$.
\end{proposition}

Now, for each $h\in [1,n]$, we introduce the notion of \emph{$h$-compatibility} between traces of  $\Ku_n$. Intuitively, this notion provides a sufficient condition to make two traces indistinguishable under the state-based semantics by means of balanced $\HS$ formulas having size at most $h$.

\begin{definition}[$h$-compatibility] Let $h\in [1,n]$. Two traces $\rho$ and $\rho'$ of $\Ku_n$ are \emph{$h$-compatible} if the following conditions hold:
\begin{itemize}
\item $N_p(\rho) = N_p(\rho')$;
  \item either $N_\emptyset(\rho) = N_\emptyset(\rho')$, or $N_\emptyset(\rho)\geq h$ and $N_\emptyset(\rho')\geq h$;
        \item either $D_p(\rho) = D_p(\rho')$, or $D_p(\rho)\geq h$ and $D_p(\rho')\geq h$.
\end{itemize}
We denote by $\tilde{\mathpzc{R}}(h)$ the binary relation over the set of traces of $\Ku_n$ such that $(\rho,\rho')\in \tilde{\mathpzc{R}}(h)$ if and only if $\rho$ and $\rho'$ are $h$-compatible. 

Notice that $\tilde{\mathpzc{R}}(h)$ is an equivalence relation, for all $h\in [1,n]$.
Moreover $\tilde{\mathpzc{R}}(h)\subseteq \tilde{\mathpzc{R}}(h-1)$, for all $h\in [2,n]$, that is, $\tilde{\mathpzc{R}}(h)$ is a refinement of $\tilde{\mathpzc{R}}(h-1)$.
\end{definition}

By construction, the following property, that will be exploited in the proof of Lemma~\ref{lemma:MainnonBranchingExpressibilityOfEventually}, can be easily shown.

\begin{proposition}\label{remark:HcompatibilityTwo} For every trace  $\rho$ of  $\Ku_n$ starting from $s_0$ (resp., $s_1$), there exists a trace $\rho'$ of
$\Ku_n$ starting from $s_1$ (resp., $s_0$) such that $(\rho,\rho')\in \tilde{\mathpzc{R}}(n)$.
\end{proposition}

The following lemma lists some useful properties of $\tilde{\mathpzc{R}}(h)$.

 \begin{lemma}\label{lemma:Hcompatibility} Let $h\in [2,n]$ and $(\rho,\rho')\in \tilde{\mathpzc{R}}(h)$. The next properties hold:
 \begin{enumerate}
  \item for each proper prefix $\sigma$ of $\rho$, there exists a proper prefix $\sigma'$ of $\rho'$ such that $(\sigma,\sigma')\in \tilde{\mathpzc{R}}(\lfloor \frac{h}{2}\rfloor)$;
   \item for each trace of the  form $\rho\cdot \sigma$, where $\sigma$ is not empty, there exists a trace of the form $\rho'\cdot\sigma'$
    such that $\sigma'$ is not empty and $(\rho\cdot\sigma,\rho'\cdot \sigma') \in \tilde{\mathpzc{R}}(\lfloor \frac{h}{2}\rfloor)$;
 \item for each proper suffix $\sigma$ of $\rho$, there exists a proper suffix $\sigma'$ of $\rho'$ such that $(\sigma,\sigma')\in \tilde{\mathpzc{R}}(h-1)$;
   \item for each trace of the form $\sigma \cdot \rho$, where $\sigma$ is not empty, there exists a trace of the form $\sigma'\cdot \rho'$ such that $\sigma'$ is not empty and $(\sigma\cdot \rho,\sigma'\cdot \rho')\in \tilde{\mathpzc{R}}(h)$.
\end{enumerate}
 \end{lemma}
 \begin{proof} We prove (1.) and (2.); (3.) and (4.) easily follow by construction and by definition of $h$-compatibility.

 (1.) We distinguish the following cases:
 \begin{enumerate}
  \item $D_p(\rho)<h$ and $N_\emptyset(\rho)<h$. Since $(\rho,\rho')\in \tilde{\mathpzc{R}}(h)$ and $h\in [2,n]$, it holds that $D_p(\rho)=D_p(\rho')$, $N_\emptyset(\rho)=N_\emptyset(\rho')$, and $N_p(\rho)=N_p(\rho')$, and thus $\rho=\rho'$.
  \item $D_p(\rho)\geq h$. Since $(\rho,\rho')\in \tilde{\mathpzc{R}}(h)$, $D_p(\rho')\geq h$, $N_p(\rho')=N_p(\rho)=0$,  and either $N_\emptyset(\rho')=N_\emptyset(\rho)$, or $N_\emptyset(\rho)\geq h$ and $N_\emptyset(\rho')\geq h$. In both cases, by construction it easily follows that for each proper prefix $\sigma$ of $\rho$,
      there exists a proper prefix $\sigma'$ of $\rho'$ such that $(\sigma,\sigma')\in \tilde{\mathpzc{R}}(h-1)\subseteq \tilde{\mathpzc{R}}(\lfloor \frac{h}{2}\rfloor)$.
  \item $D_p(\rho)<h$ and $N_\emptyset(\rho)\geq h$. Since $(\rho,\rho')\in \tilde{\mathpzc{R}}(h)$, we have that $D_p(\rho')=D_p(\rho)$ (and hence, by Proposition~\ref{remark:HcompatibilityOne}, $\lst(\rho)=\lst(\rho')$), $N_p(\rho')=N_p(\rho)$,  and $N_\emptyset(\rho')\geq h$.
  
  Let $\sigma$ be a proper prefix of $\rho$. We distinguish the following three subcases:
 \begin{enumerate}
   \item $N_\emptyset(\sigma)< \lfloor \frac{h}{2}\rfloor$. Since $N_\emptyset(\rho)\geq h$, we have that $D_p(\sigma)\geq \lfloor \frac{h}{2}\rfloor$ and $|\sigma|=N_\emptyset(\sigma)$ (and thus $N_p(\sigma)=0$). Since
    $N_\emptyset(\rho')\geq h$, by taking the proper prefix $\sigma'$ of $\rho'$ having length $N_\emptyset(\sigma)$, we obtain that
    $(\sigma,\sigma')\in  \tilde{\mathpzc{R}}(\lfloor \frac{h}{2}\rfloor)$.
   \item $N_\emptyset(\sigma)\geq  \lfloor \frac{h}{2}\rfloor$ and $D_p(\sigma)\geq  \lfloor \frac{h}{2}\rfloor$. By taking the prefix $\sigma'$ of
   $\rho'$ of length $\lfloor \frac{h}{2}\rfloor$, we get that $(\sigma,\sigma')\in  \tilde{\mathpzc{R}}(\lfloor \frac{h}{2}\rfloor)$.
   \item  $N_\emptyset(\sigma)\geq  \lfloor \frac{h}{2}\rfloor$ and $D_p(\sigma)<  \lfloor \frac{h}{2}\rfloor$. Since $\lst(\rho)=\lst(\rho')$, $N_p(\rho')=N_p(\rho)$,
   and $N_\emptyset(\rho')\geq h$,
   there exists a proper prefix $\sigma'$ of $\rho'$ such that $\lst(\sigma')=\lst(\sigma)$,  $N_p(\sigma')=N_p(\sigma)$, and
    $N_\emptyset(\sigma')\geq  \lfloor \frac{h}{2}\rfloor$. Hence $(\sigma,\sigma')\in  \tilde{\mathpzc{R}}(\lfloor \frac{h}{2}\rfloor)$.
 \end{enumerate}
\end{enumerate}
Thus, in all the cases, (1.) holds.

(2.) Let $(\rho,\rho')\in \tilde{\mathpzc{R}}(h)$ and $\sigma$ be a non-empty trace such that $\rho\cdot \sigma$ is a trace. We distinguish the following cases:
 \begin{enumerate}
  \item $D_p(\rho)<h$. Since $(\rho,\rho')\in \tilde{\mathpzc{R}}(h)$, we have that $D_p(\rho')= D_p(\rho)$, $N_p(\rho)=N_p(\rho')$, and either $N_\emptyset(\rho')=N_\emptyset(\rho)$, or $N_\emptyset(\rho)\geq h$ and $N_\emptyset(\rho')\geq h$. Hence, $\lst(\rho)\!=\!\lst(\rho')$ and, taking $\sigma'\!=\!\sigma$, we get that
        $(\rho\cdot\sigma,\rho'\cdot\sigma')\!\in\! \tilde{\mathpzc{R}}(h)\!\subseteq\! \tilde{\mathpzc{R}}(\lfloor \frac{h}{2}\rfloor)$.
  \item $D_p(\rho)\geq h$ and $D_p(\sigma)<\lfloor \frac{h}{2}\rfloor$. It follows that $N_\emptyset(\rho\cdot\sigma)\geq \lfloor \frac{h}{2}\rfloor$. Since
$D_p(\rho')\geq h$, there exists a trace of the form $\rho'\cdot \sigma'$ such that $D_p(\rho'\cdot\sigma')=D_p(\rho\cdot\sigma)$, $N_p(\rho'\cdot\sigma')=N_p(\rho\cdot\sigma)$,
and $N_\emptyset(\rho'\cdot\sigma')\geq \lfloor \frac{h}{2}\rfloor$. Hence,
$(\rho\cdot\sigma,\rho'\cdot\sigma')\in \tilde{\mathpzc{R}}(\lfloor \frac{h}{2}\rfloor)$.
\item $D_p(\rho)\geq h$ and $D_p(\sigma)\geq \lfloor \frac{h}{2}\rfloor$. Thus $D_p(\rho')\geq h$. If $N_\emptyset(\rho\cdot\sigma)<\lfloor \frac{h}{2}\rfloor$, then 
$N_\emptyset(\rho)=N_\emptyset(\rho')$. Therefore, there exists a trace of the form $\rho'\cdot\sigma'$ such that  $N_\emptyset(\rho'\cdot\sigma')=N_\emptyset(\rho\cdot\sigma)$
and $D_p(\sigma')\geq \lfloor \frac{h}{2}\rfloor$. Otherwise, $N_\emptyset(\rho\cdot\sigma)\geq \lfloor \frac{h}{2}\rfloor$ and
   there exists a trace of the form $\rho'\cdot\sigma'$ such that  $N_\emptyset(\rho'\cdot\sigma')\geq \lfloor \frac{h}{2}\rfloor$
and $D_p(\sigma')= \lfloor \frac{h}{2}\rfloor$. In both cases, $(\rho\cdot\sigma,\rho'\cdot\sigma')\in \tilde{\mathpzc{R}}(\lfloor \frac{h}{2}\rfloor)$.
\end{enumerate}
Thus, (2.) holds.
\end{proof}

By exploiting the previous lemma, we can prove the following one.
 \begin{lemma}\label{corollary:HcompatibilityOne} Let $n\in\Nat$,  $\psi$ be a balanced $\HS_\stat$ formula with $|\psi|\leq n$, and $(\rho,\rho')\in \tilde{\mathpzc{R}}(|\psi|)$. Then $\Ku_n,\rho\models \psi$ if and only if $\Ku_n,\rho'\models \psi$.
 \end{lemma}
 \begin{proof} The proof is by induction on $|\psi|$. The cases for the Boolean connectives directly follow from the inductive hypothesis and the fact that $\tilde{\mathpzc{R}}(h)\subseteq \tilde{\mathpzc{R}}(k)$, for all $h,k\in [1,n]$ with $h\geq k$. 
 
 As for the other cases, we proceed as follows:
 \begin{itemize}
  \item $\psi=p$. Since $(\rho,\rho')\in \tilde{\mathpzc{R}}(1)$, that is, either $N_{\emptyset}(\rho) = N_{\emptyset}(\rho') = 0$ or both
  $N_{\emptyset}(\rho) \geq 1$ and $N_{\emptyset}(\rho') \geq 1$, $\rho$ visits a state where $p$ does not hold if and only if $\rho'$ visits a state where $p$ does not hold, which proves the thesis.
  \item $\psi =\hsB \theta$ (resp., $\psi =\hsBt \theta$). Since $\psi$ is balanced, $\theta$ has the form $\theta=\theta_1\wedge\theta_2$, with $|\theta_1|=|\theta_2|$. Hence $|\theta_1|,|\theta_2|\leq \lfloor \frac{|\psi|}{2}\rfloor$. We focus on the case $\psi =\hsB \theta$.
  Since $\tilde{\mathpzc{R}}(|\psi|)$ is an equivalence relation, by symmetry it suffices to show that $\Ku_n,\rho\models \psi$ implies $\Ku_n,\rho'\models \psi$. If $\Ku_n,\rho\models \psi$, then there exists a proper prefix $\sigma$ of $\rho$ such that $\Ku_n,\sigma\models \theta_i$, for $i=1,2$. Since $(\rho,\rho')\in \tilde{\mathpzc{R}}(|\psi|)$, by (1.) of Lemma~\ref{lemma:Hcompatibility}, there exists a proper prefix $\sigma'$ of $\rho'$ such that  $(\sigma,\sigma')\in \tilde{\mathpzc{R}}(\lfloor \frac{|\psi|}{2}\rfloor)$. Since $\tilde{\mathpzc{R}}(\lfloor \frac{|\psi|}{2}\rfloor)\subseteq \tilde{\mathpzc{R}}(|\theta_i|)$, for $i=1,2$, by the inductive hypothesis we get that $\Ku_n,\sigma'\models \theta_i$, for $i=1,2$, thus proving that $\Ku_n,\rho'\models \psi$.
      
  The case for $\psi =\hsBt \theta$ can be dealt with similarly by exploiting (2.) of  Lemma~\ref{lemma:Hcompatibility}.
  \item $\psi =\hsE \theta$ (resp., $\psi =\hsEt \theta$). We can proceed as in the previous case by applying (3.) of Lemma~\ref{lemma:Hcompatibility} (resp., (4.) of Lemma~\ref{lemma:Hcompatibility}) and the inductive hypothesis.\qedhere
\end{itemize}
 \end{proof}

We can finally prove Lemma~\ref{lemma:MainnonBranchingExpressibilityOfEventually}.
\begin{lemma*}[\ref{lemma:MainnonBranchingExpressibilityOfEventually}]
For all $n\in\Nat^+$ and balanced $\HS_\stat$ formulas $\psi$, with $|\psi|\leq n$, 
it holds that $\Ku_n\models_\stat \psi$ if and only if $\mathpzc{M}_n\models_\stat \psi$.
\end{lemma*}
\begin{proof}
First, let us assume that $\Ku_n\not\models_\stat \psi$. Then, there exists an initial trace $\rho$ of $\Ku_n$ such that $\Ku_n,\rho \not \models_\stat \psi$. By Proposition~\ref{remark:HcompatibilityTwo}, there exists a trace $\rho'$ of $\Ku_n$, which is an initial trace for $\mathpzc{M}_n$, such that $(\rho,\rho') \in \tilde{\mathpzc{R}}(|\psi|)$. By Lemma~\ref{corollary:HcompatibilityOne}, we have 
that $\Ku_n,\rho ' \not \models_\stat \psi$. Since for any trace $\sigma$ and any $\HS_\stat$ formula $\varphi$, we have that $\Ku_n,\sigma \models_\stat \varphi$ if and only if $\mathpzc{M}_n,\sigma \models_\stat \varphi$ ($\Ku_n$ and $\mathpzc{M}_n$ feature exactly the same set of traces with exactly the same labeling; they only differ in the initial state), we can conclude that $\mathpzc{M}_n,\rho ' \not \models_\stat \psi$, and thus $\mathpzc{M}_n\not \models_\stat \psi$.

Let us now assume that $\mathpzc{M}_n\not \models_\stat \psi$. Then there exists an initial trace $\rho$ of $\mathpzc{M}_n$ such that $\mathpzc{M}_n,\rho \not \models_\stat \psi$. As in the converse direction, we have $\Ku_n,\rho \not \models_\stat \psi$, and, by Proposition~\ref{remark:HcompatibilityTwo}, we can easily find an initial trace $\rho'$ of $\Ku_n$ such that $(\rho,\rho') \in \tilde{\mathpzc{R}}(|\psi|)$. By Lemma~\ref{corollary:HcompatibilityOne} we can conclude that $\Ku_n\not\models_\stat \psi$.
\end{proof}

%% file: Chaps/Appendices/appendixGand17.tex
\chapter{Proofs and complements of Chapter~\ref{chap:Gand17}}
\minitoc\mtcskip

\section{Completion of the proof of Proposition \ref{prop:closureUnderPrefixSuffix}}\label{sec:prop:closureUnderPrefixSuffix}
\paragraph*{Language $\hsE_{\Ku}(\Lang(\Au))$.} Let us consider the $\NFA$ $\Au_{\hsE}$ over $S$ given by
\[\Au_{\hsE}=\tpl{S,(M\cup \{q'_0\})\times S ,\{q'_0\}\times S,\Delta',F},\]
where $q'_0\notin M$ is a fresh main state, and for all
$(q,s)\in (M\cup \{q'_0\})\times S$ and $s'\in S$, we have $\Delta'((q,s),s')=\emptyset$, if $s'\neq s$, and 
%
\[
\Delta((q,s),s)=  \left\{
    \begin{array}{ll}
    \Delta((q,s),s)
      &    \text{ if }  q\neq q'_0
      \\
(\{q'_0\}\times \Edges(s))\cup \{(q_0,s')\in Q_0\mid s'\in \Edges(s)\}
 &    \text{ otherwise.}
    \end{array}
  \right.
\]
%
Starting from an initial state $(q'_0,s)$, the automaton $\Au_{\hsE}$ either remains in a state whose   main component is $q'_0$, or moves to an initial state $(q_0,s')$ of $\Au$, ensuring at the same time that the portion of the input read so far is faithful to the evolution of  $\Ku$. From the state $(q_0,s')$,
 $\Au_{\hsE}$ simulates the behavior of $\Au$.  Formally, since $\Au$ is a $\Ku$-\NFA,  by construction it easily follows that $\Au_{\hsE}$ is a
   $\Ku$-\NFA\ which accepts the set of traces of $\Ku$ having a non-empty proper suffix in $\Lang(\Au)$. Hence,
   $\Lang(\Au_{\hsE})=\hsE_{\Ku}(\Lang(\Au))$.
   
\paragraph*{Language  $\hsEt_{\Ku}(\Lang(\Au))$.} Let us consider the $\NFA$ $\Au_{\hsEt}$ over $S$ given by
\[\Au_{\hsEt}=\tpl{S,(M\cup \{q_\acc\})\times S ,Q'_0,\Delta',\{q_\acc\}\times S},\] where $q_\acc\notin M$ is a fresh main state, and $Q'_0$ and $\Delta'$ are defined as follows:
\begin{itemize}
  \item the set $Q'_0$ of initial states is the set of states $(q,s)$ of $\Au$ such that there is a run of $\Au$ from some initial state to $(q,s)$ over some non-empty word.
  \item For all
$(q,s)\in (M\cup \{q_\acc\})\times S$ and $s'\in S$, we have $\Delta'((q,s),s')=\emptyset$, if $s'\neq s$, and 
\[
\Delta'((q,s),s)=  \left\{
    \begin{array}{ll}
\Delta((q,s),s)\cup \quad  \displaystyle{\smashoperator{\bigcup_{(q',s')\in F\cap \Delta((q,s),s)}}}\; \{(q_\acc,s')\}
      &    \text{ if } q\in M
      \\
    \emptyset
      &    \text{ if } q= q_\acc.
    \end{array}
  \right.
\]
\end{itemize}
Note that the set $Q'_0$ can be computed in time polynomial in the size of $\Au$. Since   $\Au_{\hsEt}$ essentially simulates $\Au$, and $\Au$ is
 a $\Ku$-\NFA, by construction we easily obtain that  $\Au_{\hsEt}$ is  a $\Ku$-$\NFA$ which accepts the set of words over $S$ which are
non-empty proper suffixes of words in $\Lang(\Au)$.
Thus, since $\Au$ is a $\Ku$-\NFA, we obtain that $\Lang(\Au_{\hsEt})=\hsEt_{\Ku}(\Lang(\Au))$.\qed

\section{Proof of Proposition~\ref{prop:invarianceLeftRightPrefix}}
\label{proof:prop:invarianceLeftRightPrefix}

\begin{proposition*}[\ref{prop:invarianceLeftRightPrefix}] Let $h\geq 0$, and $\rho$ and $\rho'$ be two $h$-prefix  bisimilar traces of  $\Ku$. Then, for all traces $\rho_L$ and $\rho_R$ of $\Ku$ such that $\rho_L \star \rho$ and  $\rho \star \rho_R$ are defined, the following properties hold:
\begin{enumerate}
    \item $\rho_L\star \rho$ and $\rho_L\star \rho'$ are $h$-prefix   bisimilar; 
    \item $\rho\star \rho_R$ and $\rho'\star \rho_R$ are $h$-prefix  bisimilar.
\end{enumerate}
\end{proposition*}

\begin{proof}
Let us note first that, since $\Summary(\rho)=\Summary(\rho')$, we have $\fst(\rho)=\fst(\rho')$ and $\lst(\rho)=\lst(\rho')$.
Hence $\rho_L\star \rho$ (resp., $\rho\star \rho_R$) is defined if and only if  $\rho_L\star \rho'$ (resp., $\rho'\star \rho_R$) is defined.
The proofs of $(1.)$ and $(2.)$ are by induction on $h\geq 0$.

$(1.)$ Since $\rho$ and $\rho'$ are $h$-prefix bisimilar, $\Summary(\rho)=\Summary(\rho')$. By Proposition~\ref{prop:Summaries},
 $\Summary(\rho_L\star \rho)=\Summary(\rho_L\star \rho')$. Thus, if $h=0$ (base case), the thesis follows. Now let $h>0$ (inductive step). Let us assume that $\nu$ is a proper prefix of
 $\rho_L\star\rho$ (the symmetric case, where we consider a proper prefix of $\rho_L\star \rho'$ is similar). We need to show that there exists a proper prefix $\nu'$ of
 $\rho_L\star\rho'$ such that $\nu$ and $\nu'$ are $(h-1)$-prefix bisimilar. If $\nu$ is a prefix of $\rho_L$, then we set $\nu'=\nu$ and the result trivially follows (note that, since $\rho$ and $\rho'$ are $h$-prefix bisimilar, it holds that $|\rho|>1$ if and only if $|\rho'|>1$). Otherwise, there is a proper prefix $\xi$ of $\rho$ such that
 $\nu=\rho_L\star \xi$. Since $\rho$ and $\rho'$ are $h$-prefix bisimilar, there exists a proper prefix $\xi'$ of $\rho'$ such that $\xi$ and $\xi'$ are $(h-1)$-prefix bisimilar.
Thus, by setting $\nu'=\rho_L\star \xi'$, by the inductive hypothesis the thesis follows.

$(2.)$ By Proposition~\ref{prop:Summaries},
 $\Summary(\rho\star \rho_R)=\Summary(\rho'\star \rho_R)$. Thus, if $h=0$, the thesis follows. Now, let us assume that $h>0$.
We proceed by a double  induction on $|\rho_R|$. As for the base case, where $|\rho_R|=1$, the result is obvious. Thus let us assume that $|\rho_R|>1$.
  Let $\nu$ be a proper prefix of
 $\rho\star\rho_R$ (the symmetric case, where we consider a proper prefix of $\rho'\star \rho_R$ is similar). We need to show that there exists a proper prefix $\nu'$ of
 $\rho'\star\rho_R$ such that $\nu$ and $\nu'$ are $(h-1)$-prefix bisimilar. If $\nu=\rho$ or $\nu$ is a proper prefix of $ \rho$, then there exists a prefix $\nu'$ of
 $ \rho'$ such that $\nu$ and $\nu'$ are $(h-1)$-prefix bisimilar. Thus, since $\nu'$ is a proper prefix of $\rho'\star\rho_R$, the result follows.
 Otherwise, there exists a proper prefix $\xi$ of $\rho_R$ such that
 $\nu=\rho\star \xi$. By setting $\nu'=\rho'\star\xi$, and considering the inductive hypothesis on $|\rho_R|$, we obtain that $\nu$ and $\nu'$ are $h$-prefix bisimilar,
 hence $(h-1)$-prefix bisimilar as well, concluding the proof.
\end{proof}

\section{Proof of Proposition~\ref{prop:fulfillmentPreservingPrefix}}
\label{proof:prop:fulfillmentPreservingPrefix}

\begin{proposition*}[\ref{prop:fulfillmentPreservingPrefix}] Let $h\geq 0$, and $\rho$ and $\rho'$ be two $h$-prefix bisimilar traces of $\Ku$. Then, for each $\AAbarBBbarEbar$
formula $\psi$ over $\SPEC$ with $\nestb(\psi)\leq h$, we have that
\[\Ku,\rho\models\psi\iff \Ku,\rho'\models\psi.\]
\end{proposition*}

\begin{proof}
We prove the proposition by a nested induction on the structure of the formula $\psi$ and on the B-nesting depth $\nestb(\psi)$. 

As for the base case, $\psi$ is a regular expression in $\SPEC$. Since $\Summary(\rho)=\Summary(\rho')$ (as $\rho$ and $\rho'$ are $h$-prefix bisimilar) the thesis holds  by Proposition~\ref{prop:Summaries}.

Let us now consider the inductive step.
The cases where the root modality of $\psi$ is a Boolean connective directly follow by the inductive hypothesis.
As for the cases where the root modality is $\hsA$ or $\hsAt$, the result follows from the fact that, being $\rho$ and $\rho'$ $h$-prefix bisimilar, we have
$\fst(\rho)=\fst(\rho')$ and $\lst(\rho)=\lst(\rho')$.
It remains to consider the cases where the root modality is  $\hsB$, $\hsBt$, or $\hsEt$. We prove the implication
$\Ku,\rho\models \psi \Longrightarrow \Ku,\rho'\models \psi$ (the converse is similar).  Let
$\Ku,\rho\models \psi$. 
\begin{itemize}
  \item $\psi=\hsB\varphi$: since $0<\nestb(\psi)\leq h$,  it holds that $h>0$.  As $\Ku,\rho\models\hsB\varphi$,   there is a proper prefix $\nu$ of $\rho$  such that $\Ku,\nu\models\varphi$.
  Since $\rho$ and $\rho'$ are $h$-prefix bisimilar, there is a proper prefix $\nu'$ of $\rho'$ such that $\nu$ and $\nu'$ are $(h-1)$-prefix bisimilar.
  Being $\nestb(\varphi)\leq h-1$, by the inductive hypothesis we obtain that  $\Ku,\nu'\models\varphi$. Hence,  $\Ku,\rho'\models\hsB\varphi$.
  \item $\psi=\hsBt\varphi$: since $\Ku,\rho\models\hsBt\varphi$, there is a trace $\rho_R$ such that $|\rho_R|>1$ and
$\Ku,\rho\star\rho_R\models\varphi$. By Proposition~\ref{prop:invarianceLeftRightPrefix}, $\rho\star\rho_R$ and $\rho'\star\rho_R$ are $h$-prefix bisimilar. By the inductive hypothesis on  the structure of the formula, we obtain that $\Ku,\rho'\star\rho_R\models\varphi$. Hence,
  $\Ku,\rho'\models\hsBt\varphi$. 
  \item    $\psi=\hsEt\varphi$: this case is similar to the previous one.\qedhere 
\end{itemize}
\end{proof}

\section{Proof of Lemma~\ref{lemma:prefixSamplingOneRegex}}\label{proof:lemma:prefixSamplingOneRegex}

\begin{lemma*}[\ref{lemma:prefixSamplingOneRegex}] For $h\geq 0$, any two traces
$\rho$ and  $\rho'$ of $\Ku$ having the same
$h$-sampling word are $h$-prefix bisimilar.
\end{lemma*}

\begin{proof}
The proof can be immediately derived by a stronger result stated in the following claim.
\begin{claim}\label{lab:clAux} Let $h\geq 0$,  $\rho$ and  $\rho'$ be two traces of $\Ku$, and $\PrefS_h$ and $\PrefS'_h$ be the two $h$-prefix samplings
of $\rho$ and $\rho'$, resp.. Assume that $\rho$ and $\rho'$ have the same $h$-sampling word, namely there is $N\geq 1$ such that
 \begin{itemize}
   \item 
   $\PrefS_h: i_1<i_2<\ldots < i_N$,
   \item 
   $\PrefS'_h: i'_1<i'_2<\ldots < i'_N$, and
   \item 
  for all $j\in [1,N]$, $\Summary(\rho(1,i_j))=\Summary(\rho'(1,i'_j))$.
 \end{itemize}

 Then, for all $j\in [1,N-1]$, $n\in [i_j+1,i_{j+1}]$ and $n'\in [i'_j+1,i'_{j+1}]$ such that $\Summary(\rho[1,n])=\Summary(\rho'[1,n'])$, it holds that
 $\rho(1,n)$ and $\rho'(1,n')$ are $h$-prefix bisimilar.
\end{claim}
\begin{proof} The proof is by induction on $h\geq 0$. For $h=0$, the result is obvious. Now let us assume that $h>0$. If $N=1$ (resp., $N=2$), then
$\rho=\rho'$  and $|\rho|=|\rho'|=N$, and the thesis trivially holds.
Now, let us assume that $N>2$. Since by hypothesis $\Summary(\rho(1,n))=\Summary(\rho'(1,n'))$, we need to show that:
 \begin{enumerate}
   \item for each $m\in [1,n-1]$, there exists $m'\in [1,n'-1]$ such that $\rho(1,m)$ and $\rho'(1,m')$ are $(h-1)$-prefix bisimilar;
   \item for each $m'\in [1,n'-1]$, there exists $m\in [1,n-1]$ such that $\rho(1,m)$ and $\rho'(1,m')$ are $(h-1)$-prefix bisimilar;
 \end{enumerate}
We only prove $(1.)$, being the proof of $(2.)$ symmetric. We exploit in the proof the following fact that can be easily shown:
let $k\in [0,h-1]$ and $1=x_1<\ldots <x_r=N$ be the subsequence of $1,\ldots,N$ such that
$i_{x_1}<\ldots<i_{x_r}$ is the $k$-prefix sampling of $\rho$. Then, $i'_{x_1}<\ldots<i'_{x_r}$ is the $k$-prefix sampling of $\rho'$.

Now we prove $(1.)$. Let $m\in [1,n-1]$. If $m=1$, we set $m'=1$, and the result follows.
Now, let us assume that $m\geq 2$. Since $h>0$, there must exist $x,y\in [1,N]$ such that $x<y$, $m\in [i_{x}+1,i_y]$, and $i_x$ and $i_y$ are two consecutive positions in the
$(h-1)$-prefix sampling of $\rho$. By the fact above, $i'_x$ and $i'_y$ are two consecutive positions in the $(h-1)$-prefix sampling of $\rho'$. We distinguish two cases:
\begin{itemize}
\item $m=i_y$. Since $n\in [i_j+1,i_{j+1}]$ and $m<n$, it holds that $i_y\leq i_j$. Hence, $i'_y\leq i'_j$ as well. Moreover, since $n'>i'_j$, it holds that $i'_y<n'$.
We set $m'=i'_y$. As $\Summary(\rho(1,i_y))=\Summary(\rho'(1,i'_y))$, $m=i_y$, $m'=i'_y$, and $i_x$ and $i_y$ (resp., $i'_x$ and $i'_y$) are two consecutive positions in
the $(h-1)$-prefix sampling of $\rho$ (resp., $\rho'$), the thesis follows by the inductive hypothesis on $h$.
\item $m\neq i_y$. Hence, $m\in [i_x+1,i_y-1]$. Since $i_x$ and $i_y$ are two consecutive positions in the $(h-1)$-prefix sampling of $\rho$, there must exist $z\in [x+1,y-1]$
such that $i_z\leq m$ and $\Summary(\rho(1,m))=\Summary(\rho(1,i_z))$. Since $i_z\leq m$, $m<n$, and $n\in [i_j+1,i_{j+1}]$, it holds that $i_z\leq i_j$.
Hence, $i'_z\leq i'_j<n'$. We set $m'=i'_z$. As $\Summary(\rho(1,i_z))=\Summary(\rho'(1,i'_z))$, we obtain that
$\Summary(\rho(1,m))=\Summary(\rho'(1,m'))$, $m\in [i_{x}+1,i_y]$ and $m'\in [i'_x+1,i'_y]$. Thus, being  $i_x$ and $i_y$ (resp., $i'_x$ and $i'_y$)   two consecutive positions in the $(h-1)$-prefix sampling of $\rho$ (resp., $\rho'$), by the inductive hypothesis on $h$ the result follows. \qedhere
\end{itemize}
\end{proof}
This concludes the proof of the claim, and of  Lemma~\ref{lemma:prefixSamplingOneRegex} as well.
\end{proof}

\section{Pseudocode of \texttt{checkFalse}}\label{sec:chkFalse}
The pseudocode of procedure $\texttt{checkFalse}$, the \lq\lq dual\rq\rq{} of $\texttt{checkTrue}$, 
is reported in Algorithm~\ref{fig-proc-checkFALSE} at page~\pageref{fig-proc-checkFALSE}.

\begin{algorithm}[p]
\caption{$\texttt{checkFalse}_{\tpl{\Ku,\varphi,\GLab}}(\WS)$}\label{fig-proc-checkFALSE}
\begin{algorithmic}[1]
\While{$\WS$ is \emph{not} universal}
    \State{deterministically select $(\psi,\rho)\in \WS$ s.t. $\psi$ is not of the form  $\hsEtu\psi'$ and $\hsBtu\psi'$}
    \State{$\WS\leftarrow \WS\setminus \{(\psi,\rho)\}$}
    \Case{ $\psi=r$  with $r\in \RE$}
        \If{$\rho\notin \Lang(r)$}
            \State{accept the input}
        \EndIf
    \case{ $\psi=\neg r$ with $r\in \RE$}
        \If{$\rho\in \Lang(r)$}
            \State{accept the input}
        \EndIf
    \case{ $\psi=\hsA \psi'$ or $\psi=\hsAu \psi'$}
        \If{$\psi\notin\GLab(\lst(\rho))$}
            \State{accept the input}
        \EndIf
    \case{ $\psi=\hsAt \psi'$ or $\psi=\hsAtu \psi'$}      \If{$\psi\notin\GLab(\fst(\rho))$}
            \State{accept the input}
        \EndIf
    \case{ $\psi=\psi_1\vee \psi_2$}
        \State{universally choose $i=1,2$}
        \State{$\WS\leftarrow \WS\cup \{(\psi_i,\rho)\}$}
        
        
    \case{ $\psi=\psi_1\wedge \psi_2$}
        \State{$\WS\leftarrow \WS\cup \{(\psi_1,\rho),(\psi_2,\rho)\}$}
    \case{ $\psi=\hsB\psi'$}
        \State{universally choose $\rho'\in \Pref(\rho)$}
        \State{$\WS\leftarrow \WS\cup \{(\psi',\rho')\}$}
    \case{ $\psi=\hsBu\psi'$}
        \State{$\WS\leftarrow \WS\cup \{(\psi',\rho')\mid \rho'\in\Pref(\rho)\}$}
    \case{ $\psi=\hsX\psi'$ with $X\in \{\overline{E},\overline{B}\}$}
        \State{universally choose an $X$-\emph{witness} $\rho'$ of $\rho$ for $(\Ku,\varphi)$}
        \State{$\mathcal{\WS}\leftarrow \mathcal{\WS}\cup \{(\psi',\rho')\}$}
    \EndCase
\EndWhile
\If{$\mathcal{\WS}=\emptyset$}
    \State{reject the input}
\Else
    \State{existentially choose $(\psi,\rho)\in \widetilde{\WS}$}
    \State{$\texttt{checkTrue}_{\tpl{\Ku,\varphi,\GLab}}(\{(\psi,\rho)\})$}
\EndIf
\end{algorithmic}
\end{algorithm}

\section{Proof of Proposition~\ref{prop:correctnessATMcheck}}\label{APP:correctnessATMcheck}

For technical convenience, given an $\AAbarBBbarEbar$ formula $\varphi$, we consider a slight variant $\Alt(\varphi)$ of $\AltN(\varphi)$. Formally, $\Alt(\varphi)$ is defined as $\AltN(\hsBt\varphi)$ (or, equivalently, as $\AltN(\hsEt\varphi)$).
Note that for each
$\AAbarBBbarEbar$ formula $\varphi$ and $X\in \{\overline{E},\overline{B}\}$, we have $\Alt(\hsXu\varphi)=\Alt(\widetilde{\hsXu\varphi})+1$.

Let $\Ku$ be a finite Kripke structure,  $\varphi$
be an $\AAbarBBbarEbar$ formula in \nnf, and $\WS$ be a well-formed set for $(\Ku,\varphi)$.
We denote by $\Alt(\WS)$ the maximum over the  alternation depths $\Alt(\psi)$, where $\psi$ is a formula occurring in $\WS$
(we set $\Alt(\WS)=0$ if $\WS=\emptyset$).
For each non-empty \emph{universal} well-formed set $\WS$ for $(\Ku,\varphi)$, we have $\Alt(\widetilde{\WS})= \Alt(\WS)- 1$.

Now, we can prove Proposition~\ref{prop:correctnessATMcheck}.
\begin{proposition*}[\ref{prop:correctnessATMcheck}] The ATM  $\texttt{check}$ is a singly exponential-time bounded ATM accepting $\FMC$ whose number of alternations on input $(\Ku,\varphi)$ is at most $\AltN(\varphi)+2$.
\end{proposition*}
\begin{proof}
Let us fix an input $(\Ku,\varphi)$, where $\varphi$
is an $\AAbarBBbarEbar$ formula in \nnf. 

Note that whenever there is a switch between the procedures $\texttt{checkTrue}$ and $\texttt{checkFalse}$, e.g., from $\texttt{checkTrue}$ to $\texttt{checkFalse}$, $(i)$~the input $\{(\psi,\rho)\}$ of the called procedure is contained in the dual $\widetilde{\WS}$ of the currently processed well-formed set $\WS$ for $(\Ku,\varphi)$, and $(ii)$~$\WS$ is non-empty and universal, hence $\Alt(\{(\psi,\rho)\})< \Alt(\WS)$. Moreover, a well-formed set $\WS$ for $(\Ku,\varphi)$ contains only formulas $\psi$ such that $\psi\in \SD(\varphi)$.

Additionally, in each iteration of the while loops of procedures $\texttt{checkTrue}$ and $\texttt{checkFalse}$, the processed pair $(\psi,\rho)$ in the current well-formed set $\WS$ is either removed from $\WS$, or it is replaced with pairs $(\psi',\rho')$ such that $\psi'$ is a strict subformula of $\psi$.
 This ensures that the algorithm always terminates. 
 
Furthermore, since the number of alternations of the ATM $\texttt{check}$  between existential choices and universal choices is evidently the number of switches between the calls to procedures $\texttt{checkTrue}$ and $\texttt{checkFalse}$ plus 2, and the top calls to $\texttt{checkTrue}$ take as input well-formed sets for $(\Ku,\varphi)$ having the form  $\{(\psi,\rho)\}$, where $\psi\in\SD(\varphi)$, we have proved the following result.

\begin{claim}
The number of alternations of the ATM \texttt{check} on input $(\Ku,\varphi)$ is at most $\AltN(\varphi)+2$.
\end{claim} 

Next, we prove the following property.
\begin{claim}\label{claim:sinExp}
The ATM $\texttt{check}$ runs in time singly exponential in the size of the input $(\Ku,\varphi)$.
\end{claim}
\begin{proof}
Let us fix an input $(\Ku,\varphi)$.  Let $T(\varphi)$ be the standard tree encoding of $\varphi$, where each node is labeled by some subformula of $\varphi$.
Let $\psi\in\SD(\varphi)$. If $\psi$ is a subformula of $\varphi$, we define $d_\psi$ as the maximum over the distances from the root in $T(\varphi)$ of $\psi$-labeled nodes. If, conversely, $\psi$ is the dual of a subformula of $\varphi$, we let
 $d_\psi = d_{\widetilde{\psi}}$. 
 
Let us denote by $H(\Ku,\varphi)$ the length of a certificate for $(\Ku,\varphi)$. Recall that   $H(\Ku,\varphi)= (|\States|\cdot 2^{(2|\SPEC|)^2})^{h+2}$,
  where $\States$ is the set of states of $\Ku$,  $\SPEC$ is the set of atomic formulas (regular expressions) occurring in $\varphi$, and $h=\nestb(\varphi)$.

By Proposition~\ref{prop:EbarBbarWitness},  it follows that each  step in an iteration of the while loops in the procedures $\texttt{checkTrue}$ and $\texttt{checkFalse}$ can be performed in time singly exponential in the size of $(\Ku,\varphi)$.
 Thus, in order to prove Claim~\ref{claim:sinExp},  it suffices to show that
for all  computations $\pi$ of the ATM $\texttt{check}$ starting from the input $(\Ku,\varphi)$, the overall number $N_\psi$ of iterations of the while loops (of procedures $\texttt{checkTrue}$ and $\texttt{checkFalse}$) along $\pi$, where the formula $\psi$ is processed, is at most $(2^{|\varphi|}\cdot H(\Ku,\varphi))^{d_\psi}$. 

The proof is done by induction on $d_\psi$. As for the base case, we have $d_\psi = 0$. Therefore, $\psi=\varphi$ or $\psi=\widetilde{\varphi}$; by construction of the algorithm, $N_\varphi$ and $N_{\widetilde{\varphi}}$ are at most equal to $1$. Thus the result holds. 

As for the inductive step, let us assume that  $d_\psi > 0$. We consider the case where $\psi$ is a subformula of $\varphi$ (the case where $\widetilde{\psi}$ is a subformula of $\varphi$ is similar). Then, the result follows from the next sequence of inequalities, where $P(\psi)$ denotes the set of  nodes of $T(\varphi)$ which are parents of the nodes labeled by $\psi$, and for each node $x$, $\textit{fo}(x)$ denotes the formula labeling $x$.
\begin{multline*}
N_\psi  \leq  \sum_{x\in P(\psi)} N_{\textit{fo}(x)} \cdot H(\Ku,\varphi) \leq \\
\leq \sum_{x\in P(\psi)} (2^{|\varphi|}\cdot H(\Ku,\varphi))^{d_{\textit{fo}(x)}} \cdot H(\Ku,\varphi) \leq \big(2^{|\varphi|}\cdot H(\Ku,\varphi)\big)^{d_\psi}.
\end{multline*}
The first inequality directly follows from the construction of the algorithm (note that if $\textit{fo}(x)=\hsBu\psi$, the processing of the subformula $\textit{fo}(x)$ in an iteration of the two while loops generates at most  $H(\Ku,\varphi)$ new ``copies'' of $\psi$). The second inequality follows by the inductive hypothesis, and the last one from the fact that $|P(\psi)|\leq 2^{|\varphi|}$  and $d_{\textit{fo}(x)}\leq d_{\psi} -1$ for all $x\in P(\psi)$. This concludes the proof of Claim~\ref{claim:sinExp}. 
\end{proof}

It remains to show that the ATM $\texttt{check}$ accepts $\FMC$.  Let us fix an input $(\Ku,\varphi)$ and let $\GLab$ be the  $\AAbar$-labeling initially and existentially guessed by
$\texttt{check}$ (at line 1).
 Evidently, after the top  calls  to $\texttt{checkTrue}$,  each configuration of the procedure $\texttt{check}$  can be described by a tuple $(\ell,\GLab, \WS, f)$, where: 
 \begin{itemize}
     \item $\WS$ is a well-formed set for $(\Ku,\varphi)$,
     \item $f=\true$ if $\WS$ is processed within $\texttt{checkTrue}$, and $f=\false$ otherwise, and
     \item $\ell$ is an instruction label corresponding to one of the instructions of the procedures $\texttt{checkTrue}$ and $\texttt{checkFalse}$.
 \end{itemize}
We denote by $\ell_0$ the label associated with the while instruction. A \emph{main configuration} is a configuration having label $\ell_0$. 

Let $\GLab_\WS$ be the restriction of $\GLab$
to the set of formulas in $\AAbar(\varphi)$ which are subformulas of formulas occurring in $\WS$. In other words, for each state $s$,
$\GLab_\WS(s)$ contains all and only the formulas $\psi\in \GLab(s)$ such that either $\psi$ or its dual $\widetilde{\psi}$
is a subformula of some formula occurring in $\WS$. $\GLab_\WS$ is said to be \emph{valid} if, for all states $s$ and $\psi\in \GLab_\WS(s)$, it holds $\Ku,s\models \psi$.

\newcommand{\Norm}[1]{\ensuremath{\|{#1}\|}}

\begin{claim}\label{claim:mainconf}
Let $\WS$ be a  well-formed set  for $(\Ku,\varphi)$ and let us assume that $\GLab_\WS$ is valid. Then
\begin{enumerate}
  \item the main configuration $(\ell_0,\GLab, \WS,\true)$ leads to acceptance \emph{if and only if} $\WS$ is valid;
  \item the main configuration $(\ell_0,\GLab, \WS,\false)$ leads to acceptance \emph{if and only if} $\WS$ is \emph{not} valid.
\end{enumerate}
\end{claim} 
\begin{proof}
 We associate with $\WS$  a natural number $\Norm{\WS}$ defined as follows. Let us fix an ordering $\psi_1,\ldots,\psi_k$ of the formulas in $\SD(\varphi)$  such that, for all $i\neq j$,  $|\psi_i|>|\psi_j|$ implies $i<j$. 
 
First, we associate with $\WS$ a $(k+1)$-tuple $\tpl{n_0,n_1,\ldots,n_k}$ of natural numbers defined as: the first component $n_0$ in the tuple is the alternation depth $\Alt(\WS)$ and,
  for all the other components $n_i$, with $1\leq i\leq k$, $n_i$ is the number of elements of $\WS$ associated with the formula $\psi_i$ (i.e., the number of elements having the form $(\psi_i,\rho)$).
 
 Then $\Norm{\WS}$ is the position of the tuple $\tpl{n_0,n_1,\ldots,n_k}$ along the total lexicographic ordering over $\Nat^{k+1}$. Note that if $\WS$ is non-empty and universal, since $\Alt(\widetilde{\WS})<\Alt(\WS)$, it holds that $\Norm{\widetilde{\WS}}<\Norm{\WS}$. Moreover, $\Norm{\WS}$ strictly decreases at each iteration of the while loops   in the procedures
$\texttt{checkTrue}$ and $\texttt{checkFalse}$ (this is because at each iteration $\Alt(\WS)$ does not increase, and  an element of $\WS$ is replaced with elements associated with smaller formulas).

The proof of Claim~\ref{claim:mainconf} is now carried out by induction on $\Norm{\WS}$. As for the base case we have  $\Norm{\WS}=0$, thus $\WS$ is empty and clearly valid. By construction $\texttt{checkTrue}$ accepts the empty set, while  $\texttt{checkFalse}$ rejects the empty set.
The result holds. 

As for the inductive step, let $\Norm{\WS}>0$, hence $\WS$ is not empty.
First, we assume that $\WS$ is universal. Recall that $\Norm{\widetilde{\WS}}<\Norm{\WS}$.  Thus,
\begin{itemize}
  \item
   $\WS$ is valid $\Longleftrightarrow$   for each $(\psi,\rho)\in \widetilde{\WS}$,
    $\{(\psi,\rho)\}$ is not valid  $\Longleftrightarrow$ (by the inductive hypothesis) for each $(\psi,\rho)\in \widetilde{\WS}$, the main configuration $(\ell_0,\GLab,\allowbreak\{(\psi,\rho)\},\false)$ leads to acceptance $\Longleftrightarrow$ (by construction of the algorithm and since $\WS$ is universal) the main configuration $(\ell_0,\GLab,\WS,\true)$ leads to acceptance.
  \item  
   $\WS$ is \emph{not}  valid $\Longleftrightarrow$   for some $(\psi,\rho)\in \widetilde{\WS}$,
    $\{(\psi,\rho)\}$ is  valid  $\Longleftrightarrow$ (by the inductive hypothesis) for some $(\psi,\rho)\in \widetilde{\WS}$, the main configuration $(\ell_0,\GLab,\allowbreak \{(\psi,\rho)\},\true)$ leads to acceptance $\Longleftrightarrow$ (by construction of the algorithm and since $\WS$ is universal) the main configuration $(\ell_0,\GLab,\WS,\false)$ leads to acceptance.
\end{itemize}
Hence, (1.) and (2.) of Claim~\ref{claim:mainconf} hold if $\WS$ is universal. 

Now, let us assume that the non-empty set  $\WS$ is not universal.
We consider (2.) of Claim~\ref{claim:mainconf} (the proof of (1.) is just the \lq\lq dual\rq\rq).
 Let $(\psi,\rho)\in \WS$ be the pair selected by the procedure $\texttt{checkFalse}$   in the iteration of the while loop   associated with the main configuration $(\ell_0,\GLab, \WS, \false)$. Here we examine the cases where either $\psi=\hsA\psi'$, or $\psi=\hsBu\psi'$, or $\psi=\hsX\psi'$  with $X\in \{\overline{B},\overline{E}\}$ (the other cases are similar or simpler).
\begin{itemize}
  \item $\psi=\hsA\psi'$. We have that $\{(\hsA\psi',\rho)\}$ is  valid if and only if $\Ku,\lst(\rho)\models \hsA\psi'$. By hypothesis, $\GLab_\WS$ is valid.
  Hence $\{(\hsA\psi',\rho)\}$ is not valid if and only if $\hsA\psi'\notin \GLab_\WS(\lst(\rho))$.
   Let $\WS' = \WS\setminus \{(\psi,\rho)\}$. Note that
      $\Norm{\WS'}< \Norm{\WS}$.
    Then
   $\WS$ is not valid $\Longleftrightarrow$  either $\hsA\psi'\notin \GLab_\WS(\lst(\rho))$ or $\WS'$ is not valid $\Longleftrightarrow$ (by the inductive hypothesis) either $\hsA\psi'\notin \GLab_\WS(\lst(\rho))$ or the main configuration $(\ell_0,\GLab,\WS',\false)$ leads to acceptance $\Longleftrightarrow$ (by construction of $\texttt{checkFalse}$) the main configuration $(\ell_0,\GLab,\WS,\false)$ leads to acceptance.
  \item $\psi=\hsBu\psi'$.  Let $\WS' = (\WS\setminus \{(\psi,\rho)\})\cup \{(\psi',\rho')\mid \rho'\in\Pref(\rho)\}$. Note that
      $\Norm{\WS'}< \Norm{\WS}$.
Then $\WS$ is not valid $\Longleftrightarrow$  $\WS'$ is not valid    $\Longleftrightarrow$ (by the inductive hypothesis) the main configuration $(\ell_0,\GLab,\WS',\false)$ leads to acceptance $\Longleftrightarrow$ (by construction of $\texttt{checkFalse}$) the main configuration $(\ell_0,\GLab,\WS,\false)$ leads to acceptance.
 \item $\psi=\hsX\psi'$  with $X\in \{\overline{B},\overline{E}\}$. By Proposition~\ref{prop:EbarBbarWitness}(1), $\Ku,\rho\models \hsX\psi'$ if and only if there exists an $X$-witness $\rho'$ of $\rho$
  for $(\Ku,\varphi)$  such that $\Ku,\rho'\models \psi'$.     Then (2.) of Claim~\ref{claim:mainconf} directly follows from the next sequence of equivalences: $\WS$ is not valid $\Longleftrightarrow$  either $\WS\setminus \{(\psi,\rho)\}$ is not valid, or  for each
  $X$-witness $\rho'$ of $\rho$
  for $(\Ku,\varphi)$,  $\{(\psi',\rho')\}$ is not valid  $\Longleftrightarrow$ for each
  $X$-witness $\rho'$ of $\rho$
  for $(\Ku,\varphi)$,   $(\WS\setminus \{(\psi,\rho)\})\cup \{(\psi',\rho')\}$ is not valid $\Longleftrightarrow$ (by the inductive hypothesis)  for each
  $X$-witness $\rho'$ of $\rho$
  for $(\Ku,\varphi)$, the main configuration $(\ell_0,\GLab,(\WS\setminus \{(\psi,\rho)\})\cup \{(\psi',\rho')\},\false)$ leads to acceptance $\Longleftrightarrow$
  (by construction of the procedure $\texttt{checkFalse}$) the main configuration  $(\ell_0,\GLab,\WS,\false)$ leads to acceptance.
\end{itemize}
This concludes the proof of Claim~\ref{claim:mainconf}.
\end{proof}

By Claim~\ref{claim:mainconf} we prove the next result, which finally concludes the proof of Proposition~\ref{prop:correctnessATMcheck}.

\begin{claim}
 The ATM $\texttt{check}$ accepts an input $(\Ku,\varphi)$ if and only if $\Ku\models \varphi$.
\end{claim}
\begin{proof}
Let us fix an input $(\Ku,\varphi)$ and an $\AAbar$-labeling $\GLab$ for $(\Ku,\varphi)$. A \emph{$\GLab$-guessing} for $(\Ku,\varphi)$
is a well-formed set $\WS$ for $(\Ku,\varphi)$  which minimally satisfies the following conditions for all states $s$ of $\Ku$:
\begin{itemize}
  \item for  all certificates $\rho$ for $(\Ku,\varphi)$ with $\fst(\rho)=\sinit$, $(\varphi,\rho)\in\WS$;
  \item for all $\hsA\psi\in \GLab(s)$ (resp., $\hsAt\psi\in \GLab(s)$), there is a certificate
  $\rho$ for $(\Ku,\varphi)$ with $\fst(\rho)=s$ (resp., $\lst(\rho)=s$) such that $(\psi,\rho)\in\WS$;
  \item for all $\hsAu\psi\in \GLab(s)$ (resp., $\hsAtu\psi\in \GLab(s)$) and for all certificates
  $\rho$ for $(\Ku,\varphi)$ with $\fst(\rho)=s$ (resp., $\lst(\rho)=s$), $(\psi,\rho)\in\WS$.
\end{itemize}
Evidently, by construction of the procedure $\texttt{check}$, for each input $(\Ku,\varphi)$, it holds:
\begin{itemize}
  \item[(*)] $\texttt{check}$ accepts $(\Ku,\varphi)$ $\Longleftrightarrow$ there are an $\AAbar$-labeling $\GLab$ and a $\GLab$-guessing $\WS$  for $(\Ku,\varphi)$
    such that, for all $(\psi,\rho)\in \WS$, the main configuration $(\ell_0,\GLab,\allowbreak \{(\psi,\rho)\},\true)$ leads to acceptance.
\end{itemize}

First we assume that $\Ku\models \varphi$.  Let $\GLab$ be the \emph{valid} $\AAbar$-labeling defined as follows for all states $s$: for all $\psi\in\AAbar(\varphi)$,
$\psi\in \GLab(s)$ if and only if $\Ku,s\models \psi$. By Theorem~\ref{theorem:singleExpTrackModel}, there exists a $\GLab$-guessing $\WS$  for $(\Ku,\varphi)$
such that for all $(\psi,\rho)\in \WS$, $\Ku,\rho\models \psi$. By Claim~\ref{claim:mainconf}, for all $(\psi,\rho)\in \WS$, the main configuration $(\ell_0,\GLab,\{(\psi,\rho)\},\true)$ leads to acceptance. Hence, by (*), $\texttt{check}$ accepts $(\Ku,\varphi)$.

For the converse direction, let us assume that $\texttt{check}$ accepts $(\Ku,\varphi)$. By (*), there exist an $\AAbar$-labeling $\GLab$ and a $\GLab$-guessing $\WS$  for $(\Ku,\varphi)$  such that, for all $(\psi,\rho)\in \WS$, the main configuration $(\ell_0,\GLab,\{(\psi,\rho)\},\true)$ leads to acceptance.
First we show that $\GLab$ is valid. 

We fix a state $s$ and a formula $\psi\in \GLab(s)$. We need to prove that $\Ku,s\models \psi$. The proof is by induction
on the nesting depth of modalities $\hsA$, $\hsAt$, $\hsAu$ and $\hsAtu$ in $\psi$. Assume that
$\psi=\hsAu\psi'$ for some $\psi'$ (the other cases, where either $\psi=\hsA\psi'$, or $\psi=\hsAt\psi'$, or   $\psi=\hsAtu\psi'$ are similar).
By definition of $\GLab$-guessing, for each certificate $\rho$   for $(\Ku,\varphi)$ with $\fst(\rho)=s$, $(\psi',\rho)\in \WS$. Moreover, by the inductive hypothesis,
one can assume that $\GLab_{\{(\psi',\rho)\}}$ is valid (note that for the base case, i.e., when $\psi'$ does not contain occurrences of modalities $\hsA$, $\hsAt$, $\hsAu$, and $\hsAtu$, $\GLab_{\{(\psi',\rho)\}}$ is trivially valid). By hypothesis, the main configuration
$(\ell_0,\GLab,\{(\psi',\rho)\},\true)$ leads  to acceptance.
By Claim~\ref{claim:mainconf}, 
for each certificate $\rho$  for $(\Ku,\varphi)$ with $\fst(\rho)=s$, it holds $\Ku,\rho\models \psi'$. Thus, by Theorem~\ref{theorem:singleExpTrackModel}, we obtain that
$\Ku,s\models \psi$. Hence $\GLab$ is valid. 

Now, by definition of $\GLab$-guessing, for each certificate $\rho$ for $(\Ku,\varphi)$ with $\fst(\rho)=\sinit$, $(\varphi,\rho)\in\WS$.
Thus, by hypothesis, by Claim~\ref{claim:mainconf}, and by Theorem~\ref{theorem:singleExpTrackModelRegex}, we have $\Ku\models \varphi$.
This concludes the proof of the claim.
\end{proof}

Proposition~\ref{prop:correctnessATMcheck} has been proved.
\end{proof}

\section{Proof of Theorem~\ref{theo:ComplexityAlternatingMT}}\label{proof:theo:ComplexityAlternatingMT}

\def\tape{\text{{\sffamily tape}}}
\def\row{\text{{\sffamily row}}}
\def\symb{\text{{\sffamily symb}}}
\def\state{\text{{\sffamily state}}}
\def\succ{\text{{\sffamily next}}}
\def\prev{\text{{\sffamily prev}}}
\def\last{\text{{\sffamily last}}}
\def\first{\text{{\sffamily first}}}
\def\Q{\text{{\sffamily Q}}}
\newcommand{\rej}{\textit{rej}}

In this section, we show that 
\begin{theorem*}[\ref{theo:ComplexityAlternatingMT}] $\!\!$The alternating multi-tiling problem is $\LINAEXPTIME\!$-complete
\end{theorem*}
Membership to $\LINAEXPTIME$ is straightforward.
As for $\LINAEXPTIME$-hardness, we establish it in two steps. In the first, we focus on a variant of the considered problem,
called \emph{TM alternation problem}, which is defined in terms of  multi-tape deterministic Turing machines, and we prove that it is
$\LINAEXPTIME$-hard. Then, in the second, we provide a polynomial-time reduction from the TM alternation problem to alternating multi-tiling.

\subsection{TM alternation problem}

Let $n\geq 1$. An \emph{$n$-ary deterministic Turing machine} (TM, for short) is a deterministic Turing machine $\mathcal{M}=\tpl{n,I,A,Q, \{q_{\textit{acc}},q_{\textit{rej}}\},q_0,\delta}$ operating on $n$ ordered semi-infinite tapes and having \emph{only one} read/write head (shared by all tapes), where: $I$ (resp., $A\supset I$) is  the input (resp., work) alphabet,   $A$ contains the blank symbol $\#\notin I$, $Q$ is the set of states, $q_{\textit{acc}}$ (resp., $q_{\textit{rej}}$) is the terminal accepting (resp., rejecting) state,
$q_0$ is the initial state, and $\delta:Q\times A \to \{\bot\}\cup  (Q\times A\times \{\leftarrow,\rightarrow\})\cup (Q\times\{\prev,\succ\})$ is the transition function, where the symbol $\bot$ is for ``undefined'', and for all $(q,a)\in Q\times A$, $\delta(q,a)=\bot$ if and only if $q\in\{q_\acc,q_\rej\}$. In each non-terminal step,
if the read/write head scans the $k$-th cell from the left of the $\ell$-th tape (for $\ell\in [1,n]$ and $k\geq 1$) and $(q,a)\in (Q\setminus\{q_{acc},q_{rej}\})\times A$ is the current pair state/scanned cell content, one of the following occurs:
\begin{itemize}
  \item $\delta(q,a)\in Q\times A\times \{\leftarrow,\rightarrow\}$ (\emph{ordinary moves}):
  $\mathcal{M}$ overwrites the tape cell being scanned, there is a change of state, and the read/write
      head moves one position to the left ($\leftarrow$) or right ($\rightarrow$) in accordance with $\delta(q,a)$.\footnote{If the read/write head points to the left-most cell of the $\ell$-th tape and $\delta(q,a)$ is of the form $(q',a,\leftarrow)$, then $\mathcal{M}$ moves to the rejecting state.}
  \item $\delta(q,a)\in Q \times\{\prev,\succ\}$ (\emph{jump moves}):  if $\delta(q,a)=(q',\prev)$ (resp., $\delta(q,a)=(q',\succ)$) for some $q'\in Q$, then the read/write head jumps to the $k$-th cell of the $(\ell-1)$-th tape (resp., $(\ell+1)$-th tape) and $\mathcal{M}$ moves to state $q'$ if $\ell>1$ (resp., $\ell<n$); otherwise, $\mathcal{M}$ moves to the rejecting state.
\end{itemize}

Initially, each tape contains a word in $I^{*}$ and the read/write
head points to the left-most cell of the first tape. Thus, an input of $\mathcal{M}$, called $n$-ary input, can be described by a tuple
 $(w_1,\ldots,w_n)\in (I^{*})^{n}$, where for all $i\in [1,n]$, $w_i$ represent the initial content of the $i$-th tape.
 $\mathcal{M}$ \emph{accepts} a $n$-ary input $(w_1,\ldots,w_n)\in (I^{*})^{n}$,  written $\mathcal{M}(w_1,\ldots,w_n)$, if the computation of $\mathcal{M}$ from $(w_1,\ldots,w_n)$  is accepting.  We consider the following problem.

\begin{definition}[TM Alternation problem] 
An instance of the problem is a tuple $(n,\mathcal{M})$, where  $n>1$  and  $\mathcal{M}$ is a  \emph{polynomial-time bounded} $n$-ary deterministic Turing machine with input alphabet $I$. The instance $(n,\mathcal{M})$ is \emph{positive} if and only if the following holds,  where,
for  $\ell\in [1,n]$,
$\Q_\ell=\exists$ if $\ell$ is odd, and $\Q_\ell=\forall$ otherwise:
\[
\Q_1 x_1\in I^{2^{n}}.\, \Q_2 x_2\in I^{2^{n}}.\,\ldots \Q_n x_n\in I^{2^{n}}. \, \mathcal{M}(x_1,\ldots,x_n).
\]
 Note that the quantifications $\Q_i$ are restricted to words over $I$ of length ${2^{n}}$. Thus, even if $\mathcal{M}$ is polynomial-time bounded, it operates on an input  whose size is exponential in $n$.
\end{definition}

\begin{proposition}\label{prop:SuitableCompleteProblem}
The TM Alternation problem is $\LINAEXPTIME$-complete.
\end{proposition}

The proof is standard; however, for  completeness, we give a proof of the hardness result, which follows from the next lemma.

\begin{lemma}\label{lemma:AlernationsPolynomial} Let $\mathcal{M}_{\mathcal{A}}$ be a singly exponential-time bounded ATM making a polynomial bounded number of alternations. Moreover, let $c\geq 1$ and $c_a\geq 1$ be  integer constants such that, for each input $\alpha$, when started on $\alpha$,
$\mathcal{M}_{\mathcal{A}}$ makes at most $|\alpha|^{c_a}$ alternations and  reaches a terminal configuration in at most $2^{|\alpha|^{c}}$ steps. Then, given an input $\alpha$, one can construct in time polynomial in $\alpha$ and in the size of  $\mathcal{M}_{\mathcal{A}}$ an instance $(2|\alpha|^{\textit{max}\{c,c_a\}}, \mathcal{M})$ of the \emph{TM Alternation problem} such that the instance is positive \emph{if and only if} $\mathcal{M}_{\mathcal{A}}$ accepts $\alpha$.
\end{lemma}
\begin{proof}
Let $\mathcal{M}_{\mathcal{A}}$, $c$, and $c_a$ be as in the above statement. Let $I_{\mathcal{A}}$ (resp., $A_{\mathcal{A}}$) be the input (resp., work) alphabet of $\mathcal{M}_{\mathcal{A}}$, where $I_{\mathcal{A}}\subset A_{\mathcal{A}}$, and $Q$ be the set of $\mathcal{M}_{\mathcal{A}}$-states.
Without loss of generality, we assume that the initial state of $\mathcal{M}_{\mathcal{A}}$ is existential.
Fix an input $\alpha\in I_{\mathcal{A}}^{*}$. We define $k= \max\{c,c_a\}$ and $n= 2|\alpha|^{k}$.

An $\alpha$-configuration is a word in $A_{\mathcal{A}}^{*}\cdot (Q\times A_{\mathcal{A}})\cdot A_{\mathcal{A}}^{*}$ of length exactly $2^{|\alpha|^{k}}$. Note that any configuration of $\mathcal{M}_{\mathcal{A}}$ reachable from the input $\alpha$ can be encoded by an $\alpha$-configuration. We denote by $C_\alpha$  the initial (existential) $\alpha$-configuration associated with the input $\alpha$. A \emph{partial computation} of $\mathcal{M}_{\mathcal{A}}$ is a finite sequence $\pi=C_1,\ldots,C_p$ of $\alpha$-configurations such that
 $p\leq 2^{|\alpha|^{k}}$ and for each $1\leq i<p$, $C_{i+1}$ is a $\mathcal{M}_{\mathcal{A}}$-successor of $C_i$ (note that a computation of $\mathcal{M}_{\mathcal{A}}$ over $\alpha$ is a partial computation).
We say that $\pi$ is \emph{uniform} if additionally one of the following holds:
\begin{itemize}
  \item $C_p$ is terminal and $\pi$ visits only existential $n$-configurations;
  \item $C_p$ is terminal and $\pi$ visits only universal $n$-configurations;
  \item $p>1$, $C_p$ is existential and for each $1\leq h<p$, $C_h$ is universal;
  \item $p>1$, $C_p$ is universal and for each $1\leq h<p$, $C_h$ is existential.
\end{itemize}

Let $\diamondsuit$ be a fresh symbol and $I=A_{\mathcal{A}}\cup\{\diamondsuit\}$. The \emph{code} of a partial computation $\pi=C_1,\ldots,C_p$ is the word over $I$ of length exactly $2^{n}$ (recall that $n=2|\alpha|^{k}$) given by $C_1,\ldots,C_p,C_{p+1}^{0},\ldots,C_{2^{|\alpha|^{k}}}^{0}$, where $C_i^{0}\in \{\diamondsuit\}^{2^{|\alpha|^{k}}}$ for all $p+1\leq i\leq 2^{|\alpha|^{k}}$. We construct a polynomial-time bounded $n$-ary deterministic Turing machine $\mathcal{M}$, which satisfies Lemma~\ref{lemma:AlernationsPolynomial} for the given input $\alpha$ of $\mathcal{M}_{\mathcal{A}}$. The input alphabet of $\mathcal{M}$ is $I$.
Given a  $n$-ary input $(w_1,\ldots,w_n)\in (I^{*})^{n}$, $\mathcal{M}$ operates in $n$-steps (macro steps). At step $i$, for $i\in [1,n]$, the behavior of $\mathcal{M}$ is as follows, where for a partial computation $\pi=C_1,\ldots,C_p$,  $\first(\pi)=C_1$ and $\last(\pi)=C_p$:
\begin{itemize}
  \item $i=1$.
  \begin{enumerate}
    \item  If $w_1\in I^{2^{n}}$ and $w_1$ encodes a uniform partial computation $\pi_1$ of $\mathcal{M}_{\mathcal{A}}$ \emph{from $C_\alpha$}, then the behavior is as follows. If $\last(\pi_1)$ is accepting (resp., rejecting), then $\mathcal{M}$ accepts (resp., rejects) the input. Conversely, if $\last(\pi_1)$ is not a terminal configuration, then $\mathcal{M}$ goes to step $i+1$.
        \item Otherwise, $\mathcal{M}$ \emph{rejects} the input.
  \end{enumerate}
  \item $i>1$.
  \begin{enumerate}
    \item  If $w_i\in I^{2^{n}}$ and $w_i$ encodes a uniform partial computation $\pi_i$ of $\mathcal{M}_{\mathcal{A}}$ such that $\first(\pi_i)=\last(\pi_{i-1})$, where $\pi_{i-1}$ is the uniform partial computation encoded by $w_{i-1}$, then the behavior is as follows. If $\last(\pi_i)$ is accepting (resp., rejecting), then $\mathcal{M}$ accepts (resp., rejects) the input. If instead $\last(\pi_i)$ is not a terminal configuration, then $\mathcal{M}$ goes to step $i+1$, if $i+1\leq n$, and rejects the input otherwise.
    \item Otherwise, if $i$ is odd (resp., even), then $\mathcal{M}$ \emph{rejects} (resp., \emph{accepts}) the input.
  \end{enumerate}
\end{itemize}

Note that (1.) in the steps above can be checked by $\mathcal{M}$ in polynomial time (in the size of the input) by using the transition function of $\mathcal{M}_{\mathcal{A}}$ and $n$-bit counters. Hence, $\mathcal{M}$ is a polynomial-time bounded $n$-ary deterministic Turing machine which can be built in time polynomial in $n$ and in the size of $\mathcal{M}_{\mathcal{A}}$. 

Now, we prove that the construction is correct, that is, $(n,\mathcal{M})$ is a positive instance of the TM Alternation problem if and only if $\mathcal{M}_{\mathcal{A}}$ accepts $\alpha$. For each $\ell\in [1,n]$, let $\Q_\ell=\exists$ if $\ell$ is odd, and $\Q_\ell=\forall$ otherwise.
Since $C_\alpha$ is existential, $\mathcal{M}_{\mathcal{A}}$ accepts $\alpha$ if and only if there is a  uniform partial  computation $\pi_1$ of  $\mathcal{M}_{\mathcal{A}}$ from $C_\alpha$ such that $\last(\pi_1)$ leads to acceptance. Moreover, for each $w_{1}\in I^{2^{n}}$, $\mathcal{M}$  accepts an input of the form $(w_1,w'_2,\ldots,w'_k)$ \emph{only if} $w_1$ encodes a non-rejecting uniform partial computation of $\mathcal{M}_{\mathcal{A}}$ from $C_\alpha$. Thus, since $\Q_1=\exists$, correctness of the construction directly follows from the following claim.

\begin{claim}
Let $\ell\in [1,n]$ and $\pi=\pi_1\ldots\pi_\ell$ be a partial computation of $\mathcal{M}_{\mathcal{A}}$ from $C_\alpha$ such that $\pi_\ell$ is uniform, and for each $1\leq j<\ell$, $\pi_j$ is non-empty and $\pi_j\cdot \first(\pi_{j+1})$ is uniform as well. Let $w_\ell\in I^{2^{n}}$ be the word encoding $\pi_\ell$, and for each $1\leq j<\ell$, $w_j\in I^{2^{n}}$ be the word encoding $\pi_j\cdot \first(\pi_{j+1})$. Then, $\last(\pi_\ell)$ leads to acceptance in $\mathcal{M}_{\mathcal{A}}$  if and only if
\begin{equation}\label{eq:Alternation}
\Q_{\ell+1} x_{\ell+1}\in I^{2^{n}}.\, \ldots \,\Q_n x_n\in I^{2^{n}}.\, \mathcal{M}(w_1,\ldots,w_\ell,x_{\ell+1},\ldots,x_n).
\end{equation}
\end{claim}
\begin{proof} The proof is by induction on $n-\ell$.

 In the base case, we have $\ell=n$. Note that, in this case, $\last(\pi_n)$ is a terminal configuration of $\mathcal{M}_{\mathcal{A}}$  (otherwise, the  number of alternations of existential and universal configurations along $\pi$ would be greater than $n-1\geq |\alpha|^{c_a}$). Thus, we need to show that $\last(\pi_n)$ is accepting if and only if $\mathcal{M}(w_1,\ldots,w_n)$. By construction, when started on the input $(w_1,\ldots,w_n)$, $\mathcal{M}$ reaches the $n$-th step and (1.) in this step is satisfied. Moreover, either $\last(\pi_n)$ is accepting and $\mathcal{M}$ accepts the input  $(w_1,\ldots,w_n)$, or
$\last(\pi_n)$ is rejecting and $\mathcal{M}$ rejects the input  $(w_1,\ldots,w_n)$. Hence, the result follows.

Let us now consider the inductive step, where $\ell<n$. First, assume that $\last(\pi_\ell)$ is a terminal configuration. By construction, on any input of the form
$(w_1,\ldots,w_\ell,w'_{\ell+1},\ldots,w'_n)$, $\mathcal{M}$ reaches the $\ell$-th step and (1.) in this step is satisfied. Moreover,
either $\last(\pi_\ell)$ is accepting and $\mathcal{M}$ accepts the input  $(w_1,\ldots,w_\ell,\allowbreak w'_{\ell+1},\ldots,w'_n)$, or
$\last(\pi_\ell)$ is rejecting and $\mathcal{M}$ rejects the input  $(w_1,\ldots,w_\ell,\linebreak w'_{\ell+1},\ldots,w'_n)$. Hence, in this case the result holds.
Now, assume that $\last(\pi_\ell)$ is not terminal. We consider the case where $\ell+1$ is even (the other case being similar).
Then, $\Q_{\ell+1}=\forall$. Since $C_\alpha$ is existential and $\last(\pi_\ell)$ is not terminal, by hypothesis,
  $\last(\pi_\ell)$ must be a universal configuration. First, assume that $\last(\pi_\ell)$ leads to acceptance. Let $w_{\ell+1}\in I^{2^{n}}$. By construction, on any input of the form
$(w_1,\ldots,w_\ell,w_{\ell+1},w'_{\ell+2}\ldots,w'_n)$, $\mathcal{M}$ reaches the $(\ell+1)$-th step. If $w_{\ell+1}$ satisfies (2.) in this step, then since $\ell+1$ is even, $\mathcal{M}$ accepts the input. Hence, $\Q_{\ell+2} x_{\ell+2}\in I^{2^{n}}.\, \ldots \Q_n x_n\in I^{2^{n}}. \, \mathcal{M}(w_1,\ldots,w_\ell,w_{\ell+1},x_{\ell+2},\ldots,x_n)$.
  Otherwise $w_{\ell+1}$ encodes a uniform partial computation $\pi_{\ell+1}$ of  $\mathcal{M}_{\mathcal{A}}$ from
$\last(\pi_\ell)$. Since $\last(\pi_\ell)$ leads to acceptance and $\last(\pi_\ell)$ is universal, $\last(\pi_{\ell+1})$ leads to acceptance as well. Thus, by applying the inductive hypothesis to the partial computation $\pi_1\ldots \pi_{\ell-1}\pi'_\ell \pi_{\ell+1}$ (where $\pi'_\ell$ is obtained from $\pi_\ell$ by removing $\last(\pi_\ell)$), it follows that $\Q_{\ell+2} x_{\ell+2}\in I^{2^{n}}.\, \ldots \Q_n x_n\in I^{2^{n}}. \, \mathcal{M}(w_1,\ldots,w_\ell,w_{\ell+1},x_{\ell+2},\ldots,$ $x_n)$. Thus the previous condition holds for each $w_{\ell+1}\in  I^{2^{n}}$. Since $\Q_{\ell+1}=\forall$, it follows that (\ref{eq:Alternation}) holds. For the converse direction, assume that (\ref{eq:Alternation}) holds. Let $\pi_{\ell+1}$ be any
uniform partial computation  of  $\mathcal{M}_{\mathcal{A}}$ from
$\last(\pi_\ell)$. We need to show that $\last(\pi_{\ell+1})$ leads to acceptance.  Since (\ref{eq:Alternation}) holds and $\Q_{\ell+1}=\forall$, we can apply the inductive hypothesis to the partial computation $\pi_1\ldots \pi_{\ell-1}\pi'_\ell \pi_{\ell+1}$ (where $\pi'_\ell$ is obtained from $\pi_\ell$ by removing $\last(\pi_\ell)$). The result follows,
 concluding the proof of the claim.
\end{proof} 
 
This concludes the proof of Lemma~\ref{lemma:AlernationsPolynomial}, as well.
\end{proof}

\subsection{$\LINAEXPTIME$-hardness of the alternating multi-tiling problem}
We prove here the $\LINAEXPTIME$-hardness of the alternating multi-tiling problem by a polynomial time reduction from the TM alternation problem.
Fix an instance $(n,\mathcal{M})$ of the TM alternation problem  where   $\mathcal{M}=\tpl{n,I,A,Q, \{q_\acc,q_{\textit{rej}}\},q_0,\delta}$
is a  polynomial-time bounded  $n$-ary deterministic Turing machine.

\begin{remark}[Assumptions on $\mathcal{M}$]\label{rem:assumptions}
In order to simplify the reduction, w.l.o.g., we can assume that $\mathcal{M}$ satisfies the following constraints:
\begin{itemize}
\item $n$ is even;
  \item for each $n$-ary input $(w_1,\ldots,w_n)\in I^{2^{n}}\times \ldots \times I^{2^{n}}$, $\mathcal{M}$ reaches a terminal configuration in exactly $2^{n}-1$ steps, and when  $\mathcal{M}$ halts, the read/write head points to a cell of the $n$-th tape;
  \item  there is no move leading to the initial state $q_0$;
  \item for all $a\in A$, $\delta(q_0,a)\in Q\times A\times \{\rightarrow\}$;
  \item for all $(q,a),(q',a')\in Q\times A$, if $\delta(q,a)\in \{q'\}\times \{\prev,\succ\}$, then $\delta(q',a')\notin Q\times \{\prev,\succ\}$.
\end{itemize}
\end{remark}

We construct, in polynomial time in the size of $(n,\mathcal{M})$, an instance $\Instance$ of the alternating multi-tiling problem
such that $(n,\mathcal{M})$ is a positive instance of the TM alternation problem if and only if $\Instance$ is a positive instance of the alternating multi-tiling problem.
By the definitions of the considered problems, it suffices to show the following.

\begin{proposition} 
One can construct in time polynomial in the size of $(n,\mathcal{M})$, an instance
$\Instance= \tpl{n,D,D_0,\allowbreak H,V, M,D_\acc}$ of the alternating multi-tiling problem such that $D_0=I$ and the following holds:
 for each $n$-ary input $(w_1,\ldots,w_n)\in I^{2^{n}}\times \ldots \times I^{2^{n}}$, $\mathcal{M}(w_1,\ldots,w_n)$ if and only if there exists a multi-tiling $F=(f_1,\ldots,f_n)$
 of $\Instance$ such that, for all $\ell\in [1,n]$, the initial condition $\Init(f_\ell)$ of the tiling $f_\ell$ is $w_\ell$.
\end{proposition}
\begin{proof} 
We adapt the well-known translation between time-space diagrams of computations (also known as computation tableaux) of a nondeterministic TM and tilings for a set of domino types entirely determined by the given TM. In such a translation, adjacent  rows in the tiled region encode successive configurations in a  computation of the  machine.

Let $\lceil A \rceil$  be a fresh copy of the work alphabet $A$ of $\mathcal{M}$.   For a word $w$ over $A$, we denote by $\lceil w \rceil $   the associated word over $\lceil A \rceil$. We define $U = A \times [1,n]$ and $\lceil U \rceil  = \lceil A \rceil \times [1,n]$.
We adopt the following set $D$ of domino types:
\[
D = I \cup U \cup \lceil U\rceil \cup   (Q \times U) \cup ((Q\times \{ \leftarrow,\rightarrow,\prev,\succ\})\times U) 
\]
Intuitively, in the encoding we keep track of the tape-indexes of $\mathcal{M}$. Moreover, the domino types in $(Q\times \{ \leftarrow,\rightarrow\})\times U$ (resp., $(Q\times \{\prev,\succ\})\times U$) are used to encode the effects of the ordinary moves (resp., jump moves). For each $d\in D$, we denote by $\symb(d)$ the associated letter in $A$ (recall that $I\subset A$).
Moreover, if $d \in D\setminus I$, we write $\tape(d)$ to mean the associated tape index $\ell\in [1,n]$.
Additionally, if $d\in (Q \times U) \cup ((Q\times \{ \leftarrow,\rightarrow,\prev,\succ\})\times U)$, we denote by $\state(d)$, the state $q\in Q$ associated with $d$. If instead $d\in I\cup U \cup \lceil U\rceil $, we set $\state(d) =\bot$ ($\bot$ is for ``undefined'').

Given $\ell\in [1,n]$ and a word $v$ over the alphabet
\[
A\cup \lceil A\rceil \cup   (Q \times A) \cup ((Q\times \{ \leftarrow,\rightarrow,\prev,\succ\})\times A)
\]
we write $v\oplus \ell$ to denote the word over alphabet $D\setminus I$ defined in the obvious way.
 
Fix an $n$-ary input $(w_1,\ldots,w_n)\in I^{2^{n}}\times \ldots \times I^{2^{n}}$ and a non-rejecting configuration $C$ of $\mathcal{M}$ reachable from the input
$(w_1,\ldots,w_n)$. Assume that in $C$, the read/write head points to the $k$-th cell of the $\ell$-th tape for some $k\geq 1$ and $\ell\in[1,n]$, and let $(q,a)\in Q \times A$ be the pair state/scanned cell content associated with $C$. Since on input  $(w_1,\ldots,w_n)$, $\mathcal{M}$ halts in $2^{n}-1$ steps, it holds that $k\leq 2^{n}$ and $C$ can be encoded by the tuples   of words in $D^{2^{n}}$ of the form $(w_1^{C}\oplus 1,\ldots,w_n^{C}\oplus n)$ defined as follows:
\begin{itemize}
\item \emph{cases $q=q_\acc$ or $\delta(q,a)\in Q\times A \times \{\leftarrow,\rightarrow\}$}:\\ for each $j\in [1,n]\setminus \{\ell\}$,
$w_j^{C}=w_j$ or $w_j^{C}=\lceil w_j \rceil$ where $w_j$ is the content of the first $2^{n}$ cells of the $j$-th tape, and one of the following holds:
\begin{itemize}
\item $q= q_\acc$: $w_\ell^{C}$ is of the form $w'\cdot (q,a) \cdot \lceil w'' \rceil$, where $w'\cdot a \cdot w''$ is the content of the first $2^{n}$ cells of the $\ell$-th tape and
$|w'|= k-1$ (the read/write head points to the $k$-th cell of the $\ell$-th tape);
\item   $\delta(q,a)  \in \{q'\}\times A \times \{\rightarrow\}$ for some $q'\in Q$:
$w_\ell^{C}$ is of the form $ w'\cdot (q, a) \cdot ((q',\rightarrow),a')\cdot \lceil w''\rceil$, where $w'\cdot a \cdot a' \cdot w''$ is the content of the first $2^{n}$ cells of the $\ell$-th tape and $|w'|= k-1$;
\item   $\delta(q,a)  \in \{q'\}\times A \times \{\leftarrow\}$ for some $q'\in Q$:
$w_\ell^{C}$ is of the form $w'\cdot ((q',\leftarrow\nolinebreak ),a')\cdot (q,a) \cdot  \lceil w''\rceil$, where $w'\cdot a' \cdot a \cdot w''$ is the content of the first $2^{n}$ cells of the $\ell$-th tape and $|w'|= k-1$;
\end{itemize}
\item \emph{case  $\delta(q,a)\in \{q'\}\times\{\prev\}$} for some $q'\in Q$:
\begin{itemize}
\item for each $j\in [1,n]\setminus \{\ell,\ell-1\}$,
     $w_j^{C}=w_j$ or $w_j^{C}=\lceil w_j \rceil$ where $w_j$ is the content of the first $2^{n}$ cells of the $j$-th tape;
\item $w_\ell^{C}$ is of the form $ w' \cdot (q,a) \cdot  \lceil w''\rceil$, where $w'\cdot a \cdot w''$ is the content of the first $2^{n}$ cells of the $\ell$-th tape and
$|w'|= k-1$;
\item if $\ell>1$, $w_{\ell-1}^{C}$ is of the form $w'\cdot ((q',\prev),a) \cdot \lceil w''\rceil$, where $w'\cdot a \cdot w''$ is the content of the first $2^{n}$ cells of the $(\ell-1)$-th tape and $|w'|= k-1$;
\end{itemize}
\item \emph{case  $\delta(q,a)\in \{q'\}\times\{\succ\}$} for some $q'\in Q$:
\begin{itemize}
\item for each $j\in [1,n]\setminus \{\ell,\ell+1\}$,
     $w_j^{C}=w_j$ or $w_j^{C}=\lceil w_j \rceil$ where $w_j$ is the content of the first $2^{n}$ cells of the $j$-th tape;
\item $w_\ell^{C}$ is of the form $w' \cdot (q, a) \cdot \lceil w''\rceil$, where $w'\cdot a \cdot w''$ is the content of the first $2^{n}$ cells of the $\ell$-th tape and
$|w'|= k-1$;
\item if $\ell<n$, $w_{\ell+1}^{C}$ is of the form $ w' \cdot ((q',\succ),a) \cdot \lceil w''\rceil$, where $w'\cdot a \cdot w''$ is the content of the first $2^{n}$ cells of the $(\ell+1)$-th tape and $|w'|= k-1$;
\end{itemize}
\end{itemize}

We construct in polynomial-time an instance $\Instance= \tpl{n,D,D_0,H,V, M,D_\acc}$ of the alternating multi-tiling problem with $D_0=I$ and $D_\acc = \{q_\acc\}\times U$
such that for each $n$-ary input $(w_1,\ldots,w_n)\in I^{2^{n}}\times \ldots \times I^{2^{n}}$,
$\mathcal{M}(w_1,\ldots,w_n)$ if and only if there exists a multi-tiling $F=(f_1,\ldots,f_n)$
 of $\Instance$ such that for all $\ell\in [1,n]$, the initial condition of the tiling $f_\ell$ is $w_\ell$. Moreover, if
$\mathcal{M}(w_1,\ldots,w_n)$, then the following holds:
\begin{itemize}
\item let $\pi = C_1 \ldots C_{2^{n}-1}$ be the accepting computation of $\mathcal{M}$ over $(w_1,\ldots,w_n)$ (by our assumptions, the
 write/read head in the accepting configuration $C_{2^{n}-1}$ points to the $n$-th tape).
 Then, there exist    multi-tilings
$F=(f_1,\ldots,f_n)$ of $\Instance$ associated with the input $(w_1,\ldots,w_n)$ such that, for all $j\in [1, 2^{n}-1]$, there is an encoding $\textit{cod}(C_j)$ 
of configuration $C_j$ so that, 
 for all $\ell\in [1,n]$, the
row of index $j$ of $f_\ell$ coincides with the $\ell$-th component of $\textit{cod}(C_j)$.
\end{itemize}

 We define the matching relations $H,V$ and $M$ in order to ensure the above conditions. In particular,  the horizontal matching relation
 $H$   guarantees that the TM configurations are correctly encoded, while the vertical matching relation is used to
 encode the ordinary moves of $\mathcal{M}$. Finally, the multi-tiling matching relation $M$ is used to encode the jump moves. Additionally, $H$, $V$ and $M$ also ensure that
 the rows of index 1 encode the initial configuration of $\mathcal{M}$ associated with the given $n$-ary input (the latter corresponds to the tuple of rows of index $0$).

 Formally $H$ is the set of pairs $(d,d')\!\in\! D\!\times\! D$ satisfying the following constraints:
\begin{itemize}
\item $d\in I$ if and only if   $d'\in I$;
\item if $d\in D\setminus I$, then $d'\in D\setminus I$ and $\tape(d)=\tape(d')$;
\item if $d,d'\in U\cup \lceil U \rceil$, then either $d,d'\in U$, or $d,d'\in \lceil U \rceil$;
\item $\state(d')\!\neq\! q_0$, and whenever $\state(d)\!=\!q_0$, then $d\!\in\!\{q_0\}\!\times\! U$ and $\tape(d)\!=\!1$;
\item if   $d\in \lceil U \rceil$ or $d\in (Q\times \{\rightarrow,\prev,\succ\})\times U$, then $d'\in \lceil U \rceil$;
\item if $d\in Q\times U$ and $\delta(\state(d),\symb(d))\notin Q\times A \times \{\rightarrow\}$, then $d'\in \lceil U \rceil$;
\item if  $d'\in U$ or $d'\in (Q\times \{\leftarrow,\prev,\succ\})\times U$, then $d\in U$;
\item if $d'\in Q\times U$ and $\delta(\state(d),\symb(d))\notin Q\times A \times \{\leftarrow\}$, then $d\in U$;
\item for $q'\in Q$,  $d\in Q\times U$ and $\delta(\state(d),\symb(d))\in \{q'\}\times A\times \{\rightarrow\}$ \emph{if and only if}  $d'\in \{(q',\rightarrow)\}\times U$;
\item for $q'\in Q$, $d'\in Q\times U$ and $\delta(\state(d),\symb(d))\in \{q'\}\times A\times \{\leftarrow\}$ \emph{if and only if} $d\in \{(q',\leftarrow)\}\times U$.
\end{itemize}

By definition of $H$ (independently of $V$ and $M$), we deduce the following.
\begin{claim}\label{cl:ccl1}
Let $f$ be a tiling of $\Instance$ and $\row$ be the content of any row of $f$. Then, either $\row \in I^{*}$ 
or $\row =\row' \oplus \ell$ for some $\ell\in [1,n]$, and $\row'$ satisfies one of the following:
\begin{itemize}
  \item $\row' \!=\! w \!\cdot\! (q,a)\!\cdot\! \lceil w' \rceil$ such that $w,\! w'\!\in\! A^{*}$ and $\delta(q,a)\!\notin\! Q\!\times\! A \!\times\! \{\leftarrow,\rightarrow\} $;
  \item $\row' = w \cdot d \cdot \lceil w' \rceil $ such that $w,w'\in A^{*}$, and $d\in  (Q \times \{\prev,\succ\})\times A$;
  \item $\row' = w$ such that $w\in A^{*}$;
  \item $\row'= \lceil w \rceil$ such that $w\in A^{*}$;
  \item $\row' = w \cdot d \cdot ((q', \rightarrow),a')\cdot \lceil w'\rceil $ such that $w,w'\in A^{*}$, $d\in Q\times A$, and $\delta(\state(d),\symb(d))\in \{q'\}\times A \times \{\rightarrow\} $;
  \item $\row' =  w  \cdot ((q', \leftarrow), a' ) \cdot d\cdot  \lceil w'\rceil $ such that $w,w'\in A^{*}$, $d\in Q\times A$,  and $\delta(\state(d),\symb(d))\in \{q'\}\times A \times \{\leftarrow\} $;
\end{itemize}
Moreover, if for some $i$, $\state(\row(i))=q_0$, then $i= 0$, $\row(0)\in \{q_0\}\times U$, and $\tape(\row(0))=1$.
\end{claim}

Now, let us define the vertical matching relation $V$. $V$ is the set of pairs $(d,d')\in D\times D$ satisfying the following constraints:
\begin{itemize}
\item if $d\!\in\! I$ then   $d'\!\in\! (\{q_0\}\times U) \cup U \cup \lceil U \rceil \cup (Q\times \{\rightarrow\})\times U$ and $\symb(d)\!=\!\symb(d')$;
\item if $d\in D\setminus I$, then $d'\in D\setminus I$ and $\tape(d)=\tape(d')$;
\item if $d \in U \cup \lceil U \rceil$ then $d'\in U \cup \lceil U \rceil \cup ((Q\times \{\prev,\succ\})\times U)$ and $\symb(d)=\symb(d')$;
\item if $d\in (Q \times \{\leftarrow,\rightarrow,\prev,\succ\})\times U$, then $d'=(\state(d),\symb(d),\tape(d))$;
\item if $d\in Q\times U$, then $\state(d)\neq q_\acc$ and one of the following holds:
\begin{itemize}
\item  $\delta(\state(d),\symb(d))\in Q\times \{\symb(d')\}\times \{\leftarrow,\rightarrow\}$ and  $d'\in U\cup \lceil U\rceil$;
\item $\delta(\state(d),\symb(d))\!\in\! Q\times \{\prev,\succ\}$, $d'\!\in\! U$ and $\symb(d')\!=\!\symb(d)$.
\end{itemize}
\end{itemize}

Finally, we define the multi-tiling matching relation $M$. $M$ is the set of pairs $(d,d')\in D\times D$ satisfying the following constraints:
\begin{itemize}
\item $d\in I$ if and only if $d'\in I$;
\item if $d\in D\setminus I$, then $d'\in D\setminus I$ and $\tape(d')=\tape(d)+1$;
\item $\state(d)\neq q_\acc$ and $\state(d')\neq q_0$;
\item for each $q\in Q$,   $d\in Q\times U$ and $\delta(\state(d),\symb(d))=(q,\succ)$ \emph{if and only if} $d'\in \{(q,\succ)\}\times U$;
\item for each $q\in Q$,   $d'\in Q\times U$ and $\delta(\state(d),\symb(d))=(q,\prev)$ \emph{if and only if} $d\in \{(q,\prev)\}\times U$;
\item if $d\in (Q \times \{\succ\})\times U$, then $\tape(d)>1$;
\item if $d'\in (Q \times \{\prev\})\times U$, then $\tape(d')<n$.
\end{itemize}

By Claim~\ref{cl:ccl1} and the definitions of the relations $V$ and $M$, one can prove that if $F$ is a multi-tiling of
$\Instance$ with initial conditions  $(w_1,\ldots,w_n)\in I^{2^{n}}\times \ldots \times I^{2^{n}}$, then $F$ encodes an accepting computation
of $\mathcal{M}$ over the $n$-ary input $(w_1,\ldots,w_n)\in I^{2^{n}}\times \ldots \times I^{2^{n}}$. Vice versa, by Remark~\ref{rem:assumptions},
if $\mathcal{M}(w_1,\ldots,w_n)$, then it follows that there is a multi-tiling encoding the accepting computation of $\mathcal{M}$ over
$(w_1,\ldots,w_n)$.
\end{proof}

\section{Proof of Proposition~\ref{proposition:wellFormdnessRegex}}\label{proof:proposition:wellFormdnessRegex}

\begin{proposition*}[\ref{proposition:wellFormdnessRegex}]
Let $\Ku=\KuDef$ be a finite Kripke structure, $\varphi$ be an $\AAbarBBbar$ formula  with $\RE$'s $r_1,\ldots ,r_k$ over $\Prop$, $\rho\in\Trk_\Ku$ be
a trace, and $(q^1,\ldots , q^k)\in Q(\varphi)$. There exists a trace $\pi\in\Trk_\Ku$, which is $(q^1,\ldots , q^k)$-well-formed w.r.t.\ $\rho$, such that $|\pi| \leq |\States|\cdot 2^{2\sum^k_{\ell=1}|r_\ell|}$.
\end{proposition*}

\begin{figure}[b]
\centering
    \scalebox{1.4}{
    \begin{tikzpicture}
    	\filldraw [gray] (0,0) circle (2pt)
    	(1.5,0) circle (2pt)
    	(2,0) circle (2pt)
    	(3.5,0) circle (2pt);
    	\filldraw [gray] (0,-0.5) circle (2pt)
    	(1.5,-0.5) circle (2pt)
    	(3,-0.5) circle (2pt);
    	\draw [red] (1.5,0) -- (2,0);
    	\draw [black]  (0,0) -- (1.5,0);
    	\draw [black] (2,0) -- (3.5,0);
    	\draw [black] (0,-0.5) -- (3,-0.5);
    	\draw [dashed, red] (0,0) -> (0,-0.5);
    	\draw [dashed, red] (1.5,0) -> (1.5,-0.5);
    	\draw [dashed, red] (2,0) -> (1.5,-0.5);
    	\draw [dashed, red] (3.5,0) -> (3,-0.5);
    	{\tiny
    		\node (a0) at (3.8,0) {$\rho$};	
    		\node (b0) at (4.1,-0.5) {$\pi=\rho(1,\! l_1)\!\star\! \rho(l_2,n)$};	
    		\node (a1) at (1.35,0.22) {$\rho(l_1)$};
    		\node (a11) at (1.72,0.22) {$=$};
    		\node (a2) at (2.1,0.22) {$\rho(l_2)$};
    		\node (a111) at (2.5,0.22) {$=$};
    		\node (a21) at (2.7,0.22) {$s$};
    		\node (a3) at (1.67,0.7) {$\forall t,\; \Du^t_{q^t}(\Lab(\rho^{l_1}))=\Du^t_{q^t}(\Lab(\rho^{l_2}))$};	
    	}
    \end{tikzpicture}}
    \caption{The contraction step of Proposition~\ref{proposition:wellFormdnessRegex}.}\label{fig:basicContrRegex}
\end{figure}
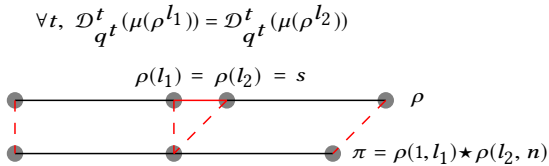

\begin{proof}
Let $\rho\in\Trk_\Ku$ with  $|\rho|=n$. If $n\leq |\States|\cdot 2^{2\sum^k_{\ell=1}|r_\ell|}$, the thesis trivially holds. 
Thus, let us assume $n > |\States|\cdot 2^{2\sum^k_{\ell=1}|r_\ell|}$. We show that there exists a trace 
which is $(q^1,\ldots , q^k)$-well-formed w.r.t.\ $\rho$, whose length is less than $n$.
The number of possible (joint) configurations of the $\DFA$s $\Du(\varphi)$ is (at most) $|Q(\varphi)|\leq 2^{2|r_1|}\cdots 2^{2|r_k|} = 2^{2\sum^k_{\ell=1}|r_\ell|}$. 
Since $n> |\States|\cdot 2^{2\sum^k_{\ell=1}|r_\ell|}$, there exists some state $s\in \States$ occurring in $\rho$ at least twice in the $\rho$-positions, say $1\leq l_1< l_2\leq |\rho|$, such that $\Du^t_{q^t}(\Lab(\rho^{l_1}))=\Du^t_{q^t}(\Lab(\rho^{l_2}))$, for all $t=1,\ldots,k$. 
Let us consider $\pi=\rho(1,l_1)\star\rho(l_2,n)$ (see Figure~\ref{fig:basicContrRegex}). 
It is easy to see that $\pi\in\Trk_\Ku$, as $\rho(l_1)=\rho(l_2)$, and $|\pi| < n$. Moreover, $\pi$ is $(q^1,\ldots , q^k)$-well-formed w.r.t.\ $\rho$ (the corresponding positions are $i_j=j$ if $j\leq l_1$, and $i_j=j+(l_2-l_1)$ otherwise). If $|\pi| \leq |\States|\cdot 2^{2\sum^k_{\ell=1}|r_\ell|}$, the thesis holds. Otherwise, the same basic step can be iterated a finite number of times, and the thesis follows by transitivity of the $(q^1,\ldots , q^k)$-well-formedness relation.
\end{proof}

\section{Proof of Theorem~\ref{theorem:expSizeModelPropertyAAbarBBbarRegex}}\label{proof:theorem:expSizeModelPropertyAAbarBBbarRegex}

\begin{theorem*}[\ref{theorem:expSizeModelPropertyAAbarBBbarRegex}, Exponential small-model for $\AAbarBBbar$]
Let $\Ku=(\Prop,\States,\allowbreak \Edges,\Lab,\sinit)$ be a finite Kripke structure, $\sigma,\rho \in \Trk_\mathpzc{K}$, and $\varphi$ be an $\AAbarBBbar$ formula in \nnf{}, with $\RE$'s  $r_1,\ldots ,r_u$ over $\Prop$, such that $\Ku,\sigma\star\rho\models \varphi$. There exists $\pi\in \Trk_\mathpzc{K}$, induced by $\rho$, such that $\Ku,\sigma\star\pi\models \varphi$ and $|\pi|\leq |\States|\cdot (|\varphi|+1) \cdot 2^{2\sum_{\ell=1}^u |r_\ell|}$.
\end{theorem*}

\begin{proof} 
Let $Wt(\varphi,\sigma\star\rho)$ be the set of witness positions of $\sigma\star\rho$ for $\varphi$.
Let $\{i_1,\ldots,i_k\}$ be the ordering of $Wt(\varphi,\sigma\star\rho)$ such that
$i_1<\ldots <i_k$. Let $i_0=1$ and $i_{k+1}=|\sigma\star\rho|$. Hence, $1=i_0\leq  i_1<\ldots <i_k < i_{k+1}=|\sigma\star\rho|$.
If the length of $\rho$ is at most $|\States|\cdot (|\varphi|+1) \cdot 2^{2\sum_{\ell=1}^u |r_\ell|}$, the thesis trivially holds.
Let us assume that $|\rho|> |\States|\cdot (|\varphi|+1) \cdot 2^{2\sum_{\ell=1}^u |r_\ell|}$. We show that there exists a trace $\pi$ induced by $\rho$, with $|\pi| < |\rho |$, such that $\Ku,\sigma\star\pi\models \varphi$. 

W.l.o.g., we can assume that $i_0\leq i_1<\ldots <i_{j-1}$, for some $j\geq 1$, are $\sigma$-positions (while $i_{j}<\ldots <i_{k+1}$ are $(\sigma\star\rho)$-positions not in $\sigma$). We claim that either $(i)$~there exists $t\in [j,k]$ such that $i_{t+1}-i_t>|\States|\cdot 2^{2\sum_{\ell=1}^u |r_\ell|}$ or $(ii)$~$|(\sigma\star\rho)(|\sigma|,i_{j})|>|\States|\cdot 2^{2\sum_{\ell=1}^u |r_\ell|}$. By way of contradiction, suppose that neither $(i)$ nor $(ii)$ holds. We need to distinguish two cases. 
If $\sigma\star\rho=\rho$, then 
$|\rho| = (i_{k+1}-i_0)+1\leq (k+1) \cdot |\States|\cdot 2^{2\sum_{\ell=1}^u |r_\ell|} +1$ (a contradiction); 
otherwise ($|\rho| < |\sigma\star\rho|$), $|\rho| = (i_{k+1} - i_{j}) + |(\sigma\star\rho)(|\sigma|,i_{j})| \leq k \cdot |\States|\cdot 2^{2\sum_{\ell=1}^u |r_\ell|} + |\States|\cdot 2^{2\sum_{\ell=1}^u |r_\ell|} \leq (k+1) \cdot |\States|\cdot 2^{2\sum_{\ell=1}^u |r_\ell|}$. The contradiction follows since $(k+1) \cdot |\States|\cdot 2^{2\sum_{\ell=1}^u |r_\ell|}+1 \leq |\varphi| \cdot |\States|\cdot 2^{2\sum_{\ell=1}^u |r_\ell|}+1 \leq |\States|\cdot (|\varphi|+1)\cdot 2^{2\sum_{\ell=1}^u |r_\ell|}$.
 
Let us define $(\alpha,\beta)=(i_t,i_{t+1})$ in case $(i)$, and $(\alpha,\beta)=(|\sigma|,i_{j})$ in case $(ii)$. Moreover let $\rho'=\rho(\alpha,\beta)$. In both the cases, we have  $|\rho'|> |\States|\cdot 2^{2\sum_{\ell=1}^u |r_\ell|}$.
By Proposition~\ref{proposition:wellFormdnessRegex}, there exists a trace $\pi'$ of $\Ku$, $(q^1,\ldots , q^u)$-well-formed with respect to $\rho'$, such that $|\pi'|\leq |\States|\cdot 2^{2\sum_{\ell=1}^u |r_\ell|} < |\rho'|$, 
where we choose $q^x=\Du^x(\Lab((\sigma\star\rho)^{\alpha-1}))$ for $x=1,\ldots , u$ (as a particular case we set $q^x$ as the initial state of $\Du^x$ if $\alpha=1$).
Let $\pi$ be the trace induced by $\rho$ obtained by replacing the subtrace $\rho'$ of $\rho$ with $\pi'$ (see Figure~\ref{fig:contr2Regex}). Since $|\pi|<|\rho|$, it remains to prove that  $\Ku,\sigma\star\pi\models \varphi$.

\begin{figure}[b]
\centering
    \resizebox{0.9\linewidth}{!}{\begin{tikzpicture}
				\filldraw [gray] (0,0) circle (2pt)
				(1.0,0) circle (2pt)
				(2.5,0) circle (2pt)
				(3.5,0) circle (2pt) 
				(5.0,0) circle (2pt)
				(6.5,0) circle (2pt);
				\filldraw [gray] (0,-0.5) circle (2pt)
				(1.0,-0.5) circle (2pt)
				(2.5,-0.5) circle (2pt)
				(3.25,-0.5) circle (2pt) 
				(4.75,-0.5) circle (2pt)
				(6.25,-0.5) circle (2pt);
				\draw [red] (2.5,0) -- (3.5,0);
				\draw [black]  (0,0) -- (2.5,0);
				\draw [black] (3.5,0) -- (6.5,0);
				\draw [black] (0,-0.5) -- (2.5,-0.5);
				\draw [red] (2.5,-0.5) -- (3.25,-0.5);
				\draw [black] (3.25,-0.5) -- (6.25,-0.5);
				\draw [dashed, red] (2.5,0) -> (2.5,-0.5);
					\draw [dashed, red] (3.5,0) -> (3.25,-0.5);
				\draw [dashed, red] (5,0) -> (4.75,-0.5);
				\draw [dashed, red] (6.5,0) -> (6.25,-0.5);
				{\tiny
					\node (a0) at (6.9,0) {$\rho$};	
					\node (b0) at (6.5,-0.8) {$\pi\!=\!\rho(1,\! i_t)\!\star\! \pi'\! \star\! \rho(i_{t+1},|\rho|)$};	
					\node (a1) at (1,0.2) {$i_1$};
					\node (a2) at (2.5,0.2) {$i_t$};
					\node (a3) at (3.5,0.2) {$i_{t+1}$};
					\node (a4) at (5,0.2) {$i_{j}$};
					\node (a1) at (1,0.55) {$\hsB\! \psi_1$};
					\node (a2) at (2.5,0.55) {$\hsB\! \psi_{t}$};
					\node (a3) at (3.5,0.55) {$\hsB\! \psi_{t+1}$};
					\node (a4) at (5,0.55) {$\hsB\! \psi_{j}$};
					\node [red] (a4) at (2.9,-0.65) {$\pi'$};
					\node [red] (a4) at (3.0,-0.15) {$\rho'$};
				}
				
			\end{tikzpicture}}
    \caption{Representation of the contraction step of Theorem~\ref{theorem:expSizeModelPropertyAAbarBBbarRegex}---case $(i)$}\label{fig:contr2Regex}
\end{figure}
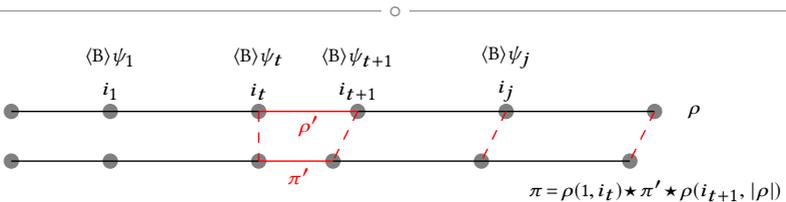

Let us denote $\sigma\star\pi$ by $\overline{\pi}$ and $\sigma\star\rho$ by $\overline{\rho}$. Moreover,
let $H:[1,|\overline{\pi}|] \rightarrow [1,|\overline{\rho}|]$ be the function mapping positions of $\overline{\pi}$ into positions of $\overline{\rho}$ in this way: positions ``outside'' $\pi'$ (i.e., outside the interval $[\alpha,\alpha+|\pi'|-1]$) are mapped into their original position in $\overline{\rho}$; positions ``inside'' $\pi'$ (i.e., in $[\alpha,\alpha+|\pi'|-1]$) are mapped to the corresponding position in 
$\rho'$ (exploiting well-formedness of $\pi'$ w.r.t.\ $\rho'$). Formally,  $H$ is defined as:
\begin{equation*}\label{equation:H}
H(m) = \begin{cases}  
	m& \text{if}\quad m<\alpha \\
\alpha+\ell_{m-\alpha+1}-1 & \mbox{if}\quad \alpha\leq m<\alpha+|\pi'| \\
m+(|\rho'| - |\pi'|) & \mbox{if}\quad m\geq \alpha+|\pi'|
\end{cases}
\end{equation*}
where  $\ell_m$ is the $\rho'$-position corresponding to the $\pi'$-position $m$.
It is easy to check that $H$ satisfies the following properties: 
\begin{enumerate}
    \item $H$ is strictly monotonic, i.e., for all $j,j'\in [1,|\overline{\pi}|]$, $j<j'\iff H(j)<H(j')$;
    \item for all $j\in [1,|\overline{\pi}|]$, $\overline{\pi}(j) = \overline{\rho}(H(j))$;
    \item $H(1)= 1$ and $H(|\overline{\pi}|)=|\overline{\rho}|$;
    \item $Wt(\varphi, \overline{\rho}) \subseteq \{H(j) \mid j \in [1,|\overline{\pi}|] \}$  i.e., all witness positions are preserved;
    \item for each $j\in [1,|\overline{\pi}|]$ and $x=1,\ldots , u$, $\Du^x(\Lab(\overline{\pi}^{j}))=\Du^x(\Lab(\overline{\rho}^{H(j)}))$.
\end{enumerate}
We only comment on Property~5.
The property holds for $j\in [1,\alpha-1]$, as $\overline{\pi}^j=\overline{\rho}^{H(j)}=\overline{\rho}^j$. 
For $j\in [\alpha,\alpha+|\pi'|-1]$, $\Du^x(\Lab(\overline{\pi}^j))=\Du^x(\Lab(\overline{\rho}^{H(j)}))$ follows from the well-formedness hypothesis.
Finally, being $\overline{\rho}(\beta,|\overline{\rho}|)=\overline{\pi}(\alpha+|\pi'|-1,|\overline{\pi}|)$ and $\Du^x(\Lab(\overline{\pi}^{\alpha+|\pi'|-1}))=\Du^x(\Lab(\overline{\rho}^\beta))$, the property holds also for $j\in [\alpha+|\pi'|,|\overline{\pi}|]$.

The fact that
$\Ku,\overline{\pi}\models \varphi$ is an immediate consequence of the following claim, considering that  $H(|\overline{\pi}|)=|\overline{\rho}|$, $\Ku,\overline{\rho}\models \varphi$,  $\overline{\rho}^{|\overline{\rho}|}=\overline{\rho}$, and $\overline{\pi}^{|\overline{\pi}|}=\overline{\pi}$.

\begin{claim}\label{claim:Hj} For all $j\in [1,|\overline{\pi}|]$, all subformulas $\psi$ of $\varphi$, and all $\xi\in\Trk_\Ku$, 
it holds that
$
\Ku,\overline{\rho}^{H(j)}\star \xi\models \psi \Longrightarrow \Ku,\overline{\pi}^{j}\star \xi\models \psi.
$
\end{claim}
\begin{proof}
Assume that $\Ku,\overline{\rho}^{H(j)}\star \xi\models \psi $. Note that $\overline{\rho}^{H(j)}\star \xi$ is defined if and only if $\overline{\pi}^{j}\star \xi$ is defined. We prove by induction on the structure of $\psi$  that
$\Ku,\overline{\pi}^{j}\star \xi\models \psi$. Since $\varphi$ is in \nnf, only the following cases can occur.

\begin{itemize}
  \item $\psi= r_t$ or $\psi=\neg r_t$ where $r_t$ is some $\RE$ over $\Prop$. By Property~5 of $H$, 
   $\Du^t(\Lab(\overline{\pi}^j))=\Du^t(\Lab(\overline{\rho}^{H(j)}))$, thus $\Du^t(\Lab(\overline{\pi}^j\star \xi))=\Du^t(\Lab(\overline{\rho}^{H(j)}\star \xi))$. It follows that
    $\Ku, \overline{\pi}^{j}\star \xi \models r_t$ if and only if $\Ku,\overline{\rho}^{H(j)}\star \xi \models r_t$, and the thesis holds.
    \item $\psi= \theta_1\wedge\theta_2$ or $\psi= \theta_1\vee\theta_2$ for some $\AAbarBBbar$ formulas $\theta_1$ and $\theta_2$. 
    The result holds by the inductive hypothesis.
   \item $\psi = \hsBu\theta$. We need to show that for each proper prefix $\eta$ of $\overline{\pi}^{j}\star \xi$, we have $\Ku,\eta\models\theta$. We distinguish two cases:
     \begin{itemize}
       \item $\eta$ is \emph{not} a proper prefix of $\overline{\pi}^{j}$. Hence, $\eta$ has the form $ \overline{\pi}^{j}\star \xi^h$ for some $h\in [1,|\xi|-1]$. Since $\Ku,\overline{\rho}^{H(j)}\star \xi\models \hsBu\theta $, then
       $\Ku,\overline{\rho}^{H(j)}\star \xi^h\models \theta $.  By the inductive hypothesis, we have $\Ku,\overline{\pi}^j\star \xi^h\models \theta $.
       \item $\eta$ is a proper prefix of $\overline{\pi}^{j}$. Hence, $\eta = \overline{\pi}^{h}$ for some $h\in [1,j-1]$.   By Property~1 of $H$, $H(h)<H(j)$, and since $\Ku,\overline{\rho}^{H(j)}\star \xi\models \hsBu\theta $, we have that $\Ku,\overline{\rho}^{H(h)}\models \theta $. By the inductive hypothesis, $\Ku,\overline{\pi}^h\models \theta $.
     \end{itemize}
    Therefore, $\Ku,\overline{\pi}^j\star \xi\models \hsBu\theta $ holds.
    \item $\psi = \hsB\theta$. We have to show that there exists a proper prefix of $\overline{\pi}^{j}\star \xi$ satisfying $\theta$.
Since $\Ku,\overline{\rho}^{H(j)}\star \xi\models \psi $, there exists a proper prefix $\eta'$ of  $\overline{\rho}^{H(j)}\star \xi$
 such that  $\Ku,\eta'\models \theta $. We distinguish two cases:
 \begin{itemize}
   \item $\eta'$ is \emph{not} a proper prefix of $\overline{\rho}^{H(j)}$. Hence, $\eta'$ is of the form $\overline{\rho}^{H(j)}\star \xi^h$ for some $h\in [1,|\xi|-1]$. By the inductive hypothesis, $\Ku,\overline{\pi}^j\star \xi^h\models \theta $, hence $\Ku,\overline{\pi}^j\star \xi\models \hsB\theta $.
   \item $\eta'$ is a proper prefix of $\overline{\rho}^{H(j)}$. Hence, $\eta' = \overline{\rho}^{i}$ for some $i\in [1,H(j)-1]$, and $\Ku,\overline{\rho}^{i}\models \theta $.
   Let $i'$ be the smallest position of $\overline{\rho}$ such that $\Ku,\overline{\rho}^{i'}\models\theta$. Hence $i'\leq i$ and $i'\in Wt(\varphi,\overline{\rho})$. By Property~4 of $H$, $i'=H(h)$ for some $\overline{\pi}$-position $h$. Since $H(h)<H(j)$, it holds that $h<j$ (Property~1). By the inductive hypothesis, $\Ku,\overline{\pi}^h\models \theta $, and thus $\Ku,\overline{\pi}^j\star \xi\models \hsB\theta $.
 \end{itemize}
 Therefore, in both cases, $\Ku,\overline{\pi}^j\star \xi\models \hsB\theta$.
    \item $\psi=\hsBtu\theta$ or $\psi=\hsBt\theta$. The thesis directly follows from the inductive hypothesis.
    \item $\psi = \hsAu\theta$, $\psi=\hsA\theta$, $\psi = \hsAtu\theta$ or $\psi=\hsAt\theta$. Since $\overline{\pi}^j\star \xi$ and $\overline{\rho}^{H(j)}\star \xi$ start at the same state and lead to the same state (by Property~2 and 3 of $H$), the result trivially follows. This concludes the proof of the claim.\qedhere
\end{itemize}
\end{proof}

We have shown that $\Ku,\overline{\pi}\models \varphi$, with $|\pi| < |\rho |$. If $|\pi|\leq |\States|\cdot (|\varphi|+1)\cdot 2^{2\sum_{\ell=1}^u |r_\ell|}$, the thesis holds. Otherwise, we can iterate the above contraction a finite number of times until the bound is reached.
\end{proof}

\section{Proof of Theorem~\ref{corrComplCheckBBbar}}\label{proof:corrComplCheckBBbar}

\begin{theorem*}[\ref{corrComplCheckBBbar}]
Let $\Phi$ be a $\B\Bbar$ formula, $\psi$ be a subformula of $\Phi$, and $\rho\in\Trk_{\Ku}$ be a trace with $s=\lst(\rho)$. Let $G$ be the subset of formulas in $\subfB(\psi)$ that hold on some proper prefix of $\rho$.
Let $\Du(\Phi)$ be the current configuration of the $\DFA$s associated with the regular expressions in $\Phi$ after reading $\mu(\rho(1,|\rho|-1))$.
%
Then $\texttt{Check}(\Ku,\psi,s,G,\Du(\Phi))=\top \!\iff\! \Ku,\rho\models \psi$.
\end{theorem*}

\begin{proof}
The proof is by induction on the structure of $\psi$. The thesis trivially follows for the cases $\psi=r$ (regular expression), $\psi=\neg\psi'$, $\psi=\psi_1\wedge\psi_2$, and $\psi=\hsB\psi'$.

Let us now assume $\psi=\hsBt\psi'$. 
\texttt{Check}$(\Ku,\psi,s,G,\Du(\Phi))=\top$ if and only if, for some $b''\in\{1,\ldots , |\States|\cdot (2|\psi'|+1)\cdot 2^{2\sum_{\ell=1}^u |r_\ell|}-1 \}$ and some $(G'',\Du(\Phi)'',s'')\in\conf(\Ku,\psi)$ ($=\conf(\Ku,\psi')$), we have \texttt{Reach}$(\Ku,\psi',(G,\Du(\Phi),s),(G'',\Du(\Phi)'', s''),\allowbreak b'')=\top$ and \texttt{Check}$(\Ku,\psi',s'',G'', \Du(\Phi)'')=\top$.
We preliminary prove the following claim.
    \begin{claim}
        Let $b\in\mathbb{N}$, $b>0$.
        Let $\tilde{\rho}\in\Trk_{\Ku}$ be a trace with $\tilde{s}=\lst(\tilde{\rho})$. Let $\tilde{G}$ be the subset of formulas in $\subfB(\psi')$ that hold on some proper prefix of $\tilde{\rho}$.
        Let $\tilde{\Du}(\Phi)$ be the current configuration of states of the $\DFA$s associated with the regular expressions in $\Phi$, reached from the initial states after reading $\Lab(\tilde{\rho}(1,|\tilde{\rho}|-1))$.
        
        For $(\tilde{G},\tilde{\Du}(\Phi),\tilde{s})$, $(G',\Du(\Phi)',s')\in\conf(\Ku,\psi')$, we have 
        \texttt{Reach}$(\Ku,\psi',\allowbreak (\tilde{G},\tilde{\Du}(\Phi),\tilde{s}),\allowbreak (G',\Du(\Phi)',s'),b)=\top$ if and only if there exists $\rho'\in\Trk_{\Ku}$ such that $\tilde{\rho}\cdot\rho'\in\Trk_\Ku$, $|\rho'|=b$, $\lst(\rho')=s'$, $G'$ is the subset of formulas in $\subfB(\psi')$ that hold on some proper prefix of $\tilde{\rho}\cdot\rho'$, and $\Du(\Phi)'$ is the current configuration of the $\DFA$s associated with the regular expressions of $\Phi$, after reading $\Lab(\tilde{\rho}\cdot\rho'(1,|\tilde{\rho}\cdot\rho'|-1))$.
    \end{claim}
    \begin{proof}
    The proof is by induction on $b\geq 1$.
    
    If $b\!=\!1$ we have \texttt{Reach}$(\Ku,\psi',(\tilde{G},\tilde{\Du}(\Phi),\tilde{s}),(G',\Du(\Phi)',s'),b)\!=\!\top$ if and only if  \texttt{Compatible}($\Ku,\psi',\allowbreak (\tilde{G},\tilde{\Du}(\Phi),\tilde{s}),(G',\Du(\Phi)',s'))\!=\!\top$; this happens if and only if: \begin{enumerate}
        \item[1.] $(\tilde{s},s')\in \Edges$, i.e., $(\tilde{s},s')$ is an edge of $\Ku$;
        \item[2.] \texttt{advance}$(\tilde{\Du}(\Phi),\Lab(\tilde{s}))=\Du(\Phi)'$;
        \item[3.] $\tilde{G}\subseteq G'$;
        \item[4.] for each $\varphi\in(G'\setminus \tilde{G})$, \texttt{Check}$(\Ku,\varphi,\tilde{s},\tilde{G}\cap \subfB(\varphi),\tilde{\Du}(\Phi))=\top$;
        \item[5.] for each $\varphi\in(\subfB(\psi')\setminus G')$, \texttt{Check}$(\Ku,\varphi,\tilde{s},\tilde{G}\cap \subfB(\varphi),\tilde{\Du}(\Phi))=\bot$.
    \end{enumerate}
    Let $\rho'=s'$. ($\Rightarrow$)
    By the inductive hypothesis (of the external theorem over $\tilde{\rho}$), by (4.) it follows that $\Ku,\tilde{\rho}\models\varphi$ for each $\varphi\in(G'\setminus \tilde{G})$. By (5.) it follows that $\Ku,\tilde{\rho}\not\models\varphi$ for each  $\varphi\in(\subfB(\psi')\setminus G')$ and the claim follows.

     ($\Leftarrow$) Conversely (1.), (2.), and (3.) easily follow. Moreover, it must hold that $\Ku,\tilde{\rho}\models\varphi$ for each $\varphi\in (G'\setminus \tilde{G})$, and $\Ku,\tilde{\rho}\not\models\varphi$ for each  $\varphi\in (\subfB(\psi')\setminus G')$ and, therefore, (4.) and (5.) follow by the inductive hypothesis (of the external theorem).
    
    If $b\geq 2$, \texttt{Reach}$(\Ku,\psi',(\tilde{G},\tilde{\Du}(\Phi),\tilde{s}),(G',\Du(\Phi)',s'),b)=\top$ if and only if, for some $(G_3,\Du(\Phi)_3,s_3)\in\conf(\Ku,\psi')$,  \texttt{Reach}($\Ku,\psi',(\tilde{G},\tilde{\Du}(\Phi),\tilde{s}),(G_3,\Du(\Phi)_3,s_3), \linebreak \lfloor b/2\rfloor)=\top$ and  \texttt{Reach}($\Ku,\psi',(G_3,\Du(\Phi)_3,s_3),(G',\Du(\Phi)',s'),b-\lfloor b/2\rfloor)=\top$.
    
    ($\Rightarrow$) By the inductive hypothesis (over $b$), there exists $\rho_3\in\Trk_{\Ku}$ such that $\tilde{\rho}\cdot\rho_3\in\Trk_\Ku$, $|\rho_3|=\lfloor b/2\rfloor$, $\lst(\rho_3)=s_3$, $G_3$ is the subset of subformulas in $\subfB(\psi')$ that hold on some proper prefix of $\tilde{\rho}\cdot\rho_3$, and $\Du(\Phi)_3$ is the current configuration of the $\DFA$s associated with the regular expressions in $\Phi$, after reading $\Lab(\tilde{\rho}\cdot\rho_3(1,|\tilde{\rho}\cdot\rho_3|-1))$.
    
    By the inductive hypothesis (over $b$, applied to the trace $\tilde{\rho}\cdot\rho_3$), there exists $\rho'\in\Trk_{\Ku}$ such that $\tilde{\rho}\cdot\rho_3\cdot \rho'\in\Trk_\Ku$, $|\rho'|=b-\lfloor b/2\rfloor$, $\lst(\rho')=s'$, $G'$ is the subset of subformulas in $\subfB(\psi')$ that hold on some proper prefix of $\tilde{\rho}\cdot\rho_3\cdot \rho'$, and $\Du(\Phi)'$ is the current configuration of the $\DFA$s associated with the regular expressions in $\Phi$, after reading $\Lab(\tilde{\rho}\cdot\rho_3\cdot \rho'(1,|\tilde{\rho}\cdot\rho_3\cdot \rho'|-1))$.
    The claim follows, as $\rho_3\cdot\rho'\in\Trk_\Ku$ and $|\rho_3\cdot\rho'|=b$.
    
    ($\Leftarrow$)
    Conversely, there exists $\rho'\in\Trk_{\Ku}$ such that $\tilde{\rho}\cdot\rho'\in\Trk_\Ku$, $|\rho'|=b\geq 2$, $\lst(\rho')=s'$, $G'$ is the subset of subformulas in $\subfB(\psi')$ that hold on some proper prefix of $\tilde{\rho}\cdot\rho'$, and $\Du(\Phi)'$ is the current configuration of the $\DFA$s associated with the regular expressions in $\Phi$, after reading $\Lab(\tilde{\rho}\cdot\rho'(1,|\tilde{\rho}\cdot\rho'|-1))$.
    Let us split $\rho'=\rho_3\cdot\rho_4$, where $|\rho_3|=\lfloor b/2\rfloor$ and $|\rho_4|=b-\lfloor b/2\rfloor$.
    Let $(G_3,\Du(\Phi)_3,s_3)\in\conf(\Ku,\psi')$ be such that $\Du(\Phi)_3$ is the current configuration of the $\DFA$s associated with the regular expressions in $\Phi$, after reading $\Lab(\tilde{\rho}\cdot\rho_3(1,|\tilde{\rho}\cdot\rho_3|-1))$, $s_3=\lst(\rho_3)$, $G_3$ is the subset of subformulas in $\subfB(\psi')$ that hold on some proper prefix of $\tilde{\rho}\cdot\rho_3$. By the inductive hypothesis (on $b$ over $\tilde{\rho}\cdot\rho_3$), 
    \texttt{Reach}($\Ku,\psi',(G_3,\Du(\Phi)_3,s_3),(G',\Du(\Phi)',s'),b-\lfloor b/2\rfloor)=\top$.
    Moreover, by the inductive hypothesis (on $b$ over $\tilde{\rho}$), we have
    \texttt{Reach}($\Ku,\psi',(\tilde{G},\tilde{\Du}(\Phi),\tilde{s}),(G_3,\Du(\Phi)_3,s_3),\allowbreak \lfloor b/2\rfloor)=\top$.
    
    Hence, both the recursive calls at line 6 return $\top$, when at line 5 $(G_3,\Du(\Phi)_3,s_3)$ is considered by the loop. Thus, \texttt{Reach}$(\Ku,\psi',(\tilde{G},\tilde{\Du}(\Phi),\tilde{s}),(G',\Du(\Phi)',s'),b)$ returns $\top$ concluding the proof of the claim.
    \end{proof}
    
    ($\Rightarrow$)
    Let us now assume that in the execution of \texttt{Check}, at lines 15--19, for some $b''\in\{1,\ldots , |\States|\cdot (2|\psi'|+1)\cdot 2^{2\sum_{\ell=1}^u |r_\ell|}-1 \}$ and some $(G'',\Du(\Phi)'',s'')\in\conf(\Ku,\psi)$ ($=\conf(\Ku,\psi')$), we have \texttt{Reach}$(\Ku,\psi',(G,\Du(\Phi),s),(G'',\Du(\Phi)'',s''),\allowbreak b'')=\top$ and \texttt{Check}$(\Ku,\psi',s'',\allowbreak G'',\Du(\Phi)'')=\top$. By the claim above, there exists $\rho''\in\Trk_{\Ku}$ such that $\rho\cdot\rho''\in\Trk_\Ku$, $\lst(\rho'')=s''$, $G''$ is the subset of subformulas in $\subfB(\psi')$ that hold on some proper prefix of $\rho\cdot\rho''$, and $\Du(\Phi)''$ is the current configuration of the $\DFA$s associated with the regular expressions of $\Phi$, after reading $\Lab(\rho\cdot\rho''(1,|\rho\cdot\rho''|-1))$.
    By the inductive hypothesis, since \texttt{Check}$(\Ku,\psi',s'',G'',\Du(\Phi)'')=\top$, we have $\Ku,\rho\cdot\rho''\models \psi'$ implying that $\Ku,\rho\models \hsBt\psi'$.
    
    ($\Leftarrow$)
    Conversely, if $\Ku,\rho\models \hsBt\psi'$, we have $\Ku,\rho\cdot\rho''\models \psi'$ for some $\rho''\in\Trk_\Ku$, with $\rho\cdot\rho''\in\Trk_\Ku$. By the exponential small-model property (Theorem~\ref{theorem:expSizeModelPropertyAAbarBBbarRegex}), there exists $\rho'\in\Trk_\Ku$ such that $\lst(\rho'')=\lst(\rho')$, $|\rho'|\leq |\States|\cdot (2|\psi'|+1)\cdot 2^{2\sum_{\ell=1}^u |r_\ell|}-1$ (recall that the factor 2 in front of $|\psi'|$ is due to the fact that 
    a formula in \nnf{} is required), $\rho\cdot\rho'\in\Trk_\Ku$ and $\Ku,\rho\cdot\rho'\models \psi'$. Let $G'$ be the subset of subformulas in $\subfB(\psi')=\subfB(\psi)$ that hold on some proper prefix of $\rho\cdot\rho'$, and $\Du(\Phi)'$ be the current configuration of the $\DFA$s associated with the regular expressions in $\Phi$, after reading $\Lab(\rho\cdot\rho'(1,|\rho\cdot\rho'|-1))$. By the inductive hypothesis (over $\rho\cdot\rho'$), \texttt{Check}$(\Ku,\psi',\lst(\rho'),G',\Du(\Phi)')=\top$.
    By the claim above, \texttt{Reach}$(\Ku,\psi',(G,\Du(\Phi),s), (G',\Du(\Phi)',\lst(\rho')),\allowbreak |\rho'|)=\top$, hence \texttt{Check}$(\Ku,\psi,s,G,\Du(\Phi))=\top$.
    This concludes the proof of the theorem.
\end{proof}

\section{Proof of Theorem~\ref{corrCheckAux}}\label{proof:corrCheckAux}

\begin{theorem*}[\ref{corrCheckAux}]
    Let $\Ku=\KuDef$ be a finite Kripke structure, and $\Phi$ be a $\B\Bbar$ formula. Then, \texttt{CheckAux}$(\Ku,\Phi)=\top\iff\Ku\models\Phi$.
\end{theorem*}

\begin{proof}
    If $\Ku\models\Phi$, then for all $\rho\in\Trk_\Ku$ with $\fst(\rho)=\sinit$, we have $\Ku,\rho\models\Phi$. Hence, we have $\Ku,\sinit\models\Phi$, and $\Ku,\sinit\cdot\rho'\models\Phi$ for all $\sinit\cdot\rho'\in\Trk_\Ku$, implying that $\Ku,\sinit\models\hsBtu\Phi$ and $\Ku,\sinit\not\models\hsBt\neg\Phi$. By Theorem~\ref{corrComplCheckBBbar}, \texttt{Check}$(\Ku,\neg\Phi,\sinit,\emptyset,\Du(\Phi)_0)=\bot$ and  \texttt{Check}$(\Ku,\hsBt\neg\Phi,\sinit,\emptyset,\Du(\Phi)_0)=\bot$ implying that \texttt{CheckAux}($\Ku,\Phi)=\top$.
   
    Conversely, if \texttt{CheckAux}($\Ku,\Phi) = \top$, it must be \texttt{Check}$(\Ku,\neg\Phi,\sinit,\emptyset,\Du(\Phi)_0)=\bot$ and  \texttt{Check}$(\Ku,\hsBt\neg\Phi,\sinit,\emptyset,\Du(\Phi)_0)=\bot$. By Theorem~\ref{corrComplCheckBBbar} applied to the trace $\rho=\sinit$, we have $\Ku,\sinit\not\models\neg\Phi$ and $\Ku,\sinit\not\models\hsBt\neg\Phi$, and thus $\Ku\models\Phi$.
\end{proof}

\section{Proof of Theorem \ref{th:hard}}\label{sec:th:hard}

\begin{theorem*}[\ref{th:hard}]
The MC problem for $\HSprop$ formulas extended with regular expressions over finite Kripke structures is $\Psp$-hard (under polynomial-time reductions).
\end{theorem*}

\begin{proof}
    Given a regular expression $r$ with $\lang(r)\subseteq \Sigma^*$,
let us define the finite Kripke structure $\Ku=(\Sigma, \{\sinit\}\cup \Sigma, \Edges, \Lab, \sinit)$, where $\sinit\not\in \Sigma$, $\Lab(\sinit)=\emptyset$, for $c\in\Sigma$,  $\Lab(c)=\{c\}$, and $\Edges=\{(\sinit,c)\mid c\in\Sigma\}\cup \{(c,c')\mid c,c'\in \Sigma\}$.

It is easy to see that \[\lang(r)=\Sigma^*\iff \Ku\models \top\cdot \overline{r},\]
where 
$\overline{r}$ is a $\RE$ over $\Sigma$, \emph{syntactically} equal to $r$
(i.e., having exactly the same structure as $r$).
Note that even though if $r$ and $\overline{r}$ are syntactically equal,  $r$ is a regular expression defining a finitary language over $\Sigma$, whereas $\overline{r}$ defines a finitary language \emph{over $2^\Sigma$} (see Section~\ref{sect:backgrRegex}). The different notations  $r$ and $\overline{r}$ are kept to avoid confusion between the two different semantics.

    We show by induction on the structure of $r$ that, for all $w\in\Sigma^*$, $w \in\lang(r)\iff\Ku,w \models \overline{r}$.
The thesis follows as $\Ku,w \models \overline{r}$ if and only if $\Ku,\sinit\cdot w \models \top\cdot \overline{r}$.
\begin{itemize}
    \item $r=\varepsilon$. We have $w \in\lang(\varepsilon)$ if and only if $w=\varepsilon$, if and only if $\Lab(w)\in\lang(\overline{\varepsilon})=\{\varepsilon\}$, if and only if $\Ku,w\models \overline{\varepsilon}$.
    \item $r=c\in\Sigma$. We have $w \in\lang(c)$ if and only if $w=c$, thus $\Lab(w)=\{c\}\in\lang(\overline{c})$, and $\Ku,w\models \overline{c}$. Conversely, if $\Ku,w\models \overline{c}$, we have $\Lab(w)\in\lang(\overline{c})=\{A\in 2^\Sigma\mid c\in A\}$. In particular $|w|=1$. Moreover, by definition of $\Lab$, $\Lab(w)$ is a singleton, hence $\Lab(w)=\{c\}$. By definition of $\Ku$, we get $w=c$, thus $w \in\lang(c)$.
    \item $r=r_1\cdot r_2$. We have $w \in\lang(r_1\cdot r_2)$ if and only if $w=w_1\cdot w_2$, with $w_1 \in\lang(r_1)$ and $w_2 \in\lang(r_2)$. By applying the inductive hypothesis, $\Ku,w_1 \models \overline{r_1}$ and $\Ku,w_2 \models \overline{r_2}$, thus $\Lab(w_1)\in\lang(\overline{r_1})$ and $\Lab(w_2)\in\lang(\overline{r_2})$. It follows that $\Lab(w)=\Lab(w_1)\cdot\Lab(w_2)\in\lang(\overline{r_1})\cdot\lang(\overline{r_2})=\lang(\overline{r_1\cdot r_2})$, namely, $\Ku,w\models \overline{r_1\cdot r_2}$.
    Conversely, $\Lab(w)\in\lang(\overline{r_1\cdot r_2})=\lang(\overline{r_1})\cdot\lang(\overline{r_2})$. Hence $\Lab(w_1)\in\lang(\overline{r_1})$ and $\Lab(w_2)\in\lang(\overline{r_2})$, for some $w_1\cdot w_2=w$. By the inductive hypothesis, $w_1\in\lang(r_1)$ and $w_2\in\lang(r_2)$, hence $w \in\lang(r_1\cdot r_2)$.
    \item $r=r_1\cup r_2$. We have $w \in\lang(r_1\cup r_2)$ if and only if $w \in\lang(r_i)$ for some $i=1,2$. By the inductive hypothesis this is true if and only if $\Ku,w \models \overline{r_i}$, if and only if $\Lab(w) \in\lang(\overline{r_i})$, if and only if $\Lab(w) \in\lang(\overline{r_1\cup r_2})$, if and only if $\Ku,w \models \overline{r_1\cup r_2}$.
    \item $r=r_1^*$. 
 The thesis trivially holds if $w=\varepsilon$. 
    Let us now assume that $w\neq\varepsilon$. We have  $w \in\lang(r_1^*)$ if and only if, for some $t\geq 1$, $w=w_1\cdots w_t$ and $w_\ell\in\lang(r_1)$ for all $1\leq \ell\leq t$. By the inductive hypothesis, $\Ku,w_\ell \models \overline{r_1}$, thus $\Lab(w_\ell)\in\lang(\overline{r_1})$, and $\Lab(w)\in\lang(\overline{r_1^*})$. We conclude that $\Ku,w \models \overline{r_1^*}$. Conversely, $\Lab(w)\in\lang(\overline{r_1^*})=(\lang(\overline{r_1}))^*$, hence it must be the case that, for some $t\geq 1$, $w=w_1\cdots w_t$ and $\Lab(w_\ell)\in\lang(\overline{r_1})$  for all $1\leq \ell\leq t$. By the inductive hypothesis, $w_\ell\in\lang(r_1)$, hence $w\in\lang(r_1^*)$.
\end{itemize}
Finally, by observing that $\Ku$ can be built in polynomial time, the thesis follows.
\end{proof}

%% file: Chaps/Appendices/appendixTimelines.tex
\chapter{Proofs and complements of Chapter~\ref{chap:Timelines}}
\minitoc\mtcskip

\section{Definition of the value transition function $T$ in the proof of Theorem~\ref{theorem:undecidability}}\label{sec:undecidabilityTrans}

The value transition function $T$ of $x_M$ is defined as follows.
 \begin{itemize}
   \item For each instruction label $\ell\in \Inc \cup \{\ell_\halt\}$, let $P_\ell=\emptyset$ if $\ell=\ell_\halt$,
   and $P_\ell=\{(\Succ(\ell),\inc_h)\}$ otherwise, where $c_h= c(\ell)$. Then, $T(\ell)$, $T((\ell,c_i))$, and $T((\ell,(c_i,\#))$, for $i=1,2$, are defined as follows:
\[
T(\ell)   =  \{(\ell,c_1),(\ell,c_2)\}\cup P_\ell 
\]

\[
T((\ell,c_1))   =  \{(\ell,c_1),(\ell,c_2)\}\cup P_\ell
\]

\[
T((\ell,c_2))   =  \{(\ell,c_2)\}\cup P_\ell 
\]
\item For each  instruction label $\ell\in\Dec$ and for each $\ell'\in \{\zero(\ell),\dec(\ell)\}$, $T((\ell,\ell'))$, $T((\ell,\ell',c_i))$, and $T((\ell,\ell',(c_i,\#))$, for $i=1,2$, are defined as:
 \[
    T((\ell,\ell'))  =  \left\{
      \begin{array}{ll}
        \{(\ell,\ell',c_2),(\ell',\zero_1)\}
        &    \text{if }  c(\ell) =c_1 , \ell'=\zero(\ell)
        \\
        \{(\ell,\ell',c_1),(\ell',\zero_2)\}
        &    \text{if }  c(\ell) =c_2 , \ell'=\zero(\ell)
        \\
        \{(\ell,\ell',(c_1,\#))\}
        &    \text{if }  c(\ell) =c_1 , \ell'=\dec(\ell)
        \\
        \{(\ell,\ell', c_1),(\ell,\ell',(c_2,\#))\}
        &    \text{otherwise }
       \end{array}
    \right.
  \]
  \[
    T((\ell,\ell',c_1))  =  \left\{
      \begin{array}{ll}
        \emptyset
        &    \text{if }  c(\ell)\! =\!c_1 , \ell'\!=\!\zero(\ell)
        \\
        \{(\ell,\ell',c_1),(\ell',\zero_2)\}
        &    \text{if }  c(\ell)\! =\!c_2 , \ell'\!=\!\zero(\ell)
        \\
        \{(\ell,\ell',c_1),(\ell,\ell',c_2),(\ell',\dec_1)\}
        &    \text{if }  c(\ell)\! =\!c_1 , \ell'\!=\!\dec(\ell)
        \\
        \{(\ell,\ell', c_1),(\ell,\ell',(c_2,\#))\}
        &    \text{otherwise }
       \end{array}
    \right.
  \]
  \[
    T((\ell,\ell',c_2))  =  \left\{
      \begin{array}{ll}
      \{(\ell,\ell',c_2),(\ell',\zero_1)\}
        &    \text{if }  c(\ell) =c_1 , \ell'=\zero(\ell)
        \\
       \emptyset
        &    \text{if }  c(\ell) =c_2 , \ell'=\zero(\ell)
        \\
        \{ (\ell,\ell',c_2),(\ell',\dec_1)\}
        &    \text{if }  c(\ell) =c_1 , \ell'=\dec(\ell)
        \\
        \{ (\ell,\ell',c_2),(\ell',\dec_2)\}
        &    \text{otherwise }
       \end{array}
    \right.
  \]
   \[
    T((\ell,\ell',(c_1,\#)))  =  \left\{
      \begin{array}{ll}
        \{(\ell,\ell'\!,c_1),\!(\ell,\ell'\!,c_2),\!(\ell'\!,\dec_1)\}
        &    \text{if }  c(\ell) \!=\! c_1 , \ell'\!\!=\!\dec(\ell)
        \\
         \emptyset
        &    \text{otherwise }
       \end{array}
    \right.
  \]
   \[
    T((\ell,\ell',(c_2,\#)))  =  \left\{
      \begin{array}{ll}
        \{ (\ell,\ell',c_2),(\ell',\dec_2)\}
        &    \text{if }  c(\ell) = c_2 , \ell'=\dec(\ell)
        \\
         \emptyset
        &    \text{otherwise }
       \end{array}
    \right.
  \]

 \item For each label $\ell\in\InstructLab$ and operation $\op\in\{\inc_1,\inc_2,\zero_1,\zero_2,\dec_1,\allowbreak \dec_2\}$, $T((\ell,\op))$, $T((\ell,\op,c_i))$, and $T((\ell,\op,(c_i,\#))$, for $i=1,2$, are defined as follows,
 where $S_\ell=\{(\ell,\zero(\ell)),(\ell,\dec(\ell))\}$ if $\ell\in\Dec$, and $S_\ell=\{\ell\}$ otherwise:
\[
    T((\ell,\op))  =  \left\{
      \begin{array}{ll}
        \{(\ell,\op,c_2)\}\cup S_\ell
        &    \text{if }  \op = \zero_1 , \ell\neq \ell_\init
        \\
        \{(\ell,\op,c_1)\} \cup S_\ell
        &    \text{if }  \op = \zero_2 , \ell\neq \ell_\init
        \\
        \{(\ell,\op,c_1),(\ell,\op,c_2)\} \cup S_\ell
        &    \text{if }  \op \in \{\dec_1,\dec_2\} , \ell\neq \ell_\init
        \\
        \{(\ell,\op,(c_1,\#))\}
        &    \text{if }  \op = \inc_1 , \ell\neq \ell_\init
        \\
        \{(\ell,\op,c_1),(\ell,\op,(c_2,\#))\}
        &    \text{if }  \op = \inc_2 , \ell\neq \ell_\init\\
        \{\ell_\init\}  &    \text{if }  \op = \zero_1 , \ell= \ell_\init\\
         \emptyset
        &    \text{otherwise}
      \end{array}
    \right.
  \]
\[
    T((\ell,\op,c_1))  =  \left\{
      \begin{array}{ll}
        \emptyset
        &    \text{if }   \op = \zero_1 \text{ or } \ell =\ell_\init
        \\
        \{(\ell,\op,c_1)\} \cup S_\ell
        &    \text{if }  \op = \zero_2 , \ell\neq \ell_\init
        \\
        \{(\ell,\op,c_1),(\ell,\op,c_2)\} \cup S_\ell
        &    \text{if }  \op \in \{\dec_1,\dec_2,\inc_1\} , \\ & \phantom{if }\ell\neq \ell_\init
        \\
         \{(\ell,\op,c_1),(\ell,\op,(c_2,\#))\}
        &    \text{if }  \op = \inc_2 , \ell\neq \ell_\init
      \end{array}
    \right.
  \]
\[
    T((\ell,\op,c_2))  =  \left\{
      \begin{array}{ll}
        \emptyset
        &    \text{if }   \op = \zero_2 \text{ or } \ell =\ell_\init
        \\
         \{(\ell,\op,c_2)\} \cup S_\ell
        &    \text{otherwise }
      \end{array}
    \right.
  \]
  \[
    T((\ell,\op,(c_1,\#)))  =  \left\{
      \begin{array}{ll}
        \emptyset
        &    \text{if }   \op \neq \inc_1 \text{ or } \ell =\ell_\init
        \\
        \{(\ell,\op,c_1),(\ell,\op,c_2)\} \cup S_\ell
        & \text{otherwise }
      \end{array}
    \right.
  \]
  \[
    T((\ell,\op,(c_2,\#)))  =  \left\{
      \begin{array}{ll}
        \emptyset
        &    \text{if }   \op \neq \inc_2 \text{ or } \ell =\ell_\init
        \\
        \{(\ell,\op,c_2)\} \cup S_\ell
        & \text{otherwise }
      \end{array}
    \right.
  \]
  \end{itemize}

This concludes the definition of $T$ of $x_M$.

\section{Non-primitive recursive-hardness of future \allowbreak TP}\label{sec:NPRHardness}

In this section, we establish the following result.
\begin{theorem*}[\ref{theorem:NPRHardness}]
The future TP problem, even with \emph{one state variable}, is non-primitive recursive-hard also under one of the following two assumptions: \emph{either} $(1)$~the trigger rules are simple,
\emph{or} $(2)$~the intervals are in $\Intv_{(0,\infty)}$. %
\end{theorem*}

Theorem~\ref{theorem:NPRHardness} is proved by a polynomial-time reduction from the halting problem for \emph{gainy counter machines}~\cite{DemriL09}, a variant of standard Minsky machines, whose counters may erroneously  increase. Such a machine is a tuple $M = \tpl{Q,q_\init,q_\halt,n, \Delta}$,
where:
\begin{itemize}
  \item  $Q$ is a finite set of (control) locations/states, $q_\init\in Q$ is the initial location, and $q_\halt\in Q$ is the halting location,
  \item  $n\in\Nat\setminus\{0\}$ is the number of counters of $M$, and
  \item $\Delta \subseteq Q\times \InstL \times Q$ is a transition relation over the instruction set $\InstL= \{\inc,\dec,\allowbreak \zero\}\times \{1,\ldots,n\}$.
\end{itemize}
We adopt the following notational conventions.
 For an instruction
$\instr\in \InstL$, let $c(\instr)\in\{1,\ldots,n\}$ be the counter associated with $\instr$.
For a transition $\delta\in \Delta$ of the form $\delta=(q,\instr,q')$, we define $\From(\delta)= q$, $\instr(\delta)=\instr$, $c(\delta)= c(\instr)$,
and $\To(\delta)= q'$.  We denote by $\instr_\init$ the instruction $(\zero,1)$.
W.l.o.g., we make these assumptions:
\begin{itemize}
  \item for each transition $\delta\in \Delta$, $\From(\delta)\neq q_\halt$ and $\To(\delta )\neq q_\init$, and
    \item there is exactly one transition in $\Delta$, denoted $\delta_\init$, having as source the initial location $q_\init$.
\end{itemize}

An \emph{$M$-configuration} is a pair $(q,\nu)$ consisting of a location $q\in Q$ and a counter valuation $\nu: \{1,\ldots,n\}\to \Nat$. Given
two valuations $\nu$ and $\nu'$, we write $\nu\geq \nu'$ if and only if $\nu(c)\geq \nu'(c)$ for all $c\in\{1,\ldots,n\}$.

Under the \emph{exact semantics} (with no errors), $M$ induces a transition relation, denoted by $\longrightarrow$, over pairs of $M$-configurations and instructions, defined as follows:
 for configurations $(q,\nu)$ and $(q',\nu')$, and instructions $\instr \in \InstL$, we have $(q,\nu) \der{\instr} (q',\nu')$ if the following holds, where $c\in \{1,\ldots,n\}$ is the counter associated with the instruction
 $\instr$:
\begin{itemize}
  \item  $(q,\instr,q')\in \Delta$ and $\nu'(c')= \nu(c')$ for all $c'\in \{1,\ldots,n\}\setminus\{c\}$;
  \item  $\nu'(c)= \nu(c) +1$ if $\instr=(\inc,c)$;
  \item $\nu'(c)= \nu(c) -1$ if $\instr=(\dec,c)$ (in particular, it has to be $v(c)>0$);
   \item  $\nu'(c)= \nu(c)=0$ if $\instr=(\zero,c)$.
\end{itemize}

The \emph{gainy semantics} is obtained from the exact one by allowing \emph{increment} errors.
Formally, 
$M$  induces a transition relation, denoted by $\longrightarrow_\gainy$, defined as follows:
for configurations $(q,\nu)$ and $(q',\nu')$, and instructions $\instr \in \InstL$, we have $(q,\nu) \derG{\instr} (q',\nu')$ if the following holds, where $c=c(\instr)$ is the counter associated with the instruction
$\instr$:
$(q,\nu) \derG{\instr} (q',\nu')$ iff there are valuations $\nu_+$ and $\nu'_+$ such that
$\nu_+ \geq \nu$, $(q,\nu_+) \der{\instr} (q',\nu'_+)$, and $\nu' \geq \nu'_+$. Equivalently, $(q,\nu) \derG{\instr} (q',\nu')$ iff
the following conditions hold: 
%
\begin{itemize}
  \item  $(q,\instr,q')\in \Delta$ and $\nu'(c')\geq  \nu(c')$ for all $c'\in \{1,\ldots,n\}\setminus\{c\}$;
  \item  $\nu'(c)\geq  \nu(c) +1$ if $\instr=(\inc,c)$;
  \item $\nu'(c)\geq  \nu(c) -1$ if $\instr=(\dec,c)$;
   \item  $\nu(c)=0$ if $\instr=(\zero,c)$.
\end{itemize}

A (gainy) \emph{$M$-computation} is a finite sequence of the form:
\[
(q_0,\nu_0) \derG{\instr_0} (q_1,\nu_1) \derG{\instr_1} \cdots  \derG{\instr_{k-1}} (q_k,\nu_k).
\]
$M$ \emph{halts} if there exists an $M$-computation starting at the \emph{initial} configuration $(q_\init, \nu_\init)$, where $\nu_\init(c) = 0$ for all $c\in \{1,\ldots,n\}$, and leading to some
halting configuration
$(q_{\halt}, \nu)$. 
Given a gainy counter machine $M$,
\emph{the halting problem for $M$} is to decide whether $M$ halts, and it was shown to be decidable and non-primitive recursive~\cite{DemriL09}.

We now prove the following result, from which Theorem~\ref{theorem:NPRHardness} directly follows.

\begin{proposition}\label{prop:NPRHardness}
One can construct in polynomial time a TP domain $P=(\{x_M\},R_M)$ where the trigger rules in $R_M$ are simple (resp., the intervals in $P$ are in $\Intv_{(0,\infty)}$)
such that $M$ halts iff there is a future plan for $P$.
\end{proposition}
\begin{proof}
We focus on the reduction where the intervals in $P$ are in $\Intv_{(0,\infty)}$. At the end of the proof, we show how to adapt the construction for 
the case of simple trigger rules with arbitrary intervals.

\paragraph*{Encoding of $M$-computations.}
First, we define a suitable encoding of a computation of $M$ as a timeline for $x_M$. For this,
we exploit the finite set of symbols $V= V_{\main}\cup V_{\cont}\cup V_{\dummy}$ corresponding to the finite domain of the state variable $x_M$.
The   set of \emph{main} values $V_{\main}$ is given by
\[
V_\main = \{(\delta,\instr)\in\Delta\times\InstL\mid \instr\neq (\inc,c) \text{ if }\instr(\delta)=(\zero,c)\}.
\]
Intuitively, in the encoding of an $M$-computation, a main value $(\delta,\instr)$ keeps track of the transition $\delta$ used in the current step of the computation, while
$\instr$ represents the instruction exploited in the previous computation step (if any).

The set of \emph{secondary} values $V_{\cont}$ is defined as
\[
 V_\cont = V_\main \times \{1,\ldots,n\} \times 2^{\{\#_\inc,\#_\dec\}},
\]
where $\#_\inc$ and $\#_\dec$
are two special symbols used as markers. $V_{\cont}$ is used for encoding counter values, as shown later.
Finally, the set of \emph{dummy values} is $V_{\dummy}=(V_{\main}\cup V_{\cont})\times \{\dummy\}$;
their use will be clear when we introduce synchronization rules:
they are used to specify punctual time constraints by means of
non-simple trigger rules over intervals in
$\Intv_{(0,\infty)}$. 

Given a word $w\in V^{*}$, we denote by $||w||$ the length of the word
obtained from $w$ by removing dummy symbols.

 For $c\in \{1,\ldots,n\}$ and $v_\main = (\delta,\instr)\in V_\main$,
the \emph{set $\Tag(c,v_\main)$ of markers of counter $c$ for the main value $v_\main$} is the subset of  $\{\#_\inc,\#_\dec\}$ defined as follows:
\begin{itemize}
  \item $\#_\inc\in \Tag(c,v_\main)$ iff $\instr = (\inc,c)$;
  \item $\#_\dec\in \Tag(c,v_\main)$ iff $\instr(\delta) = (\dec,c)$;
\end{itemize}

A \emph{$c$-code for the main value $v_\main= (\delta,\instr)$} is a  finite word $w_c$ over $V$ such that
\emph{either} $(i)$ $w_c$ is empty and $\#_\inc\notin\Tag(c,v_\main)$, \emph{or} $(ii)$ $\instr(\delta)\neq (\zero,c)$ and $w_c=(v_\main,c,\Tag(c,v_\main))(v_\main,c,\emptyset,\dummy)^{h_0}\cdot (v_\main,c,\emptyset)\cdot (v_\main,c,\emptyset,\dummy)^{h_1}\!\cdots\allowbreak (v_\main,c,\emptyset)\cdot (v_\main,c,\emptyset,\dummy)^{h_n}$ for some $n\geq 0$ and $h_0,h_1,\ldots,\allowbreak h_n\geq 0$. The $c$-code $w_c$ encodes the value for the counter $c$
given by $||w_c||$.
Intuitively, $w_c$ can be seen as an interleaving of secondary values with dummy ones, the latter being present only for technical aspects, but not encoding any counter value.

A \emph{configuration-code $w$  for a main value $v_\main=(\delta,\instr)\in V_\main$} is a finite word over $V$
of the form $w= v_\main \cdot (v_\main,\dummy)^{h}\cdot w_1\cdots w_n$, where $h\geq 0$ and for each counter $c\in \{1,\ldots,n\}$, $w_c$ is a $c$-code
for the main value $v_\main$. The configuration-code $w$ encodes the $M$-configuration $(\From(\delta),\nu)$, where $\nu(c)=||w_c||$
for all $c\in \{1,\ldots,n\}$. Note that if $\instr(\delta)=(\zero,c)$, then $\nu(c)=0$ and $\instr\neq (\inc,c)$. 

The marker $\#_\inc$ occurs in $w$ iff $\instr$ is an increment instruction, and in such a case
 $\#_\inc$ marks the \emph{first} symbol of the encoding $w_{c(\instr)}$ of counter $c(\instr)$. Intuitively, if the operation performed in the previous step
 of the computation increments counter $c$, then the tag $\#_\inc$ \lq\lq marks\rq\rq\ the unit of the counter $c$ in the current configuration which has been added by the increment.

The marker $\#_\dec$ occurs in $w$ iff $\delta$ is a decrement instruction and the value of counter $c(\delta)$ in $w$ is non-zero; in such a case,
 $\#_\dec$ marks the \emph{first} symbol of the encoding $w_{c(\delta)}$ of counter $c(\delta)$. Intuitively, if the operation to be performed in the current step
 decrements counter $c$ and the current value of 
 $c$ is non-zero, then the tag $\#_\dec$  marks  the unit of the counter $c$  in the current configuration which has to be removed by the decrement.

A \emph{computation}-code is a sequence of configuration-codes $\pi= w_{(\delta_0 ,\instr_0)} \cdots\allowbreak w_{(\delta_k,\instr_k)}$, where,  for all $0\leq i\leq k$, $w_{(\delta_i,\instr_i)}$ is a configuration-code with main value $(\delta_i,\instr_i)$, and whenever
  $i<k$, it holds that $\To(\delta_i)=\From(\delta_{i+1})$ and $\instr(\delta_i)=\instr_{i+1}$. Note that by our assumptions $\To(\delta_i)\neq q_\halt$ for all $0\leq i<k$, and
  $\delta_j\neq \delta_\init$ for all $0<j\leq k$.
  The computation-code $\pi$ is \emph{initial} if  the first configuration-code $w_{(\delta_0 ,\instr_0)}$ is $(\delta_\init,\instr_\init)$ (which encodes the initial configuration), and it is \emph{halting} if
  for the last  configuration-code $w_{(\delta_k,\instr_k)}$ in $\pi$, it holds that $\To(\delta_k)=q_\halt$.
For all $0\leq i\leq k$, let $(q_i,\nu_i)$ be the $M$-configuration encoded by the configuration-code $w_{(\delta_i,\instr_i)}$ and $c_i= c(\delta_i)$.
 The computation-code $\pi$ is \emph{well-formed} if, additionally,   for all $0\leq j\leq k-1$, the following conditions hold:
\begin{itemize}
  \item $\nu_{j+1}(c)\geq \nu_j(c)$ for all $c\in \{1,\ldots,n\}\setminus \{c_j\}$ (\emph{gainy monotonicity});
\item $\nu_{j+1}(c_j)\geq \nu_j(c_j)+1$ if $\instr(\delta_j)= (\inc,c_j)$ (\emph{increment requirement});
 \item $\nu_{j+1}(c_j)\geq \nu_j(c_j)-1$ if $\instr(\delta_j)= (\dec,c_j)$ (\emph{decrement requirement}).
\end{itemize}
Clearly, 
$M$ halts \emph{iff} there is an initial and halting well-formed computation-code. 

\paragraph*{Definition of $x_M$ and $R_M$.} We now define a state variable $x_M$ and a set $R_M$ of  synchronization rules for $x_M$ with intervals in $\Intv_{(0,\infty)}$ such that the untimed part of any \emph{future plan} for $P=(\{x_M\},R_M)$
is an initial and halting well-formed computation-code. Thus, $M$ halts if and only if there is a future plan of $P$.

Formally, the state 
variable $x_M$ is given by $x_M= (V,T,D)$ where, for each $v\in V$,
$D(v)=\mathopen]0,\infty\mathclose[$ if $v\notin V_{\dummy}$, and $D(v)=\mathopen[0,\infty\mathclose[$ otherwise: we require that the duration of a non-dummy token 
is always greater than zero (\emph{strict time monotonicity}).

The value transition function $T$ of $x_M$ ensures the following requirement.
\begin{claim}\label{ref:claimGainy}
The untimed part of any timeline for $x_M$ whose first token has value $(\delta_\init,\instr_\init)$ corresponds
 to a prefix of some initial computation-code. Moreover, $(\delta_\init,\instr_\init)\notin T(v)$ for all $v\in V$.
\end{claim}

$T$ can be built by adapting the construction of Appendix~\ref{sec:undecidabilityTrans}. 

Let $V_\halt=\{(\delta,\instr)\in V_\main\mid \To(\delta)=q_\halt\}$.
 By Claim~\ref{ref:claimGainy} and the assumption that  $\From(\delta)\neq q_\halt$ for each transition $\delta\in \Delta$, to ensure the initialization and halting requirements,
 it suffices  to enforce the timeline to feature a token with value $(\delta_\init,\instr_\init)$ and a token with value in $V_\halt$. This is captured by the trigger-less rules
   \[
   \true \rightarrow \exists   o[x_M=(\delta_\init,\instr_\init)].  \true
   \]
   and  
   \[
   \true \rightarrow \bigvee_{v\in V_\halt} \exists   o[x_M=v].  \true .
   \]

The crucial well-formedness requirement is captured by the trigger rules in $R_M$ which express the following punctual time constraints.
Note that we take advantage of the dense temporal domain to allow
for the encoding of arbitrarily large values of counters in two time units.
 \begin{itemize}
   \item \emph{2-Time distance between consecutive main values:} the overall duration of the sequence of tokens corresponding to a configuration-code  amounts exactly to 2 time units. 
By Claim~\ref{ref:claimGainy}, strict time monotonicity, and the halting requirement, it suffices to ensure that each token $tk$ having a  main value in $V_\main \setminus V_\halt$ is eventually followed by a token $tk'$  such that $tk'$ has a  main value and $\startTime(tk')-\startTime(tk)=2$. To this aim, for each  $v\in V_\main \setminus V_\halt$,  we have the following non-simple trigger rule with intervals in $\Intv_{(0,\infty)}$ which uses a dummy token for capturing the punctual time constraint:
\begin{multline*}
o[x_M=v] \rightarrow \bigvee_{u\in V_\main}\bigvee_{u_d\in V_\dummy} \exists  o'[x_M= u]\exists  o_d[x_M= u_d].  o\leq^{\start,\start}_{[1,+\infty[} o_d \,\wedge\\ o_d\leq^{\start,\start}_{[1,+\infty[}o'\, \wedge\, o\leq^{\start,\start}_{[0,2]} o'.
\end{multline*}
   \item 
   For a counter $c\in \{1,\ldots,n\}$, let us denote as $V_c\subseteq V_\cont$ the set of secondary values given
    by $V_\main \times \{c\} \times 2^{\{\#_\inc,\#_\dec\}}$. We require that  each token $tk$  with
    a $V_{c}$-value of the form $((\delta,\instr),c,\Tag)$  such that $c\neq c(\delta)$ and $\To(\delta)\neq q_\halt$ is eventually followed by a token $tk'$ with a $V_{c}$-value such that  $\startTime(tk')-\startTime(tk)=2$.
Note that our encoding, Claim~\ref{ref:claimGainy}, strict time monotonicity, and 2-Time distance between consecutive main values guarantee  that the previous requirement captures \emph{gainy monotonicity}.
 Thus, for each counter $c$ and $v\in V_{c}$ such that $v$ is of the form $((\delta,\instr),c,\Tag)$, where $c\neq c(\delta)$ and $\To(\delta)\neq q_\halt$, we have the following non-simple trigger rule over $\Intv_{(0,\infty)}$:\\
\begin{multline*}
o[x_M=v] \rightarrow \bigvee_{u\in V_{c}}\bigvee_{u_d\in V_\dummy} \exists  o'[x_M= u]\exists  o_d[x_M= u_d].  o\leq^{\start,\start}_{[1,+\infty[} o_d \,\wedge\\ o_d\leq^{\start,\start}_{[1,+\infty[}o'\, \wedge\, o\leq^{\start,\start}_{[0,2]} o'.
\end{multline*}
  \item For capturing the increment and decrement requirements, by construction, it suffices to enforce that: 
  \begin{enumerate}
    \item each token $tk$  with
    a $V_{c}$-value of the form $((\delta,\instr),c,\Tag)$  such that $\To(\delta)\neq q_\halt$ and $\delta=(\inc,c)$ is eventually followed by a token $tk'$ with a $V_{c}$-value which is \emph{not} marked by $\#_\inc$ such that  $\startTime(tk')-\startTime(tk)=2$; 
    \item
  each token $tk$  with
    a $V_{c}$-value of the form $((\delta,\instr),c,\Tag)$  such that $\To(\delta)\neq q_\halt$, $\delta=(\dec,c)$, and  $\#_\dec\notin \Tag$ is eventually followed by a token $tk'$ with a $V_{c}$-value  such that  $\startTime(tk')-\startTime(tk)=2$. These requirements can be expressed by non-simple trigger rules with intervals in $\Intv_{(0,\infty)}$ similar to the previous ones.
  \end{enumerate}    
\end{itemize}
Finally, to prove Proposition~\ref{prop:NPRHardness} for the case of simple trigger rules with arbitrary intervals, it suffices to remove the dummy values and replace the conjunction
  $o\leq^{\start,\start}_{[1,+\infty[} o_d \,\wedge\, o_d\leq^{\start,\start}_{[1,+\infty[}o'\, \wedge\, o\leq^{\start,\start}_{[0,2]} o'$ in the previous trigger
  rules with the \lq\lq punctual\rq\rq\ atom $ o\leq^{\start,\start}_{[2,2]} o'$, whose interval at the subscript is singular.

  This concludes the proof of Proposition~\ref{prop:NPRHardness}.
\end{proof}

\section[Future TP, simple trigger rules, non-sing.\ intervals: $\EXPSPACE$-hardness]{Future TP with simple trigger rules and non-singular intervals: $\EXPSPACE$-hardness}\label{sec:EXPSPHardFutTP}

Here we show that the future TP problem with simple trigger rules and non-singular intervals
is $\EXPSPACE$-hard. 
The claim is proved by a polynomial-time reduction from the domino-tiling problem for grids with rows of single exponential length~\cite{harel92} (see Section~\ref{sec:BEhard} for the definition and notation). 
Hardness holds also when only a single state variable is involved.

\begin{theorem*}[\ref{theorem:EXPSPlowerBound}]
The future TP problem, even with \emph{one state variable}, with simple trigger rules and non-singular intervals is $\EXPSPACE$-hard (under polynomial-time reductions).
\end{theorem*}
\begin{proof}
For the sake of the reduction, 
we define the state variable $y=(V,T,D)$ where (we recall that $\Delta$ represents the set of domino-types and $d,d',d_\Final\in\Delta$): 
\begin{itemize}
    \item $V=\{\$,\$'\}\cup\Delta$ (with $\$,\$'\notin\Delta$),
    \item $T(\$)=\Delta$ and $T(\$')=\{\$'\}$,
    \item for $d\in \Delta\setminus\{d_\Final\}$, $T(d)=\{\$\}\cup\{d'\in\Delta\mid [d]_{\Right}=[d']_{\Left} \}$,
    \item $T(d_\Final)=\{\$,\$'\}\cup\{d'\in\Delta\mid [d_\Final]_{\Right}=[d']_{\Left} \}$,
    \item for all $v\in\ V$, $D(v)=[2,+\infty[$.
\end{itemize}
Basically, the domain of the state variable $y$ contains all domino-types, as well as two auxiliary symbols $\$$ and $\$'$. 
The idea is encoding a tiling by the concatenation of its rows, separated by an occurrence of $\$$. The last row is terminated by $\$'$.

More precisely,
each cell of the grid is encoded by (the value of) a token having duration 2.
A row of the grid is then represented by the sequence of tokens of its cells, ordered by increasing column index.
Finally, a full tiling is just given by the timeline for $y$ obtained by concatenating the sequences of tokens of all rows, ordered by increasing row index.
See Figure~\ref{fig:rowEXPhardTP} for an example.
\begin{figure}[tb]
    \centering
    \includegraphics[scale=0.42]{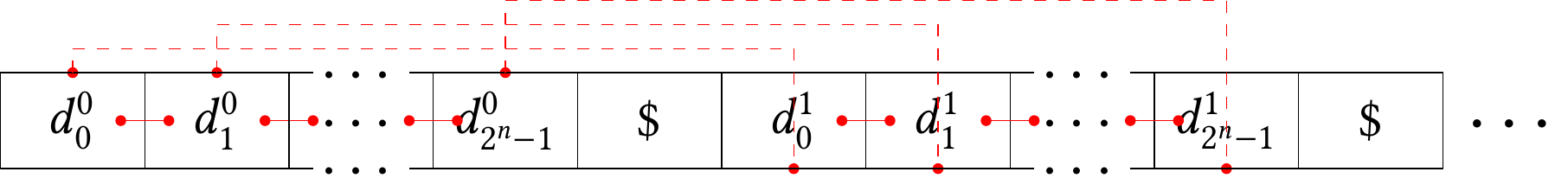}\hspace{1pt}
    \includegraphics[scale=0.42]{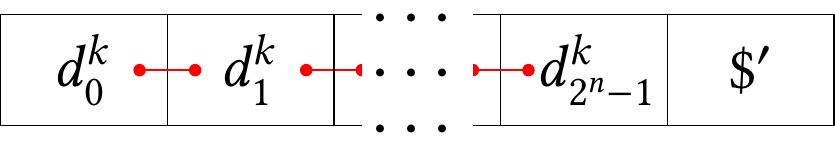}
    \caption{A timeline encoding the ordered concatenation of the rows of a tiling. Red lines represent the horizontal and vertical constraints among domino-types.}\label{fig:rowEXPhardTP}
\end{figure}

We observe that $T$ guarantees the horizontal constraint among domino-types, and that it allows only occurrences of $\$'$ after the first $\$'$. 

We start with the next simple trigger rules, one for each $v\in V$: 
\[o[y=v]\to o\leq^{\mathsf{s},\mathsf{e}}_{[0,2]} o.\] 
These, paired with the constraint function $D$, enforce
all tokens' durations to be exactly 2. This is done for technical convenience: intuitively, since we exclude singular intervals, requiring, for instance, that a token $o'$ starts  $t$ instants of time after the end of $o$, with $t\in [\ell,\ell+1]$ and even $\ell\in\Nat$, boils down to $o'$ starting \emph{exactly} $\ell$ instants after the end of $o$. We also observe that, given the constant token duration, in this proof the density of the time domain does not play any role.

We now define the following synchronization rules (of which all trigger ones are simple and future). The next ones state (together) that the \emph{first occurrence} of (a token having value) $\$$ starts exactly at $2\cdot 2^n$:
\begin{equation}\label{equat:1}
    \true \to \exists o[y=\$]. o\geq^{\mathsf{s}}_{[0,1]} 2\cdot 2^n ,
\end{equation}
and 
\begin{equation}\label{eq:2}
    o[y=\$] \to  o\geq^{\mathsf{s}}_{[0,+\infty[} 2\cdot 2^n.
\end{equation}
Thus, all tokens before such a first occurrence of $\$$ have a value in $\Delta$.

Every occurrence of $\$$ must be followed, after exactly $2\cdot 2^n$ instants of time (namely, after $2^n$ tokens), by another occurrence of $\$$ or of $\$'$.
\begin{multline}\label{eq:3}
o[y=\$]\to \\ (\exists o'[y=\$]. o\leq ^{\mathsf{e},\mathsf{s}}_{[2\cdot 2^n,2\cdot 2^n+1]} o') \vee (\exists o''[y=\$']. o\leq ^{\mathsf{e},\mathsf{s}}_{[2\cdot 2^n,2\cdot 2^n+1]} o'').
\end{multline}

Now we force every token with value $d\in \Delta$ either $(i)$ to be followed, after $2\cdot 2^n$ instants, by another token with value  $d'\in\Delta$, in particular, satisfying the vertical requirement, i.e., $[d]_{\Up}=[d']_{\Down}$, or $(ii)$ to be in the last row (which is terminated by $\$'$). For each $d\in \Delta$, 
\begin{multline}\label{eq:4}
o[y=d]\to \\ \Big(\smashoperator{\bigvee_{d'\in\Delta, \,[d]_{\Up}=[d']_{\Down}}} \exists o'[y=d']. o\leq ^{\mathsf{e},\mathsf{s}}_{[2\cdot 2^n,2\cdot 2^n+1]} o'\Big) \vee (\exists o''[y=\$']. o\leq ^{\mathsf{e},\mathsf{s}}_{[0,2\cdot 2^n-2]} o'').
\end{multline}

It is straightforward to check that rules (\ref{equat:1}), (\ref{eq:2}), (\ref{eq:3}), and (\ref{eq:4}), along with the horizontal constraint guaranteed by the function $T$, enforce the following property.
\begin{proposition}
There exists $k'\in\Nat^+$ such that
all tokens with value $\$$ end at all and only times $k\cdot 2(2^n+1)$, for $1\leq k< k'$. Moreover the first token with value $\$'$ ends at time $k'\cdot 2(2^n+1)$. Finally, all other tokens having end time less than $k'\cdot 2(2^n+1)$ have value in $\Delta$ and satisfy the horizontal and vertical constraints.
\end{proposition}

Finally, we settle the initialization and acceptance requirements by means of the following pair of trigger-less rules:
\[
    \true \to \exists o[y=d_\Init]. o\geq^{\mathsf{s}}_{[0,1]} 0,
\]
\[
    \true \to \exists o[y=d_\Final]\exists o'[y=\$']. o\leq^{\mathsf{e},\mathsf{s}}_{[0,1]} o'.
\]
The former rule states that a token with value $d_\Init$ must start at $t=0$, the latter that a
token with value $d_\Final$ must occur just before the terminator of the last row $\$'$.

To conclude the proof, we observe that the state variable $y=(V,T,D)$ as well as all synchronization rules can be generated in polynomial time in the size of the instance $\Instance$ of the domino-tiling problem (in particular, note that all interval bounds and time constants of time-point atoms have a value, encoded in binary, which is in $O(2^n)$).
\end{proof}

%% file: ms.bbl
\newcommand{\etalchar}[1]{$^{#1}$}
\begin{thebibliography}{BDMG{\etalchar{+}}08}

\bibitem[AAB{\etalchar{+}}07]{AlurABEIL07}
R.~Alur, M.~Arenas, P.~Barcel{\'o}, K.~Etessami, N.~Immerman,
  and L.~Libkin.
\newblock First-order and temporal logics for nested words.
\newblock In {\em Proceedings of LICS}, pages 151--160. IEEE Computer Society,
  2007.
\newblock \href {http://dx.doi.org/10.1109/LICS.2007.19}
  {\path{doi:10.1109/LICS.2007.19}}.

\bibitem[ACC07]{4271662}
A.~Armando, R.~Carbone, and L.~Compagna.
\newblock {LTL} {M}odel {C}hecking for {S}ecurity {P}rotocols.
\newblock In {\em Proceedings of CSF}, pages 385--396. IEEE Computer Society,
  2007.
\newblock \href {http://dx.doi.org/10.1109/CSF.2007.24}
  {\path{doi:10.1109/CSF.2007.24}}.

\bibitem[ACZ06]{AlurCZ06}
R.~Alur, P.~Cern{\'y}, and S.~Zdancewic.
\newblock Preserving secrecy under refinement.
\newblock In {\em Proceedings of ICALP}, pages 107--118. Springer, 2006.
\newblock \href {http://dx.doi.org/10.1007/11787006_10}
  {\path{doi:10.1007/11787006_10}}.

\bibitem[AD94]{ALUR1994183}
R.~Alur and D.~L. Dill.
\newblock A theory of timed automata.
\newblock {\em Theoretical Computer Science}, 126(2):183--235, 1994.
\newblock \href {http://dx.doi.org/10.1016/0304-3975(94)90010-8}
  {\path{doi:10.1016/0304-3975(94)90010-8}}.

\bibitem[AEF{\etalchar{+}}05]{Arons2005}
T.~Arons, E.~Elster, L.~Fix, S.~Mador-Haim, M.~Mishaeli, J.~Shalev,
  E.~Singerman, A.~Tiemeyer, M.~Y. Vardi, and L.e~D. Zuck.
\newblock {\em Formal Verification of Backward Compatibility of Microcode},
  pages 185--198.
\newblock Springer, 2005.
\newblock \href {http://dx.doi.org/10.1007/11513988_20}
  {\path{doi:10.1007/11513988_20}}.

\bibitem[AEM04]{AlurEM04}
R.~Alur, K.~Etessami, and P.~Madhusudan.
\newblock A temporal logic of nested calls and returns.
\newblock In {\em Proceedings of TACAS}, pages 467--481. Springer, 2004.
\newblock \href {http://dx.doi.org/10.1007/978-3-540-24730-2_35}
  {\path{doi:10.1007/978-3-540-24730-2_35}}.

\bibitem[AFH96]{Alur:1996}
R.~Alur, T.~Feder, and T.~A. Henzinger.
\newblock The benefits of relaxing punctuality.
\newblock {\em Journal of the ACM}, 43(1):116--146, 1996.
\newblock \href {http://dx.doi.org/10.1145/227595.227602}
  {\path{doi:10.1145/227595.227602}}.

\bibitem[AH93]{AlurH93}
R.~Alur and T.~A. Henzinger.
\newblock {Real-Time Logics: Complexity and Expressiveness}.
\newblock {\em Information and Computation}, 104(1):35--77, 1993.
\newblock \href {http://dx.doi.org/10.1006/inco.1993.1025}
  {\path{doi:10.1006/inco.1993.1025}}.

\bibitem[All83]{All83}
J.~F. Allen.
\newblock Maintaining knowledge about temporal intervals.
\newblock {\em Communications of the ACM}, 26(11):832--843, 1983.
\newblock \href {http://dx.doi.org/10.1145/182.358434}
  {\path{doi:10.1145/182.358434}}.

\bibitem[BBD{\etalchar{+}}12]{barreiro2012europa}
J.~Barreiro, M.~Boyce, M.~Do, J.~Frank, M.~Iatauro, T.~Kichkaylo, P.~Morris,
  J.~Ong, E.~Remolina, T.~Smith, and D.~Smith.
\newblock {EUROPA: A Platform for AI Planning, Scheduling, Constraint
  Programming, and Optimization}.
\newblock In {\em Proceedings of ICKEPS}, 2012.

\bibitem[BCC{\etalchar{+}}99]{biere1999symbolic}
A.~Biere, A.~Cimatti, E.~M. Clarke, M.~Fujita, and Y.~Zhu.
\newblock Symbolic model checking using sat procedures instead of bdds.
\newblock In {\em Proceedings of DAC}, pages 317--320. ACM, 1999.
\newblock \href {http://dx.doi.org/10.1145/309847.309942}
  {\path{doi:10.1145/309847.309942}}.

\bibitem[BCC{\etalchar{+}}03]{biere2003bounded}
A.~Biere, A.~Cimatti, E.~M. Clarke, O.~Strichman, and Y.~Zhu.
\newblock Bounded model checking.
\newblock {\em Advances in computers}, 58:117--148, 2003.
\newblock \href {http://dx.doi.org/10.1016/S0065-2458(03)58003-2}
  {\path{doi:10.1016/S0065-2458(03)58003-2}}.

\bibitem[BCM{\etalchar{+}}90]{bur90}
J.~R.~Burch, E.~M.~Clarke, K.~L.~McMillan, D.~L.~Dill, and L.~J.~Hwang.
\newblock Symbolic model checking: 10\textsuperscript{20} states and beyond.
\newblock In {\em Proceedings of LICS}, pages 428--439. IEEE Computer Society,
  1990.
\newblock \href {http://dx.doi.org/10.1109/LICS.1990.113767}
  {\path{doi:10.1109/LICS.1990.113767}}.

\bibitem[BCM15]{basin2015model}
D.~Basin, C.~Cremers, and C.~Meadows.
\newblock Model checking security protocols, 2015.
\newblock URL:
  \url{http://www-oldurls.inf.ethz.ch/personal/basin/pubs/security-modelchecking.pdf}.

\bibitem[BCMD90]{Burch:1991}
J.~R. Burch, E.~M. Clarke, K.~L. McMillan, and David~L. Dill.
\newblock Sequential circuit verification using symbolic model checking.
\newblock In {\em Proceedings of DAC}, pages 46--51. ACM, 1990.
\newblock \href {http://dx.doi.org/10.1145/123186.123223}
  {\path{doi:10.1145/123186.123223}}.

\bibitem[BDG{\etalchar{+}}09]{DBLP:conf/time/BresolinMGMS09}
D.~Bresolin, D.~{Della Monica}, V.~Goranko, A.~Montanari, and G.~Sciavicco.
\newblock Undecidability of interval temporal logics with the overlap modality.
\newblock In {\em Proceedings of {TIME}}, pages 88--95. IEEE Computer Society,
  2009.
\newblock \href {http://dx.doi.org/10.1109/TIME.2009.24}
  {\path{doi:10.1109/TIME.2009.24}}.

\bibitem[BDG{\etalchar{+}}10]{DBLP:conf/ecai/BresolinMGMS10}
D.~Bresolin, D.~{Della Monica}, V.~Goranko, A.~Montanari, and G.~Sciavicco.
\newblock Metric propositional neighborhood logics: Expressiveness,
  decidability, and undecidability.
\newblock In {\em Proceedings of {ECAI}}, pages 695--700. ACM, 2010.
\newblock \href {http://dx.doi.org/10.3233/978-1-60750-606-5-695}
  {\path{doi:10.3233/978-1-60750-606-5-695}}.

\bibitem[BDG{\etalchar{+}}13]{DBLP:journals/sosym/BresolinMGMS13}
D.~Bresolin, D.~{Della Monica}, V.~Goranko, A.~Montanari, and G.~Sciavicco.
\newblock Metric propositional neighborhood logics on natural numbers.
\newblock {\em Software and System Modeling}, 12(2):245--264, 2013.
\newblock \href {http://dx.doi.org/10.1007/s10270-011-0195-y}
  {\path{doi:10.1007/s10270-011-0195-y}}.

\bibitem[BDG{\etalchar{+}}14]{DBLP:journals/amai/BresolinMGMS14}
D.~Bresolin, D.~{Della Monica}, V.~Goranko, A.~Montanari, and G.~Sciavicco.
\newblock The dark side of interval temporal logic: marking the undecidability
  border.
\newblock {\em Annals of Mathematics and Artificial Intelligence},
  71(1--3):41--83, 2014.
\newblock \href {http://dx.doi.org/10.1007/s10472-013-9376-4}
  {\path{doi:10.1007/s10472-013-9376-4}}.

\bibitem[BDMG{\etalchar{+}}08]{Bresolin2008}
D.~Bresolin, D.~Della~Monica, V.~Goranko, A.~Montanari, and G.~Sciavicco.
\newblock {\em Decidable and Undecidable Fragments of Halpern and Shoham's
  Interval Temporal Logic: Towards a Complete Classification}, pages 590--604.
\newblock Springer, 2008.
\newblock \href {http://dx.doi.org/10.1007/978-3-540-89439-1_41}
  {\path{doi:10.1007/978-3-540-89439-1_41}}.

\bibitem[BGG{\etalchar{+}}17]{BGGMM17}
M.~Benerecetti, R.~De Guglielmo, U.~Gentile, S.~Marrone, N.~Mazzocca,
  R.~Nardone, A.~Peron, L.~Velardi, and V.~Vittorini.
\newblock Dynamic state machines for modelling railway control systems.
\newblock {\em Science of Computer Programming}, 133:116--153, 2017.
\newblock \href {http://dx.doi.org/10.1016/j.scico.2016.09.002}
  {\path{doi:10.1016/j.scico.2016.09.002}}.

\bibitem[BGMS09]{DBLP:journals/apal/BresolinGMS09}
D.~Bresolin, V.~Goranko, A.~Montanari, and G.~Sciavicco.
\newblock Propositional interval neighborhood logics: Expressiveness,
  decidability, and undecidable extensions.
\newblock {\em Annals of Pure and Applied Logic}, 161(3):289--304, 2009.
\newblock \href {http://dx.doi.org/10.1016/j.apal.2009.07.003}
  {\path{doi:10.1016/j.apal.2009.07.003}}.

\bibitem[BGMS10]{DBLP:journals/logcom/BresolinGMS10}
D.~Bresolin, V.~Goranko, A.~Montanari, and P.~Sala.
\newblock Tableaux for logics of subinterval structures over dense orderings.
\newblock {\em Journal of Logic and Computation}, 20(1):133--166, 2010.
\newblock \href {http://dx.doi.org/10.1093/logcom/exn063}
  {\path{doi:10.1093/logcom/exn063}}.

\bibitem[BH91]{buss1991}
S.~Buss and L.~Hay.
\newblock {On truth-table reducibility to SAT}.
\newblock {\em Information and Computation}, 102:86--102, 1991.
\newblock \href {http://dx.doi.org/10.1109/SCT.1988.5282}
  {\path{doi:10.1109/SCT.1988.5282}}.

\bibitem[BK08]{Baier2008}
C.~Baier and J.~P. Katoen.
\newblock {\em Principles of Model Checking}.
\newblock Representation and Mind Series. The MIT Press, 2008.

\bibitem[BMM{\etalchar{+}}16a]{ictcs16}
L.~Bozzelli, A.~Molinari, A.~Montanari, A.~Peron, and P.~Sala.
\newblock Interval temporal logic model checking based on track bisimilarity
  and prefix sampling.
\newblock In {\em Proceedings of ICTCS}, pages 49--61. CEUR Workshop
  Proceedings, 2016.

\bibitem[BMM{\etalchar{+}}16b]{ijcar16}
L.~Bozzelli, A.~Molinari, A.~Montanari, A.~Peron, and P.~Sala.
\newblock Interval temporal logic model checking: The border between good and
  bad {HS} fragments.
\newblock In {\em Proceedings of IJCAR}, pages 389--405. Springer, 2016.
\newblock \href {http://dx.doi.org/10.1007/978-3-319-40229-1_27}
  {\path{doi:10.1007/978-3-319-40229-1_27}}.

\bibitem[BMM{\etalchar{+}}16c]{fsttcs16}
L.~Bozzelli, A.~Molinari, A.~Montanari, A.~Peron, and P.~Sala.
\newblock Interval vs. point temporal logic model checking: an expressiveness
  comparison.
\newblock In {\em Proceedings of FSTTCS}, pages 26:1--26:14. Schloss
  Dagstuhl--Leibniz-Zentrum fuer Informatik, 2016.
\newblock \href {http://dx.doi.org/10.4230/LIPIcs.FSTTCS.2016.26}
  {\path{doi:10.4230/LIPIcs.FSTTCS.2016.26}}.

\bibitem[BMM{\etalchar{+}}16d]{gandalf16}
L.~Bozzelli, A.~Molinari, A.~Montanari, A.~Peron, and P.~Sala.
\newblock Model checking the logic of allen's relations meets and started-by is
  p\({}^{\mbox{np}}\)-complete.
\newblock In {\em Proceedings of GandALF}, pages 76--90. EPTCS, 2016.
\newblock \href {http://dx.doi.org/10.4204/EPTCS.226.6}
  {\path{doi:10.4204/EPTCS.226.6}}.

\bibitem[BMM{\etalchar{+}}17]{icalp17}
L.~Bozzelli, A.~Molinari, A.~Montanari, A.~Peron, and P.~Sala.
\newblock {Satisfiability and Model Checking for the Logic of Sub-Intervals
  under the Homogeneity Assumption}.
\newblock In {\em Proceedings of ICALP}, pages 120:1--120:14. Schloss
  Dagstuhl--Leibniz-Zentrum fuer Informatik, 2017.
\newblock \href {http://dx.doi.org/10.4230/LIPIcs.ICALP.2017.120}
  {\path{doi:10.4230/LIPIcs.ICALP.2017.120}}.

\bibitem[BMM{\etalchar{+}}18a]{ictcs18}
L.~Bozzelli, A.~Molinari, A.~Montanari, A.~Peron, and G.~Woeginger.
\newblock Timeline-based planning over dense temporal domains with trigger-less
  rules is {NP}-complete.
\newblock In {\em Proceedings of ICTCS}. CEUR Workshop Proceedings, 2018.

\bibitem[BMM{\etalchar{+}}18b]{tocl18}
L.~Bozzelli, A.~Molinari, A.~Montanari, A.~Peron, and P.~Sala.
\newblock Interval vs.\ point temporal logic model checking: an expressiveness
  comparison.
\newblock {\em Transactions on Computational Logic}, 2018.

\bibitem[BMM{\etalchar{+}}18c]{BOZZELLI2018IC}
L.~Bozzelli, A.~Molinari, A.~Montanari, A.~Peron, and P.~Sala.
\newblock Model checking for fragments of the interval temporal logic {HS} at
  the low levels of the polynomial time hierarchy.
\newblock {\em Information and Computation}, 262:241--264, 2018.
\newblock \href {http://dx.doi.org/10.1016/j.ic.2018.09.006}
  {\path{doi:10.1016/j.ic.2018.09.006}}.

\bibitem[BMM{\etalchar{+}}18d]{BOZZELLI2018}
L.~Bozzelli, A.~Molinari, A.~Montanari, A.~Peron, and P.~Sala.
\newblock Which fragments of the interval temporal logic {HS} are tractable in
  model checking?
\newblock {\em Theoretical Computer Science}, 2018.
\newblock \href {http://dx.doi.org/10.1016/j.tcs.2018.04.011}
  {\path{doi:10.1016/j.tcs.2018.04.011}}.

\bibitem[BMMP17a]{sefm17}
L.~Bozzelli, A.~Molinari, A.~Montanari, and A.~Peron.
\newblock An in-depth investigation of interval temporal logic model checking
  with regular expressions.
\newblock In {\em Proceedings of SEFM}, pages 104--119. Springer, 2017.
\newblock \href {http://dx.doi.org/10.1007/978-3-319-66197-1_7}
  {\path{doi:10.1007/978-3-319-66197-1_7}}.

\bibitem[BMMP17b]{gandalf17}
L.~Bozzelli, A.~Molinari, A.~Montanari, and A.~Peron.
\newblock On the complexity of model checking for syntactically maximal
  fragments of the interval temporal logic {HS} with regular expressions.
\newblock In {\em Proceedings of GandALF}, pages 31--45. EPTCS, 2017.
\newblock \href {http://dx.doi.org/10.4204/EPTCS.256.3}
  {\path{doi:10.4204/EPTCS.256.3}}.

\bibitem[BMMP18a]{gand18}
L.~Bozzelli, A.~Molinari, A.~Montanari, and A.~Peron.
\newblock {Complexity of timeline-based planning over dense temporal domains:
  exploring the middle ground}.
\newblock In {\em Proceedings of {GandALF}}, pages 191--205. EPTCS, 2018.
\newblock \href {http://dx.doi.org/10.4204/EPTCS.277.14}
  {\path{doi:10.4204/EPTCS.277.14}}.

\bibitem[BMMP18b]{kr18}
L.~Bozzelli, A.~Molinari, A.~Montanari, and A.~Peron.
\newblock {Decidability and Complexity of Timeline-based Planning over Dense
  Temporal Domains}.
\newblock In {\em Proceedings of KR}. AAAI Press, 2018.
\newblock Full version at
  \url{https://www.uniud.it/it/ateneo-uniud/ateneo-uniud-organizzazione/dipartimenti/dmif/assets/preprints/1-2018-molinari}.

\bibitem[BMMP18c]{icRegex}
L.~Bozzelli, A.~Molinari, A.~Montanari, and A.~Peron.
\newblock Model checking interval temporal logics with regular expressions.
\newblock {\em Information and Computation}, 2018.

\bibitem[BMP09]{batzold2009}
T.~Batzold, G.~Morin, and J.~Pecoraro.
\newblock {Completeness in the $\Sigma^p_2$ Hierarchy. A Compendium}, 2009.

\bibitem[BMSS11a]{DBLP:journals/corr/abs-1106-1241}
D.~Bresolin, A.~Montanari, P.~Sala, and G.~Sciavicco.
\newblock An optimal decision procedure for {MPNL} over the integers.
\newblock In {\em Proceedings of GandALF}, pages 192--206. EPTCS, 2011.
\newblock \href {http://dx.doi.org/10.4204/EPTCS.54.14}
  {\path{doi:10.4204/EPTCS.54.14}}.

\bibitem[BMSS11b]{DBLP:conf/tableaux/BresolinMSS11INS}
D.~Bresolin, A.~Montanari, P.~Sala, and G.~Sciavicco.
\newblock Optimal tableau systems for propositional neighborhood logic over
  all, dense, and discrete linear orders.
\newblock In {\em Proceedings of TABLEAUX}, pages 73--87. Springer, 2011.
\newblock \href {http://dx.doi.org/10.1007/978-3-642-22119-4_8}
  {\path{doi:10.1007/978-3-642-22119-4_8}}.

\bibitem[BMSS11c]{BMSS11}
D.~Bresolin, A.~Montanari, P.~Sala, and G.~Sciavicco.
\newblock What's decidable about {H}alpern and {S}hoham's interval logic? {T}he
  maximal fragment {$\mathsf{AB \overline{BL}}$}.
\newblock In {\em Proceedings of LICS}, pages 387--396. IEEE Computer Society,
  2011.
\newblock \href {http://dx.doi.org/10.1109/LICS.2011.35}
  {\path{doi:10.1109/LICS.2011.35}}.

\bibitem[BPR01]{Ball2001}
T.~Ball, A.~Podelski, and S.~K. Rajamani.
\newblock {\em Boolean and Cartesian Abstraction for Model Checking C
  Programs}, pages 268--283.
\newblock Springer, 2001.
\newblock \href {http://dx.doi.org/10.1007/3-540-45319-9_19}
  {\path{doi:10.1007/3-540-45319-9_19}}.

\bibitem[Bra11]{Bradley2011}
A.~R. Bradley.
\newblock {\em SAT-Based Model Checking without Unrolling}, pages 70--87.
\newblock Springer, 2011.
\newblock \href {http://dx.doi.org/10.1007/978-3-642-18275-4_7}
  {\path{doi:10.1007/978-3-642-18275-4_7}}.

\bibitem[Bra12]{Bradley2012}
A.~R. Bradley.
\newblock {\em Understanding IC3}, pages 1--14.
\newblock Springer, 2012.
\newblock \href {http://dx.doi.org/10.1007/978-3-642-31612-8_1}
  {\path{doi:10.1007/978-3-642-31612-8_1}}.

\bibitem[BS96]{BS98}
P.~Blackburn and J.~Seligman.
\newblock What are hybrid languages?
\newblock In {\em Proceedings of AiML}, pages 41--62. CSLI Publications, 1996.

\bibitem[BT03]{DBLP:journals/logcom/BowmanT03}
H.~Bowman and S.~J. Thompson.
\newblock A decision procedure and complete axiomatization of finite interval
  temporal logic with projection.
\newblock {\em Journal of Logic and Computation}, 13(2):195--239, 2003.
\newblock \href {http://dx.doi.org/10.1093/logcom/13.2.195}
  {\path{doi:10.1093/logcom/13.2.195}}.

\bibitem[BvDP15]{tcs15l}
L.~Bozzelli, H.~van Ditmarsch, and S.~Pinchinat.
\newblock The complexity of one-agent refinement modal logic.
\newblock {\em Theoretical Computer Science}, 603(C):58--83, 2015.
\newblock \href {http://dx.doi.org/10.1016/j.tcs.2015.07.015}
  {\path{doi:10.1016/j.tcs.2015.07.015}}.

\bibitem[CAB{\etalchar{+}}98]{708566}
W.~Chan, R.~J. Anderson, P.~Beame, S.~Burns, F.~Modugno, D.~Notkin, and J.~D.
  Reese.
\newblock Model checking large software specifications.
\newblock {\em Transactions on Software Engineering}, 24(7):498--520, 1998.
\newblock \href {http://dx.doi.org/10.1109/32.708566}
  {\path{doi:10.1109/32.708566}}.

\bibitem[CCF{\etalchar{+}}07]{CestaCFOP07}
A.~Cesta, G.~Cortellessa, S.~Fratini, A.~Oddi, and N.~Policella.
\newblock {A}n {I}nnovative {P}roduct for {S}pace {M}ission {P}lanning: {A}n
  {A} {P}osteriori {E}valuation.
\newblock In {\em Proceedings of ICAPS}, pages 57--64. AAAI Press, 2007.

\bibitem[CCGR00]{Cimatti2000nusmv}
A.~Cimatti, E.~Clarke, F.~Giunchiglia, and M.~Roveri.
\newblock {NuSMV}: a new symbolic model checker.
\newblock {\em International Journal on Software Tools for Technology
  Transfer}, 2(4):410--425, 2000.
\newblock \href {http://dx.doi.org/10.1007/s100090050046}
  {\path{doi:10.1007/s100090050046}}.

\bibitem[CE81]{CE81}
E.~M. Clarke and E.~A. Emerson.
\newblock {Design and Synthesis of Synchronization Skeletons Using
  Branching-Time Temporal Logic}.
\newblock In {\em Proceedings of LP}, pages 52--71. Springer, 1981.
\newblock \href {http://dx.doi.org/10.1007/BFb0025774}
  {\path{doi:10.1007/BFb0025774}}.

\bibitem[CES86]{Clarke:1986}
E.~M. Clarke, E.~A. Emerson, and A.~P. Sistla.
\newblock Automatic verification of finite-state concurrent systems using
  temporal logic specifications.
\newblock {\em Transactions on Programming Languages and Systems},
  8(2):244--263, 1986.
\newblock \href {http://dx.doi.org/10.1145/5397.5399}
  {\path{doi:10.1145/5397.5399}}.

\bibitem[CG12]{Cimatti2012}
A.~Cimatti and A.~Griggio.
\newblock {\em Software Model Checking via IC3}, pages 277--293.
\newblock Springer, 2012.
\newblock \href {http://dx.doi.org/10.1007/978-3-642-31424-7_23}
  {\path{doi:10.1007/978-3-642-31424-7_23}}.

\bibitem[CGMT14]{Cimatti2014}
A.~Cimatti, A.~Griggio, S.~Mover, and S.~Tonetta.
\newblock {\em IC3 Modulo Theories via Implicit Predicate Abstraction}, pages
  46--61.
\newblock Springer, 2014.
\newblock \href {http://dx.doi.org/10.1007/978-3-642-54862-8_4}
  {\path{doi:10.1007/978-3-642-54862-8_4}}.

\bibitem[CGMT16]{Cimatti2016}
A.~Cimatti, A.~Griggio, S.~Mover, and S.~Tonetta.
\newblock Infinite-state invariant checking with ic3 and predicate abstraction.
\newblock {\em Formal Methods in System Design}, 49(3):190--218, 2016.
\newblock \href {http://dx.doi.org/10.1007/s10703-016-0257-4}
  {\path{doi:10.1007/s10703-016-0257-4}}.

\bibitem[CGP02]{CGP02}
E.~M. Clarke, O.~Grumberg, and D.~A. Peled.
\newblock {\em {Model Checking}}.
\newblock Cyber Physical Systems Series. MIT Press, 2002.

\bibitem[CH04]{DBLP:series/eatcs/ChaochenH04}
Z.~Chaochen and M.~R. Hansen.
\newblock {\em Duration Calculus---{A} Formal Approach to Real-Time Systems}.
\newblock Monographs in Theoretical Computer Science. An {EATCS} Series.
  Springer, 2004.

\bibitem[CHR91]{CHAOCHEN1991269}
Z.~Chaochen, C.~A.~R. Hoare, and A.~P. Ravn.
\newblock A calculus of durations.
\newblock {\em Information Processing Letters}, 40(5):269--276, 1991.
\newblock \href {http://dx.doi.org/10.1016/0020-0190(91)90122-X}
  {\path{doi:10.1016/0020-0190(91)90122-X}}.

\bibitem[Cim01]{Cimatti2001}
A.~Cimatti.
\newblock {\em Industrial Applications of Model Checking}, pages 153--168.
\newblock Springer, 2001.
\newblock \href {http://dx.doi.org/10.1007/3-540-45510-8_6}
  {\path{doi:10.1007/3-540-45510-8_6}}.

\bibitem[CKS81]{CKS81}
A.~K. Chandra, D.~C. Kozen, and L.~J. Stockmeyer.
\newblock Alternation.
\newblock {\em Journal of the ACM}, 28(1):114--133, 1981.
\newblock \href {http://dx.doi.org/10.1145/322234.322243}
  {\path{doi:10.1145/322234.322243}}.

\bibitem[CM90]{cou90}
O.~Coudert and J.C. Madre.
\newblock A unified framework for the formal verification of sequential
  circuits.
\newblock In {\em Proceedings of ICCAD}, pages 126--129. IEEE Computer Society,
  1990.
\newblock \href {http://dx.doi.org/10.1109/ICCAD.1990.129859}
  {\path{doi:10.1109/ICCAD.1990.129859}}.

\bibitem[COU16]{MayerOU16}
M.~{Cialdea Mayer}, A.~Orlandini, and A.~Umbrico.
\newblock {P}lanning and {E}xecution with {F}lexible {T}imelines: a {F}ormal
  {A}ccount.
\newblock {\em Acta Informatica}, 53(6--8):649--680, 2016.
\newblock \href {http://dx.doi.org/10.1007/s00236-015-0252-z}
  {\path{doi:10.1007/s00236-015-0252-z}}.

\bibitem[CS92]{castro92}
J.~Castro and C.~Seara.
\newblock Characterizations of some complexity classes between {$\Theta_2^p$}
  and {$\Delta_2^p$}.
\newblock In {\em Proceedings of STACS}, pages 303--317. Springer, 1992.

\bibitem[CTR{\etalchar{+}}10]{aspen2010}
S.~Chien, D.~Tran, G.~Rabideau, S.R. Schaffer, D.~Mandl, and S.~Frye.
\newblock Timeline-based space operations scheduling with external constraints.
\newblock In {\em Proceedings of ICAPS}, pages 34--41. AAAI Press, 2010.

\bibitem[DGL16]{DBLP:books/cu/Demri2016}
S.~Demri, V.~Goranko, and M.~Lange.
\newblock {\em Temporal Logics in Computer Science: Finite-State Systems}.
\newblock Cambridge Tracts in Theoretical Computer Science. Cambridge
  University Press, 2016.
\newblock \href {http://dx.doi.org/10.1017/CBO9781139236119}
  {\path{doi:10.1017/CBO9781139236119}}.

\bibitem[DGMS11]{DBLP:journals/eatcs/MonicaGMS11}
D.~{Della Monica}, V.~Goranko, A.~Montanari, and G.~Sciavicco.
\newblock Interval temporal logics: a journey.
\newblock {\em Bulletin of the {EATCS}}, 105:73--99, 2011.

\bibitem[DL09]{DemriL09}
S.~Demri and R.~Lazic.
\newblock {LTL} with the freeze quantifier and register automata.
\newblock {\em Transactions on Computational Logic}, 10(3):16:1--16:30, 2009.
\newblock \href {http://dx.doi.org/10.1145/1507244.1507246}
  {\path{doi:10.1145/1507244.1507246}}.

\bibitem[DM11]{DarioDM:phd_thesis}
D.~Della~Monica.
\newblock {\em Expressiveness, decidability, and undecidability of Interval
  Temporal Logic}.
\newblock PhD thesis, University of Udine, 2011.

\bibitem[DMRT06]{DONINI200619}
F.~Donini, M.~Mongiello, M.~Ruta, and R.~Totaro.
\newblock A model checking-based method for verifying web application design.
\newblock {\em Electronic Notes in Theoretical Computer Science},
  151(2):19--32, 2006.
\newblock \href {http://dx.doi.org/10.1016/j.entcs.2005.07.034}
  {\path{doi:10.1016/j.entcs.2005.07.034}}.

\bibitem[DP14]{daniel_et_al}
J.~Daniel and P.~Par{\'i}zek.
\newblock {Predicate Abstraction in Program Verification: Survey and Current
  Trends}.
\newblock In {\em Imperial College Computing Student Workshop}, pages 27--35.
  Schloss Dagstuhl--Leibniz-Zentrum fuer Informatik, 2014.
\newblock \href {http://dx.doi.org/10.4230/OASIcs.ICCSW.2014.27}
  {\path{doi:10.4230/OASIcs.ICCSW.2014.27}}.

\bibitem[DV13]{DBLP:conf/ijcai/GiacomoV13}
G.~{De Giacomo} and M.~Y. Vardi.
\newblock Linear temporal logic and linear dynamic logic on finite traces.
\newblock In {\em Proceedings of {IJCAI}}, pages 854--860. AAAI Press, 2013.

\bibitem[EH86]{EH86}
E.~A. Emerson and J.~Y. Halpern.
\newblock {``Sometimes'' and ``Not Never'' Revisited: On Branching Versus
  Linear Time}.
\newblock {\em Journal of the ACM}, 33(1):151--178, 1986.
\newblock \href {http://dx.doi.org/10.1145 / 4904.4999} {\path{doi:10.1145 /
  4904.4999}}.

\bibitem[EL85]{DBLP:conf/popl/EmersonL85}
E.~A. Emerson and C.~Lei.
\newblock Modalities for model checking: Branching time strikes back.
\newblock In {\em Proceedings of {PoPL}}, pages 84--96. Elsevier, 1985.
\newblock \href {http://dx.doi.org/10.1145/318593.318620}
  {\path{doi:10.1145/318593.318620}}.

\bibitem[Fis85]{intvgraphs}
P.~C. Fishburn.
\newblock {\em Interval Orders and Interval Graphs: A Study of Partially
  Ordered Sets}.
\newblock Wiley-Interscience Series in Discrete Mathematics. John Wiley \&
  Sons, 1985.

\bibitem[Fix08]{fix2008}
L.~Fix.
\newblock Fifteen years of formal property verification in intel.
\newblock {\em 25 Years of Model Checking}, pages 139--144, 2008.
\newblock \href {http://dx.doi.org/10.1007/978-3-540-69850-0_8}
  {\path{doi:10.1007/978-3-540-69850-0_8}}.

\bibitem[FJ03]{FrankJ03}
J.~Frank and A.~J{\'o}nsson.
\newblock Constraint-based {A}ttribute and {I}nterval {P}lanning.
\newblock {\em Constraints}, 8(4):339--364, 2003.
\newblock \href {http://dx.doi.org/10.1023/A:1025842019552}
  {\path{doi:10.1023/A:1025842019552}}.

\bibitem[FL79]{FL79}
M.~J. Fischer and R.~E. Ladner.
\newblock {Propositional Dynamic Logic of Regular Programs}.
\newblock {\em Journal of Computer and System Sciences}, 18(2):194--211, 1979.
\newblock \href {http://dx.doi.org/10.1016/0022-0000(79)90046-1}
  {\path{doi:10.1016/0022-0000(79)90046-1}}.

\bibitem[FR75]{FR75}
J.~Ferrante and C.~Rackoff.
\newblock {A Decision Procedure for the First Order Theory of Real Addition
  with Order}.
\newblock {\em SIAM Journal of Computation}, 4(1):69--76, 1975.
\newblock \href {http://dx.doi.org/10.1137/0204006}
  {\path{doi:10.1137/0204006}}.

\bibitem[Gab87]{Gabbay87}
D.~M. Gabbay.
\newblock The declarative past and imperative future: Executable temporal logic
  for interactive systems.
\newblock In {\em Temporal Logic in Specification}, pages 409--448. Springer,
  1987.
\newblock \href {http://dx.doi.org/10.1007/3-540-51803-7_36}
  {\path{doi:10.1007/3-540-51803-7_36}}.

\bibitem[GJ79]{Garey79}
M.~Garey and D.~Johnson.
\newblock {\em Computers and Intractability: A Guide to the Theory of
  NP-Completeness}.
\newblock Series of Books in the Mathematical Sciences. W. H. Freeman and
  Company, 1979.

\bibitem[GM13]{Gligoric2013}
M.~Gligoric and R.~Majumdar.
\newblock {\em Model Checking Database Applications}, pages 549--564.
\newblock Springer, 2013.
\newblock \href {http://dx.doi.org/10.1007/978-3-642-36742-7_40}
  {\path{doi:10.1007/978-3-642-36742-7_40}}.

\bibitem[GMCO16]{GiganteMCO16}
N.~Gigante, A.~Montanari, M.~{Cialdea Mayer}, and A.~Orlandini.
\newblock {T}imelines are {E}xpressive {E}nough to {C}apture {A}ction-based
  {T}emporal {P}lanning.
\newblock In {\em Proceedings of TIME}, pages 100--109. IEEE Computer Society,
  2016.
\newblock \href {http://dx.doi.org/10.1109/TIME.2016.18}
  {\path{doi:10.1109/TIME.2016.18}}.

\bibitem[GMCO17]{GiganteMCO17}
N.~Gigante, A.~Montanari, M.~{Cialdea Mayer}, and A.~Orlandini.
\newblock Complexity of timeline-based planning.
\newblock In {\em Proceedings of ICAPS}, pages 116--124. AAAI Press, 2017.

\bibitem[GMS04]{roadmap_intervals}
V.~Goranko, A.~Montanari, and G.~Sciavicco.
\newblock A road map of interval temporal logics and duration calculi.
\newblock {\em Journal of Applied Non-Classical Logics}, 14(1--2):9--54, 2004.
\newblock \href {http://dx.doi.org/10.3166/jancl.14.9-54}
  {\path{doi:10.3166/jancl.14.9-54}}.

\bibitem[God97]{Godefroid:1997}
P.~Godefroid.
\newblock Model checking for programming languages using verisoft.
\newblock In {\em Proceedings of POPL}, pages 174--186. ACM, 1997.
\newblock \href {http://dx.doi.org/10.1145/263699.263717}
  {\path{doi:10.1145/263699.263717}}.

\bibitem[Got95]{gottlob1995}
G.~Gottlob.
\newblock {NP} {T}rees and {C}arnap's {M}odal {L}ogic.
\newblock {\em Journal of the ACM}, 42(2):421--457, 1995.
\newblock \href {http://dx.doi.org/10.1145/201019.201031}
  {\path{doi:10.1145/201019.201031}}.

\bibitem[GT99]{DBLP:conf/ecp/GiunchigliaT99}
F.~Giunchiglia and P.~Traverso.
\newblock Planning as model checking.
\newblock In {\em Proceedings of ECP}, pages 1--20. Springer, 1999.
\newblock \href {http://dx.doi.org/10.1007/10720246_1}
  {\path{doi:10.1007/10720246_1}}.

\bibitem[GTB{\etalchar{+}}06]{Giordano06}
L.~Giordano, P.~Terenziani, A.~Bottrighi, S.~Montani, and L.~Donzella.
\newblock Model checking for clinical guidelines: an agent-based approach.
\newblock In {\em Proceedings of AMIA}, pages 289--293, 2006.

\bibitem[Har12]{harel92}
D.~Harel.
\newblock {\em Algorithmics: The spirit of computing}.
\newblock Springer, 3rd edition, 2012.

\bibitem[HD01]{Hatcliff2001}
J.~Hatcliff and M.~Dwyer.
\newblock {\em Using the Bandera Tool Set to Model-Check Properties of
  Concurrent Java Software}, pages 39--58.
\newblock Springer, 2001.
\newblock \href {http://dx.doi.org/10.1007/3-540-44685-0_5}
  {\path{doi:10.1007/3-540-44685-0_5}}.

\bibitem[HJMS02]{Henzinger02}
T.~A. Henzinger, R.~Jhala, R.~Majumdar, and G.~Sutre.
\newblock Lazy abstraction.
\newblock In {\em Proceedings of POPL}, pages 58--70. ACM, 2002.
\newblock \href {http://dx.doi.org/10.1145/503272.503279}
  {\path{doi:10.1145/503272.503279}}.

\bibitem[HK11]{holzer}
M.~Holzer and M.~Kutrib.
\newblock Descriptional and computational complexity of finite automata---a
  survey.
\newblock {\em Information and Computation}, 209(3):456--470, 2011.
\newblock \href {http://dx.doi.org/10.1016/j.ic.2010.11.013}
  {\path{doi:10.1016/j.ic.2010.11.013}}.

\bibitem[Hol94]{holzmann1994}
G.~J. Holzmann.
\newblock The theory and practice of a formal method: Newcore.
\newblock In {\em Proceedings of the IFIP World Computer Congress}, 1994.

\bibitem[HP00]{Havelund2000}
K.~Havelund and T.~Pressburger.
\newblock Model checking java programs using java pathfinder.
\newblock {\em International Journal on Software Tools for Technology
  Transfer}, 2(4):366--381, 2000.
\newblock \href {http://dx.doi.org/10.1007/s100090050043}
  {\path{doi:10.1007/s100090050043}}.

\bibitem[HS91]{HS91}
J.~Y. Halpern and Y.~Shoham.
\newblock A propositional modal logic of time intervals.
\newblock {\em Journal of the ACM}, 38(4):935--962, 1991.
\newblock \href {http://dx.doi.org/10.1145/115234.115351}
  {\path{doi:10.1145/115234.115351}}.

\bibitem[HVH07]{Hansen2007}
M.~R. Hansen and D.~Van~Hung.
\newblock {\em A Theory of Duration Calculus with Application}, pages 119--176.
\newblock Springer, 2007.
\newblock \href {http://dx.doi.org/10.1007/978-3-540-74964-6_3}
  {\path{doi:10.1007/978-3-540-74964-6_3}}.

\bibitem[JM97]{jazequel1997}
J.~M. Jazequel and B.~Meyer.
\newblock Design by contract: The lessons of ariane.
\newblock {\em Computer}, 30(1):129--130, 1997.
\newblock \href {http://dx.doi.org/10.1109/2.562936}
  {\path{doi:10.1109/2.562936}}.

\bibitem[JMM{\etalchar{+}}00]{JonssonMMRS00}
A.~K. J{\'{o}}nsson, P.~H. Morris, N.~Muscettola, K.~Rajan, and B.~D. Smith.
\newblock {P}lanning in {I}nterplanetary {S}pace: {T}heory and {P}ractice.
\newblock In {\em Proceedings of ICAPS}, pages 177--186. AAAI Press, 2000.

\bibitem[Jun13]{Jung}
D.~Jungnickel.
\newblock {\em Graphs, Networks and Algorithms}.
\newblock Algorithms and Computation in Mathematics. Springer, 2013.
\newblock \href {http://dx.doi.org/10.1007/978-3-642-32278-5}
  {\path{doi:10.1007/978-3-642-32278-5}}.

\bibitem[Kam68]{Kamp}
H.~Kamp.
\newblock {\em Tense Logic and the Theory of Linear Order}.
\newblock PhD thesis, UCLA, 1968.

\bibitem[Koy90]{Koymans90}
R.~Koymans.
\newblock Specifying real-time properties with metric temporal logic.
\newblock {\em Real-Time Systems}, 2(4):255--299, 1990.
\newblock \href {http://dx.doi.org/10.1007/BF01995674}
  {\path{doi:10.1007/BF01995674}}.

\bibitem[KPV09]{DBLP:journals/fmsd/KupfermanPV09}
O.~Kupferman, N.~Piterman, and M.~Y. Vardi.
\newblock From liveness to promptness.
\newblock {\em Formal Methods in System Design}, 34(2):83--103, 2009.
\newblock \href {http://dx.doi.org/10.1007/s10703-009-0067-z}
  {\path{doi:10.1007/s10703-009-0067-z}}.

\bibitem[KPV12]{jcss/KupfermanPV12}
O.~Kupferman, A.~Pnueli, and M.~Y. Vardi.
\newblock Once and for all.
\newblock {\em Journal of Computer and System Sciences}, 78(3):981--996, 2012.
\newblock \href {http://dx.doi.org/10.1016/j.jcss.2011.08.006}
  {\path{doi:10.1016/j.jcss.2011.08.006}}.

\bibitem[Lan06]{Lan06}
M.~Lange.
\newblock Model checking propositional dynamic logic with all extras.
\newblock {\em Journal of Applied Logic}, 4(1):39--49, 2006.
\newblock \href {http://dx.doi.org/10.1016/j.jal.2005.08.002}
  {\path{doi:10.1016/j.jal.2005.08.002}}.

\bibitem[LM13]{LM13}
A.~Lomuscio and J.~Michaliszyn.
\newblock An epistemic {H}alpern-{S}hoham logic.
\newblock In {\em Proceedings of IJCAI}. AAAI Press, 2013.

\bibitem[LM14]{LM14}
A.~Lomuscio and J.~Michaliszyn.
\newblock Decidability of model checking multi-agent systems against a class of
  {EHS} specifications.
\newblock In {\em Proceedings of {ECAI}}, pages 543--548. IOS, 2014.
\newblock \href {http://dx.doi.org/10.3233/978-1-61499-419-0-543}
  {\path{doi:10.3233/978-1-61499-419-0-543}}.

\bibitem[LM16]{lm16}
A.~Lomuscio and J.~Michaliszyn.
\newblock Model checking multi-agent systems against epistemic {HS}
  specifications with regular expressions.
\newblock In {\em Proceedings of {KR}}, pages 298--308. AAAI Press, 2016.

\bibitem[LMP10]{Lmp10}
F.~Laroussinie, A.~Meyer, and E.~Petonnet.
\newblock {\em Counting {CTL}}, pages 206--220.
\newblock Springer, 2010.
\newblock \href {http://dx.doi.org/10.1007/978-3-642-12032-9_15}
  {\path{doi:10.1007/978-3-642-12032-9_15}}.

\bibitem[LMS01]{LMS01}
F.~Laroussinie, N.~Markey, and P.~Schnoebelen.
\newblock {Model Checking CTL+ and FCTL is Hard}.
\newblock In {\em Proceedings of FOSSACS}, pages 318--331. Springer, 2001.
\newblock \href {http://dx.doi.org/10.1007/3-540-45315-6_21}
  {\path{doi:10.1007/3-540-45315-6_21}}.

\bibitem[LMS02]{LMS02}
F.~Laroussinie, N.~Markey, and P.~Schnoebelen.
\newblock {Temporal Logic with Forgettable Past}.
\newblock In {\em Proceedings of LICS}, pages 383--392. IEEE Computer Society,
  2002.
\newblock \href {http://dx.doi.org/10.1109/LICS.2002.1029846}
  {\path{doi:10.1109/LICS.2002.1029846}}.

\bibitem[Lod00]{DBLP:conf/asian/Lodaya00}
K.~Lodaya.
\newblock Sharpening the undecidability of interval temporal logic.
\newblock In {\em Proceedings of ASIAN}, pages 290--298. Springer, 2000.
\newblock \href {http://dx.doi.org/10.1007/3-540-44464-5_21}
  {\path{doi:10.1007/3-540-44464-5_21}}.

\bibitem[Low96]{Lowe1996}
G.~Lowe.
\newblock {\em Breaking and fixing the Needham-Schroeder Public-Key Protocol
  using FDR}, pages 147--166.
\newblock Springer, 1996.
\newblock \href {http://dx.doi.org/10.1007/3-540-61042-1_43}
  {\path{doi:10.1007/3-540-61042-1_43}}.

\bibitem[LP00]{DBLP:journals/igpl/LichtensteinP00}
O.~Lichtenstein and A.~Pnueli.
\newblock Propositional temporal logics: Decidability and completeness.
\newblock {\em Logic Journal of the {IGPL}}, 8(1):55--85, 2000.
\newblock \href {http://dx.doi.org/10.1093/jigpal/8.1.55}
  {\path{doi:10.1093/jigpal/8.1.55}}.

\bibitem[LPY97]{UPP}
G.~K.~Larsen, P.~Pettersson, and W.~Yi.
\newblock {UPPAAL} in a nutshell.
\newblock {\em International Journal on Software Tools for Technology
  Transfer}, 1:134--152, 1997.
\newblock \href {http://dx.doi.org/10.1007/s100090050010}
  {\path{doi:10.1007/s100090050010}}.

\bibitem[LR06]{DBLP:conf/tacas/LomuscioR06}
A.~Lomuscio and F.~Raimondi.
\newblock {MCMAS:} {A} model checker for multi-agent systems.
\newblock In {\em Proceedings of {TACAS}}, pages 450--454. Springer, 2006.
\newblock \href {http://dx.doi.org/10.1007/11691372_31}
  {\path{doi:10.1007/11691372_31}}.

\bibitem[LS95]{LS95}
F.~Laroussinie and P.~Schnoebelen.
\newblock A hierarchy of temporal logics with past.
\newblock {\em Theoretical Computer Science}, 148(2):303--324, 1995.
\newblock \href {http://dx.doi.org/10.1016/0304-3975(95)00035-U}
  {\path{doi:10.1016/0304-3975(95)00035-U}}.

\bibitem[LS07]{Leucker2007}
M.~Leucker and C.~S{\'a}nchez.
\newblock Regular linear temporal logic.
\newblock In {\em Proceedings of ICTAC}, pages 291--305. Springer, 2007.
\newblock \href {http://dx.doi.org/10.1007/978-3-540-75292-9_20}
  {\path{doi:10.1007/978-3-540-75292-9_20}}.

\bibitem[MBT00]{cadence}
A.~A. Mir, S.~Balakrishnan, and S.~Tahar.
\newblock Modeling and verification of embedded systems using {C}adence {SMV}.
\newblock In {\em Proceedings of CCECE}, pages 179--1831. IEEE Computer
  Society, 2000.
\newblock \href {http://dx.doi.org/10.1109/CCECE.2000.849694}
  {\path{doi:10.1109/CCECE.2000.849694}}.

\bibitem[McM93]{mcm93}
K.~L. McMillan.
\newblock {\em Symbolic Model Checking}.
\newblock Kluwer, 1993.

\bibitem[McM06]{McMillan2006}
K.~L. McMillan.
\newblock {\em Lazy Abstraction with Interpolants}, pages 123--136.
\newblock Springer, 2006.
\newblock \href {http://dx.doi.org/10.1007/11817963_14}
  {\path{doi:10.1007/11817963_14}}.

\bibitem[Min67]{Minsky67}
M.~L. Minsky.
\newblock {\em Computation: Finite and Infinite Machines}.
\newblock Automatic Computation. Prentice-Hall, Inc., 1967.

\bibitem[MK12]{CPE:CPE1876}
A.~Mentis and P.~Katsaros.
\newblock Model checking and code generation for transaction processing
  software.
\newblock {\em Concurrency and Computation: Practice and Experience},
  24(7):711--722, 2012.
\newblock \href {http://dx.doi.org/10.1002/cpe.1876}
  {\path{doi:10.1002/cpe.1876}}.

\bibitem[MM14]{DBLP:journals/fuin/MarcinkowskiM14}
J.~Marcinkowski and J.~Michaliszyn.
\newblock The undecidability of the logic of subintervals.
\newblock {\em Fundamenta Informaticae}, 131(2):217--240, 2014.
\newblock \href {http://dx.doi.org/10.3233/FI-2014-1011}
  {\path{doi:10.3233/FI-2014-1011}}.

\bibitem[MMDdJ11]{MATEESCU20112854}
R.~Mateescu, P.~T. Monteiro, E.~Dumas, and H.~de~Jong.
\newblock {CTRL}: Extension of {CTL} with regular expressions and fairness
  operators to verify genetic regulatory networks.
\newblock {\em Theoretical Computer Science}, 412(26):2854--2883, 2011.
\newblock \href {http://dx.doi.org/10.1016/j.tcs.2010.05.009}
  {\path{doi:10.1016/j.tcs.2010.05.009}}.

\bibitem[MMM{\etalchar{+}}16]{MMMPP15}
A.~Molinari, A.~Montanari, A.~Murano, G.~Perelli, and
  A.~Peron.
\newblock Checking interval properties of computations.
\newblock {\em Acta Informatica}, 53(6):587--619, 2016.
\newblock \href {http://dx.doi.org/10.1007/s00236-015-0250-1}
  {\path{doi:10.1007/s00236-015-0250-1}}.

\bibitem[MMP15a]{MMP15B}
A.~Molinari, A.~Montanari, and A.~Peron.
\newblock {Complexity of ITL model checking: some well-behaved fragments of the
  interval logic HS}.
\newblock In {\em Proceedings of TIME}, pages 90--100. IEEE Computer Society,
  2015.
\newblock \href {http://dx.doi.org/10.1109/TIME.2015.12}
  {\path{doi:10.1109/TIME.2015.12}}.

\bibitem[MMP15b]{csl15}
A.~Molinari, A.~Montanari, and A.~Peron.
\newblock A model checking procedure for interval temporal logics based on
  track representatives.
\newblock In {\em Proceedings of CSL}, pages 193--210. Schloss
  Dagstuhl--Leibniz-Zentrum fuer Informatik, 2015.
\newblock \href {http://dx.doi.org/10.4230/LIPIcs.CSL.2015.193}
  {\path{doi:10.4230/LIPIcs.CSL.2015.193}}.

\bibitem[MMP17]{MMP15}
A.~Molinari, A.~Montanari, and A.~Peron.
\newblock Model checking for fragments of {H}alpern and {S}hoham's interval
  temporal logic based on track representatives.
\newblock {\em Information and Computation}, 259:412--443, 2017.
\newblock \href {http://dx.doi.org/10.1016/j.ic.2017.08.011}
  {\path{doi:10.1016/j.ic.2017.08.011}}.

\bibitem[MMPS16]{kr16}
A.~Molinari, A.~Montanari, A.~Peron, and P.~Sala.
\newblock Model checking well-behaved fragments of {HS:} the (almost) final
  picture.
\newblock In {\em Proceedings of KR}, pages 473--483. AAAI Press, 2016.

\bibitem[MMTV08]{MEIER2008201}
A.~Meier, M.~Mundhenk, M.~Thomas, and H.~Vollmer.
\newblock The complexity of satisfiability for fragments of ctl and ctl*.
\newblock {\em Electronic Notes in Theoretical Computer Science}, 223:201--213,
  2008.
\newblock \href {http://dx.doi.org/10.1016/j.entcs.2008.12.040}
  {\path{doi:10.1016/j.entcs.2008.12.040}}.

\bibitem[Mos83]{digital_circuits_thesis}
B.~Moszkowski.
\newblock {\em Reasoning About Digital Circuits}.
\newblock PhD thesis, Stanford University, 1983.

\bibitem[MPS10a]{DBLP:conf/time/MontanariPS10}
A.~Montanari, I.~Pratt{-}Hartmann, and P.~Sala.
\newblock Decidability of the logics of the reflexive sub-interval and
  super-interval relations over finite linear orders.
\newblock In {\em Proceedings of TIME}, pages 27--34. IEEE Computer Society,
  2010.
\newblock \href {http://dx.doi.org/10.1109/TIME.2010.18}
  {\path{doi:10.1109/TIME.2010.18}}.

\bibitem[MPS10b]{MPS10}
A.~Montanari, G.~Puppis, and P.~Sala.
\newblock Maximal decidable fragments of {H}alpern and {S}hoham's modal logic
  of intervals.
\newblock In {\em Proceedings of ICALP}, pages 345--356. Springer, 2010.
\newblock \href {http://dx.doi.org/10.1007/978-3-642-14162-1_29}
  {\path{doi:10.1007/978-3-642-14162-1_29}}.

\bibitem[MPS14]{DBLP:conf/mfcs/MontanariPS14Ratio}
A.~Montanari, G.~Puppis, and P.~Sala.
\newblock Decidability of the interval temporal logic
  $\mathsf{A,\overline{A},B,\overline{B}}$ over the rationals.
\newblock In {\em Proceedings of {MFCS}}, pages 451--463. Springer, 2014.
\newblock \href {http://dx.doi.org/10.1007/978-3-662-44522-8_38}
  {\path{doi:10.1007/978-3-662-44522-8_38}}.

\bibitem[MPSS10]{MPSS10}
A.~Montanari, G.~Puppis, P.~Sala, and G.~Sciavicco.
\newblock Decidability of the interval temporal logic
  $\mathsf{\overline{a},b,\overline{b}}$ over the natural numbers.
\newblock In {\em Proceedings of STACS}, pages 597--608. Schloss
  Dagstuhl--Leibniz-Zentrum fuer Informatik, 2010.
\newblock \href {http://dx.doi.org/10.4230/LIPIcs.STACS.2010.2488}
  {\path{doi:10.4230/LIPIcs.STACS.2010.2488}}.

\bibitem[MS12]{DBLP:conf/time/MontanariS12INS}
A.~Montanari and P.~Sala.
\newblock An optimal tableau system for the logic of temporal neighborhood over
  the reals.
\newblock In {\em Proceedings of TIME}, pages 39--46. IEEE Computer Society,
  2012.
\newblock \href {http://dx.doi.org/10.1109/TIME.2012.18}
  {\path{doi:10.1109/TIME.2012.18}}.

\bibitem[MS14]{DBLP:journals/corr/MontanariS14}
A.~Montanari and P.~Sala.
\newblock Interval-based synthesis.
\newblock In {\em Proceedings of GandALF}, pages 102--115. EPTCS, 2014.
\newblock \href {http://dx.doi.org/10.4204/EPTCS.161.11}
  {\path{doi:10.4204/EPTCS.161.11}}.

\bibitem[Mus94]{Muscettola94}
N.~Muscettola.
\newblock {{HSTS}}: {I}ntegrating {P}lanning and {S}cheduling.
\newblock In {\em Intelligent Scheduling}, pages 169--212. Morgan Kaufmann,
  1994.

\bibitem[NGB{\etalchar{+}}16]{Nardone2016}
R.~Nardone, U.~Gentile, M.~Benerecetti, A.~Peron, V.~Vittorini, S.~Marrone, and
  N.~Mazzocca.
\newblock {\em Modeling Railway Control Systems in Promela}, pages 121--136.
\newblock Springer, 2016.
\newblock \href {http://dx.doi.org/10.1007/978-3-319-29510-7_7}
  {\path{doi:10.1007/978-3-319-29510-7_7}}.

\bibitem[Ott01]{DBLP:journals/jsyml/Otto01}
M.~Otto.
\newblock Two variable first-order logic over ordered domains.
\newblock {\em Journal of Symbolic Logic}, 66(2):685--702, 2001.
\newblock \href {http://dx.doi.org/10.2307/2695037}
  {\path{doi:10.2307/2695037}}.

\bibitem[OW07]{OuaknineW07}
J.~Ouaknine and J.~Worrell.
\newblock On the decidability and complexity of metric temporal logic over
  finite words.
\newblock {\em Logical Methods in Computer Science}, 3(1), 2007.
\newblock \href {http://dx.doi.org/10.2168/LMCS-3(1:8)2007}
  {\path{doi:10.2168/LMCS-3(1:8)2007}}.

\bibitem[OW08]{Ouaknine08}
J.~Ouaknine and J.~Worrell.
\newblock Some recent results in metric temporal logic.
\newblock In {\em Proceedings of FORMATS}, pages 1--13. Springer, 2008.
\newblock \href {http://dx.doi.org/10.1007/978-3-540-85778-5_1}
  {\path{doi:10.1007/978-3-540-85778-5_1}}.

\bibitem[Pap94]{Pap94}
C.~H. Papadimitriou.
\newblock {\em Computational complexity}.
\newblock Addison-Wesley, 1994.

\bibitem[Pel93]{pored93}
D.~Peled.
\newblock All from one, one for all: on model checking using representatives.
\newblock In {\em Proceedings of CAV}, pages 409--423. Springer, 1993.
\newblock \href {http://dx.doi.org/10.1007/3-540-56922-7_34}
  {\path{doi:10.1007/3-540-56922-7_34}}.

\bibitem[Pnu77]{Pnu77}
A.~Pnueli.
\newblock {The Temporal Logic of Programs}.
\newblock In {\em Proceedings of FOCS}, pages 46--57. IEEE Computer Society,
  1977.
\newblock \href {http://dx.doi.org/10.1109/SFCS.1977.32}
  {\path{doi:10.1109/SFCS.1977.32}}.

\bibitem[Pra95]{pratt1995}
V.~Pratt.
\newblock Anatomy of the pentium bug.
\newblock In {\em Proceedings of TAPSOFT}, pages 97--107. Springer, 1995.
\newblock \href {http://dx.doi.org/10.1007/3-540-59293-8_189}
  {\path{doi:10.1007/3-540-59293-8_189}}.

\bibitem[Pra05]{DBLP:journals/ai/Pratt-Hartmann05}
I.~Pratt{-}Hartmann.
\newblock Temporal prepositions and their logic.
\newblock {\em Artificial Intelligence}, 166(1--2):1--36, 2005.
\newblock \href {http://dx.doi.org/10.1016/j.artint.2005.04.003}
  {\path{doi:10.1016/j.artint.2005.04.003}}.

\bibitem[PZ82]{Papadimitriou82}
C.~H. Papadimitriou and S.~K. Zachos.
\newblock Two remarks on the power of counting.
\newblock {\em Theoretical Computer Science}, pages 269--275, 1982.
\newblock \href {http://dx.doi.org/10.1007/BFb0036487}
  {\path{doi:10.1007/BFb0036487}}.

\bibitem[QS82]{Queille1982}
J.~P. Queille and J.~Sifakis.
\newblock {\em Specification and verification of concurrent systems in CESAR},
  pages 337--351.
\newblock Springer, 1982.
\newblock \href {http://dx.doi.org/10.1007/3-540-11494-7_22}
  {\path{doi:10.1007/3-540-11494-7_22}}.

\bibitem[Roe80]{Roe80}
P.~Roeper.
\newblock Intervals and tenses.
\newblock {\em Journal of Philosophical Logic}, 9:451--469, 1980.
\newblock \href {http://dx.doi.org/10.1007/BF00262866}
  {\path{doi:10.1007/BF00262866}}.

\bibitem[SC85]{Sistla:1985}
A.~P. Sistla and E.~M. Clarke.
\newblock The complexity of propositional linear temporal logics.
\newblock {\em Journal of the ACM}, 32(3):733--749, 1985.
\newblock \href {http://dx.doi.org/10.1145/3828.3837}
  {\path{doi:10.1145/3828.3837}}.

\bibitem[Sch03]{schnoebelen2003}
P.~Schnoebelen.
\newblock Oracle circuits for branching-time model checking.
\newblock In {\em Proceedings of ICALP}, pages 790--801. Springer, 2003.
\newblock \href {http://dx.doi.org/10.1007/3-540-45061-0_62}
  {\path{doi:10.1007/3-540-45061-0_62}}.

\bibitem[Sch16]{Schmitz:2016}
S.~Schmitz.
\newblock Complexity hierarchies beyond elementary.
\newblock {\em Transactions on Computation Theory}, 8(1):3:1--3:36, 2016.
\newblock \href {http://dx.doi.org/10.1145/2858784}
  {\path{doi:10.1145/2858784}}.

\bibitem[Sha04]{DBLP:conf/aiml/Shapirovsky04}
I.~Shapirovsky.
\newblock On {PSPACE}-decidability in transitive modal logic.
\newblock In {\em Proceedings of AiML}, pages 269--287. CSLI Publications,
  2004.

\bibitem[Sip12]{Sip12}
M.~Sipser.
\newblock {\em Introduction to the Theory of Computation}.
\newblock International Thomson Publishing, 3rd edition, 2012.

\bibitem[SM73]{Stockmeyer:1973}
L.~J. Stockmeyer and A.~R. Meyer.
\newblock Word problems requiring exponential time (preliminary report).
\newblock In {\em Proceedings of STOC}, pages 1--9. ACM, 1973.
\newblock \href {http://dx.doi.org/10.1145/800125.804029}
  {\path{doi:10.1145/800125.804029}}.

\bibitem[SS05]{DBLP:journals/logcom/ShapirovskyS05}
I.~Shapirovsky and V.~B. Shehtman.
\newblock {Modal Logics of Regions and Minkowski Spacetime}.
\newblock {\em Journal of Logic and Computation}, 15(4):559--574, 2005.
\newblock \href {http://dx.doi.org/10.1093/logcom/exi039}
  {\path{doi:10.1093/logcom/exi039}}.

\bibitem[Sto74]{stockmeyer1974}
L.~J.~Stockmeyer.
\newblock {\em The complexity of decision problems in automata theory and
  logic}.
\newblock PhD thesis, Massachusetts Institute of Technology, 1974.

\bibitem[Sto76]{stockmeyer1976}
L.~J. Stockmeyer.
\newblock {The polynomial-time hierarchy}.
\newblock {\em Theoretical Computer Science}, 3(1):1--22, 1976.
\newblock \href {http://dx.doi.org/10.1016/0304-3975(76)90061-X}
  {\path{doi:10.1016/0304-3975(76)90061-X}}.

\bibitem[Tho94]{thomas1994}
M.~Thomas.
\newblock A proof of incorrectness using the lp theorem prover: the editing
  problem in the therac-25.
\newblock {\em High Integrity Systems Journal}, 1(1):35--48, 1994.

\bibitem[{van}91]{DBLP:journals/jphil/Benthem91}
J.~{van Benthem}.
\newblock Language in action.
\newblock {\em Journal of Philosophical Logic}, 20(3):225--263, 1991.
\newblock \href {http://dx.doi.org/10.1007/BF00250539}
  {\path{doi:10.1007/BF00250539}}.

\bibitem[Var96]{vardi1996automata}
M.~Y. Vardi.
\newblock An automata-theoretic approach to linear temporal logic.
\newblock In {\em Logics for concurrency}, pages 238--266. Springer, 1996.
\newblock \href {http://dx.doi.org/10.1007/3-540-60915-6_6}
  {\path{doi:10.1007/3-540-60915-6_6}}.

\bibitem[Ven90]{venema1990}
Y.~Venema.
\newblock Expressiveness and completeness of an interval tense logic.
\newblock {\em Notre Dame Journal of Formal Logic}, 31(4):529--547, 1990.
\newblock \href {http://dx.doi.org/10.1305/ndjfl/1093635589}
  {\path{doi:10.1305/ndjfl/1093635589}}.

\bibitem[Ven91]{chopping_intervals}
Y.~Venema.
\newblock A modal logic for chopping intervals.
\newblock {\em Journal of Logic and Computation}, 1(4):453--476, 1991.
\newblock \href {http://dx.doi.org/10.1093/logcom/1.4.453}
  {\path{doi:10.1093/logcom/1.4.453}}.

\bibitem[VW86]{VW86b}
M.Y. Vardi and P.~Wolper.
\newblock {An Automata-Theoretic Approach to Automatic Program Verification}.
\newblock In {\em Proceedings of LICS}, pages 332--344. IEEE Computer Society,
  1986.

\bibitem[Wag87]{WAGNER87}
K.~W. Wagner.
\newblock More complicated questions about maxima and minima, and some closures
  of {NP}.
\newblock {\em Theoretical Computer Science}, 51(1):53--80, 1987.
\newblock \href {http://dx.doi.org/10.1016/0304-3975(87)90049-1}
  {\path{doi:10.1016/0304-3975(87)90049-1}}.

\bibitem[Wag90]{wagner90}
K.~W. Wagner.
\newblock Bounded query classes.
\newblock {\em SIAM Journal of Computation}, 19(5):833--846, 1990.
\newblock \href {http://dx.doi.org/10.1137/0219058}
  {\path{doi:10.1137/0219058}}.

\bibitem[WBKW07]{Witkowski:2007}
T.~Witkowski, N.~Blanc, D.~Kroening, and G.~Weissenbacher.
\newblock Model checking concurrent {L}inux device drivers.
\newblock In {\em Proceedings of ASE}, pages 501--504. ACM, 2007.
\newblock \href {http://dx.doi.org/10.1145/1321631.1321719}
  {\path{doi:10.1145/1321631.1321719}}.

\bibitem[Wil99]{stacs/Wilke99}
T.~Wilke.
\newblock Classifying discrete temporal properties.
\newblock In {\em Proceedings of {STACS}}, pages 32--46. Springer, 1999.
\newblock \href {http://dx.doi.org/10.1007/3-540-49116-3_3}
  {\path{doi:10.1007/3-540-49116-3_3}}.

\end{thebibliography}
